 \documentclass[final,authoryear,5p,times,twocolumn]{elsarticle}

\usepackage{graphicx}
\usepackage{pstricks}

\graphicspath{{app-figs/}}

\usepackage{amssymb}

\usepackage[switch,pagewise,running]{lineno}

\journal{PSS}

  \bibpunct{(}{)}{;}{a}{}{,}
\usepackage{hyperref}
\hypersetup{
    unicode=false,          
    pdftoolbar=true,        
    pdfmenubar=true,        
    pdffitwindow=false,     
    pdftitle={Density of asteroids},
    pdfauthor={Benoit Carry},
    pdfsubject={Planetary Science},
    pdfkeywords={},         
    pdfnewwindow=true,      
    colorlinks=true,        
    linkcolor=gray,         
    citecolor=blue,         
    filecolor=gray,         
    urlcolor=gray           
}

\usepackage{ulem}

\newcommand{\numb}[1]{\textcolor{red}{\textbf{#1}}}

\newcommand{\ie}{\textsl{i.e.}}
\newcommand{\eg}{\textsl{e.g.}}

\newcommand{\macro}{\ensuremath{\mathcal{P}}}

\renewcommand{\numb}[1]{#1}

\usepackage{longtable}

\begin{document}

\begin{frontmatter}



\title{Density of asteroids}



\author{B. Carry}
\ead{benoit.carry@esa.int}
\address{European Space Astronomy Centre, ESA, P.O. Box 78, 28691 Villanueva de la Ca{\~n}ada, Madrid, Spain}

\begin{abstract}
%
%
  The small bodies of our solar system are the remnants of the
  early stages of planetary formation.
  A considerable amount of information regarding
  the processes that occurred during the
  accretion of the early planetesimals is still present 
  among this population. 
  A review of our current knowledge of the density of small bodies is
  presented here.
  Density is indeed a fundamental property for the understanding of
  their composition and internal structure.
  Intrinsic physical properties of small
  bodies are sought by searching for relationships between
  the dynamical and taxonomic classes, size, and density.
  Mass and volume estimates for \numb{287} small bodies (asteroids, comets,
  and transneptunian objects) are collected from the literature.
  The accuracy and biases affecting the methods used
  to estimate these quantities are discussed and 
  best-estimates are strictly selected. %
  Bulk densities are subsequently computed and compared with meteorite
  density, allowing to estimate the macroporosity
  (\ie, amount of voids) within these bodies.
  Dwarf-planets apparently have no macroporosity, while smaller bodies
  ($<$400\,km) can have large voids.
  This trend is apparently correlated with size: C and S-complex
  asteroids tends to have larger density with increasing diameter.
  The average density of each Bus-DeMeo taxonomic classes is computed
  (DeMeo et al., 2009, Icarus 202).
  S-complex asteroids are more dense on average than those in
  the C-complex that in turn have a larger macroporosity, although
  both complexes partly overlap.
  Within the C-complex asteroids, B-types stand out
  in albedo, reflectance spectra, and density, indicating a 
  unique composition and structure.
  Asteroids in the X-complex span a wide range of densities,
  suggesting that many compositions are included in the complex.
  Comets and TNOs have high macroporosity and low density, supporting
  the current models of internal structures made of icy aggregates.
  Although the number of density estimates sky-rocketed during last
  decade from a handful to \numb{287}, only a third of the estimates
  are more precise than 20\%.
  Several lines of investigation to refine this statistic are
  contemplated, including observations of multiple systems, 3-D shape
  modeling, and orbital analysis from Gaia astrometry.
\end{abstract}

\begin{keyword}
  Minor planets \sep Mass \sep Volume \sep Density \sep Porosity 


\end{keyword}

\end{frontmatter}


\section{Small bodies as remnants of planetesimals\label{sec: int1}}
  \indent The small bodies of our solar System are the left-overs of
  the building blocks that accreted to form the planets, some 4.6 Gyr ago.
  They represent the most direct witnesses of the conditions that
  reigned in the proto-planetary nebula
  \citep{2002-AsteroidsIII-1-Bottke}.
  Indeed, terrestrial planets have thermally evolved and in some cases
  suffered erosion (\eg, plate tectonic, volcanism) erasing evidence
  of their primitive composition. 
  For most small bodies, however, their small diameter limited the
  amount of radiogenic nuclides in their interior, and thus the
  amount of energy for internal heating.
  The evolution of small bodies is therefore mainly exogenous,
  through eons of collisions, external heating, and bombardment
  by high energy particles.\\ 
  \indent A detailed study of the composition of small bodies can
  be achieved in the laboratory, by analyzing their terrestrial
  counterparts: meteorites.
  The distribution of elements, isotopes in meteorites,
  together with the level of heating and aqueous alteration they
  experienced tell us about the 
  temperature, elemental abundance, and timescales during the
  accretion stages \citep[\eg,][]{2006-MESS2-Halliday}. 
  The connection of this information with specific locations in
  the Solar System constrains the formation scenarios of our
  Solar System. This requires the identification of links between
  the meteorites and the different populations of small bodies.\\
  \indent Indeed, if meteorites are samples from the Solar System, 
  several questions are raised.
  Is this sampling complete? Is this sampling homogeneous?
  Some of the identified asteroid types (see Sect.~\ref{sec: intaxo}) 
  lack of a terrestrial analog.
  The most flagrant example are the O-type asteroids
  (3628) Bo{\v z}n{\v e}mcov{\' a} and 
  (7472) Kumakiri
  that appear unlike any measured meteorite assemblage
  \citep{2011-LPI-Burbine}.
  Coupled mineralogical and dynamical studies have shown that
  meteorites come from specific locations. Other regions of
  the Solar System may therefore be unrepresented in our meteorite
  collection
  \citep[see the discussions in][for instance]{
    2002-AsteroidsIII-5.2-Burbine, 
    2002-AsteroidsIII-4.1-Bottke, 2008-Nature-454-Vernazza}.\\
  \indent Additionally, the current orbits of small bodies may be
  different from the place they originally formed.
  For instance, it has been suggested that the giant
  planets migrated to their current orbits
  \citep[the Nice model, see][]{2005-Nature-435-Tsiganis}, injecting
  material from the Kuiper Belt into the inner Solar System
  \citep{2009-Nature-460-Levison}.
  Similarly, gravitational interaction among planetary embryos may
  have caused 
  outward migration of planetesimals from Earth's vicinity into the
  main belt \citep{2006-Nature-439-Bottke}.
  Current distribution of small bodies may therefore not reflect the
  original distribution of material in the Solar System. It however
  tells us about the dynamical processes that occurred over history. 
  Analysis of the composition of meteorites in the laboratory, of
  small bodies from remote-sensing, and of their
  distribution in the Solar System are therefore pre-requisites to
  understanding the formation and evolution of our Solar System.

\section{Linking small bodies with meteorites\label{sec: intaxo}}
  \indent Most of our knowledge on the mineralogy of asteroids has
  been derived by analysis of their reflectance spectra
  in the visible and near-infrared (VNIR). 
  The shape of these spectra has been used to classify the
  asteroids into broad groups, following
  several classification schemes called taxonomies.
  In what follows, I refer to
  the taxonomy by \citet{2009-Icarus-202-DeMeo}, based on the largest
  wavelength range (0.4--2.4\,$\mu$m). 
  It encloses 15 classes grouped into three \textsl{complexes}
  (C, S, and X), with 9 additional classes called \textsl{end-members}
  \citep[see][for a detailed description of the
    classes]{2009-Icarus-202-DeMeo}.  
  Mineralogical interpretations and links with meteorites have 
  been proposed for several classes. \\
  \indent Asteroids belonging to the S complex (S, Sa, Sq, Sr, and Sv)
  and to the Q class have been successfully linked to the most common
  meteorites, the ordinary chondrites (OCs).
  This link had been suggested for years based on the 
  presence of two deep absorption bands in their spectra,
  around 1 and 2 microns,
  similar to that of OCs and
  characteristic of a mixture of olivines and pyroxenes
  \citep[see for instance][among many
    others]{1996-MPS-31-Chapman, 2006-Icarus-184-Brunetto}. 
  The analysis of the sample from the S-type asteroid Itokawa
  returned by the Hayabusa spacecraft confirmed this link
  \citep{2011-Science-333-Yurimoto}.
  The two end-member classes A and V have a mineralogy 
  related to the S-complex.
  A-types are asteroids made of almost pure olivine, 
  which possible analogs are
  the achondrite meteorites of the Brachinite and Pallasite groups 
  \citep[see, \eg,][]{1989-AsteroidsII-4-Bell, 2004-AA-422-deLeon}. 
  In opposition, V-types are made of pure pyroxenes and are related to
  the HED achondrite meteorites
  \citep[\eg,][]{1970-Science-168-McCord}.
  A- and V-types are believed to 
  correspond to the mantle and the crust
  of differentiated parent bodies 
  \citep{1996-MPS-31-Burbine}.\\
  \indent The link between the hydrated carbonaceous chondrites (CCs) CI
  and CM and the asteroids in the C-complex seems well
  established \citep{2011-Icarus-212-Cloutis,
    2011-Icarus-216-Cloutis}.
  The anhydrous CV/CO carbonaceous chondrites have also been linked with
  B-types \citep{2010-JGR-115-Clark}.
  The scarcity and low contrast of absorption features in the VNIR
  prevents a detailed description of the mineralogy and
  association with meteorites of these asteroid types (B, C, Cb, Cg, Cgh, Ch).
  Spectroscopy in the 2.5--4\,$\mu$m wavelength range, however, revealed
  the presence of hydration features \citep{1978-MNRAS-182-Lebofsky,
    1990-Icarus-88-Jones, 2002-AsteroidsIII-2.2-Rivkin}.
  These features were interpreted as evidences for aqueous
  alteration, similar to that experienced by CI/CM parent bodies
  \citep{2011-Icarus-212-Cloutis, 2011-Icarus-216-Cloutis}. 
  Due to their similar composition to that of the solar photosphere, CI
  meteorites are often considered the most primitive material in the
  Solar System \citep[see][for an overview of meteorite
    classes]{2006-MESS2-Weisberg}.
  This has made the compositional study of these
  so-called \textsl{primitive asteroids} a primary goal in planetary
  science. \\
  \indent The VNIR spectra of asteroids in the X-complex are 
  devoid of strong absorption bands.
  However, several weak features (\eg, around 0.9\,$\mu$m) have been
  identified and used to discriminate sub-classes
  \citep{2004-AJ-128-Clark, 2010-Icarus-210-Ockert-Bell,
    2011-Icarus-214-Fornasier}. 
  Proposed meteorite analogs for X, Xc, Xe, and Xk asteroids virtually cover 
  the entire meteorite collection:
  the anhydrous CV/CO carbonaceous chondrites
  \citep{2005-AA-430-Barucci, 2012-PSS-Barucci},
  enstatite chondrites and aubrites \citep{2009-Icarus-202-Vernazza,
     2011-Icarus-216-Vernazza, 2010-Icarus-210-Ockert-Bell},
  mesosiderites \citep{2009-Icarus-202-Vernazza},
  stony-iron \citep{2010-Icarus-210-Ockert-Bell}, and iron meteorites 
  \citep{2011-Icarus-214-Fornasier}.
  The mineralogy represented in the X-complex is therefore probably
  more diverse than in the S- and C-complexes, due to the limits of the
  taxonomy based on spectral features only. 
  In is worth noting that in former taxonomies
  \citep[\eg,][]{1989-AsteroidsII-Tholen}, the X-complex was divided
  into three main groups, E, M, and P, distinguished by albedo. \\
  \indent L-types have been suggested to be the most ancient asteroids 
  that currently exist. From the comparison of their VNIR spectra with
  laboratory material, a fraction of 30\,$\pm$\,10\% of
  Calcium- and aluminum-rich inclusions was proposed
  \citep{2008-Science-320-Sunshine}.
  This value is significantly higher than that of meteorites.
  This suggests a very early accretion together with a low degree of
  alteration while crossing the entire history of the Solar System.
  With a similar spectral shape, K-types have often be described as
  intermediates between S- and C-like material
  \citep{2009-Icarus-202-DeMeo}. Most of the K-type are associated
  with the Eos dynamical family in the outer Main Belt. 
  They have been tentatively linked with the anhydrous CO, CV, and CK,
  and hydrated but metal-rich CR carbonaceous chondrites meteorites
  \citep{1989-AsteroidsII-4-Bell, 1998-Icarus-131-Doressoundiram, 
    2009-Icarus-202-Clark}. \\
  \indent The mineralogy of the remaining end-members classes is
  more uncertain, owing 
  to the apparent absence of strong spectral features (D and T) or 
  to the mismatch of features with any known material (O and R).
  It has been suggested that T-types contain a high fraction of
  metallic contents, and may be related to the iron cores of
  differentiated asteroids, hence iron meteorites
  \citep{1992-Metic-Britt}.
  D-types are among the reddest objects in the Solar System, not
  unlike that of comet nuclei and some transneptunian objects
  \citep{2008-SSBN-3-Barucci}.
  Their emission spectra in the mid-infrared indeed show
  striking similarities with that of comet nuclei
  \citep{2006-Icarus-182-Emery, 2011-AJ-141-Emery}.
  Both O and R classes were defined to describe the spectral shape of
  a single object, 
  (3628) Bo{\v z}n{\v e}mcov{\' a} and 
  (349) Dembowska respectively.
  Both types display broad absorption bands around 1 and 2 microns.
  These bands are however unlike those of S-types or any type of
  pyroxenes and olivines in our sample collection
  \citep{2011-LPI-Burbine}. \\
  \indent Comets and transneptunian objects (TNOs) are volatile-rich
  bodies. These two populations are dynamically linked, the later
  being one of the reservoir of periodic comets 
  \citep{2004-CometsII-7-Jewitt}.
  Several compositional groups have been identified among TNOs: 
  water ice dominated spectra, 
  methane-rich spectra, 
  and featureless spectra similar to that of comet nuclei
  \citep{2008-SSBN-3-Barucci}.
  There is no evidence for a meteorite sample from these dynamic
  classes, although 
  the delivery from Kuiper Belt material to Earth should be possible 
  \citep{2008-SSBN-9-Gounelle}. \\ 
  \indent As seen from this short summary,
  asteroid-meteorites connections and detailed mineralogy
  remain open questions in many cases:
  only about half of the 24 classes defining the
  taxonomy by \citet{2009-Icarus-202-DeMeo}
  have a mineralogical interpretation.
  Expanding the taxonomy toward longer wavelengths
  (2--5 and 5--40\,$\mu$m range) will help in that respect
  \citep[\eg,][]{1995-Icarus-117-Rivkin, 2002-AsteroidsIII-2.2-Rivkin,
    2006-Icarus-182-Emery}.
  Additional constraints must however be used to refine current
  mineralogy interpretations, especially for objects with featureless
  spectra. 
  Visible and radar albedos, thermal inertia, and density provide
  valuable constraints on the composition of these objects
  \citep[\eg,][]{2011-Icarus-214-Fornasier}. 
  Among these, the most fundamental property to 
  understand the composition and internal structure is perhaps the
  density \citep{2002-AsteroidsIII-4.2-Britt, 
    2008-ChEG-68-Consolmagno}.

\section{The density: a fundamental property\label{sec: int2}}
  \indent As described above, from the analysis of the surface
  properties such as reflectance spectra or albedo,
  it is possible to make inferences on composition.
  These observables however tell us about surface composition only, 
  which may or may not be reflective of the bulk composition of the
  body \citep{2011-EPSL-305-Elkins-Tanton}. 
  For instance, the surface of Earth, the \textsl{Blue Planet},
  is covered by water while its overall composition is
  totally different.
  Earth's density is indeed indicative of a rocky
  composition with a core of denser material. 
  Densities of small bodies are much more subtle, but still 
  contain critical information. \\
  \indent From the compilation of the density of about 20 asteroids, 
  \citet{2002-AsteroidsIII-4.2-Britt} already showed that differences
  are visible among that population.
  In a more recent review including 40 small bodies,
  \citet{2008-ChEG-68-Consolmagno}
  highlighted four trends in macroporosity (hereafter \macro).
  The macroporosity reflects the amount of voids 
  larger than the typical micrometer-sized cracks of
  meteorites.
  The largest asteroids (mass above 10$^{20}$\,kg) are
  apparently compact bodies without any macroporosity. This contrasts
  strongly with all the other less massive small bodies that have
  20\% or more macroporosity. The fraction of voids increases
  dramatically for icy bodies (comets and TNOs).
  Finally, \textsl{primitive} C-type asteroids tends to have larger
  macroporosity than the \textsl{basaltic} S-type. \\
  \indent Macroporosity, if present to a large extend, may have strong
  consequences on certain physical properties such as gravity field,
  thermal diffusivity, seismic velocity, and of course on
  collisional lifetimes
  \citep[see the review by][]{2002-AsteroidsIII-4.2-Britt}.  
  Macroporosity can also help in understanding the collisional history: 
  intact bodies are expected to have low-to-no macroporosity, while
  heavily impacted objects may have large cracks, fractures
  (\ie, moderate \macro), or be
  gravitational re-accumulation of material (\ie,
  rubble-piles, characterized by high values of \macro).

\section{Determination of density\label{sec: compil}}
  \indent Direct measurement of the \textsl{bulk} density ($\rho$)
  involves the independent measures of the mass ($M$) and 
  volume ($V$): $\rho$\,$=$\,$M/V$.
  Indirect determination of the density are also possible by modeling the 
  mutual eclipses of a binary system \citep[\eg,][]{2006-AA-446-Behrend} or
  the non-gravitational forces on a comet nucleus
  \citep[\eg,][]{2007-Icarus-187-Davidsson}.
  This study aims at deriving constraints on the intrinsic physical
  properties of small bodies by searching for relationships between,
  the dynamical and taxonomic classes, size, and density. 
  An extensive compilation of the mass, volume, and resulting density
  estimates available in the literature is therefore presented here. \\
  \indent There are \numb{994} published mass estimates for \numb{267} small
  bodies (Sect.~\ref{ssec: mass}).
  For each object, the volume determinations are also compiled here,
  resulting in \numb{1454}
  independent estimates (Sect.~\ref{ssec: diam}). Finally, 
  the density of \numb{24} small bodies has also been indirectly
  determined (Sect.~\ref{ssec: dens}).
  In total, \numb{287} density estimates are available, 
  for small bodies pertaining to all the dynamical classes:
  \numb{17} near-Earth asteroids (NEAs), 
  \numb{230} Main-Belt (MBAs) and Trojan asteroids, 
  \numb{12} comets, and
  \numb{28} transneptunian objects (TNOs).
  There is however a large spread among the independent estimates of  
  the mass and volume estimates of these objects.
  Additionally, several estimates lead to obvious non-physical
  densities such as 0.05 or 20, the respective densities of Aerogel
  and Platinum. 
  A rigorous selection of the different estimates is therefore needed.
  Some specifics of mass and diameter estimates are discussed below,
  together with selection criteria.

  \subsection{Mass estimates\label{ssec: mass}}
%
\begin{figure}[!t]
\centering
  \includegraphics[width=.45\textwidth]{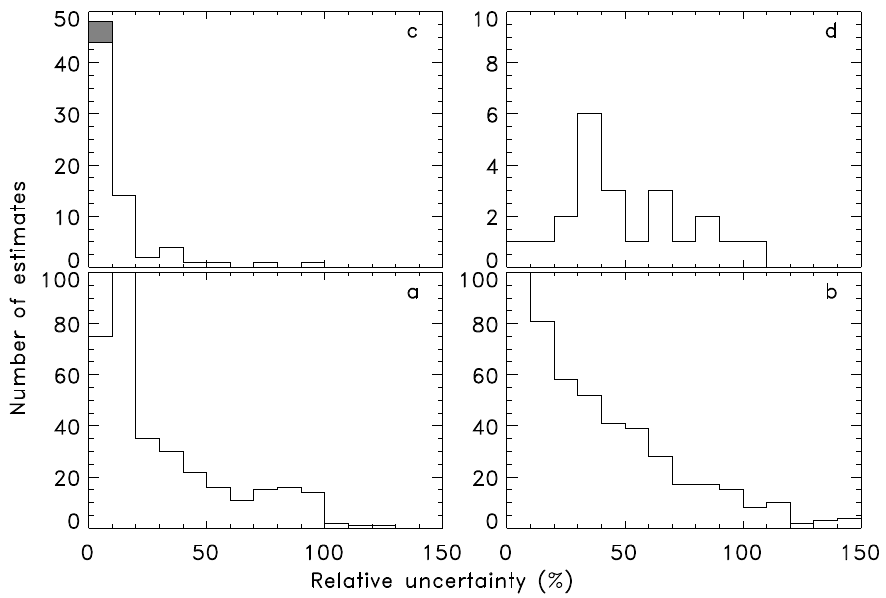}
  \caption[Statistics on precision of mass estimates]{%
    Distribution of the relative accuracy of mass estimates obtained
    with four different methods (see text):
    (a) orbit deflection during close encounters, 
    (b) planetary ephemeris, 
    (c) orbit of natural satellites or spacecrafts (gray bar), and
    (d) indirect determination of density (Sect.~\ref{ssec: dens})
    converted into mass.    
    \label{fig: statM}
  }
\end{figure}
%
%
    \indent The determination of the mass of a minor planet
    relies on the analysis of its gravitational effects on other
    objects \citep[see the review by][for
      instance]{2002-AsteroidsIII-2.2-Hilton}. 
    The \numb{994} mass estimates for \numb{267} small bodies 
    listed in~\ref{app: mass} can be divided in 4 categories, owing to
    the gravitational effects that were analyzed:
    \begin{enumerate}
        \vspace*{-.5em}
      \item \textbf{Orbit deflection during close encounters:}
        The mass of small bodies is several order of magnitude lower than
        that of planets. Asteroids can nevertheless slightly influence the orbit of
        other smaller asteroids \citep[\eg,][]{2000-AA-360-Michalak,
          2001-AA-374-Michalak} and of Mars
        \cite[\eg,][]{2001-AA-371-Pitjeva, 2009-AA-508-Mouret} during
        \textsl{close} encounters.
        This method has been widely used, resulting in \numb{547} mass
        estimates. An accuracy of few percent can be reached for the
        most massive asteroids such as (1) Ceres, (2) Pallas, or (4)
        Vesta \citep[\eg,][]{2006-Icarus-182-Konopliv,
          2011-AJ-142-Zielenbach}.
        The accuracy however drops for smaller asteroids, and about a
        third have uncertainties cruder than 100\% 
        \citep[see, for instance][and
          Fig.~\ref{fig: statM}.a]{2010-PSS-58-Somenzi,
          2011-AJ-142-Zielenbach}.         
      \item \textbf{Planetary ephemeris:} 
        Numerical models have been developed to describe and predict
        the position of planets and minor planets around the Sun.
        In addition to the Sun and the planets, the gravitational
        influence of several asteroids must
        be taken into account to properly describe the observed
        position of planets, satellites, and spacecrafts
        \citep[see][for details]{2008-CeMDA-100-Baer,
          2011-AJ-141-Baer, 
          2008-AA-477-Fienga, 2009-AA-507-Fienga,
          2010-SciNote-Fienga, 2009-SciNote-Folkner}.
        In that respect, this method is similar to the analysis of
        close encounters. There is however a strong philosophical
        difference between these two methods:
        analysis of close encounters consists of considering N times
        a 1-to-1 gravitational interaction, while 
        planetary ephemeris are conceptually closer to a N-to-1
        interaction. 
        Similarly to the results obtained from close encounters, the
        best accuracy is achieved for largest asteroids and becomes
        cruder for smaller objects. The mean accuracy is of 45\%,
        but values are distributed up to 100\%
        (Fig.~\ref{fig: statM}.b).
      \item \textbf{Spacecraft tracking:}
        The Doppler shifts of the radio signals sent by spacecraft
        around an asteroid can be used to determine its orbit or
        the deflection of its trajectory during a flyby.
        These frequency shifts are imposed by the gravitational
        perturbation and are related to the mass of the asteroid
        \citep{1997-Science-278-Yeomans, 2000-Science-289-Yeomans, 
          2006-Science-312-Fujiwara, 2011-Science-334-Paetzold}.
        It is by far the most precise technique with a typical
        accuracy of a couple of percent (Fig.~\ref{fig: statM}.c).
        It will however remain
        limited to a handful of small bodies (only four to date).
      \item \textbf{Orbit of a satellite:}
        From optical or radar
        images of the components of the system, their mutual orbit
        can be determined and the mass derived with Kepler's third
        law \citep[see, for instance,][]{1997-Icarus-130-Petit,
          1999-Nature-401-Merline,
          2002-AsteroidsIII-2.2-Merline, 
          2002-Science-296-Margot,
          2005-Icarus-178-Marchis, 2008-Icarus-196-Marchis,
          2008-Icarus-195-Marchis, 
          2005-ApJ-632-Brown, 2010-AJ-139-Brown, 
          2011-AA-534-Carry, 2011-AJ-141-Fang}. 
        The \numb{28} mass estimates available for TNOs were derived from
        optical imaging with the Hubble space telescope or large
        ground-based telescopes equipped with adaptive-optics cameras
        \citep[\eg,][]{2009-Icarus-200-Grundy,
          2011-AA-528-Dumas}.
        Similarly, the \numb{17} mass estimates for NEAs
        were all derived from radar \citep[\eg,][]{2006-Science-314-Ostro,
        2006-Icarus-184-Shepard}, with the exception of Itokawa
        which was the target of the Hayabusa sample-return mission
        \citep{2006-Science-312-Fujiwara}. 
        Additionally, the mass of \numb{26} MBAs was determined
        by optical imaging.
        In total, \numb{68} mass estimates have
        been derived by analyzing the orbit of a satellite. It is the
        second most-precise technique with a typical accuracy
        of about 10--15\% (Fig.~\ref{fig: statM}.c).
        It is the most productive method of \textsl{accurate}
        mass determinations. With currently more than
        200 known binaries, many mass estimates are still to come. 
    \end{enumerate} 

\begin{figure}[!t]
\centering
  \includegraphics[width=.45\textwidth]{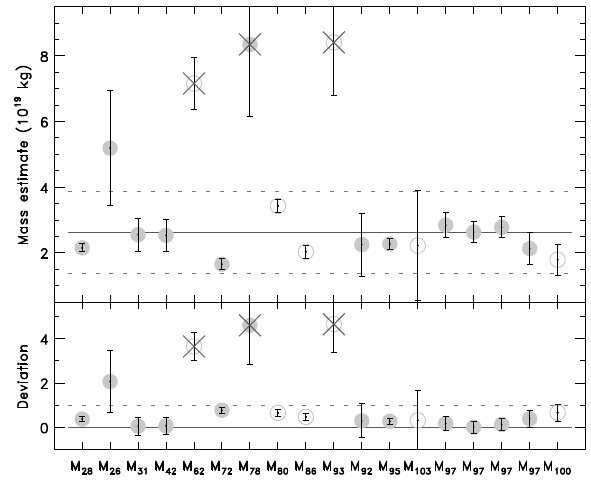}
  \caption[Mass estimates for (52) Europa]{%
    The \numb{18} mass estimates for (52) Europa
    (see \ref{app: biblio} for the references).
    \textbf{Top:} The different mass estimates M$_i$, in 10$^{19}$\,kg.
    Symbols indicate the method used to determine the mass:
    deflections (gray disk) or 
    planetary ephemeris (open circle).
    Crossed estimates were discarded from the analysis (see text).
    Horizontal solid and dashed lines are respectively 
    the weighted average ($\mu$)
    and standard deviation ($\sigma$)
    of the mass estimates before selection.
    \textbf{Bottom:} Same as above, but plotted as
    a function of the distance to the average value, in units of
    deviation: (M$_i$\,-\,$\mu$)/ $\sigma$.
    Similar plots for each of the \numb{140} small bodies with
    multiple mass estimates are provided in \ref{app: mass}.
   \label{fig: mex52}
  }
\end{figure}

\begin{figure}
\centering
  \includegraphics[width=.45\textwidth]{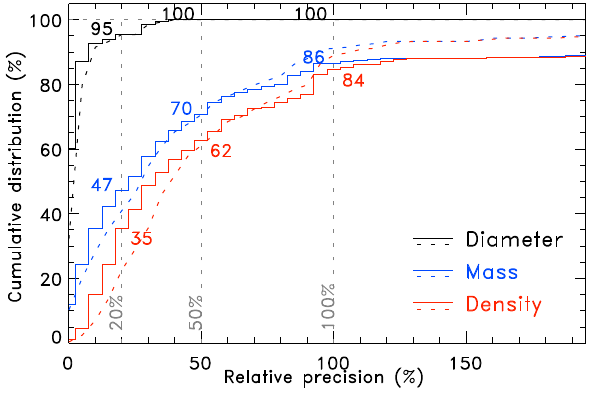}
  \caption[Statistics on precision]{%
    Cumulative distribution of the accuracy on the
    diameter (black), mass (blue), and density (red) estimates.
    Dashed and solid lines represent the distributions before and
    after selection of best estimates (see text for details).
    Three reference levels for the relative accuracy are drawn:
    20\%, 50\% and 100\%,
    with the fraction of targets with a better accuracy
    reported for each estimate (after selection only).
    \label{fig: prec}
  }
\end{figure}

    \indent Based on these considerations and a close 
    inspection of the different mass estimates available
    (\eg, Fig.~\ref{fig: mex52}), 
    the following criteria for selecting mass
    estimates were applied:
    Mass estimates derived from either the
    third or the fourth method (spacecraft or satellite)
    prevail upon the first two methods (deflection and ephemeris).
    Mass estimates leading to non-physical densities are discarded.
    Mass estimates that do not agree within uncertainties with the range
    drawn by the weighted average and standard deviation are
    discarded. The weighted average and standard deviation are
    subsequently recomputed. 
    The \numb{994} mass estimates are provided in \ref{app: mass}
    together with bibliographic references and
    notes on selection. \\
%
%
%
%
    \indent A summary of the precision achieved on mass estimates is
    presented in Fig.~\ref{fig: prec}. 
    The contribution provided here is
    illustrated by the difference between the cumulative distribution
    of relative precision before (dashed line) and after (solid line)
    the selection (about 20\% of the estimates were discarded).
    For estimates with a relative uncertainty below 50\% , the
    selection of estimates slightly improves the final accuracy,
    increasing the number of accurate estimates by 5 to 10\%.
    The apparent \textsl{degradation} introduced by the selection for
    low-precision estimates is due to rejection of about 10\% of these
    estimates. In other words, these estimates lead
    to unrealistic densities and should not be considered.
    Furthermore, the distribution presented in Fig.~\ref{fig: prec} is
    based on the uncertainties reported by the different authors.
    The discrepancy between estimates however often reaches 
    disconcerting levels. For instance, the estimates 
    M$_{\rm 28}$ \citep{2001-IAA-Krasinsky}, 
    M$_{\rm 72}$ \citep{2008-DPS-40-Baer}, 
    M$_{\rm 80}$ \citep{2009-AA-507-Fienga}, and 
    M$_{\rm 86}$ \citep{2009-SciNote-Folkner} of the mass of (52) Europa
    fall within the range drawn by the weighted mean
    and deviation (Fig.~\ref{fig: mex52}).
    They nevertheless strongly disagree: the different values are
    between 4 and 11\,$\sigma$ one from each other. \\
    \indent Such differences are indicative of underestimated
    uncertainties. Accuracy is often reported as the formal standard
    deviation ($\sigma$), which in some cases may be small compared to
    systematics. 
    The uncertainties on the mass determinations should therefore be
    considered as lower limits, to which some systematics could be
    added.
    As a result, the cumulative distribution of the relative precision 
    presented in Fig.~\ref{fig: prec} is optimistic and gives an upper
    limit to the amount of accurate estimates. 
    Therefore, even with mass estimates available for more than 250
    small bodies, our knowledge is still very limited:
    Only about half of the estimates are more accurate than 20\%, and
    no more than 70\% of the estimates are more accurate than 50\%
    (higher uncertainties preventing any firm conclusion).
%
%
%

  \subsection{Volume estimates\label{ssec: diam}}
%
\begin{figure}[!t]
\centering
  \includegraphics[width=.45\textwidth]{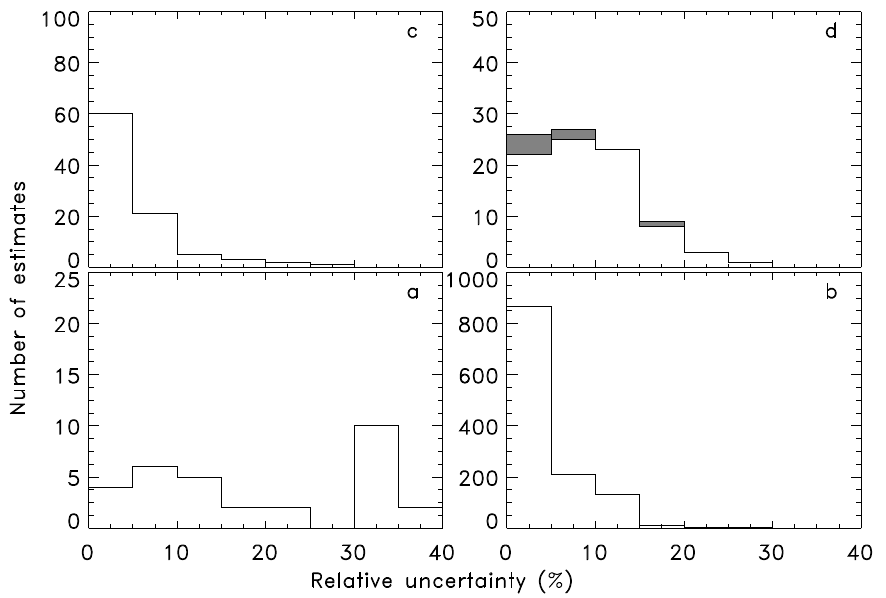}
  \caption[Statistics on precision of diameter estimates]{%
    Distribution of the relative accuracy of diameter estimates obtained
    with four classes of different methods (see text):
    (a) crude estimates from absolute magnitude, 
    (b) thermal radiometry, 
    (c) direct measurement limited to a single geometry, and
    (d) shape modeling based on several geometries (gray bars
    represent the diameters derived from spacecraft encounters).
    Although estimates in sub-plot (d) are expected to be the most
    precise, it is not reflected in their relative uncertainty
    distribution. The possible underestimation of biases in other
    techniques may be the cause (see text).    
    \label{fig: statD}
  }
\end{figure}
%
%
    \indent As already noted by several authors, 
    the most problematic part of determining the density of a small
    body is to measure any mass at all
    \citep[\eg,][]{2002-AsteroidsIII-2.2-Merline,
      2008-ChEG-68-Consolmagno}. 
    The number of density estimates
    presented here is limited by the number of mass
    estimates, and not by the number of volume estimates (generally
    reported as volume-equivalent diameter, hereafter $\phi$).
    Many different observing techniques and methods of
    analysis have been used to evaluate the diameter
    of small bodies \citep[see the review by][]{2012-PSS--Carry}.
    The \numb{1454} diameter estimates listed in \ref{app: diam} 
    were derived with 15 different methods, that can be grouped into 4
    categories: 

\begin{figure}[!t]
\centering
  \includegraphics[width=.45\textwidth]{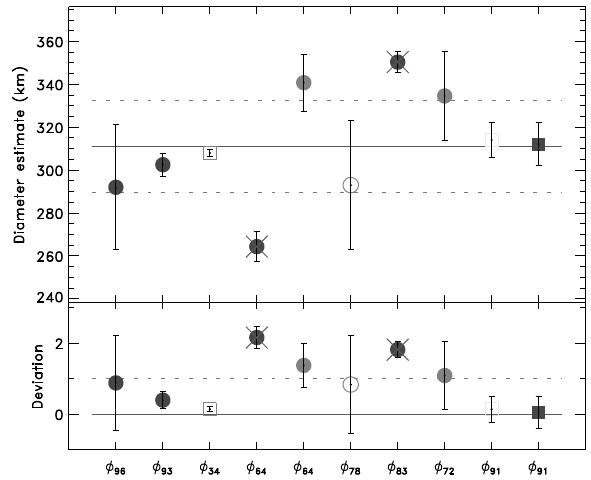}
  \caption[diameter estimates for (52) Europa]{%
    The \numb{10} diameter estimates for (52) Europa
    (see \ref{app: biblio} for the references).
    \textbf{Top:} The different diameter estimates $\phi_i$, in km.
    Symbols indicate the method used to determine the diameter:
    mid-infrared radiometry modeled using
    the Standard Thermal Model
       (STM: $\phi_{96}$, $\phi_{93}$, $\phi_{64}$, and $\phi_{83}$)
    and the near-Earth asteroid thermal model
    (NEATM: $\phi_{64}$ and $\phi_{72}$), 
    disk-resolved imaging on a single epoch ($\phi_{34}$),
    combination of lightcurves and stellar occultations ($\phi_{78}$), 
    or shape modeling ($\phi_{91}$).
    See \ref{app: diam} for a
    complete description of the symbols.
    Crossed estimates were discarded from the analysis (see text).
    Horizontal solid and dashed lines are respectively 
    the weighted average ($\mu$)
    and standard deviation ($\sigma$)
    of the diameter estimates before selection.
    \textbf{Bottom:} Same as above, but plotted as
    a function of the distance to the average value, in units of
    deviation: ($\phi_i$\,-\,$\mu$)/ $\sigma$.
    Similar plots for each of the \numb{246} small bodies with
    multiple diameter estimates are provided in \ref{app: diam}.
   \label{fig: dex52}
  }
\end{figure}
%
    \begin{enumerate}
        \vspace*{-.5em}
      \item \textbf{Absolute magnitude:} It could almost be considered
        an \textsl{absence} of size estimate. It is the crudest method
        to evaluate the diameter of a small body (Fig.~\ref{fig: statD}.a).
        From the absolute magnitude $H$ and an
        \textsl{assumed} geometric albedo $p$, the diameter is given
        by $\phi\,({\rm km}) = 1329\, p^{-0.5} 10^{-0.2 H}$
        \citep[][and references therein]{2007-Icarus-190-Pravec}.
        The diameter of \numb{29} small bodies presented here were
        derived using their absolute magnitude, in absence of any
        other estimates. This particularly applies to TNOs.
      \item \textbf{Thermal modeling of mid-infrared radiometry:}
        It is by far the main provider of diameter estimates:
        \numb{1233} diameter estimates out of the 
        \numb{1454} listed in \ref{app: diam} (\ie, $\approx$85\%).
        Asteroids are indeed among the brightest sources in the sky at 
        mid-infrared wavelengths (5--20\,$\mu$m), so
        infrared satellites (IRAS, ISO, AKARI, Spitzer, and WISE) have 
        been able to acquire observations of a vast number of these
        objects \citep[see][]{2002-AJ-123-Tedesco-a, 2010-AJ-140-Ryan, 
          2011-PASJ-63-Usui, 2011-ApJ-741-Masiero,
          2011-AJ-141-Mueller}.
        The diameter and albedo of the colder TNOs have also been
        studied at longer wavelengths with Spitzer and Herschel
        \citep[\eg,][]{2008-SSBN-3-Stansberry, 2009-EMP-105-Muller}. 
        As visible in Fig.~\ref{fig: statD}.b, the typical uncertainty
        is of only few percent. In many case, however, the
        different estimates from thermal modeling disagree above
        their respective quoted uncertainty
        \citep[see Table 3 in][illustrating the
          issue]{2009-PSS-57-Delbo}.
        For instance, in the case of Europa (Fig.~\ref{fig: dex52}),
        both diameter estimates
        $\phi_{64}$ \citep{2010-AJ-140-Ryan} where based on the same
        data, but used two different thermal modeling, and disagree
        at more than 6\,$\sigma$.
        Such differences are again indicative of underestimated
        uncertainties. Accuracy is often reported as the formal standard
        deviation ($\sigma$), which in some cases may be small compared to
        systematics. In the present case, the simplified 
        standard thermal model \citep{1986-Icarus-68-Lebofsky}
        and near-Earth asteroid thermal model
        \citep{1998-Icarus-131-Harris} 
        widely used do not take into account the spin and shape of the
        small body into account, and can therefore be strongly biased.
        A more realistic level of accuracy is about 10\%
        \citep{2010-AA-518-Lim}, at which these estimates are still
        highly valuable given the
        huge number of small bodies that have been studied that way.
      \item \textbf{Direct measurements of a single geometry:}
        Stellar occultations or disk-resolved images can provide an
        extremely precise measure of the apparent size and shape of a
        small body
        \citep[\eg,][]{2004-AJ-127-Brown, 2006-ApJ-643-Brown, 
          2006-Icarus-185-Marchis, 2008-Icarus-196-Marchis, PDSSBN-OCC}.
        When these direct measurements are limited to a
        single geometry, however, the evaluation of the diameter may
        be biased. The volume is 3-D while a single geometry only
        provides 2-D constraints. The typical accuracy of 5\%
        (Fig.~\ref{fig: statD}.c) may therefore be optimistic.
        Nevertheless, these estimates are highly valuable, being
        based on direct measurements.
      \item \textbf{Shape modeling based on several geometries:}
        The least numerous but most precise diameter estimates are
        derived when the spin and 3-D shape of the objects are
        modeled, thus limiting the 2-D to 3-D related biases
        (Fig.~\ref{fig: statD}.d).
        Small bodies can be modeled as smooth tri-axial ellipsoids
        \citep[\eg,][]{2005-Nature-437-Thomas,
          2009-Science-326-Schmidt,
          2009-Icarus-202-Drummond, 2010-AA-523-Drummond}, 
        convex shapes
        \citep{2007-Icarus-187-Descamps, 2011-Icarus-214-Durech}, 
        or realistic 3-D shapes
        \citep{2000-Science-289-Veverka, 
          2006-Science-314-Ostro, 2010-Icarus-207-Ostro, 
          2010-Icarus-205-Carry-a, 2010-AA-523-Carry, 
          2011-Science-334-Sierks}.
        In particular, spacecraft encounters with
        (25\,143) Itokawa and (21) Lutetia have shown that
        multi-data approaches provide reliable and precise
        diameter estimates:
        \eg, lightcurve-derived shape model with thermal radiometry
        \citep{2006-AA-447-Mueller} or
        combined inversion of disk-resolved imaging and lightcurves 
        \citep{2011-IPI-5-Kaasalainen, 2010-AA-523-Carry,
          2012-PSS--Carry}.
    \end{enumerate}

    \indent As visible in Figs.~\ref{fig: prec} and~\ref{fig: statD},
    the diameter estimates are generally intrinsically much
    more precise than the mass determination: all the estimates 
    are known to better than 50\% relative precision, and a large
    majority to better than 10\%. 
    Diameter estimates from different techniques moreover
    generally agree, suggesting that systematics are commensurable
    with formal uncertainties. 
    The same selection criteria than for mass estimates were applied
    here, and about 15\% of the estimates were discarded.
    Paradoxically, once the mass is determined, the
    uncertainty on the volume ($\delta V/V$) often becomes the major
    source of uncertainty on the density ($\rho$). Indeed,

    \begin{equation}
      \frac{\delta \rho}{\rho} = \sqrt{ \left(\frac{\delta M}{M}\right)^2  + 
                                        \left(\frac{\delta V}{V}\right)^2 }
                               = \sqrt{ \left(\frac{\delta M}{M}\right)^2  + 
                                      9 \left(\frac{\delta \phi}{\phi}\right)^2 }
      \label{eq: prec}                     
    \end{equation}

    \noindent The contribution of the uncertainty on the diameter
    ($\delta \phi/\phi$) therefore easily overwhelms that of the mass
    ($\delta M/M$) . 
    In the compilation presented here, however, the mass is the
    limiting factor for \numb{61}\% of the objects, contributing to 
    $\approx$\numb{72}\% of the density uncertainty.
    This is mainly due to the high number of non-precise mass estimates 
    (Fig.~\ref{fig: prec}).
    If only the density estimates with a relative precision better
    than 20\% are considered, then the situation is reversed:
    the diameter is the
    limiting factor for \numb{75}\% of the objects, contributing to 
    $\approx$\numb{68}\% of the density uncertainty.
    For these reasons, the mass should therefore be
    considered the limiting factor in most of the cases.
    As already discussed elsewhere, however, when a reliable mass
    estimate is available (\ie, usually from the presence of a
    satellite), the precision on the volume generally limits the
    accuracy on the density \citep{2002-AsteroidsIII-2.2-Merline, 
    2002-AsteroidsIII-4.2-Britt, 2008-ChEG-68-Consolmagno}.

  \subsection{Indirect density estimates\label{ssec: dens}}
    \indent For small bodies with diameters of a few to tens of kilometers
    the methods to estimate their mass listed above
    (Sect.~\ref{ssec: mass}) cannot be used.
    The gravitational influence of these very small bodies is too tiny
    to be measured. 
    Even in the case of binary systems, their angular extent
    is generally too small to be imaged with current technology.
    The only exception are the small binary NEAs that can be imaged
    with radar during close approaches with Earth. 
    Yet, a large fraction of the currently known binaries are
    small-sized systems discovered by studying their lightcurves
    \citep[\numb{86} out of \numb{207}, \eg,][]{2000-Icarus-146-Mottola, 
      2002-AsteroidsIII-2.2-Pravec, 2006-Icarus-181-Pravec}.
    Indeed, photometric observations of the mutual eclipses of a
    system provide many constraints, for instance, on
    the ratio between the diameters of the two components
    or between the primary diameter and the orbit semi-major axis
    \citep[see][]{2009-Icarus-200-Scheirich}. \\
    \indent Nevertheless, these parameters are dimensionless from
    lightcurve observations only.
    The \textsl{absolute} scale, hence semi-major
    axis and thus mass, cannot be derived.
    Usually, both components are \textsl{assumed} to have the same
    bulk density to bypass this restriction 
    \citep[\eg,][]{2009-Icarus-200-Scheirich}.
    These estimates are indirect, being derived without measuring the
    mass nor the size.
    The accuracy reached greatly depends on each system, and 
    ranges from a few percent to 100\% (Fig.~\ref{fig: statM}.d).
    It is worth noting that if small-sized binaries are formed by
    rotational breakup \citep{2008-Nature-454-Walsh}
    as suggested by the fast rotations of the primaries
    \citep{2002-AsteroidsIII-2.2-Pravec, 2006-Icarus-181-Pravec,
      2010-Nature-466-Pravec}, the porosity, hence density,
    of the components may be significantly different.
    These density estimates may therefore be biased, in absence of an
    independent measure of the scale of the systems. \\
    \indent Measuring the mass of comets is another challenge.
    With diameters typically smaller than 10\,km, comets have very
    small masses. In absence of a satellite, studying their
    gravitational effect on other objects is hopeless.
    The activity of their nucleus
    however provides an indirect way to estimate their mass.
    Indeed, the forces resulting from the gas jets slowly change the
    orbit of the nucleus around the Sun. 
    Modeling this non-gravitational effect provides the mass of
    the nucleus \citep[\eg,][]{2004-Icarus-168-Davidsson, 
      2005-Icarus-176-Davidsson, 2006-Icarus-180-Davidsson, 
      2007-Icarus-187-Davidsson, 2009-MNRAS-393-Sosa}.
    The masses of \numb{11} comets have been derived using this
    approach. 
    \citet{2007-Icarus-190-Richardson}
    have also studied the expansion of ejecta created by the
    Deep Impact experiment on
    the comet 9P/Tempel1. This is the most
    direct measurement of the mass of a comet, independent of
    the non-gravitational effect. \\

  \indent A summary of the mass, volume-equivalent diameter and bulk
  density of the \numb{287} small bodies compiled here is provided in
  Table~\ref{tabSM: dens}. The values listed are the weighted average
  and standard deviation of all the selected estimates
  (see \ref{app: mass}, \ref{app: diam}, and \ref{app: indens}).
  The density is given normalized to that of liquid water
  (1\,000\,kg\,m$^{-3}$), \ie, dimensionless. 
  The estimates have been ranked from A to E, owing to the level of
  relative accuracy achieved on the density:
  B better than 20\%, 
  C between 20 and 50\%, 
  D between 50 and 100\%, and
  E cruder than 100\%.
  A stands for reliable estimates (more precise than 20\%),
  based on more than
  5 mass estimates and 5 diameter estimates,
  or a spacecraft encounter.
  Irrelevant densities are tagged with a cross (\ding{55}).
  Only about a third of the \numb{287} density estimates have a
  relative precision better than 20\% (Fig.~\ref{fig: prec}),
  and two third better than 50\%, above which level nothing
  relevant can be derived.

  The fraction of volume occupied by voids,
  the macroporosity \macro,
  is also reported, computed as:
  \begin{equation}
    \macro\,(\%) = 100 ~\left( 1 - \frac{\rho}{\rho_m}
    \right)\label{eq: poro}
  \end{equation}
  \noindent with $\rho$ the asteroid bulk density and $\rho_m$ the bulk density
  of the associated meteorite (Table~\ref{tabSM: met}).
  The macroporosity is the least constrained of all the quantities
  discussed here. Indeed, it is affected by the uncertainties and
  possible biases on the diameter and mass estimates and also from the
  possible ambiguous links with meteorites (Sect.~\ref{sec: intaxo}
  and Table~\ref{tab: taxodens}).
  Depending on the meteorite association, the macroporosity may change
  by 30--40\%. For instance, while (16) Psyche was the most porous
  asteroid listed by \citet{2002-AsteroidsIII-4.2-Britt}
  and \citet{2008-ChEG-68-Consolmagno}
  with a macroporosity of about 70\%, it stands in the low
  macroporosity range (about 18\%).
  A low macroporosity is actually more consistent with the link
  between Psyche and iron meteorites than the very high value of
  $\sim$75\% found previously.

\begin{table*}[!ht]
\setcounter{table}{0}
\centering
\caption[Density estimates]{%
  Compilation of the average mass ($M$) and volume-equivalent diameter ($\phi$) estimates 
  (see~\ref{app: mass}, \ref{app: diam}, and~\ref{app: indens}), 
  and resulting bulk density ($\rho$) and macroporosity ($\mathcal{P}$)
  for \numb{287} objects, with their associated uncertainties.
  For each object, the dynamical class is listed (Dyn.), together with the 
  taxonomic class (Tax., for asteroids only) and associated meteorite (Met.).
  The density estimates are ranked A to E, owing to the level 
  of confidence at which they are determined (see text). 
  Unrealistic density estimates are marked with a cross (\ding{55}) and 
  uncertainties on the macroporosity larger than 100\% are listed as $\infty$.
  \textbf{References}:
    (1) \citet{2010-JGR-115-Clark}, 
    (2) \citet{2010-Icarus-210-Ockert-Bell}, and
    (3) \citet{2011-Icarus-214-Fornasier}.
\label{tabSM: dens}
}
\begin{tabular}{r@{ }l c@{\hspace{.75em}}l@{\hspace{.75em}}l c@{\,$\pm$\,}c@{\,$\times$\,}l@{\hspace{.5em}}l r@{.}l@{\,$\pm$\,}r@{.}l@{\hspace{.5em}}l r@{.}l@{\,$\pm$\,}r@{.}l r@{\,$\pm$\,}rc}
 \hline
 \hline
 \multicolumn{2}{c}{Designation} & 
 \multicolumn{3}{c}{Classification} & 
 \multicolumn{4}{c}{Masses (kg)} & 
 \multicolumn{5}{c}{Diameter (km)} & 
 \multicolumn{4}{c}{Density} & 
 \multicolumn{2}{c}{Porosity} & 
 Rank \\ 
 \multicolumn{1}{c}{\#} & \multicolumn{1}{c}{Name} & 
 Dyn. & Tax. & Met. &
 \multicolumn{1}{c}{$M$} & \multicolumn{1}{c}{$\delta M$} & & Fig. & 
 \multicolumn{2}{c}{$\phi$} & \multicolumn{2}{c}{$\delta \phi$} & Fig. & 
 \multicolumn{2}{c}{$\rho$} & \multicolumn{2}{c}{$\delta \rho$} & 
 \multicolumn{1}{c}{$\mathcal{P}$} & $\delta \mathcal{P}$ \\ 
 \hline
     1 & Ceres                & MBA & C   & CM        & 9.44 & 0.06 & $10^{20}$ & \ref{fap: mass000001}       &  944 & 79 &   22 & 99 & \ref{fap: diam000001}       &   2 & 13 &   0 & 15 &   4 &  7 &         A \\
     2 & Pallas               & MBA & B   & CK $^{1}$ & 2.04 & 0.04 & $10^{20}$ & \ref{fap: mass000002}       &  514 & 41 &   19 & 12 & \ref{fap: diam000002}       &   2 & 86 &   0 & 32 &   0 & 11 &         A \\
     3 & Juno                 & MBA & Sq  & OC        & 2.73 & 0.29 & $10^{19}$ & \ref{fap: mass000003}       &  241 & 79 &   10 & 58 & \ref{fap: diam000003}       &   3 & 68 &   0 & 62 &   0 & 16 &         A \\
     4 & Vesta                & MBA & V   & HED       & 2.63 & 0.05 & $10^{20}$ & \ref{fap: mass000004}       &  519 & 33 &    6 & 84 & \ref{fap: diam000004}       &   3 & 58 &   0 & 15 &   0 &  4 &         A \\
     5 & Astraea              & MBA & S   & OC        & 2.64 & 0.44 & $10^{18}$ & \ref{fap: mass000005}       &  113 & 41 &    3 & 53 & \ref{fap: diam000005}       &   3 & 45 &   0 & 66 &   0 & 19 &         B \\
\hline
\end{tabular}
\end{table*}

\begin{table*}[!ht]
\setcounter{table}{0}
\centering
\caption{Continued}
\begin{tabular}{r@{ }l c@{\hspace{.75em}}l@{\hspace{.75em}}l c@{\,$\pm$\,}c@{\,$\times$\,}l@{\hspace{.5em}}l r@{.}l@{\,$\pm$\,}r@{.}l@{\hspace{.5em}}l r@{.}l@{\,$\pm$\,}r@{.}l r@{\,$\pm$\,}rc}
 \hline
 \hline
 \multicolumn{2}{c}{Designation} & 
 \multicolumn{3}{c}{Class} & 
 \multicolumn{4}{c}{Masses (kg)} & 
 \multicolumn{5}{c}{Diameter (km)} & 
 \multicolumn{4}{c}{Density} & 
 \multicolumn{2}{c}{Porosity} & 
 Rank \\ 
 \multicolumn{1}{c}{\#} & \multicolumn{1}{c}{Name} & 
 Dyn. & Tax. & Met. & 
 \multicolumn{1}{c}{$M$} & \multicolumn{1}{c}{$\delta M$} & & Fig. & 
 \multicolumn{2}{c}{$\phi$} & \multicolumn{2}{c}{$\delta \phi$} & Fig. & 
 \multicolumn{2}{c}{$\rho$} & \multicolumn{2}{c}{$\delta \rho$} & 
 \multicolumn{1}{c}{$\mathcal{P}$} & $\delta \mathcal{P}$ \\ 
 \hline
    59 & Elpis                & MBA & B   & CV        & 3.00 & 0.50 & $10^{18}$ & \ref{fap: mass000059}       &  163 & 61 &    6 & 50 & \ref{fap: diam000059}       &   1 & 30 &   0 & 26 &  53 & 20 &         C \\
    60 & Echo                 & MBA & S   & OC        & 3.15 & 0.32 & $10^{17}$ & \ref{fap: mass000060}       &   60 & 00 &    1 & 33 & \ref{fap: diam000060}       &   2 & 78 &   0 & 33 &  16 & 12 &         B \\
    61 & Danae                & MBA & S   & OC        & 2.89 & 2.78 & $10^{18}$ &                             &   82 & 52 &    2 & 73 & \ref{fap: diam000061}       &   9 & 81 &   9 & 49 &   0 & 96 &         D \\
    63 & Ausonia              & MBA & S   & OC        & 1.53 & 0.15 & $10^{18}$ & \ref{fap: mass000063}       &   94 & 45 &    7 & 15 & \ref{fap: diam000063}       &   3 & 46 &   0 & 86 &   0 & 24 &         C \\
    65 & Cybele               & MBA & Xk  & Mes       & 1.36 & 0.31 & $10^{19}$ & \ref{fap: mass000065}       &  248 & 29 &   17 & 59 & \ref{fap: diam000065}       &   1 & 70 &   0 & 52 &  59 & 30 &         C \\
\hline
\end{tabular}
\end{table*}

\begin{table*}[!ht]
\setcounter{table}{0}
\centering
\caption{Continued}
\begin{tabular}{r@{ }l c@{\hspace{.75em}}l@{\hspace{.75em}}l c@{\,$\pm$\,}c@{\,$\times$\,}l@{\hspace{.5em}}l r@{.}l@{\,$\pm$\,}r@{.}l@{\hspace{.5em}}l r@{.}l@{\,$\pm$\,}r@{.}l r@{\,$\pm$\,}rc}
 \hline
 \hline
 \multicolumn{2}{c}{Designation} & 
 \multicolumn{3}{c}{Class} & 
 \multicolumn{4}{c}{Masses (kg)} & 
 \multicolumn{5}{c}{Diameter (km)} & 
 \multicolumn{4}{c}{Density} & 
 \multicolumn{2}{c}{Porosity} & 
 Rank \\ 
 \multicolumn{1}{c}{\#} & \multicolumn{1}{c}{Name} & 
 Dyn. & Tax. & Met. & 
 \multicolumn{1}{c}{$M$} & \multicolumn{1}{c}{$\delta M$} & & Fig. & 
 \multicolumn{2}{c}{$\phi$} & \multicolumn{2}{c}{$\delta \phi$} & Fig. & 
 \multicolumn{2}{c}{$\rho$} & \multicolumn{2}{c}{$\delta \rho$} & 
 \multicolumn{1}{c}{$\mathcal{P}$} & $\delta \mathcal{P}$ \\ 
 \hline
   164 & Eva                  & MBA & X   & CV        & 9.29 & 7.76 & $10^{17}$ &                             &  101 & 77 &    3 & 61 & \ref{fap: diam000164}       &   1 & 68 &   1 & 41 &  39 & 84 &         D \\
   165 & Loreley              & MBA & C   & CM        & 1.91 & 0.19 & $10^{19}$ & \ref{fap: mass000165}       &  164 & 92 &    8 & 14 & \ref{fap: diam000165}       &   8 & 14 &   1 & 46 &   0 & 17 & \ding{55} \\
   168 & Sibylla              & MBA & Ch  & CM        & 3.92 & 1.80 & $10^{18}$ & \ref{fap: mass000168}       &  149 & 06 &    4 & 29 & \ref{fap: diam000168}       &   2 & 26 &   1 & 05 &   0 & 46 &         C \\
   173 & Ino                  & MBA & X   & CV        & 4.79 & 3.11 & $10^{18}$ & \ref{fap: mass000173}       &  160 & 07 &    6 & 04 & \ref{fap: diam000173}       &   2 & 23 &   1 & 47 &  20 & 65 &         D \\
   179 & Klytaemnestra        & MBA & S   & OC        & 2.49 & 1.19 & $10^{17}$ &                             &   75 & 02 &    3 & 21 & \ref{fap: diam000179}       &   1 & 12 &   0 & 55 &  66 & 49 &         C \\
\hline
\end{tabular}
\end{table*}

\begin{table*}[!ht]
\setcounter{table}{0}
\centering
\caption{Continued}
\begin{tabular}{r@{ }l c@{\hspace{.75em}}l@{\hspace{.75em}}l c@{\,$\pm$\,}c@{\,$\times$\,}l@{\hspace{.5em}}l r@{.}l@{\,$\pm$\,}r@{.}l@{\hspace{.5em}}l r@{.}l@{\,$\pm$\,}r@{.}l r@{\,$\pm$\,}rc}
 \hline
 \hline
 \multicolumn{2}{c}{Designation} & 
 \multicolumn{3}{c}{Class} & 
 \multicolumn{4}{c}{Masses (kg)} & 
 \multicolumn{5}{c}{Diameter (km)} & 
 \multicolumn{4}{c}{Density} & 
 \multicolumn{2}{c}{Porosity} & 
 Rank \\ 
 \multicolumn{1}{c}{\#} & \multicolumn{1}{c}{Name} & 
 Dyn. & Tax. & Met. & 
 \multicolumn{1}{c}{$M$} & \multicolumn{1}{c}{$\delta M$} & & Fig. & 
 \multicolumn{2}{c}{$\phi$} & \multicolumn{2}{c}{$\delta \phi$} & Fig. & 
 \multicolumn{2}{c}{$\rho$} & \multicolumn{2}{c}{$\delta \rho$} & 
 \multicolumn{1}{c}{$\mathcal{P}$} & $\delta \mathcal{P}$ \\ 
 \hline
   410 & Chloris              & MBA & Ch  & CM        & 6.24 & 0.30 & $10^{18}$ & \ref{fap: mass000410}       &  115 & 55 &    8 & 22 & \ref{fap: diam000410}       &   7 & 72 &   1 & 69 &   0 & 21 & \ding{55} \\
   416 & Vaticana             & MBA & S   & OC        & 3.27 & 3.10 & $10^{18}$ &                             &   87 & 10 &    2 & 57 & \ref{fap: diam000416}       &   9 & 44 &   8 & 99 &   0 & 95 &         D \\
   419 & Aurelia              & MBA & Cb  & CM $^{1}$ & 1.72 & 0.34 & $10^{18}$ & \ref{fap: mass000419}       &  124 & 47 &    3 & 08 & \ref{fap: diam000419}       &   1 & 70 &   0 & 35 &  24 & 21 &         C \\
   420 & Bertholda            & MBA & X   & CV        & 1.48 & 0.09 & $10^{19}$ &                             &  141 & 54 &    2 & 08 & \ref{fap: diam000420}       &   9 & 96 &   0 & 75 &   0 &  7 & \ding{55} \\
   423 & Diotima              & MBA & C   & CM        & 6.91 & 1.93 & $10^{18}$ & \ref{fap: mass000423}       &  211 & 64 &   16 & 02 & \ref{fap: diam000423}       &   1 & 39 &   0 & 50 &  38 & 35 &         C \\
\hline
\end{tabular}
\end{table*}

\begin{table*}[!ht]
\setcounter{table}{0}
\centering
\caption{Continued}
\begin{tabular}{r@{ }l c@{\hspace{.75em}}l@{\hspace{.75em}}l c@{\,$\pm$\,}c@{\,$\times$\,}l@{\hspace{.5em}}l r@{.}l@{\,$\pm$\,}r@{.}l@{\hspace{.5em}}l r@{.}l@{\,$\pm$\,}r@{.}l r@{\,$\pm$\,}rc}
 \hline
 \hline
 \multicolumn{2}{c}{Designation} & 
 \multicolumn{3}{c}{Class} & 
 \multicolumn{4}{c}{Masses (kg)} & 
 \multicolumn{5}{c}{Diameter (km)} & 
 \multicolumn{4}{c}{Density} & 
 \multicolumn{2}{c}{Porosity} & 
 Rank \\ 
 \multicolumn{1}{c}{\#} & \multicolumn{1}{c}{Name} & 
 Dyn. & Tax. & Met. & 
 \multicolumn{1}{c}{$M$} & \multicolumn{1}{c}{$\delta M$} & & Fig. & 
 \multicolumn{2}{c}{$\phi$} & \multicolumn{2}{c}{$\delta \phi$} & Fig. & 
 \multicolumn{2}{c}{$\rho$} & \multicolumn{2}{c}{$\delta \rho$} & 
 \multicolumn{1}{c}{$\mathcal{P}$} & $\delta \mathcal{P}$ \\ 
 \hline
   790 & Pretoria             & MBA & X   & CV        & 4.58 & 0.28 & $10^{18}$ & \ref{fap: mass000790}       &  160 & 98 &   11 & 16 & \ref{fap: diam000790}       &   2 & 09 &   0 & 45 &  24 & 21 &         C \\
   804 & Hispania             & MBA & C   & CM        & 5.00 & 1.78 & $10^{18}$ & \ref{fap: mass000804}       &  148 & 25 &    4 & 08 & \ref{fap: diam000804}       &   2 & 93 &   1 & 06 &   0 & 36 &         C \\
   809 & Lundia               & MBA & V   & HED       & 9.27 & 3.09 & $10^{14}$ &                             &   10 & 26 &    0 & 07 &                             &   1 & 64 &   0 & 10 &  49 &  6 &         B \\
   854 & Frostia              & MBA & V   & HED       & 1.06 & 0.95 & $10^{15}$ &                             &    8 & 39 &    1 & 27 & \ref{fap: diam000854}       &   0 & 88 &   0 & 13 &  72 & 14 &         B \\
   895 & Helio                & MBA & B   & CV        & 9.87 & 6.05 & $10^{18}$ & \ref{fap: mass000895}       &  148 & 43 &    5 & 02 & \ref{fap: diam000895}       &   5 & 76 &   3 & 58 &   0 & 62 &         D \\
\hline
\end{tabular}
\end{table*}

\begin{table*}[!ht]
\setcounter{table}{0}
\centering
\caption{Continued}
\begin{tabular}{r@{ }l c@{\hspace{.75em}}l@{\hspace{.75em}}l c@{\,$\pm$\,}c@{\,$\times$\,}l@{\hspace{.5em}}l r@{.}l@{\,$\pm$\,}r@{.}l@{\hspace{.5em}}l r@{.}l@{\,$\pm$\,}r@{.}l r@{\,$\pm$\,}rc}
 \hline
 \hline
 \multicolumn{2}{c}{Designation} & 
 \multicolumn{3}{c}{Class} & 
 \multicolumn{4}{c}{Masses (kg)} & 
 \multicolumn{5}{c}{Diameter (km)} & 
 \multicolumn{4}{c}{Density} & 
 \multicolumn{2}{c}{Porosity} & 
 Rank \\ 
 \multicolumn{1}{c}{\#} & \multicolumn{1}{c}{Name} & 
 Dyn. & Tax. & Met. & 
 \multicolumn{1}{c}{$M$} & \multicolumn{1}{c}{$\delta M$} & & Fig. & 
 \multicolumn{2}{c}{$\phi$} & \multicolumn{2}{c}{$\delta \phi$} & Fig. & 
 \multicolumn{2}{c}{$\rho$} & \multicolumn{2}{c}{$\delta \rho$} & 
 \multicolumn{1}{c}{$\mathcal{P}$} & $\delta \mathcal{P}$ \\ 
 \hline
       & 1999 OJ4             & TNO &     & Ice       & 3.91 & 0.22 & $10^{17}$ &                             &  130 & 00 &   45 & 00 &                             &   0 & 33 &   0 & 35 &  66 & $\infty$ &         E \\
       & 2000 CF105           & TNO &     & Ice       & 1.85 & 0.12 & $10^{17}$ &                             &  188 & 00 &   38 & 00 &                             &   0 & 05 &   0 & 03 &  94 & 60 & \ding{55} \\
       & 2000 QL251           & TNO &     & Ice       & 3.11 & 0.05 & $10^{18}$ &                             &  150 & 00 &   50 & 00 &                             &   1 & 75 &   1 & 76 &   0 & $\infty$ &         E \\
       & 2000 UG11            & NEA &     & CM        & 9.35 & 1.59 & $10^{ 9}$ &                             &    0 & 30 &    0 & 10 &                             &   0 & 66 &   0 & 67 &  70 & $\infty$ &         E \\
       & 2001 QC298           & TNO &     & Ice       & 1.08 & 0.07 & $10^{19}$ &                             &  244 & 00 &   55 & 00 &                             &   1 & 41 &   0 & 96 &   0 & 67 &         D \\
\hline
\end{tabular}
\end{table*}

\begin{table}[!t]
\centering
\caption[Density and porosity of meteorites]{%
  Average bulk density ($\rho$) 
  measured on $\mathcal{N}_s$ sample of $\mathcal{N}_m$ meteorites
  used in Table~\ref{tabSM: dens}:
  Ordinary chondrites (OC: H, L, and LL),
  Carbonaceous chondrites (CC: CI, CM, CR, CO, CV, and CK),
  Enstatites chondrites (EH and EL), 
  Achondrites HED (\ie, average of Howardites, Eucrites, and Diogenites), 
  Stony-Iron (Pallasites, Mesosiderites, and Steinbach), 
  and Iron meteorites (Ataxites and Hexahedrites).
  Terrestrial weathering has a strong effect on the porosity of found OCs
  with respect to fallen OCs \citep{2008-ChEG-68-Consolmagno}. Only
  measurements on falls are therefore used here.
  For the other meteorite classes, both finds and falls are used.
  The density of liquid water of 1.00\,$\pm$\,0.10 is used as a proxy for the
  volatiles that compose icy bodies.
  \textbf{References}:
    (1) \citet{1998-MPS-33-Consolmagno}.
    (2) \citet{2003-MPS-38-Britt}, 
    (3) \citet{2008-ChEG-68-Consolmagno}, 
    (4) \citet{2010-MPS-45-Macke}, and
    (5) \citet{2011-MPS-46-Macke}.
\label{tabSM: met}
}
\begin{tabular}{llccrrc}
 \hline
 \hline
 \multicolumn{2}{c}{Meteorite} & $\rho$ &
 $\mathcal{N}_s$ & $\mathcal{N}_m$ & Refs.\\
 \hline
 Ord. chondrites   & H   & 3.42 $\pm$ 0.18 & 265 & 157 & 2,3 \\
 Ord. chondrites   & L   & 3.36 $\pm$ 0.16 & 277 & 160 & 2,3 \\
 Ord. chondrites   & LL  & 3.22 $\pm$ 0.22 & 149 &  39 & 2,3 \\
 \noalign{\smallskip}
 Carb. Chondrites  & CI  & 1.60 $\pm$ 0.03 &  14 &   4 & 2,3 \\
 Carb. Chondrites  & CM  & 2.25 $\pm$ 0.08 &  33 &  18 & 2,3 \\
 Carb. Chondrites  & CR  & 3.10            &   7 &   3 & 2 \\
 Carb. Chondrites  & CO  & 3.03 $\pm$ 0.19 &  22 &   8 & 2,3 \\
 Carb. Chondrites  & CV  & 2.79 $\pm$ 0.06 &  51 &  10 & 2,3 \\
 Carb. Chondrites  & CK  & 2.85 $\pm$ 0.08 &   3 &   3 & 3 \\
 \noalign{\smallskip}
 Enstatites        & EH  & 3.47 $\pm$ 0.21 &  16 &   9 & 4 \\
 Enstatites        & EL  & 3.46 $\pm$ 0.32 &  25 &  14 & 4 \\
 \noalign{\smallskip}
 Achondrites       & HED & 3.25 $\pm$ 0.26 &  96 &  56 & 5 \\
 \noalign{\smallskip}
 Stony-Iron        & Pal & 4.76 $\pm$ 0.10 &  10 &   5 & 2 \\
 Stony-Iron        & Mes & 4.35 $\pm$ 0.02 &   8 &   3 & 2 \\
 Stony-Iron        & Ste & 4.18 $\pm$ 0.10 &   2 &   1 & 2 \\
 \noalign{\smallskip}
 Iron              & Ata & 4.01 $\pm$ 0.04 &  1 &   1 & 1 \\
 Iron              & Hex & 7.37 $\pm$ 0.14 &  2 &   2 & 1 \\
 Iron              & Oct & 7.14 $\pm$ 0.13 &  5 &   5 & 1 \\
 \hline
\end{tabular}
\end{table}

\section{Density and macroporosity of small bodies\label{sec: disc}}
%
%
\begin{table*}[!ht]
\centering
\caption[Density of taxonomic classes]{
  Average density $\rho_i$ for each asteroid taxonomic type
  \citep{2009-Icarus-202-DeMeo}, based on 
  $\mathcal{N}_i$ estimates. The $i$ indices stand for the level of
  accuracy considered: more accurate than 20\%, 50\%, and no restriction on
  precision ($\infty$).
  For each class, the associated meteorite (Met., see
  Table~\ref{tabSM: met}) and number of asteroids observed by
  \citeauthor{2009-Icarus-202-DeMeo} with the corresponding fraction
  represented by the class are reported. 
  The average density for transneptunian objects and comets are also reported.
\label{tab: taxodens}}
\begin{tabular}{lc rr rc rc rc}
 \hline
 \hline
 Type & Met. & \multicolumn{2}{c}{Taxonomy} & \multicolumn{6}{c}{Average density for each class}\\
 && (\#) & (\%) &
     $\mathcal{N}_{\infty}$ & $\rho_{\infty}$ & 
     $\mathcal{N}_{50}$ & $\rho_{50}$ & 
     $\mathcal{N}_{20}$ & $\rho_{20}$ \\
\hline
 S   & OC  & 144 & 38  &  50 & 2.66 $\pm$ 1.29  &  28 & 2.70 $\pm$ 0.69  &  11 & 2.72 $\pm$ 0.54  \\ 
 Sa  & OC  &   2 &$<$1 &   1 & 1.07 $\pm$ 0.25  &   1 & 1.07 $\pm$ 0.25  & &\multicolumn{1}{c}{--}\\ 
 Sq  & OC  &  29 &  7  &   5 & 2.78 $\pm$ 0.85  &   4 & 2.78 $\pm$ 0.81  &   2 & 3.43 $\pm$ 0.20  \\ 
 Sr  & OC  &  22 &  5  & &\multicolumn{1}{c}{--}& &\multicolumn{1}{c}{--}& &\multicolumn{1}{c}{--}\\ 
 Sv  & OC  &   2 &$<$1 & &\multicolumn{1}{c}{--}& &\multicolumn{1}{c}{--}& &\multicolumn{1}{c}{--}\\ 
 \noalign{\smallskip}
 B   & CV  &   4 &  1  &  10 & 2.19 $\pm$ 1.00  &   4 & 2.15 $\pm$ 0.74  &   2 & 2.38 $\pm$ 0.45  \\ 
 C   & CM  &  13 &  3  &  33 & 1.57 $\pm$ 1.38  &  19 & 1.41 $\pm$ 0.69  &   5 & 1.33 $\pm$ 0.58  \\ 
 Cb  & CM  &   3 &$<$1 &  13 & 1.88 $\pm$ 2.09  &   6 & 1.43 $\pm$ 0.74  &   3 & 1.25 $\pm$ 0.21  \\ 
 Cg  & CM  &   1 &$<$1 &   1 & 0.96 $\pm$ 0.27  &   1 & 0.96 $\pm$ 0.27  & &\multicolumn{1}{c}{--}\\ 
 Cgh & CM  &  10 &  2  &   5 & 2.64 $\pm$ 1.35  &   1 & 3.48 $\pm$ 1.06  & &\multicolumn{1}{c}{--}\\ 
 Ch  & CM  &  18 &  4  &  47 & 1.96 $\pm$ 1.65  &  27 & 1.70 $\pm$ 1.10  &   9 & 1.41 $\pm$ 0.29  \\ 
 \noalign{\smallskip}
 X   & CV  &   4 &  1  &  26 & 2.87 $\pm$ 2.59  &  15 & 1.99 $\pm$ 0.99  &   8 & 1.85 $\pm$ 0.81  \\ 
 Xc  & Mes &   3 &$<$1 &   9 & 4.96 $\pm$ 2.39  &   3 & 4.63 $\pm$ 0.76  &   2 & 4.86 $\pm$ 0.81  \\ 
 Xe  & EH  &   7 &  1  &   4 & 2.94 $\pm$ 0.85  &   2 & 2.91 $\pm$ 0.65  &   1 & 2.60 $\pm$ 0.20  \\ 
 Xk  & Mes &  18 &  4  &  13 & 3.85 $\pm$ 1.27  &   9 & 3.79 $\pm$ 1.18  &   3 & 4.22 $\pm$ 0.65  \\ 
 \noalign{\smallskip}
 D   & CM  &  16 &  4  &   3 & 9.56 $\pm$ 0.22  & &\multicolumn{1}{c}{--}& &\multicolumn{1}{c}{--}\\ 
 K   & CV  &  16 &  4  &   2 & 4.25 $\pm$ 2.03  &   1 & 3.54 $\pm$ 0.21  &   1 & 3.54 $\pm$ 0.21  \\ 
 L   & CO  &  22 &  5  &   4 & 3.24 $\pm$ 1.03  &   3 & 3.22 $\pm$ 0.97  & &\multicolumn{1}{c}{--}\\ 
 T   & Ata &   4 &  1  &   1 & 2.61 $\pm$ 2.54  & &\multicolumn{1}{c}{--}& &\multicolumn{1}{c}{--}\\ 
 \noalign{\smallskip}
 A   & Pal &   6 &  1  &   1 & 3.73 $\pm$ 1.40  &   1 & 3.73 $\pm$ 1.40  & &\multicolumn{1}{c}{--}\\ 
 O   & OC  &   1 &$<$1 & &\multicolumn{1}{c}{--}& &\multicolumn{1}{c}{--}& &\multicolumn{1}{c}{--}\\ 
 Q   & OC  &   8 &  2  & &\multicolumn{1}{c}{--}& &\multicolumn{1}{c}{--}& &\multicolumn{1}{c}{--}\\ 
 R   & OC  &   1 &$<$1 &   1 & 2.23 $\pm$ 1.02  &   1 & 2.23 $\pm$ 1.02  & &\multicolumn{1}{c}{--}\\ 
 V   & HED &  17 &  4  &   3 & 1.93 $\pm$ 1.07  &   3 & 1.93 $\pm$ 1.07  &   3 & 1.93 $\pm$ 1.07  \\ 
\hline
\multicolumn{4}{c}{Transneptunian objects}&22&0.77 $\pm$ 0.80  &  10 & 1.06 $\pm$ 0.80  &   6 & 1.06 $\pm$ 0.75  \\ 
\multicolumn{4}{c}{Comets}&12&0.47 $\pm$ 0.25  &   4 & 0.56 $\pm$ 0.14  &   3 & 0.54 $\pm$ 0.09  \\ 
\hline
\end{tabular}
\end{table*}
%
%
\begin{figure}
  \includegraphics[width=.49\textwidth]{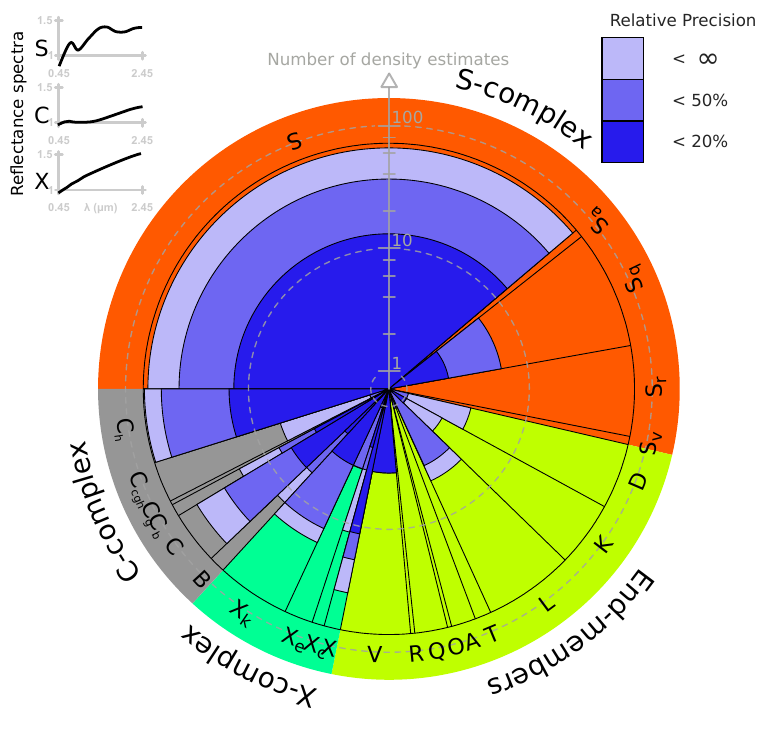}
  \caption[Bus-DeMeo taxonomy and number of density estimates]{%
    Pie chart showing the fraction of asteroids within each class of
    the taxonomy by \citet{2009-Icarus-202-DeMeo}, based on 371
    objects. Complexes (C, S, and X) and end-members are displayed in 
    gray, red, green and yellow respectively. Typical
    reflectance spectra of the complexes are also reported (top left).
    For each class, the number of density estimates, with a relative
    precision better than 20\%, 50\%, and regardless to the precision
    ($\infty$), are drawn in blue wedges.
    \label{fig: taxo}
  }
\end{figure}
%
%
%
%
%
%
  \indent The density and macroporosity of small bodies and their
  relationships with asteroid taxonomy, dynamical class, and diameter
  are discussed here. \\
  \indent For asteroids, the distribution of density estimates over 
  taxonomic classes is presented in Fig.~\ref{fig: taxo}.
  The taxonomy is based on a limited sample
  \citep[371 objects, see][]{2009-Icarus-202-DeMeo} and
  the relative part represented by each class in the whole population 
  may be substantially different \citep{1999-PhD-Bus} but this discussion
  is beyond the scope of present analysis.
  Density estimates are available for the three complexes: 
  \numb{109} for C complex, 
  and \numb{50} for both S and X complexes.
  End-members are less represented: only \numb{15} density estimates
  are available, although end-members represent about 20\% of the
  asteroids. For density estimates with relative accuracy better 20\%
  only, the statistic is however based on low-numbers (see
  Table~\ref{tab: taxodens}). The situation is particularly dramatic
  for end-members: only K-type and V-types have reliable estimates. 
  The number of density estimates for comets and TNOs also drops
  with increasing levels of relative precision
  (Table~\ref{tab: taxodens}). \\
  \indent The density estimates are plotted in Fig.~\ref{fig: dens}, 
  regrouped into 6 categories:
  TNOs, comets, and four asteroid groups:
  S, C, and X complexes, and end-members. 
  Macroporosity estimates (Eq.~\ref{eq: poro}) 
  are similarly plotted in Fig.~\ref{fig: poro}. 
  Several trends can be observed:%
  \begin{itemize}
        \vspace*{-.5em}
        \setlength{\itemsep}{0em}
        \setlength{\parskip}{.25em}
    \item[$\bullet$] Asteroids in the S-complex are more dense than
      those in the C-complex
      \citep[confirming][findings]{2002-AsteroidsIII-4.2-Britt}.
    \item[$\bullet$] Asteroids in the C-complex seem to have
      larger macroporosity than those in the S-complex.
    \item[$\bullet$] The density of asteroids from both the S-complex and
      the C-complex seems to increase with the mass, apparently
      resulting from a decreasing macroporosity.
    \item[$\bullet$] In both C and S-complex, NEAs seem to have a lower
      density than MBAs, following the trend between mass and density
      observed for MBAs.
    \item[$\bullet$] At comparable sizes, 
      B-types appear significantly denser
      ($\rho$\,$\sim$\,2.4) than the other types of the C-complex
      that gather around $\rho$\,$\sim$\,1.4. 

    \item[$\bullet$] The density of the X-complex asteroids covers a large
      range, from the most dense Xc-types with $\rho$\,$\sim$\,4.9
      to X-types with $\rho$\,$\sim$\,1.8.
    \item[$\bullet$] Comets have very low densities
      ($\rho$\,$\sim$\,0.5), low even 
      considering their volatile-rich composition
      \citep[in agreement with spacecraft observations,
        see][]{2007-Icarus-190-Richardson}.
    \item[$\bullet$] The density of TNOs covers a large range, from 
      comet-like ($\rho$\,$\sim$\,0.5)
      to the rocky (50\,000) Quaoar ($\rho$\,$\sim$\,3.6). 
    \item[$\bullet$] Dwarf-planets apparently have no macroporosity,
      contrary to small bodies whose masses are inferior to
      $\approx 10^{20}$\,kg.
    \item[$\bullet$] For each type of small body, the dispersion
      in density and macroporosity is huge. 
  \end{itemize}

  \indent These trends are discussed below.
  The large dispersion of values is however attributed to
  observational and methodological biases, rather than to genuine
  physical effects. Indeed, when considering different levels of
  accuracy, the distributions narrow with precision.
  In other words, biased estimates artificially spread the density
  distribution, hence the need for \textsl{realistic} evaluation of
  uncertainties.

%
\begin{figure*}[!t]
\centering
  \includegraphics[width=.9\textwidth]{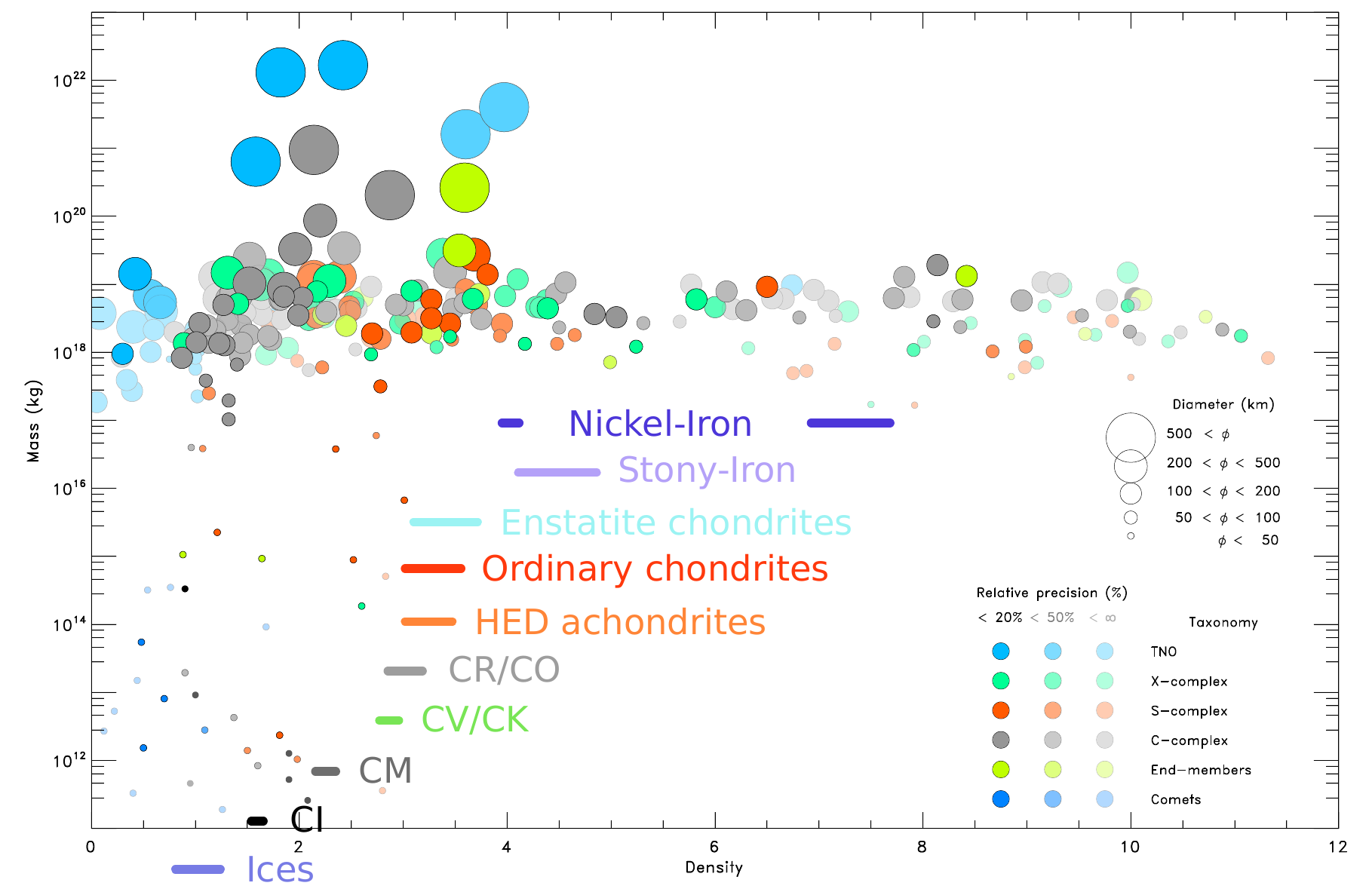}
  \caption[Density vs. Mass]{Density vs. Mass.
    Small bodies are divided into 6 categories:
    TNOs (light blue), 
    comets (blue),
    and asteroids (all dynamic class together) divided into four
    taxonomic groups:
    S-complex in red,
    C-complex in grey, 
    X-complex in green, and
    end-members in yellow (similar to Fig.~\ref{fig: taxo}).
    Asteroids which taxonomy is unknown are plotted in black.
    The size of the symbols is a function of the object diameters,
    and the three different levels of contrast correspond to
    three cuts of relative accuracy: $<$\,20\%, $<$\,50\%, and
    regardless to the precision ($<$\,$\infty$).
    The density of the different class of meteorites is also
    drawn, at arbitrary masses (Table~\ref{tabSM: met}).
    \label{fig: dens}
  }
\end{figure*}

\begin{figure*}[!ht]
\centering
  \includegraphics[width=.9\textwidth]{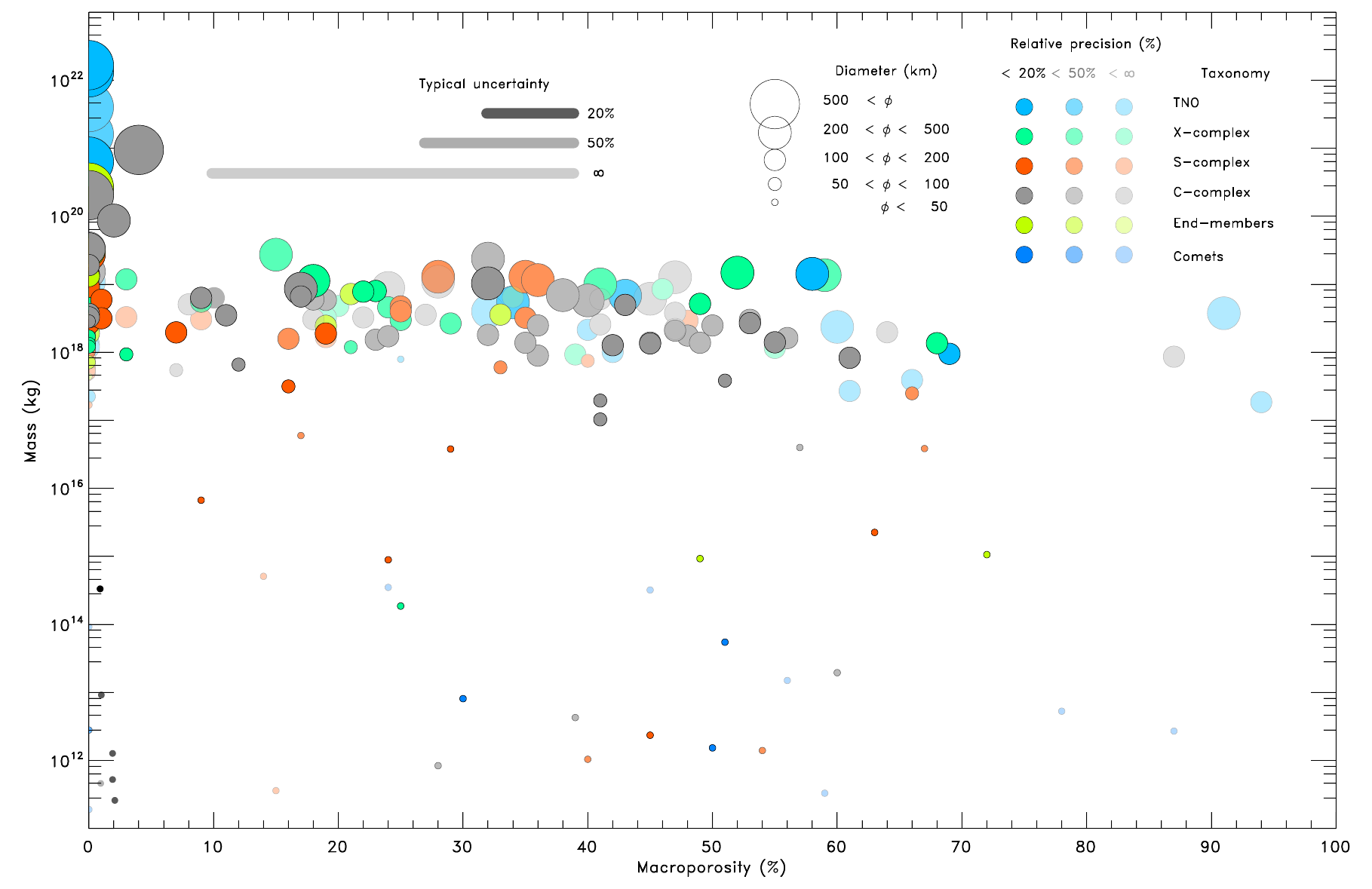}
  \caption[Macroporosity vs. Mass]{Macroporosity vs. Mass.
    The color and size of the symbols are similar to
    Fig.~\ref{fig: dens}.
    Macroporosity is obtained from Eq.~\ref{eq: poro} and the 
    asteroid-meteorite links listed in~\ref{tab: taxodens}.
    The typical uncertainty in macroporosity for three precision level
    on density are displayed (20\%, 50\%, and regardless to the
    precision: $\infty$).
    Additionally, an erroneous asteroid-meteorite link can shift any
    value by 30--40\%.
    \label{fig: poro}
  }
\end{figure*}

  \subsection{C-complex and sub-groups\label{ssec: ctype}}
    \indent Most of the asteroids in the C-complex have densities
    ranging from the highly porous
    (253) Mathilde ($\rho$\,$\sim$\,1.3) to the dense
    (2) Pallas ($\rho$\,$\sim$\,2.9).
    This interval overlaps with CCs meteorites, and the structure of 
    these asteroids ranges from large, compact, bodies
    ($\mathcal{P}$\,$\sim$\,0\%) to rubble-piles 
    ($\mathcal{P}$\,$\sim$\,40--60\%).
    This trend for large bodies to present a zero macroporosity can be
    explained by the high pressure of their interiors. 
    Following \citet[][and references
      therein]{2002-AsteroidsIII-4.2-Britt}, silicate grains
    start to fracture when the pressure reaches $\sim$10$^7$\,Pa.
    This threshold is reached within the first few
    kilometers from the surface of large bodies,
    allowing a thin layer only to host macroporosity. 
    Because large-scale \textsl{grains} (\ie, rubble) are expected to
    grind at much smaller pressures, the transition from compact
    to fractured bodies is expected to be smooth. \\
    \indent Indeed, these different structures are apparently
    correlated with the mass of the asteroids (Fig.~\ref{fig: Vrho}).
    The correlation coefficient between density and diameter is
    \numb{68}\% and this trend seems real although the sample is still
    size-limited.
    From this trend (the linear regression in Fig.~\ref{fig: Vrho}),
    the mass of hypothetical asteroids made of each type of CCs
    meteorites, without macroporosity, are all within
    $10^{19}$--$10^{20}$\,kg, corresponding to the observed transition
    between compact and fractured asteroids.
    This suggests that large C-complex asteroids ($\phi$\,$\geq$\,300\,km)
    have intact structures, while smaller asteroids have porous
    interiors because the internal pressure never reaches the threshold
    for silicate compaction.
    This is consistent with the current vision of the dynamical history
    of the Main Belt:
    large asteroids survived intact throughout the history of the Solar System,
    while most of the material was removed or grinded into pieces
    \citep{2009-Icarus-204-Morbidelli}. 
    This is also supported by the apparent lower density of
    about \numb{1.2} for the \numb{7} NEAs, with respect to 
    about \numb{2} for the \numb{53} MBAs. \\
    \indent Among the C-complex, B-types have distinct surface properties:
    negative spectral slope in the visible and higher albedo
    \citep[see the compilation of albedo per taxonomic class from][]{
      2010-AJ-140-Ryan, 2011-PASJ-63-Usui, 2011-ApJ-741-Masiero}. 
    From a comprehensive comparison of 22 C-complex asteroids with
    laboratory spectra of meteorites, 
    \citet{2010-JGR-115-Clark} indeed found that 
    spectra of C-types were best matched by 
    aqueous-altered CI/CM carbonaceous chondrites
    while those of B-types by other CCs sub-groups
    (mainly CO, CV, but also CK and CR).
    This is supported by the density estimates
    (Table~\ref{tab: taxodens}):
    B-types are significantly denser
    ($\rho$\,$\sim$\,2.4) than the other 
    types of the C-complex, following the trend observed in
    meteorites. 
    Although only two B-types have density estimates more accurate
    than 20\%, (2) Pallas and (704) Interamnia, this trend of a larger
    density is constantly found at different levels of precision
    (Table~\ref{tab: taxodens}) and
    diameters (Table~\ref{tabSM: dens}). 
    B-types are thus intrinsically more dense than the other C-types,
    independently from the mass-density trend observed among C-complex
    asteroids (see above).
    Therefore, in addition to albedo and reflectance spectra that
    point toward different surface properties/composition, 
    density suggests that there are fundamental differences in the
    composition and internal structures of B-types.
    The recent recovery of the Almahata Sitta meteorite, originating
    from the impact of asteroid 2008 TC$_3$ on Earth in October 2008,
    indeed indicated that B-type could be associated with unusual
    Ureilite achondrites \citep{2009-Nature-458-Jenniskens}.
    Based on a comparison of the densities of (1) Ceres
    and (2) Pallas 
    (used as archetypes for the definition of C and B taxonomic classes), 
    \citet{2010-Icarus-205-Carry-a} had suggested that B-types were
    less hydrated than C-types; a hypothesis supported by the lack of
    signature of organic or icy material in their spectra
    \citep{1990-Icarus-88-Jones}. \\
    \indent Finally, the three D-types have density estimates around
    9. These estimates were discarded from the analysis, as their
    uncertainty range does not overlap with meteorites, even the
    highly dense iron hexahedrites (Table~\ref{tabSM: met}).

  \subsection{S-complex and related end-members\label{ssec: stype}}
    \indent The density of S-complex asteroids is distributed in a
    narrow interval (about 2 to 3), slightly below the density of their
    associated meteorites, the ordinary chondrites.
    The resulting macroporosity is generally smaller than 30\%, \ie, 
    these asteroids may present cracks and fractures but are
    still coherent (not rubble-piles).
    This highlights intrinsic differences with the
    C-complex.
    The higher density is revelatory of the difference in composition: 
    \textsl{basaltic} ordinary chondrites vs.
    \textsl{primitive} CI/CM carbonaceous chondrites. 
    The lower macroporosity suggests a difference in formation and
    response to shocks. 
    S-complex asteroids are made of \textsl{igneous} rocks, \ie, they
    experienced a stage of high temperatures and were partly or
    entirely melted. 
    If S-types acquired some cohesion in the process, 
    subsequent impacts would have either not enough energy to overpass 
    this cohesion barrier, leaving them with cracks and fractures
    only, or enough energy to break their structure 
    and destroy them \citep[``battered to
      bits'':][]{1996-MPS-31-Burbine}.
    The current S-complex asteroids would therefore be the few
    remnants of an originally much larger population
    \citep{2009-Icarus-204-Morbidelli}. \\ 
    \indent There are only four density estimates of asteroids
    belonging to the A and V classes.
    The only A-type, (354) Eleonora, has a density of 
    3.7\,$\pm$\,1.4, much higher than S-types.
    This value is in agreement with the density of terrestrial
    olivines and stony-iron Pallasites meteorites
    (Sect.~\ref{sec: intaxo} and Table~\ref{tabSM: met}),
    although the rough relative accuracy allows a wide range of
    possibilities. 
    Average density of the three V-types is surprisingly low:
    $\approx$1.9. A close inspection however reveals that 
    (4) Vesta has a high density of 3.6 while the two
    10\,km-sized (809) Lundia and (854) Frostia have low densities of 
    1.6 and 0.9 respectively.
    These density measurement are hardly comparable.
    Vesta is a differentiated asteroid
    with a pyroxene-rich crust, analog to the HED meteorites, 
    and a denser olivine-rich mantle
    \citep[\eg,][]{1970-Science-168-McCord, 1997-Icarus-128-Binzel}. 
    The low density of Lundia and Frostia implies a high
    macroporosity, above 50\%, in the rubble-pile regime.
    Owing to their small size, they are the product of the collisional
    disruption of a larger parent body, and such a porous structure is
    not so surprising.

  \subsection{X complex, or X melting pot?\label{ssec: xtype}}
    \indent The large spread in density and macroporosity of asteroids
    in the X-complex does not reduce with increasing levels of
    accuracy, contrary to the other groups of small bodies.
    This suggests that multiple compositions are present in
    the complex.
    This is supported by the many different proposed analog meteorites
    (see Sect.~\ref{sec: intaxo}) and wider distribution of
    albedo with respect to C and S complexes
    \citep[Fig.~9 by][]{2010-AJ-140-Ryan}.
    The current definition of the X-complex
    \citep{2009-Icarus-202-DeMeo} indeed encompass the former 
    E, M, and P groups that were 
    distinguished owing to their albedo
    \citep[][]{1989-AsteroidsII-Tholen}. \\
    \indent Both Xc and Xk class have densities above 4, in the range
    of stony-iron and iron meteorites (Table~\ref{tabSM: met}).
    The density of X-types and Xe-types is
    lower, at about 1.8 and 2.6 respectively, closer to the proposed 
    CV carbonaceous chondrites and enstatite chondrites meteorites.
    These asteroids have been grouped together in the taxonomy by
    \citet{2009-Icarus-202-DeMeo} owing to    
    their spectra similarity. 
    Given the low-contrast of their reflectance spectra, however, this
    grouping may be artificial.
    Many different compositions are likely to be represented among
    the X-complex. Further understanding and classification of these
    asteroids will benefit using a larger wavelength range
    \citep[\eg,][]{2011-Icarus-216-Vernazza} and albedo
    \citep[\eg,][]{2010-Icarus-210-Ockert-Bell,
      2011-Icarus-214-Fornasier}.

%
%
%

%
\begin{figure}{!t}
\centering
  \includegraphics[width=.45\textwidth]{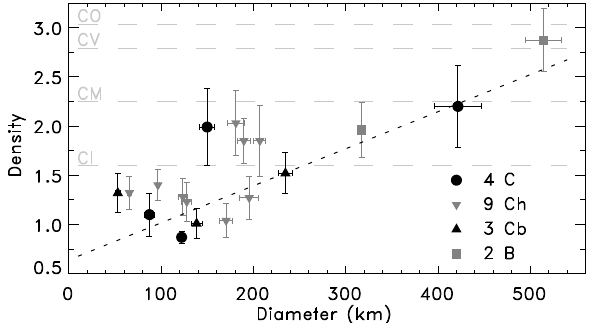}\\
  \includegraphics[width=.45\textwidth]{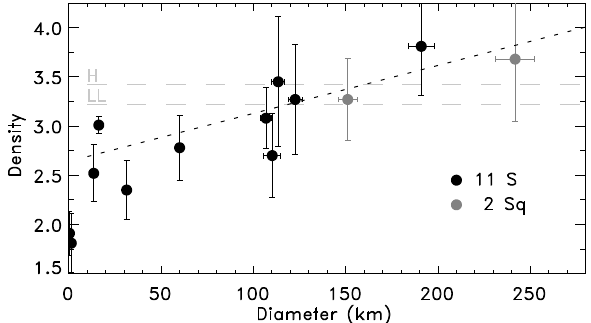}
  \caption[Density vs. Diameter]{%
    Density \textsl{vs.} diameter.
    \textbf{Top:} The \numb{18} asteroids in the C-complex
    (20\% relative precision only, without Ceres). 
    Average density for the CI, CM, CV, and CV carbonaceous chondrites
    are also reported (light grey horizontal dashed lines).
    The oblique dotted line is a linear regression on this sample with
    a correlation coefficient of \numb{68}\%.
    \textbf{Bottom:} The \numb{13} asteroids in the S-complex
    (20\% relative precision only). 
    Average density for the LL and H ordinary chondrites
    are also reported (light grey horizontal dashed lines).
    The oblique dotted line is a linear regression on this sample with
    a correlation coefficient of \numb{51}\%.
    \label{fig: Vrho}
  }
\end{figure}

  \subsection{Dwarf planets and small bodies\label{ssec: size}}
    \indent There are only 8 small bodies
    more massive than 10$^{20}$\,kg:
    Ceres, Pallas, Vesta, Quaoar, Orcus, Pluto, Haumea,
    and Eris.
    These objects have diameters larger than 500\,km and can be
    considered dwarf planets. Their density is high, between 2
    and 4, above that of their analog meteorites.
    This population particularly stands out in Fig.~\ref{fig: poro}, 
    where the dwarf planets ($M$\,$\geq$\,10$^{20}$\,kg) are all
    packed near the $\mathcal{P}$\,$\approx$\,0 axis, and the other
    small bodies below 10$^{20}$\,kg are spread over the entire graph.
    This suggests that these bodies are differentiated, with the
    presence of higher density material below the surface, 
    \eg, silicate or iron cores
    \citep{2010-ApJ-714-Fraser, 2010-Icarus-205-Castillo-Rogez}. \\
    \indent The majority (\numb{75}\%) of the small bodies in the
    sample compiled here are main-belt and Trojan asteroids
    with masses between 10$^{17}$ and 10$^{20}$\,kg.
    These asteroids have diameters between 50 and 400\,km, 
    densities between \numb{0.9} and \numb{5.8}, and macroporosities up to 70\%, 
    from the highly porous (90) Antiope to the very compact
    asteroid (46) Hestia. 
    The pressure inside an object with a mass lower
    than $\approx$10$^{20}$\,kg never reaches 10$^7$\,Pa
    \citep[the threshold for silicate grain compaction, see
      Sect.~\ref{ssec: ctype}, Fig.~\ref{fig: Vrho},
      and][]{2008-ChEG-68-Consolmagno}.
    These high levels of macroporosity are therefore not unexpected. 
    The broad range of densities is more surprising.
    It is partly due to different compositions of these objects, 
    but also to the often large biases affecting the density
    estimates. 
    Indeed, the fraction of asteroids with densities lower
    than 4, decreases from \numb{56}\% to \numb{16}\% by considering the
    estimates more precise than 20\% only.
    Said differently, most of the small bodies with a density
    larger than $\approx$4 suffer from low-precision estimates with
    underestimated volume and/or overestimated mass. \\
    \indent This is supported by the distribution of density and
    macroporosity among NEAs ($M$\,$\leq$\,$10^{17}$\,kg).
    If the macroporosity of NEA also spans a similar range up to 70\%,
    the most dense NEA is (433) Eros, with $\rho$\,$\sim$\,3
    \textsl{only}.
    By opposition to MBAs, all the mass estimates available for NEA
    were derived from a spacecraft encounter or from the orbit of a
    satellite, the most-precise techniques
    (Fig.~\ref{fig: statM}).
    The accuracy on their density is therefore limited by the relative
    precision of their volume-equivalent diameters, which is generally
    less affected by biases (Sect.~\ref{sec: compil}).
    The distribution of density among NEAs may therefore be more
    representative of the \textsl{real} density distribution than
    what we now observe for main-belt asteroids. 
    Because NEAs only represent \numb{6}\% of the sample presented
    here, strong efforts to improve the mass estimates of MBAs must
    be undertaken.

  \subsection{Transneptunian objects\label{ssec: tnos}}
    \indent This population includes a wide range of sizes, from
    dwarf-planets such as Pluto with
    diameters above 2\,000\,km
    down to small bodies of a few tens of kilometers.
    All of the \numb{28} TNOs listed in Table~\ref{tabSM: dens}
    have satellites and the main source of uncertainty is the
    precision on volume estimates, similarly to NEAs.
    The situation is however worse for TNOs.
    Indeed, volume estimates from 
    thermal radiometry \citep[\eg,][]{2010-AA-518-Lellouch}, 
    stellar occultation \citep[\eg,][]{2011-Nature-478-Sicardy} or 
    direct imaging \citep[\eg,][]{2010-ApJ-714-Fraser}
    are available for the few larger TNOs only.
    The diameter of \numb{11} TNOs was roughly estimated from
    their apparent magnitude.
    Given the lack of knowledge on their albedo and the 20\%
    uncertainty affecting albedo estimates
    \citep[see][]{2010-AA-518-Lim},
    only crude diameter estimates can be derived
    (Fig.~\ref{fig: statD}.a).
    The diameters of 
    \numb{7} additional TNOs have been estimated from an
    \textsl{assumed} density of 1.0 
    \citep{2011-ApJ-743-Parker, 2011-arXiv-Sheppard}. 
    Only \numb{10} density estimates were therefore determined from
    \textsl{direct} measurements. Of these, only \numb{5} have a
    relative precision better than 20\%:
    1999 TC$_{36}$, Typhon, Orcus, Pluto, and Eris.\\
    \indent The \numb{5} TNOs larger than $\approx$1\,000\,km have
    densities above 1.5, indicating differentiated interiors as
    described before (Sect.~\ref{ssec: size}).
    On the contrary, the \numb{5} other 100\,km-sized TNOs have
    densities around 0.5, indicative of highly porous structures
    ($\mathcal{P}$\,$\geq$\,50\%).
    The increase in macroporosity for smaller objects is similar to
    that observed for asteroids. 
    Current asteroid and TNO populations are the result of
    collisions over History and such similarities are therefore
    expected.

  \subsection{Comets\label{ssec: comet}}
    \indent The comets are the least massive objects listed here, from
    $10^{14}$ to $10^{17}$\,kg. 
    With a diameter of typically a few hundreds meters to a couple of
    kilometers and a very bright coma with respect to the nucleus
    itself as soon as they are active, observations of comet nuclei
    are very difficult. 
    Current knowledge of the physical properties of comet nuclei is
    therefore still limited \citep{2004-CometsII-4-Lamy}. \\
    \indent The comets have a very low density: \numb{9} of the
    \numb{12} comets listed here have a density below 1.
    The weighted average density of all \numb{12} comets is
    0.47\,$\pm$\,0.25 only, marginally below the limit value of
    0.6 inferred from rotation properties
    \citep[\eg,][]{2004-CometsII-4-Lamy, 2006-MNRAS-373-Snodgrass}.
    The resulting macroporosity is generally high
    ($\mathcal{P}$\,$\geq$\,30--50\%), consistent with our current
    understanding of the structure of a comet nucleus: a highly porous
    assemblage of ices and silicates
    \citep[see][for a review]{2004-CometsII-4-Weissman}. 
    These values of density and macroporosity are consistent with
    those of the small-sized TNOs (Sect.~\ref{ssec: tnos}). This is
    reassuring given that TNOs are thought to be the reservoir of
    Jupiter-family comets \citep{2004-CometsII-7-Jewitt}.

\section{Perspectives\label{sec: persp}}
  \indent Our knowledge on the density and macroporosity of small
  bodies has seen a revolution in the last 10 years, from
  17 objects listed by \citet{2002-AsteroidsIII-4.2-Britt},
  to 40 by \citet{2008-ChEG-68-Consolmagno},
  to \numb{287} here.
  If the sample has increased by about an order of
  magnitude, only a third of the density estimates have a relative
  precision better than 20\%. 
  Improving the accuracy of mass and volume estimates is
  therefore necessary.
  Several lines of investigations are still required to
  improve our understanding of asteroids composition and internal
  structure.

  \subsection{Asteroid-meteorite link\label{ssec: plink}}
    \indent As briefly described in Sect.~\ref{sec: intaxo}, 
    only half of the 24 classes of the asteroid
    taxonomy have mineralogy interpretations
    \citep{2009-Icarus-202-DeMeo}.
    Together with the dynamic of asteroids, it is one of the
    fundamental knowledge required to 
    constrain the models of planetary formation
    \citep[\eg,][]{2005-Nature-435-Morbidelli, 
      2011-Nature-475-Walsh}.
    Efforts to determine the surface properties must be continued.
    Irradiation experiments in the laboratory have allowed to
    understand the space weathering 
    processes on the surfaces of olivines and pyroxenes-rich S-complex
    asteroids \citep[see][among many others]{1996-MPS-31-Chapman,
      2005-Icarus-174-Strazzulla,  
      2006-AA-451-Vernazza, 2009-Nature-458-Vernazza}, including the
    related end-members A and V types 
    \citep{2007-AA-472-Brunetto, 2012-AA-537-Fulvio}.
    The influence of the space weathering on the reflectance spectra
    of most meteorite types is however still unknown, apart from 
    some experiments on enstatite chondrites and mesosiderites
    \citep{2009-Icarus-202-Vernazza}. \\
    \indent Mid-infrared spectroscopy (2--5 and 5--40\,$\mu$m range) will 
    also help refining the mineralogy
    \citep[\eg,][]{2002-AsteroidsIII-2.2-Rivkin,
    2006-Icarus-182-Emery}, providing the regolith packing can be
    reproduced in the laboratory
    \citep[\eg,][]{2010-Icarus-207-Vernazza, 2011-LPI-Vernazza, 
    2011-LPI-King}.
    Albedo measurements should also be used
    \citep{2011-Icarus-214-Fornasier}, although the typical
    uncertainty of about 20\% that can be expected from simple thermal
    models \citep{2010-AA-518-Lim}
    may preclude strong conclusions for the time being.
    Density can also greatly help in that respect.
    The comparison of bulk density resulting from the possible
    composition with the asteroid bulk density may confirm or
    invalidate the composition \citep{2011-Science-334-Sierks}.
    Refining the asteroid-meteorite links will allow to secure the
    macroporosity estimates, hence our knowledge of the interior of
    small bodies.

  \subsection{Accurate mass estimates\label{ssec: pmass}}
    \indent Estimating any mass at all is the limiting factor in
    determining the density of small bodies
    \citep[\eg,][]{2008-ChEG-68-Consolmagno}.
    Furthermore, in most of the cases, the density accuracy is hampered
    by the large uncertainty of mass estimates
    (Sect.~\ref{ssec: diam}).
    Improving the number and accuracy of mass estimates is therefore
    required. \\ 
    \indent The study of binary systems is highly relevant in that
    respect. It is the most productive method to determine accurate
    mass estimates (Fig.~\ref{fig: statM}.c).
    However, only a third of the 200 known binaries have a mass
    estimate (Sect.~\ref{ssec: mass}).
    Most of the binaries were indeed discovered from
    lightcurves, and their angular separation is too small to be
    resolved. 
    Upcoming facilities such as the ALMA interferometer or the E-ELT
    will provide the angular resolution required to resolve these
    systems, and many more accurate mass determinations should be
    available in few years.
    Additional optical and radar imaging observations, together with
    lightcurves of mutual events of known binaries,
    will also help improving the
    current mass estimates \citep[\eg,][]{2008-Icarus-196-Descamps}. \\
    \indent In parallel, the astrometry observations by Gaia will
    provide additional mass determinations. 
    Around 350\,000 small bodies are expected to be observed during the
    5 years mission, with an average of 50 to 60 epochs on each
    \citep{2007-EMP-101-Mignard}.
    The micro-arcsecond precision of Gaia's astrometry will allow to
    refine the accuracy on the orbit of asteroids by several orders of 
    magnitude.
    Such a precision will have a snowball effect on subsequent mass
    estimates from planetary ephemeris and orbit deflections.
    Close encounters between asteroids will also be observed
    during the mission and the mass of about 50
    asteroids with an expected relative precision better than
    10\% will be determined \citep{2007-AA-472-Mouret}.
    Although most of these objects are most likely already
    listed in~\ref{app: mass}, the mass estimates
    are expected to be less affected by biases, owing to 
    the unprecedented completeness of Gaia catalog.
    The number of mass estimates and their level of accuracy is
    therefore expected to improve significantly at the 2020 horizon.

  \subsection{Accurate volume estimates\label{ssec: pvol}}
    \indent As described in Sect.~\ref{sec: compil}, the 
    contributions of the mass and diameter uncertainties to the density
    uncertainty are not even.
    The precision on the diameter is indeed
    the limiting factor of the most accurate density estimates
    (see also Fig.~\ref{fig: prec}).
    Relative precision on the \textsl{volume} below 10--15\% are
    required to take advantage of any mass determination.
    The accuracy on the \textsl{diameter} should therefore be of a few
    percent at most.
    Thanks to improved observing facilities and from improved methods
    of analysis, our understanding of the physical properties
    of asteroids as seen a revolution in
    last decade, making such a goal achievable. \\
    \indent Many different observing techniques and methods of
    analysis can be used to evaluate the diameter
    of small bodies. In particular, multi-data approaches have been
    proven successful in determining the 3-D shape, size, and spin axis of
    small bodies (see Sect.~\ref{ssec: diam}).
    The recent flyby of asteroid (21) Lutetia by the ESA Rosetta
    mission showed that the diameter estimate derived before the flyby
    from optical lightcurves and disk-resolved images 
    was accurate to 2\%
    \citep[using the KOALA 3-D shape modeling algorithm,
      see][]{2011-IPI-5-Kaasalainen, 2010-AA-523-Carry,
      2012-PSS--Carry}. 
    Besides, 3-D shape models offer the possibility to analyze
    thermal radiometry data with more advanced thermal models 
    \citep[\eg,][]{1996-AA-310-Lagerros, 1997-AA-325-Lagerros,
      2005-AA-443-Muller, 2006-AA-447-Mueller, 
      2009-PSS-57-Delbo, 2011-MNRAS-415-Rozitis,
      2012-PSS--Rourke}.
    Such models allow to derive several surface properties such as the
    albedo and thermal inertia. These quantities can in turn be used 
    to help constraining the asteroid-meteorite
    links (Sect.~\ref{ssec: plink}).
    Large observing programs (\eg, lightcurves,
    adaptive-optics disk-resolved images on large telescopes, stellar
    occultation campaigns) to derive 3-D shape models of all the small
    bodies listed in Table~\ref{tabSM: dens} have therefore
    far-reaching implications.

\section{Conclusion\label{sec: conclu}}
  \indent An extensive review of current knowledge on the density and
  macroporosity of small bodies is presented.
  The density estimates of \numb{287} small bodies are presented,
  computed from 
  \numb{994} mass estimates, 
  \numb{1454} volume-equivalent diameter estimates, and
  \numb{24} indirect density estimates.
  All the dynamical classes are represented in the sample:
  \numb{17} near-Earth asteroids, 
  \numb{230} Main-Belt and Trojan asteroids, 
  \numb{12} comets, and
  \numb{28} transneptunian objects.
  The accuracy and biases affecting mass and diameter estimates are
  discussed and best-estimates are strictly selected.
  Bulk densities are computed and compared with meteorite density,
  allowing to estimate the macroporosity.
  Although the sample still suffers from large uncertainties and often
  biases (Sect.~\ref{sec: compil} and~\ref{sec: disc}), several
  trends can be identified:
  \begin{enumerate}
    \item Dwarf-planets apparently have no macroporosity,
      contrary to small bodies which mass is inferior to
      $\approx 10^{20}$\,kg.
    \item Asteroids in the S-complex are more dense than those in the
      C-complex that in turn present a larger macroporosity.
    \item There is a trend of increasing density with mass for asteroids in
      both S and C complexes. This trends is also visible from the
      lower density of NEAs with respect to MBAs.
    \item B-types seem structurally different from other C-complex
      asteroids (albedo, reflectance spectra, density).
    \item The X-complex encompasses many different compositions and
      should be revised using additional data (\eg, albedo).
    \item Comets and TNOs have similar low density and high macroporosity,
      consistent with a structure of porous icy agglomerates.
  \end{enumerate}

  \indent Several lines of investigations to improve the number and
  accuracy of density estimates are discussed.
  The search for binary asteroids and subsequent orbital analysis,
  together with detailed 3-D shape modeling from multi-data inversion
  techniques stand out as key programs.

\section*{Acknowledgments}
  \indent A big thank-you to P.\,Tanga and D.\,Hestroffer
  for poking me about
  coming to the Gaia GREAT meeting in Pisa, without them
  I wouldn't have started this painful (but fruitful!) task of compiling
  masses.
  Dankesch{\"o}n, merci to T.\,M{\" u}ller, F. Marchis and A. Fienga
  for sharing their results ahead of publication.
  Thanks to F.\,DeMeo for constructive discussions.
  Thank you to the two anonymous referees for their constructive
  comments. As a result, the present manuscript includes a significantly
  higher amount of material in the introductory sections.
  Gracias R.\,Soja and E.\,Treguier for all our discussions about
  this topic and for the fun in the office.
  This research made heavy use of NASA's Astrophysics Data System
  Abstract Service (ADS) and Data Archive, thanks to the developers
  and maintainers.

\bibliographystyle{model2-names}
\bibliography{biblio}

\begin{thebibliography}{257}
\expandafter\ifx\csname natexlab\endcsname\relax\def\natexlab#1{#1}\fi
\expandafter\ifx\csname url\endcsname\relax
  \def\url#1{\texttt{#1}}\fi
\expandafter\ifx\csname urlprefix\endcsname\relax\def\urlprefix{URL }\fi
\providecommand{\eprint}[2][]{\url{#2}}
\providecommand{\bibinfo}[2]{#2}
\ifx\xfnm\relax \def\xfnm[#1]{\unskip,\space#1}\fi
\bibitem[{A'Hearn et~al.(2005)A'Hearn, Belton, Delamere, Kissel, Klaasen,
  McFadden, Meech, Melosh, Schultz, Sunshine, Thomas, Veverka, Yeomans, Baca,
  Busko, Crockett, Collins, Desnoyer, Eberhardy, Ernst, Farnham, Feaga,
  Groussin, Hampton, Ipatov, Li, Lindler, Lisse, Mastrodemos, Owen, Richardson,
  Wellnitz and White}]{2005-Science-310-AHearn}
\bibinfo{author}{A'Hearn, M.F.}, \bibinfo{author}{Belton, M.J.S.},
  \bibinfo{author}{Delamere, W.A.}, \bibinfo{author}{Kissel, J.},
  \bibinfo{author}{Klaasen, K.P.}, \bibinfo{author}{McFadden, L.A.},
  \bibinfo{author}{Meech, K.J.}, \bibinfo{author}{Melosh, H.J.},
  \bibinfo{author}{Schultz, P.H.}, \bibinfo{author}{Sunshine, J.M.},
  \bibinfo{author}{Thomas, P.C.}, \bibinfo{author}{Veverka, J.},
  \bibinfo{author}{Yeomans, D.K.}, \bibinfo{author}{Baca, M.W.},
  \bibinfo{author}{Busko, I.}, \bibinfo{author}{Crockett, C.J.},
  \bibinfo{author}{Collins, S.M.}, \bibinfo{author}{Desnoyer, M.},
  \bibinfo{author}{Eberhardy, C.A.}, \bibinfo{author}{Ernst, C.M.},
  \bibinfo{author}{Farnham, T.L.}, \bibinfo{author}{Feaga, L.},
  \bibinfo{author}{Groussin, O.}, \bibinfo{author}{Hampton, D.},
  \bibinfo{author}{Ipatov, S.I.}, \bibinfo{author}{Li, J.Y.},
  \bibinfo{author}{Lindler, D.}, \bibinfo{author}{Lisse, C.M.},
  \bibinfo{author}{Mastrodemos, N.}, \bibinfo{author}{Owen, W.M.},
  \bibinfo{author}{Richardson, J.E.}, \bibinfo{author}{Wellnitz, D.D.},
  \bibinfo{author}{White, R.L.}, \bibinfo{year}{2005}.
\newblock \bibinfo{title}{{Deep Impact: Excavating Comet Tempel 1}}.
\newblock \bibinfo{journal}{Science} \bibinfo{volume}{310},
  \bibinfo{pages}{258--264}.
\bibitem[{Altenhoff et~al.(2004)Altenhoff, Bertoldi and
  Menten}]{2004-AA-415-Altenhoff}
\bibinfo{author}{Altenhoff, W.J.}, \bibinfo{author}{Bertoldi, F.},
  \bibinfo{author}{Menten, K.M.}, \bibinfo{year}{2004}.
\newblock \bibinfo{title}{{Size estimates of some optically bright KBOs}}.
\newblock \bibinfo{journal}{Astronomy and Astrophysics} \bibinfo{volume}{415},
  \bibinfo{pages}{771--775}.
\bibitem[{Archinal et~al.(2011)Archinal, A'Hearn, Bowell, Conrad, Consolmagno,
  Courtin, Fukushima, Hestroffer, Hilton, Krasinsky, Neumann, Oberst,
  Seidelmann, Stooke, Tholen, Thomas and Williams}]{2011-CeMDA-109-Archinal}
\bibinfo{author}{Archinal, B.A.}, \bibinfo{author}{A'Hearn, M.F.},
  \bibinfo{author}{Bowell, E.}, \bibinfo{author}{Conrad, A.},
  \bibinfo{author}{Consolmagno, G.J.}, \bibinfo{author}{Courtin, R.},
  \bibinfo{author}{Fukushima, T.}, \bibinfo{author}{Hestroffer, D.},
  \bibinfo{author}{Hilton, J.L.}, \bibinfo{author}{Krasinsky, G.A.},
  \bibinfo{author}{Neumann, G.}, \bibinfo{author}{Oberst, J.},
  \bibinfo{author}{Seidelmann, P.K.}, \bibinfo{author}{Stooke, P.},
  \bibinfo{author}{Tholen, D.J.}, \bibinfo{author}{Thomas, P.C.},
  \bibinfo{author}{Williams, I.P.}, \bibinfo{year}{2011}.
\newblock \bibinfo{title}{{Report of the IAU Working Group on Cartographic
  Coordinates and Rotational Elements: 2009}}.
\newblock \bibinfo{journal}{Celestial Mechanics and Dynamical Astronomy}
  \bibinfo{volume}{109}, \bibinfo{pages}{101--135}.
\bibitem[{Aslan et~al.(2007)Aslan, Gumerov, Hudkova, Ivantsov, Khamitov and
  Pinigin}]{2007-ASPC-370-Aslan}
\bibinfo{author}{Aslan, Z.}, \bibinfo{author}{Gumerov, R.},
  \bibinfo{author}{Hudkova, L.}, \bibinfo{author}{Ivantsov, A.},
  \bibinfo{author}{Khamitov, I.}, \bibinfo{author}{Pinigin, G.},
  \bibinfo{year}{2007}.
\newblock \bibinfo{title}{{Mass Determination of Small Solar System Bodies with
  Ground-based Observations}}, in: \bibinfo{editor}{{O.~Demircan, S.~O.~Selam,
  \& B.~Albayrak}} (Ed.), \bibinfo{booktitle}{Solar and Stellar Physics Through
  Eclipses}, pp. \bibinfo{pages}{52--58}.
\bibitem[{Baer and Chesley(2008)}]{2008-CeMDA-100-Baer}
\bibinfo{author}{Baer, J.}, \bibinfo{author}{Chesley, S.R.},
  \bibinfo{year}{2008}.
\newblock \bibinfo{title}{{Astrometric masses of 21 asteroids, and an
  integrated asteroid ephemeris}}.
\newblock \bibinfo{journal}{Celestial Mechanics and Dynamical Astronomy}
  \bibinfo{volume}{100}, \bibinfo{pages}{27--42}.
\bibitem[{Baer et~al.(2011)Baer, Chesley and Matson}]{2011-AJ-141-Baer}
\bibinfo{author}{Baer, J.}, \bibinfo{author}{Chesley, S.R.},
  \bibinfo{author}{Matson, R.D.}, \bibinfo{year}{2011}.
\newblock \bibinfo{title}{{Astrometric Masses of 26 Asteroids and Observations
  on Asteroid Porosity}}.
\newblock \bibinfo{journal}{Astronomical Journal} \bibinfo{volume}{141},
  \bibinfo{pages}{143--155}.
\bibitem[{Baer et~al.(2008)Baer, Milani, Chesley and Matson}]{2008-DPS-40-Baer}
\bibinfo{author}{Baer, J.}, \bibinfo{author}{Milani, A.},
  \bibinfo{author}{Chesley, S.R.}, \bibinfo{author}{Matson, R.D.},
  \bibinfo{year}{2008}.
\newblock \bibinfo{title}{{An Observational Error Model, and Application to
  Asteroid Mass Determination}}, in: \bibinfo{booktitle}{Bulletin of the
  American Astronomical Society}, p. \bibinfo{pages}{493}.
\bibitem[{Bange(1998)}]{1998-AA-340-Bange}
\bibinfo{author}{Bange, J.}, \bibinfo{year}{1998}.
\newblock \bibinfo{title}{{An estimation of the mass of asteroid 20-Massalia
  derived from the HIPPARCOS minor planets data}}.
\newblock \bibinfo{journal}{Astronomy and Astrophysics} \bibinfo{volume}{340},
  \bibinfo{pages}{L1--L4}.
\bibitem[{Bange and Bec-Borsenberger(1997)}]{1997-ESAP-402-Bange}
\bibinfo{author}{Bange, J.F.}, \bibinfo{author}{Bec-Borsenberger, A.},
  \bibinfo{year}{1997}.
\newblock \bibinfo{title}{{Determination of the Masses of Minor Planets}}, in:
  \bibinfo{editor}{{R.~M.~Bonnet, E.~H{\o}g, P.~L.~Bernacca, L.~Emiliani,
  A.~Blaauw, C.~Turon, J.~Kovalevsky, L.~Lindegren, H.~Hassan, M.~Bouffard,
  B.~Strim, D.~Heger, M.~A.~C.~Perryman, \& L.~Woltjer}} (Ed.),
  \bibinfo{booktitle}{Hipparcos - Venice '97}, pp. \bibinfo{pages}{169--172}.
\bibitem[{Barucci et~al.(2012)Barucci, Belskaya, Fornasier, Fulchignoni, Clark,
  Coradini, Capaccioni, Dotto, Birlan, Leyrat, Sierks, Thomas and
  Vincent}]{2012-PSS-Barucci}
\bibinfo{author}{Barucci, A.}, \bibinfo{author}{Belskaya, I.},
  \bibinfo{author}{Fornasier, S.}, \bibinfo{author}{Fulchignoni, M.},
  \bibinfo{author}{Clark, B.E.}, \bibinfo{author}{Coradini, A.},
  \bibinfo{author}{Capaccioni, F.}, \bibinfo{author}{Dotto, E.},
  \bibinfo{author}{Birlan, M.}, \bibinfo{author}{Leyrat, C.},
  \bibinfo{author}{Sierks, H.}, \bibinfo{author}{Thomas, N.},
  \bibinfo{author}{Vincent, J.B.}, \bibinfo{year}{2012}.
\newblock \bibinfo{title}{{Overview of Lutetia's surface composition}}.
\newblock \bibinfo{journal}{Planetary and Space Science} .
\bibitem[{Barucci et~al.(2008)Barucci, Brown, Emery and
  Merlin}]{2008-SSBN-3-Barucci}
\bibinfo{author}{Barucci, M.A.}, \bibinfo{author}{Brown, M.E.},
  \bibinfo{author}{Emery, J.P.}, \bibinfo{author}{Merlin, F.},
  \bibinfo{year}{2008}.
\newblock \bibinfo{title}{{Composition and Surface Properties of Transneptunian
  Objects and Centaurs}}.
\newblock \bibinfo{journal}{The Solar System Beyond Neptune} ,
  \bibinfo{pages}{143--160}.
\bibitem[{Barucci et~al.(2005)Barucci, Fulchignoni, Fornasier, Dotto, Vernazza,
  Birlan, Binzel, Carvano, Merlin, Barbieri and Belskaya}]{2005-AA-430-Barucci}
\bibinfo{author}{Barucci, M.A.}, \bibinfo{author}{Fulchignoni, M.},
  \bibinfo{author}{Fornasier, S.}, \bibinfo{author}{Dotto, E.},
  \bibinfo{author}{Vernazza, P.}, \bibinfo{author}{Birlan, M.},
  \bibinfo{author}{Binzel, R.P.}, \bibinfo{author}{Carvano, J.M.},
  \bibinfo{author}{Merlin, F.}, \bibinfo{author}{Barbieri, C.},
  \bibinfo{author}{Belskaya, I.N.}, \bibinfo{year}{2005}.
\newblock \bibinfo{title}{{Asteroid target selection for the new Rosetta
  mission baseline. 21 Lutetia and 2867 Steins}}.
\newblock \bibinfo{journal}{Astronomy and Astrophysics} \bibinfo{volume}{430},
  \bibinfo{pages}{313--317}.
\bibitem[{Behrend et~al.(2006)Behrend, Bernasconi, Roy, Klotz, Colas, Antonini,
  Aoun, Augustesen, Barbotin, Berger, Berrouachdi, Brochard, Cazenave,
  Cavadore, Coloma, Cotrez, Deconihout, Demeautis, Dorseuil, Dubos, Durkee,
  Frappa, Hormuth, Itkonen, Jacques, Kurtze, Laffont, Lavayssi{\`e}re,
  Lecacheux, Leroy, Manzini, Masi, Matter, Michelsen, Nomen, Oksanen,
  P{\"a}{\"a}kk{\"o}nen, Peyrot, Pimentel, Pray, Rinner, Sanchez, Sonnenberg,
  Sposetti, Starkey, Stoss, Teng, Vignand and Waelchli}]{2006-AA-446-Behrend}
\bibinfo{author}{Behrend, R.}, \bibinfo{author}{Bernasconi, L.},
  \bibinfo{author}{Roy, R.}, \bibinfo{author}{Klotz, A.},
  \bibinfo{author}{Colas, F.}, \bibinfo{author}{Antonini, P.},
  \bibinfo{author}{Aoun, R.}, \bibinfo{author}{Augustesen, K.},
  \bibinfo{author}{Barbotin, E.}, \bibinfo{author}{Berger, N.},
  \bibinfo{author}{Berrouachdi, H.}, \bibinfo{author}{Brochard, E.},
  \bibinfo{author}{Cazenave, A.}, \bibinfo{author}{Cavadore, C.},
  \bibinfo{author}{Coloma, J.}, \bibinfo{author}{Cotrez, V.},
  \bibinfo{author}{Deconihout, S.}, \bibinfo{author}{Demeautis, C.},
  \bibinfo{author}{Dorseuil, J.}, \bibinfo{author}{Dubos, G.},
  \bibinfo{author}{Durkee, R.}, \bibinfo{author}{Frappa, E.},
  \bibinfo{author}{Hormuth, F.}, \bibinfo{author}{Itkonen, T.},
  \bibinfo{author}{Jacques, C.}, \bibinfo{author}{Kurtze, L.},
  \bibinfo{author}{Laffont, A.}, \bibinfo{author}{Lavayssi{\`e}re, M.},
  \bibinfo{author}{Lecacheux, J.}, \bibinfo{author}{Leroy, A.},
  \bibinfo{author}{Manzini, F.}, \bibinfo{author}{Masi, G.},
  \bibinfo{author}{Matter, D.}, \bibinfo{author}{Michelsen, R.},
  \bibinfo{author}{Nomen, J.}, \bibinfo{author}{Oksanen, A.},
  \bibinfo{author}{P{\"a}{\"a}kk{\"o}nen, P.}, \bibinfo{author}{Peyrot, A.},
  \bibinfo{author}{Pimentel, E.}, \bibinfo{author}{Pray, D.P.},
  \bibinfo{author}{Rinner, C.}, \bibinfo{author}{Sanchez, S.},
  \bibinfo{author}{Sonnenberg, K.}, \bibinfo{author}{Sposetti, S.},
  \bibinfo{author}{Starkey, D.}, \bibinfo{author}{Stoss, R.},
  \bibinfo{author}{Teng, J.P.}, \bibinfo{author}{Vignand, M.},
  \bibinfo{author}{Waelchli, N.}, \bibinfo{year}{2006}.
\newblock \bibinfo{title}{{Four new binary minor planets: (854) Frostia, (1089)
  Tama, (1313) Berna, (4492) Debussy}}.
\newblock \bibinfo{journal}{Astronomy and Astrophysics} \bibinfo{volume}{446},
  \bibinfo{pages}{1177--1184}.
\bibitem[{Bell et~al.(1989)Bell, Davis, Hartmann and
  Gaffey}]{1989-AsteroidsII-4-Bell}
\bibinfo{author}{Bell, J.F.}, \bibinfo{author}{Davis, D.R.},
  \bibinfo{author}{Hartmann, W.K.}, \bibinfo{author}{Gaffey, M.J.},
  \bibinfo{year}{1989}.
\newblock \bibinfo{title}{{Asteroids - The big picture}}.
\newblock \bibinfo{journal}{Asteroids II} , \bibinfo{pages}{921--945}.
\bibitem[{Benner et~al.(2010)Benner, Margot, Nolan, Giorgini, Brozovic,
  Scheeres, Magri and Ostro}]{2010-DPS-42-Benner}
\bibinfo{author}{Benner, L.A.M.}, \bibinfo{author}{Margot, J.},
  \bibinfo{author}{Nolan, M.C.}, \bibinfo{author}{Giorgini, J.D.},
  \bibinfo{author}{Brozovic, M.}, \bibinfo{author}{Scheeres, D.J.},
  \bibinfo{author}{Magri, C.}, \bibinfo{author}{Ostro, S.J.},
  \bibinfo{year}{2010}.
\newblock \bibinfo{title}{{Radar Imaging and a Physical Model of Binary
  Asteroid 65803 Didymos}}, in: \bibinfo{booktitle}{AAS/Division for Planetary
  Sciences Meeting Abstracts \#42}, p. \bibinfo{pages}{1056}.
\bibitem[{Bertoldi et~al.(2006)Bertoldi, Altenhoff, Weiss, Menten and
  Thum}]{2006-Nature-439-Bertoldi}
\bibinfo{author}{Bertoldi, F.}, \bibinfo{author}{Altenhoff, W.},
  \bibinfo{author}{Weiss, A.}, \bibinfo{author}{Menten, K.M.},
  \bibinfo{author}{Thum, C.}, \bibinfo{year}{2006}.
\newblock \bibinfo{title}{{The trans-neptunian object UB$_{313}$ is larger than
  Pluto}}.
\newblock \bibinfo{journal}{Nature} \bibinfo{volume}{439},
  \bibinfo{pages}{563--564}.
\bibitem[{Binzel et~al.(1997)Binzel, Gaffey, Thomas, Zellner, Storrs and
  Wells}]{1997-Icarus-128-Binzel}
\bibinfo{author}{Binzel, R.P.}, \bibinfo{author}{Gaffey, M.J.},
  \bibinfo{author}{Thomas, P.C.}, \bibinfo{author}{Zellner, B.H.},
  \bibinfo{author}{Storrs, A.D.}, \bibinfo{author}{Wells, E.N.},
  \bibinfo{year}{1997}.
\newblock \bibinfo{title}{{Geologic Mapping of Vesta from 1994 Hubble Space
  Telescope Images}}.
\newblock \bibinfo{journal}{Icarus} \bibinfo{volume}{128},
  \bibinfo{pages}{95--103}.
\bibitem[{Bottke et~al.(2002a)Bottke, Cellino, Paolicchi and
  Binzel}]{2002-AsteroidsIII-1-Bottke}
\bibinfo{author}{Bottke, Jr., W.F.}, \bibinfo{author}{Cellino, A.},
  \bibinfo{author}{Paolicchi, P.}, \bibinfo{author}{Binzel, R.P.},
  \bibinfo{year}{2002}a.
\newblock \bibinfo{title}{{An Overview of the Asteroids: The Asteroids III
  Perspective}}.
\newblock \bibinfo{journal}{Asteroids III} , \bibinfo{pages}{3--15}.
\bibitem[{Bottke et~al.(2006)Bottke, Nesvorn{\'y}, Grimm, Morbidelli and
  O'Brien}]{2006-Nature-439-Bottke}
\bibinfo{author}{Bottke, Jr., W.F.}, \bibinfo{author}{Nesvorn{\'y}, D.},
  \bibinfo{author}{Grimm, R.E.}, \bibinfo{author}{Morbidelli, A.},
  \bibinfo{author}{O'Brien, D.P.}, \bibinfo{year}{2006}.
\newblock \bibinfo{title}{{Iron meteorites as remnants of planetesimals formed
  in the terrestrial planet region}}.
\newblock \bibinfo{journal}{Nature} \bibinfo{volume}{439},
  \bibinfo{pages}{821--824}.
\bibitem[{Bottke et~al.(2002b)Bottke, Vokrouhlick{\'y}, Rubincam and
  Broz}]{2002-AsteroidsIII-4.1-Bottke}
\bibinfo{author}{Bottke, Jr., W.F.}, \bibinfo{author}{Vokrouhlick{\'y}, D.},
  \bibinfo{author}{Rubincam, D.P.}, \bibinfo{author}{Broz, M.},
  \bibinfo{year}{2002}b.
\newblock \bibinfo{title}{{The Effect of Yarkovsky Thermal Forces on the
  Dynamical Evolution of Asteroids and Meteoroids}}.
\newblock \bibinfo{journal}{Asteroids III} , \bibinfo{pages}{395--408}.
\bibitem[{Britt et~al.(1992)Britt, Bell, Haack and Scott}]{1992-Metic-Britt}
\bibinfo{author}{Britt, D.T.}, \bibinfo{author}{Bell, J.F.},
  \bibinfo{author}{Haack, H.}, \bibinfo{author}{Scott, E.R.D.},
  \bibinfo{year}{1992}.
\newblock \bibinfo{title}{{The Reflectance Spectrum of Troilite and the T-Type
  Asteroids}}.
\newblock \bibinfo{journal}{Meteoritics} \bibinfo{volume}{27},
  \bibinfo{pages}{207}.
\bibitem[{Britt and Consolmagno(2003)}]{2003-MPS-38-Britt}
\bibinfo{author}{Britt, D.T.}, \bibinfo{author}{Consolmagno, G.J.},
  \bibinfo{year}{2003}.
\newblock \bibinfo{title}{{Stony meteorite porosities and densities: A review
  of the data through 2001}}.
\newblock \bibinfo{journal}{Meteoritics and Planetary Science}
  \bibinfo{volume}{38}, \bibinfo{pages}{1161--1180}.
\bibitem[{Britt et~al.(2002)Britt, Yeomans, Housen and
  Consolmagno}]{2002-AsteroidsIII-4.2-Britt}
\bibinfo{author}{Britt, D.T.}, \bibinfo{author}{Yeomans, D.K.},
  \bibinfo{author}{Housen, K.R.}, \bibinfo{author}{Consolmagno, G.J.},
  \bibinfo{year}{2002}.
\newblock \bibinfo{title}{{Asteroid Density, Porosity, and Structure}}.
\newblock \bibinfo{journal}{Asteroids III} , \bibinfo{pages}{485--500}.
\bibitem[{Brooks(2006)}]{2006-AAS-209-Brooks}
\bibinfo{author}{Brooks, H.E.}, \bibinfo{year}{2006}.
\newblock \bibinfo{title}{{Orbits of Binary Near-Earth Asteroids from Radar
  Observations}}, in: \bibinfo{booktitle}{AAS/Division for Planetary Sciences
  Meeting Abstracts \#38}, pp. \bibinfo{pages}{934--+}.
\bibitem[{Brown et~al.(2005)Brown, Bouchez, Rabinowitz, Sari, Trujillo, van
  Dam, Campbell, Chin, Hartman, Johansson, Lafon, Le~Mignant, Stomski~Jr.,
  Summers and Wizinowich}]{2005-ApJ-632-Brown}
\bibinfo{author}{Brown, M.E.}, \bibinfo{author}{Bouchez, A.H.},
  \bibinfo{author}{Rabinowitz, D.L.}, \bibinfo{author}{Sari, R.},
  \bibinfo{author}{Trujillo, C.A.}, \bibinfo{author}{van Dam, M.A.},
  \bibinfo{author}{Campbell, R.D.}, \bibinfo{author}{Chin, J.C.Y.},
  \bibinfo{author}{Hartman, S.K.}, \bibinfo{author}{Johansson, E.M.},
  \bibinfo{author}{Lafon, R.E.}, \bibinfo{author}{Le~Mignant, D.},
  \bibinfo{author}{Stomski~Jr., P.J.}, \bibinfo{author}{Summers, D.M.},
  \bibinfo{author}{Wizinowich, P.L.}, \bibinfo{year}{2005}.
\newblock \bibinfo{title}{{Keck Observatory Laser Guide Star Adaptive Optics
  Discovery and Characterization of a Satellite to the Large Kuiper Belt Pbject
  2003 EL$_{61}$}}.
\newblock \bibinfo{journal}{Astrophysical Journal} \bibinfo{volume}{632},
  \bibinfo{pages}{L45--L48}.
\bibitem[{Brown et~al.(2010)Brown, Ragozzine, Stansberry and
  Fraser}]{2010-AJ-139-Brown}
\bibinfo{author}{Brown, M.E.}, \bibinfo{author}{Ragozzine, D.},
  \bibinfo{author}{Stansberry, J.}, \bibinfo{author}{Fraser, W.C.},
  \bibinfo{year}{2010}.
\newblock \bibinfo{title}{{The Size, Density, and Formation of the Orcus-Vanth
  System in the Kuiper Belt}}.
\newblock \bibinfo{journal}{Astronomical Journal} \bibinfo{volume}{139},
  \bibinfo{pages}{2700--2705}.
\bibitem[{Brown and Schaller(2007)}]{2007-Science-316-Brown}
\bibinfo{author}{Brown, M.E.}, \bibinfo{author}{Schaller, E.L.},
  \bibinfo{year}{2007}.
\newblock \bibinfo{title}{{The Mass of Dwarf Planet Eris}}.
\newblock \bibinfo{journal}{Science} \bibinfo{volume}{316},
  \bibinfo{pages}{1585--}.
\bibitem[{Brown et~al.(2006)Brown, Schaller, Roe, Rabinowitz and
  Trujillo}]{2006-ApJ-643-Brown}
\bibinfo{author}{Brown, M.E.}, \bibinfo{author}{Schaller, E.L.},
  \bibinfo{author}{Roe, H.G.}, \bibinfo{author}{Rabinowitz, D.L.},
  \bibinfo{author}{Trujillo, C.A.}, \bibinfo{year}{2006}.
\newblock \bibinfo{title}{{Direct Measurement of the Size of 2003 UB313 from
  the Hubble Space Telescope}}.
\newblock \bibinfo{journal}{Astrophysical Journal} \bibinfo{volume}{643},
  \bibinfo{pages}{L61--L63}.
\bibitem[{Brown and Trujillo(2004)}]{2004-AJ-127-Brown}
\bibinfo{author}{Brown, M.E.}, \bibinfo{author}{Trujillo, C.A.},
  \bibinfo{year}{2004}.
\newblock \bibinfo{title}{{Direct Measurement of the Size of the Large Kuiper
  Belt Object (50000) Quaoar}}.
\newblock \bibinfo{journal}{Astronomical Journal} \bibinfo{volume}{127},
  \bibinfo{pages}{2413--2417}.
\bibitem[{Brownlee et~al.(2004)Brownlee, Horz, Newburn, Zolensky, Duxbury,
  Sandford, Sekanina, Tsou, Hanner, Clark, Green and
  Kissel}]{2004-Science-304-Brownlee}
\bibinfo{author}{Brownlee, D.E.}, \bibinfo{author}{Horz, F.},
  \bibinfo{author}{Newburn, R.L.}, \bibinfo{author}{Zolensky, M.},
  \bibinfo{author}{Duxbury, T.C.}, \bibinfo{author}{Sandford, S.},
  \bibinfo{author}{Sekanina, Z.}, \bibinfo{author}{Tsou, P.},
  \bibinfo{author}{Hanner, M.S.}, \bibinfo{author}{Clark, B.C.},
  \bibinfo{author}{Green, S.F.}, \bibinfo{author}{Kissel, J.},
  \bibinfo{year}{2004}.
\newblock \bibinfo{title}{{Surface of Young Jupiter Family Comet 81 P/Wild 2:
  View from the Stardust Spacecraft}}.
\newblock \bibinfo{journal}{Science} \bibinfo{volume}{304},
  \bibinfo{pages}{1764--1769}.
\bibitem[{Brozovi{\'c} et~al.(2011)Brozovi{\'c}, Benner, Taylor, Nolan, Howell,
  Magri, Scheeres, Giorgini, Pollock, Pravec, Gal{\'a}d, Fang, Margot, Busch,
  Shepard, Reichart, Ivarsen, Haislip, Lacluyze, Jao, Slade, Lawrence and
  Hicks}]{2011-Icarus-216-Brozovic}
\bibinfo{author}{Brozovi{\'c}, M.}, \bibinfo{author}{Benner, L.A.M.},
  \bibinfo{author}{Taylor, P.A.}, \bibinfo{author}{Nolan, M.C.},
  \bibinfo{author}{Howell, E.S.}, \bibinfo{author}{Magri, C.},
  \bibinfo{author}{Scheeres, D.J.}, \bibinfo{author}{Giorgini, J.D.},
  \bibinfo{author}{Pollock, J.T.}, \bibinfo{author}{Pravec, P.},
  \bibinfo{author}{Gal{\'a}d, A.}, \bibinfo{author}{Fang, J.},
  \bibinfo{author}{Margot, J.L.}, \bibinfo{author}{Busch, M.W.},
  \bibinfo{author}{Shepard, M.K.}, \bibinfo{author}{Reichart, D.E.},
  \bibinfo{author}{Ivarsen, K.M.}, \bibinfo{author}{Haislip, J.B.},
  \bibinfo{author}{Lacluyze, A.P.}, \bibinfo{author}{Jao, J.},
  \bibinfo{author}{Slade, M.A.}, \bibinfo{author}{Lawrence, K.J.},
  \bibinfo{author}{Hicks, M.D.}, \bibinfo{year}{2011}.
\newblock \bibinfo{title}{{Radar and optical observations and physical modeling
  of triple near-Earth Asteroid (136617) 1994 CC}}.
\newblock \bibinfo{journal}{Icarus} \bibinfo{volume}{216},
  \bibinfo{pages}{241--256}.
\bibitem[{Brucker et~al.(2009)Brucker, Grundy, Stansberry, Spencer, Sheppard,
  Chiang and Buie}]{2009-Icarus-201-Brucker}
\bibinfo{author}{Brucker, M.J.}, \bibinfo{author}{Grundy, W.M.},
  \bibinfo{author}{Stansberry, J.A.}, \bibinfo{author}{Spencer, J.R.},
  \bibinfo{author}{Sheppard, S.S.}, \bibinfo{author}{Chiang, E.I.},
  \bibinfo{author}{Buie, M.W.}, \bibinfo{year}{2009}.
\newblock \bibinfo{title}{{High albedos of low inclination Classical Kuiper
  belt objects}}.
\newblock \bibinfo{journal}{Icarus} \bibinfo{volume}{201},
  \bibinfo{pages}{284--294}.
\bibitem[{Brunetto et~al.(2007)Brunetto, de~Le{\'o}n and
  Licandro}]{2007-AA-472-Brunetto}
\bibinfo{author}{Brunetto, R.}, \bibinfo{author}{de~Le{\'o}n, J.},
  \bibinfo{author}{Licandro, J.}, \bibinfo{year}{2007}.
\newblock \bibinfo{title}{{Testing space weathering models on A-type asteroid
  (1951) Lick}}.
\newblock \bibinfo{journal}{Astronomy and Astrophysics} \bibinfo{volume}{472},
  \bibinfo{pages}{653--656}.
\bibitem[{Brunetto et~al.(2006)Brunetto, Vernazza, Marchi, Birlan, Fulchignoni,
  Orofino and Strazzulla}]{2006-Icarus-184-Brunetto}
\bibinfo{author}{Brunetto, R.}, \bibinfo{author}{Vernazza, P.},
  \bibinfo{author}{Marchi, S.}, \bibinfo{author}{Birlan, M.},
  \bibinfo{author}{Fulchignoni, M.}, \bibinfo{author}{Orofino, V.},
  \bibinfo{author}{Strazzulla, G.}, \bibinfo{year}{2006}.
\newblock \bibinfo{title}{{Modeling asteroid surfaces from observations and
  irradiation experiments: The case of 832 Karin}}.
\newblock \bibinfo{journal}{Icarus} \bibinfo{volume}{184},
  \bibinfo{pages}{327--337}.
\bibitem[{Buie et~al.(2006)Buie, Grundy, Young, Young and
  Stern}]{2006-AJ-132-Buie}
\bibinfo{author}{Buie, M.W.}, \bibinfo{author}{Grundy, W.M.},
  \bibinfo{author}{Young, E.F.}, \bibinfo{author}{Young, L.A.},
  \bibinfo{author}{Stern, S.A.}, \bibinfo{year}{2006}.
\newblock \bibinfo{title}{{Orbits and Photometry of Pluto's Satellites: Charon,
  S/2005 P1, and S/2005 P2}}.
\newblock \bibinfo{journal}{Astronomical Journal} \bibinfo{volume}{132},
  \bibinfo{pages}{290--298}.
\bibitem[{Burbine et~al.(2011)Burbine, Duffard, Buchanan, Cloutis and
  Binzel}]{2011-LPI-Burbine}
\bibinfo{author}{Burbine, T.H.}, \bibinfo{author}{Duffard, R.},
  \bibinfo{author}{Buchanan, P.C.}, \bibinfo{author}{Cloutis, E.A.},
  \bibinfo{author}{Binzel, R.P.}, \bibinfo{year}{2011}.
\newblock \bibinfo{title}{{Spectroscopy of O-Type Asteroids}}, in:
  \bibinfo{booktitle}{Lunar and Planetary Institute Science Conference
  Abstracts}, p. \bibinfo{pages}{2483}.
\bibitem[{Burbine et~al.(2002)Burbine, McCoy, Meibom, Gladman and
  Keil}]{2002-AsteroidsIII-5.2-Burbine}
\bibinfo{author}{Burbine, T.H.}, \bibinfo{author}{McCoy, T.J.},
  \bibinfo{author}{Meibom, A.}, \bibinfo{author}{Gladman, B.},
  \bibinfo{author}{Keil, K.}, \bibinfo{year}{2002}.
\newblock \bibinfo{title}{{Meteoritic Parent Bodies: Their Number and
  Identification}}.
\newblock \bibinfo{journal}{Asteroids III} , \bibinfo{pages}{653--667}.
\bibitem[{Burbine et~al.(1996)Burbine, Meibom and Binzel}]{1996-MPS-31-Burbine}
\bibinfo{author}{Burbine, T.H.}, \bibinfo{author}{Meibom, A.},
  \bibinfo{author}{Binzel, R.P.}, \bibinfo{year}{1996}.
\newblock \bibinfo{title}{{Mantle material in the main belt: Battered to
  bits?}}
\newblock \bibinfo{journal}{Meteoritics and Planetary Science}
  \bibinfo{volume}{31}, \bibinfo{pages}{607--620}.
\bibitem[{Bus(1999)}]{1999-PhD-Bus}
\bibinfo{author}{Bus, S.J.}, \bibinfo{year}{1999}.
\newblock \bibinfo{title}{{Compositional structure in the asteroid belt:
  Results of a spectroscopic survey}}.
\newblock Ph.D. thesis. MASSACHUSETTS INSTITUTE OF TECHNOLOGY.
\bibitem[{Carpino and Knezevic(1996)}]{1996-IAUS-172-Carpino}
\bibinfo{author}{Carpino, M.}, \bibinfo{author}{Knezevic, Z.},
  \bibinfo{year}{1996}.
\newblock \bibinfo{title}{{Asteroid mass determination: (1) Ceres}}, in:
  \bibinfo{editor}{{S.~Ferraz-Mello, B.~Morando, \& J.-E.~Arlot}} (Ed.),
  \bibinfo{booktitle}{Dynamics, Ephemerides, and Astrometry of the Solar
  System}, pp. \bibinfo{pages}{203--+}.
\bibitem[{Carry et~al.(2008)Carry, Dumas, Fulchignoni, Merline, Berthier,
  Hestroffer, Fusco and Tamblyn}]{2008-AA-478-Carry}
\bibinfo{author}{Carry, B.}, \bibinfo{author}{Dumas, C.},
  \bibinfo{author}{Fulchignoni, M.}, \bibinfo{author}{Merline, W.J.},
  \bibinfo{author}{Berthier, J.}, \bibinfo{author}{Hestroffer, D.},
  \bibinfo{author}{Fusco, T.}, \bibinfo{author}{Tamblyn, P.},
  \bibinfo{year}{2008}.
\newblock \bibinfo{title}{{Near-Infrared Mapping and Physical Properties of the
  Dwarf-Planet Ceres}}.
\newblock \bibinfo{journal}{Astronomy and Astrophysics} \bibinfo{volume}{478},
  \bibinfo{pages}{235--244}.
\bibitem[{Carry et~al.(2010a)Carry, Dumas, Kaasalainen, Berthier, Merline,
  Erard, Conrad, Drummond, Hestroffer, Fulchignoni and
  Fusco}]{2010-Icarus-205-Carry-a}
\bibinfo{author}{Carry, B.}, \bibinfo{author}{Dumas, C.},
  \bibinfo{author}{Kaasalainen, M.}, \bibinfo{author}{Berthier, J.},
  \bibinfo{author}{Merline, W.J.}, \bibinfo{author}{Erard, S.},
  \bibinfo{author}{Conrad, A.R.}, \bibinfo{author}{Drummond, J.D.},
  \bibinfo{author}{Hestroffer, D.}, \bibinfo{author}{Fulchignoni, M.},
  \bibinfo{author}{Fusco, T.}, \bibinfo{year}{2010}a.
\newblock \bibinfo{title}{{Physical properties of (2) Pallas}}.
\newblock \bibinfo{journal}{Icarus} \bibinfo{volume}{205},
  \bibinfo{pages}{460--472}.
\bibitem[{Carry et~al.(2011)Carry, Hestroffer, DeMeo, Thirouin, Berthier,
  Lacerda, Sicardy, Doressoundiram, Dumas, Farrelly and
  M{\"u}ller}]{2011-AA-534-Carry}
\bibinfo{author}{Carry, B.}, \bibinfo{author}{Hestroffer, D.},
  \bibinfo{author}{DeMeo, F.E.}, \bibinfo{author}{Thirouin, A.},
  \bibinfo{author}{Berthier, J.}, \bibinfo{author}{Lacerda, P.},
  \bibinfo{author}{Sicardy, B.}, \bibinfo{author}{Doressoundiram, A.},
  \bibinfo{author}{Dumas, C.}, \bibinfo{author}{Farrelly, D.},
  \bibinfo{author}{M{\"u}ller, T.G.}, \bibinfo{year}{2011}.
\newblock \bibinfo{title}{{Integral-field spectroscopy of (90482)
  Orcus-Vanth}}.
\newblock \bibinfo{journal}{Astronomy and Astrophysics} \bibinfo{volume}{534},
  \bibinfo{pages}{A115}.
\bibitem[{Carry et~al.(2010b)Carry, Kaasalainen, Leyrat, Merline, Drummond,
  Conrad, Weaver, Tamblyn, Chapman, Dumas, Colas, Christou, Dotto, Perna,
  Fornasier, Bernasconi, Behrend, Vachier, Kryszczynska, Polinska, Fulchignoni,
  Roy, Naves, Poncy and Wiggins}]{2010-AA-523-Carry}
\bibinfo{author}{Carry, B.}, \bibinfo{author}{Kaasalainen, M.},
  \bibinfo{author}{Leyrat, C.}, \bibinfo{author}{Merline, W.J.},
  \bibinfo{author}{Drummond, J.D.}, \bibinfo{author}{Conrad, A.R.},
  \bibinfo{author}{Weaver, H.A.}, \bibinfo{author}{Tamblyn, P.M.},
  \bibinfo{author}{Chapman, C.R.}, \bibinfo{author}{Dumas, C.},
  \bibinfo{author}{Colas, F.}, \bibinfo{author}{Christou, J.C.},
  \bibinfo{author}{Dotto, E.}, \bibinfo{author}{Perna, D.},
  \bibinfo{author}{Fornasier, S.}, \bibinfo{author}{Bernasconi, L.},
  \bibinfo{author}{Behrend, R.}, \bibinfo{author}{Vachier, F.},
  \bibinfo{author}{Kryszczynska, A.}, \bibinfo{author}{Polinska, M.},
  \bibinfo{author}{Fulchignoni, M.}, \bibinfo{author}{Roy, R.},
  \bibinfo{author}{Naves, R.}, \bibinfo{author}{Poncy, R.},
  \bibinfo{author}{Wiggins, P.}, \bibinfo{year}{2010}b.
\newblock \bibinfo{title}{{Physical properties of the ESA Rosetta target
  asteroid (21) Lutetia. II. Shape and flyby geometry}}.
\newblock \bibinfo{journal}{Astronomy and Astrophysics} \bibinfo{volume}{523},
  \bibinfo{pages}{A94}.
\bibitem[{Carry et~al.(2012)Carry, Kaasalainen, Merline, M{\"u}ller, Jorda,
  Drummond, Berthier, O'Rourke, {\v D}urech, K{\"u}ppers, Conrad, Dumas, Sierks
  and the OSIRIS~Team}]{2012-PSS--Carry}
\bibinfo{author}{Carry, B.}, \bibinfo{author}{Kaasalainen, M.},
  \bibinfo{author}{Merline, W.J.}, \bibinfo{author}{M{\"u}ller, T.G.},
  \bibinfo{author}{Jorda, L.}, \bibinfo{author}{Drummond, J.D.},
  \bibinfo{author}{Berthier, J.}, \bibinfo{author}{O'Rourke, L.},
  \bibinfo{author}{{\v D}urech, J.}, \bibinfo{author}{K{\"u}ppers, M.},
  \bibinfo{author}{Conrad, A.R.}, \bibinfo{author}{Dumas, C.},
  \bibinfo{author}{Sierks, H.}, \bibinfo{author}{the OSIRIS~Team},
  \bibinfo{year}{2012}.
\newblock \bibinfo{title}{{KOALA shape modeling technique validated at (21)
  Lutetia by ESA Rosetta mission}}.
\newblock \bibinfo{journal}{Planetary and Space Science} \bibinfo{volume}{in
  press}.
\bibitem[{Castillo-Rogez and McCord(2010)}]{2010-Icarus-205-Castillo-Rogez}
\bibinfo{author}{Castillo-Rogez, J.C.}, \bibinfo{author}{McCord, T.B.},
  \bibinfo{year}{2010}.
\newblock \bibinfo{title}{{Ceres: evolution and present state constrained by
  shape data}}.
\newblock \bibinfo{journal}{Icarus} \bibinfo{volume}{205},
  \bibinfo{pages}{443--459}.
\bibitem[{Chapman(1996)}]{1996-MPS-31-Chapman}
\bibinfo{author}{Chapman, C.R.}, \bibinfo{year}{1996}.
\newblock \bibinfo{title}{{S-Type Asteroids, Ordinary Chondrites, and Space
  Weathering: The Evidence from Galileo's Fly-bys of Gaspra and Ida}}.
\newblock \bibinfo{journal}{Meteoritics and Planetary Science}
  \bibinfo{volume}{31}, \bibinfo{pages}{699--725}.
\bibitem[{Chernetenko and Kochetova(2002)}]{2002-ACM-Chernetenko}
\bibinfo{author}{Chernetenko, Y.A.}, \bibinfo{author}{Kochetova, O.M.},
  \bibinfo{year}{2002}.
\newblock \bibinfo{title}{{Masses of some large minor planets}}, in:
  \bibinfo{editor}{{B.~Warmbein}} (Ed.), \bibinfo{booktitle}{Asteroids, Comets,
  and Meteors: ACM 2002}, pp. \bibinfo{pages}{437--440}.
\bibitem[{Chesley et~al.(2005)Chesley, Owen, Hayne, Sullivan, Dumas, Giorgini,
  Chamberlin, Synnott and Vazquez}]{2005-DDA-36-Chesley}
\bibinfo{author}{Chesley, S.R.}, \bibinfo{author}{Owen, Jr., W.M.},
  \bibinfo{author}{Hayne, E.W.}, \bibinfo{author}{Sullivan, A.M.},
  \bibinfo{author}{Dumas, R.C.}, \bibinfo{author}{Giorgini, J.D.},
  \bibinfo{author}{Chamberlin, A.B.}, \bibinfo{author}{Synnott, S.P.},
  \bibinfo{author}{Vazquez, C.S.}, \bibinfo{year}{2005}.
\newblock \bibinfo{title}{{The Mass of Asteroid 10 Hygiea}}, in:
  \bibinfo{booktitle}{AAS/Division of Dynamical Astronomy Meeting \#36}, p.
  \bibinfo{pages}{524}.
\bibitem[{Clark et~al.(2004)Clark, Bus, Rivkin, Shepard and
  Shah}]{2004-AJ-128-Clark}
\bibinfo{author}{Clark, B.E.}, \bibinfo{author}{Bus, S.J.},
  \bibinfo{author}{Rivkin, A.S.}, \bibinfo{author}{Shepard, M.K.},
  \bibinfo{author}{Shah, S.}, \bibinfo{year}{2004}.
\newblock \bibinfo{title}{{Spectroscopy of X-Type Asteroids}}.
\newblock \bibinfo{journal}{Astronomical Journal} \bibinfo{volume}{128},
  \bibinfo{pages}{3070--3081}.
\bibitem[{Clark et~al.(2009)Clark, Ockert-Bell, Cloutis, Nesvorny,
  Moth{\'e}-Diniz and Bus}]{2009-Icarus-202-Clark}
\bibinfo{author}{Clark, B.E.}, \bibinfo{author}{Ockert-Bell, M.E.},
  \bibinfo{author}{Cloutis, E.A.}, \bibinfo{author}{Nesvorny, D.},
  \bibinfo{author}{Moth{\'e}-Diniz, T.}, \bibinfo{author}{Bus, S.J.},
  \bibinfo{year}{2009}.
\newblock \bibinfo{title}{{Spectroscopy of K-complex asteroids: Parent bodies
  of carbonaceous meteorites?}}
\newblock \bibinfo{journal}{Icarus} \bibinfo{volume}{202},
  \bibinfo{pages}{119--133}.
\bibitem[{Clark et~al.(2010)Clark, Ziffer, Nesvorn{\'y}, Campins, Rivkin,
  Hiroi, Barucci, Fulchignoni, Binzel, Fornasier, DeMeo, Ockert-Bell, Licandro
  and Moth{\'e}-Diniz}]{2010-JGR-115-Clark}
\bibinfo{author}{Clark, B.E.}, \bibinfo{author}{Ziffer, J.},
  \bibinfo{author}{Nesvorn{\'y}, D.}, \bibinfo{author}{Campins, H.},
  \bibinfo{author}{Rivkin, A.S.}, \bibinfo{author}{Hiroi, T.},
  \bibinfo{author}{Barucci, M.A.}, \bibinfo{author}{Fulchignoni, M.},
  \bibinfo{author}{Binzel, R.P.}, \bibinfo{author}{Fornasier, S.},
  \bibinfo{author}{DeMeo, F.}, \bibinfo{author}{Ockert-Bell, M.E.},
  \bibinfo{author}{Licandro, J.}, \bibinfo{author}{Moth{\'e}-Diniz, T.},
  \bibinfo{year}{2010}.
\newblock \bibinfo{title}{{Spectroscopy of B-type asteroids: Subgroups and
  meteorite analogs}}.
\newblock \bibinfo{journal}{Journal of Geophysical Research (Planets)}
  \bibinfo{volume}{115}, \bibinfo{pages}{E06005}.
\bibitem[{Cloutis et~al.(2011a)Cloutis, Hiroi, Gaffey, Alexander and
  Mann}]{2011-Icarus-212-Cloutis}
\bibinfo{author}{Cloutis, E.A.}, \bibinfo{author}{Hiroi, T.},
  \bibinfo{author}{Gaffey, M.J.}, \bibinfo{author}{Alexander, C.M.O.},
  \bibinfo{author}{Mann, P.}, \bibinfo{year}{2011}a.
\newblock \bibinfo{title}{{Spectral reflectance properties of carbonaceous
  chondrites: 1. CI chondrites}}.
\newblock \bibinfo{journal}{Icarus} \bibinfo{volume}{212},
  \bibinfo{pages}{180--209}.
\bibitem[{Cloutis et~al.(2011b)Cloutis, Hudon, Hiroi, Gaffey and
  Mann}]{2011-Icarus-216-Cloutis}
\bibinfo{author}{Cloutis, E.A.}, \bibinfo{author}{Hudon, P.},
  \bibinfo{author}{Hiroi, T.}, \bibinfo{author}{Gaffey, M.J.},
  \bibinfo{author}{Mann, P.}, \bibinfo{year}{2011}b.
\newblock \bibinfo{title}{{Spectral reflectance properties of carbonaceous
  chondrites: 2. CM chondrites}}.
\newblock \bibinfo{journal}{Icarus} \bibinfo{volume}{216},
  \bibinfo{pages}{309--346}.
\bibitem[{Conrad et~al.(2007)Conrad, Dumas, Merline, Drummond, Campbell,
  Goodrich, Le~Mignant, Chaffee, Fusco, Kwok and
  Knight}]{2007-Icarus-191-Conrad}
\bibinfo{author}{Conrad, A.R.}, \bibinfo{author}{Dumas, C.},
  \bibinfo{author}{Merline, W.J.}, \bibinfo{author}{Drummond, J.D.},
  \bibinfo{author}{Campbell, R.D.}, \bibinfo{author}{Goodrich, R.W.},
  \bibinfo{author}{Le~Mignant, D.}, \bibinfo{author}{Chaffee, F.H.},
  \bibinfo{author}{Fusco, T.}, \bibinfo{author}{Kwok, S.H.},
  \bibinfo{author}{Knight, R.I.}, \bibinfo{year}{2007}.
\newblock \bibinfo{title}{{Direct measurement of the size, shape, and pole of
  511 Davida with Keck AO in a single night}}.
\newblock \bibinfo{journal}{Icarus} \bibinfo{volume}{191},
  \bibinfo{pages}{616--627}.
\bibitem[{Consolmagno et~al.(2008)Consolmagno, Britt and
  Macke}]{2008-ChEG-68-Consolmagno}
\bibinfo{author}{Consolmagno, G.}, \bibinfo{author}{Britt, D.},
  \bibinfo{author}{Macke, R.}, \bibinfo{year}{2008}.
\newblock \bibinfo{title}{{The significance of meteorite density and
  porosity}}.
\newblock \bibinfo{journal}{Chemie der Erde / Geochemistry}
  \bibinfo{volume}{68}, \bibinfo{pages}{1--29}.
\bibitem[{Consolmagno and Britt(1998)}]{1998-MPS-33-Consolmagno}
\bibinfo{author}{Consolmagno, G.J.}, \bibinfo{author}{Britt, D.T.},
  \bibinfo{year}{1998}.
\newblock \bibinfo{title}{{The density and porosity of meteorites from the
  Vatican collection}}.
\newblock \bibinfo{journal}{Meteoritics and Planetary Science}
  \bibinfo{volume}{33}, \bibinfo{pages}{1231--1241}.
\bibitem[{Davidsson and Guti{\'e}rrez(2004)}]{2004-Icarus-168-Davidsson}
\bibinfo{author}{Davidsson, B.J.R.}, \bibinfo{author}{Guti{\'e}rrez, P.J.},
  \bibinfo{year}{2004}.
\newblock \bibinfo{title}{{Estimating the nucleus density of Comet
  19P/Borrelly}}.
\newblock \bibinfo{journal}{Icarus} \bibinfo{volume}{168},
  \bibinfo{pages}{392--408}.
\bibitem[{Davidsson and Guti{\'e}rrez(2005)}]{2005-Icarus-176-Davidsson}
\bibinfo{author}{Davidsson, B.J.R.}, \bibinfo{author}{Guti{\'e}rrez, P.J.},
  \bibinfo{year}{2005}.
\newblock \bibinfo{title}{{Nucleus properties of Comet 67P/Churyumov
  Gerasimenko estimated from non-gravitational force modeling}}.
\newblock \bibinfo{journal}{Icarus} \bibinfo{volume}{176},
  \bibinfo{pages}{453--477}.
\bibitem[{Davidsson and Guti{\'e}rrez(2006)}]{2006-Icarus-180-Davidsson}
\bibinfo{author}{Davidsson, B.J.R.}, \bibinfo{author}{Guti{\'e}rrez, P.J.},
  \bibinfo{year}{2006}.
\newblock \bibinfo{title}{{Non-gravitational force modeling of Comet 81P/Wild
  2. I. A nucleus bulk density estimate}}.
\newblock \bibinfo{journal}{Icarus} \bibinfo{volume}{180},
  \bibinfo{pages}{224--242}.
\bibitem[{Davidsson et~al.(2007)Davidsson, Guti{\'e}rrez and
  Rickman}]{2007-Icarus-187-Davidsson}
\bibinfo{author}{Davidsson, B.J.R.}, \bibinfo{author}{Guti{\'e}rrez, P.J.},
  \bibinfo{author}{Rickman, H.}, \bibinfo{year}{2007}.
\newblock \bibinfo{title}{{Nucleus properties of Comet 9P/Tempel 1 estimated
  from non-gravitational force modeling}}.
\newblock \bibinfo{journal}{Icarus} \bibinfo{volume}{187},
  \bibinfo{pages}{306--320}.
\bibitem[{{de Le{\'o}n} et~al.(2004){de Le{\'o}n}, Duffard, Licandro and
  Lazzaro}]{2004-AA-422-deLeon}
\bibinfo{author}{{de Le{\'o}n}, J.}, \bibinfo{author}{Duffard, R.},
  \bibinfo{author}{Licandro, J.}, \bibinfo{author}{Lazzaro, D.},
  \bibinfo{year}{2004}.
\newblock \bibinfo{title}{{Mineralogical characterization of A-type asteroid
  (1951) Lick}}.
\newblock \bibinfo{journal}{Astronomy and Astrophysics} \bibinfo{volume}{422},
  \bibinfo{pages}{L59--L62}.
\bibitem[{Delbo(2004)}]{PDSSBN-DELBO}
\bibinfo{author}{Delbo, M.}, \bibinfo{year}{2004}.
\newblock \bibinfo{title}{{Thermal Infrared Asteroid Diameters and Albedo}}.
\newblock \bibinfo{howpublished}{NASA Planetary Data System}.
\newblock \bibinfo{note}{MSX-A-SPIRIT3-5-SBN0003-MIMPS-V1.0}.
\bibitem[{Delbo et~al.(2009)Delbo, Ligori, Matter, Cellino and
  Berthier}]{2009-ApJ-694-Delbo}
\bibinfo{author}{Delbo, M.}, \bibinfo{author}{Ligori, S.},
  \bibinfo{author}{Matter, A.}, \bibinfo{author}{Cellino, A.},
  \bibinfo{author}{Berthier, J.}, \bibinfo{year}{2009}.
\newblock \bibinfo{title}{{First VLTI-MIDI Direct Determinations of Asteroid
  Sizes}}.
\newblock \bibinfo{journal}{Astrophysical Journal} \bibinfo{volume}{694},
  \bibinfo{pages}{1228--1236}.
\bibitem[{Delbo and Tanga(2009)}]{2009-PSS-57-Delbo}
\bibinfo{author}{Delbo, M.}, \bibinfo{author}{Tanga, P.}, \bibinfo{year}{2009}.
\newblock \bibinfo{title}{{Thermal inertia of main belt asteroids smaller than
  100 km from IRAS data}}.
\newblock \bibinfo{journal}{Planetary and Space Science} \bibinfo{volume}{57},
  \bibinfo{pages}{259--265}.
\bibitem[{DeMeo et~al.(2009)DeMeo, Binzel, Slivan and
  Bus}]{2009-Icarus-202-DeMeo}
\bibinfo{author}{DeMeo, F.E.}, \bibinfo{author}{Binzel, R.P.},
  \bibinfo{author}{Slivan, S.M.}, \bibinfo{author}{Bus, S.J.},
  \bibinfo{year}{2009}.
\newblock \bibinfo{title}{{An extension of the Bus asteroid taxonomy into the
  near-infrared}}.
\newblock \bibinfo{journal}{Icarus} \bibinfo{volume}{202},
  \bibinfo{pages}{160--180}.
\bibitem[{Descamps et~al.(2011)Descamps, Marchis, Berthier, Emery, Duch{\^e}ne,
  de~Pater, Wong, Lim, Hammel, Vachier, Wiggins, Teng-Chuen-Yu, Peyrot,
  Pollock, Assafin, Vieira-Martins, Camargo, Braga-Ribas and
  Macomber}]{2011-Icarus-211-Descamps}
\bibinfo{author}{Descamps, P.}, \bibinfo{author}{Marchis, F.},
  \bibinfo{author}{Berthier, J.}, \bibinfo{author}{Emery, J.P.},
  \bibinfo{author}{Duch{\^e}ne, G.}, \bibinfo{author}{de~Pater, I.},
  \bibinfo{author}{Wong, M.H.}, \bibinfo{author}{Lim, L.},
  \bibinfo{author}{Hammel, H.B.}, \bibinfo{author}{Vachier, F.},
  \bibinfo{author}{Wiggins, P.}, \bibinfo{author}{Teng-Chuen-Yu, J.P.},
  \bibinfo{author}{Peyrot, A.}, \bibinfo{author}{Pollock, J.},
  \bibinfo{author}{Assafin, M.}, \bibinfo{author}{Vieira-Martins, R.},
  \bibinfo{author}{Camargo, J.I.B.}, \bibinfo{author}{Braga-Ribas, F.},
  \bibinfo{author}{Macomber, B.}, \bibinfo{year}{2011}.
\newblock \bibinfo{title}{{Triplicity and physical characteristics of Asteroid
  (216) Kleopatra}}.
\newblock \bibinfo{journal}{Icarus} \bibinfo{volume}{211},
  \bibinfo{pages}{1022--1033}.
\bibitem[{Descamps et~al.(2009)Descamps, Marchis, Durech, Emery, Harris,
  Kaasalainen, Berthier, Teng-Chuen-Yu, Peyrot, Hutton, Greene, Pollock,
  Assafin, Vieira-Martins, Camargo, Braga-Ribas, Vachier, Reichart, Ivarsen,
  Crain, Nysewander, Lacluyze, Haislip, Behrend, Colas, Lecacheux, Bernasconi,
  Roy, Baudouin, Brunetto, Sposetti and Manzini}]{2009-Icarus-203-Descamps}
\bibinfo{author}{Descamps, P.}, \bibinfo{author}{Marchis, F.},
  \bibinfo{author}{Durech, J.}, \bibinfo{author}{Emery, J.P.},
  \bibinfo{author}{Harris, A.W.}, \bibinfo{author}{Kaasalainen, M.},
  \bibinfo{author}{Berthier, J.}, \bibinfo{author}{Teng-Chuen-Yu, J.P.},
  \bibinfo{author}{Peyrot, A.}, \bibinfo{author}{Hutton, L.},
  \bibinfo{author}{Greene, J.}, \bibinfo{author}{Pollock, J.},
  \bibinfo{author}{Assafin, M.}, \bibinfo{author}{Vieira-Martins, R.},
  \bibinfo{author}{Camargo, J.I.B.}, \bibinfo{author}{Braga-Ribas, F.},
  \bibinfo{author}{Vachier, F.}, \bibinfo{author}{Reichart, D.E.},
  \bibinfo{author}{Ivarsen, K.M.}, \bibinfo{author}{Crain, J.A.},
  \bibinfo{author}{Nysewander, M.C.}, \bibinfo{author}{Lacluyze, A.P.},
  \bibinfo{author}{Haislip, J.B.}, \bibinfo{author}{Behrend, R.},
  \bibinfo{author}{Colas, F.}, \bibinfo{author}{Lecacheux, J.},
  \bibinfo{author}{Bernasconi, L.}, \bibinfo{author}{Roy, R.},
  \bibinfo{author}{Baudouin, P.}, \bibinfo{author}{Brunetto, L.},
  \bibinfo{author}{Sposetti, S.}, \bibinfo{author}{Manzini, F.},
  \bibinfo{year}{2009}.
\newblock \bibinfo{title}{{New insights on the binary Asteroid 121 Hermione}}.
\newblock \bibinfo{journal}{Icarus} \bibinfo{volume}{203},
  \bibinfo{pages}{88--101}.
\bibitem[{Descamps et~al.(2007a)Descamps, Marchis, Michalowski, Colas,
  Berthier, Vachier, Teng-Chuen-Yu, Peyrot, Payet, Dorseuil, L{\'e}onie,
  Dijoux, Berrouachdi, {Chion Hock} and Benard}]{2007-Icarus-189-Descamps}
\bibinfo{author}{Descamps, P.}, \bibinfo{author}{Marchis, F.},
  \bibinfo{author}{Michalowski, T.}, \bibinfo{author}{Colas, F.},
  \bibinfo{author}{Berthier, J.}, \bibinfo{author}{Vachier, F.},
  \bibinfo{author}{Teng-Chuen-Yu, J.P.}, \bibinfo{author}{Peyrot, A.},
  \bibinfo{author}{Payet, B.}, \bibinfo{author}{Dorseuil, J.},
  \bibinfo{author}{L{\'e}onie, Y.}, \bibinfo{author}{Dijoux, T.},
  \bibinfo{author}{Berrouachdi, H.}, \bibinfo{author}{{Chion Hock}, C.},
  \bibinfo{author}{Benard, F.}, \bibinfo{year}{2007}a.
\newblock \bibinfo{title}{{Nature of the small main belt Asteroid 3169 Ostro}}.
\newblock \bibinfo{journal}{Icarus} \bibinfo{volume}{189},
  \bibinfo{pages}{362--369}.
\bibitem[{Descamps et~al.(2007b)Descamps, Marchis, Michalowski, Vachier, Colas,
  Berthier, Assafin, Dunckel, Polinska, Pych, Hestroffer, Miller,
  Vieira-Martins, Birlan, Teng-Chuen-Yu, Peyrot, Payet, Dorseuil, L{\'e}onie
  and Dijoux}]{2007-Icarus-187-Descamps}
\bibinfo{author}{Descamps, P.}, \bibinfo{author}{Marchis, F.},
  \bibinfo{author}{Michalowski, T.}, \bibinfo{author}{Vachier, F.},
  \bibinfo{author}{Colas, F.}, \bibinfo{author}{Berthier, J.},
  \bibinfo{author}{Assafin, M.}, \bibinfo{author}{Dunckel, P.B.},
  \bibinfo{author}{Polinska, M.}, \bibinfo{author}{Pych, W.},
  \bibinfo{author}{Hestroffer, D.}, \bibinfo{author}{Miller, K.P.M.},
  \bibinfo{author}{Vieira-Martins, R.}, \bibinfo{author}{Birlan, M.},
  \bibinfo{author}{Teng-Chuen-Yu, J.P.}, \bibinfo{author}{Peyrot, A.},
  \bibinfo{author}{Payet, B.}, \bibinfo{author}{Dorseuil, J.},
  \bibinfo{author}{L{\'e}onie, Y.}, \bibinfo{author}{Dijoux, T.},
  \bibinfo{year}{2007}b.
\newblock \bibinfo{title}{{Figure of the double Asteroid 90 Antiope from
  adaptive optics and lightcurve observations}}.
\newblock \bibinfo{journal}{Icarus} \bibinfo{volume}{187},
  \bibinfo{pages}{482--499}.
\bibitem[{Descamps et~al.(2008)Descamps, Marchis, Pollock, Berthier, Vachier,
  Birlan, Kaasalainen, Harris, Wong, Romanishin, Cooper, Kettner, Wiggins,
  Kryszczynska, Polinska, Coliac, Devyatkin, Verestchagina and
  Gorshanov}]{2008-Icarus-196-Descamps}
\bibinfo{author}{Descamps, P.}, \bibinfo{author}{Marchis, F.},
  \bibinfo{author}{Pollock, J.}, \bibinfo{author}{Berthier, J.},
  \bibinfo{author}{Vachier, F.}, \bibinfo{author}{Birlan, M.},
  \bibinfo{author}{Kaasalainen, M.}, \bibinfo{author}{Harris, A.W.},
  \bibinfo{author}{Wong, M.H.}, \bibinfo{author}{Romanishin, W.J.},
  \bibinfo{author}{Cooper, E.M.}, \bibinfo{author}{Kettner, K.A.},
  \bibinfo{author}{Wiggins, P.}, \bibinfo{author}{Kryszczynska, A.},
  \bibinfo{author}{Polinska, M.}, \bibinfo{author}{Coliac, J.F.},
  \bibinfo{author}{Devyatkin, A.}, \bibinfo{author}{Verestchagina, I.},
  \bibinfo{author}{Gorshanov, D.}, \bibinfo{year}{2008}.
\newblock \bibinfo{title}{{New determination of the size and bulk density of
  the binary Asteroid 22 Kalliope from observations of mutual eclipses}}.
\newblock \bibinfo{journal}{Icarus} \bibinfo{volume}{196},
  \bibinfo{pages}{578--600}.
\bibitem[{Doressoundiram et~al.(1998)Doressoundiram, Barucci, Fulchignoni and
  Florczak}]{1998-Icarus-131-Doressoundiram}
\bibinfo{author}{Doressoundiram, A.}, \bibinfo{author}{Barucci, M.A.},
  \bibinfo{author}{Fulchignoni, M.}, \bibinfo{author}{Florczak, M.},
  \bibinfo{year}{1998}.
\newblock \bibinfo{title}{{EOS Family: A Spectroscopic Study}}.
\newblock \bibinfo{journal}{Icarus} \bibinfo{volume}{131},
  \bibinfo{pages}{15--31}.
\bibitem[{Drummond and Christou(2008)}]{2008-Icarus-197-Drummond}
\bibinfo{author}{Drummond, J.D.}, \bibinfo{author}{Christou, J.C.},
  \bibinfo{year}{2008}.
\newblock \bibinfo{title}{{Triaxial ellipsoid dimensions and rotational poles
  of seven asteroids from Lick Observatory adaptive optics images, and of
  Ceres}}.
\newblock \bibinfo{journal}{Icarus} \bibinfo{volume}{197},
  \bibinfo{pages}{480--496}.
\bibitem[{Drummond et~al.(2009)Drummond, Christou and
  Nelson}]{2009-Icarus-202-Drummond}
\bibinfo{author}{Drummond, J.D.}, \bibinfo{author}{Christou, J.C.},
  \bibinfo{author}{Nelson, J.}, \bibinfo{year}{2009}.
\newblock \bibinfo{title}{{Triaxial ellipsoid dimensions and poles of asteroids
  from AO observations at the Keck-II telescope}}.
\newblock \bibinfo{journal}{Icarus} \bibinfo{volume}{202},
  \bibinfo{pages}{147--159}.
\bibitem[{Drummond et~al.(2010)Drummond, Conrad, Merline, Carry, Chapman,
  Weaver, Tamblyn, Christou and Dumas}]{2010-AA-523-Drummond}
\bibinfo{author}{Drummond, J.D.}, \bibinfo{author}{Conrad, A.},
  \bibinfo{author}{Merline, W.J.}, \bibinfo{author}{Carry, B.},
  \bibinfo{author}{Chapman, C.R.}, \bibinfo{author}{Weaver, H.A.},
  \bibinfo{author}{Tamblyn, P.M.}, \bibinfo{author}{Christou, J.C.},
  \bibinfo{author}{Dumas, C.}, \bibinfo{year}{2010}.
\newblock \bibinfo{title}{{Physical properties of the ESA Rosetta target
  asteroid (21) Lutetia. I. The triaxial ellipsoid dimensions, rotational pole,
  and bulk density}}.
\newblock \bibinfo{journal}{Astronomy and Astrophysics} \bibinfo{volume}{523},
  \bibinfo{pages}{A93}.
\bibitem[{Drummond et~al.(2011)Drummond, Merline, Conrad, Christou, Tamblyn and
  Carry}]{2011-DPS-Drummond}
\bibinfo{author}{Drummond, J.D.}, \bibinfo{author}{Merline, W.J.},
  \bibinfo{author}{Conrad, A.}, \bibinfo{author}{Christou, J.},
  \bibinfo{author}{Tamblyn, P.}, \bibinfo{author}{Carry, B.},
  \bibinfo{year}{2011}.
\newblock \bibinfo{title}{{Asteroid (19) Fortuna: Triaxial Ellipsoid Dimensions
  and Rotational Pole with AO at Gemini North}}, in:
  \bibinfo{booktitle}{AAS/Division for Planetary Sciences Meeting Abstracts
  \#41}, pp.~\bibinfo{pages}{--}.
\bibitem[{Dumas et~al.(2011)Dumas, Carry, Hestroffer and
  Merlin}]{2011-AA-528-Dumas}
\bibinfo{author}{Dumas, C.}, \bibinfo{author}{Carry, B.},
  \bibinfo{author}{Hestroffer, D.}, \bibinfo{author}{Merlin, F.},
  \bibinfo{year}{2011}.
\newblock \bibinfo{title}{{High-contrast observations of (136108) Haumea. A
  crystalline water-ice multiple system}}.
\newblock \bibinfo{journal}{Astronomy and Astrophysics} \bibinfo{volume}{528},
  \bibinfo{pages}{A105}.
\bibitem[{Dunham et~al.(2011)Dunham, Herald, Frappa, Hayamizu, Talbot and
  Timerson}]{PDSSBN-OCC}
\bibinfo{author}{Dunham, D.W.}, \bibinfo{author}{Herald, D.},
  \bibinfo{author}{Frappa, E.}, \bibinfo{author}{Hayamizu, T.},
  \bibinfo{author}{Talbot, J.}, \bibinfo{author}{Timerson, B.},
  \bibinfo{year}{2011}.
\newblock \bibinfo{title}{{Asteroid Occultations}}.
\newblock \bibinfo{howpublished}{NASA Planetary Data System}.
\newblock \bibinfo{note}{{EAR-A-3-RDR-OCCULTATIONS-V9.0}}.
\bibitem[{{\v D}urech et~al.(2011){\v D}urech, Kaasalainen, Herald, Dunham,
  Timerson, Hanu{\v s}, Frappa, Talbot, Hayamizu, Warner, Pilcher and
  Gal{\'a}d}]{2011-Icarus-214-Durech}
\bibinfo{author}{{\v D}urech, J.}, \bibinfo{author}{Kaasalainen, M.},
  \bibinfo{author}{Herald, D.}, \bibinfo{author}{Dunham, D.},
  \bibinfo{author}{Timerson, B.}, \bibinfo{author}{Hanu{\v s}, J.},
  \bibinfo{author}{Frappa, E.}, \bibinfo{author}{Talbot, J.},
  \bibinfo{author}{Hayamizu, T.}, \bibinfo{author}{Warner, B.D.},
  \bibinfo{author}{Pilcher, F.}, \bibinfo{author}{Gal{\'a}d, A.},
  \bibinfo{year}{2011}.
\newblock \bibinfo{title}{{Combining asteroid models derived by lightcurve
  inversion with asteroidal occultation silhouettes}}.
\newblock \bibinfo{journal}{Icarus} \bibinfo{volume}{214},
  \bibinfo{pages}{652--670}.
\bibitem[{Elkins-Tanton et~al.(2011)Elkins-Tanton, Weiss and
  Zuber}]{2011-EPSL-305-Elkins-Tanton}
\bibinfo{author}{Elkins-Tanton, L.T.}, \bibinfo{author}{Weiss, B.P.},
  \bibinfo{author}{Zuber, M.T.}, \bibinfo{year}{2011}.
\newblock \bibinfo{title}{{Chondrites as samples of differentiated
  planetesimals}}.
\newblock \bibinfo{journal}{Earth and Planetary Science Letters}
  \bibinfo{volume}{305}, \bibinfo{pages}{1--10}.
\bibitem[{Emery et~al.(2011)Emery, Burr and Cruikshank}]{2011-AJ-141-Emery}
\bibinfo{author}{Emery, J.P.}, \bibinfo{author}{Burr, D.M.},
  \bibinfo{author}{Cruikshank, D.P.}, \bibinfo{year}{2011}.
\newblock \bibinfo{title}{{Near-infrared Spectroscopy of Trojan Asteroids:
  Evidence for Two Compositional Groups}}.
\newblock \bibinfo{journal}{Astronomical Journal} \bibinfo{volume}{141},
  \bibinfo{pages}{25}.
\bibitem[{Emery et~al.(2006)Emery, Cruikshank and van
  Cleve}]{2006-Icarus-182-Emery}
\bibinfo{author}{Emery, J.P.}, \bibinfo{author}{Cruikshank, D.P.},
  \bibinfo{author}{van Cleve, J.}, \bibinfo{year}{2006}.
\newblock \bibinfo{title}{{Thermal emission spectroscopy (5.2 38 {$\mu$}m) of
  three Trojan asteroids with the Spitzer Space Telescope: Detection of
  fine-grained silicates}}.
\newblock \bibinfo{journal}{Icarus} \bibinfo{volume}{182},
  \bibinfo{pages}{496--512}.
\bibitem[{Fang et~al.(2011)Fang, Margot, Brozovic, Nolan, Benner and
  Taylor}]{2011-AJ-141-Fang}
\bibinfo{author}{Fang, J.}, \bibinfo{author}{Margot, J.L.},
  \bibinfo{author}{Brozovic, M.}, \bibinfo{author}{Nolan, M.C.},
  \bibinfo{author}{Benner, L.A.M.}, \bibinfo{author}{Taylor, P.A.},
  \bibinfo{year}{2011}.
\newblock \bibinfo{title}{{Orbits of Near-Earth Asteroid Triples 2001 SN263 and
  1994 CC: Properties, Origin, and Evolution}}.
\newblock \bibinfo{journal}{Astronomical Journal} \bibinfo{volume}{141},
  \bibinfo{pages}{154--169}.
\bibitem[{Fern{\'a}ndez et~al.(2003)Fern{\'a}ndez, Sheppard and
  Jewitt}]{2003-AJ-126-Fernandez}
\bibinfo{author}{Fern{\'a}ndez, Y.R.}, \bibinfo{author}{Sheppard, S.S.},
  \bibinfo{author}{Jewitt, D.C.}, \bibinfo{year}{2003}.
\newblock \bibinfo{title}{{The Albedo Distribution of Jovian Trojan
  Asteroids}}.
\newblock \bibinfo{journal}{Astronomical Journal} \bibinfo{volume}{126},
  \bibinfo{pages}{1563--1574}.
\bibitem[{Fienga et~al.(2011)Fienga, Kuchynka, Laskar, Manche and
  Gastineau}]{2011-DPS-Fienga}
\bibinfo{author}{Fienga, A.}, \bibinfo{author}{Kuchynka, P.},
  \bibinfo{author}{Laskar, J.}, \bibinfo{author}{Manche, H.},
  \bibinfo{author}{Gastineau, M.}, \bibinfo{year}{2011}.
\newblock \bibinfo{title}{{Asteroid mass determinations with INPOP planetary
  ephemerides}}.
\newblock \bibinfo{journal}{EPSC-DPS Joint Meeting 2011} ,
  \bibinfo{pages}{1879}.
\bibitem[{Fienga et~al.(2009)Fienga, Laskar, Morley, Manche, Kuchynka,
  Le~Poncin-Lafitte, Budnik, Gastineau and Somenzi}]{2009-AA-507-Fienga}
\bibinfo{author}{Fienga, A.}, \bibinfo{author}{Laskar, J.},
  \bibinfo{author}{Morley, T.}, \bibinfo{author}{Manche, H.},
  \bibinfo{author}{Kuchynka, P.}, \bibinfo{author}{Le~Poncin-Lafitte, C.},
  \bibinfo{author}{Budnik, F.}, \bibinfo{author}{Gastineau, M.},
  \bibinfo{author}{Somenzi, L.}, \bibinfo{year}{2009}.
\newblock \bibinfo{title}{{INPOP08, a 4-D planetary ephemeris: from asteroid
  and time-scale computations to ESA Mars Express and Venus Express
  contributions}}.
\newblock \bibinfo{journal}{Astronomy and Astrophysics} \bibinfo{volume}{507},
  \bibinfo{pages}{1675--1686}.
\bibitem[{Fienga et~al.(2010)Fienga, Manche, Kuchynka, Laskar and
  Gastineau}]{2010-SciNote-Fienga}
\bibinfo{author}{Fienga, A.}, \bibinfo{author}{Manche, H.},
  \bibinfo{author}{Kuchynka, P.}, \bibinfo{author}{Laskar, J.},
  \bibinfo{author}{Gastineau, M.}, \bibinfo{year}{2010}.
\newblock \bibinfo{title}{{INPOP10a}}.
\newblock \bibinfo{journal}{Scientific Notes} .
\bibitem[{Fienga et~al.(2008)Fienga, Manche, Laskar and
  Gastineau}]{2008-AA-477-Fienga}
\bibinfo{author}{Fienga, A.}, \bibinfo{author}{Manche, H.},
  \bibinfo{author}{Laskar, J.}, \bibinfo{author}{Gastineau, M.},
  \bibinfo{year}{2008}.
\newblock \bibinfo{title}{{INPOP06: a new numerical planetary ephemeris}}.
\newblock \bibinfo{journal}{Astronomy and Astrophysics} \bibinfo{volume}{477},
  \bibinfo{pages}{315--327}.
\bibitem[{Folkner et~al.(2009)Folkner, Williams and
  Boggs}]{2009-SciNote-Folkner}
\bibinfo{author}{Folkner, W.M.}, \bibinfo{author}{Williams, J.G.},
  \bibinfo{author}{Boggs, D.H.}, \bibinfo{year}{2009}.
\newblock \bibinfo{title}{The planetary and lunar ephemeris de 421}.
\newblock \bibinfo{journal}{IPN Progress Report} \bibinfo{volume}{42},
  \bibinfo{pages}{1--34}.
\bibitem[{Fornasier et~al.(2011)Fornasier, Clark and
  Dotto}]{2011-Icarus-214-Fornasier}
\bibinfo{author}{Fornasier, S.}, \bibinfo{author}{Clark, B.E.},
  \bibinfo{author}{Dotto, E.}, \bibinfo{year}{2011}.
\newblock \bibinfo{title}{{Spectroscopic survey of X-type asteroids}}.
\newblock \bibinfo{journal}{Icarus} \bibinfo{volume}{214},
  \bibinfo{pages}{131--146}.
\bibitem[{Fraser and Brown(2010)}]{2010-ApJ-714-Fraser}
\bibinfo{author}{Fraser, W.C.}, \bibinfo{author}{Brown, M.E.},
  \bibinfo{year}{2010}.
\newblock \bibinfo{title}{{Quaoar: A Rock in the Kuiper Belt}}.
\newblock \bibinfo{journal}{Astrophysical Journal} \bibinfo{volume}{714},
  \bibinfo{pages}{1547--1550}.
\bibitem[{Fujiwara et~al.(2006)Fujiwara, Kawaguchi, Yeomans, Abe, Mukai, Okada,
  Saito, Yano, Yoshikawa, Scheeres, Barnouin-Jha, Cheng, Demura, Gaskell,
  Hirata, Ikeda, Kominato, Miyamoto, Nakamura, Sasaki and
  Uesugi}]{2006-Science-312-Fujiwara}
\bibinfo{author}{Fujiwara, A.}, \bibinfo{author}{Kawaguchi, J.},
  \bibinfo{author}{Yeomans, D.K.}, \bibinfo{author}{Abe, M.},
  \bibinfo{author}{Mukai, T.}, \bibinfo{author}{Okada, T.},
  \bibinfo{author}{Saito, J.}, \bibinfo{author}{Yano, H.},
  \bibinfo{author}{Yoshikawa, M.}, \bibinfo{author}{Scheeres, D.J.},
  \bibinfo{author}{Barnouin-Jha, O.S.}, \bibinfo{author}{Cheng, A.F.},
  \bibinfo{author}{Demura, H.}, \bibinfo{author}{Gaskell, G.W.},
  \bibinfo{author}{Hirata, N.}, \bibinfo{author}{Ikeda, H.},
  \bibinfo{author}{Kominato, T.}, \bibinfo{author}{Miyamoto, H.},
  \bibinfo{author}{Nakamura, R.}, \bibinfo{author}{Sasaki, S.},
  \bibinfo{author}{Uesugi, K.}, \bibinfo{year}{2006}.
\newblock \bibinfo{title}{{The Rubble-Pile Asteroid Itokawa as Observed by
  Hayabusa}}.
\newblock \bibinfo{journal}{Science} \bibinfo{volume}{312},
  \bibinfo{pages}{1330--1334}.
\bibitem[{Fulvio et~al.(2012)Fulvio, Brunetto, Vernazza and
  Strazzulla}]{2012-AA-537-Fulvio}
\bibinfo{author}{Fulvio, D.}, \bibinfo{author}{Brunetto, R.},
  \bibinfo{author}{Vernazza, P.}, \bibinfo{author}{Strazzulla, G.},
  \bibinfo{year}{2012}.
\newblock \bibinfo{title}{{Space weathering of Vesta and V-type asteroids: new
  irradiation experiments on HED meteorites}}.
\newblock \bibinfo{journal}{Astronomy and Astrophysics} \bibinfo{volume}{537},
  \bibinfo{pages}{L11}.
\bibitem[{Goffin(2001)}]{2001-AA-365-Goffin}
\bibinfo{author}{Goffin, E.}, \bibinfo{year}{2001}.
\newblock \bibinfo{title}{{New determination of the mass of Pallas}}.
\newblock \bibinfo{journal}{Astronomy and Astrophysics} \bibinfo{volume}{365},
  \bibinfo{pages}{627--630}.
\bibitem[{Gounelle et~al.(2008)Gounelle, Morbidelli, Bland, Spurny, Young and
  Sephton}]{2008-SSBN-9-Gounelle}
\bibinfo{author}{Gounelle, M.}, \bibinfo{author}{Morbidelli, A.},
  \bibinfo{author}{Bland, P.A.}, \bibinfo{author}{Spurny, P.},
  \bibinfo{author}{Young, E.D.}, \bibinfo{author}{Sephton, M.},
  \bibinfo{year}{2008}.
\newblock \bibinfo{title}{{Meteorites from the Outer Solar System?}}
\newblock \bibinfo{journal}{The Solar System Beyond Neptune} ,
  \bibinfo{pages}{525--541}.
\bibitem[{Grav et~al.(2011)Grav, Mainzer, Bauer, Masiero, Spahr, McMillan,
  Walker, Cutri, Wright, Eisenhardt, Blauvelt, DeBaun, Elsbury, Gautier,
  Gomillion, Hand and Wilkins}]{2011-ApJ-742-Grav}
\bibinfo{author}{Grav, T.}, \bibinfo{author}{Mainzer, A.K.},
  \bibinfo{author}{Bauer, J.}, \bibinfo{author}{Masiero, J.},
  \bibinfo{author}{Spahr, T.}, \bibinfo{author}{McMillan, R.S.},
  \bibinfo{author}{Walker, R.}, \bibinfo{author}{Cutri, R.},
  \bibinfo{author}{Wright, E.}, \bibinfo{author}{Eisenhardt, P.R.M.},
  \bibinfo{author}{Blauvelt, E.}, \bibinfo{author}{DeBaun, E.},
  \bibinfo{author}{Elsbury, D.}, \bibinfo{author}{Gautier, IV, T.},
  \bibinfo{author}{Gomillion, S.}, \bibinfo{author}{Hand, E.},
  \bibinfo{author}{Wilkins, A.}, \bibinfo{year}{2011}.
\newblock \bibinfo{title}{{WISE/NEOWISE Observations of the Jovian Trojans:
  Preliminary Results}}.
\newblock \bibinfo{journal}{Astrophysical Journal} \bibinfo{volume}{742},
  \bibinfo{pages}{40}.
\bibitem[{Grundy et~al.(2009)Grundy, Noll, Buie, Benecchi, Stephens and
  Levison}]{2009-Icarus-200-Grundy}
\bibinfo{author}{Grundy, W.M.}, \bibinfo{author}{Noll, K.S.},
  \bibinfo{author}{Buie, M.W.}, \bibinfo{author}{Benecchi, S.D.},
  \bibinfo{author}{Stephens, D.C.}, \bibinfo{author}{Levison, H.F.},
  \bibinfo{year}{2009}.
\newblock \bibinfo{title}{{Mutual orbits and masses of six transneptunian
  binaries}}.
\newblock \bibinfo{journal}{Icarus} \bibinfo{volume}{200},
  \bibinfo{pages}{627--635}.
\bibitem[{Grundy et~al.(2008)Grundy, Noll, Virtanen, Muinonen, Kern, Stephens,
  Stansberry, Levison and Spencer}]{2008-Icarus-197-Grundy}
\bibinfo{author}{Grundy, W.M.}, \bibinfo{author}{Noll, K.S.},
  \bibinfo{author}{Virtanen, J.}, \bibinfo{author}{Muinonen, K.},
  \bibinfo{author}{Kern, S.D.}, \bibinfo{author}{Stephens, D.C.},
  \bibinfo{author}{Stansberry, J.A.}, \bibinfo{author}{Levison, H.F.},
  \bibinfo{author}{Spencer, J.R.}, \bibinfo{year}{2008}.
\newblock \bibinfo{title}{{(42355) Typhon Echidna: Scheduling observations for
  binary orbit determination}}.
\newblock \bibinfo{journal}{Icarus} \bibinfo{volume}{197},
  \bibinfo{pages}{260--268}.
\bibitem[{Grundy et~al.(2007)Grundy, Stansberry, Noll, Stephens, Trilling,
  Kern, Spencer, Cruikshank and Levison}]{2007-Icarus-191-Grundy}
\bibinfo{author}{Grundy, W.M.}, \bibinfo{author}{Stansberry, J.A.},
  \bibinfo{author}{Noll, K.S.}, \bibinfo{author}{Stephens, D.C.},
  \bibinfo{author}{Trilling, D.E.}, \bibinfo{author}{Kern, S.D.},
  \bibinfo{author}{Spencer, J.R.}, \bibinfo{author}{Cruikshank, D.P.},
  \bibinfo{author}{Levison, H.F.}, \bibinfo{year}{2007}.
\newblock \bibinfo{title}{{The orbit, mass, size, albedo, and density of
  (65489) Ceto/Phorcys: A tidally-evolved binary Centaur}}.
\newblock \bibinfo{journal}{Icarus} \bibinfo{volume}{191},
  \bibinfo{pages}{286--297}.
\bibitem[{Halliday and Kleine(2006)}]{2006-MESS2-Halliday}
\bibinfo{author}{Halliday, A.N.}, \bibinfo{author}{Kleine, T.},
  \bibinfo{year}{2006}.
\newblock \bibinfo{title}{{Meteorites and the Timing, Mechanisms, and
  Conditions of Terrestrial Planet Accretion and Early Differentiation}}.
\newblock \bibinfo{journal}{Meteorites and the Early Solar System II} ,
  \bibinfo{pages}{775--801}.
\bibitem[{Harris(1998)}]{1998-Icarus-131-Harris}
\bibinfo{author}{Harris, A.W.}, \bibinfo{year}{1998}.
\newblock \bibinfo{title}{{A Thermal Model for Near-Earth Asteroids}}.
\newblock \bibinfo{journal}{Icarus} \bibinfo{volume}{131},
  \bibinfo{pages}{291--301}.
\bibitem[{Harris and Davies(1999)}]{1999-Icarus-142-Harris}
\bibinfo{author}{Harris, A.W.}, \bibinfo{author}{Davies, J.K.},
  \bibinfo{year}{1999}.
\newblock \bibinfo{title}{{Physical Characteristics of Near-Earth Asteroids
  from Thermal Infrared Spectrophotometry}}.
\newblock \bibinfo{journal}{Icarus} \bibinfo{volume}{142},
  \bibinfo{pages}{464--475}.
\bibitem[{Hilton(1999)}]{1999-AJ-117-Hilton}
\bibinfo{author}{Hilton, J.L.}, \bibinfo{year}{1999}.
\newblock \bibinfo{title}{{US Naval Observatory Ephemerides of the Largest
  Asteroids}}.
\newblock \bibinfo{journal}{Astronomical Journal} \bibinfo{volume}{117},
  \bibinfo{pages}{1077--1086}.
\bibitem[{Hilton(2002)}]{2002-AsteroidsIII-2.2-Hilton}
\bibinfo{author}{Hilton, J.L.}, \bibinfo{year}{2002}.
\newblock \bibinfo{title}{{Asteroid Masses and Densities}}.
\newblock \bibinfo{journal}{Asteroids III} , \bibinfo{pages}{103--112}.
\bibitem[{Ivantsov(2008)}]{2008-PSS-56-Ivantsov}
\bibinfo{author}{Ivantsov, A.}, \bibinfo{year}{2008}.
\newblock \bibinfo{title}{{Asteroid mass determination at Nikolaev
  Observatory}}.
\newblock \bibinfo{journal}{Planetary and Space Science} \bibinfo{volume}{56},
  \bibinfo{pages}{1857--1861}.
\bibitem[{Jenniskens et~al.(2009)Jenniskens, Shaddad, Numan, Elsir, Kudoda,
  Zolensky, Le, Robinson, Friedrich, Rumble, Steele, Chesley, Fitzsimmons,
  Duddy, Hsieh, Ramsay, Brown, Edwards, Tagliaferri, Boslough, Spalding,
  Dantowitz, Kozubal, Pravec, Borovicka, Charvat, Vaubaillon, Kuiper, Albers,
  Bishop, Mancinelli, Sandford, Milam, Nuevo and
  P.}]{2009-Nature-458-Jenniskens}
\bibinfo{author}{Jenniskens, P.}, \bibinfo{author}{Shaddad, M.H.},
  \bibinfo{author}{Numan, D.}, \bibinfo{author}{Elsir, S.},
  \bibinfo{author}{Kudoda, A.M.}, \bibinfo{author}{Zolensky, M.E.},
  \bibinfo{author}{Le, L.}, \bibinfo{author}{Robinson, G.A.},
  \bibinfo{author}{Friedrich, J.M.}, \bibinfo{author}{Rumble, D.},
  \bibinfo{author}{Steele, A.}, \bibinfo{author}{Chesley, S.R.},
  \bibinfo{author}{Fitzsimmons, A.}, \bibinfo{author}{Duddy, S.},
  \bibinfo{author}{Hsieh, H.H.}, \bibinfo{author}{Ramsay, G.},
  \bibinfo{author}{Brown, P.G.}, \bibinfo{author}{Edwards, W.N.},
  \bibinfo{author}{Tagliaferri, E.}, \bibinfo{author}{Boslough, M.B.},
  \bibinfo{author}{Spalding, R.E.}, \bibinfo{author}{Dantowitz, R.},
  \bibinfo{author}{Kozubal, M.}, \bibinfo{author}{Pravec, P.},
  \bibinfo{author}{Borovicka, J.}, \bibinfo{author}{Charvat, Z.},
  \bibinfo{author}{Vaubaillon, J.}, \bibinfo{author}{Kuiper, J.},
  \bibinfo{author}{Albers, J.}, \bibinfo{author}{Bishop, J.L.},
  \bibinfo{author}{Mancinelli, R.L.}, \bibinfo{author}{Sandford, S.A.},
  \bibinfo{author}{Milam, S.N.}, \bibinfo{author}{Nuevo, M.},
  \bibinfo{author}{P., W.S.}, \bibinfo{year}{2009}.
\newblock \bibinfo{title}{{The impact and recovery of asteroid 2008 TC$_{3}$}}.
\newblock \bibinfo{journal}{Nature} \bibinfo{volume}{458},
  \bibinfo{pages}{485--488}.
\bibitem[{Jewitt(2004)}]{2004-CometsII-7-Jewitt}
\bibinfo{author}{Jewitt, D.C.}, \bibinfo{year}{2004}.
\newblock \bibinfo{title}{{From cradle to grave: the rise and demise of the
  comets}} , \bibinfo{pages}{659--676}.
\bibitem[{Jones et~al.(1990)Jones, Lebofsky, Lewis and
  Marley}]{1990-Icarus-88-Jones}
\bibinfo{author}{Jones, T.D.}, \bibinfo{author}{Lebofsky, L.A.},
  \bibinfo{author}{Lewis, J.S.}, \bibinfo{author}{Marley, M.S.},
  \bibinfo{year}{1990}.
\newblock \bibinfo{title}{{The composition and Origin of the C,P and D
  Asteroids: Water as Tracer of Thermal Evolution in the Outer Belt}}.
\newblock \bibinfo{journal}{Icarus} \bibinfo{volume}{88},
  \bibinfo{pages}{172--193}.
\bibitem[{Kaasalainen(2011)}]{2011-IPI-5-Kaasalainen}
\bibinfo{author}{Kaasalainen, M.}, \bibinfo{year}{2011}.
\newblock \bibinfo{title}{{Maximum compatibility estimates and shape
  reconstruction with boundary curves and volumes of generalized projections}}.
\newblock \bibinfo{journal}{Inverse Problems and Imaging} \bibinfo{volume}{5},
  \bibinfo{pages}{37--57}.
\bibitem[{Kern and Elliot(2005)}]{2005-DPS-37-Kern}
\bibinfo{author}{Kern, S.D.}, \bibinfo{author}{Elliot, J.L.},
  \bibinfo{year}{2005}.
\newblock \bibinfo{title}{{Studies of Kuiper Belt Binaries}}, in:
  \bibinfo{booktitle}{AAS/Division for Planetary Sciences Meeting Abstracts
  \#37}, p. \bibinfo{pages}{738}.
\bibitem[{Kern and Elliot(2006)}]{2006-Icarus-183-Kern}
\bibinfo{author}{Kern, S.D.}, \bibinfo{author}{Elliot, J.L.},
  \bibinfo{year}{2006}.
\newblock \bibinfo{title}{{Discovery and characteristics of the Kuiper belt
  binary 2003QY90}}.
\newblock \bibinfo{journal}{Icarus} \bibinfo{volume}{183},
  \bibinfo{pages}{179--185}.
\bibitem[{{King} et~al.(2011){King}, {Izawa}, {Vernazza}, {McCutcheon},
  {Berger} and {Dunn}}]{2011-LPI-King}
\bibinfo{author}{{King}, P.L.}, \bibinfo{author}{{Izawa}, M.R.M.},
  \bibinfo{author}{{Vernazza}, P.}, \bibinfo{author}{{McCutcheon}, W.A.},
  \bibinfo{author}{{Berger}, J.A.}, \bibinfo{author}{{Dunn}, T.},
  \bibinfo{year}{2011}.
\newblock \bibinfo{title}{{Salt: A Critical Material to Consider when Exploring
  the Solar System}}.
\newblock \bibinfo{journal}{Lunar and Planetary Institute Science Conference
  Abstracts} \bibinfo{volume}{42}, \bibinfo{pages}{1985}.
\bibitem[{Kochetova(2004)}]{2004-SoSyR-38-Kochetova}
\bibinfo{author}{Kochetova, O.M.}, \bibinfo{year}{2004}.
\newblock \bibinfo{title}{{Determination of Large Asteroid Masses by the
  Dynamical Method}}.
\newblock \bibinfo{journal}{Solar System Research} \bibinfo{volume}{38},
  \bibinfo{pages}{66--75}.
\bibitem[{Konopliv et~al.(2011)Konopliv, Asmar, Folkner, Karatekin, Nunes,
  Smrekar, Yoder and Zuber}]{2011-Icarus-211-Konopliv}
\bibinfo{author}{Konopliv, A.S.}, \bibinfo{author}{Asmar, S.W.},
  \bibinfo{author}{Folkner, W.M.}, \bibinfo{author}{Karatekin, {\"O}.},
  \bibinfo{author}{Nunes, D.C.}, \bibinfo{author}{Smrekar, S.E.},
  \bibinfo{author}{Yoder, C.F.}, \bibinfo{author}{Zuber, M.T.},
  \bibinfo{year}{2011}.
\newblock \bibinfo{title}{{Mars high resolution gravity fields from MRO, Mars
  seasonal gravity, and other dynamical parameters}}.
\newblock \bibinfo{journal}{Icarus} \bibinfo{volume}{211},
  \bibinfo{pages}{401--428}.
\bibitem[{Konopliv et~al.(2002)Konopliv, Miller, Owen, Yeomans, Giorgini,
  Garmier and Barriot}]{2002-Icarus-160-Konopliv}
\bibinfo{author}{Konopliv, A.S.}, \bibinfo{author}{Miller, J.K.},
  \bibinfo{author}{Owen, W.M.}, \bibinfo{author}{Yeomans, D.K.},
  \bibinfo{author}{Giorgini, J.D.}, \bibinfo{author}{Garmier, R.},
  \bibinfo{author}{Barriot, J.P.}, \bibinfo{year}{2002}.
\newblock \bibinfo{title}{{A Global Solution for the Gravity Field, Rotation,
  Landmarks, and Ephemeris of Eros}}.
\newblock \bibinfo{journal}{Icarus} \bibinfo{volume}{160},
  \bibinfo{pages}{289--299}.
\bibitem[{Konopliv et~al.(2006)Konopliv, Yoder, Standish, Yuan and
  Sjogren}]{2006-Icarus-182-Konopliv}
\bibinfo{author}{Konopliv, A.S.}, \bibinfo{author}{Yoder, C.F.},
  \bibinfo{author}{Standish, E.M.}, \bibinfo{author}{Yuan, D.N.},
  \bibinfo{author}{Sjogren, W.L.}, \bibinfo{year}{2006}.
\newblock \bibinfo{title}{{A global solution for the Mars static and seasonal
  gravity, Mars orientation, Phobos and Deimos masses, and Mars ephemeris}}.
\newblock \bibinfo{journal}{Icarus} \bibinfo{volume}{182},
  \bibinfo{pages}{23--50}.
\bibitem[{Kova{\v c}evi{\'c}(2005)}]{2005-AA-430-Kovacevic}
\bibinfo{author}{Kova{\v c}evi{\'c}, A.}, \bibinfo{year}{2005}.
\newblock \bibinfo{title}{{Determination of the mass of (4) Vesta based on new
  close approaches}}.
\newblock \bibinfo{journal}{Astronomy and Astrophysics} \bibinfo{volume}{430},
  \bibinfo{pages}{319--325}.
\bibitem[{Kova{\v c}evi{\'c} and Kuzmanoski(2007)}]{2007-EMP-100-Kovacevic}
\bibinfo{author}{Kova{\v c}evi{\'c}, A.}, \bibinfo{author}{Kuzmanoski, M.},
  \bibinfo{year}{2007}.
\newblock \bibinfo{title}{{A New Determination of the Mass of (1) Ceres}}.
\newblock \bibinfo{journal}{Earth Moon and Planets} \bibinfo{volume}{100},
  \bibinfo{pages}{117--123}.
\bibitem[{Krasinsky et~al.(2001)Krasinsky, Pitjeva, Vasiliev and
  Yagudina}]{2001-IAA-Krasinsky}
\bibinfo{author}{Krasinsky, G.A.}, \bibinfo{author}{Pitjeva, E.V.},
  \bibinfo{author}{Vasiliev, M.V.}, \bibinfo{author}{Yagudina, E.I.},
  \bibinfo{year}{2001}.
\newblock \bibinfo{title}{{Estimating masses of asteroids}}, in:
  \bibinfo{booktitle}{Communications of IAA of RAS}.
\bibitem[{Kryszczy{\'n}ska et~al.(2009)Kryszczy{\'n}ska, Colas, Descamps,
  Bartczak, Poli{\'n}ska, Kwiatkowski, Lecacheux, Hirsch, Fagas, Kami{\'n}ski,
  Micha{\l}owski and Marciniak}]{2009-AA-501-Kryszczynska}
\bibinfo{author}{Kryszczy{\'n}ska, A.}, \bibinfo{author}{Colas, F.},
  \bibinfo{author}{Descamps, P.}, \bibinfo{author}{Bartczak, P.},
  \bibinfo{author}{Poli{\'n}ska, M.}, \bibinfo{author}{Kwiatkowski, T.},
  \bibinfo{author}{Lecacheux, J.}, \bibinfo{author}{Hirsch, R.},
  \bibinfo{author}{Fagas, M.}, \bibinfo{author}{Kami{\'n}ski, K.},
  \bibinfo{author}{Micha{\l}owski, T.}, \bibinfo{author}{Marciniak, A.},
  \bibinfo{year}{2009}.
\newblock \bibinfo{title}{{New binary asteroid 809 Lundia. I. Photometry and
  modelling}}.
\newblock \bibinfo{journal}{Astronomy and Astrophysics} \bibinfo{volume}{501},
  \bibinfo{pages}{769--776}.
\bibitem[{Kuzmanoski(1996)}]{1996-IAUS-172-Kuzmanoski}
\bibinfo{author}{Kuzmanoski, M.}, \bibinfo{year}{1996}.
\newblock \bibinfo{title}{{A method for asteroid mass determination}}, in:
  \bibinfo{editor}{{S.~Ferraz-Mello, B.~Morando, \& J.-E.~Arlot}} (Ed.),
  \bibinfo{booktitle}{Dynamics, Ephemerides, and Astrometry of the Solar
  System}, pp. \bibinfo{pages}{207--+}.
\bibitem[{Kuzmanoski et~al.(2010)Kuzmanoski, Apostolovska and
  Novakovi{\'c}}]{2010-AJ-140-Kuzmanoski}
\bibinfo{author}{Kuzmanoski, M.}, \bibinfo{author}{Apostolovska, G.},
  \bibinfo{author}{Novakovi{\'c}, B.}, \bibinfo{year}{2010}.
\newblock \bibinfo{title}{{The Mass of (4) Vesta Derived from its Largest
  Gravitational Effects}}.
\newblock \bibinfo{journal}{Astronomical Journal} \bibinfo{volume}{140},
  \bibinfo{pages}{880--886}.
\bibitem[{Kuzmanoski and Kova{\v c}evi{\'c}(2002)}]{2002-AA-395-Kuzmanoski}
\bibinfo{author}{Kuzmanoski, M.}, \bibinfo{author}{Kova{\v c}evi{\'c}, A.},
  \bibinfo{year}{2002}.
\newblock \bibinfo{title}{{Motion of the asteroid (13206) 1997GC22 and the mass
  of (16) Psyche}}.
\newblock \bibinfo{journal}{Astronomy and Astrophysics} \bibinfo{volume}{395},
  \bibinfo{pages}{L17--L19}.
\bibitem[{Lagerros(1996)}]{1996-AA-310-Lagerros}
\bibinfo{author}{Lagerros, J.S.V.}, \bibinfo{year}{1996}.
\newblock \bibinfo{title}{{Thermal physics of asteroids. I. Effects of shape,
  heat conduction and beaming}}.
\newblock \bibinfo{journal}{Astronomy and Astrophysics} \bibinfo{volume}{310},
  \bibinfo{pages}{1011--1020}.
\bibitem[{Lagerros(1997)}]{1997-AA-325-Lagerros}
\bibinfo{author}{Lagerros, J.S.V.}, \bibinfo{year}{1997}.
\newblock \bibinfo{title}{{Thermal physics of asteroids. III. Irregular shapes
  and albedo variegations.}}
\newblock \bibinfo{journal}{Astronomy and Astrophysics} \bibinfo{volume}{325},
  \bibinfo{pages}{1226--1236}.
\bibitem[{Lamy et~al.(2004)Lamy, Toth, Fernandez and
  Weaver}]{2004-CometsII-4-Lamy}
\bibinfo{author}{Lamy, P.L.}, \bibinfo{author}{Toth, I.},
  \bibinfo{author}{Fernandez, Y.R.}, \bibinfo{author}{Weaver, H.A.},
  \bibinfo{year}{2004}.
\newblock \bibinfo{title}{{The sizes, shapes, albedos, and colors of cometary
  nuclei}}.
\newblock \bibinfo{journal}{Comets II} , \bibinfo{pages}{223--264}.
\bibitem[{Lamy et~al.(2008)Lamy, Toth, Groussin, Jorda, Kelley and
  Stansberry}]{2008-AA-489-Lamy}
\bibinfo{author}{Lamy, P.L.}, \bibinfo{author}{Toth, I.},
  \bibinfo{author}{Groussin, O.}, \bibinfo{author}{Jorda, L.},
  \bibinfo{author}{Kelley, M.S.}, \bibinfo{author}{Stansberry, J.A.},
  \bibinfo{year}{2008}.
\newblock \bibinfo{title}{{Spitzer Space Telescope observations of the nucleus
  of comet 67P/Churyumov-Gerasimenko}}.
\newblock \bibinfo{journal}{Astronomy and Astrophysics} \bibinfo{volume}{489},
  \bibinfo{pages}{777--785}.
\bibitem[{Lamy et~al.(2006)Lamy, Toth, Weaver, Jorda, Kaasalainen and
  Guti{\'e}rrez}]{2006-AA-458-Lamy}
\bibinfo{author}{Lamy, P.L.}, \bibinfo{author}{Toth, I.},
  \bibinfo{author}{Weaver, H.A.}, \bibinfo{author}{Jorda, L.},
  \bibinfo{author}{Kaasalainen, M.}, \bibinfo{author}{Guti{\'e}rrez, P.J.},
  \bibinfo{year}{2006}.
\newblock \bibinfo{title}{{Hubble Space Telescope observations of the nucleus
  and inner coma of comet 67P/Churyumov-Gerasimenko}}.
\newblock \bibinfo{journal}{Astronomy and Astrophysics} \bibinfo{volume}{458},
  \bibinfo{pages}{669--678}.
\bibitem[{Landgraf(1992)}]{1992-IAUS-152-Landgraf}
\bibinfo{author}{Landgraf, W.}, \bibinfo{year}{1992}.
\newblock \bibinfo{title}{{A Determination of the Mass of (704) Interamnia from
  Observations of (993) Moultona}}, in: \bibinfo{editor}{{S.~Ferraz-Mello}}
  (Ed.), \bibinfo{booktitle}{Chaos, Resonance, and Collective Dynamical
  Phenomena in the Solar System}, pp. \bibinfo{pages}{179--183}.
\bibitem[{Lebofsky(1978)}]{1978-MNRAS-182-Lebofsky}
\bibinfo{author}{Lebofsky, L.A.}, \bibinfo{year}{1978}.
\newblock \bibinfo{title}{{Asteroid 1 Ceres - Evidence for water of
  hydration}}.
\newblock \bibinfo{journal}{Monthly Notices of the Royal Astronomical Society}
  \bibinfo{volume}{182}, \bibinfo{pages}{17--21}.
\bibitem[{Lebofsky et~al.(1986)Lebofsky, Sykes, Tedesco, Veeder, Matson, Brown,
  Gradie, Feierberg and Rudy}]{1986-Icarus-68-Lebofsky}
\bibinfo{author}{Lebofsky, L.A.}, \bibinfo{author}{Sykes, M.V.},
  \bibinfo{author}{Tedesco, E.F.}, \bibinfo{author}{Veeder, G.J.},
  \bibinfo{author}{Matson, D.L.}, \bibinfo{author}{Brown, R.H.},
  \bibinfo{author}{Gradie, J.C.}, \bibinfo{author}{Feierberg, M.A.},
  \bibinfo{author}{Rudy, R.J.}, \bibinfo{year}{1986}.
\newblock \bibinfo{title}{{A refined 'standard' thermal model for asteroids
  based on observations of 1 Ceres and 2 Pallas}}.
\newblock \bibinfo{journal}{Icarus} \bibinfo{volume}{68},
  \bibinfo{pages}{239--251}.
\bibitem[{Lellouch et~al.(2010)Lellouch, Kiss, Santos-Sanz, M{\"u}ller,
  Fornasier, Groussin, Lacerda, Ortiz, Thirouin, Delsanti, Duffard, Harris,
  Henry, Lim, Moreno, Mommert, Mueller, Protopapa, Stansberry, Trilling,
  Vilenius, Barucci, Crovisier, Doressoundiram, Dotto, Guti{\'e}rrez, Hainaut,
  Hartogh, Hestroffer, Horner, Jorda, Kidger, Lara, Rengel, Swinyard and
  Thomas}]{2010-AA-518-Lellouch}
\bibinfo{author}{Lellouch, E.}, \bibinfo{author}{Kiss, C.},
  \bibinfo{author}{Santos-Sanz, P.}, \bibinfo{author}{M{\"u}ller, T.G.},
  \bibinfo{author}{Fornasier, S.}, \bibinfo{author}{Groussin, O.},
  \bibinfo{author}{Lacerda, P.}, \bibinfo{author}{Ortiz, J.L.},
  \bibinfo{author}{Thirouin, A.}, \bibinfo{author}{Delsanti, A.},
  \bibinfo{author}{Duffard, R.}, \bibinfo{author}{Harris, A.W.},
  \bibinfo{author}{Henry, F.}, \bibinfo{author}{Lim, T.},
  \bibinfo{author}{Moreno, R.}, \bibinfo{author}{Mommert, M.},
  \bibinfo{author}{Mueller, M.}, \bibinfo{author}{Protopapa, S.},
  \bibinfo{author}{Stansberry, J.}, \bibinfo{author}{Trilling, D.},
  \bibinfo{author}{Vilenius, E.}, \bibinfo{author}{Barucci, A.},
  \bibinfo{author}{Crovisier, J.}, \bibinfo{author}{Doressoundiram, A.},
  \bibinfo{author}{Dotto, E.}, \bibinfo{author}{Guti{\'e}rrez, P.J.},
  \bibinfo{author}{Hainaut, O.R.}, \bibinfo{author}{Hartogh, P.},
  \bibinfo{author}{Hestroffer, D.}, \bibinfo{author}{Horner, J.},
  \bibinfo{author}{Jorda, L.}, \bibinfo{author}{Kidger, M.},
  \bibinfo{author}{Lara, L.M.}, \bibinfo{author}{Rengel, M.},
  \bibinfo{author}{Swinyard, B.M.}, \bibinfo{author}{Thomas, N.},
  \bibinfo{year}{2010}.
\newblock \bibinfo{title}{{``TNOs are cool'': A survey of the trans-Neptunian
  region. II. The thermal lightcurve of (136108) Haumea}}.
\newblock \bibinfo{journal}{Astronomy and Astrophysics} \bibinfo{volume}{518},
  \bibinfo{pages}{L147}.
\bibitem[{Levison et~al.(2009)Levison, Bottke, Gounelle, Morbidelli,
  Nesvorn{\'y} and Tsiganis}]{2009-Nature-460-Levison}
\bibinfo{author}{Levison, H.F.}, \bibinfo{author}{Bottke, W.F.},
  \bibinfo{author}{Gounelle, M.}, \bibinfo{author}{Morbidelli, A.},
  \bibinfo{author}{Nesvorn{\'y}, D.}, \bibinfo{author}{Tsiganis, K.},
  \bibinfo{year}{2009}.
\newblock \bibinfo{title}{{Contamination of the asteroid belt by primordial
  trans-Neptunian objects}}.
\newblock \bibinfo{journal}{Nature} \bibinfo{volume}{460},
  \bibinfo{pages}{364--366}.
\bibitem[{Lim et~al.(2010)Lim, Stansberry, M{\"u}ller, Mueller, Lellouch, Kiss,
  Santos-Sanz, Vilenius, Protopapa, Moreno, Delsanti, Duffard, Fornasier,
  Groussin, Harris, Henry, Horner, Lacerda, Mommert, Ortiz, Rengel, Thirouin,
  Trilling, Barucci, Crovisier, Doressoundiram, Dotto,
  Guti{\'e}rrez~Buenestado, Hainaut, Hartogh, Hestroffer, Kidger, Lara,
  Swinyard and Thomas}]{2010-AA-518-Lim}
\bibinfo{author}{Lim, T.L.}, \bibinfo{author}{Stansberry, J.},
  \bibinfo{author}{M{\"u}ller, T.G.}, \bibinfo{author}{Mueller, M.},
  \bibinfo{author}{Lellouch, E.}, \bibinfo{author}{Kiss, C.},
  \bibinfo{author}{Santos-Sanz, P.}, \bibinfo{author}{Vilenius, E.},
  \bibinfo{author}{Protopapa, S.}, \bibinfo{author}{Moreno, R.},
  \bibinfo{author}{Delsanti, A.}, \bibinfo{author}{Duffard, R.},
  \bibinfo{author}{Fornasier, S.}, \bibinfo{author}{Groussin, O.},
  \bibinfo{author}{Harris, A.W.}, \bibinfo{author}{Henry, F.},
  \bibinfo{author}{Horner, J.}, \bibinfo{author}{Lacerda, P.},
  \bibinfo{author}{Mommert, M.}, \bibinfo{author}{Ortiz, J.L.},
  \bibinfo{author}{Rengel, M.}, \bibinfo{author}{Thirouin, A.},
  \bibinfo{author}{Trilling, D.}, \bibinfo{author}{Barucci, A.},
  \bibinfo{author}{Crovisier, J.}, \bibinfo{author}{Doressoundiram, A.},
  \bibinfo{author}{Dotto, E.}, \bibinfo{author}{Guti{\'e}rrez~Buenestado,
  P.J.}, \bibinfo{author}{Hainaut, O.R.}, \bibinfo{author}{Hartogh, P.},
  \bibinfo{author}{Hestroffer, D.}, \bibinfo{author}{Kidger, M.},
  \bibinfo{author}{Lara, L.M.}, \bibinfo{author}{Swinyard, B.M.},
  \bibinfo{author}{Thomas, N.}, \bibinfo{year}{2010}.
\newblock \bibinfo{title}{{``TNOs are Cool'': A survey of the trans-Neptunian
  region . III. Thermophysical properties of 90482 Orcus and 136472 Makemake}}.
\newblock \bibinfo{journal}{Astronomy and Astrophysics} \bibinfo{volume}{518},
  \bibinfo{pages}{L148}.
\bibitem[{Lopez~Garcia et~al.(1997)Lopez~Garcia, Medvedev and
  Morano~Fernendez}]{1997-IAU-LopezGarcia}
\bibinfo{author}{Lopez~Garcia, A.}, \bibinfo{author}{Medvedev, Y.D.},
  \bibinfo{author}{Morano~Fernendez, J.A.}, \bibinfo{year}{1997}.
\newblock \bibinfo{title}{{Using Close Encounters of Minor Planets for the
  Improvement of their Masses}}, in: \bibinfo{editor}{{I.~M.~Wytrzyszczak,
  J.~H.~Lieske, \& R.~A.~Feldman}} (Ed.), \bibinfo{booktitle}{IAU Colloq. 165:
  Dynamics and Astrometry of Natural and Artificial Celestial Bodies}, p.
  \bibinfo{pages}{199}.
\bibitem[{Macke et~al.(2011)Macke, Britt and Consolmagno}]{2011-MPS-46-Macke}
\bibinfo{author}{Macke, R.J.}, \bibinfo{author}{Britt, D.T.},
  \bibinfo{author}{Consolmagno, G.J.}, \bibinfo{year}{2011}.
\newblock \bibinfo{title}{{Density, porosity, and magnetic susceptibility of
  achondritic meteorites}}.
\newblock \bibinfo{journal}{Meteoritics and Planetary Science}
  \bibinfo{volume}{46}, \bibinfo{pages}{311--326}.
\bibitem[{Macke et~al.(2010)Macke, Consolmagno, Britt and
  Hutson}]{2010-MPS-45-Macke}
\bibinfo{author}{Macke, R.J.}, \bibinfo{author}{Consolmagno, G.J.},
  \bibinfo{author}{Britt, D.T.}, \bibinfo{author}{Hutson, M.L.},
  \bibinfo{year}{2010}.
\newblock \bibinfo{title}{{Enstatite chondrite density, magnetic
  susceptibility, and porosity}}.
\newblock \bibinfo{journal}{Meteoritics and Planetary Science}
  \bibinfo{volume}{45}, \bibinfo{pages}{1513--1526}.
\bibitem[{Marchis et~al.(2008a)Marchis, Descamps, Baek, Harris, Kaasalainen,
  Berthier, Hestroffer and Vachier}]{2008-Icarus-196-Marchis}
\bibinfo{author}{Marchis, F.}, \bibinfo{author}{Descamps, P.},
  \bibinfo{author}{Baek, M.}, \bibinfo{author}{Harris, A.W.},
  \bibinfo{author}{Kaasalainen, M.}, \bibinfo{author}{Berthier, J.},
  \bibinfo{author}{Hestroffer, D.}, \bibinfo{author}{Vachier, F.},
  \bibinfo{year}{2008}a.
\newblock \bibinfo{title}{{Main belt binary asteroidal systems with circular
  mutual orbits}}.
\newblock \bibinfo{journal}{Icarus} \bibinfo{volume}{196},
  \bibinfo{pages}{97--118}.
\bibitem[{Marchis et~al.(2008b)Marchis, Descamps, Berthier, Hestroffer,
  Vachier, Baek, Harris and Nesvorn{\'y}}]{2008-Icarus-195-Marchis}
\bibinfo{author}{Marchis, F.}, \bibinfo{author}{Descamps, P.},
  \bibinfo{author}{Berthier, J.}, \bibinfo{author}{Hestroffer, D.},
  \bibinfo{author}{Vachier, F.}, \bibinfo{author}{Baek, M.},
  \bibinfo{author}{Harris, A.W.}, \bibinfo{author}{Nesvorn{\'y}, D.},
  \bibinfo{year}{2008}b.
\newblock \bibinfo{title}{{Main belt binary asteroidal systems with eccentric
  mutual orbits}}.
\newblock \bibinfo{journal}{Icarus} \bibinfo{volume}{195},
  \bibinfo{pages}{295--316}.
\bibitem[{Marchis et~al.(2011)Marchis, Descamps, Dalba, Enriquez, Durech,
  Emery, Berthier, Vachier, Merlbourne, Stockton, Fassnacht and
  Dupuy}]{2011-DPS-Marchis}
\bibinfo{author}{Marchis, F.}, \bibinfo{author}{Descamps, P.},
  \bibinfo{author}{Dalba, P.}, \bibinfo{author}{Enriquez, J.E.},
  \bibinfo{author}{Durech, J.}, \bibinfo{author}{Emery, J.P.},
  \bibinfo{author}{Berthier, J.}, \bibinfo{author}{Vachier, F.},
  \bibinfo{author}{Merlbourne, J.}, \bibinfo{author}{Stockton, A.N.},
  \bibinfo{author}{Fassnacht, C.D.}, \bibinfo{author}{Dupuy, T.J.},
  \bibinfo{year}{2011}.
\newblock \bibinfo{title}{{A Detailed Picture of the (93) Minerva Triple
  System}}, in: \bibinfo{booktitle}{EPSC-DPS Joint Meeting 2011}, p.
  \bibinfo{pages}{653}.
\bibitem[{Marchis et~al.(2005a)Marchis, Descamps, Hestroffer and
  Berthier}]{2005-Nature-436-Marchis}
\bibinfo{author}{Marchis, F.}, \bibinfo{author}{Descamps, P.},
  \bibinfo{author}{Hestroffer, D.}, \bibinfo{author}{Berthier, J.},
  \bibinfo{year}{2005}a.
\newblock \bibinfo{title}{{Discovery of the triple asteroidal system 87
  Sylvia}}.
\newblock \bibinfo{journal}{Nature} \bibinfo{volume}{436},
  \bibinfo{pages}{822--824}.
\bibitem[{Marchis et~al.(2006a)Marchis, Hestroffer, Descamps, Berthier,
  Bouchez, Campbell, Chin, van Dam, Hartman, Johansson, Lafon, Le~Mignant,
  de~Pater, Stomski~Jr., Summers, Vachier, Wizinowich and
  Wong}]{2006-Nature-439-Marchis}
\bibinfo{author}{Marchis, F.}, \bibinfo{author}{Hestroffer, D.},
  \bibinfo{author}{Descamps, P.}, \bibinfo{author}{Berthier, J.},
  \bibinfo{author}{Bouchez, A.H.}, \bibinfo{author}{Campbell, R.D.},
  \bibinfo{author}{Chin, J.C.Y.}, \bibinfo{author}{van Dam, M.A.},
  \bibinfo{author}{Hartman, S.K.}, \bibinfo{author}{Johansson, E.M.},
  \bibinfo{author}{Lafon, R.E.}, \bibinfo{author}{Le~Mignant, D.},
  \bibinfo{author}{de~Pater, I.}, \bibinfo{author}{Stomski~Jr., P.J.},
  \bibinfo{author}{Summers, D.M.}, \bibinfo{author}{Vachier, F.},
  \bibinfo{author}{Wizinowich, P.L.}, \bibinfo{author}{Wong, M.H.},
  \bibinfo{year}{2006}a.
\newblock \bibinfo{title}{{A low density of 0.8gcm-3 for the Trojan binary
  asteroid 617 Patroclus}}.
\newblock \bibinfo{journal}{Nature} \bibinfo{volume}{439},
  \bibinfo{pages}{565--567}.
\bibitem[{Marchis et~al.(2005b)Marchis, Hestroffer, Descamps, Berthier, Laver
  and de~Pater}]{2005-Icarus-178-Marchis}
\bibinfo{author}{Marchis, F.}, \bibinfo{author}{Hestroffer, D.},
  \bibinfo{author}{Descamps, P.}, \bibinfo{author}{Berthier, J.},
  \bibinfo{author}{Laver, C.}, \bibinfo{author}{de~Pater, I.},
  \bibinfo{year}{2005}b.
\newblock \bibinfo{title}{{Mass and density of Asteroid 121 Hermione from an
  analysis of its companion orbit}}.
\newblock \bibinfo{journal}{Icarus} \bibinfo{volume}{178},
  \bibinfo{pages}{450--464}.
\bibitem[{Marchis et~al.(2006b)Marchis, Kaasalainen, Hom, Berthier, Enriquez,
  Hestroffer, Le~Mignant and de~Pater}]{2006-Icarus-185-Marchis}
\bibinfo{author}{Marchis, F.}, \bibinfo{author}{Kaasalainen, M.},
  \bibinfo{author}{Hom, E.F.Y.}, \bibinfo{author}{Berthier, J.},
  \bibinfo{author}{Enriquez, J.}, \bibinfo{author}{Hestroffer, D.},
  \bibinfo{author}{Le~Mignant, D.}, \bibinfo{author}{de~Pater, I.},
  \bibinfo{year}{2006}b.
\newblock \bibinfo{title}{{Shape, size and multiplicity of main-belt
  asteroids}}.
\newblock \bibinfo{journal}{Icarus} \bibinfo{volume}{185},
  \bibinfo{pages}{39--63}.
\bibitem[{Margot and Brown(2001)}]{2001-DPS-33-Margot}
\bibinfo{author}{Margot, J.L.}, \bibinfo{author}{Brown, M.E.},
  \bibinfo{year}{2001}.
\newblock \bibinfo{title}{{Discovery and characterization of binary asteroids
  22 Kalliope and 87 Sylvia}}, in: \bibinfo{booktitle}{AAS/Division for
  Planetary Sciences Meeting Abstracts \#33}, p. \bibinfo{pages}{1133}.
\bibitem[{Margot and Brown(2003)}]{2003-Science-300-Margot}
\bibinfo{author}{Margot, J.L.}, \bibinfo{author}{Brown, M.E.},
  \bibinfo{year}{2003}.
\newblock \bibinfo{title}{{A Low-Density M-type Asteroid in the Main Belt}}.
\newblock \bibinfo{journal}{Science} \bibinfo{volume}{300},
  \bibinfo{pages}{1939--1942}.
\bibitem[{Margot et~al.(2004)Margot, Brown, Trujillo and
  Sari}]{2004-DPS-36-Margot}
\bibinfo{author}{Margot, J.L.}, \bibinfo{author}{Brown, M.E.},
  \bibinfo{author}{Trujillo, C.A.}, \bibinfo{author}{Sari, R.},
  \bibinfo{year}{2004}.
\newblock \bibinfo{title}{{HST observations of Kuiper Belt binaries}}, in:
  \bibinfo{booktitle}{AAS/Division for Planetary Sciences Meeting Abstracts
  \#36}, p. \bibinfo{pages}{1081}.
\bibitem[{Margot et~al.(2002)Margot, Nolan, Benner, Ostro, Jurgens, Giorgini,
  Slade and Campbell}]{2002-Science-296-Margot}
\bibinfo{author}{Margot, J.L.}, \bibinfo{author}{Nolan, M.C.},
  \bibinfo{author}{Benner, L.A.M.}, \bibinfo{author}{Ostro, S.J.},
  \bibinfo{author}{Jurgens, R.F.}, \bibinfo{author}{Giorgini, J.D.},
  \bibinfo{author}{Slade, M.A.}, \bibinfo{author}{Campbell, D.B.},
  \bibinfo{year}{2002}.
\newblock \bibinfo{title}{{Binary Asteroids in the Near-Earth Object
  Population}}.
\newblock \bibinfo{journal}{Science} \bibinfo{volume}{296},
  \bibinfo{pages}{1445--1448}.
\bibitem[{Masiero et~al.(2011)Masiero, Mainzer, Grav, Bauer, Cutri, Dailey,
  Eisenhardt, McMillan, Spahr, Skrutskie, Tholen, Walker, Wright, DeBaun,
  Elsbury, Gautier, Gomillion and Wilkins}]{2011-ApJ-741-Masiero}
\bibinfo{author}{Masiero, J.R.}, \bibinfo{author}{Mainzer, A.K.},
  \bibinfo{author}{Grav, T.}, \bibinfo{author}{Bauer, J.M.},
  \bibinfo{author}{Cutri, R.M.}, \bibinfo{author}{Dailey, J.},
  \bibinfo{author}{Eisenhardt, P.R.M.}, \bibinfo{author}{McMillan, R.S.},
  \bibinfo{author}{Spahr, T.B.}, \bibinfo{author}{Skrutskie, M.F.},
  \bibinfo{author}{Tholen, D.}, \bibinfo{author}{Walker, R.G.},
  \bibinfo{author}{Wright, E.L.}, \bibinfo{author}{DeBaun, E.},
  \bibinfo{author}{Elsbury, D.}, \bibinfo{author}{Gautier, IV, T.},
  \bibinfo{author}{Gomillion, S.}, \bibinfo{author}{Wilkins, A.},
  \bibinfo{year}{2011}.
\newblock \bibinfo{title}{{Main Belt Asteroids with WISE/NEOWISE. I.
  Preliminary Albedos and Diameters}}.
\newblock \bibinfo{journal}{Astrophysical Journal} \bibinfo{volume}{741},
  \bibinfo{pages}{68}.
\bibitem[{Matter et~al.(2011)Matter, Delbo, Ligori, Crouzet and
  Tanga}]{2011-Icarus-215-Matter}
\bibinfo{author}{Matter, A.}, \bibinfo{author}{Delbo, M.},
  \bibinfo{author}{Ligori, S.}, \bibinfo{author}{Crouzet, N.},
  \bibinfo{author}{Tanga, P.}, \bibinfo{year}{2011}.
\newblock \bibinfo{title}{{Determination of physical properties of the asteroid
  (41) Daphne from interferometric observations in the thermal infrared}}.
\newblock \bibinfo{journal}{Icarus} \bibinfo{volume}{215},
  \bibinfo{pages}{47--56}.
\bibitem[{McCord et~al.(1970)McCord, Adams and
  Johnson}]{1970-Science-168-McCord}
\bibinfo{author}{McCord, T.B.}, \bibinfo{author}{Adams, J.B.},
  \bibinfo{author}{Johnson, T.V.}, \bibinfo{year}{1970}.
\newblock \bibinfo{title}{{Asteroid Vesta: Spectral Reflectivity and
  Compositional Implications}}.
\newblock \bibinfo{journal}{Science} \bibinfo{volume}{168},
  \bibinfo{pages}{1445--1447}.
\bibitem[{Merline et~al.(1999)Merline, Close, Dumas, Chapman, Roddier,
  M\'{e}nard, Slater, Duvert, Shelton and Morgan}]{1999-Nature-401-Merline}
\bibinfo{author}{Merline, W.J.}, \bibinfo{author}{Close, L.M.},
  \bibinfo{author}{Dumas, C.}, \bibinfo{author}{Chapman, C.R.},
  \bibinfo{author}{Roddier, F.}, \bibinfo{author}{M\'{e}nard, F.},
  \bibinfo{author}{Slater, D.C.}, \bibinfo{author}{Duvert, G.},
  \bibinfo{author}{Shelton, C.}, \bibinfo{author}{Morgan, T.},
  \bibinfo{year}{1999}.
\newblock \bibinfo{title}{{Discovery of a moon orbiting the asteroid 45
  Eugenia}}.
\newblock \bibinfo{journal}{Nature} \bibinfo{volume}{401},
  \bibinfo{pages}{565--568}.
\bibitem[{Merline et~al.(2002)Merline, Weidenschilling, Durda, Margot, Pravec
  and Storrs}]{2002-AsteroidsIII-2.2-Merline}
\bibinfo{author}{Merline, W.J.}, \bibinfo{author}{Weidenschilling, S.J.},
  \bibinfo{author}{Durda, D.D.}, \bibinfo{author}{Margot, J.L.},
  \bibinfo{author}{Pravec, P.}, \bibinfo{author}{Storrs, A.D.},
  \bibinfo{year}{2002}.
\newblock \bibinfo{title}{{Asteroids Do Have Satellites}}.
\newblock \bibinfo{journal}{Asteroids III} , \bibinfo{pages}{289--312}.
\bibitem[{Michalak(2000)}]{2000-AA-360-Michalak}
\bibinfo{author}{Michalak, G.}, \bibinfo{year}{2000}.
\newblock \bibinfo{title}{{Determination of asteroid masses --- I. (1) Ceres,
  (2) Pallas and (4) Vesta}}.
\newblock \bibinfo{journal}{Astronomy and Astrophysics} \bibinfo{volume}{360},
  \bibinfo{pages}{363--374}.
\bibitem[{Michalak(2001)}]{2001-AA-374-Michalak}
\bibinfo{author}{Michalak, G.}, \bibinfo{year}{2001}.
\newblock \bibinfo{title}{{Determination of asteroid masses. II. (6) Hebe, (10)
  Hygiea, (15) Eunomia, (52) Europa, (88) Thisbe, (444) Typtis, (511) Davida
  and (704) Interamnia}}.
\newblock \bibinfo{journal}{Astronomy and Astrophysics} \bibinfo{volume}{374},
  \bibinfo{pages}{703--711}.
\bibitem[{Mignard et~al.(2007)Mignard, Cellino, Muinonen, Tanga, Delbo,
  Dell'Oro, Granvik, Hestroffer, Mouret, Thuillot and
  Virtanen}]{2007-EMP-101-Mignard}
\bibinfo{author}{Mignard, F.}, \bibinfo{author}{Cellino, A.},
  \bibinfo{author}{Muinonen, K.}, \bibinfo{author}{Tanga, P.},
  \bibinfo{author}{Delbo, M.}, \bibinfo{author}{Dell'Oro, A.},
  \bibinfo{author}{Granvik, M.}, \bibinfo{author}{Hestroffer, D.},
  \bibinfo{author}{Mouret, S.}, \bibinfo{author}{Thuillot, W.},
  \bibinfo{author}{Virtanen, J.}, \bibinfo{year}{2007}.
\newblock \bibinfo{title}{{The Gaia Mission: Expected Applications to Asteroid
  Science}}.
\newblock \bibinfo{journal}{Earth Moon and Planets} \bibinfo{volume}{101},
  \bibinfo{pages}{97--125}.
\bibitem[{Mommert et~al.(2012)Mommert, Harris, Kiss, P{\'a}l, Santos~Sanz,
  Stansberry, Delsanti, Vilenius, M{\"u}ller, Peixinho, Lellouch, Szalai,
  B{\"o}nhardt, Duffard, Fornasier, Hartogh, Mueller, Ortiz, Protopapa, Rengel
  and Thirouin}]{2012-AA--Mommert}
\bibinfo{author}{Mommert, M.}, \bibinfo{author}{Harris, A.W.},
  \bibinfo{author}{Kiss, C.}, \bibinfo{author}{P{\'a}l, A.},
  \bibinfo{author}{Santos~Sanz, P.}, \bibinfo{author}{Stansberry, J.},
  \bibinfo{author}{Delsanti, A.}, \bibinfo{author}{Vilenius, E.},
  \bibinfo{author}{M{\"u}ller, T.}, \bibinfo{author}{Peixinho, N.},
  \bibinfo{author}{Lellouch, E.}, \bibinfo{author}{Szalai, N. amd~Henry, F.},
  \bibinfo{author}{B{\"o}nhardt, H.}, \bibinfo{author}{Duffard, R.},
  \bibinfo{author}{Fornasier, S.}, \bibinfo{author}{Hartogh, P.},
  \bibinfo{author}{Mueller, M.}, \bibinfo{author}{Ortiz, J.L.},
  \bibinfo{author}{Protopapa, S.}, \bibinfo{author}{Rengel, M.},
  \bibinfo{author}{Thirouin, A.}, \bibinfo{year}{2012}.
\newblock \bibinfo{title}{{TNOs are Cool: A survey of the trans-Neptunian
  region V. Physical characterization of 18 Plutinos using Herschel PACS
  observations}}.
\newblock \bibinfo{journal}{Astronomy and Astrophysics} \bibinfo{volume}{in
  press}.
\bibitem[{Morbidelli et~al.(2009)Morbidelli, Bottke, Nesvorn{\'y} and
  Levison}]{2009-Icarus-204-Morbidelli}
\bibinfo{author}{Morbidelli, A.}, \bibinfo{author}{Bottke, W.F.},
  \bibinfo{author}{Nesvorn{\'y}, D.}, \bibinfo{author}{Levison, H.F.},
  \bibinfo{year}{2009}.
\newblock \bibinfo{title}{{Asteroids were born big}}.
\newblock \bibinfo{journal}{Icarus} \bibinfo{volume}{204},
  \bibinfo{pages}{558--573}.
\bibitem[{Morbidelli et~al.(2005)Morbidelli, Levison, Tsiganis and
  Gomes}]{2005-Nature-435-Morbidelli}
\bibinfo{author}{Morbidelli, A.}, \bibinfo{author}{Levison, H.F.},
  \bibinfo{author}{Tsiganis, K.}, \bibinfo{author}{Gomes, R.},
  \bibinfo{year}{2005}.
\newblock \bibinfo{title}{{Chaotic capture of Jupiter's Trojan asteroids in the
  early Solar System}}.
\newblock \bibinfo{journal}{Nature} \bibinfo{volume}{435},
  \bibinfo{pages}{462--465}.
\bibitem[{Mottola and Lahulla(2000)}]{2000-Icarus-146-Mottola}
\bibinfo{author}{Mottola, S.}, \bibinfo{author}{Lahulla, F.},
  \bibinfo{year}{2000}.
\newblock \bibinfo{title}{{Mutual Eclipse Events in Asteroidal Binary System
  1996 FG$_{3}$: Observations and a Numerical Model}}.
\newblock \bibinfo{journal}{Icarus} \bibinfo{volume}{146},
  \bibinfo{pages}{556--567}.
\bibitem[{Mouret et~al.(2007)Mouret, Hestroffer and
  Mignard}]{2007-AA-472-Mouret}
\bibinfo{author}{Mouret, S.}, \bibinfo{author}{Hestroffer, D.},
  \bibinfo{author}{Mignard, F.}, \bibinfo{year}{2007}.
\newblock \bibinfo{title}{{Asteroid masses and improvement with GAIA}}.
\newblock \bibinfo{journal}{Astronomy and Astrophysics} \bibinfo{volume}{472},
  \bibinfo{pages}{1017--1027}.
\bibitem[{Mouret et~al.(2009)Mouret, Simon, Mignard and
  Hestroffer}]{2009-AA-508-Mouret}
\bibinfo{author}{Mouret, S.}, \bibinfo{author}{Simon, J.L.},
  \bibinfo{author}{Mignard, F.}, \bibinfo{author}{Hestroffer, D.},
  \bibinfo{year}{2009}.
\newblock \bibinfo{title}{{The list of asteroids perturbing the Mars orbit to
  be seen during future space missions}}.
\newblock \bibinfo{journal}{Astronomy and Astrophysics} \bibinfo{volume}{508},
  \bibinfo{pages}{479--489}.
\bibitem[{Mueller et~al.(2011)Mueller, Delbo, Hora, Trilling, Bhattacharya,
  Bottke, Chesley, Emery, Fazio, Harris, Mainzer, Mommert, Penprase, Smith,
  Spahr, Stansberry and Thomas}]{2011-AJ-141-Mueller}
\bibinfo{author}{Mueller, M.}, \bibinfo{author}{Delbo, M.},
  \bibinfo{author}{Hora, J.L.}, \bibinfo{author}{Trilling, D.E.},
  \bibinfo{author}{Bhattacharya, B.}, \bibinfo{author}{Bottke, W.F.},
  \bibinfo{author}{Chesley, S.}, \bibinfo{author}{Emery, J.P.},
  \bibinfo{author}{Fazio, G.}, \bibinfo{author}{Harris, A.W.},
  \bibinfo{author}{Mainzer, A.}, \bibinfo{author}{Mommert, M.},
  \bibinfo{author}{Penprase, B.}, \bibinfo{author}{Smith, H.A.},
  \bibinfo{author}{Spahr, T.B.}, \bibinfo{author}{Stansberry, J.A.},
  \bibinfo{author}{Thomas, C.A.}, \bibinfo{year}{2011}.
\newblock \bibinfo{title}{{ExploreNEOs. III. Physical Characterization of 65
  Potential Spacecraft Target Asteroids}}.
\newblock \bibinfo{journal}{Astronomical Journal} \bibinfo{volume}{141},
  \bibinfo{pages}{109}.
\bibitem[{Mueller et~al.(2006)Mueller, Harris, Bus, Hora, Kassis and
  Adams}]{2006-AA-447-Mueller}
\bibinfo{author}{Mueller, M.}, \bibinfo{author}{Harris, A.W.},
  \bibinfo{author}{Bus, S.J.}, \bibinfo{author}{Hora, J.L.},
  \bibinfo{author}{Kassis, M.}, \bibinfo{author}{Adams, J.D.},
  \bibinfo{year}{2006}.
\newblock \bibinfo{title}{{The size and albedo of Rosetta fly-by target 21
  Lutetia from new IRTF measurements and thermal modeling}}.
\newblock \bibinfo{journal}{Astronomy and Astrophysics} \bibinfo{volume}{447},
  \bibinfo{pages}{1153--1158}.
\bibitem[{Mueller et~al.(2010)Mueller, Marchis, Emery, Harris, Mottola,
  Hestroffer, Berthier and di~Martino}]{2010-Icarus-205-Mueller}
\bibinfo{author}{Mueller, M.}, \bibinfo{author}{Marchis, F.},
  \bibinfo{author}{Emery, J.P.}, \bibinfo{author}{Harris, A.W.},
  \bibinfo{author}{Mottola, S.}, \bibinfo{author}{Hestroffer, D.},
  \bibinfo{author}{Berthier, J.}, \bibinfo{author}{di~Martino, M.},
  \bibinfo{year}{2010}.
\newblock \bibinfo{title}{{Eclipsing binary Trojan asteroid Patroclus: Thermal
  inertia from Spitzer observations}}.
\newblock \bibinfo{journal}{Icarus} \bibinfo{volume}{205},
  \bibinfo{pages}{505--515}.
\newblock \eprint{0908.4198}.
\bibitem[{M{\"u}ller and Blommaert(2004)}]{2004-AA-418-Muller}
\bibinfo{author}{M{\"u}ller, T.G.}, \bibinfo{author}{Blommaert, J.A.D.L.},
  \bibinfo{year}{2004}.
\newblock \bibinfo{title}{{65 Cybele in the thermal infrared: Multiple
  observations and thermophysical analysis}}.
\newblock \bibinfo{journal}{Astronomy and Astrophysics} \bibinfo{volume}{418},
  \bibinfo{pages}{347--356}.
\bibitem[{M{\"u}ller et~al.(2009)M{\"u}ller, Lellouch, B{\"o}hnhardt,
  Stansberry, Barucci, Crovisier, Delsanti, Doressoundiram, Dotto, Duffard,
  Fornasier, Groussin, Guti{\'e}rrez, Hainaut, Harris, Hartogh, Hestroffer,
  Horner, Jewitt, Kidger, Kiss, Lacerda, Lara, Lim, Mueller, Moreno, Ortiz,
  Rengel, Santos-Sanz, Swinyard, Thomas, Thirouin and
  Trilling}]{2009-EMP-105-Muller}
\bibinfo{author}{M{\"u}ller, T.G.}, \bibinfo{author}{Lellouch, E.},
  \bibinfo{author}{B{\"o}hnhardt, H.}, \bibinfo{author}{Stansberry, J.},
  \bibinfo{author}{Barucci, A.}, \bibinfo{author}{Crovisier, J.},
  \bibinfo{author}{Delsanti, A.}, \bibinfo{author}{Doressoundiram, A.},
  \bibinfo{author}{Dotto, E.}, \bibinfo{author}{Duffard, R.},
  \bibinfo{author}{Fornasier, S.}, \bibinfo{author}{Groussin, O.},
  \bibinfo{author}{Guti{\'e}rrez, P.J.}, \bibinfo{author}{Hainaut, O.R.},
  \bibinfo{author}{Harris, A.W.}, \bibinfo{author}{Hartogh, P.},
  \bibinfo{author}{Hestroffer, D.}, \bibinfo{author}{Horner, J.},
  \bibinfo{author}{Jewitt, D.C.}, \bibinfo{author}{Kidger, M.},
  \bibinfo{author}{Kiss, C.}, \bibinfo{author}{Lacerda, P.},
  \bibinfo{author}{Lara, L.M.}, \bibinfo{author}{Lim, T.},
  \bibinfo{author}{Mueller, M.}, \bibinfo{author}{Moreno, R.},
  \bibinfo{author}{Ortiz, J.L.}, \bibinfo{author}{Rengel, M.},
  \bibinfo{author}{Santos-Sanz, P.}, \bibinfo{author}{Swinyard, B.},
  \bibinfo{author}{Thomas, N.}, \bibinfo{author}{Thirouin, A.},
  \bibinfo{author}{Trilling, D.}, \bibinfo{year}{2009}.
\newblock \bibinfo{title}{{TNOs are Cool: A Survey of the Transneptunian
  Region}}.
\newblock \bibinfo{journal}{Earth Moon and Planets} \bibinfo{volume}{105},
  \bibinfo{pages}{209--219}.
\bibitem[{M{\"u}ller et~al.(2005)M{\"u}ller, Sekiguchi, Kaasalainen, Abe and
  Hasegawa}]{2005-AA-443-Muller}
\bibinfo{author}{M{\"u}ller, T.G.}, \bibinfo{author}{Sekiguchi, T.},
  \bibinfo{author}{Kaasalainen, M.}, \bibinfo{author}{Abe, M.},
  \bibinfo{author}{Hasegawa, S.}, \bibinfo{year}{2005}.
\newblock \bibinfo{title}{{Thermal infrared observations of the Hayabusa
  spacecraft target asteroid 25143 Itokawa}}.
\newblock \bibinfo{journal}{Astronomy and Astrophysics} \bibinfo{volume}{443},
  \bibinfo{pages}{347--355}.
\bibitem[{Neish et~al.(2003)Neish, Nolan, Howell and
  Rivkin}]{2003-DPS-35-Neish}
\bibinfo{author}{Neish, C.D.}, \bibinfo{author}{Nolan, M.C.},
  \bibinfo{author}{Howell, E.S.}, \bibinfo{author}{Rivkin, A.S.},
  \bibinfo{year}{2003}.
\newblock \bibinfo{title}{{Radar Observations of Binary Asteroid 5381
  Sekhmet}}, in: \bibinfo{booktitle}{American Astronomical Society Meeting
  Abstracts}, p. \bibinfo{pages}{134.02}.
\bibitem[{Nolan et~al.(2004)Nolan, Howell and Miranda}]{2004-DPS-36-Nolan}
\bibinfo{author}{Nolan, M.C.}, \bibinfo{author}{Howell, E.S.},
  \bibinfo{author}{Miranda, G.}, \bibinfo{year}{2004}.
\newblock \bibinfo{title}{{Radar Images of Binary Asteroid 2003 YT1}}, in:
  \bibinfo{booktitle}{AAS/Division for Planetary Sciences Meeting Abstracts
  \#36}, p. \bibinfo{pages}{1132}.
\bibitem[{Noll et~al.(2004a)Noll, Stephens, Grundy and
  Griffin}]{2004-Icarus-172-Noll}
\bibinfo{author}{Noll, K.S.}, \bibinfo{author}{Stephens, D.C.},
  \bibinfo{author}{Grundy, W.M.}, \bibinfo{author}{Griffin, I.},
  \bibinfo{year}{2004}a.
\newblock \bibinfo{title}{{The orbit, mass, and albedo of transneptunian binary
  (66652) 1999 RZ$_{253}$}}.
\newblock \bibinfo{journal}{Icarus} \bibinfo{volume}{172},
  \bibinfo{pages}{402--407}.
\bibitem[{Noll et~al.(2002)Noll, Stephens, Grundy, Millis, Spencer, Buie,
  Tegler, Romanishin and Cruikshank}]{2002-AJ-124-Noll}
\bibinfo{author}{Noll, K.S.}, \bibinfo{author}{Stephens, D.C.},
  \bibinfo{author}{Grundy, W.M.}, \bibinfo{author}{Millis, R.L.},
  \bibinfo{author}{Spencer, J.}, \bibinfo{author}{Buie, M.W.},
  \bibinfo{author}{Tegler, S.C.}, \bibinfo{author}{Romanishin, W.},
  \bibinfo{author}{Cruikshank, D.P.}, \bibinfo{year}{2002}.
\newblock \bibinfo{title}{{Detection of Two Binary Trans-Neptunian Objects,
  1997 CQ$_{29}$ and 2000 CF$_{105}$, with the Hubble Space Telescope}}.
\newblock \bibinfo{journal}{Astronomical Journal} \bibinfo{volume}{124},
  \bibinfo{pages}{3424--3429}.
\bibitem[{Noll et~al.(2004b)Noll, Stephens, Grundy, Osip and
  Griffin}]{2004-AJ-128-Noll}
\bibinfo{author}{Noll, K.S.}, \bibinfo{author}{Stephens, D.C.},
  \bibinfo{author}{Grundy, W.M.}, \bibinfo{author}{Osip, D.J.},
  \bibinfo{author}{Griffin, I.}, \bibinfo{year}{2004}b.
\newblock \bibinfo{title}{{The Orbit and Albedo of Trans-Neptunian Binary
  (58534) 1997 CQ$_{29}$}}.
\newblock \bibinfo{journal}{Astronomical Journal} \bibinfo{volume}{128},
  \bibinfo{pages}{2547--2552}.
\bibitem[{Ockert-Bell et~al.(2010)Ockert-Bell, Clark, Shepard, Isaacs, Cloutis,
  Fornasier and Bus}]{2010-Icarus-210-Ockert-Bell}
\bibinfo{author}{Ockert-Bell, M.E.}, \bibinfo{author}{Clark, B.E.},
  \bibinfo{author}{Shepard, M.K.}, \bibinfo{author}{Isaacs, R.A.},
  \bibinfo{author}{Cloutis, E.A.}, \bibinfo{author}{Fornasier, S.},
  \bibinfo{author}{Bus, S.J.}, \bibinfo{year}{2010}.
\newblock \bibinfo{title}{{The composition of M-type asteroids: Synthesis of
  spectroscopic and radar observations}}.
\newblock \bibinfo{journal}{Icarus} \bibinfo{volume}{210},
  \bibinfo{pages}{674--692}.
\bibitem[{O'Rourke et~al.(2012)O'Rourke, M{\"u}ller, Valtchanov, Altieri,
  González-Garcia, Bhattacharya, Jorda, Carry, K{\"u}ppers, Groussin, Altwegg,
  Barucci, Bockel{\'e}e-Morvan, Crovisier, Dotto, Garcia-Lario, Kidger,
  Llorente, Llorente, Marstona, Sanchez~Portal, Schulz, Sierra, Teyssiera and
  Vavreka}]{2012-PSS--Rourke}
\bibinfo{author}{O'Rourke, L.}, \bibinfo{author}{M{\"u}ller, T.G.},
  \bibinfo{author}{Valtchanov, I.}, \bibinfo{author}{Altieri, B.},
  \bibinfo{author}{González-Garcia, B.}, \bibinfo{author}{Bhattacharya, B.},
  \bibinfo{author}{Jorda, L.}, \bibinfo{author}{Carry, B.},
  \bibinfo{author}{K{\"u}ppers, M.}, \bibinfo{author}{Groussin, O.},
  \bibinfo{author}{Altwegg, K.}, \bibinfo{author}{Barucci, A.},
  \bibinfo{author}{Bockel{\'e}e-Morvan, D.}, \bibinfo{author}{Crovisier, J.},
  \bibinfo{author}{Dotto, E.}, \bibinfo{author}{Garcia-Lario, P.},
  \bibinfo{author}{Kidger, M.}, \bibinfo{author}{Llorente, A.},
  \bibinfo{author}{Llorente, R.}, \bibinfo{author}{Marstona, A.P.},
  \bibinfo{author}{Sanchez~Portal, M.}, \bibinfo{author}{Schulz, R.},
  \bibinfo{author}{Sierra, M.}, \bibinfo{author}{Teyssiera, D.},
  \bibinfo{author}{Vavreka, R.}, \bibinfo{year}{2012}.
\newblock \bibinfo{title}{{Thermal \& Shape properties of Asteroid (21) Lutetia
  from Herschel Observations around the Rosetta Flyby}}.
\newblock \bibinfo{journal}{Planetary and Space Science} \bibinfo{volume}{in
  press}.
\bibitem[{Osip et~al.(2003)Osip, Kern and Elliot}]{2003-EMP-92-Osip}
\bibinfo{author}{Osip, D.J.}, \bibinfo{author}{Kern, S.D.},
  \bibinfo{author}{Elliot, J.L.}, \bibinfo{year}{2003}.
\newblock \bibinfo{title}{{Physical Characterization of the Binary
  Edgeworth-Kuiper Belt Object 2001 QT$_{297}$}}.
\newblock \bibinfo{journal}{Earth Moon and Planets} \bibinfo{volume}{92},
  \bibinfo{pages}{409--421}.
\bibitem[{Ostro et~al.(2005)Ostro, Benner, Magri, Giorgini, Rose, Jurgens,
  Yeomans, Hine, Nolan, Scheeres, Broschart, Kaasalainen and
  Margot}]{2005-MPS-40-Ostro}
\bibinfo{author}{Ostro, S.J.}, \bibinfo{author}{Benner, L.A.M.},
  \bibinfo{author}{Magri, C.}, \bibinfo{author}{Giorgini, J.D.},
  \bibinfo{author}{Rose, R.}, \bibinfo{author}{Jurgens, R.F.},
  \bibinfo{author}{Yeomans, D.K.}, \bibinfo{author}{Hine, A.A.},
  \bibinfo{author}{Nolan, M.C.}, \bibinfo{author}{Scheeres, D.J.},
  \bibinfo{author}{Broschart, S.B.}, \bibinfo{author}{Kaasalainen, M.},
  \bibinfo{author}{Margot, J.}, \bibinfo{year}{2005}.
\newblock \bibinfo{title}{{Radar observations of Itokawa in 2004 and improved
  shape estimation}}.
\newblock \bibinfo{journal}{Meteoritics and Planetary Science}
  \bibinfo{volume}{40}, \bibinfo{pages}{1563--1574}.
\bibitem[{Ostro et~al.(2000)Ostro, Hudson, Nolan, Margot, Scheeres, Campbell,
  Magri, Giorgini and Yeomans}]{2000-Science-288-Ostro}
\bibinfo{author}{Ostro, S.J.}, \bibinfo{author}{Hudson, R.S.},
  \bibinfo{author}{Nolan, M.C.}, \bibinfo{author}{Margot, J.L.},
  \bibinfo{author}{Scheeres, D.J.}, \bibinfo{author}{Campbell, D.B.},
  \bibinfo{author}{Magri, C.}, \bibinfo{author}{Giorgini, J.D.},
  \bibinfo{author}{Yeomans, D.K.}, \bibinfo{year}{2000}.
\newblock \bibinfo{title}{{Radar Observations of Asteroid 216 Kleopatra}}.
\newblock \bibinfo{journal}{Science} \bibinfo{volume}{288},
  \bibinfo{pages}{836--839}.
\bibitem[{Ostro et~al.(2010)Ostro, Magri, Benner, Giorgini, Nolan, Hine, Busch
  and Margot}]{2010-Icarus-207-Ostro}
\bibinfo{author}{Ostro, S.J.}, \bibinfo{author}{Magri, C.},
  \bibinfo{author}{Benner, L.A.M.}, \bibinfo{author}{Giorgini, J.D.},
  \bibinfo{author}{Nolan, M.C.}, \bibinfo{author}{Hine, A.A.},
  \bibinfo{author}{Busch, M.W.}, \bibinfo{author}{Margot, J.L.},
  \bibinfo{year}{2010}.
\newblock \bibinfo{title}{{Radar imaging of Asteroid 7 Iris}}.
\newblock \bibinfo{journal}{Icarus} \bibinfo{volume}{207},
  \bibinfo{pages}{285--294}.
\bibitem[{Ostro et~al.(2006)Ostro, Margot, Benner, Giorgini, Scheeres,
  Fahnestock, Broschart, Bellerose, Nolan, Magri, Pravec, Scheirich, Rose,
  Jurgens, De~Jong and Suzuki}]{2006-Science-314-Ostro}
\bibinfo{author}{Ostro, S.J.}, \bibinfo{author}{Margot, J.L.},
  \bibinfo{author}{Benner, L.A.M.}, \bibinfo{author}{Giorgini, J.D.},
  \bibinfo{author}{Scheeres, D.J.}, \bibinfo{author}{Fahnestock, E.G.},
  \bibinfo{author}{Broschart, S.B.}, \bibinfo{author}{Bellerose, J.},
  \bibinfo{author}{Nolan, M.C.}, \bibinfo{author}{Magri, C.},
  \bibinfo{author}{Pravec, P.}, \bibinfo{author}{Scheirich, P.},
  \bibinfo{author}{Rose, R.}, \bibinfo{author}{Jurgens, R.F.},
  \bibinfo{author}{De~Jong, E.M.}, \bibinfo{author}{Suzuki, S.},
  \bibinfo{year}{2006}.
\newblock \bibinfo{title}{{Radar Imaging of Binary Near-Earth Asteroid (66391)
  1999 KW4}}.
\newblock \bibinfo{journal}{Science} \bibinfo{volume}{314},
  \bibinfo{pages}{1276--1280}.
\bibitem[{Parker et~al.(2011)Parker, Kavelaars, Petit, Jones, Gladman and
  Parker}]{2011-ApJ-743-Parker}
\bibinfo{author}{Parker, A.H.}, \bibinfo{author}{Kavelaars, J.J.},
  \bibinfo{author}{Petit, J.M.}, \bibinfo{author}{Jones, L.},
  \bibinfo{author}{Gladman, B.}, \bibinfo{author}{Parker, J.},
  \bibinfo{year}{2011}.
\newblock \bibinfo{title}{{Characterization of Seven Ultra-wide Trans-Neptunian
  Binaries}}.
\newblock \bibinfo{journal}{Astrophysical Journal} \bibinfo{volume}{743},
  \bibinfo{pages}{1}.
\bibitem[{P{\"a}tzold et~al.(2011)P{\"a}tzold, Andert, Asmar, Anderson,
  Barriot, Bird, Häusler, Hahn, Tellmann, Sierks, Lamy and
  Weiss}]{2011-Science-334-Paetzold}
\bibinfo{author}{P{\"a}tzold, M.}, \bibinfo{author}{Andert, T.},
  \bibinfo{author}{Asmar, S.W.}, \bibinfo{author}{Anderson, J.D.},
  \bibinfo{author}{Barriot, J.P.}, \bibinfo{author}{Bird, M.K.},
  \bibinfo{author}{Häusler, B.}, \bibinfo{author}{Hahn, M.},
  \bibinfo{author}{Tellmann, S.}, \bibinfo{author}{Sierks, H.},
  \bibinfo{author}{Lamy, P.}, \bibinfo{author}{Weiss, B.P.},
  \bibinfo{year}{2011}.
\newblock \bibinfo{title}{{Asteroid 21 Lutetia: Low Mass, High Density}}.
\newblock \bibinfo{journal}{Science} \bibinfo{volume}{334},
  \bibinfo{pages}{491}.
\bibitem[{Petit et~al.(1997)Petit, Durda, Greenberg, Hurford and
  Geissler}]{1997-Icarus-130-Petit}
\bibinfo{author}{Petit, J.M.}, \bibinfo{author}{Durda, D.D.},
  \bibinfo{author}{Greenberg, R.}, \bibinfo{author}{Hurford, T.A.},
  \bibinfo{author}{Geissler, P.E.}, \bibinfo{year}{1997}.
\newblock \bibinfo{title}{{The Long-Term Dynamics of Dactyl's Orbit}}.
\newblock \bibinfo{journal}{Icarus} \bibinfo{volume}{130},
  \bibinfo{pages}{177--197}.
\bibitem[{Pitjeva(2001)}]{2001-AA-371-Pitjeva}
\bibinfo{author}{Pitjeva, E.V.}, \bibinfo{year}{2001}.
\newblock \bibinfo{title}{{Progress in the determination of some astronomical
  constants from radiometric observations of planets and spacecraft}}.
\newblock \bibinfo{journal}{Astronomy and Astrophysics} \bibinfo{volume}{371},
  \bibinfo{pages}{760--765}.
\bibitem[{Pitjeva(2004)}]{2004-COSPAR-35-Pitjeva}
\bibinfo{author}{Pitjeva, E.V.}, \bibinfo{year}{2004}.
\newblock \bibinfo{title}{{Estimations of masses of the largest asteroids and
  the main asteroid belt from ranging to planets, Mars orbiters and landers}},
  in: \bibinfo{editor}{{J.-P.~Paill{\'e}}} (Ed.), \bibinfo{booktitle}{35th
  COSPAR Scientific Assembly}, p. \bibinfo{pages}{2014}.
\bibitem[{Pitjeva(2005)}]{2005-SoSyR-39-Pitjeva}
\bibinfo{author}{Pitjeva, E.V.}, \bibinfo{year}{2005}.
\newblock \bibinfo{title}{{High-Precision Ephemerides of Planets - EPM and
  Determination of Some Astronomical Constants}}.
\newblock \bibinfo{journal}{Solar System Research} \bibinfo{volume}{39},
  \bibinfo{pages}{176--186}.
\bibitem[{Pravec and Harris(2007)}]{2007-Icarus-190-Pravec}
\bibinfo{author}{Pravec, P.}, \bibinfo{author}{Harris, A.W.},
  \bibinfo{year}{2007}.
\newblock \bibinfo{title}{{Binary asteroid population. 1. Angular momentum
  content}}.
\newblock \bibinfo{journal}{Icarus} \bibinfo{volume}{190},
  \bibinfo{pages}{250--259}.
\bibitem[{Pravec et~al.(2002)Pravec, Harris and
  Michalowski}]{2002-AsteroidsIII-2.2-Pravec}
\bibinfo{author}{Pravec, P.}, \bibinfo{author}{Harris, A.W.},
  \bibinfo{author}{Michalowski, T.}, \bibinfo{year}{2002}.
\newblock \bibinfo{title}{{Asteroid Rotations}}.
\newblock \bibinfo{journal}{Asteroids III} , \bibinfo{pages}{113--122}.
\bibitem[{Pravec et~al.(2006)Pravec, Scheirich, Ku{\v s}nir{\'a}k, {\v
  S}arounov{\'a}, Mottola, Hahn, Brown, Esquerdo, Kaiser, Krzeminski, Pray,
  Warner, Harris, Nolan, Howell, Benner, Margot, Gal{\'a}d, Holliday, Hicks,
  Krugly, Tholen, Whiteley, Marchis, Degraff, Grauer, Larson, Velichko, Cooney,
  Stephens, Zhu, Kirsch, Dyvig, Snyder, Reddy, Moore, Gajdo{\v s}, Vil{\'a}gi,
  Masi, Higgins, Funkhouser, Knight, Slivan, Behrend, Grenon, Burki, Roy,
  Demeautis, Matter, Waelchli, Revaz, Klotz, Rieugn{\'e}, Thierry, Cotrez,
  Brunetto and Kober}]{2006-Icarus-181-Pravec}
\bibinfo{author}{Pravec, P.}, \bibinfo{author}{Scheirich, P.},
  \bibinfo{author}{Ku{\v s}nir{\'a}k, P.}, \bibinfo{author}{{\v S}arounov{\'a},
  L.}, \bibinfo{author}{Mottola, S.}, \bibinfo{author}{Hahn, G.},
  \bibinfo{author}{Brown, P.G.}, \bibinfo{author}{Esquerdo, G.A.},
  \bibinfo{author}{Kaiser, N.}, \bibinfo{author}{Krzeminski, Z.},
  \bibinfo{author}{Pray, D.P.}, \bibinfo{author}{Warner, B.D.},
  \bibinfo{author}{Harris, A.W.}, \bibinfo{author}{Nolan, M.C.},
  \bibinfo{author}{Howell, E.S.}, \bibinfo{author}{Benner, L.A.M.},
  \bibinfo{author}{Margot, J.L.}, \bibinfo{author}{Gal{\'a}d, A.},
  \bibinfo{author}{Holliday, W.}, \bibinfo{author}{Hicks, M.D.},
  \bibinfo{author}{Krugly, Y.N.}, \bibinfo{author}{Tholen, D.J.},
  \bibinfo{author}{Whiteley, R.J.}, \bibinfo{author}{Marchis, F.},
  \bibinfo{author}{Degraff, D.R.}, \bibinfo{author}{Grauer, A.},
  \bibinfo{author}{Larson, S.}, \bibinfo{author}{Velichko, F.P.},
  \bibinfo{author}{Cooney, W.R.}, \bibinfo{author}{Stephens, R.},
  \bibinfo{author}{Zhu, J.}, \bibinfo{author}{Kirsch, K.},
  \bibinfo{author}{Dyvig, R.}, \bibinfo{author}{Snyder, L.},
  \bibinfo{author}{Reddy, V.}, \bibinfo{author}{Moore, S.},
  \bibinfo{author}{Gajdo{\v s}, {\v S}.}, \bibinfo{author}{Vil{\'a}gi, J.},
  \bibinfo{author}{Masi, G.}, \bibinfo{author}{Higgins, D.},
  \bibinfo{author}{Funkhouser, G.}, \bibinfo{author}{Knight, B.},
  \bibinfo{author}{Slivan, S.M.}, \bibinfo{author}{Behrend, R.},
  \bibinfo{author}{Grenon, M.}, \bibinfo{author}{Burki, G.},
  \bibinfo{author}{Roy, R.}, \bibinfo{author}{Demeautis, C.},
  \bibinfo{author}{Matter, D.}, \bibinfo{author}{Waelchli, N.},
  \bibinfo{author}{Revaz, Y.}, \bibinfo{author}{Klotz, A.},
  \bibinfo{author}{Rieugn{\'e}, M.}, \bibinfo{author}{Thierry, P.},
  \bibinfo{author}{Cotrez, V.}, \bibinfo{author}{Brunetto, L.},
  \bibinfo{author}{Kober, G.}, \bibinfo{year}{2006}.
\newblock \bibinfo{title}{{Photometric survey of binary near-Earth asteroids}}.
\newblock \bibinfo{journal}{Icarus} \bibinfo{volume}{181},
  \bibinfo{pages}{63--93}.
\bibitem[{Pravec et~al.(2010)Pravec, Vokrouhlick{\'y}, Polishook, Scheeres,
  Harris, Gal{\'a}d, Vaduvescu, Pozo, Barr, Longa, Vachier, Colas, Pray,
  Pollock, Reichart, Ivarsen, Haislip, Lacluyze, Ku{\v s}nir{\'a}k, Henych,
  Marchis, Macomber, Jacobson, Krugly, Sergeev and
  Leroy}]{2010-Nature-466-Pravec}
\bibinfo{author}{Pravec, P.}, \bibinfo{author}{Vokrouhlick{\'y}, D.},
  \bibinfo{author}{Polishook, D.}, \bibinfo{author}{Scheeres, D.J.},
  \bibinfo{author}{Harris, A.W.}, \bibinfo{author}{Gal{\'a}d, A.},
  \bibinfo{author}{Vaduvescu, O.}, \bibinfo{author}{Pozo, F.},
  \bibinfo{author}{Barr, A.}, \bibinfo{author}{Longa, P.},
  \bibinfo{author}{Vachier, F.}, \bibinfo{author}{Colas, F.},
  \bibinfo{author}{Pray, D.P.}, \bibinfo{author}{Pollock, J.},
  \bibinfo{author}{Reichart, D.}, \bibinfo{author}{Ivarsen, K.},
  \bibinfo{author}{Haislip, J.}, \bibinfo{author}{Lacluyze, A.},
  \bibinfo{author}{Ku{\v s}nir{\'a}k, P.}, \bibinfo{author}{Henych, T.},
  \bibinfo{author}{Marchis, F.}, \bibinfo{author}{Macomber, B.},
  \bibinfo{author}{Jacobson, S.A.}, \bibinfo{author}{Krugly, Y.N.},
  \bibinfo{author}{Sergeev, A.V.}, \bibinfo{author}{Leroy, A.},
  \bibinfo{year}{2010}.
\newblock \bibinfo{title}{{Formation of asteroid pairs by rotational fission}}.
\newblock \bibinfo{journal}{Nature} \bibinfo{volume}{466},
  \bibinfo{pages}{1085--1088}.
\bibitem[{Rabinowitz et~al.(2006)Rabinowitz, Barkume, Brown, Roe, Schwartz,
  Tourtellotte and Trujillo}]{2006-ApJ-639-Rabinowitz}
\bibinfo{author}{Rabinowitz, D.L.}, \bibinfo{author}{Barkume, K.M.},
  \bibinfo{author}{Brown, M.E.}, \bibinfo{author}{Roe, H.G.},
  \bibinfo{author}{Schwartz, M.}, \bibinfo{author}{Tourtellotte, S.W.},
  \bibinfo{author}{Trujillo, C.A.}, \bibinfo{year}{2006}.
\newblock \bibinfo{title}{{Photometric Observations Constraining the Size,
  Shape, and Albedo of 2003 EL61, a Rapidly Rotating, Pluto-sized Object in the
  Kuiper Belt}}.
\newblock \bibinfo{journal}{Astrophysical Journal} \bibinfo{volume}{639},
  \bibinfo{pages}{1238--1251}.
\bibitem[{Ragozzine and Brown(2009)}]{2009-AJ-137-Ragozzine}
\bibinfo{author}{Ragozzine, D.}, \bibinfo{author}{Brown, M.E.},
  \bibinfo{year}{2009}.
\newblock \bibinfo{title}{{Orbits and Masses of the Satellites of the Dwarf
  Planet Haumea (2003 EL61)}}.
\newblock \bibinfo{journal}{Astronomical Journal} \bibinfo{volume}{137},
  \bibinfo{pages}{4766--4776}.
\bibitem[{Richardson et~al.(2007)Richardson, Melosh, Lisse and
  Carcich}]{2007-Icarus-190-Richardson}
\bibinfo{author}{Richardson, J.E.}, \bibinfo{author}{Melosh, H.J.},
  \bibinfo{author}{Lisse, C.M.}, \bibinfo{author}{Carcich, B.},
  \bibinfo{year}{2007}.
\newblock \bibinfo{title}{{A ballistics analysis of the Deep Impact ejecta
  plume: Determining Comet Tempel 1's gravity, mass, and density}}.
\newblock \bibinfo{journal}{Icarus} \bibinfo{volume}{190},
  \bibinfo{pages}{357--390}.
\bibitem[{Rivkin et~al.(1995)Rivkin, Howell, Britt, Lebofsky, Nolan and
  Branston}]{1995-Icarus-117-Rivkin}
\bibinfo{author}{Rivkin, A.S.}, \bibinfo{author}{Howell, E.S.},
  \bibinfo{author}{Britt, D.T.}, \bibinfo{author}{Lebofsky, L.A.},
  \bibinfo{author}{Nolan, M.C.}, \bibinfo{author}{Branston, D.D.},
  \bibinfo{year}{1995}.
\newblock \bibinfo{title}{{Three-micron spectrometric survey of M-and E-class
  asteroids.}}
\newblock \bibinfo{journal}{Icarus} \bibinfo{volume}{117},
  \bibinfo{pages}{90--100}.
\bibitem[{Rivkin et~al.(2002)Rivkin, Howell, Vilas and
  Lebofsky}]{2002-AsteroidsIII-2.2-Rivkin}
\bibinfo{author}{Rivkin, A.S.}, \bibinfo{author}{Howell, E.S.},
  \bibinfo{author}{Vilas, F.}, \bibinfo{author}{Lebofsky, L.A.},
  \bibinfo{year}{2002}.
\newblock \bibinfo{title}{{Hydrated Minerals on Asteroids: The Astronomical
  Record}}.
\newblock \bibinfo{journal}{Asteroids III} , \bibinfo{pages}{235--253}.
\bibitem[{Rojo and Margot(2011)}]{2011-ApJ-727-Rojo}
\bibinfo{author}{Rojo, P.}, \bibinfo{author}{Margot, J.L.},
  \bibinfo{year}{2011}.
\newblock \bibinfo{title}{{Mass and Density of the B-type Asteroid (702)
  Alauda}}.
\newblock \bibinfo{journal}{Astrophysical Journal} \bibinfo{volume}{727},
  \bibinfo{pages}{69--74}.
\bibitem[{Rozitis and Green(2011)}]{2011-MNRAS-415-Rozitis}
\bibinfo{author}{Rozitis, B.}, \bibinfo{author}{Green, S.F.},
  \bibinfo{year}{2011}.
\newblock \bibinfo{title}{{Directional characteristics of thermal-infrared
  beaming from atmosphereless planetary surfaces - a new thermophysical
  model}}.
\newblock \bibinfo{journal}{Monthly Notices of the Royal Astronomical Society}
  \bibinfo{volume}{415}, \bibinfo{pages}{2042--2062}.
\bibitem[{Ryan and Woodward(2010)}]{2010-AJ-140-Ryan}
\bibinfo{author}{Ryan, E.L.}, \bibinfo{author}{Woodward, C.E.},
  \bibinfo{year}{2010}.
\newblock \bibinfo{title}{{Rectified Asteroid Albedos and Diameters from IRAS
  and MSX Photometry Catalogs}}.
\newblock \bibinfo{journal}{Astronomical Journal} \bibinfo{volume}{140},
  \bibinfo{pages}{933--943}.
\bibitem[{Santos~Sanz et~al.(2012)Santos~Sanz, Lellouch, Fornasier, Kiss,
  P{\'a}l, M{\"u}ller, Vilenius, Stansberry, Mommert, Delsanti, Mueller,
  B{\"o}nhardt, Peixinho, Henry, Ortiz, Thirouin, Protopapa, Duffard, Szalai,
  Lim, Ejeta, Hartogh, Harris and Rengel}]{2012-AA--SantoSanz}
\bibinfo{author}{Santos~Sanz, P.}, \bibinfo{author}{Lellouch, E.},
  \bibinfo{author}{Fornasier, S.}, \bibinfo{author}{Kiss, C.},
  \bibinfo{author}{P{\'a}l, A.}, \bibinfo{author}{M{\"u}ller, T.},
  \bibinfo{author}{Vilenius, E.}, \bibinfo{author}{Stansberry, J.},
  \bibinfo{author}{Mommert, M.}, \bibinfo{author}{Delsanti, A.},
  \bibinfo{author}{Mueller, M.}, \bibinfo{author}{B{\"o}nhardt, H.},
  \bibinfo{author}{Peixinho, N.}, \bibinfo{author}{Henry, F.},
  \bibinfo{author}{Ortiz, J.L.}, \bibinfo{author}{Thirouin, A.},
  \bibinfo{author}{Protopapa, S.}, \bibinfo{author}{Duffard, R.},
  \bibinfo{author}{Szalai, N.}, \bibinfo{author}{Lim, T.},
  \bibinfo{author}{Ejeta, C.}, \bibinfo{author}{Hartogh, P.},
  \bibinfo{author}{Harris, A.W.}, \bibinfo{author}{Rengel, M.},
  \bibinfo{year}{2012}.
\newblock \bibinfo{title}{{TNOs are Cool: A Survey of the Transneptunian Region
  IV. Size/albedo characterization of 15 scattered disk and detached objects
  observed with Herschel Space Observatory-PACS}}.
\newblock \bibinfo{journal}{Astronomy and Astrophysics} \bibinfo{volume}{in
  press}.
\bibitem[{Scheirich and Pravec(2009)}]{2009-Icarus-200-Scheirich}
\bibinfo{author}{Scheirich, P.}, \bibinfo{author}{Pravec, P.},
  \bibinfo{year}{2009}.
\newblock \bibinfo{title}{{Modeling of lightcurves of binary asteroids}}.
\newblock \bibinfo{journal}{Icarus} \bibinfo{volume}{200},
  \bibinfo{pages}{531--547}.
\bibitem[{Schmidt et~al.(2009)Schmidt, Thomas, Bauer, Li, McFadden, Mutchler,
  Radcliffe, Rivkin, Russell, Parker and Stern}]{2009-Science-326-Schmidt}
\bibinfo{author}{Schmidt, B.E.}, \bibinfo{author}{Thomas, P.C.},
  \bibinfo{author}{Bauer, J.M.}, \bibinfo{author}{Li, J.},
  \bibinfo{author}{McFadden, L.A.}, \bibinfo{author}{Mutchler, M.J.},
  \bibinfo{author}{Radcliffe, S.C.}, \bibinfo{author}{Rivkin, A.S.},
  \bibinfo{author}{Russell, C.T.}, \bibinfo{author}{Parker, J.W.},
  \bibinfo{author}{Stern, S.A.}, \bibinfo{year}{2009}.
\newblock \bibinfo{title}{{The Shape and Surface Variation of 2 Pallas from the
  Hubble Space Telescope}}.
\newblock \bibinfo{journal}{Science} \bibinfo{volume}{326},
  \bibinfo{pages}{275--278}.
\bibitem[{Shepard et~al.(2008)Shepard, Clark, Nolan, Howell, Magri, Giorgini,
  Benner, Ostro, Harris, Warner, Pray, Pravec, Fauerbach, Bennett, Klotz,
  Behrend, Correia, Coloma, Casulli and Rivkin}]{2008-Icarus-195-Shepard}
\bibinfo{author}{Shepard, M.K.}, \bibinfo{author}{Clark, B.E.},
  \bibinfo{author}{Nolan, M.C.}, \bibinfo{author}{Howell, E.S.},
  \bibinfo{author}{Magri, C.}, \bibinfo{author}{Giorgini, J.D.},
  \bibinfo{author}{Benner, L.A.M.}, \bibinfo{author}{Ostro, S.J.},
  \bibinfo{author}{Harris, A.W.}, \bibinfo{author}{Warner, B.D.},
  \bibinfo{author}{Pray, D.P.}, \bibinfo{author}{Pravec, P.},
  \bibinfo{author}{Fauerbach, M.}, \bibinfo{author}{Bennett, T.},
  \bibinfo{author}{Klotz, A.}, \bibinfo{author}{Behrend, R.},
  \bibinfo{author}{Correia, H.}, \bibinfo{author}{Coloma, J.},
  \bibinfo{author}{Casulli, S.}, \bibinfo{author}{Rivkin, A.S.},
  \bibinfo{year}{2008}.
\newblock \bibinfo{title}{{A radar survey of M- and X-class asteroids}}.
\newblock \bibinfo{journal}{Icarus} \bibinfo{volume}{195},
  \bibinfo{pages}{184--205}.
\bibitem[{Shepard et~al.(2006)Shepard, Margot, Magri, Nolan, Schlieder, Estes,
  Bus, Volquardsen, Rivkin, Benner, Giorgini, Ostro and
  Busch}]{2006-Icarus-184-Shepard}
\bibinfo{author}{Shepard, M.K.}, \bibinfo{author}{Margot, J.L.},
  \bibinfo{author}{Magri, C.}, \bibinfo{author}{Nolan, M.C.},
  \bibinfo{author}{Schlieder, J.}, \bibinfo{author}{Estes, B.},
  \bibinfo{author}{Bus, S.J.}, \bibinfo{author}{Volquardsen, E.L.},
  \bibinfo{author}{Rivkin, A.S.}, \bibinfo{author}{Benner, L.A.M.},
  \bibinfo{author}{Giorgini, J.D.}, \bibinfo{author}{Ostro, S.J.},
  \bibinfo{author}{Busch, M.W.}, \bibinfo{year}{2006}.
\newblock \bibinfo{title}{{Radar and infrared observations of binary near-Earth
  Asteroid 2002 CE26}}.
\newblock \bibinfo{journal}{Icarus} \bibinfo{volume}{184},
  \bibinfo{pages}{198--210}.
\bibitem[{Sheppard et~al.(2011)Sheppard, Ragozzine and
  Trujillo}]{2011-arXiv-Sheppard}
\bibinfo{author}{Sheppard, S.S.}, \bibinfo{author}{Ragozzine, D.},
  \bibinfo{author}{Trujillo, C.}, \bibinfo{year}{2011}.
\newblock \bibinfo{title}{{2007 TY430: A Cold Classical Kuiper Belt Type Binary
  in the Plutino Population}}.
\newblock \bibinfo{journal}{Accepted for publication in the Astronomical
  Journal} .
\bibitem[{Sicardy et~al.(2011)Sicardy, Ortiz, Assafin, Jehin, Maury, Lellouch,
  Hutton, Braga-Ribas, Colas, Hestroffer, Lecacheux, Roques, Santos-Sanz,
  Widemann, Morales, Duffard, Thirouin, Castro-Tirado, Jel{\'{\i}}nek,
  Kub{\'a}nek, Sota, S{\'a}nchez-Ram{\'{\i}}rez, Andrei, Camargo,
  da~Silva~Neto, Gomes, Martins, Gillon, Manfroid, Tozzi, Harlingten, Saravia,
  Behrend, Mottola, {Melendo}, Peris, Fabregat, Madiedo, Cuesta, Eibe,
  Ull{\'a}n, Organero, Pastor, de~Los~Reyes, Pedraz, Castro, de~La~Cueva,
  Muler, Steele, Cebri{\'a}n, Monta{\~n}{\'e}s-Rodr{\'{\i}}guez, Oscoz, Weaver,
  Jacques, Corradi, Santos, Reis, Milone, Emilio, Guti{\'e}rrez, V{\'a}zquez
  and Hern{\'a}ndez-Toledo}]{2011-Nature-478-Sicardy}
\bibinfo{author}{Sicardy, B.}, \bibinfo{author}{Ortiz, J.L.},
  \bibinfo{author}{Assafin, M.}, \bibinfo{author}{Jehin, E.},
  \bibinfo{author}{Maury, A.}, \bibinfo{author}{Lellouch, E.},
  \bibinfo{author}{Hutton, R.G.}, \bibinfo{author}{Braga-Ribas, F.},
  \bibinfo{author}{Colas, F.}, \bibinfo{author}{Hestroffer, D.},
  \bibinfo{author}{Lecacheux, J.}, \bibinfo{author}{Roques, F.},
  \bibinfo{author}{Santos-Sanz, P.}, \bibinfo{author}{Widemann, T.},
  \bibinfo{author}{Morales, N.}, \bibinfo{author}{Duffard, R.},
  \bibinfo{author}{Thirouin, A.}, \bibinfo{author}{Castro-Tirado, A.J.},
  \bibinfo{author}{Jel{\'{\i}}nek, M.}, \bibinfo{author}{Kub{\'a}nek, P.},
  \bibinfo{author}{Sota, A.}, \bibinfo{author}{S{\'a}nchez-Ram{\'{\i}}rez, R.},
  \bibinfo{author}{Andrei, A.H.}, \bibinfo{author}{Camargo, J.I.B.},
  \bibinfo{author}{da~Silva~Neto, D.N.}, \bibinfo{author}{Gomes, A.R.},
  \bibinfo{author}{Martins, R.V.}, \bibinfo{author}{Gillon, M.},
  \bibinfo{author}{Manfroid, J.}, \bibinfo{author}{Tozzi, G.P.},
  \bibinfo{author}{Harlingten, C.}, \bibinfo{author}{Saravia, S.},
  \bibinfo{author}{Behrend, R.}, \bibinfo{author}{Mottola, S.},
  \bibinfo{author}{{Melendo}, E.G.}, \bibinfo{author}{Peris, V.},
  \bibinfo{author}{Fabregat, J.}, \bibinfo{author}{Madiedo, J.M.},
  \bibinfo{author}{Cuesta, L.}, \bibinfo{author}{Eibe, M.T.},
  \bibinfo{author}{Ull{\'a}n, A.}, \bibinfo{author}{Organero, F.},
  \bibinfo{author}{Pastor, S.}, \bibinfo{author}{de~Los~Reyes, J.A.},
  \bibinfo{author}{Pedraz, S.}, \bibinfo{author}{Castro, A.},
  \bibinfo{author}{de~La~Cueva, I.}, \bibinfo{author}{Muler, G.},
  \bibinfo{author}{Steele, I.A.}, \bibinfo{author}{Cebri{\'a}n, M.},
  \bibinfo{author}{Monta{\~n}{\'e}s-Rodr{\'{\i}}guez, P.},
  \bibinfo{author}{Oscoz, A.}, \bibinfo{author}{Weaver, D.},
  \bibinfo{author}{Jacques, C.}, \bibinfo{author}{Corradi, W.J.B.},
  \bibinfo{author}{Santos, F.P.}, \bibinfo{author}{Reis, W.},
  \bibinfo{author}{Milone, A.}, \bibinfo{author}{Emilio, M.},
  \bibinfo{author}{Guti{\'e}rrez, L.}, \bibinfo{author}{V{\'a}zquez, R.},
  \bibinfo{author}{Hern{\'a}ndez-Toledo, H.}, \bibinfo{year}{2011}.
\newblock \bibinfo{title}{{A Pluto-like radius and a high albedo for the dwarf
  planet Eris from an occultation}}.
\newblock \bibinfo{journal}{Nature} \bibinfo{volume}{478},
  \bibinfo{pages}{493--496}.
\bibitem[{Sierks et~al.(2011)Sierks, Lamy, Barbieri, Koschny, Rickman, Rodrigo,
  A'Hearn, Angrilli, Barucci, Bertaux, Bertini, Besse, Carry, Cremonese,
  Da~Deppo, Davidsson, Debei, De~Cecco, De~Leon, Ferri, Fornasier, Fulle,
  Hviid, Gaskell, Groussin, Gutierrez, Jorda, Kaasalainen, Keller, Knollenberg,
  Kramm, K{\"u}hrt, K{\"u}ppers, Lara, Lazzarin, Leyrat, Lopez~Moreno, Magrin,
  Marchi, Marzari, Massironi, Michalik, Moissl, Naletto, Preusker, Sabau,
  Sabolo, Scholten, Snodgrass, Thomas, Tubiana, Vernazza, Vincent, Wenzel,
  Andert, P{\"a}tzold and Weiss}]{2011-Science-334-Sierks}
\bibinfo{author}{Sierks, H.}, \bibinfo{author}{Lamy, P.},
  \bibinfo{author}{Barbieri, C.}, \bibinfo{author}{Koschny, D.},
  \bibinfo{author}{Rickman, H.}, \bibinfo{author}{Rodrigo, R.},
  \bibinfo{author}{A'Hearn, M.F.}, \bibinfo{author}{Angrilli, F.},
  \bibinfo{author}{Barucci, A.}, \bibinfo{author}{Bertaux, J.L.},
  \bibinfo{author}{Bertini, I.}, \bibinfo{author}{Besse, S.},
  \bibinfo{author}{Carry, B.}, \bibinfo{author}{Cremonese, G.},
  \bibinfo{author}{Da~Deppo, V.}, \bibinfo{author}{Davidsson, B.},
  \bibinfo{author}{Debei, S.}, \bibinfo{author}{De~Cecco, M.},
  \bibinfo{author}{De~Leon, J.}, \bibinfo{author}{Ferri, F.},
  \bibinfo{author}{Fornasier, S.}, \bibinfo{author}{Fulle, M.},
  \bibinfo{author}{Hviid, S.F.}, \bibinfo{author}{Gaskell, G.W.},
  \bibinfo{author}{Groussin, O.}, \bibinfo{author}{Gutierrez, P.J.},
  \bibinfo{author}{Jorda, L.}, \bibinfo{author}{Kaasalainen, M.},
  \bibinfo{author}{Keller, H.U.}, \bibinfo{author}{Knollenberg, J.},
  \bibinfo{author}{Kramm, J.R.}, \bibinfo{author}{K{\"u}hrt, E.},
  \bibinfo{author}{K{\"u}ppers, M.}, \bibinfo{author}{Lara, L.M.},
  \bibinfo{author}{Lazzarin, M.}, \bibinfo{author}{Leyrat, C.},
  \bibinfo{author}{Lopez~Moreno, J.L.}, \bibinfo{author}{Magrin, S.},
  \bibinfo{author}{Marchi, S.}, \bibinfo{author}{Marzari, F.},
  \bibinfo{author}{Massironi, M.}, \bibinfo{author}{Michalik, H.},
  \bibinfo{author}{Moissl, R.}, \bibinfo{author}{Naletto, G.},
  \bibinfo{author}{Preusker, F.}, \bibinfo{author}{Sabau, L.},
  \bibinfo{author}{Sabolo, W.}, \bibinfo{author}{Scholten, F.},
  \bibinfo{author}{Snodgrass, C.}, \bibinfo{author}{Thomas, N.},
  \bibinfo{author}{Tubiana, C.}, \bibinfo{author}{Vernazza, P.},
  \bibinfo{author}{Vincent, J.B.}, \bibinfo{author}{Wenzel, K.P.},
  \bibinfo{author}{Andert, T.}, \bibinfo{author}{P{\"a}tzold, M.},
  \bibinfo{author}{Weiss, B.P.}, \bibinfo{year}{2011}.
\newblock \bibinfo{title}{{Images of asteroid (21) Lutetia: A remnant
  planetesimal from the early Solar System}}.
\newblock \bibinfo{journal}{Science} \bibinfo{volume}{334},
  \bibinfo{pages}{487--490}.
\bibitem[{Sitarski and Todorovic-Juchniewicz(1992)}]{1992-AcA-42-Sitarski}
\bibinfo{author}{Sitarski, G.}, \bibinfo{author}{Todorovic-Juchniewicz, B.},
  \bibinfo{year}{1992}.
\newblock \bibinfo{title}{{Determination of the mass of (1) Ceres from
  perturbations on (203) Pompeja and (348) May}}.
\newblock \bibinfo{journal}{Acta Astronomica} \bibinfo{volume}{42},
  \bibinfo{pages}{139--144}.
\bibitem[{Sitarski and Todorovic-Juchniewicz(1995)}]{1995-AcA-45-Sitarski}
\bibinfo{author}{Sitarski, G.}, \bibinfo{author}{Todorovic-Juchniewicz, B.},
  \bibinfo{year}{1995}.
\newblock \bibinfo{title}{{Determination of Masses of Ceres and Vesta from
  Their Perturbations on Four Asteroids}}.
\newblock \bibinfo{journal}{Acta Astronomica} \bibinfo{volume}{45},
  \bibinfo{pages}{673--677}.
\bibitem[{Snodgrass et~al.(2006)Snodgrass, Lowry and
  Fitzsimmons}]{2006-MNRAS-373-Snodgrass}
\bibinfo{author}{Snodgrass, C.}, \bibinfo{author}{Lowry, S.C.},
  \bibinfo{author}{Fitzsimmons, A.}, \bibinfo{year}{2006}.
\newblock \bibinfo{title}{{Photometry of cometary nuclei: rotation rates,
  colours and a comparison with Kuiper Belt Objects}}.
\newblock \bibinfo{journal}{Monthly Notices of the Royal Astronomical Society}
  \bibinfo{volume}{373}, \bibinfo{pages}{1590--1602}.
\bibitem[{Solem(1994)}]{1994-Nature-370-Solem}
\bibinfo{author}{Solem, J.C.}, \bibinfo{year}{1994}.
\newblock \bibinfo{title}{{Density and size of comet Shoemaker-Levy 9 deduced
  from a tidal breakup model}}.
\newblock \bibinfo{journal}{Nature} \bibinfo{volume}{370},
  \bibinfo{pages}{349--351}.
\bibitem[{Somenzi et~al.(2010)Somenzi, Fienga, Laskar and
  Kuchynka}]{2010-PSS-58-Somenzi}
\bibinfo{author}{Somenzi, L.}, \bibinfo{author}{Fienga, A.},
  \bibinfo{author}{Laskar, J.}, \bibinfo{author}{Kuchynka, P.},
  \bibinfo{year}{2010}.
\newblock \bibinfo{title}{{Determination of asteroid masses from their close
  encounters with Mars}}.
\newblock \bibinfo{journal}{Planetary and Space Science} \bibinfo{volume}{58},
  \bibinfo{pages}{858--863}.
\bibitem[{Sosa and Fern{\'a}ndez(2009)}]{2009-MNRAS-393-Sosa}
\bibinfo{author}{Sosa, A.}, \bibinfo{author}{Fern{\'a}ndez, J.A.},
  \bibinfo{year}{2009}.
\newblock \bibinfo{title}{{Cometary masses derived from non-gravitational
  forces}}.
\newblock \bibinfo{journal}{Monthly Notices of the Royal Astronomical Society}
  \bibinfo{volume}{393}, \bibinfo{pages}{192--214}.
\bibitem[{Spencer et~al.(2006)Spencer, Stansberry, Grundy and
  Noll}]{2006-DPS-38-Spencer}
\bibinfo{author}{Spencer, J.R.}, \bibinfo{author}{Stansberry, J.A.},
  \bibinfo{author}{Grundy, W.M.}, \bibinfo{author}{Noll, K.S.},
  \bibinfo{year}{2006}.
\newblock \bibinfo{title}{{A Low Density for Binary Kuiper Belt Object (26308)
  1998 SM165}}, in: \bibinfo{booktitle}{AAS/Division for Planetary Sciences
  Meeting Abstracts \#38}, pp. \bibinfo{pages}{546--+}.
\bibitem[{Standish(2001)}]{2001-JPL-Standish}
\bibinfo{author}{Standish, E.M.}, \bibinfo{year}{2001}.
\newblock \bibinfo{title}{Suggested GM values for Ceres, Pallas, and Vesta}.
\newblock \bibinfo{type}{Technical Report}. JPL Interoffice Memorandum.
\bibitem[{Stansberry et~al.(2008)Stansberry, Grundy, Brown, Cruikshank,
  Spencer, Trilling and Margot}]{2008-SSBN-3-Stansberry}
\bibinfo{author}{Stansberry, J.}, \bibinfo{author}{Grundy, W.},
  \bibinfo{author}{Brown, M.E.}, \bibinfo{author}{Cruikshank, D.P.},
  \bibinfo{author}{Spencer, J.R.}, \bibinfo{author}{Trilling, D.},
  \bibinfo{author}{Margot, J.L.}, \bibinfo{year}{2008}.
\newblock \bibinfo{title}{{Physical Properties of Kuiper Belt and Centaur
  Objects: Constraints from the Spitzer Space Telescope}}.
\newblock \bibinfo{journal}{The Solar System Beyond Neptune} ,
  \bibinfo{pages}{161--179}.
\bibitem[{Stansberry et~al.(2006)Stansberry, Grundy, Margot, Cruikshank, Emery,
  Rieke and Trilling}]{2006-ApJ-643-Stansberry}
\bibinfo{author}{Stansberry, J.A.}, \bibinfo{author}{Grundy, W.M.},
  \bibinfo{author}{Margot, J.L.}, \bibinfo{author}{Cruikshank, D.P.},
  \bibinfo{author}{Emery, J.P.}, \bibinfo{author}{Rieke, G.H.},
  \bibinfo{author}{Trilling, D.E.}, \bibinfo{year}{2006}.
\newblock \bibinfo{title}{{The Albedo, Size, and Density of Binary Kuiper Belt
  Object (47171) 1999 TC$_{36}$}}.
\newblock \bibinfo{journal}{Astrophysical Journal} \bibinfo{volume}{643},
  \bibinfo{pages}{556--566}.
\bibitem[{Strazzulla et~al.(2005)Strazzulla, Dotto, Binzel, Brunetto, Barucci,
  Blanco and Orofino}]{2005-Icarus-174-Strazzulla}
\bibinfo{author}{Strazzulla, G.}, \bibinfo{author}{Dotto, E.},
  \bibinfo{author}{Binzel, R.P.}, \bibinfo{author}{Brunetto, R.},
  \bibinfo{author}{Barucci, M.A.}, \bibinfo{author}{Blanco, A.},
  \bibinfo{author}{Orofino, V.}, \bibinfo{year}{2005}.
\newblock \bibinfo{title}{{Spectral alteration of the Meteorite Epinal (H5)
  induced by heavy ion irradiation: a simulation of space weathering effects on
  near-Earth asteroids}}.
\newblock \bibinfo{journal}{Icarus} \bibinfo{volume}{174},
  \bibinfo{pages}{31--35}.
\bibitem[{Sunshine et~al.(2008)Sunshine, Connolly, McCoy, Bus and
  La~Croix}]{2008-Science-320-Sunshine}
\bibinfo{author}{Sunshine, J.M.}, \bibinfo{author}{Connolly, H.C.},
  \bibinfo{author}{McCoy, T.J.}, \bibinfo{author}{Bus, S.J.},
  \bibinfo{author}{La~Croix, L.M.}, \bibinfo{year}{2008}.
\newblock \bibinfo{title}{{Ancient Asteroids Enriched in Refractory
  Inclusions}}.
\newblock \bibinfo{journal}{Science} \bibinfo{volume}{320},
  \bibinfo{pages}{514--}.
\bibitem[{Tancredi et~al.(2006)Tancredi, Fern{\'a}ndez, Rickman and
  Licandro}]{2006-Icarus-182-Tancredi}
\bibinfo{author}{Tancredi, G.}, \bibinfo{author}{Fern{\'a}ndez, J.A.},
  \bibinfo{author}{Rickman, H.}, \bibinfo{author}{Licandro, J.},
  \bibinfo{year}{2006}.
\newblock \bibinfo{title}{{Nuclear magnitudes and the size distribution of
  Jupiter family comets}}.
\newblock \bibinfo{journal}{Icarus} \bibinfo{volume}{182},
  \bibinfo{pages}{527--549}.
\bibitem[{Taylor et~al.(2008)Taylor, Margot, Nolan, Benner, Ostro, Giorgini and
  Magri}]{2008-LPI-Taylor}
\bibinfo{author}{Taylor, P.A.}, \bibinfo{author}{Margot, J.L.},
  \bibinfo{author}{Nolan, M.C.}, \bibinfo{author}{Benner, L.A.M.},
  \bibinfo{author}{Ostro, S.J.}, \bibinfo{author}{Giorgini, J.D.},
  \bibinfo{author}{Magri, C.}, \bibinfo{year}{2008}.
\newblock \bibinfo{title}{{The Shape, Mutual Orbit, and Tidal Evolution of
  Binary Near-Earth Asteroid 2004 DC}}.
\newblock \bibinfo{journal}{LPI Contributions} \bibinfo{volume}{1405},
  \bibinfo{pages}{8322}.
\bibitem[{Tedesco et~al.(2004a)Tedesco, Egan and Price}]{PDSSBN-MSX}
\bibinfo{author}{Tedesco, E.F.}, \bibinfo{author}{Egan, M.P.},
  \bibinfo{author}{Price, S.D.}, \bibinfo{year}{2004}a.
\newblock \bibinfo{title}{{MSX Infrared Minor Planet Survey}}.
\newblock \bibinfo{howpublished}{NASA Planetary Data System}.
\newblock \bibinfo{note}{{MSX-A-SPIRIT3-5-SBN0003-MIMPS-V1.0}}.
\bibitem[{Tedesco et~al.(2002)Tedesco, Noah, Noah and
  Price}]{2002-AJ-123-Tedesco-a}
\bibinfo{author}{Tedesco, E.F.}, \bibinfo{author}{Noah, P.V.},
  \bibinfo{author}{Noah, M.C.}, \bibinfo{author}{Price, S.D.},
  \bibinfo{year}{2002}.
\newblock \bibinfo{title}{{The Supplemental IRAS Minor Planet Survey}}.
\newblock \bibinfo{journal}{Astronomical Journal} \bibinfo{volume}{123},
  \bibinfo{pages}{1056--1085}.
\bibitem[{Tedesco et~al.(2004b)Tedesco, Noah, Noah and Price}]{PDSSBN-IRAS}
\bibinfo{author}{Tedesco, E.F.}, \bibinfo{author}{Noah, P.V.},
  \bibinfo{author}{Noah, M.C.}, \bibinfo{author}{Price, S.D.},
  \bibinfo{year}{2004}b.
\newblock \bibinfo{title}{{IRAS Minor Planet Survey}}.
\newblock \bibinfo{howpublished}{NASA Planetary Data System}.
\newblock \bibinfo{note}{{IRAS-A-FPA-3-RDR-IMPS-V6.0}}.
\bibitem[{Tholen and Barucci(1989)}]{1989-AsteroidsII-Tholen}
\bibinfo{author}{Tholen, D.J.}, \bibinfo{author}{Barucci, M.A.},
  \bibinfo{year}{1989}.
\newblock \bibinfo{title}{{Asteroid taxonomy}}.
\newblock \bibinfo{journal}{Asteroids II} , \bibinfo{pages}{298--315}.
\bibitem[{Thomas et~al.(1997)Thomas, Binzel, Gaffey, Zellner, Storrs and
  Wells}]{1997-Icarus-128-Thomas}
\bibinfo{author}{Thomas, P.C.}, \bibinfo{author}{Binzel, R.P.},
  \bibinfo{author}{Gaffey, M.J.}, \bibinfo{author}{Zellner, B.H.},
  \bibinfo{author}{Storrs, A.D.}, \bibinfo{author}{Wells, E.N.},
  \bibinfo{year}{1997}.
\newblock \bibinfo{title}{{Vesta: Spin Pole, Size, and Shape from HST Images}}.
\newblock \bibinfo{journal}{Icarus} \bibinfo{volume}{128},
  \bibinfo{pages}{88--94}.
\bibitem[{Thomas et~al.(2005)Thomas, Parker, McFadden, Russell, Stern, Sykes
  and Young}]{2005-Nature-437-Thomas}
\bibinfo{author}{Thomas, P.C.}, \bibinfo{author}{Parker, J.W.},
  \bibinfo{author}{McFadden, L.A.}, \bibinfo{author}{Russell, C.T.},
  \bibinfo{author}{Stern, S.A.}, \bibinfo{author}{Sykes, M.V.},
  \bibinfo{author}{Young, E.F.}, \bibinfo{year}{2005}.
\newblock \bibinfo{title}{{Differentiation of the asteroid Ceres as revealed by
  its shape}}.
\newblock \bibinfo{journal}{Nature} \bibinfo{volume}{437},
  \bibinfo{pages}{224--226}.
\bibitem[{Tsiganis et~al.(2005)Tsiganis, Gomes, Morbidelli and
  Levison}]{2005-Nature-435-Tsiganis}
\bibinfo{author}{Tsiganis, K.}, \bibinfo{author}{Gomes, R.},
  \bibinfo{author}{Morbidelli, A.}, \bibinfo{author}{Levison, H.F.},
  \bibinfo{year}{2005}.
\newblock \bibinfo{title}{{Origin of the orbital architecture of the giant
  planets of the Solar System}}.
\newblock \bibinfo{journal}{Nature} \bibinfo{volume}{435},
  \bibinfo{pages}{459--461}.
\bibitem[{Usui et~al.(2011)Usui, Kuroda, M{\"u}ller, Hasegawa, Ishiguro,
  Ootsubo, Ishihara, Kataza, Takita, Oyabu, Ueno, Matsuhara and
  Onaka}]{2011-PASJ-63-Usui}
\bibinfo{author}{Usui, F.}, \bibinfo{author}{Kuroda, D.},
  \bibinfo{author}{M{\"u}ller, T.G.}, \bibinfo{author}{Hasegawa, S.},
  \bibinfo{author}{Ishiguro, M.}, \bibinfo{author}{Ootsubo, T.},
  \bibinfo{author}{Ishihara, D.}, \bibinfo{author}{Kataza, H.},
  \bibinfo{author}{Takita, S.}, \bibinfo{author}{Oyabu, S.},
  \bibinfo{author}{Ueno, M.}, \bibinfo{author}{Matsuhara, H.},
  \bibinfo{author}{Onaka, T.}, \bibinfo{year}{2011}.
\newblock \bibinfo{title}{{Asteroid Catalog Using Akari: AKARI/IRC Mid-Infrared
  Asteroid Survey}}.
\newblock \bibinfo{journal}{Publications of the Astronomical Society of Japan}
  \bibinfo{volume}{63}, \bibinfo{pages}{1117--1138}.
\bibitem[{Vasiliev and Yagudina(1999)}]{1999-IAA-Vasiliev}
\bibinfo{author}{Vasiliev, M.V.}, \bibinfo{author}{Yagudina, E.I.},
  \bibinfo{year}{1999}.
\newblock \bibinfo{title}{{Determination of masses for 26 selected minor
  planets from analysis of observations their mutual encounters with asteroids
  of lesser mass}}, in: \bibinfo{booktitle}{Communications of IAA of RAS}.
\bibitem[{Vernazza et~al.(2009a)Vernazza, Binzel, Rossi, Fulchignoni and
  Birlan}]{2009-Nature-458-Vernazza}
\bibinfo{author}{Vernazza, P.}, \bibinfo{author}{Binzel, R.P.},
  \bibinfo{author}{Rossi, A.}, \bibinfo{author}{Fulchignoni, M.},
  \bibinfo{author}{Birlan, M.}, \bibinfo{year}{2009}a.
\newblock \bibinfo{title}{{Solar wind as the origin of rapid reddening of
  asteroid surfaces}}.
\newblock \bibinfo{journal}{Nature} \bibinfo{volume}{458},
  \bibinfo{pages}{993--995}.
\bibitem[{Vernazza et~al.(2008)Vernazza, Binzel, Thomas, DeMeo, Bus, Rivkin and
  Tokunaga}]{2008-Nature-454-Vernazza}
\bibinfo{author}{Vernazza, P.}, \bibinfo{author}{Binzel, R.P.},
  \bibinfo{author}{Thomas, C.A.}, \bibinfo{author}{DeMeo, F.E.},
  \bibinfo{author}{Bus, S.J.}, \bibinfo{author}{Rivkin, A.S.},
  \bibinfo{author}{Tokunaga, A.T.}, \bibinfo{year}{2008}.
\newblock \bibinfo{title}{{Compositional differences between meteorites and
  near-Earth asteroids}}.
\newblock \bibinfo{journal}{Nature} \bibinfo{volume}{454},
  \bibinfo{pages}{858--860}.
\bibitem[{Vernazza et~al.(2009b)Vernazza, Brunetto, Binzel, Perron, Fulvio,
  Strazzulla and Fulchignoni}]{2009-Icarus-202-Vernazza}
\bibinfo{author}{Vernazza, P.}, \bibinfo{author}{Brunetto, R.},
  \bibinfo{author}{Binzel, R.P.}, \bibinfo{author}{Perron, C.},
  \bibinfo{author}{Fulvio, D.}, \bibinfo{author}{Strazzulla, G.},
  \bibinfo{author}{Fulchignoni, M.}, \bibinfo{year}{2009}b.
\newblock \bibinfo{title}{{Plausible parent bodies for enstatite chondrites and
  mesosiderites: Implications for Lutetia's fly-by}}.
\newblock \bibinfo{journal}{Icarus} \bibinfo{volume}{202},
  \bibinfo{pages}{477--486}.
\bibitem[{Vernazza et~al.(2006)Vernazza, Brunetto, Strazzulla, Fulchignoni,
  Rochette, Meyer-Vernet and Zouganelis}]{2006-AA-451-Vernazza}
\bibinfo{author}{Vernazza, P.}, \bibinfo{author}{Brunetto, R.},
  \bibinfo{author}{Strazzulla, G.}, \bibinfo{author}{Fulchignoni, M.},
  \bibinfo{author}{Rochette, P.}, \bibinfo{author}{Meyer-Vernet, N.},
  \bibinfo{author}{Zouganelis, I.}, \bibinfo{year}{2006}.
\newblock \bibinfo{title}{{Asteroid colors: a novel tool for magnetic field
  detection? The case of Vesta}}.
\newblock \bibinfo{journal}{Astronomy and Astrophysics} \bibinfo{volume}{451},
  \bibinfo{pages}{43--46}.
\bibitem[{Vernazza et~al.(2010)Vernazza, Carry, Emery, Hora, Cruikshank,
  Binzel, Jackson, Helbert and Maturilli}]{2010-Icarus-207-Vernazza}
\bibinfo{author}{Vernazza, P.}, \bibinfo{author}{Carry, B.},
  \bibinfo{author}{Emery, J.P.}, \bibinfo{author}{Hora, J.L.},
  \bibinfo{author}{Cruikshank, D.P.}, \bibinfo{author}{Binzel, R.P.},
  \bibinfo{author}{Jackson, J.}, \bibinfo{author}{Helbert, J.},
  \bibinfo{author}{Maturilli, A.}, \bibinfo{year}{2010}.
\newblock \bibinfo{title}{{Mid-infrared spectral variability for
  compositionally similar asteroids: Implications for asteroid particle size
  distributions}}.
\newblock \bibinfo{journal}{Icarus} \bibinfo{volume}{207},
  \bibinfo{pages}{800--809}.
\bibitem[{Vernazza et~al.(2011a)Vernazza, King, Izawa, Maturilli, Helbert,
  Cruikshank, Brunetto, Marchis, Binzel and Flemming}]{2011-LPI-Vernazza}
\bibinfo{author}{Vernazza, P.}, \bibinfo{author}{King, P.L.},
  \bibinfo{author}{Izawa, M.R.M.}, \bibinfo{author}{Maturilli, A.},
  \bibinfo{author}{Helbert, J.}, \bibinfo{author}{Cruikshank, D.},
  \bibinfo{author}{Brunetto, R.}, \bibinfo{author}{Marchis, F.},
  \bibinfo{author}{Binzel, R.P.}, \bibinfo{author}{Flemming, R.L.},
  \bibinfo{year}{2011}a.
\newblock \bibinfo{title}{{Opening the Mid-IR Window on Asteroid Physical
  Properties}}.
\newblock \bibinfo{journal}{Lunar and Planetary Institute Science Conference
  Abstracts} \bibinfo{volume}{42}, \bibinfo{pages}{1344}.
\bibitem[{Vernazza et~al.(2011b)Vernazza, Lamy, Groussin, Hiroi, Jorda, King,
  Izawa, Marchis, Birlan and Brunetto}]{2011-Icarus-216-Vernazza}
\bibinfo{author}{Vernazza, P.}, \bibinfo{author}{Lamy, P.},
  \bibinfo{author}{Groussin, O.}, \bibinfo{author}{Hiroi, T.},
  \bibinfo{author}{Jorda, L.}, \bibinfo{author}{King, P.L.},
  \bibinfo{author}{Izawa, M.R.M.}, \bibinfo{author}{Marchis, F.},
  \bibinfo{author}{Birlan, M.}, \bibinfo{author}{Brunetto, R.},
  \bibinfo{year}{2011}b.
\newblock \bibinfo{title}{{Asteroid (21) Lutetia as a remnant of Earth's
  precursor planetesimals}}.
\newblock \bibinfo{journal}{Icarus} \bibinfo{volume}{216},
  \bibinfo{pages}{650--659}.
\bibitem[{Veverka et~al.(2000)Veverka, Robinson, Thomas, Murchie, Bell,
  Izenberg, Chapman, Harch, Bell, Carcich, Cheng, Clark, Domingue, Dunham,
  Farquhar, Gaffey, Hawkins, Joseph, Kirk, Li, Lucey, Malin, Martin, McFadden,
  Merline, Miller, Owen, Peterson, Prockter, Warren, Wellnitz, Williams and
  Yeomans}]{2000-Science-289-Veverka}
\bibinfo{author}{Veverka, J.}, \bibinfo{author}{Robinson, M.},
  \bibinfo{author}{Thomas, P.}, \bibinfo{author}{Murchie, S.},
  \bibinfo{author}{Bell, J.F.}, \bibinfo{author}{Izenberg, N.},
  \bibinfo{author}{Chapman, C.}, \bibinfo{author}{Harch, A.},
  \bibinfo{author}{Bell, M.}, \bibinfo{author}{Carcich, B.},
  \bibinfo{author}{Cheng, A.}, \bibinfo{author}{Clark, B.},
  \bibinfo{author}{Domingue, D.}, \bibinfo{author}{Dunham, D.},
  \bibinfo{author}{Farquhar, R.}, \bibinfo{author}{Gaffey, M.J.},
  \bibinfo{author}{Hawkins, E.}, \bibinfo{author}{Joseph, J.},
  \bibinfo{author}{Kirk, R.}, \bibinfo{author}{Li, H.}, \bibinfo{author}{Lucey,
  P.}, \bibinfo{author}{Malin, M.}, \bibinfo{author}{Martin, P.},
  \bibinfo{author}{McFadden, L.}, \bibinfo{author}{Merline, W.J.},
  \bibinfo{author}{Miller, J.K.}, \bibinfo{author}{Owen, W.M.},
  \bibinfo{author}{Peterson, C.}, \bibinfo{author}{Prockter, L.},
  \bibinfo{author}{Warren, J.}, \bibinfo{author}{Wellnitz, D.},
  \bibinfo{author}{Williams, B.G.}, \bibinfo{author}{Yeomans, D.K.},
  \bibinfo{year}{2000}.
\newblock \bibinfo{title}{{NEAR at Eros: Imaging and Spectral Results}}.
\newblock \bibinfo{journal}{Science} \bibinfo{volume}{289},
  \bibinfo{pages}{2088--2097}.
\bibitem[{Viateau(2000)}]{2000-AA-354-Viateau}
\bibinfo{author}{Viateau, B.}, \bibinfo{year}{2000}.
\newblock \bibinfo{title}{{Mass and density of asteroids (16) Psyche and (121)
  Hermione}}.
\newblock \bibinfo{journal}{Astronomy and Astrophysics} \bibinfo{volume}{354},
  \bibinfo{pages}{725--731}.
\bibitem[{Viateau and Rapaport(1995)}]{1995-AA-111-Viateau}
\bibinfo{author}{Viateau, B.}, \bibinfo{author}{Rapaport, M.},
  \bibinfo{year}{1995}.
\newblock \bibinfo{title}{{The orbit of (2) Pallas.}}
\newblock \bibinfo{journal}{Astronomy and Astrophysics} \bibinfo{volume}{111},
  \bibinfo{pages}{305--+}.
\bibitem[{Viateau and Rapaport(1997a)}]{1997-ESASP-402-Viateau}
\bibinfo{author}{Viateau, B.}, \bibinfo{author}{Rapaport, M.},
  \bibinfo{year}{1997}a.
\newblock \bibinfo{title}{{Improvement of the Orbits of Asteroids and the Mass
  of (1) Ceres}}, in: \bibinfo{editor}{{R.~M.~Bonnet, E.~H{\o}g,
  P.~L.~Bernacca, L.~Emiliani, A.~Blaauw, C.~Turon, J.~Kovalevsky,
  L.~Lindegren, H.~Hassan, M.~Bouffard, B.~Strim, D.~Heger, M.~A.~C.~Perryman,
  \& L.~Woltjer}} (Ed.), \bibinfo{booktitle}{Hipparcos - Venice '97}, pp.
  \bibinfo{pages}{91--94}.
\bibitem[{Viateau and Rapaport(1997b)}]{1997-AA-320-Viateau}
\bibinfo{author}{Viateau, B.}, \bibinfo{author}{Rapaport, M.},
  \bibinfo{year}{1997}b.
\newblock \bibinfo{title}{{The Bordeaux meridian observations of asteroids.
  First determination of the mass of (11) Parthenope.}}
\newblock \bibinfo{journal}{Astronomy} \bibinfo{volume}{320},
  \bibinfo{pages}{652--658}.
\bibitem[{Viateau and Rapaport(1998)}]{1998-AA-334-Viateau}
\bibinfo{author}{Viateau, B.}, \bibinfo{author}{Rapaport, M.},
  \bibinfo{year}{1998}.
\newblock \bibinfo{title}{{The mass of (1) Ceres from its gravitational
  perturbations on the orbits of 9 asteroids}}.
\newblock \bibinfo{journal}{Astronomy and Astrophysics} \bibinfo{volume}{334},
  \bibinfo{pages}{729--735}.
\bibitem[{Viateau and Rapaport(2001)}]{2001-AA-370-Viateau}
\bibinfo{author}{Viateau, B.}, \bibinfo{author}{Rapaport, M.},
  \bibinfo{year}{2001}.
\newblock \bibinfo{title}{{Mass and density of asteroids (4) Vesta and (11)
  Parthenope}}.
\newblock \bibinfo{journal}{Astronomy and Astrophysics} \bibinfo{volume}{370},
  \bibinfo{pages}{602--609}.
\bibitem[{Vitagliano and Stoss(2006)}]{2006-AA-455-Vitagliano}
\bibinfo{author}{Vitagliano, A.}, \bibinfo{author}{Stoss, R.M.},
  \bibinfo{year}{2006}.
\newblock \bibinfo{title}{{New mass determination of (15) Eunomia based on a
  very close encounter with (50278) 2000CZ12}}.
\newblock \bibinfo{journal}{Astronomy and Astrophysics} \bibinfo{volume}{455},
  \bibinfo{pages}{L29--L31}.
\bibitem[{Walsh et~al.(2012)Walsh, Delbo, Mueller, Binzel and
  DeMeo}]{2012-ApJ-186-Walsh}
\bibinfo{author}{Walsh, K.J.}, \bibinfo{author}{Delbo, M.},
  \bibinfo{author}{Mueller, M.}, \bibinfo{author}{Binzel, R.P.},
  \bibinfo{author}{DeMeo, F.}, \bibinfo{year}{2012}.
\newblock \bibinfo{title}{{Physical characterization and origin of binary
  near-earth asteroid (175706) 1996 FG3}}.
\newblock \bibinfo{journal}{Astrophysical Journal} \bibinfo{volume}{186},
  \bibinfo{pages}{498}.
\bibitem[{Walsh et~al.(2011)Walsh, Morbidelli, Raymond, O'Brien and
  Mandell}]{2011-Nature-475-Walsh}
\bibinfo{author}{Walsh, K.J.}, \bibinfo{author}{Morbidelli, A.},
  \bibinfo{author}{Raymond, S.N.}, \bibinfo{author}{O'Brien, D.P.},
  \bibinfo{author}{Mandell, A.M.}, \bibinfo{year}{2011}.
\newblock \bibinfo{title}{{A low mass for Mars from Jupiter's early gas-driven
  migration}}.
\newblock \bibinfo{journal}{Nature} \bibinfo{volume}{475},
  \bibinfo{pages}{206--209}.
\bibitem[{Walsh et~al.(2008)Walsh, Richardson and
  Michel}]{2008-Nature-454-Walsh}
\bibinfo{author}{Walsh, K.J.}, \bibinfo{author}{Richardson, D.C.},
  \bibinfo{author}{Michel, P.}, \bibinfo{year}{2008}.
\newblock \bibinfo{title}{{Rotational breakup as the origin of small binary
  asteroids}}.
\newblock \bibinfo{journal}{Nature} \bibinfo{volume}{454},
  \bibinfo{pages}{188--191}.
\bibitem[{Weisberg et~al.(2006)Weisberg, McCoy and Krot}]{2006-MESS2-Weisberg}
\bibinfo{author}{Weisberg, M.K.}, \bibinfo{author}{McCoy, T.J.},
  \bibinfo{author}{Krot, A.N.}, \bibinfo{year}{2006}.
\newblock \bibinfo{title}{{Systematics and Evaluation of Meteorite
  Classification}}.
\newblock \bibinfo{journal}{Meteorites and the Early Solar System II} ,
  \bibinfo{pages}{19--52}.
\bibitem[{Weissman et~al.(2004)Weissman, Asphaug and
  Lowry}]{2004-CometsII-4-Weissman}
\bibinfo{author}{Weissman, P.R.}, \bibinfo{author}{Asphaug, E.},
  \bibinfo{author}{Lowry, S.C.}, \bibinfo{year}{2004}.
\newblock \bibinfo{title}{{Structure and density of cometary nuclei}}.
\newblock \bibinfo{journal}{Comets II} , \bibinfo{pages}{337--357}.
\bibitem[{Williams(1992)}]{1992-ACM-Williams}
\bibinfo{author}{Williams, G.V.}, \bibinfo{year}{1992}.
\newblock \bibinfo{title}{{The mass of (1) Ceres from perturbations on (348)
  May}}, in: \bibinfo{booktitle}{Asteroids, Comets, Meteors 1991}, pp.
  \bibinfo{pages}{641--643}.
\bibitem[{Wolters et~al.(2008)Wolters, Green, McBride and
  Davies}]{2008-Icarus-193-Wolters}
\bibinfo{author}{Wolters, S.D.}, \bibinfo{author}{Green, S.F.},
  \bibinfo{author}{McBride, N.}, \bibinfo{author}{Davies, J.K.},
  \bibinfo{year}{2008}.
\newblock \bibinfo{title}{{Thermal infrared and optical observations of four
  near-Earth asteroids}}.
\newblock \bibinfo{journal}{Icarus} \bibinfo{volume}{193},
  \bibinfo{pages}{535--552}.
\bibitem[{Wolters et~al.(2011)Wolters, Rozitis, Duddy, Lowry, Green, Snodgrass,
  Hainaut and Weissman}]{2011-MNRAS-418-Wolters}
\bibinfo{author}{Wolters, S.D.}, \bibinfo{author}{Rozitis, B.},
  \bibinfo{author}{Duddy, S.R.}, \bibinfo{author}{Lowry, S.C.},
  \bibinfo{author}{Green, S.F.}, \bibinfo{author}{Snodgrass, C.},
  \bibinfo{author}{Hainaut, O.R.}, \bibinfo{author}{Weissman, P.},
  \bibinfo{year}{2011}.
\newblock \bibinfo{title}{{Physical characterization of low delta-V asteroid
  (175706) 1996 FG3}}.
\newblock \bibinfo{journal}{Monthly Notices of the Royal Astronomical Society}
  \bibinfo{volume}{418}, \bibinfo{pages}{1246--1257}.
\bibitem[{Yeomans et~al.(2000)Yeomans, Antreasian, Barriot, Chesley, Dunham,
  Farquhar, Giorgini, Helfrich, Konopliv, McAdams, Miller, Owen, Scheeres,
  Thomas, Veverka and Williams}]{2000-Science-289-Yeomans}
\bibinfo{author}{Yeomans, D.K.}, \bibinfo{author}{Antreasian, P.G.},
  \bibinfo{author}{Barriot, J.P.}, \bibinfo{author}{Chesley, S.R.},
  \bibinfo{author}{Dunham, D.W.}, \bibinfo{author}{Farquhar, R.W.},
  \bibinfo{author}{Giorgini, J.D.}, \bibinfo{author}{Helfrich, C.E.},
  \bibinfo{author}{Konopliv, A.S.}, \bibinfo{author}{McAdams, J.V.},
  \bibinfo{author}{Miller, J.K.}, \bibinfo{author}{Owen, W.M.},
  \bibinfo{author}{Scheeres, D.J.}, \bibinfo{author}{Thomas, P.C.},
  \bibinfo{author}{Veverka, J.}, \bibinfo{author}{Williams, B.G.},
  \bibinfo{year}{2000}.
\newblock \bibinfo{title}{{Radio Science Results During the NEAR-Shoemaker
  Spacecraft Rendezvous with Eros}}.
\newblock \bibinfo{journal}{Science} \bibinfo{volume}{289},
  \bibinfo{pages}{2085--2088}.
\bibitem[{Yeomans et~al.(1997)Yeomans, Barriot, Dunham, Farquhar, Giorgini,
  Helfrich, Konopliv, McAdams, Miller, Owen, Scheeres, Synnott and
  Williams}]{1997-Science-278-Yeomans}
\bibinfo{author}{Yeomans, D.K.}, \bibinfo{author}{Barriot, J.P.},
  \bibinfo{author}{Dunham, D.W.}, \bibinfo{author}{Farquhar, R.W.},
  \bibinfo{author}{Giorgini, J.D.}, \bibinfo{author}{Helfrich, C.E.},
  \bibinfo{author}{Konopliv, A.S.}, \bibinfo{author}{McAdams, J.V.},
  \bibinfo{author}{Miller, J.K.}, \bibinfo{author}{Owen, Jr., W.M.},
  \bibinfo{author}{Scheeres, D.J.}, \bibinfo{author}{Synnott, S.P.},
  \bibinfo{author}{Williams, B.G.}, \bibinfo{year}{1997}.
\newblock \bibinfo{title}{{Estimating the Mass of Asteroid 253 Mathilde from
  Tracking Data During the NEAR Flyby}}.
\newblock \bibinfo{journal}{Science} \bibinfo{volume}{278},
  \bibinfo{pages}{2106}.
\bibitem[{Yurimoto et~al.(2011)Yurimoto, Abe, Abe, Ebihara, Fujimura,
  Hashiguchi, Hashizume, Ireland, Itoh, Katayama, Kato, Kawaguchi, Kawasaki,
  Kitajima, Kobayashi, Meike, Mukai, Nagao, Nakamura, Naraoka, Noguchi,
  Okazaki, Park, Sakamoto, Seto, Takei, Tsuchiyama, Uesugi, Wakaki, Yada,
  Yamamoto, Yoshikawa and Zolensky}]{2011-Science-333-Yurimoto}
\bibinfo{author}{Yurimoto, H.}, \bibinfo{author}{Abe, K.i.},
  \bibinfo{author}{Abe, M.}, \bibinfo{author}{Ebihara, M.},
  \bibinfo{author}{Fujimura, A.}, \bibinfo{author}{Hashiguchi, M.},
  \bibinfo{author}{Hashizume, K.}, \bibinfo{author}{Ireland, T.R.},
  \bibinfo{author}{Itoh, S.}, \bibinfo{author}{Katayama, J.},
  \bibinfo{author}{Kato, C.}, \bibinfo{author}{Kawaguchi, J.},
  \bibinfo{author}{Kawasaki, N.}, \bibinfo{author}{Kitajima, F.},
  \bibinfo{author}{Kobayashi, S.}, \bibinfo{author}{Meike, T.},
  \bibinfo{author}{Mukai, T.}, \bibinfo{author}{Nagao, K.},
  \bibinfo{author}{Nakamura, T.}, \bibinfo{author}{Naraoka, H.},
  \bibinfo{author}{Noguchi, T.}, \bibinfo{author}{Okazaki, R.},
  \bibinfo{author}{Park, C.}, \bibinfo{author}{Sakamoto, N.},
  \bibinfo{author}{Seto, Y.}, \bibinfo{author}{Takei, M.},
  \bibinfo{author}{Tsuchiyama, A.}, \bibinfo{author}{Uesugi, M.},
  \bibinfo{author}{Wakaki, S.}, \bibinfo{author}{Yada, T.},
  \bibinfo{author}{Yamamoto, K.}, \bibinfo{author}{Yoshikawa, M.},
  \bibinfo{author}{Zolensky, M.E.}, \bibinfo{year}{2011}.
\newblock \bibinfo{title}{{Oxygen Isotopic Compositions of Asteroidal Materials
  Returned from Itokawa by the Hayabusa Mission}}.
\newblock \bibinfo{journal}{Science} \bibinfo{volume}{333},
  \bibinfo{pages}{1116--1118}.
\bibitem[{Zielenbach(2010)}]{2010-AJ-139-Zielenbach}
\bibinfo{author}{Zielenbach, W.}, \bibinfo{year}{2010}.
\newblock \bibinfo{title}{{The Mass of (15) Eunomia from 923 Test Bodies}}.
\newblock \bibinfo{journal}{Astronomical Journal} \bibinfo{volume}{139},
  \bibinfo{pages}{816--824}.
\bibitem[{Zielenbach(2011)}]{2011-AJ-142-Zielenbach}
\bibinfo{author}{Zielenbach, W.}, \bibinfo{year}{2011}.
\newblock \bibinfo{title}{{Mass Determination Studies of 104 Large Asteroids}}.
\newblock \bibinfo{journal}{Astronomical Journal} \bibinfo{volume}{142},
  \bibinfo{pages}{120--128}.

\end{thebibliography}


\appendix
\clearpage

\section{Compilation of mass estimates\label{app: mass}}
  \indent The \numb{994} mass estimates gathered
  in the literature are listed in Table~\ref{tabSM: mass}.
  For objects with more than a single mass determination, 
  Fig.~\ref{fap: mass000001} to
  Fig.~\ref{fap: mass9P/Tempel1} presents a comparison of the mass
  estimates, with additional information on discarded values.
  See~\ref{app: biblio} for the references, and
  Fig.~\ref{fap: mass9P/Tempel1} for symbols key.

\setcounter{figure}{0}
\setcounter{table}{0}
\begin{table}[!ht]
\centering
\caption[Mass estimates]{%
  Compilation of mass estimates ($M$, in kg) for 287 objects,
  with their associated uncertainty ($\delta M$),
  bibliographic references (see~\ref{app: biblio}), and method of analysis:
  \textsl{Deflec}: orbital deflection, 
  \textsl{Ephem}: planetary ephemeris, 
  \textsl{PheMu}: mutual eclipsing phenomena in binary systems, 
  \textsl{BinImg}: binary imaged at optical wavelength, 
  \textsl{BinRad}: binary imaged with radar, 
  \textsl{FlyBy}: radio experiment for spacecraft flyby or orbit, and, 
 for comet nucleii
  \textsl{CNGF}: non-gravitational forces, and
  \textsl{BkUp}: break-up modeling. 
  Estimates marked with a dagger ($\dagger$) were rejected from average mass computation.
\label{tabSM: mass}
}
\begin{tabular}{r@{ }lccll}
 \hline
 \hline
 \# & Designation & $M$ & $\delta M$ & Method & Refs.\\
 \hline
     1 & Ceres                & 9.55E+20 & 4.38E+19 & Deflec     & M$_{1}$            \\ 
     1 & Ceres                & 9.54E+20 & 1.69E+19 & Deflec     & M$_{2}$            \\ 
     1 & Ceres                & 9.94E+20 & 3.98E+19 & Deflec     & M$_{5}$$^\dagger$  \\ 
     1 & Ceres                & 9.19E+20 & 1.41E+19 & Deflec     & M$_{6}$            \\ 
     1 & Ceres                & 9.29E+20 & 1.79E+19 & Deflec     & M$_{7}$            \\ 
     1 & Ceres                & 8.27E+20 & 3.78E+19 & Deflec     & M$_{8}$$^\dagger$  \\ 
     1 & Ceres                & 9.52E+20 & 7.76E+18 & Deflec     & M$_{11}$           \\
\hline
\end{tabular}
 \end{table}
\setcounter{table}{0}
\begin{table}[!ht]
\caption{Continued}
\begin{tabular}{r@{ }lccll}
 \hline
 \hline
 \# & Designation & $M$ & $\delta M$ & Method & Refs.\\
 \hline
     3 & Juno                 & 4.16E+19 & 6.96E+18 & Deflec     & M$_{31}$$^\dagger$     \\ 
     3 & Juno                 & 4.16E+19 & 6.96E+18 & Deflec     & M$_{42}$$^\dagger$     \\ 
     3 & Juno                 & 2.82E+19 & 1.19E+18 & Deflec     & M$_{39}$               \\ 
     3 & Juno                 & 3.00E+19 & 5.97E+17 & Deflec     & M$_{49}$               \\ 
     3 & Juno                 & 2.96E+19 & 2.98E+18 & Deflec     & M$_{56}$               \\ 
     3 & Juno                 & 2.67E+19 & 4.57E+18 & Deflec     & M$_{72}$               \\ 
     3 & Juno                 & 2.30E+19 & 2.30E+18 & Ephem      & M$_{86}$               \\
\hline
\end{tabular}
 \end{table}
\setcounter{table}{0}
\begin{table}[!ht]
\caption{Continued}
\begin{tabular}{r@{ }lccll}
 \hline
 \hline
 \# & Designation & $M$ & $\delta M$ & Method & Refs.\\
 \hline
     6 & Hebe                 & 1.37E+19 & 4.38E+18 & Deflec     & M$_{26}$               \\ 
     6 & Hebe                 & 1.37E+19 & 1.79E+18 & Deflec     & M$_{42}$               \\ 
     6 & Hebe                 & 1.28E+19 & 6.36E+17 & Deflec     & M$_{72}$               \\ 
     6 & Hebe                 & 3.18E+17 & 2.19E+17 & Ephem      & M$_{80}$$^\dagger$     \\ 
     6 & Hebe                 & 9.07E+18 & 9.07E+17 & Ephem      & M$_{86}$$^\dagger$     \\ 
     6 & Hebe                 & 1.41E+19 & 2.41E+18 & Ephem      & M$_{93}$               \\ 
     6 & Hebe                 & 1.34E+19 & 3.26E+18 & Ephem      & M$_{103}$              \\
\hline
\end{tabular}
 \end{table}
\setcounter{table}{0}
\begin{table}[!ht]
\caption{Continued}
\begin{tabular}{r@{ }lccll}
 \hline
 \hline
 \# & Designation & $M$ & $\delta M$ & Method & Refs.\\
 \hline
    10 & Hygiea               & 1.55E+20 & 4.97E+19 & Deflec     & M$_{18}$$^\dagger$     \\ 
    10 & Hygiea               & 5.57E+19 & 1.99E+18 & Deflec     & M$_{28}$$^\dagger$     \\ 
    10 & Hygiea               & 1.11E+20 & 1.39E+19 & Deflec     & M$_{26}$$^\dagger$     \\ 
    10 & Hygiea               & 9.96E+19 & 8.15E+18 & Deflec     & M$_{31}$               \\ 
    10 & Hygiea               & 9.97E+19 & 8.16E+18 & Deflec     & M$_{42}$               \\ 
    10 & Hygiea               & 9.03E+19 & 2.59E+18 & Deflec     & M$_{45}$               \\ 
    10 & Hygiea               & 4.18E+19 & 5.97E+18 & Ephem      & M$_{62}$$^\dagger$     \\
\hline
\end{tabular}
 \end{table}
\setcounter{table}{0}
\begin{table}[!ht]
\caption{Continued}
\begin{tabular}{r@{ }lccll}
 \hline
 \hline
 \# & Designation & $M$ & $\delta M$ & Method & Refs.\\
 \hline
    15 & Eunomia              & 1.99E+19 & 7.96E+18 & Deflec     & M$_{18}$               \\ 
    15 & Eunomia              & 2.78E+19 & 3.98E+18 & Deflec     & M$_{28}$               \\ 
    15 & Eunomia              & 2.51E+19 & 5.97E+18 & Deflec     & M$_{26}$               \\ 
    15 & Eunomia              & 2.43E+19 & 3.18E+18 & Deflec     & M$_{31}$$^\dagger$     \\ 
    15 & Eunomia              & 2.11E+19 & 3.18E+18 & Deflec     & M$_{42}$$^\dagger$     \\ 
    15 & Eunomia              & 3.26E+19 & 1.19E+18 & Deflec     & M$_{51}$               \\ 
    15 & Eunomia              & 3.12E+19 & 3.98E+17 & Deflec     & M$_{72}$               \\
\hline
\end{tabular}
 \end{table}
\setcounter{table}{0}
\begin{table}[!ht]
\caption{Continued}
\begin{tabular}{r@{ }lccll}
 \hline
 \hline
 \# & Designation & $M$ & $\delta M$ & Method & Refs.\\
 \hline
    19 & Fortuna              & 1.27E+19 & 4.97E+17 & Deflec     & M$_{72}$$^\dagger$     \\ 
    19 & Fortuna              & 4.02E+18 & 3.98E+17 & Ephem      & M$_{80}$$^\dagger$     \\ 
    19 & Fortuna              & 6.94E+18 & 6.94E+17 & Ephem      & M$_{86}$               \\ 
    19 & Fortuna              & 6.37E+18 & 2.90E+18 & Deflec     & M$_{92}$               \\ 
    19 & Fortuna              & 8.31E+18 & 7.16E+17 & Deflec     & M$_{95}$               \\ 
    19 & Fortuna              & 6.37E+18 & 1.05E+18 & Ephem      & M$_{103}$              \\ 
    19 & Fortuna              & 1.00E+19 & 1.08E+18 & Deflec     & M$_{97}$               \\
\hline
\end{tabular}
 \end{table}
\setcounter{table}{0}
\begin{table}[!ht]
\caption{Continued}
\begin{tabular}{r@{ }lccll}
 \hline
 \hline
 \# & Designation & $M$ & $\delta M$ & Method & Refs.\\
 \hline
    27 & Euterpe              & 1.26E+18 & 1.26E+17 & Ephem      & M$_{86}$               \\ 
    27 & Euterpe              & 4.35E+18 & 1.98E+18 & Deflec     & M$_{97}$               \\ 
    27 & Euterpe              & 2.09E+18 & 1.47E+18 & Deflec     & M$_{97}$               \\ 
    27 & Euterpe              & 2.20E+18 & 1.46E+18 & Deflec     & M$_{97}$               \\ 
    27 & Euterpe              & 5.59E+18 & 3.07E+18 & Deflec     & M$_{97}$               \\ 
\noalign{\smallskip}
    28 & Bellona              & 2.78E+18 & 2.78E+17 & Deflec     & M$_{28}$               \\ 
    28 & Bellona              & 1.39E+19 & 3.98E+18 & Deflec     & M$_{78}$$^\dagger$     \\
\hline
\end{tabular}
 \end{table}
\setcounter{table}{0}
\begin{table}[!ht]
\caption{Continued}
\begin{tabular}{r@{ }lccll}
 \hline
 \hline
 \# & Designation & $M$ & $\delta M$ & Method & Refs.\\
 \hline
    41 & Daphne               & 1.05E+19 & 9.95E+17 & Ephem      & M$_{80}$$^\dagger$     \\ 
    41 & Daphne               & 7.90E+18 & 7.90E+17 & Ephem      & M$_{86}$$^\dagger$     \\ 
    41 & Daphne               & 1.83E+19 & 5.23E+18 & Ephem      & M$_{93}$$^\dagger$     \\ 
    41 & Daphne               & 8.43E+18 & 3.52E+18 & Ephem      & M$_{103}$$^\dagger$    \\ 
    41 & Daphne               & 6.31E+18 & 1.10E+17 & BinImg     & M$_{107}$              \\ 
    41 & Daphne               & 1.82E+19 & 7.20E+18 & Deflec     & M$_{97}$$^\dagger$     \\ 
    41 & Daphne               & 3.02E+17 & 5.67E+18 & Deflec     & M$_{97}$$^\dagger$     \\
\hline
\end{tabular}
 \end{table}
\setcounter{table}{0}
\begin{table}[!ht]
\caption{Continued}
\begin{tabular}{r@{ }lccll}
 \hline
 \hline
 \# & Designation & $M$ & $\delta M$ & Method & Refs.\\
 \hline
    51 & Nemausa              & 2.16E+18 & 2.16E+17 & Ephem      & M$_{86}$               \\ 
    51 & Nemausa              & 1.44E+18 & 8.37E+17 & Ephem      & M$_{93}$               \\ 
    51 & Nemausa              & 4.55E+18 & 2.72E+18 & Deflec     & M$_{97}$               \\ 
    51 & Nemausa              & 3.39E+18 & 1.62E+18 & Deflec     & M$_{97}$               \\ 
    51 & Nemausa              & 3.36E+18 & 1.62E+18 & Deflec     & M$_{97}$               \\ 
    51 & Nemausa              & 4.25E+18 & 2.08E+18 & Deflec     & M$_{97}$               \\ 
    51 & Nemausa              & 5.63E+18 & 1.30E+18 & Ephem      & M$_{100}$$^\dagger$    \\
\hline
\end{tabular}
 \end{table}
\setcounter{table}{0}
\begin{table}[!ht]
\caption{Continued}
\begin{tabular}{r@{ }lccll}
 \hline
 \hline
 \# & Designation & $M$ & $\delta M$ & Method & Refs.\\
 \hline
    65 & Cybele               & 1.15E+19 & 2.98E+18 & Deflec     & M$_{31}$               \\ 
    65 & Cybele               & 1.15E+19 & 2.98E+18 & Deflec     & M$_{42}$               \\ 
    65 & Cybele               & 1.78E+19 & 1.21E+18 & Deflec     & M$_{72}$               \\ 
    65 & Cybele               & 1.04E+19 & 1.04E+18 & Ephem      & M$_{86}$               \\ 
    65 & Cybele               & 1.43E+19 & 8.47E+18 & Ephem      & M$_{93}$               \\ 
    65 & Cybele               & 1.05E+19 & 1.91E+18 & Deflec     & M$_{95}$               \\ 
    65 & Cybele               & 1.62E+19 & 3.68E+18 & Deflec     & M$_{97}$               \\
\hline
\end{tabular}
 \end{table}
\setcounter{table}{0}
\begin{table}[!ht]
\caption{Continued}
\begin{tabular}{r@{ }lccll}
 \hline
 \hline
 \# & Designation & $M$ & $\delta M$ & Method & Refs.\\
 \hline
    88 & Thisbe               & 1.47E+19 & 2.59E+18 & Deflec     & M$_{26}$               \\ 
    88 & Thisbe               & 1.17E+19 & 2.39E+18 & Deflec     & M$_{42}$               \\ 
    88 & Thisbe               & 1.05E+19 & 1.79E+18 & Deflec     & M$_{72}$               \\ 
    88 & Thisbe               & 1.72E+19 & 5.17E+17 & Ephem      & M$_{80}$               \\ 
    88 & Thisbe               & 1.83E+19 & 1.09E+18 & Deflec     & M$_{95}$               \\ 
    88 & Thisbe               & 9.53E+18 & 3.20E+18 & Deflec     & M$_{97}$               \\ 
    88 & Thisbe               & 7.61E+18 & 2.27E+18 & Deflec     & M$_{97}$$^\dagger$     \\
\hline
\end{tabular}
 \end{table}
 \clearpage
\setcounter{table}{0}
\begin{table}[!ht]
\caption{Continued}
\begin{tabular}{r@{ }lccll}
 \hline
 \hline
 \# & Designation & $M$ & $\delta M$ & Method & Refs.\\
 \hline
   117 & Lomia                & 1.72E+19 & 1.99E+17 & Ephem      & M$_{93}$$^\dagger$     \\ 
   117 & Lomia                & 6.36E+18 & 4.35E+18 & Deflec     & M$_{97}$               \\ 
   117 & Lomia                & 5.85E+18 & 3.59E+18 & Deflec     & M$_{97}$               \\ 
   117 & Lomia                & 6.73E+18 & 3.55E+18 & Deflec     & M$_{97}$               \\ 
   117 & Lomia                & 4.88E+18 & 6.61E+18 & Deflec     & M$_{97}$               \\ 
\noalign{\smallskip}
   121 & Hermione             & 9.35E+18 & 1.59E+18 & Deflec     & M$_{20}$$^\dagger$     \\ 
   121 & Hermione             & 5.38E+18 & 2.98E+17 & BinImg     & M$_{47}$               \\
\hline
\end{tabular}
 \end{table}
\setcounter{table}{0}
\begin{table}[!ht]
\caption{Continued}
\begin{tabular}{r@{ }lccll}
 \hline
 \hline
 \# & Designation & $M$ & $\delta M$ & Method & Refs.\\
 \hline
   144 & Vibilia              & 9.08E+18 & 5.92E+18 & Ephem      & M$_{93}$               \\ 
   144 & Vibilia              & 4.63E+18 & 2.84E+18 & Deflec     & M$_{97}$               \\ 
   144 & Vibilia              & 5.32E+18 & 1.86E+18 & Deflec     & M$_{97}$               \\ 
   144 & Vibilia              & 4.95E+18 & 1.85E+18 & Deflec     & M$_{97}$               \\ 
   144 & Vibilia              & 4.62E+18 & 3.06E+18 & Deflec     & M$_{97}$               \\ 
\noalign{\smallskip}
   145 & Adeona               & 2.26E+18 & 2.26E+17 & Ephem      & M$_{86}$               \\ 
   145 & Adeona               & 1.73E+17 & 2.39E+18 & Deflec     & M$_{97}$               \\
\hline
\end{tabular}
 \end{table}
\setcounter{table}{0}
\begin{table}[!ht]
\caption{Continued}
\begin{tabular}{r@{ }lccll}
 \hline
 \hline
 \# & Designation & $M$ & $\delta M$ & Method & Refs.\\
 \hline
   196 & Philomela            & 6.39E+18 & 5.54E+18 & Deflec     & M$_{97}$               \\ 
   196 & Philomela            & 3.16E+18 & 1.61E+18 & Deflec     & M$_{97}$               \\ 
   196 & Philomela            & 3.30E+18 & 1.61E+18 & Deflec     & M$_{97}$               \\ 
   196 & Philomela            & 7.59E+18 & 6.87E+18 & Deflec     & M$_{97}$               \\ 
\noalign{\smallskip}
   200 & Dynamene             & 1.07E+19 & 1.61E+18 & Ephem      & M$_{93}$               \\ 
\noalign{\smallskip}
   204 & Kallisto             & 6.01E+17 & 1.81E+18 & Ephem      & M$_{93}$               \\ 
\noalign{\smallskip}
   209 & Dido                 & 4.59E+18 & 7.42E+18 & Deflec     & M$_{97}$               \\
\hline
\end{tabular}
 \end{table}
\setcounter{table}{0}
\begin{table}[!ht]
\caption{Continued}
\begin{tabular}{r@{ }lccll}
 \hline
 \hline
 \# & Designation & $M$ & $\delta M$ & Method & Refs.\\
 \hline
   283 & Emma                 & 1.38E+18 & 3.00E+16 & BinImg     & M$_{73}$               \\ 
   283 & Emma                 & 1.00E+19 & 5.16E+18 & Deflec     & M$_{97}$$^\dagger$     \\ 
   283 & Emma                 & 8.38E+18 & 3.68E+18 & Deflec     & M$_{97}$$^\dagger$     \\ 
   283 & Emma                 & 8.31E+18 & 3.65E+18 & Deflec     & M$_{97}$$^\dagger$     \\ 
   283 & Emma                 & 1.63E+19 & 6.03E+18 & Deflec     & M$_{97}$$^\dagger$     \\ 
\noalign{\smallskip}
   304 & Olga                 & 1.15E+18 & 1.12E+18 & Ephem      & M$_{93}$               \\ 
\noalign{\smallskip}
   306 & Unitas               & 5.33E+17 & 5.77E+17 & Ephem      & M$_{93}$               \\
\hline
\end{tabular}
 \end{table}
\setcounter{table}{0}
\begin{table}[!ht]
\caption{Continued}
\begin{tabular}{r@{ }lccll}
 \hline
 \hline
 \# & Designation & $M$ & $\delta M$ & Method & Refs.\\
 \hline
   375 & Ursula               & 2.64E+18 & 5.67E+18 & Deflec     & M$_{97}$               \\ 
   375 & Ursula               & 7.14E+18 & 3.93E+18 & Deflec     & M$_{97}$               \\ 
   375 & Ursula               & 7.44E+18 & 3.89E+18 & Deflec     & M$_{97}$               \\ 
   375 & Ursula               & 1.88E+19 & 6.40E+18 & Ephem      & M$_{100}$              \\ 
\noalign{\smallskip}
   379 & Huenna               & 3.83E+17 & 1.99E+16 & BinImg     & M$_{73}$               \\ 
\noalign{\smallskip}
   381 & Myrrha               & 9.18E+18 & 7.96E+17 & Ephem      & M$_{93}$               \\ 
\noalign{\smallskip}
   386 & Siegena              & 2.35E+19 & 4.18E+17 & Ephem      & M$_{93}$$^\dagger$     \\
\hline
\end{tabular}
 \end{table}
\setcounter{table}{0}
\begin{table}[!ht]
\caption{Continued}
\begin{tabular}{r@{ }lccll}
 \hline
 \hline
 \# & Designation & $M$ & $\delta M$ & Method & Refs.\\
 \hline
   455 & Bruchsalia           & 1.19E+18 & 1.19E+17 & Ephem      & M$_{93}$               \\ 
\noalign{\smallskip}
   469 & Argentina            & 7.25E+18 & 6.25E+18 & Ephem      & M$_{93}$$^\dagger$     \\ 
   469 & Argentina            & 4.53E+18 & 1.76E+18 & Ephem      & M$_{100}$              \\ 
\noalign{\smallskip}
   471 & Papagena             & 1.26E+19 & 7.16E+17 & Ephem      & M$_{93}$$^\dagger$     \\ 
   471 & Papagena             & 2.31E+18 & 4.88E+18 & Deflec     & M$_{97}$               \\ 
   471 & Papagena             & 2.08E+18 & 4.78E+18 & Deflec     & M$_{97}$               \\ 
   471 & Papagena             & 6.52E+18 & 9.75E+18 & Deflec     & M$_{97}$               \\
\hline
\end{tabular}
 \end{table}
\setcounter{table}{0}
\begin{table}[!ht]
\caption{Continued}
\begin{tabular}{r@{ }lccll}
 \hline
 \hline
 \# & Designation & $M$ & $\delta M$ & Method & Refs.\\
 \hline
   536 & Merapi               & 2.15E+19 & 2.44E+19 & Deflec     & M$_{97}$               \\ 
   536 & Merapi               & 3.10E+19 & 2.55E+19 & Deflec     & M$_{97}$               \\ 
\noalign{\smallskip}
   554 & Peraga               & 6.59E+17 & 6.59E+16 & Ephem      & M$_{86}$               \\ 
   554 & Peraga               & 3.13E+18 & 2.54E+18 & Ephem      & M$_{93}$$^\dagger$     \\ 
\noalign{\smallskip}
   582 & Olympia              & 4.28E+17 & 1.17E+18 & Ephem      & M$_{93}$               \\ 
\noalign{\smallskip}
   584 & Semiramis            & 8.23E+17 & 5.77E+17 & Ephem      & M$_{93}$               \\ 
\noalign{\smallskip}
   602 & Marianna             & 1.02E+19 & 4.77E+17 & Ephem      & M$_{93}$               \\
\hline
\end{tabular}
 \end{table}
\setcounter{table}{0}
\begin{table}[!ht]
\caption{Continued}
\begin{tabular}{r@{ }lccll}
 \hline
 \hline
 \# & Designation & $M$ & $\delta M$ & Method & Refs.\\
 \hline
   747 & Winchester           & 7.96E+15 & 3.98E+15 & Ephem      & M$_{80}$$^\dagger$     \\ 
   747 & Winchester           & 2.94E+18 & 2.94E+17 & Ephem      & M$_{86}$               \\ 
   747 & Winchester           & 1.20E+19 & 4.59E+18 & Ephem      & M$_{93}$               \\ 
   747 & Winchester           & 8.10E+18 & 5.86E+18 & Deflec     & M$_{97}$               \\ 
   747 & Winchester           & 4.31E+18 & 4.65E+18 & Deflec     & M$_{97}$               \\ 
   747 & Winchester           & 5.25E+18 & 4.47E+18 & Deflec     & M$_{97}$               \\ 
   747 & Winchester           & 1.33E+19 & 1.47E+18 & Ephem      & M$_{100}$$^\dagger$    \\
\hline
\end{tabular}
 \end{table}
\setcounter{table}{0}
\begin{table}[!ht]
\caption{Continued}
\begin{tabular}{r@{ }lccll}
 \hline
 \hline
 \# & Designation & $M$ & $\delta M$ & Method & Refs.\\
 \hline
  1669 & Dagmar               & 4.18E+19 & 3.18E+19 & Deflec     & M$_{18}$$^\dagger$     \\ 
  1669 & Dagmar               & 3.98E+16 & 7.96E+15 & Deflec     & M$_{28}$               \\ 
  1669 & Dagmar               & 2.78E+19 & 3.98E+18 & Deflec     & M$_{78}$$^\dagger$     \\ 
\noalign{\smallskip}
  1686 & De Sitter            & 6.76E+18 & 3.18E+18 & Deflec     & M$_{78}$               \\ 
\noalign{\smallskip}
  3169 & Ostro                & 1.86E+14 & 6.20E+13 & PheMu      & M$_{66}$               \\ 
\noalign{\smallskip}
  3671 & Dionysus             & 8.38E+11 & 2.79E+11 & PheMu      & M$_{55}$               \\ 
\noalign{\smallskip}
  3749 & Balam                & 5.09E+14 & 1.99E+13 & BinImg     & M$_{73}$               \\
\hline
\end{tabular}
 \end{table}
\setcounter{table}{0}
\begin{table}[!ht]
\caption{Continued}
\begin{tabular}{r@{ }lccll}
 \hline
 \hline
 \# & Designation & $M$ & $\delta M$ & Method & Refs.\\
 \hline
       & 1999 OJ4             & 3.91E+17 & 2.20E+16 & BinImg     & M$_{82}$               \\ 
       & 2000 QL251           & 3.11E+18 & 5.10E+16 & BinImg     & M$_{82}$               \\ 
       & 2000 UG11            & 9.35E+09 & 1.59E+09 & BinRad     & M$_{32}$               \\ 
       & 2001 QC298           & 1.08E+19 & 7.00E+17 & BinImg     & M$_{40}$               \\ 
       & 2001 XR254           & 4.00E+18 & 1.70E+17 & BinImg     & M$_{82}$               \\ 
       & 2003 QY90            & 1.01E+18 & 7.85E+17 & BinImg     & M$_{57}$               \\ 
       & 2003 TJ58            & 2.25E+17 & 1.50E+16 & BinImg     & M$_{82}$               \\
\hline
 \end{tabular}
 \end{table}

\clearpage
    \begin{figure}[!ht]
  \centering
  \includegraphics[width=.49\textwidth]{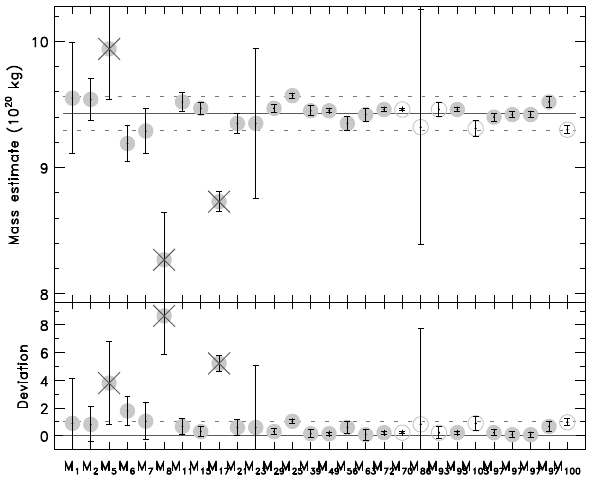}
\caption[Mass estimates for (1) Ceres]{%
  \label{fap: mass000001}
  Mass estimates for (1) Ceres.
}
\end{figure}

  \begin{figure}[!ht]
  \centering
  \includegraphics[width=.49\textwidth]{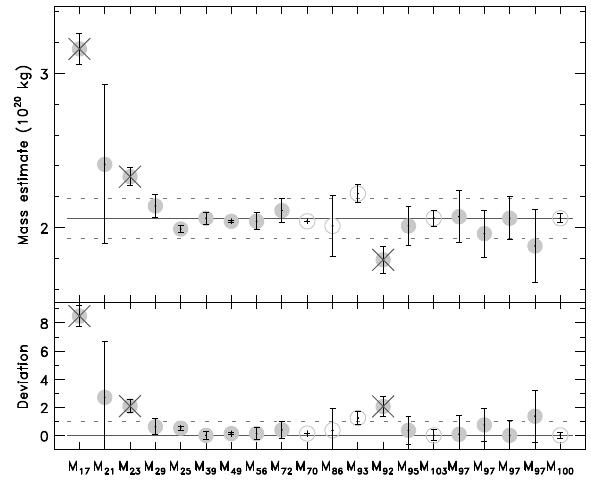}
\caption[Mass estimates for (2) Pallas]{%
  \label{fap: mass000002}
  Mass estimates for (2) Pallas.
}
\end{figure}

  \begin{figure}[!ht]
  \centering
  \includegraphics[width=.49\textwidth]{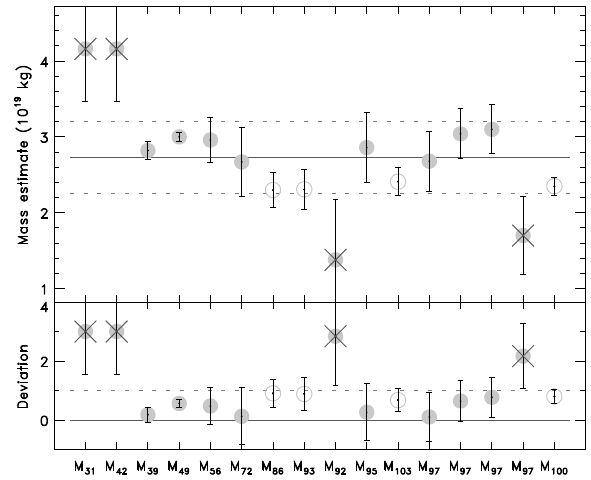}
\caption[Mass estimates for (3) Juno]{%
  \label{fap: mass000003}
  Mass estimates for (3) Juno.
}
\end{figure}

  \begin{figure}[!ht]
  \centering
  \includegraphics[width=.49\textwidth]{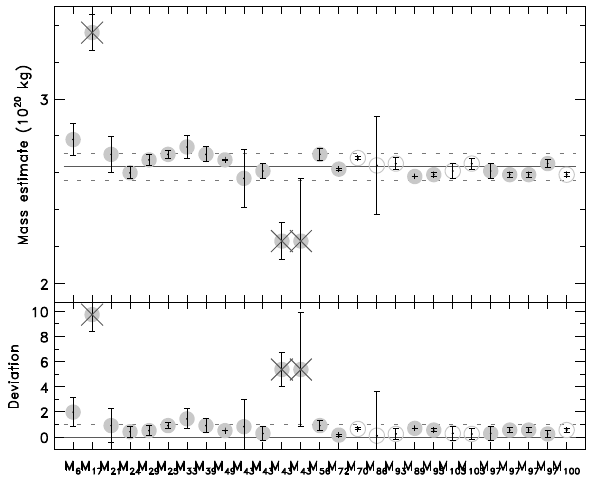}
\caption[Mass estimates for (4) Vesta]{%
  \label{fap: mass000004}
  Mass estimates for (4) Vesta.
}
\end{figure}

  \begin{figure}[!ht]
  \centering
  \includegraphics[width=.49\textwidth]{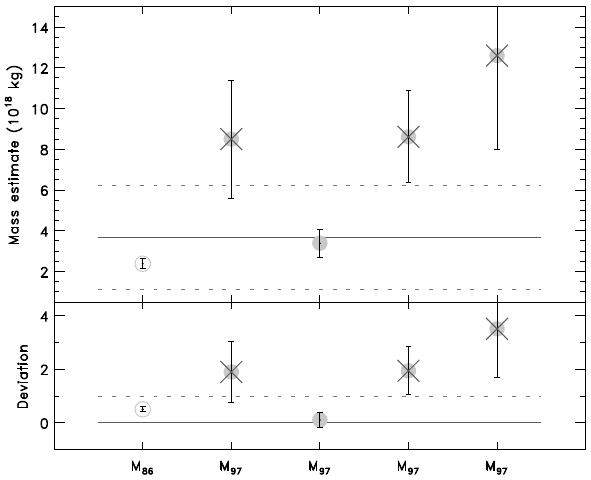}
\caption[Mass estimates for (5) Astraea]{%
  \label{fap: mass000005}
  Mass estimates for (5) Astraea.
}
\end{figure}

  \begin{figure}[!ht]
  \centering
  \includegraphics[width=.49\textwidth]{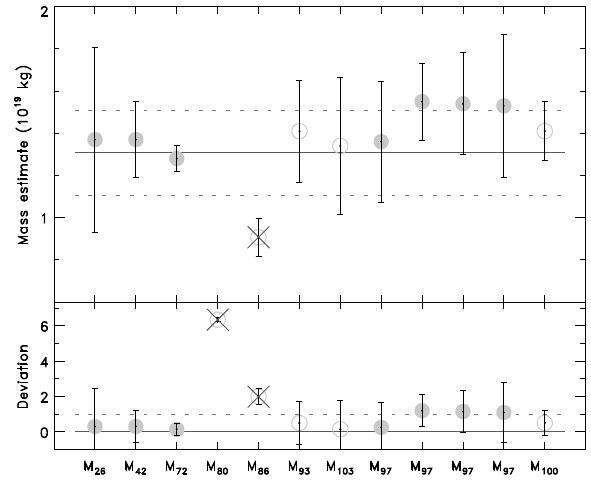}
\caption[Mass estimates for (6) Hebe]{%
  \label{fap: mass000006}
  Mass estimates for (6) Hebe.
}
\end{figure}

  \begin{figure}[!ht]
  \centering
  \includegraphics[width=.49\textwidth]{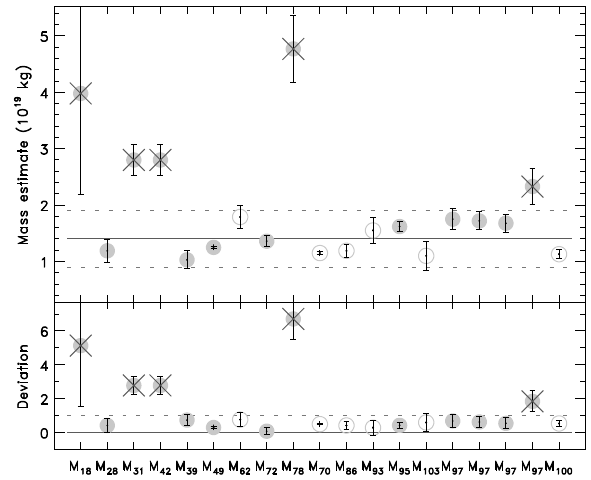}
\caption[Mass estimates for (7) Iris]{%
  \label{fap: mass000007}
  Mass estimates for (7) Iris.
}
\end{figure}

  \begin{figure}[!ht]
  \centering
  \includegraphics[width=.49\textwidth]{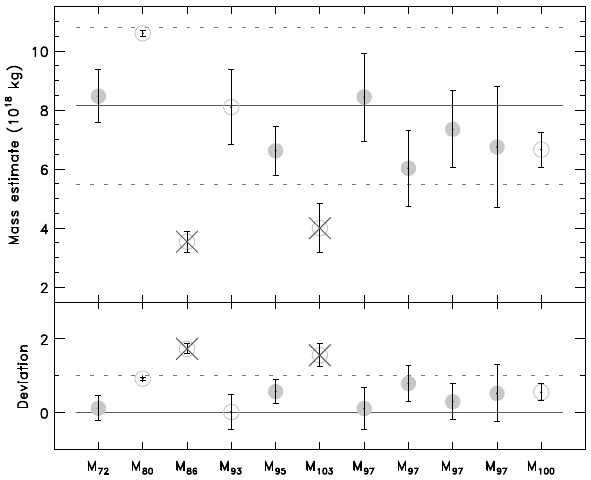}
\caption[Mass estimates for (8) Flora]{%
  \label{fap: mass000008}
  Mass estimates for (8) Flora.
}
\end{figure}

  \begin{figure}[!ht]
  \centering
  \includegraphics[width=.49\textwidth]{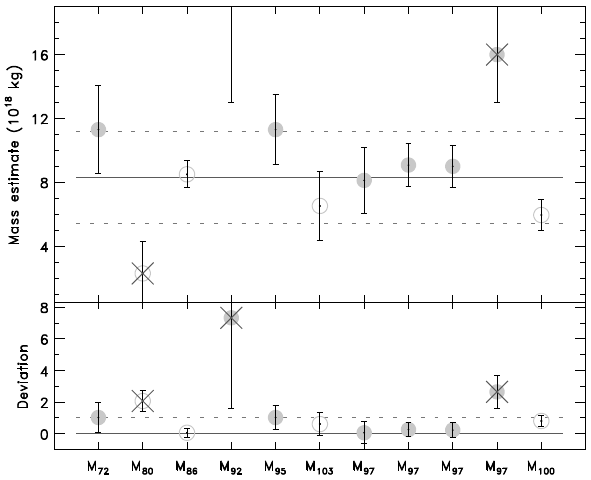}
\caption[Mass estimates for (9) Metis]{%
  \label{fap: mass000009}
  Mass estimates for (9) Metis.
}
\end{figure}

  \begin{figure}[!ht]
  \centering
  \includegraphics[width=.49\textwidth]{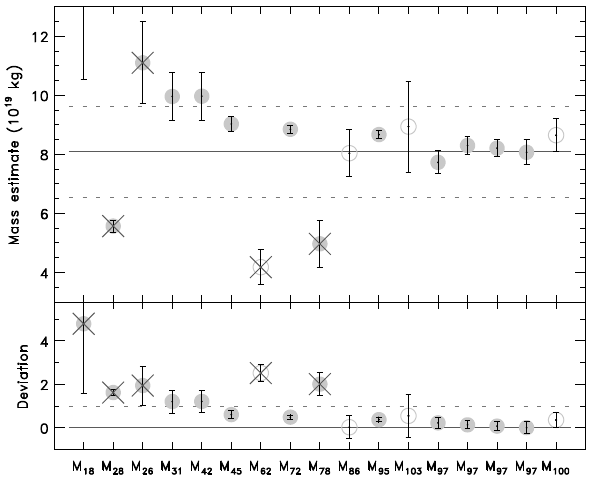}
\caption[Mass estimates for (10) Hygiea]{%
  \label{fap: mass000010}
  Mass estimates for (10) Hygiea.
}
\end{figure}

  \begin{figure}[!ht]
  \centering
  \includegraphics[width=.49\textwidth]{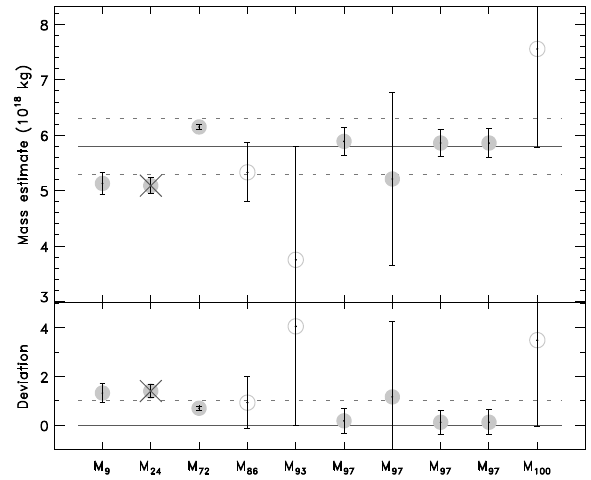}
\caption[Mass estimates for (11) Parthenope]{%
  \label{fap: mass000011}
  Mass estimates for (11) Parthenope.
}
\end{figure}

  \begin{figure}[!ht]
  \centering
  \includegraphics[width=.49\textwidth]{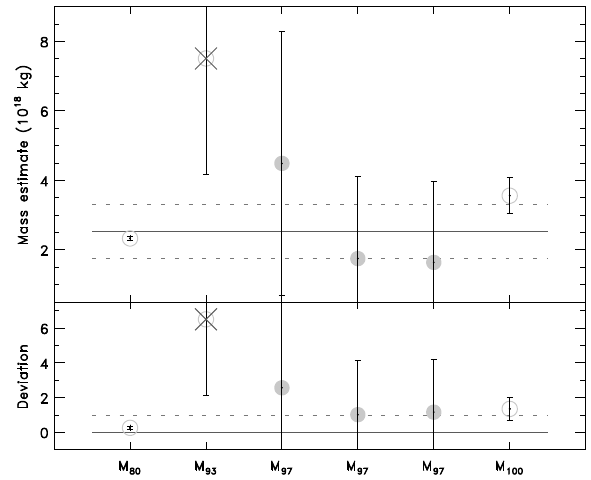}
\caption[Mass estimates for (12) Victoria]{%
  \label{fap: mass000012}
  Mass estimates for (12) Victoria.
}
\end{figure}

  \begin{figure}[!ht]
  \centering
  \includegraphics[width=.49\textwidth]{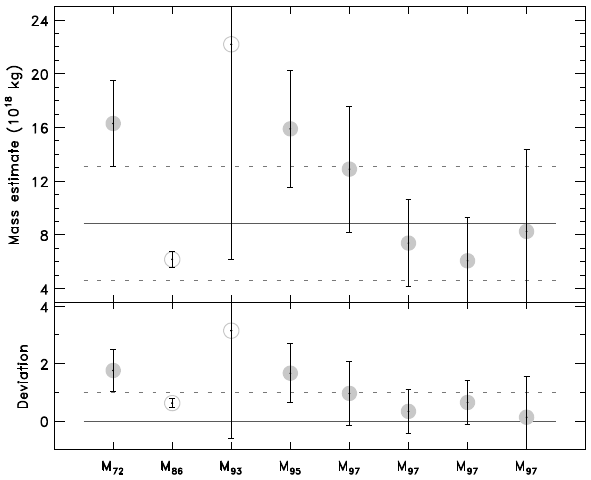}
\caption[Mass estimates for (13) Egeria]{%
  \label{fap: mass000013}
  Mass estimates for (13) Egeria.
}
\end{figure}

  \begin{figure}[!ht]
  \centering
  \includegraphics[width=.49\textwidth]{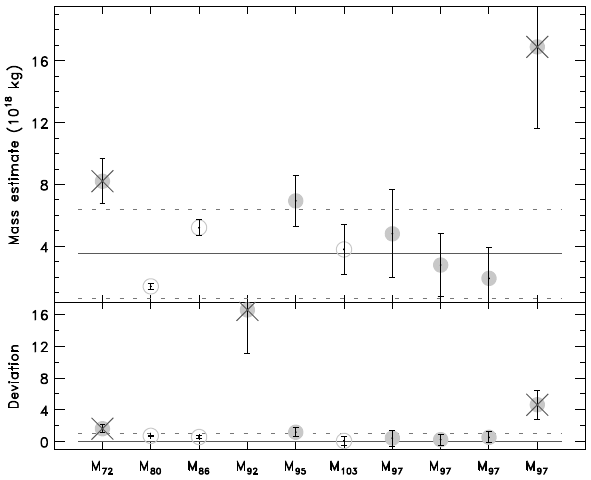}
\caption[Mass estimates for (14) Irene]{%
  \label{fap: mass000014}
  Mass estimates for (14) Irene.
}
\end{figure}

  \begin{figure}[!ht]
  \centering
  \includegraphics[width=.49\textwidth]{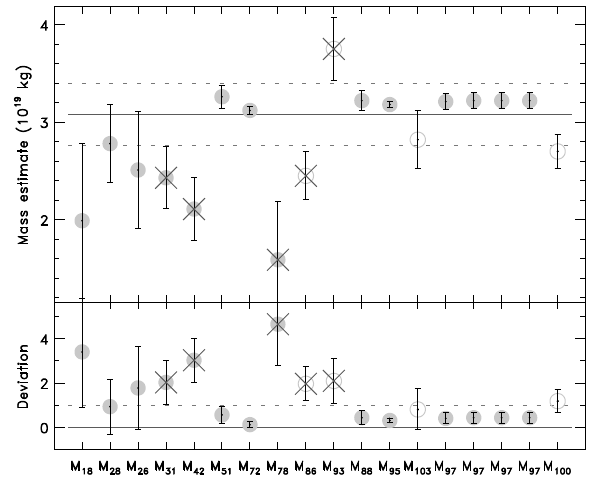}
\caption[Mass estimates for (15) Eunomia]{%
  \label{fap: mass000015}
  Mass estimates for (15) Eunomia.
}
\end{figure}

  \begin{figure}[!ht]
  \centering
  \includegraphics[width=.49\textwidth]{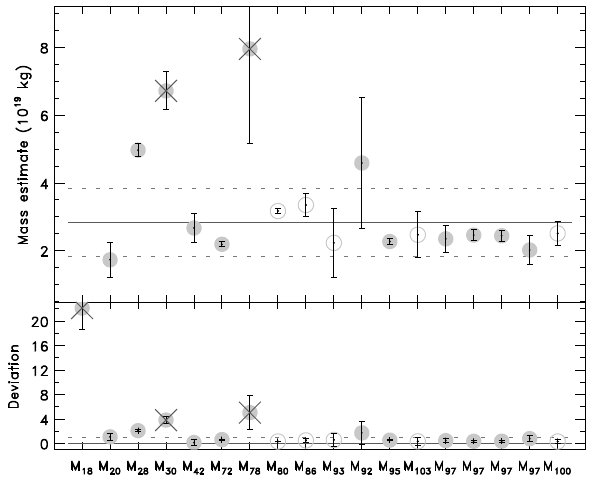}
\caption[Mass estimates for (16) Psyche]{%
  \label{fap: mass000016}
  Mass estimates for (16) Psyche.
}
\end{figure}

  \begin{figure}[!ht]
  \centering
  \includegraphics[width=.49\textwidth]{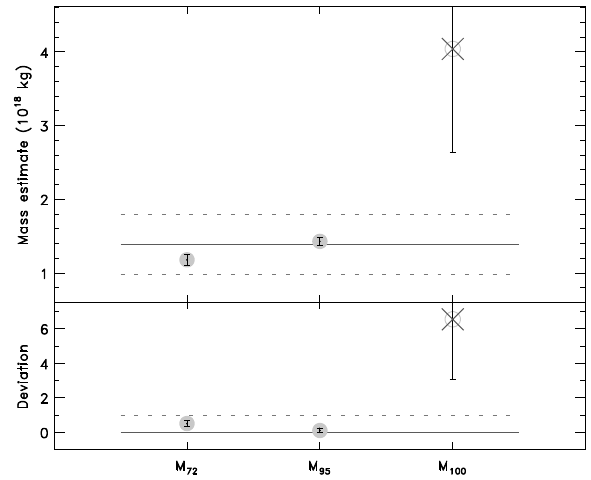}
\caption[Mass estimates for (17) Thetis]{%
  \label{fap: mass000017}
  Mass estimates for (17) Thetis.
}
\end{figure}

  \begin{figure}[!ht]
  \centering
  \includegraphics[width=.49\textwidth]{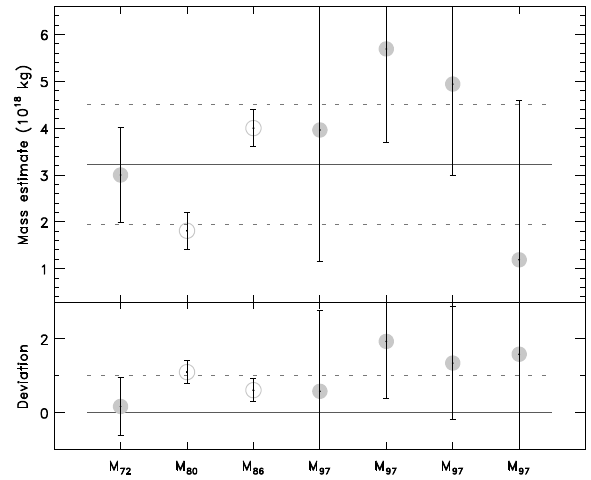}
\caption[Mass estimates for (18) Melpomene]{%
  \label{fap: mass000018}
  Mass estimates for (18) Melpomene.
}
\end{figure}

\clearpage
  \begin{figure}[!ht]
  \centering
  \includegraphics[width=.49\textwidth]{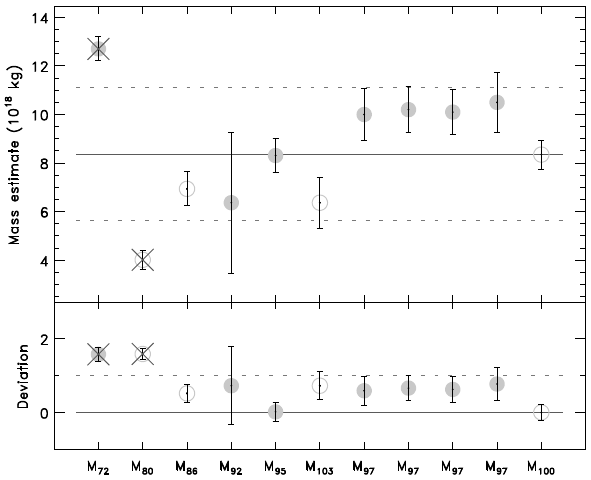}
\caption[Mass estimates for (19) Fortuna]{%
  \label{fap: mass000019}
  Mass estimates for (19) Fortuna.
}
\end{figure}

  \begin{figure}[!ht]
  \centering
  \includegraphics[width=.49\textwidth]{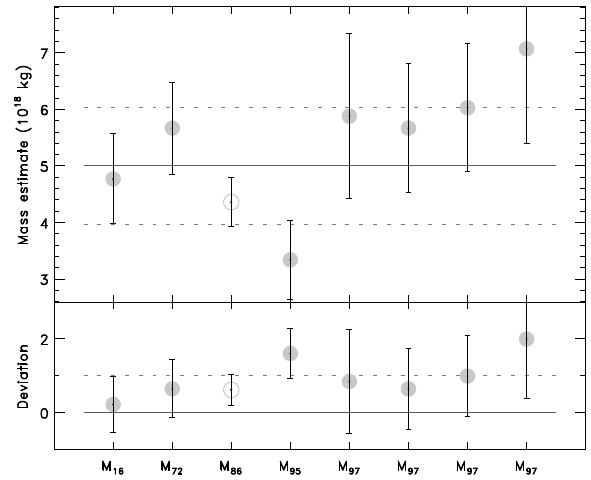}
\caption[Mass estimates for (20) Massalia]{%
  \label{fap: mass000020}
  Mass estimates for (20) Massalia.
}
\end{figure}

  \begin{figure}[!ht]
  \centering
  \includegraphics[width=.49\textwidth]{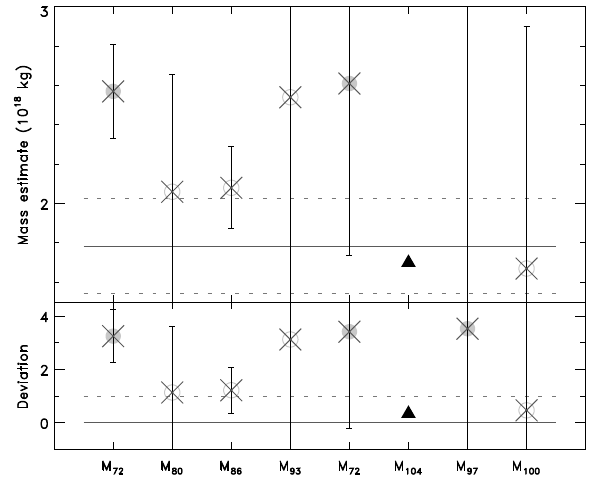}
\caption[Mass estimates for (21) Lutetia]{%
  \label{fap: mass000021}
  Mass estimates for (21) Lutetia.
  Only the flyby estimate (M$_{104}$) is used here. 
}
\end{figure}

  \begin{figure}[!ht]
  \centering
  \includegraphics[width=.49\textwidth]{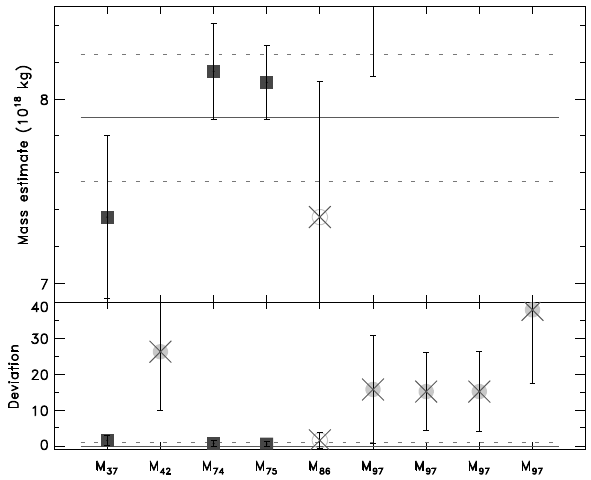}
\caption[Mass estimates for (22) Kalliope]{%
  \label{fap: mass000022}
  Mass estimates for (22) Kalliope.
  Only the estimates based on direct imaging of the system are used here (M$_{37}$, M$_{74}$, and M$_{75}$). Apart from M$_{86}$, all the indirect determinations are far from the solution derived from  the analysis of the satellite's orbit
}
\end{figure}

  \begin{figure}[!ht]
  \centering
  \includegraphics[width=.49\textwidth]{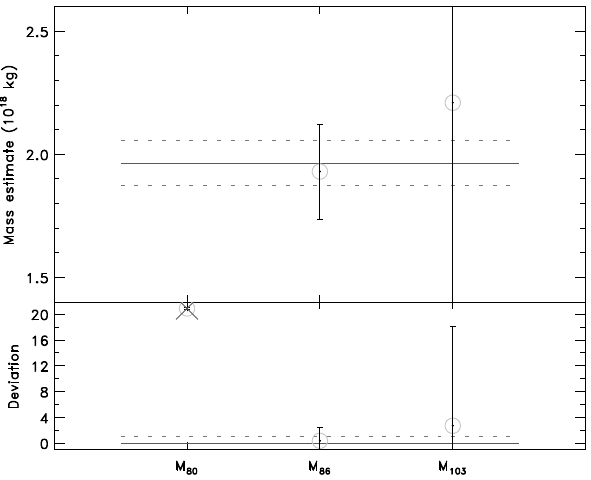}
\caption[Mass estimates for (23) Thalia]{%
  \label{fap: mass000023}
  Mass estimates for (23) Thalia.
  The mass estimate from M$_{80}$ gives an unrealistic density of 0.09\,$\pm$\,0.03 if used alone, and is therefore discarded.
}
\end{figure}

  \begin{figure}[!ht]
  \centering
  \includegraphics[width=.49\textwidth]{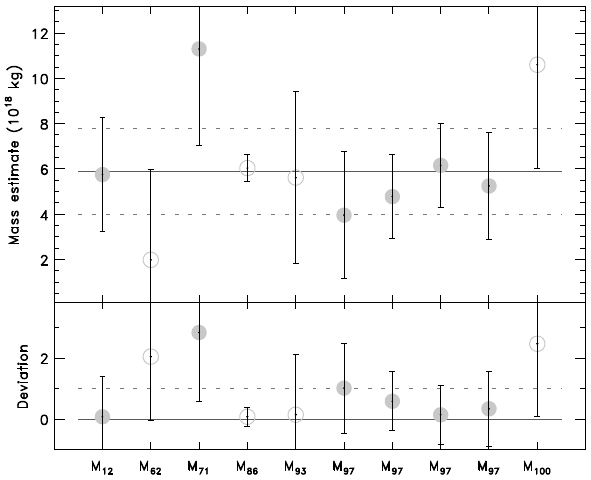}
\caption[Mass estimates for (24) Themis]{%
  \label{fap: mass000024}
  Mass estimates for (24) Themis.
}
\end{figure}

  \begin{figure}[!ht]
  \centering
  \includegraphics[width=.49\textwidth]{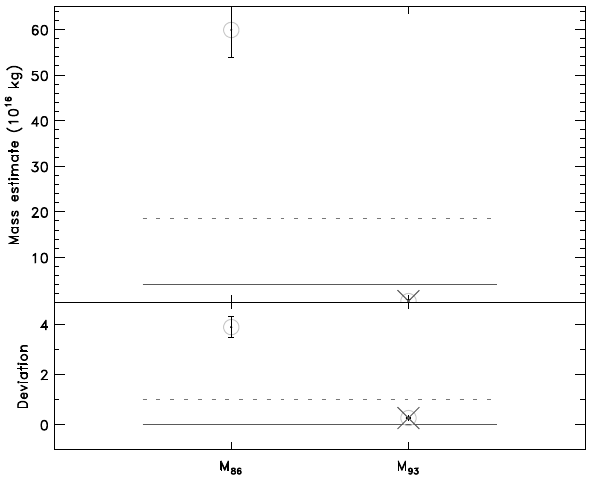}
\caption[Mass estimates for (25) Phocaea]{%
  \label{fap: mass000025}
  Mass estimates for (25) Phocaea.
  Although the mass estimate from M$_{86}$ is outside the weighted average range, this is due to the very small uncertainty associated with the determination from M$_{93}$ which gives an unrealistic density of 0.02\,$\pm$\,0.02 if used alone. M$_{93}$ estimate is therefore discarded.
}
\end{figure}

  \begin{figure}[!ht]
  \centering
  \includegraphics[width=.49\textwidth]{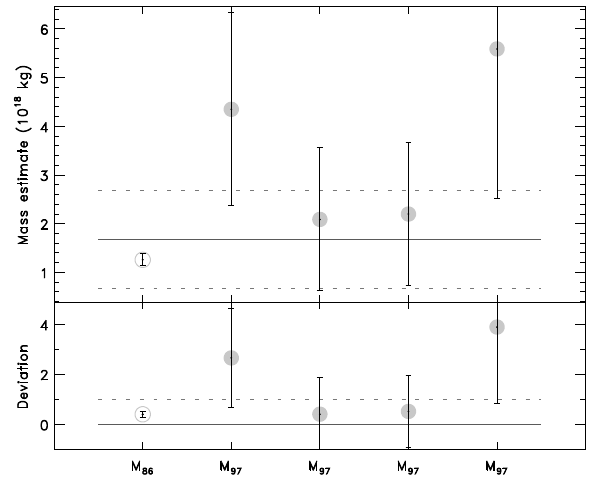}
\caption[Mass estimates for (27) Euterpe]{%
  \label{fap: mass000027}
  Mass estimates for (27) Euterpe.
}
\end{figure}

  \begin{figure}[!ht]
  \centering
  \includegraphics[width=.49\textwidth]{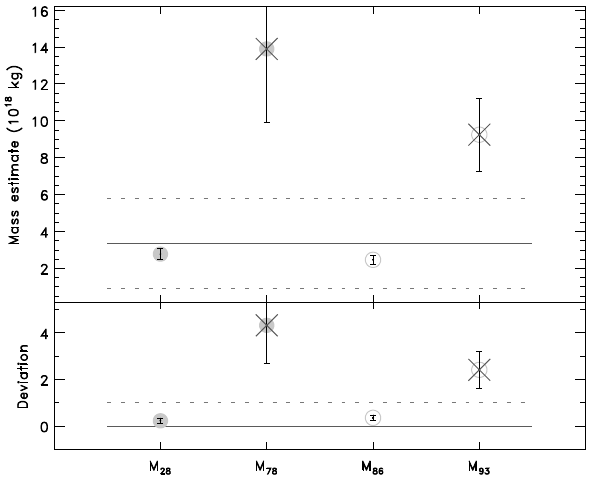}
\caption[Mass estimates for (28) Bellona]{%
  \label{fap: mass000028}
  Mass estimates for (28) Bellona.
  The mass estimates from M$_{78}$ and M$_{93}$ give unrealistic densities of  18\,$\pm$\,7 and 12\,$\pm$\,4 respectively, and are therefore discarded.
}
\end{figure}

  \begin{figure}[!ht]
  \centering
  \includegraphics[width=.49\textwidth]{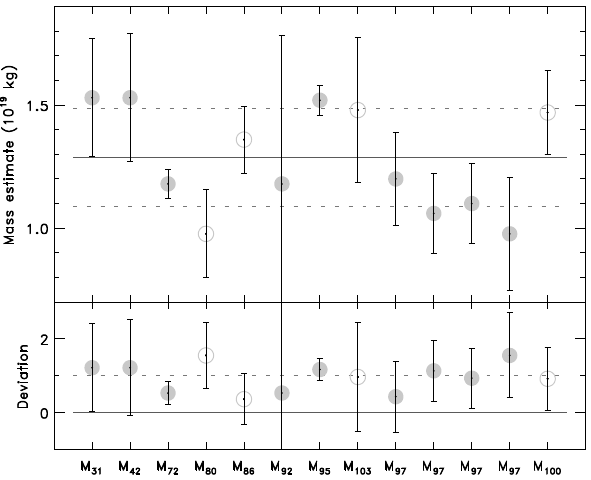}
\caption[Mass estimates for (29) Amphitrite]{%
  \label{fap: mass000029}
  Mass estimates for (29) Amphitrite.
}
\end{figure}

  \begin{figure}[!ht]
  \centering
  \includegraphics[width=.49\textwidth]{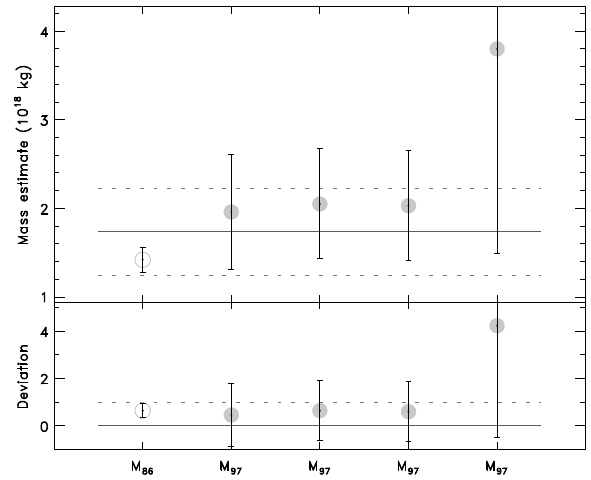}
\caption[Mass estimates for (30) Urania]{%
  \label{fap: mass000030}
  Mass estimates for (30) Urania.
}
\end{figure}

  \begin{figure}[!ht]
  \centering
  \includegraphics[width=.49\textwidth]{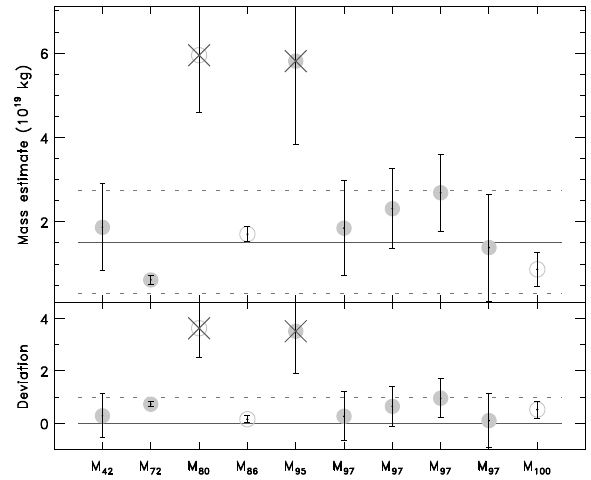}
\caption[Mass estimates for (31) Euphrosyne]{%
  \label{fap: mass000031}
  Mass estimates for (31) Euphrosyne.
}
\end{figure}

  \begin{figure}[!ht]
  \centering
  \includegraphics[width=.49\textwidth]{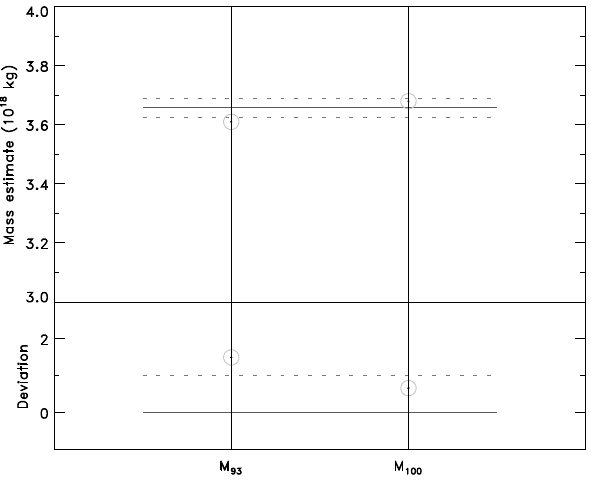}
\caption[Mass estimates for (34) Circe]{%
  \label{fap: mass000034}
  Mass estimates for (34) Circe.
}
\end{figure}

  \begin{figure}[!ht]
  \centering
  \includegraphics[width=.49\textwidth]{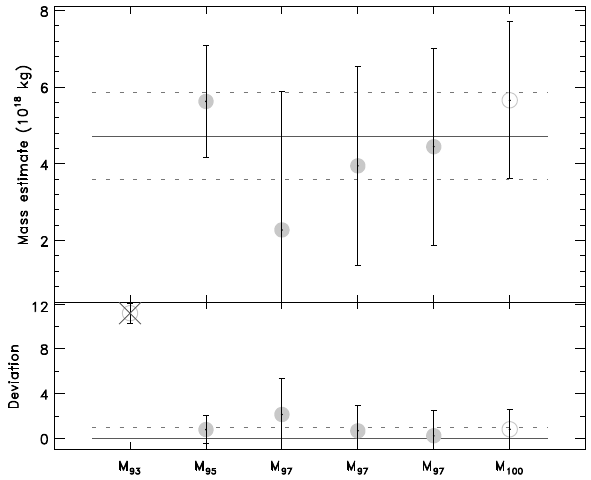}
\caption[Mass estimates for (39) Laetitia]{%
  \label{fap: mass000039}
  Mass estimates for (39) Laetitia.
  The mass estimate from M$_{93}$ gives an unrealistic density of 8.9\,$\pm$\,0.9 if used alone, and is therefore discarded.
}
\end{figure}

  \begin{figure}[!ht]
  \centering
  \includegraphics[width=.49\textwidth]{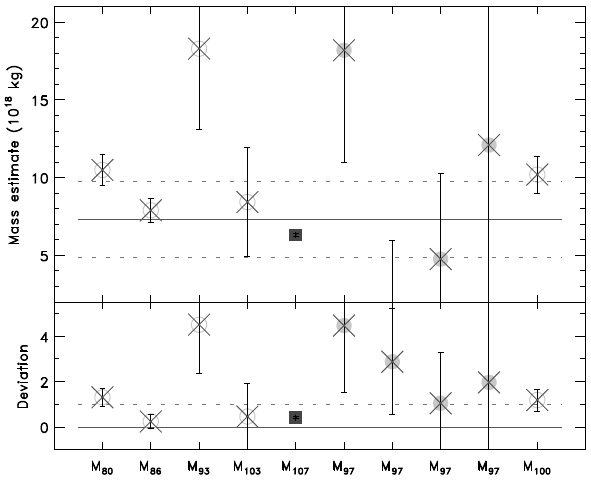}
\caption[Mass estimates for (41) Daphne]{%
  \label{fap: mass000041}
  Mass estimates for (41) Daphne.
}
\end{figure}

  \begin{figure}[!ht]
  \centering
  \includegraphics[width=.49\textwidth]{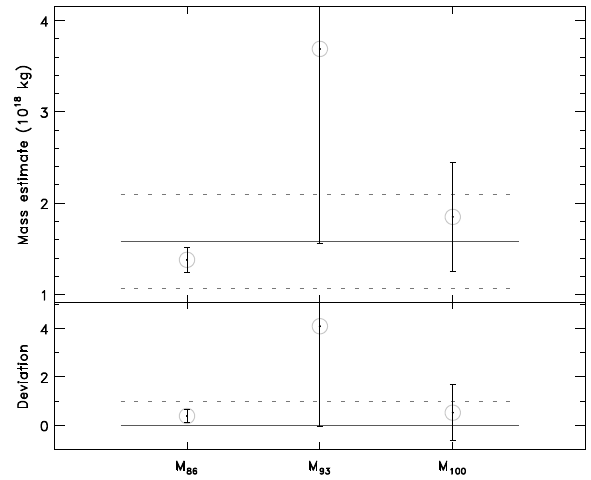}
\caption[Mass estimates for (42) Isis]{%
  \label{fap: mass000042}
  Mass estimates for (42) Isis.
}
\end{figure}

  \begin{figure}[!ht]
  \centering
  \includegraphics[width=.49\textwidth]{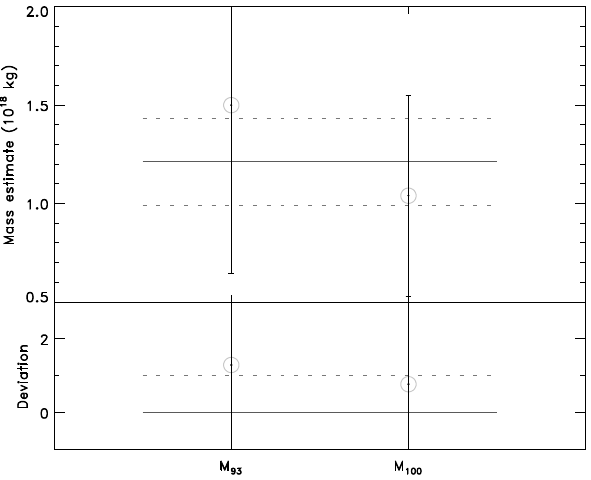}
\caption[Mass estimates for (43) Ariadne]{%
  \label{fap: mass000043}
  Mass estimates for (43) Ariadne.
}
\end{figure}

  \begin{figure}[!ht]
  \centering
  \includegraphics[width=.49\textwidth]{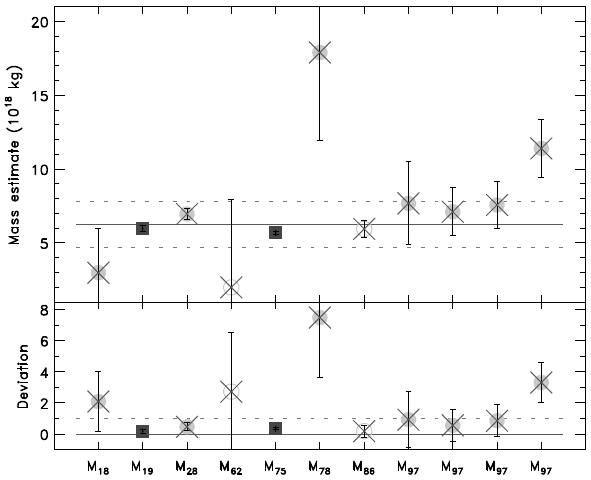}
\caption[Mass estimates for (45) Eugenia]{%
  \label{fap: mass000045}
  Mass estimates for (45) Eugenia.
  Only the mass estimates based on direct imaging of the system are used here (M$_{19}$ and M$_{75}$). 
}
\end{figure}

\clearpage
  \begin{figure}[!ht]
  \centering
  \includegraphics[width=.49\textwidth]{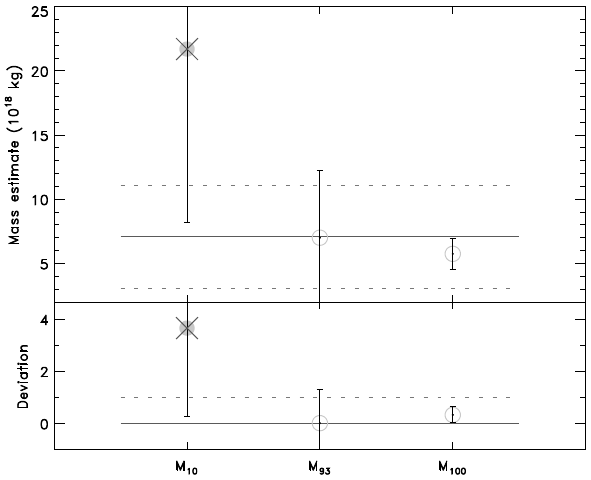}
\caption[Mass estimates for (46) Hestia]{%
  \label{fap: mass000046}
  Mass estimates for (46) Hestia.
  The mass estimate from M$_{10}$ gives an unrealistic density of 19\,$\pm$\,12 if used alone, and is therefore discarded.
}
\end{figure}

  \begin{figure}[!ht]
  \centering
  \includegraphics[width=.49\textwidth]{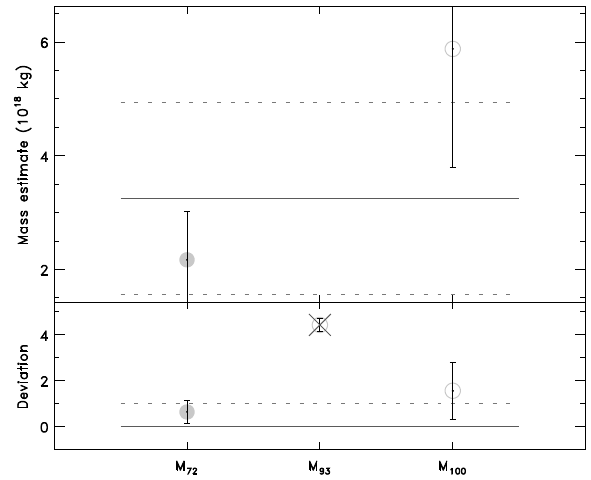}
\caption[Mass estimates for (47) Aglaja]{%
  \label{fap: mass000047}
  Mass estimates for (47) Aglaja.
  The mass estimate from M$_{93}$ gives an unrealistic density of 10.3\,$\pm$\,4.3 if used alone, and is therefore discarded.
}
\end{figure}

  \begin{figure}[!ht]
  \centering
  \includegraphics[width=.49\textwidth]{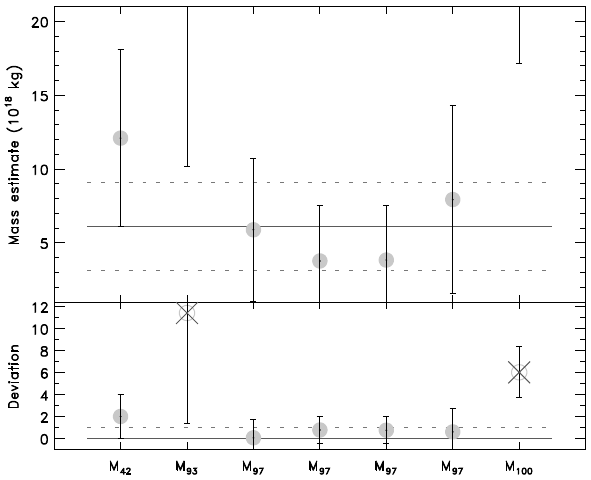}
\caption[Mass estimates for (48) Doris]{%
  \label{fap: mass000048}
  Mass estimates for (48) Doris.
}
\end{figure}

  \begin{figure}[!ht]
  \centering
  \includegraphics[width=.49\textwidth]{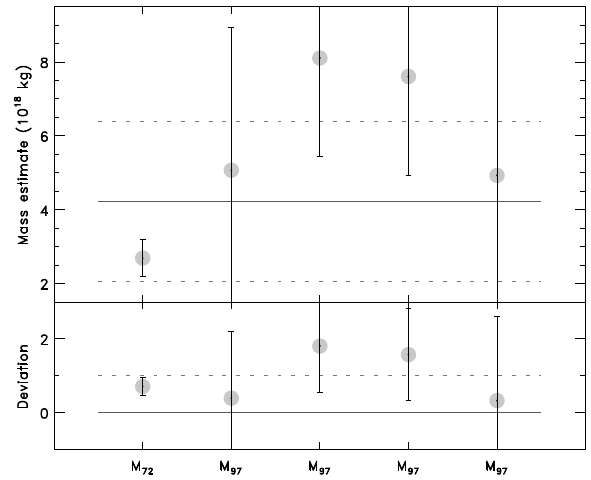}
\caption[Mass estimates for (49) Pales]{%
  \label{fap: mass000049}
  Mass estimates for (49) Pales.
}
\end{figure}

  \begin{figure}[!ht]
  \centering
  \includegraphics[width=.49\textwidth]{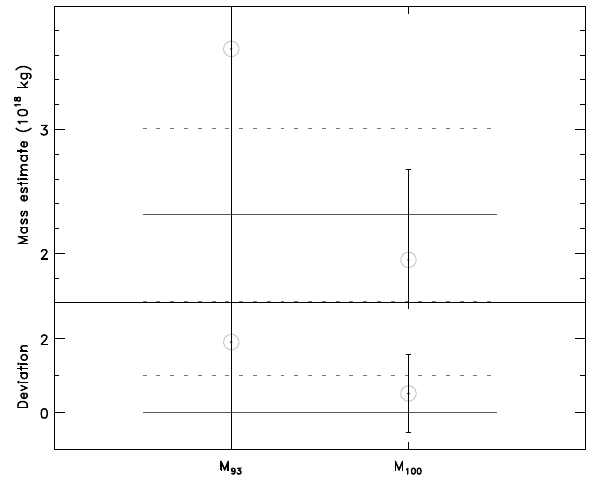}
\caption[Mass estimates for (50) Virginia]{%
  \label{fap: mass000050}
  Mass estimates for (50) Virginia.
}
\end{figure}

  \begin{figure}[!ht]
  \centering
  \includegraphics[width=.49\textwidth]{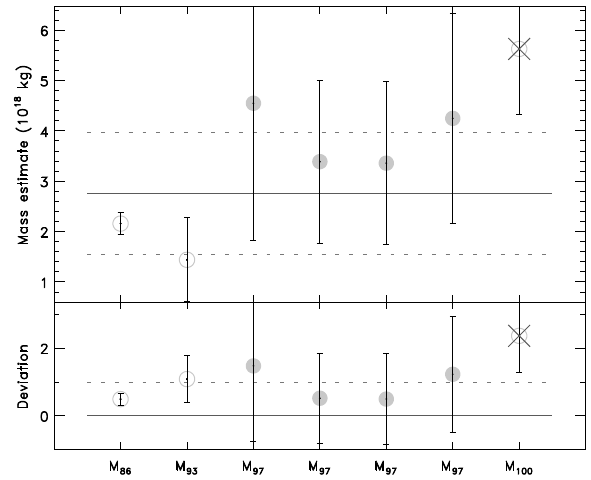}
\caption[Mass estimates for (51) Nemausa]{%
  \label{fap: mass000051}
  Mass estimates for (51) Nemausa.
}
\end{figure}

  \begin{figure}[!ht]
  \centering
  \includegraphics[width=.49\textwidth]{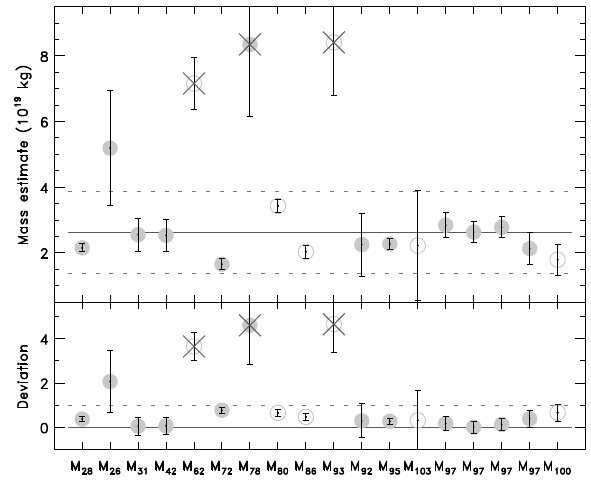}
\caption[Mass estimates for (52) Europa]{%
  \label{fap: mass000052}
  Mass estimates for (52) Europa.
}
\end{figure}

  \begin{figure}[!ht]
  \centering
  \includegraphics[width=.49\textwidth]{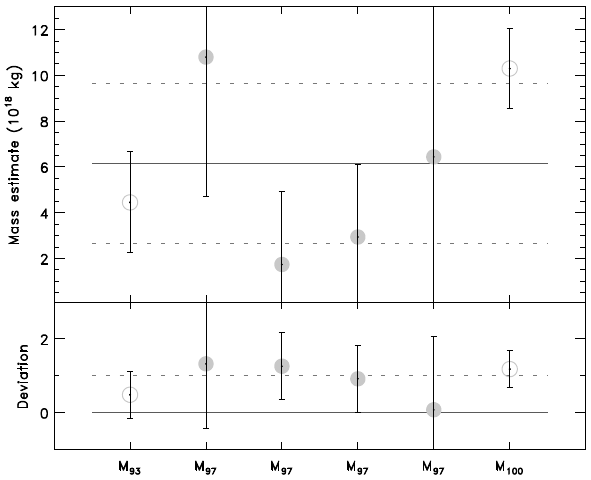}
\caption[Mass estimates for (54) Alexandra]{%
  \label{fap: mass000054}
  Mass estimates for (54) Alexandra.
}
\end{figure}

  \begin{figure}[!ht]
  \centering
  \includegraphics[width=.49\textwidth]{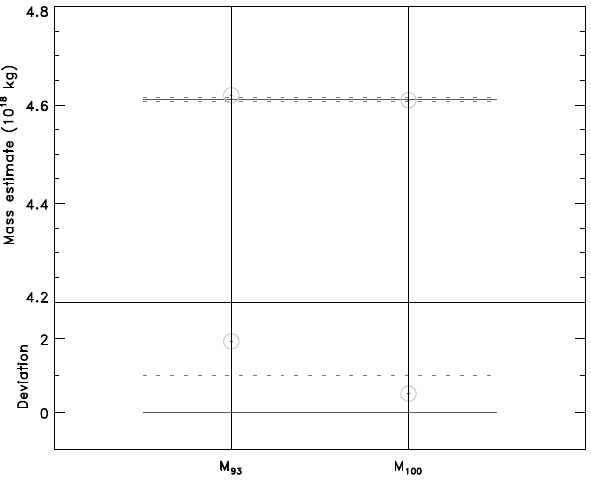}
\caption[Mass estimates for (56) Melete]{%
  \label{fap: mass000056}
  Mass estimates for (56) Melete.
}
\end{figure}

  \begin{figure}[!ht]
  \centering
  \includegraphics[width=.49\textwidth]{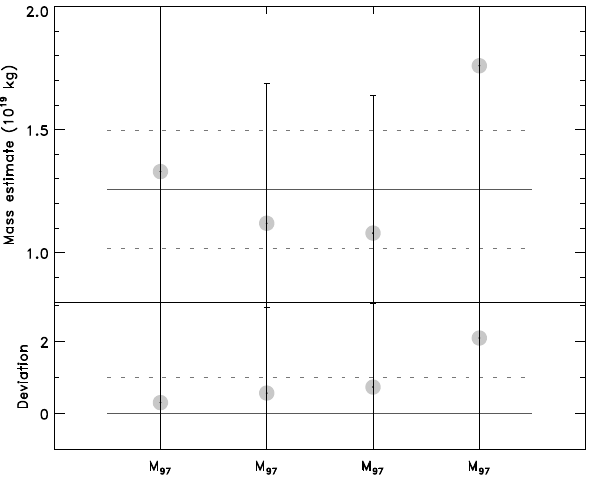}
\caption[Mass estimates for (57) Mnemosyne]{%
  \label{fap: mass000057}
  Mass estimates for (57) Mnemosyne.
}
\end{figure}

  \begin{figure}[!ht]
  \centering
  \includegraphics[width=.49\textwidth]{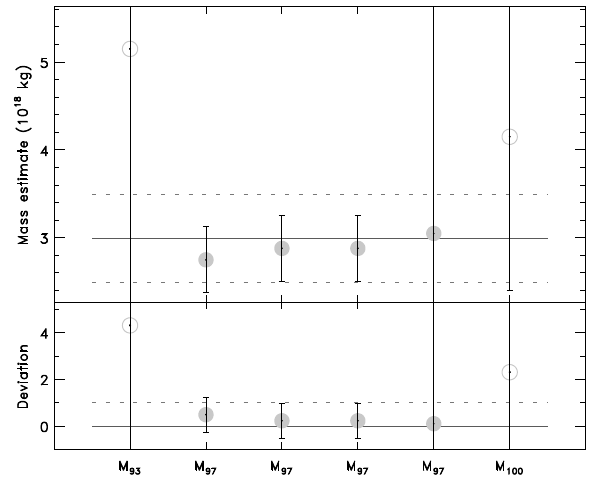}
\caption[Mass estimates for (59) Elpis]{%
  \label{fap: mass000059}
  Mass estimates for (59) Elpis.
}
\end{figure}

  \begin{figure}[!ht]
  \centering
  \includegraphics[width=.49\textwidth]{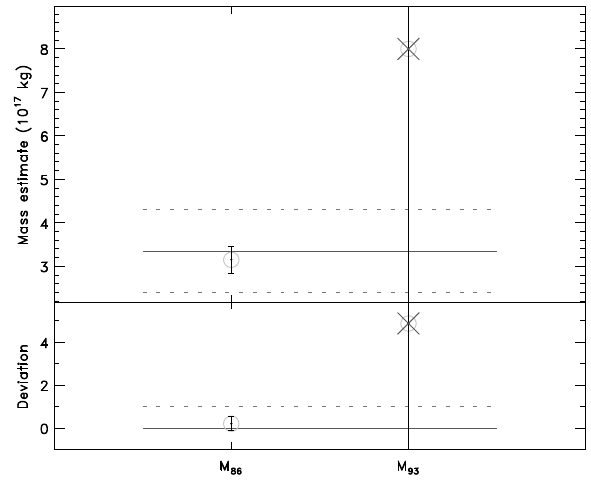}
\caption[Mass estimates for (60) Echo]{%
  \label{fap: mass000060}
  Mass estimates for (60) Echo.
  The mass estimate from M$_{93}$ gives a low-constrained density of 6.8\,$\pm$\,6.4 if used alone.  Only the estimate from M$_{86}$ is used.
}
\end{figure}

  \begin{figure}[!ht]
  \centering
  \includegraphics[width=.49\textwidth]{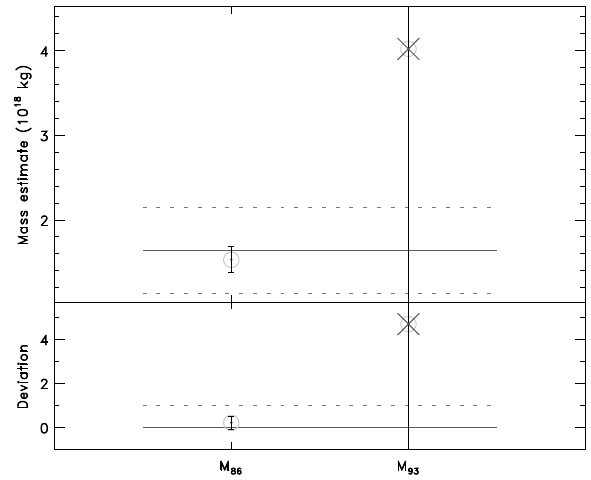}
\caption[Mass estimates for (63) Ausonia]{%
  \label{fap: mass000063}
  Mass estimates for (63) Ausonia.
  The mass estimate from M$_{93}$ gives a low-constrained density of 9.7\,$\pm$\,8.3 if used alone.  Only the estimate from M$_{86}$ is used.
}
\end{figure}

  \begin{figure}[!ht]
  \centering
  \includegraphics[width=.49\textwidth]{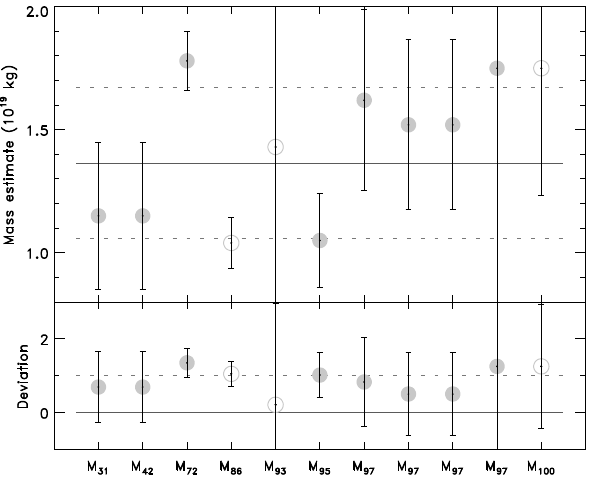}
\caption[Mass estimates for (65) Cybele]{%
  \label{fap: mass000065}
  Mass estimates for (65) Cybele.
}
\end{figure}

  \begin{figure}[!ht]
  \centering
  \includegraphics[width=.49\textwidth]{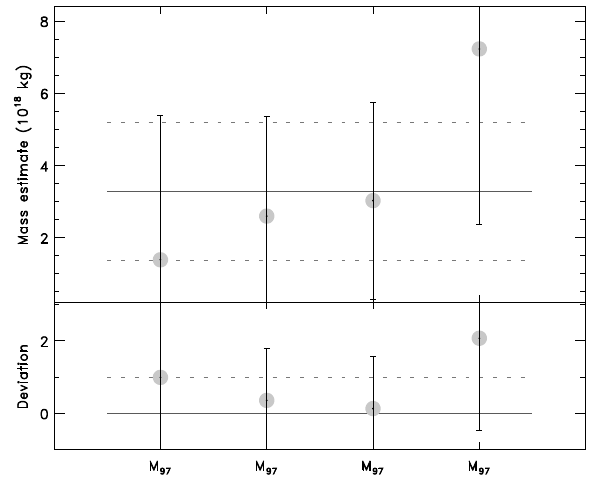}
\caption[Mass estimates for (68) Leto]{%
  \label{fap: mass000068}
  Mass estimates for (68) Leto.
}
\end{figure}

  \begin{figure}[!ht]
  \centering
  \includegraphics[width=.49\textwidth]{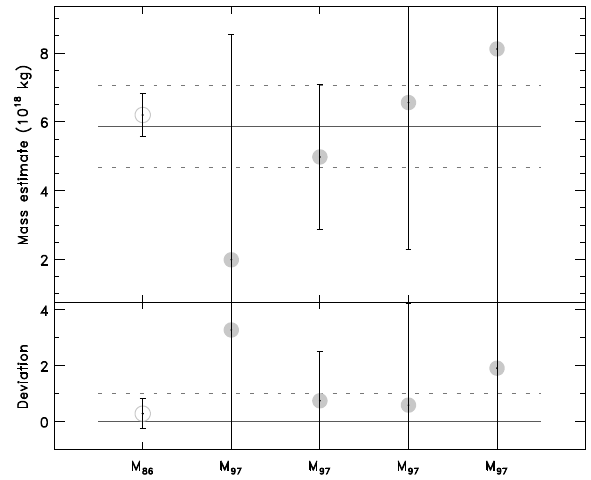}
\caption[Mass estimates for (69) Hesperia]{%
  \label{fap: mass000069}
  Mass estimates for (69) Hesperia.
}
\end{figure}

  \begin{figure}[!ht]
  \centering
  \includegraphics[width=.49\textwidth]{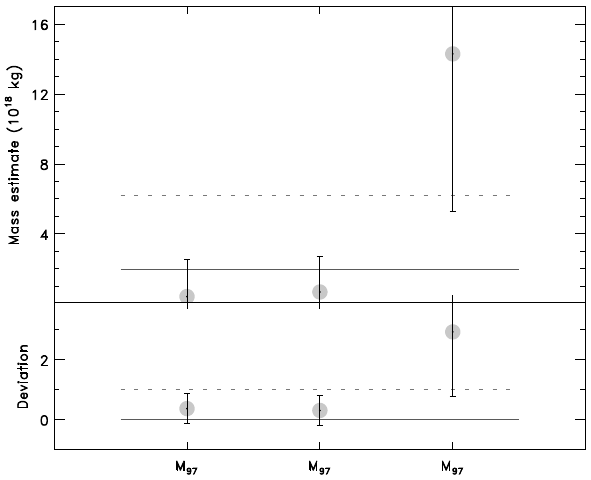}
\caption[Mass estimates for (76) Freia]{%
  \label{fap: mass000076}
  Mass estimates for (76) Freia.
}
\end{figure}

  \begin{figure}[!ht]
  \centering
  \includegraphics[width=.49\textwidth]{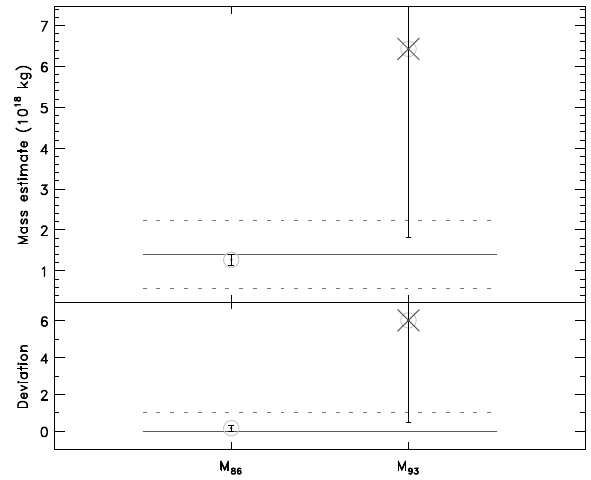}
\caption[Mass estimates for (78) Diana]{%
  \label{fap: mass000078}
  Mass estimates for (78) Diana.
  The mass estimate from M$_{93}$ gives a low-constrained density of 6.9\,$\pm$\,5.0 if used alone.  Only the estimate from M$_{86}$ is used.
}
\end{figure}

\clearpage
  \begin{figure}[!ht]
  \centering
  \includegraphics[width=.49\textwidth]{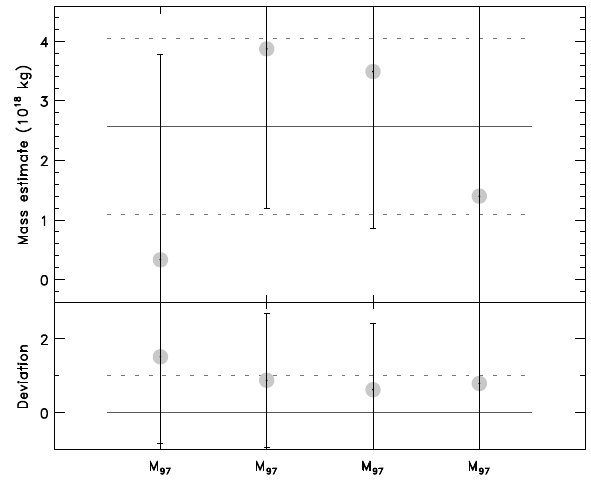}
\caption[Mass estimates for (85) Io]{%
  \label{fap: mass000085}
  Mass estimates for (85) Io.
}
\end{figure}

  \begin{figure}[!ht]
  \centering
  \includegraphics[width=.49\textwidth]{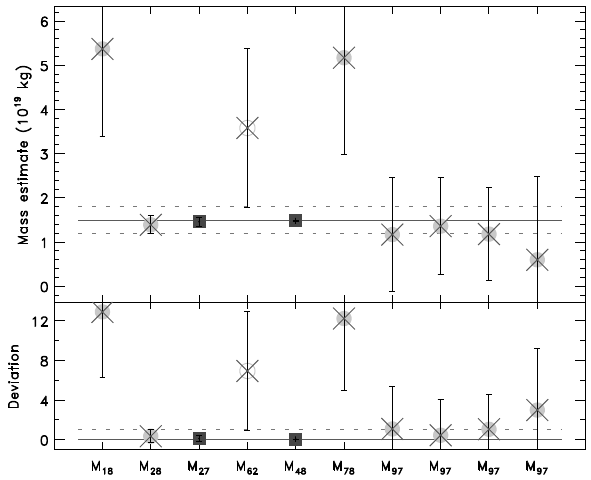}
\caption[Mass estimates for (87) Sylvia]{%
  \label{fap: mass000087}
  Mass estimates for (87) Sylvia.
  Only the mass estimates based on direct imaging of the system are used here (M$_{27}$ and M$_{48}$). 
}
\end{figure}

  \begin{figure}[!ht]
  \centering
  \includegraphics[width=.49\textwidth]{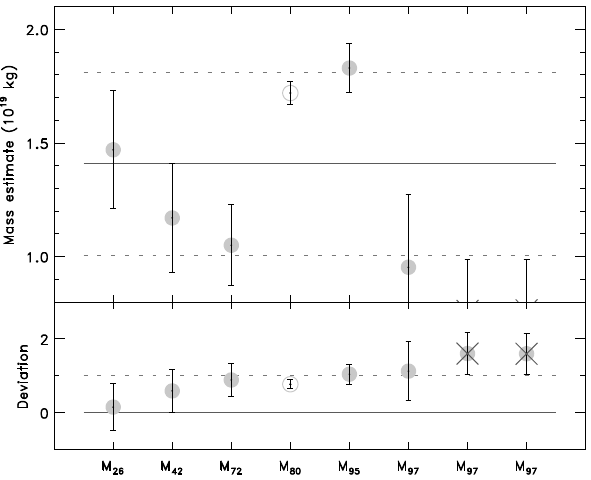}
\caption[Mass estimates for (88) Thisbe]{%
  \label{fap: mass000088}
  Mass estimates for (88) Thisbe.
}
\end{figure}

  \begin{figure}[!ht]
  \centering
  \includegraphics[width=.49\textwidth]{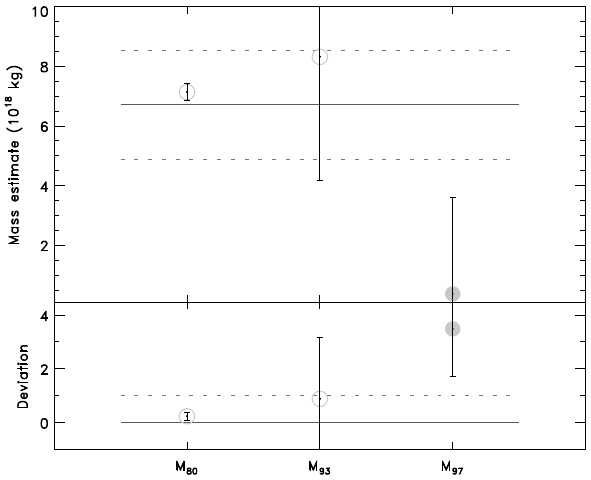}
\caption[Mass estimates for (89) Julia]{%
  \label{fap: mass000089}
  Mass estimates for (89) Julia.
}
\end{figure}

  \begin{figure}[!ht]
  \centering
  \includegraphics[width=.49\textwidth]{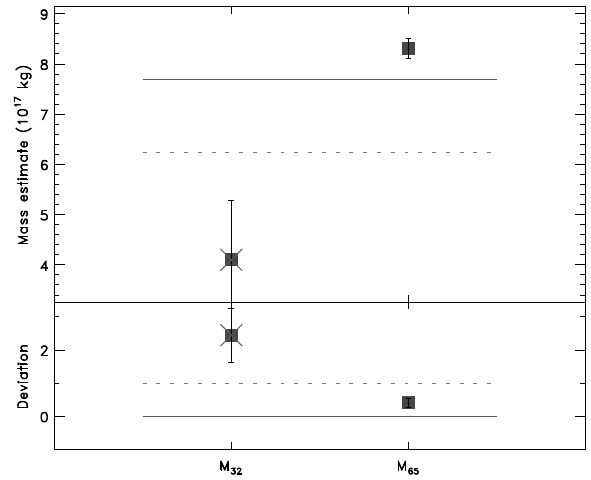}
\caption[Mass estimates for (90) Antiope]{%
  \label{fap: mass000090}
  Mass estimates for (90) Antiope.
  The mass estimate from M$_{32}$ was based on few discovery images, and the estimate from M$_{65}$ is preferred.
}
\end{figure}

  \begin{figure}[!ht]
  \centering
  \includegraphics[width=.49\textwidth]{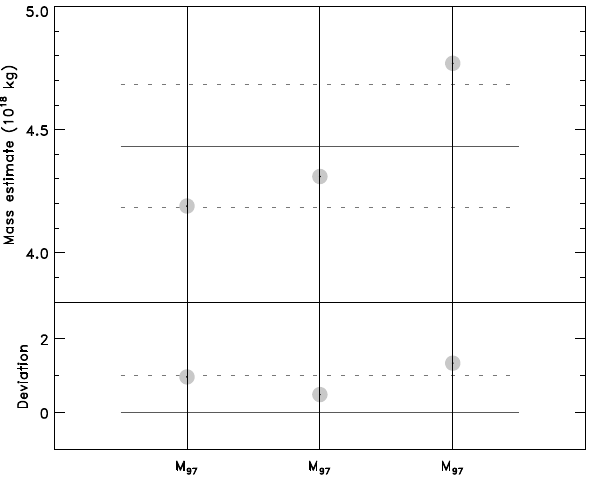}
\caption[Mass estimates for (92) Undina]{%
  \label{fap: mass000092}
  Mass estimates for (92) Undina.
}
\end{figure}

  \begin{figure}[!ht]
  \centering
  \includegraphics[width=.49\textwidth]{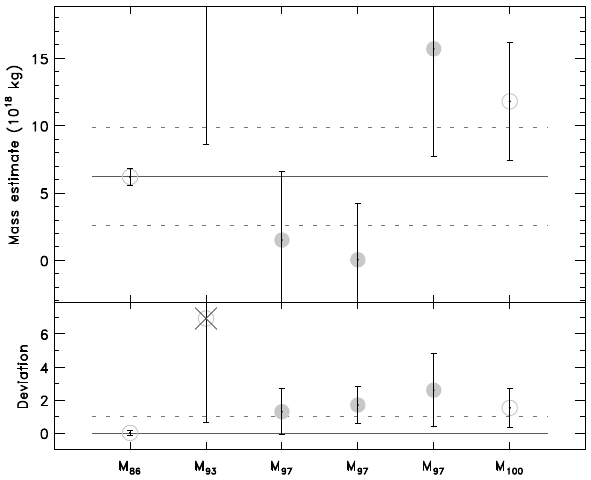}
\caption[Mass estimates for (94) Aurora]{%
  \label{fap: mass000094}
  Mass estimates for (94) Aurora.
}
\end{figure}

  \begin{figure}[!ht]
  \centering
  \includegraphics[width=.49\textwidth]{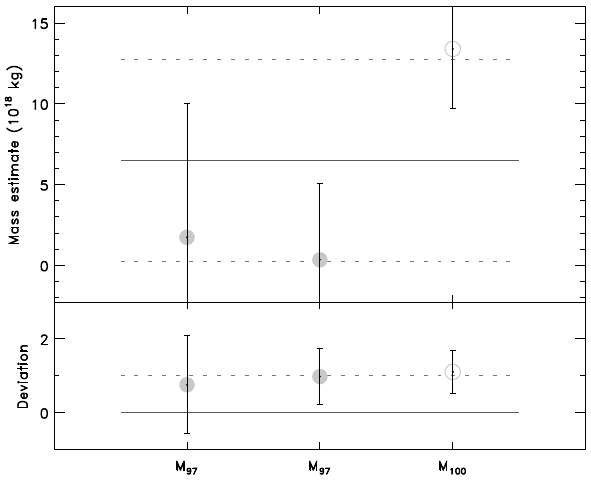}
\caption[Mass estimates for (96) Aegle]{%
  \label{fap: mass000096}
  Mass estimates for (96) Aegle.
}
\end{figure}

  \begin{figure}[!ht]
  \centering
  \includegraphics[width=.49\textwidth]{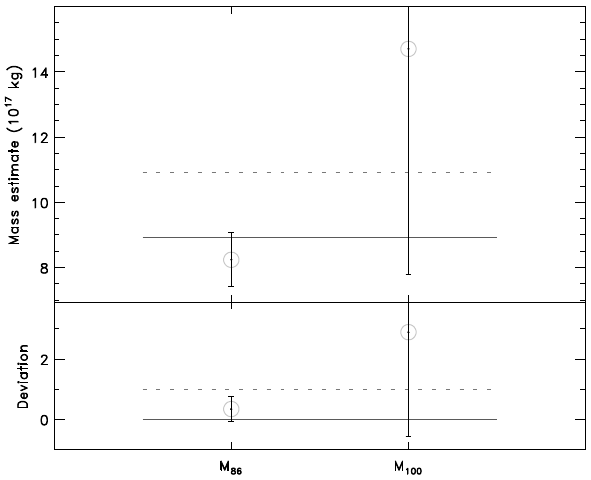}
\caption[Mass estimates for (98) Ianthe]{%
  \label{fap: mass000098}
  Mass estimates for (98) Ianthe.
}
\end{figure}

  \begin{figure}[!ht]
  \centering
  \includegraphics[width=.49\textwidth]{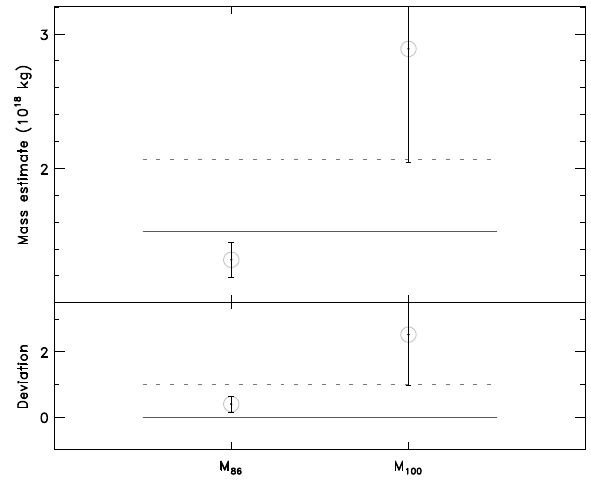}
\caption[Mass estimates for (105) Artemis]{%
  \label{fap: mass000105}
  Mass estimates for (105) Artemis.
}
\end{figure}

  \begin{figure}[!ht]
  \centering
  \includegraphics[width=.49\textwidth]{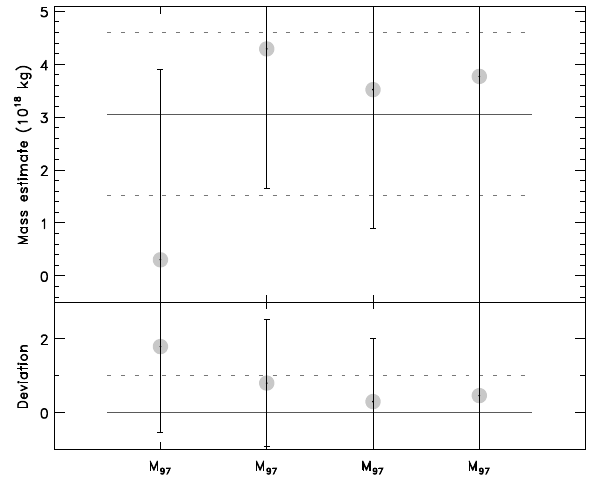}
\caption[Mass estimates for (106) Dione]{%
  \label{fap: mass000106}
  Mass estimates for (106) Dione.
}
\end{figure}

  \begin{figure}[!ht]
  \centering
  \includegraphics[width=.49\textwidth]{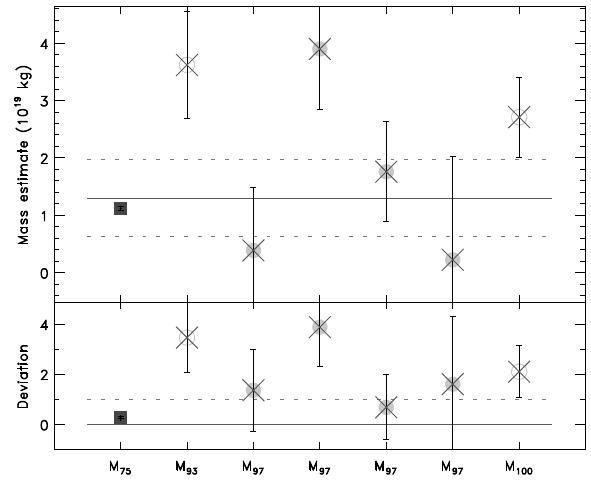}
\caption[Mass estimates for (107) Camilla]{%
  \label{fap: mass000107}
  Mass estimates for (107) Camilla.
  The mass estimate from M$_{75}$ based on direct imaging of the binary system   is preferred over the other estimates.
}
\end{figure}

  \begin{figure}[!ht]
  \centering
  \includegraphics[width=.49\textwidth]{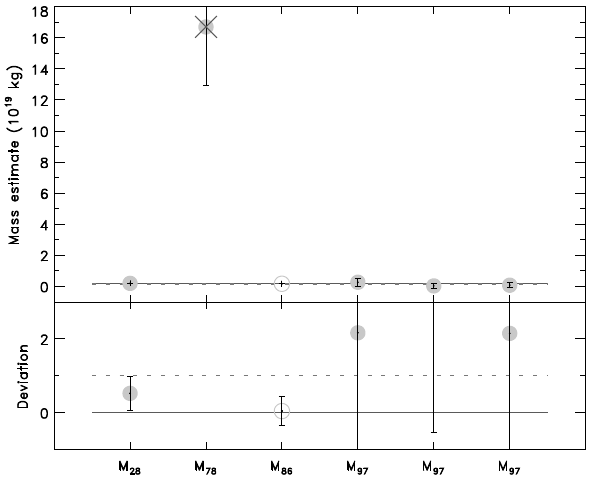}
\caption[Mass estimates for (111) Ate]{%
  \label{fap: mass000111}
  Mass estimates for (111) Ate.
  The mass estimate from M$_{78}$ gives an unrealistic density of 113\,$\pm$\,30 if used alone, and is therefore discarded.
}
\end{figure}

  \begin{figure}[!ht]
  \centering
  \includegraphics[width=.49\textwidth]{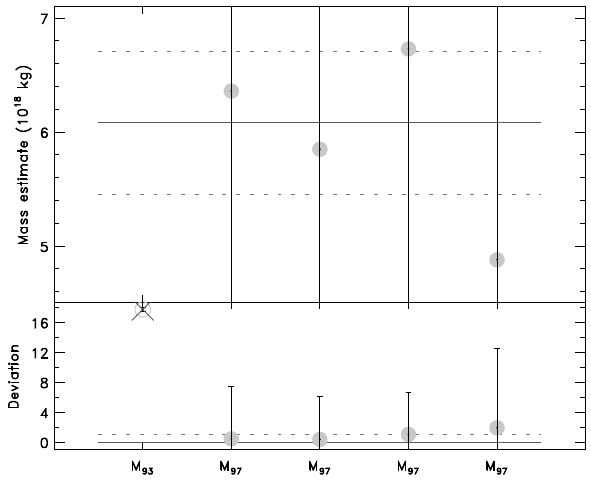}
\caption[Mass estimates for (117) Lomia]{%
  \label{fap: mass000117}
  Mass estimates for (117) Lomia.
}
\end{figure}

  \begin{figure}[!ht]
  \centering
  \includegraphics[width=.49\textwidth]{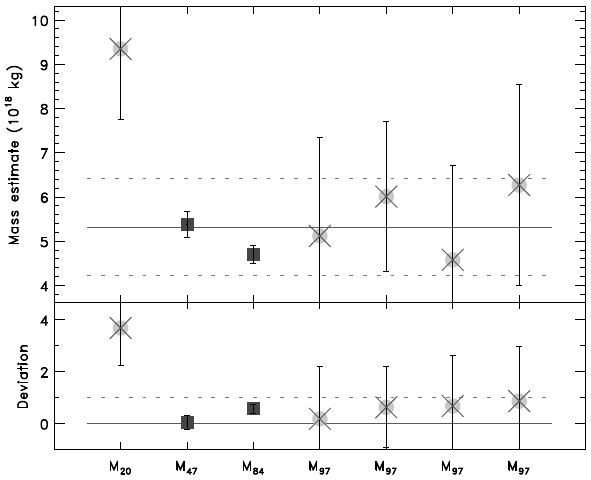}
\caption[Mass estimates for (121) Hermione]{%
  \label{fap: mass000121}
  Mass estimates for (121) Hermione.
  Only the mass estimates based on direct imaging of the system are used here (M$_{47}$ and M$_{84}$). 
}
\end{figure}

  \begin{figure}[!ht]
  \centering
  \includegraphics[width=.49\textwidth]{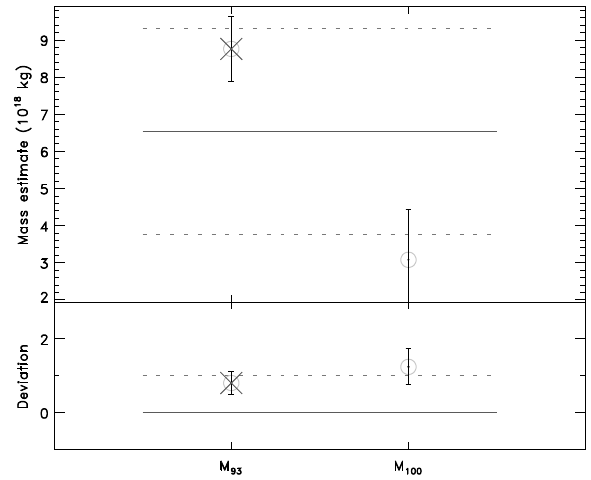}
\caption[Mass estimates for (127) Johanna]{%
  \label{fap: mass000127}
  Mass estimates for (127) Johanna.
}
\end{figure}

  \begin{figure}[!ht]
  \centering
  \includegraphics[width=.49\textwidth]{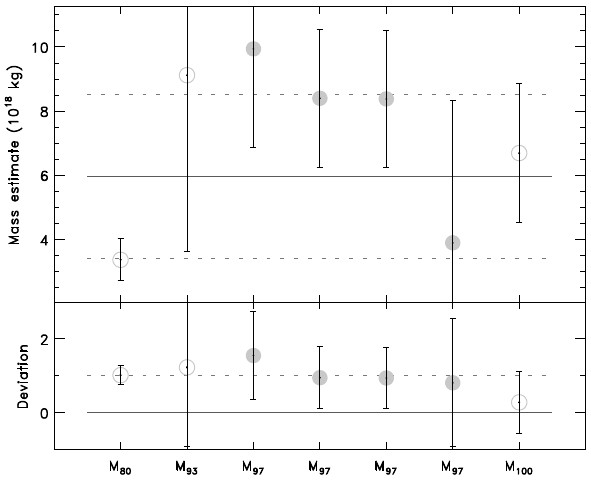}
\caption[Mass estimates for (128) Nemesis]{%
  \label{fap: mass000128}
  Mass estimates for (128) Nemesis.
}
\end{figure}

  \begin{figure}[!ht]
  \centering
  \includegraphics[width=.49\textwidth]{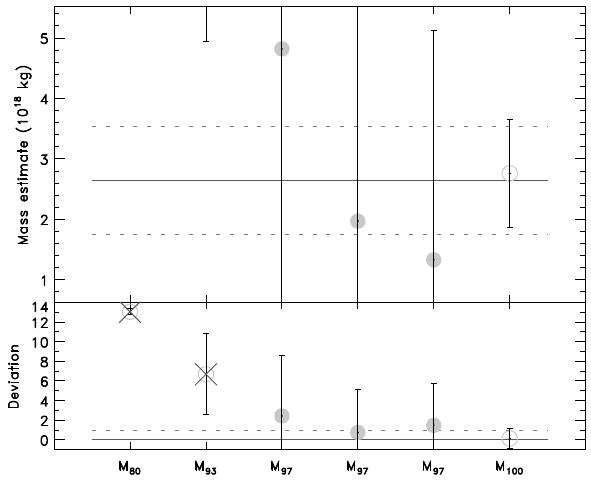}
\caption[Mass estimates for (129) Antigone]{%
  \label{fap: mass000129}
  Mass estimates for (129) Antigone.
  Both mass estimates from ephemeris (M$_{80}$ and M$_{93}$) result in unrealistic high densities, and are thus discarded.
}
\end{figure}

\clearpage
  \begin{figure}[!ht]
  \centering
  \includegraphics[width=.49\textwidth]{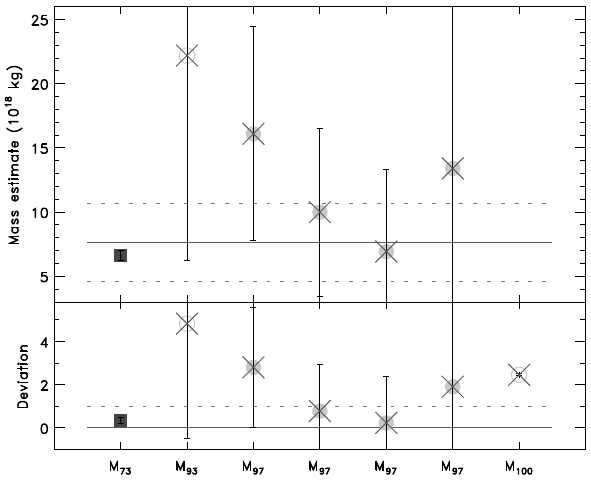}
\caption[Mass estimates for (130) Elektra]{%
  \label{fap: mass000130}
  Mass estimates for (130) Elektra.
 Only the estimate from M$_{73}$, based on direct imaging of the binary system is used.
}
\end{figure}

  \begin{figure}[!ht]
  \centering
  \includegraphics[width=.49\textwidth]{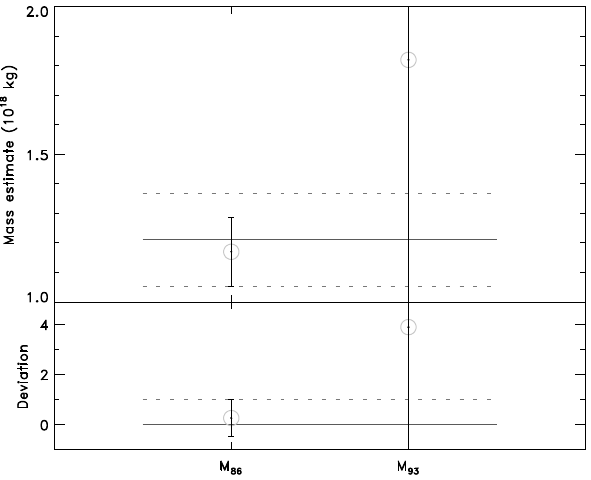}
\caption[Mass estimates for (135) Hertha]{%
  \label{fap: mass000135}
  Mass estimates for (135) Hertha.
}
\end{figure}

  \begin{figure}[!ht]
  \centering
  \includegraphics[width=.49\textwidth]{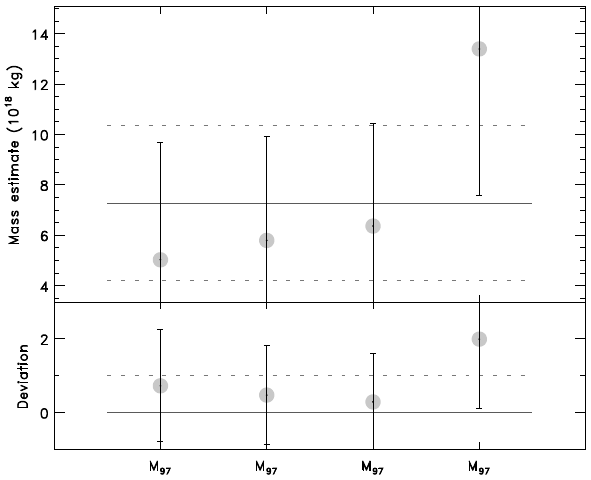}
\caption[Mass estimates for (137) Meliboea]{%
  \label{fap: mass000137}
  Mass estimates for (137) Meliboea.
}
\end{figure}

  \begin{figure}[!ht]
  \centering
  \includegraphics[width=.49\textwidth]{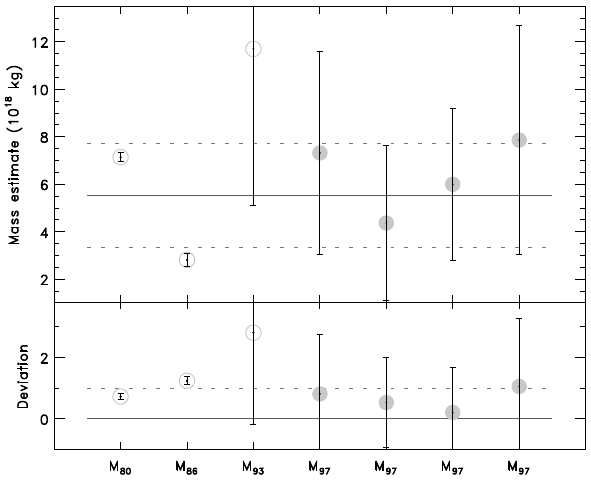}
\caption[Mass estimates for (139) Juewa]{%
  \label{fap: mass000139}
  Mass estimates for (139) Juewa.
}
\end{figure}

  \begin{figure}[!ht]
  \centering
  \includegraphics[width=.49\textwidth]{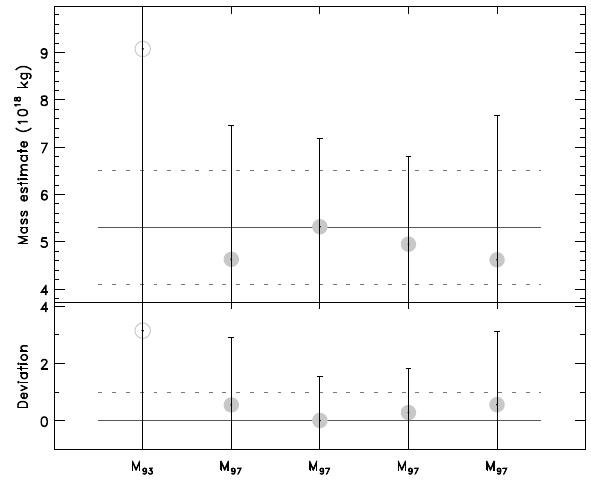}
\caption[Mass estimates for (144) Vibilia]{%
  \label{fap: mass000144}
  Mass estimates for (144) Vibilia.
}
\end{figure}

  \begin{figure}[!ht]
  \centering
  \includegraphics[width=.49\textwidth]{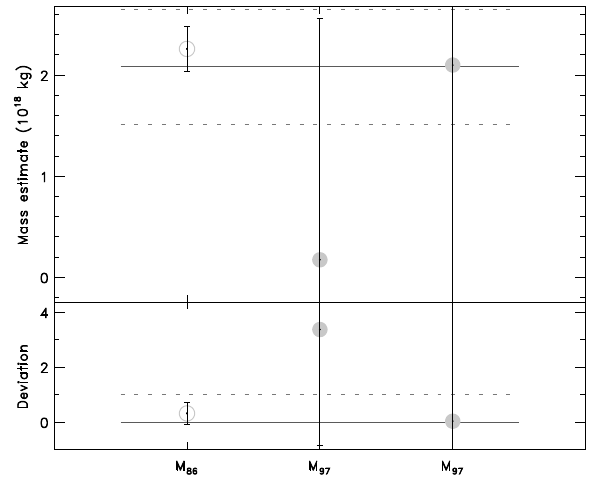}
\caption[Mass estimates for (145) Adeona]{%
  \label{fap: mass000145}
  Mass estimates for (145) Adeona.
}
\end{figure}

  \begin{figure}[!ht]
  \centering
  \includegraphics[width=.49\textwidth]{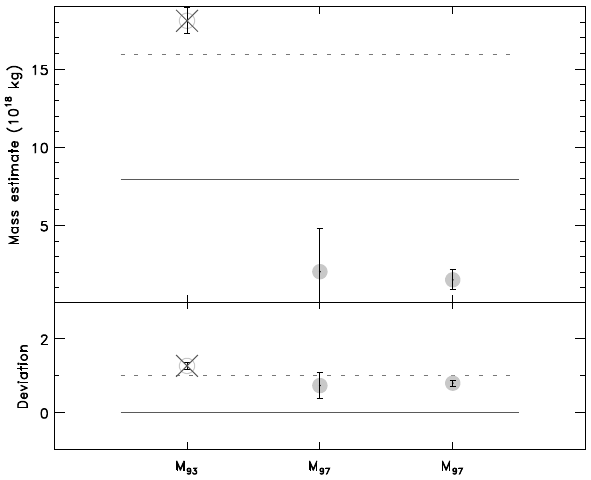}
\caption[Mass estimates for (150) Nuwa]{%
  \label{fap: mass000150}
  Mass estimates for (150) Nuwa.
 The mass estimate M$_{93}$ gives an unrealistic density of 10.4\,$\pm$\,1.9 if used alone, and is therefore discarded.
}
\end{figure}

  \begin{figure}[!ht]
  \centering
  \includegraphics[width=.49\textwidth]{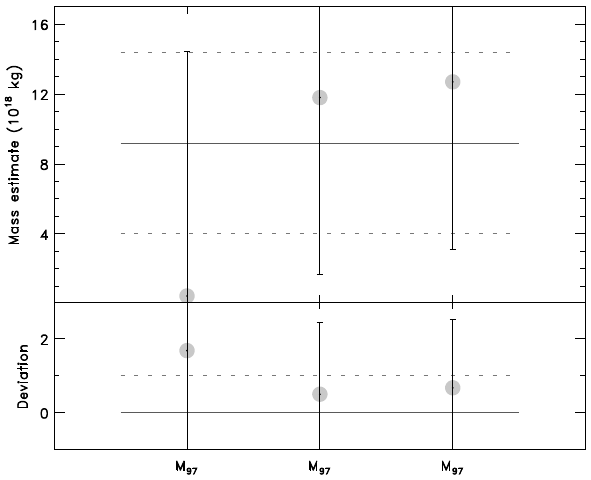}
\caption[Mass estimates for (154) Bertha]{%
  \label{fap: mass000154}
  Mass estimates for (154) Bertha.
}
\end{figure}

  \begin{figure}[!ht]
  \centering
  \includegraphics[width=.49\textwidth]{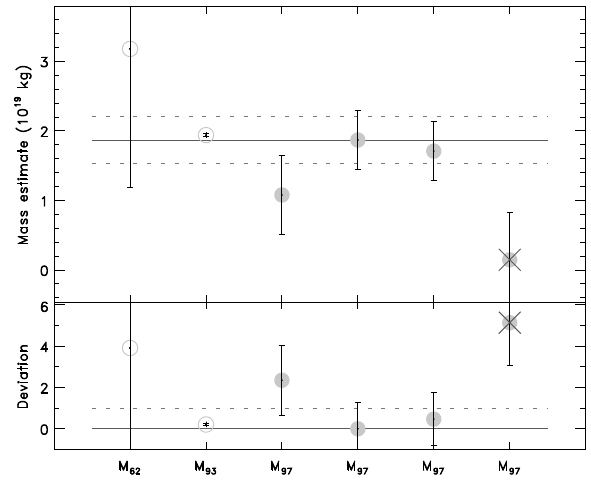}
\caption[Mass estimates for (165) Loreley]{%
  \label{fap: mass000165}
  Mass estimates for (165) Loreley.
}
\end{figure}

  \begin{figure}[!ht]
  \centering
  \includegraphics[width=.49\textwidth]{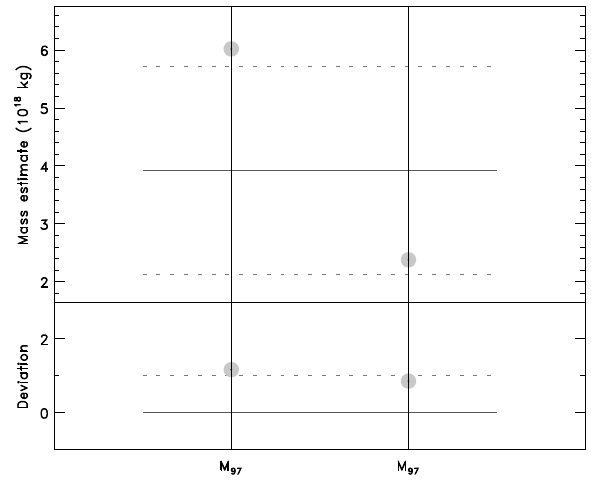}
\caption[Mass estimates for (168) Sibylla]{%
  \label{fap: mass000168}
  Mass estimates for (168) Sibylla.
}
\end{figure}

  \begin{figure}[!ht]
  \centering
  \includegraphics[width=.49\textwidth]{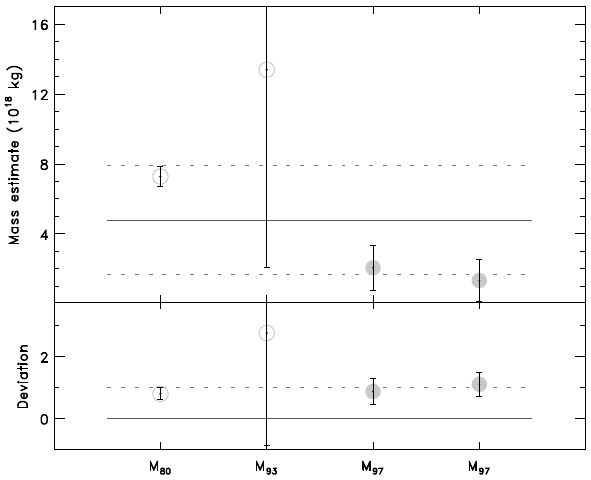}
\caption[Mass estimates for (173) Ino]{%
  \label{fap: mass000173}
  Mass estimates for (173) Ino.
}
\end{figure}

  \begin{figure}[!ht]
  \centering
  \includegraphics[width=.49\textwidth]{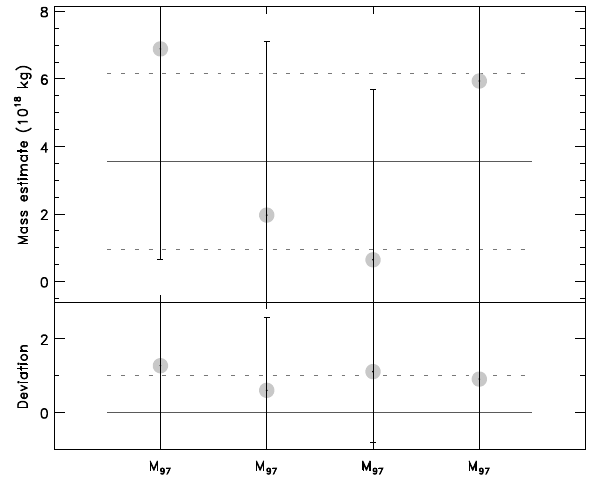}
\caption[Mass estimates for (185) Eunike]{%
  \label{fap: mass000185}
  Mass estimates for (185) Eunike.
}
\end{figure}

  \begin{figure}[!ht]
  \centering
  \includegraphics[width=.49\textwidth]{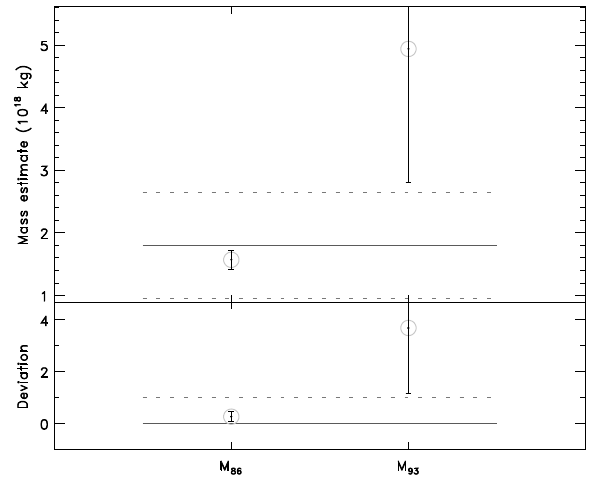}
\caption[Mass estimates for (187) Lamberta]{%
  \label{fap: mass000187}
  Mass estimates for (187) Lamberta.
}
\end{figure}

  \begin{figure}[!ht]
  \centering
  \includegraphics[width=.49\textwidth]{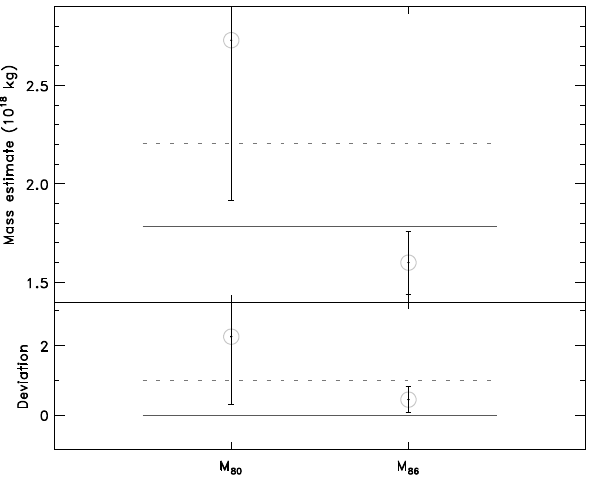}
\caption[Mass estimates for (192) Nausikaa]{%
  \label{fap: mass000192}
  Mass estimates for (192) Nausikaa.
}
\end{figure}

  \begin{figure}[!ht]
  \centering
  \includegraphics[width=.49\textwidth]{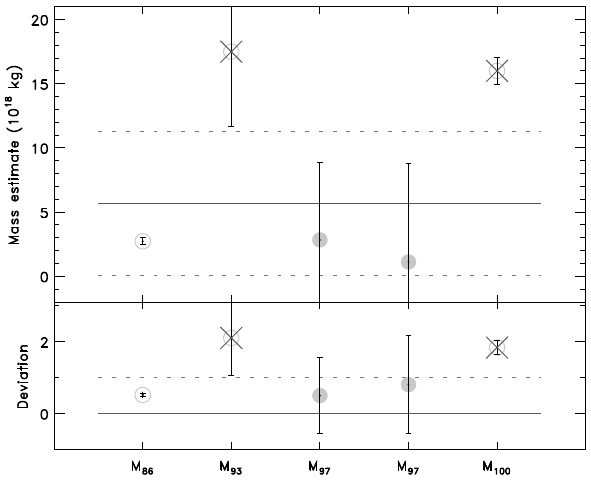}
\caption[Mass estimates for (194) Prokne]{%
  \label{fap: mass000194}
  Mass estimates for (194) Prokne.
}
\end{figure}

  \begin{figure}[!ht]
  \centering
  \includegraphics[width=.49\textwidth]{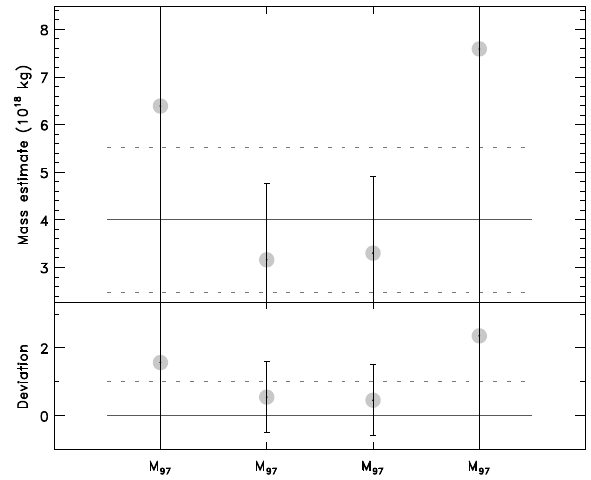}
\caption[Mass estimates for (196) Philomela]{%
  \label{fap: mass000196}
  Mass estimates for (196) Philomela.
}
\end{figure}

  \begin{figure}[!ht]
  \centering
  \includegraphics[width=.49\textwidth]{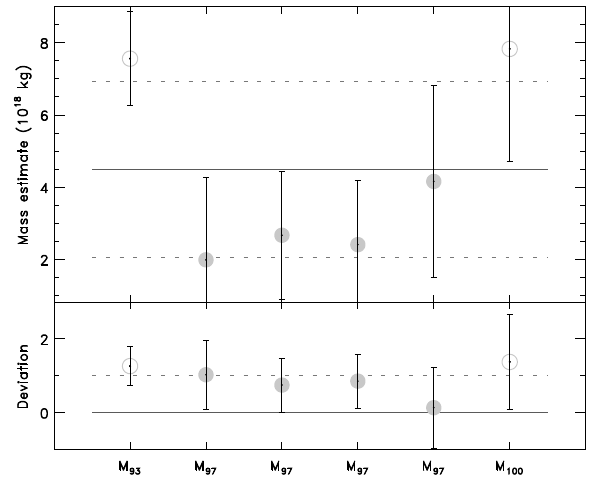}
\caption[Mass estimates for (211) Isolda]{%
  \label{fap: mass000211}
  Mass estimates for (211) Isolda.
}
\end{figure}

  \begin{figure}[!ht]
  \centering
  \includegraphics[width=.49\textwidth]{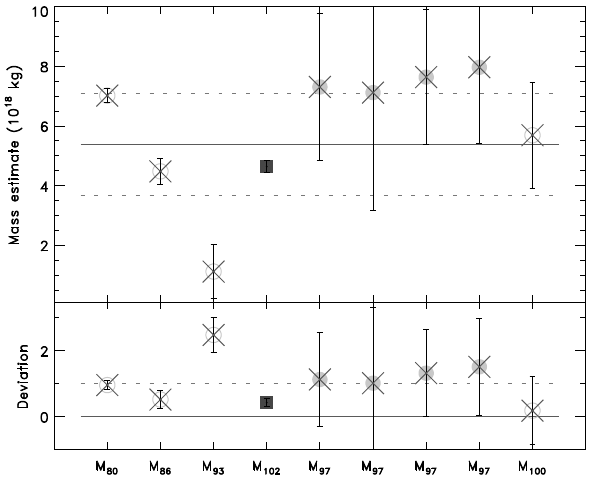}
\caption[Mass estimates for (216) Kleopatra]{%
  \label{fap: mass000216}
  Mass estimates for (216) Kleopatra.
  Only the mass estimate based on direct imaging of the system is used here (M$_{102}$). 
}
\end{figure}

\clearpage
  \begin{figure}[!ht]
  \centering
  \includegraphics[width=.49\textwidth]{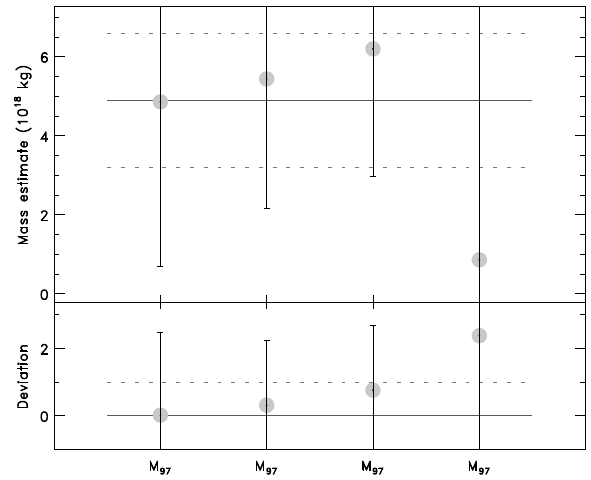}
\caption[Mass estimates for (238) Hypatia]{%
  \label{fap: mass000238}
  Mass estimates for (238) Hypatia.
}
\end{figure}

  \begin{figure}[!ht]
  \centering
  \includegraphics[width=.49\textwidth]{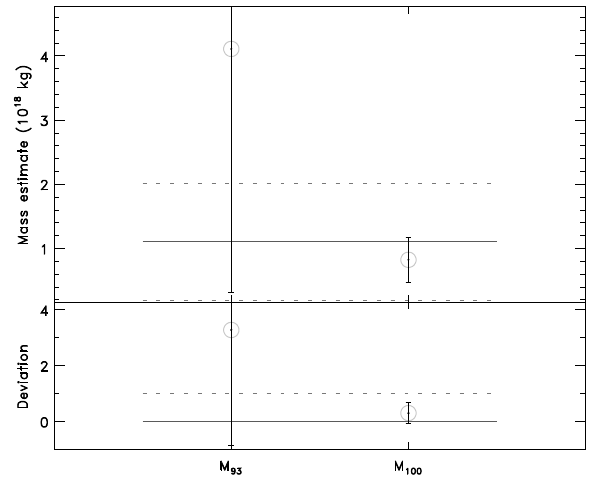}
\caption[Mass estimates for (240) Vanadis]{%
  \label{fap: mass000240}
  Mass estimates for (240) Vanadis.
}
\end{figure}

  \begin{figure}[!ht]
  \centering
  \includegraphics[width=.49\textwidth]{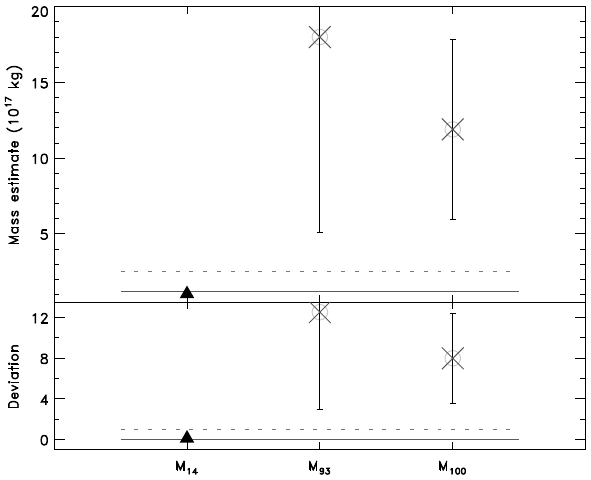}
\caption[Mass estimates for (253) Mathilde]{%
  \label{fap: mass000253}
  Mass estimates for (253) Mathilde.
  Only the flyby estimate (M$_{14}$) is used here.
}
\end{figure}

  \begin{figure}[!ht]
  \centering
  \includegraphics[width=.49\textwidth]{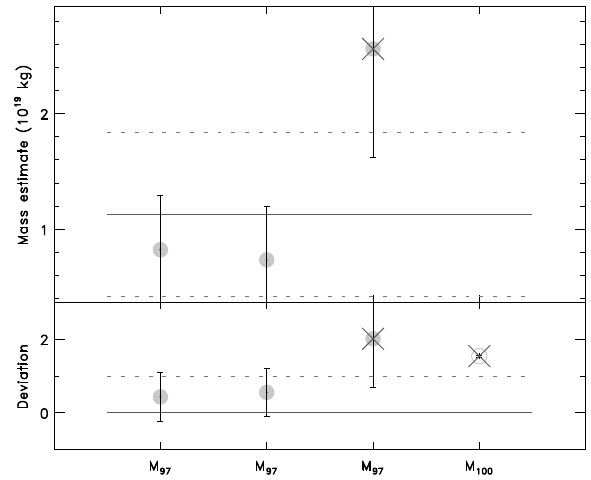}
\caption[Mass estimates for (259) Aletheia]{%
  \label{fap: mass000259}
  Mass estimates for (259) Aletheia.
}
\end{figure}

  \begin{figure}[!ht]
  \centering
  \includegraphics[width=.49\textwidth]{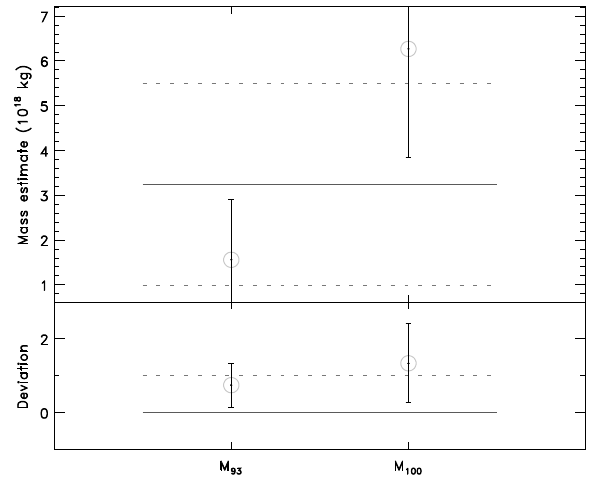}
\caption[Mass estimates for (268) Adorea]{%
  \label{fap: mass000268}
  Mass estimates for (268) Adorea.
}
\end{figure}

  \begin{figure}[!ht]
  \centering
  \includegraphics[width=.49\textwidth]{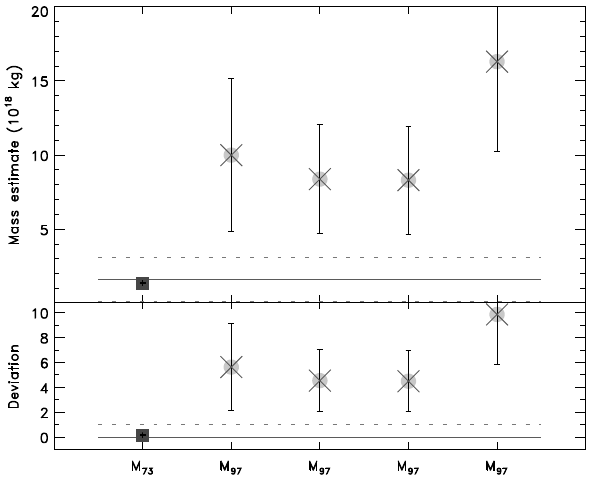}
\caption[Mass estimates for (283) Emma]{%
  \label{fap: mass000283}
  Mass estimates for (283) Emma.
  Only the mass estimate based on direct imaging of the system is used here (M$_{73}$). 
}
\end{figure}

  \begin{figure}[!ht]
  \centering
  \includegraphics[width=.49\textwidth]{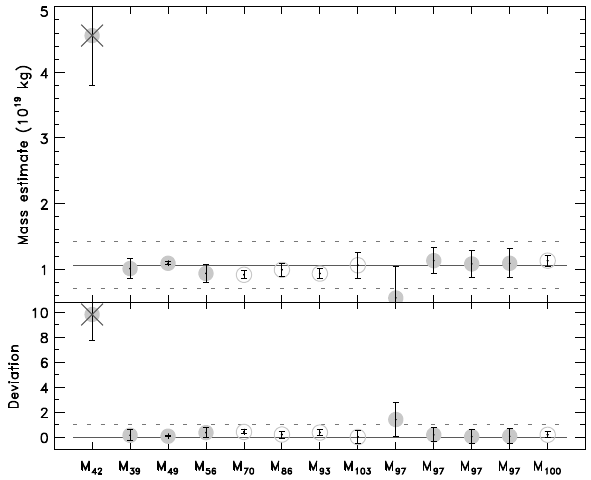}
\caption[Mass estimates for (324) Bamberga]{%
  \label{fap: mass000324}
  Mass estimates for (324) Bamberga.
}
\end{figure}

  \begin{figure}[!ht]
  \centering
  \includegraphics[width=.49\textwidth]{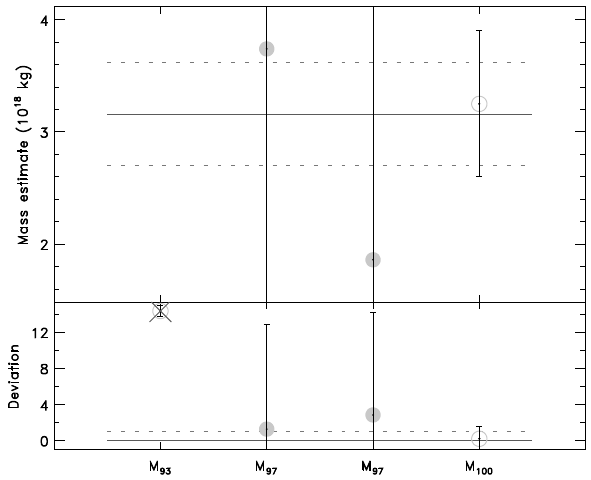}
\caption[Mass estimates for (328) Gudrun]{%
  \label{fap: mass000328}
  Mass estimates for (328) Gudrun.
}
\end{figure}

  \begin{figure}[!ht]
  \centering
  \includegraphics[width=.49\textwidth]{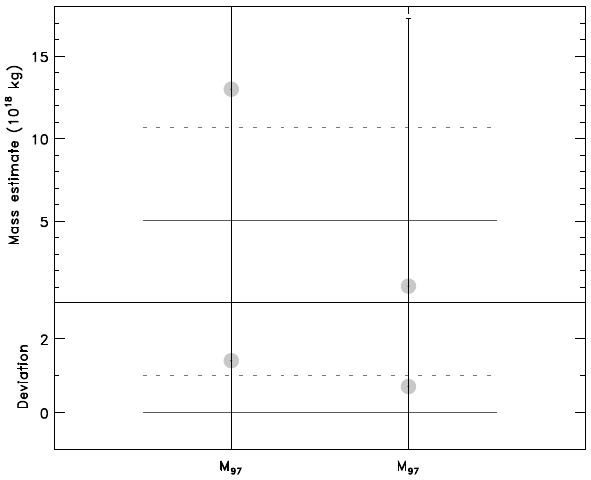}
\caption[Mass estimates for (334) Chicago]{%
  \label{fap: mass000334}
  Mass estimates for (334) Chicago.
}
\end{figure}

  \begin{figure}[!ht]
  \centering
  \includegraphics[width=.49\textwidth]{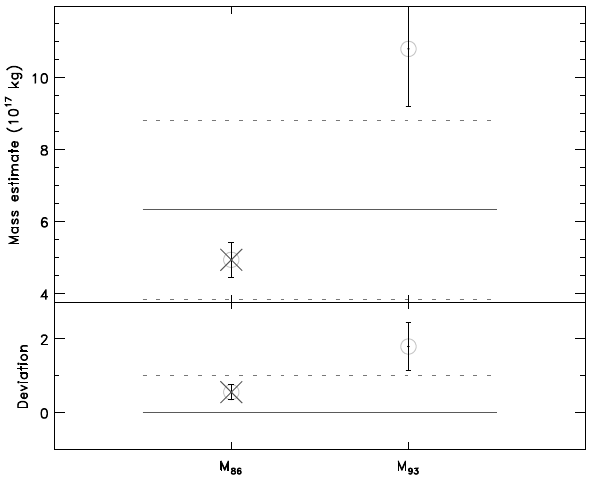}
\caption[Mass estimates for (337) Devosa]{%
  \label{fap: mass000337}
  Mass estimates for (337) Devosa.
  \citet{2011-Icarus-214-Fornasier} found that Hexahedrite iron meteorites  provided the best match to (337) Devosa's reflectance spectrum. Mass estimate M$_{86}$ leads to a density of 3.5\,$\pm$\,1.4 if used alone, far from the density of these meteorites ($\sim$7.4).  The estimate from M$_{93}$ results in a density of 7.7\,$\pm$\,3.2, in better agreement  with the meteorite.
}
\end{figure}

  \begin{figure}[!ht]
  \centering
  \includegraphics[width=.49\textwidth]{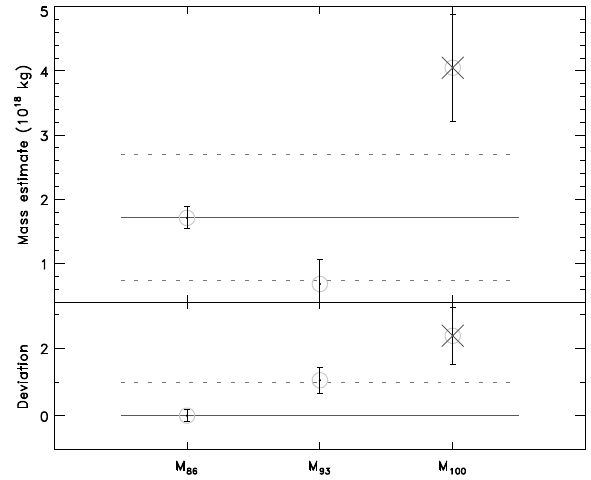}
\caption[Mass estimates for (344) Desiderata]{%
  \label{fap: mass000344}
  Mass estimates for (344) Desiderata.
}
\end{figure}

  \begin{figure}[!ht]
  \centering
  \includegraphics[width=.49\textwidth]{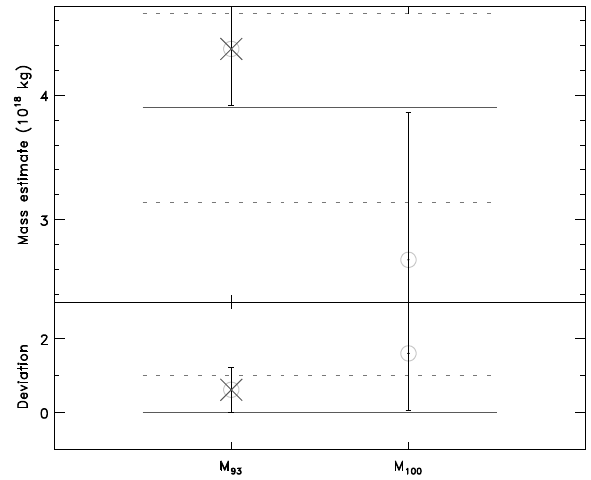}
\caption[Mass estimates for (345) Tercidina]{%
  \label{fap: mass000345}
  Mass estimates for (345) Tercidina.
  The mass estimate from M$_{93}$ gives a unrealistic high density of 8.6\,$\pm$\,1.1 if used alone, and is therefore discarded.
}
\end{figure}

  \begin{figure}[!ht]
  \centering
  \includegraphics[width=.49\textwidth]{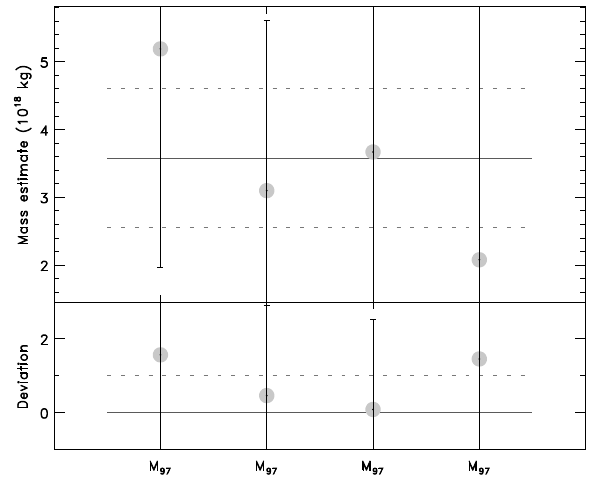}
\caption[Mass estimates for (349) Dembowska]{%
  \label{fap: mass000349}
  Mass estimates for (349) Dembowska.
}
\end{figure}

  \begin{figure}[!ht]
  \centering
  \includegraphics[width=.49\textwidth]{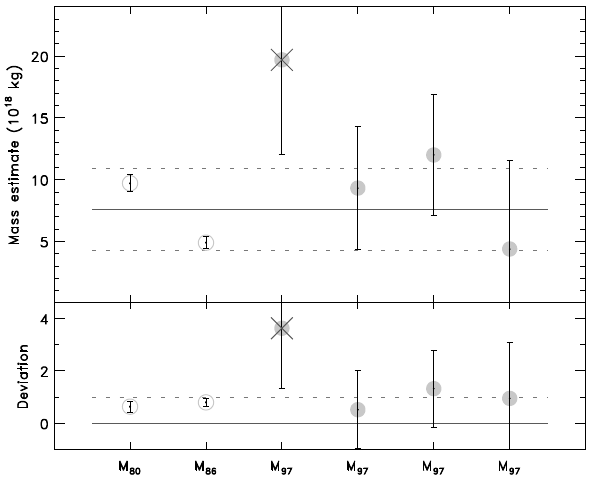}
\caption[Mass estimates for (354) Eleonora]{%
  \label{fap: mass000354}
  Mass estimates for (354) Eleonora.
}
\end{figure}

  \begin{figure}[!ht]
  \centering
  \includegraphics[width=.49\textwidth]{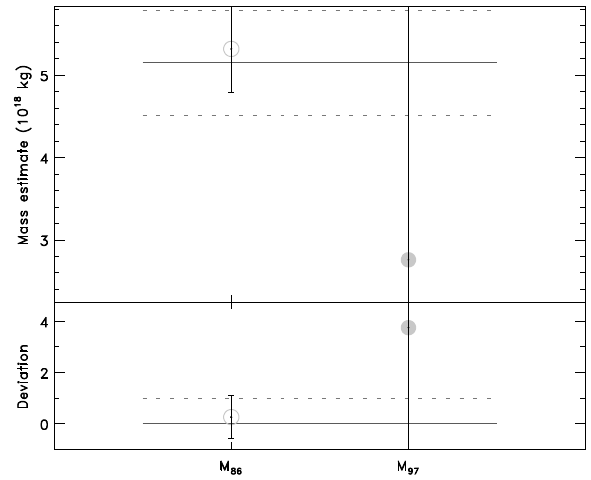}
\caption[Mass estimates for (372) Palma]{%
  \label{fap: mass000372}
  Mass estimates for (372) Palma.
}
\end{figure}

  \begin{figure}[!ht]
  \centering
  \includegraphics[width=.49\textwidth]{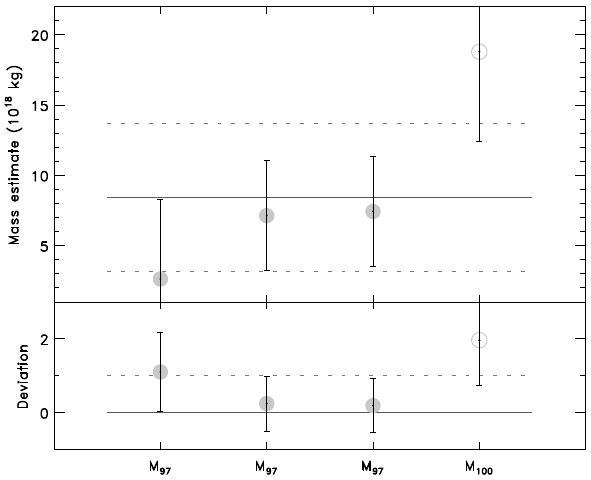}
\caption[Mass estimates for (375) Ursula]{%
  \label{fap: mass000375}
  Mass estimates for (375) Ursula.
}
\end{figure}

  \begin{figure}[!ht]
  \centering
  \includegraphics[width=.49\textwidth]{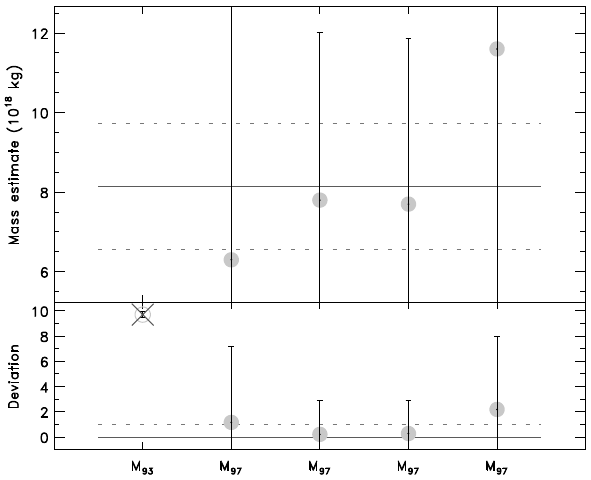}
\caption[Mass estimates for (386) Siegena]{%
  \label{fap: mass000386}
  Mass estimates for (386) Siegena.
}
\end{figure}

  \begin{figure}[!ht]
  \centering
  \includegraphics[width=.49\textwidth]{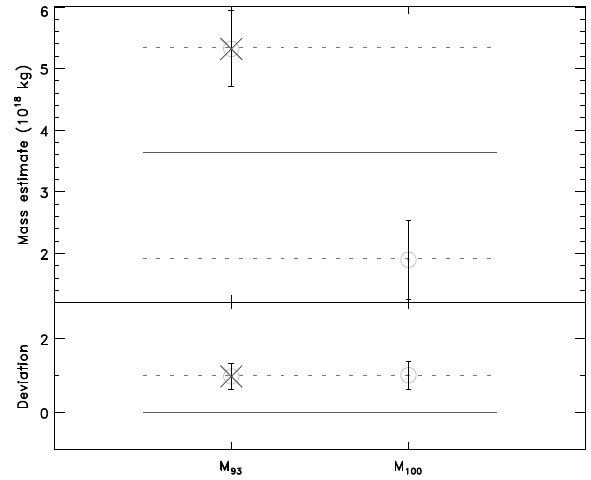}
\caption[Mass estimates for (387) Aquitania]{%
  \label{fap: mass000387}
  Mass estimates for (387) Aquitania.
}
\end{figure}

\clearpage
  \begin{figure}[!ht]
  \centering
  \includegraphics[width=.49\textwidth]{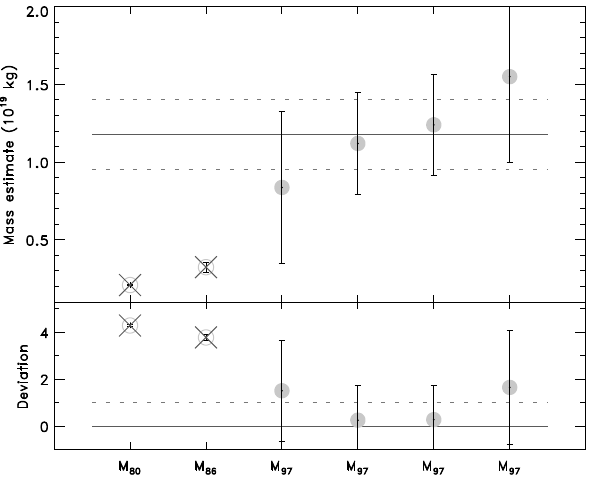}
\caption[Mass estimates for (409) Aspasia]{%
  \label{fap: mass000409}
  Mass estimates for (409) Aspasia.
}
\end{figure}

  \begin{figure}[!ht]
  \centering
  \includegraphics[width=.49\textwidth]{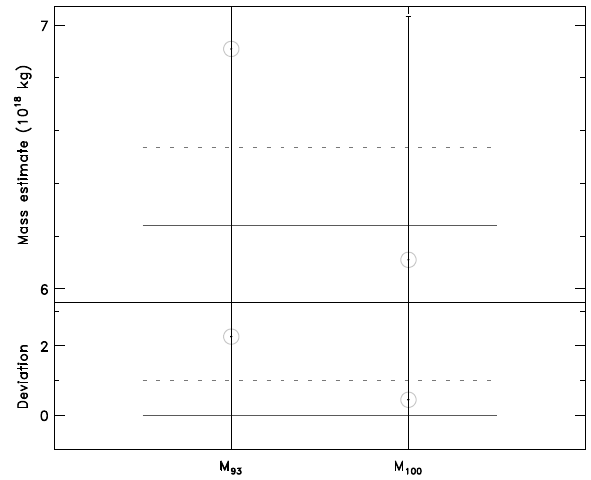}
\caption[Mass estimates for (410) Chloris]{%
  \label{fap: mass000410}
  Mass estimates for (410) Chloris.
}
\end{figure}

  \begin{figure}[!ht]
  \centering
  \includegraphics[width=.49\textwidth]{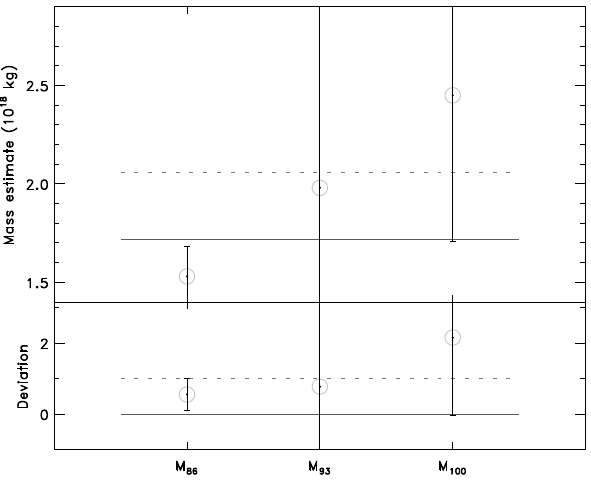}
\caption[Mass estimates for (419) Aurelia]{%
  \label{fap: mass000419}
  Mass estimates for (419) Aurelia.
}
\end{figure}

  \begin{figure}[!ht]
  \centering
  \includegraphics[width=.49\textwidth]{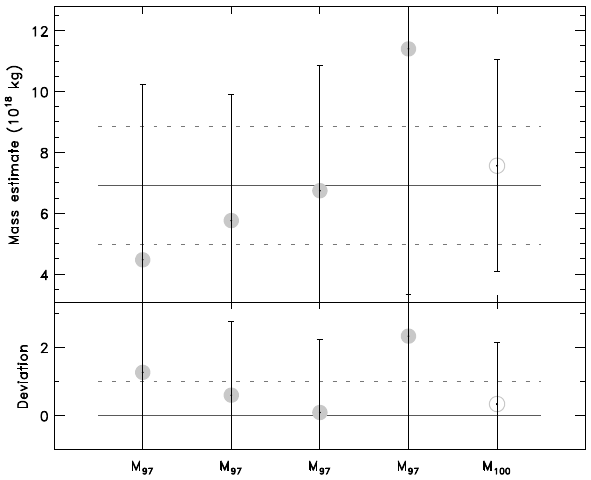}
\caption[Mass estimates for (423) Diotima]{%
  \label{fap: mass000423}
  Mass estimates for (423) Diotima.
}
\end{figure}

  \begin{figure}[!ht]
  \centering
  \includegraphics[width=.49\textwidth]{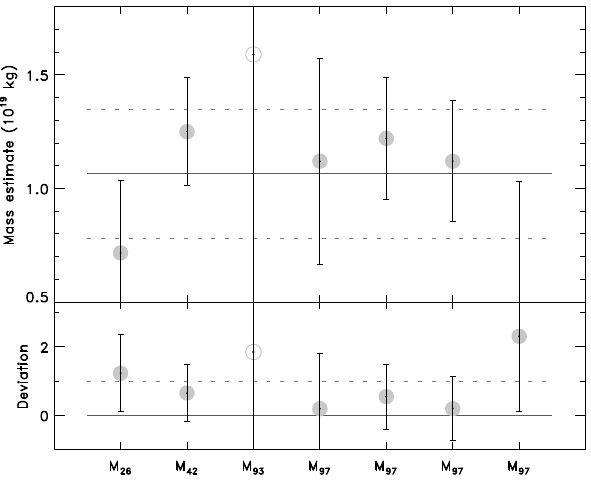}
\caption[Mass estimates for (444) Gyptis]{%
  \label{fap: mass000444}
  Mass estimates for (444) Gyptis.
}
\end{figure}

  \begin{figure}[!ht]
  \centering
  \includegraphics[width=.49\textwidth]{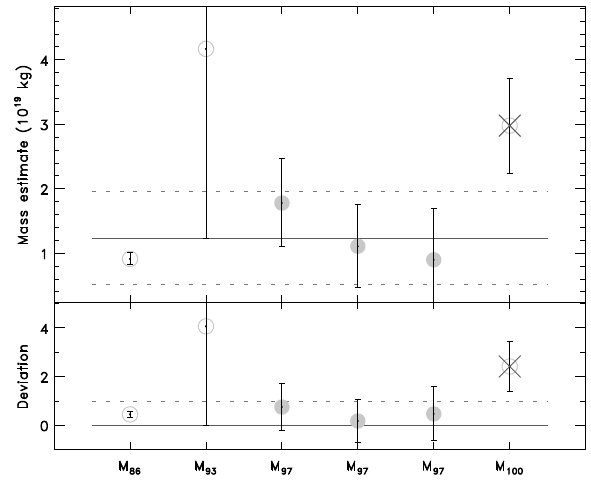}
\caption[Mass estimates for (451) Patientia]{%
  \label{fap: mass000451}
  Mass estimates for (451) Patientia.
}
\end{figure}

  \begin{figure}[!ht]
  \centering
  \includegraphics[width=.49\textwidth]{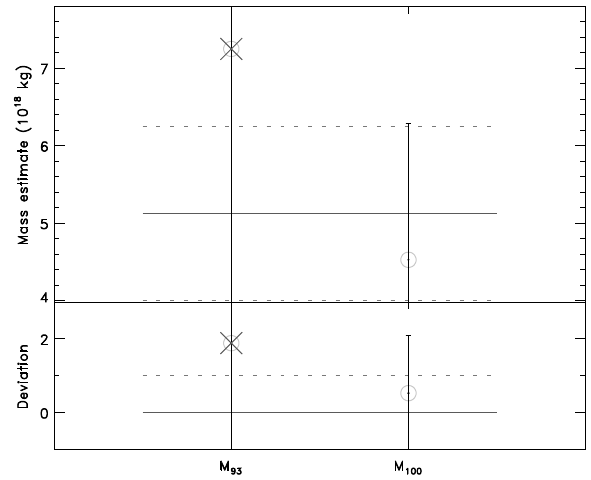}
\caption[Mass estimates for (469) Argentina]{%
  \label{fap: mass000469}
  Mass estimates for (469) Argentina.
}
\end{figure}

  \begin{figure}[!ht]
  \centering
  \includegraphics[width=.49\textwidth]{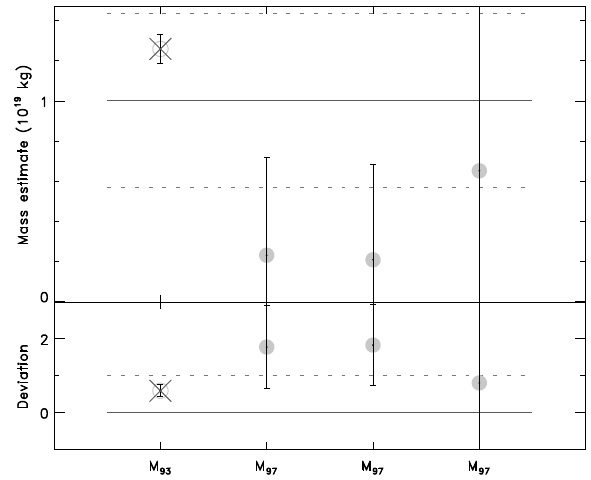}
\caption[Mass estimates for (471) Papagena]{%
  \label{fap: mass000471}
  Mass estimates for (471) Papagena.
  The mass estimate from M$_{93}$ gives a unrealistic high density of 12.4\,$\pm$\,2.5 if used alone, and is therefore discarded.
}
\end{figure}

  \begin{figure}[!ht]
  \centering
  \includegraphics[width=.49\textwidth]{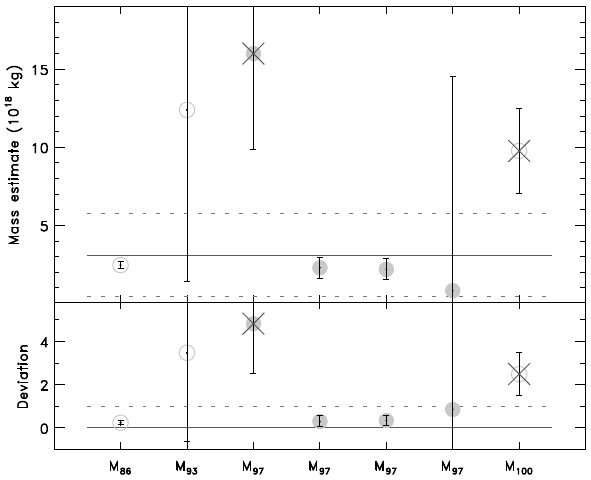}
\caption[Mass estimates for (488) Kreusa]{%
  \label{fap: mass000488}
  Mass estimates for (488) Kreusa.
}
\end{figure}

  \begin{figure}[!ht]
  \centering
  \includegraphics[width=.49\textwidth]{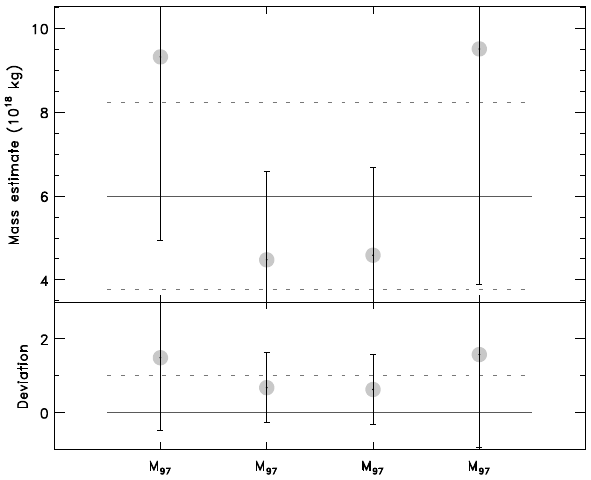}
\caption[Mass estimates for (490) Veritas]{%
  \label{fap: mass000490}
  Mass estimates for (490) Veritas.
}
\end{figure}

  \begin{figure}[!ht]
  \centering
  \includegraphics[width=.49\textwidth]{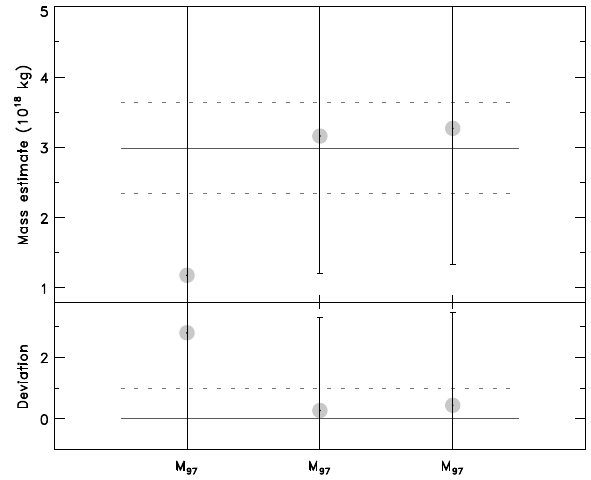}
\caption[Mass estimates for (508) Princetonia]{%
  \label{fap: mass000508}
  Mass estimates for (508) Princetonia.
}
\end{figure}

  \begin{figure}[!ht]
  \centering
  \includegraphics[width=.49\textwidth]{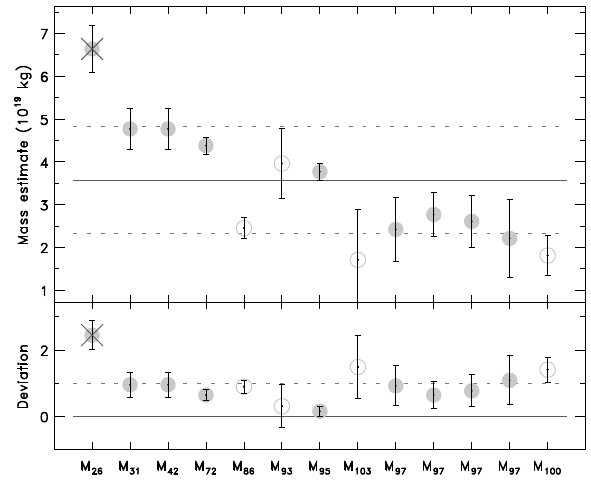}
\caption[Mass estimates for (511) Davida]{%
  \label{fap: mass000511}
  Mass estimates for (511) Davida.
}
\end{figure}

  \begin{figure}[!ht]
  \centering
  \includegraphics[width=.49\textwidth]{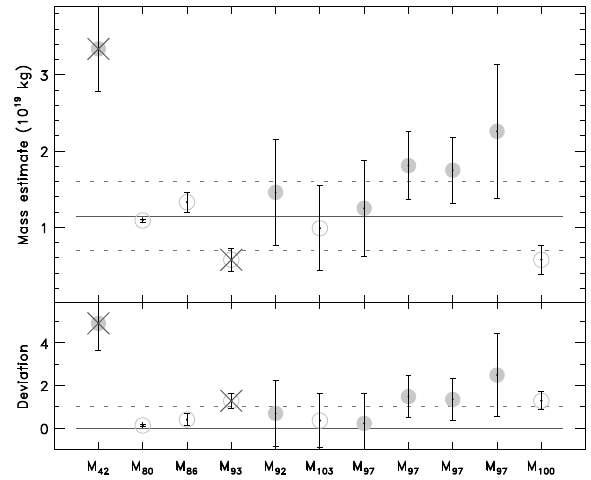}
\caption[Mass estimates for (532) Herculina]{%
  \label{fap: mass000532}
  Mass estimates for (532) Herculina.
}
\end{figure}

\clearpage
  \begin{figure}[!ht]
  \centering
  \includegraphics[width=.49\textwidth]{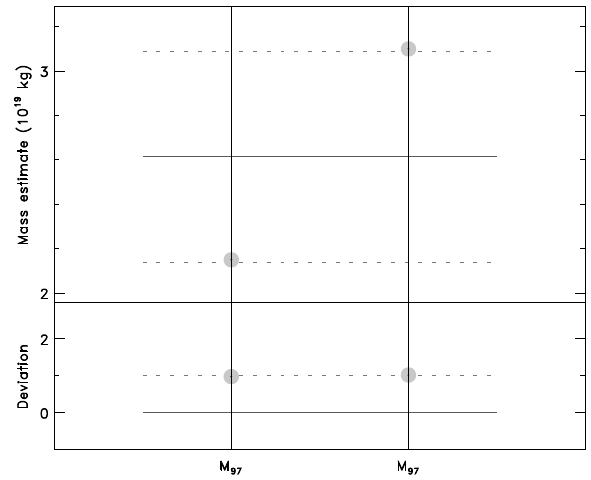}
\caption[Mass estimates for (536) Merapi]{%
  \label{fap: mass000536}
  Mass estimates for (536) Merapi.
}
\end{figure}

  \begin{figure}[!ht]
  \centering
  \includegraphics[width=.49\textwidth]{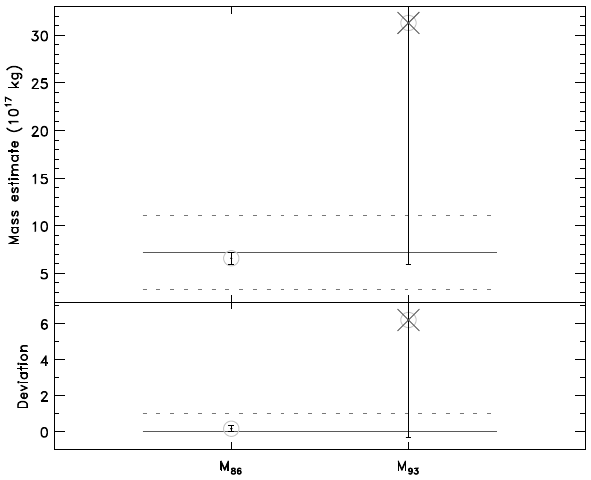}
\caption[Mass estimates for (554) Peraga]{%
  \label{fap: mass000554}
  Mass estimates for (554) Peraga.
  The mass estimate from M$_{93}$ gives a low-constrained density of 6.2\,$\pm$\,5.2 if used alone, far from the CI meteorite analogue.  Only the estimate from M$_{86}$ is used.
}
\end{figure}

  \begin{figure}[!ht]
  \centering
  \includegraphics[width=.49\textwidth]{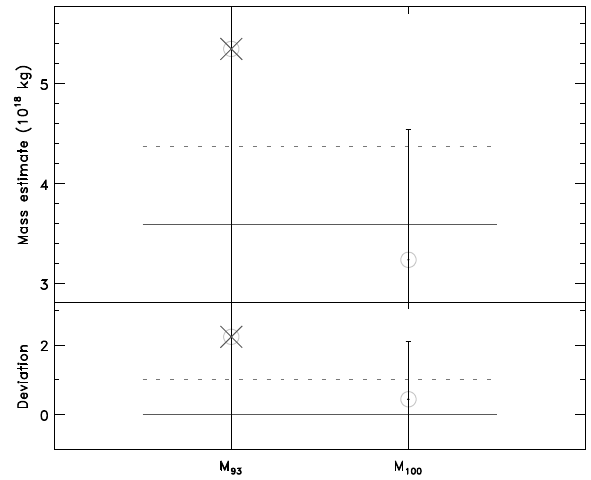}
\caption[Mass estimates for (626) Notburga]{%
  \label{fap: mass000626}
  Mass estimates for (626) Notburga.
}
\end{figure}

  \begin{figure}[!ht]
  \centering
  \includegraphics[width=.49\textwidth]{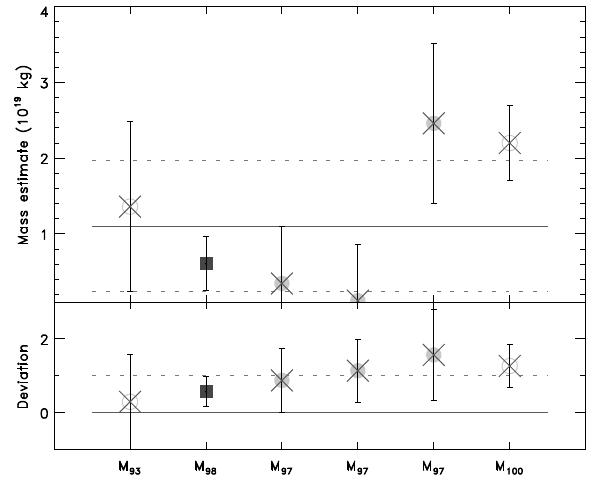}
\caption[Mass estimates for (702) Alauda]{%
  \label{fap: mass000702}
  Mass estimates for (702) Alauda.
 The estimate from M$_{98}$ obtained by imaging the binary system is preferred.
}
\end{figure}

  \begin{figure}[!ht]
  \centering
  \includegraphics[width=.49\textwidth]{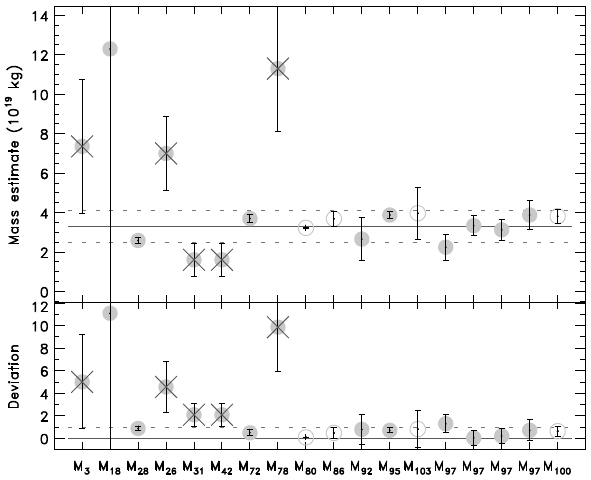}
\caption[Mass estimates for (704) Interamnia]{%
  \label{fap: mass000704}
  Mass estimates for (704) Interamnia.
}
\end{figure}

  \begin{figure}[!ht]
  \centering
  \includegraphics[width=.49\textwidth]{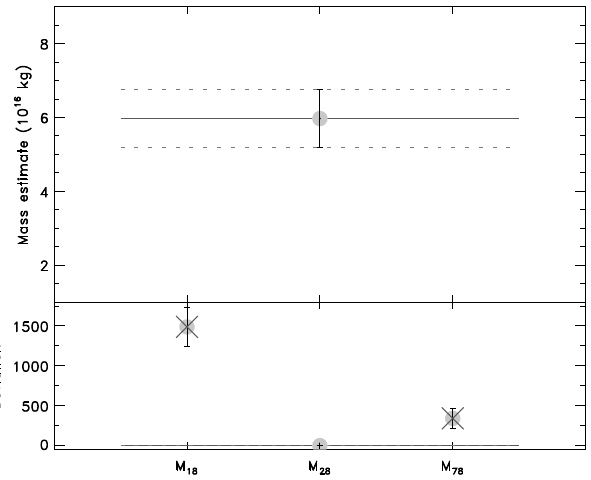}
\caption[Mass estimates for (720) Bohlinia]{%
  \label{fap: mass000720}
  Mass estimates for (720) Bohlinia.
}
\end{figure}

  \begin{figure}[!ht]
  \centering
  \includegraphics[width=.49\textwidth]{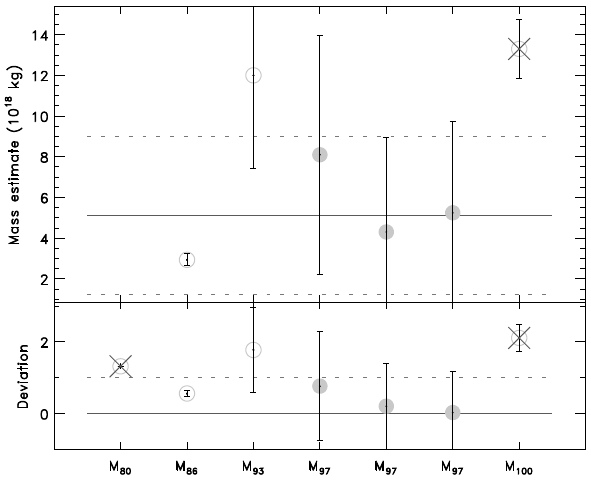}
\caption[Mass estimates for (747) Winchester]{%
  \label{fap: mass000747}
  Mass estimates for (747) Winchester.
  The mass estimate from M$_{80}$ was listed as unrealistic by the authors, and is discarded here.
}
\end{figure}

  \begin{figure}[!ht]
  \centering
  \includegraphics[width=.49\textwidth]{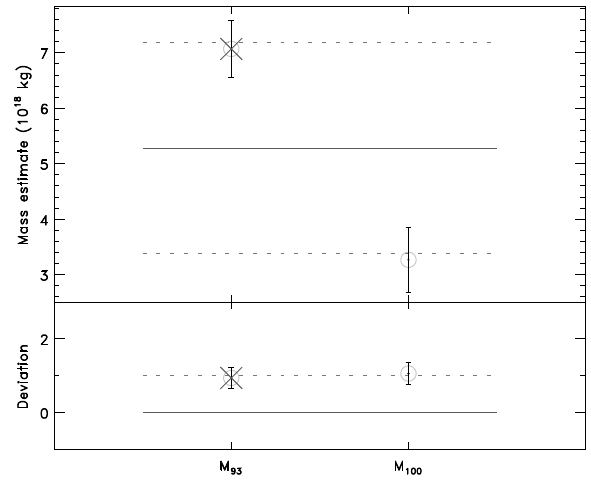}
\caption[Mass estimates for (751) Faina]{%
  \label{fap: mass000751}
  Mass estimates for (751) Faina.
}
\end{figure}

  \begin{figure}[!ht]
  \centering
  \includegraphics[width=.49\textwidth]{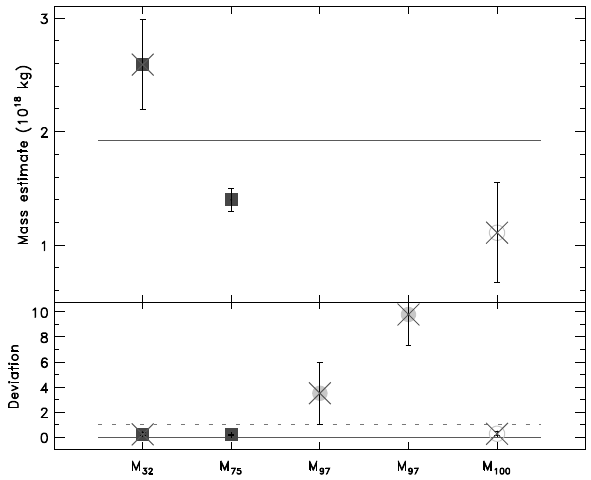}
\caption[Mass estimates for (762) Pulcova]{%
  \label{fap: mass000762}
  Mass estimates for (762) Pulcova.
  The mass estimate from M$_{32}$ was based on few discovery images, and the estimate from M$_{75}$ is preferred. 
}
\end{figure}

  \begin{figure}[!ht]
  \centering
  \includegraphics[width=.49\textwidth]{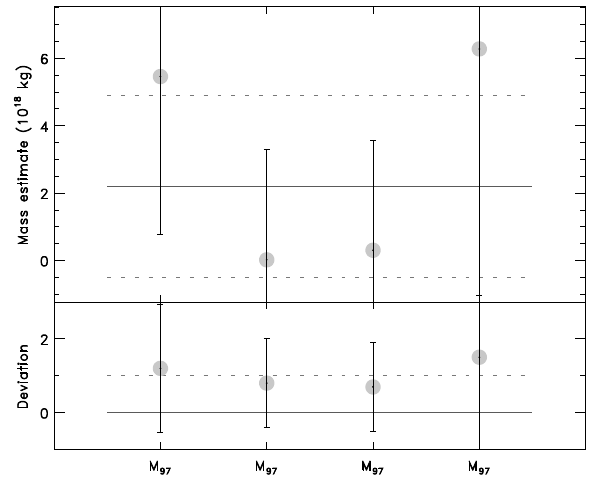}
\caption[Mass estimates for (776) Berbericia]{%
  \label{fap: mass000776}
  Mass estimates for (776) Berbericia.
}
\end{figure}

  \begin{figure}[!ht]
  \centering
  \includegraphics[width=.49\textwidth]{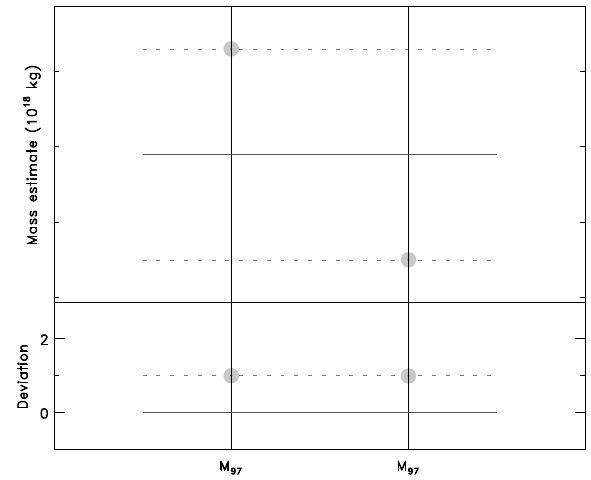}
\caption[Mass estimates for (790) Pretoria]{%
  \label{fap: mass000790}
  Mass estimates for (790) Pretoria.
}
\end{figure}

  \begin{figure}[!ht]
  \centering
  \includegraphics[width=.49\textwidth]{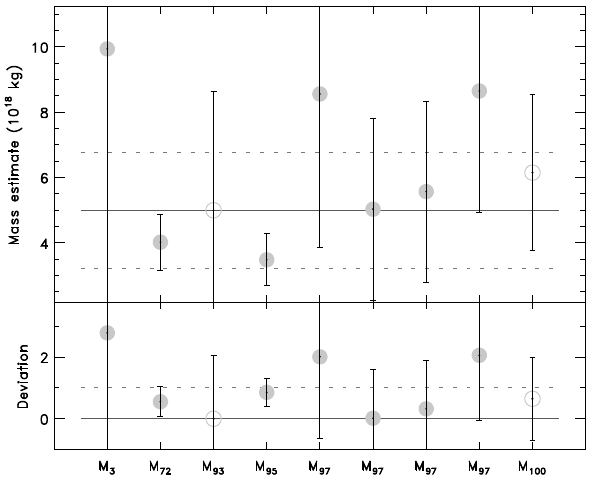}
\caption[Mass estimates for (804) Hispania]{%
  \label{fap: mass000804}
  Mass estimates for (804) Hispania.
}
\end{figure}

  \begin{figure}[!ht]
  \centering
  \includegraphics[width=.49\textwidth]{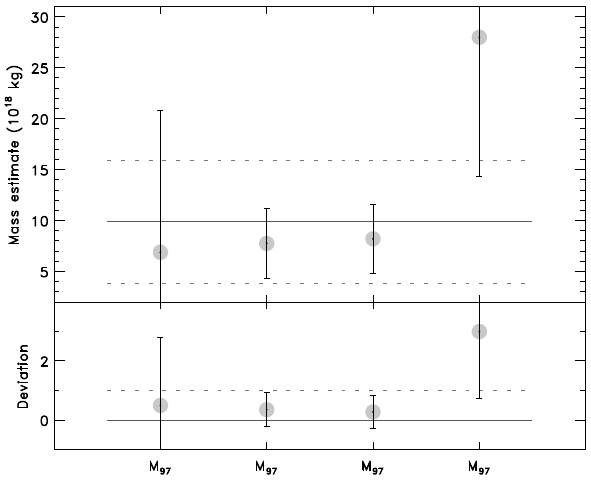}
\caption[Mass estimates for (895) Helio]{%
  \label{fap: mass000895}
  Mass estimates for (895) Helio.
}
\end{figure}

  \begin{figure}[!ht]
  \centering
  \includegraphics[width=.49\textwidth]{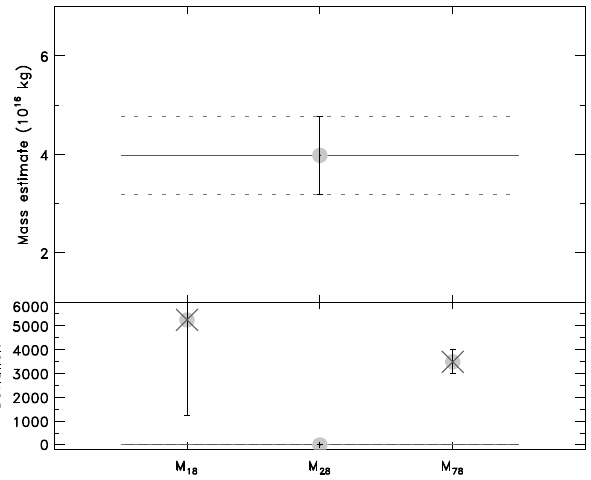}
\caption[Mass estimates for (1669) Dagmar]{%
  \label{fap: mass001669}
  Mass estimates for (1669) Dagmar.
}
\end{figure}

  \begin{figure}[!ht]
  \centering
  \includegraphics[width=.49\textwidth]{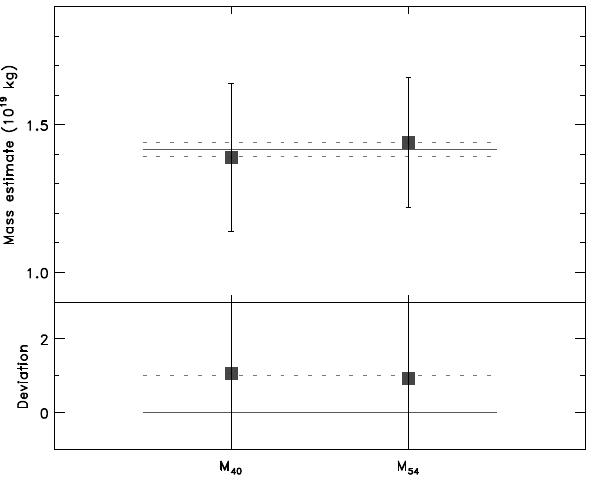}
\caption[Mass estimates for (47171) 1999 TC36]{%
  \label{fap: mass047171}
  Mass estimates for (47171) 1999 TC36.
}
\end{figure}

  \begin{figure}[!ht]
  \centering
  \includegraphics[width=.49\textwidth]{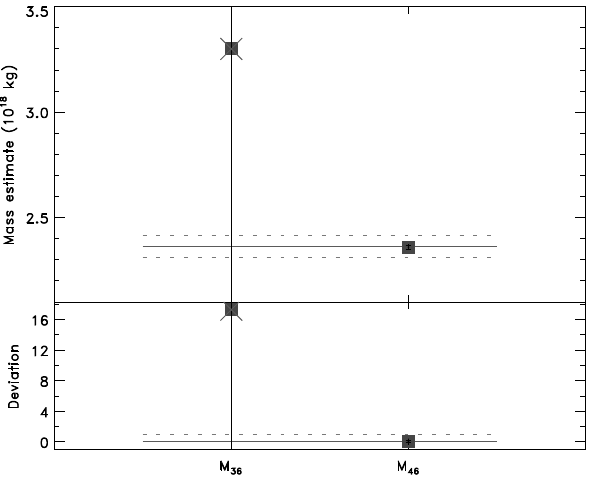}
\caption[Mass estimates for (88611) 2001 QT297]{%
  \label{fap: mass088611}
  Mass estimates for (88611) 2001 QT297.
  The mass estimate from M$_{36}$ has a crude relative precision (90\%). The estimate from M$_{46}$ is preferred.
}
\end{figure}

  \begin{figure}[!ht]
  \centering
  \includegraphics[width=.49\textwidth]{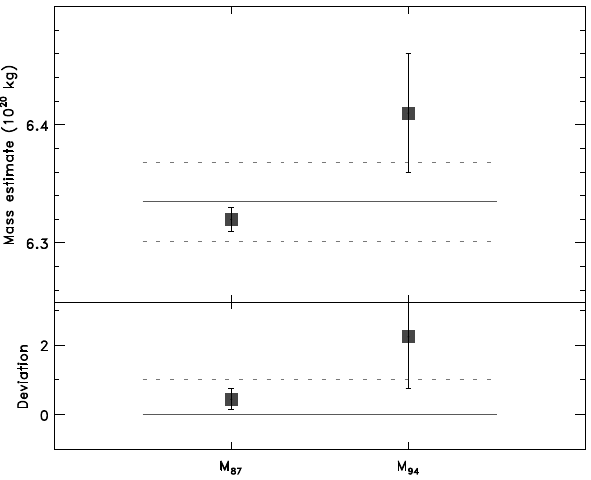}
\caption[Mass estimates for (90482) Orcus]{%
  \label{fap: mass090482}
  Mass estimates for (90482) Orcus.
}
\end{figure}

\clearpage
  \begin{figure}[!ht]
  \centering
  \includegraphics[width=.49\textwidth]{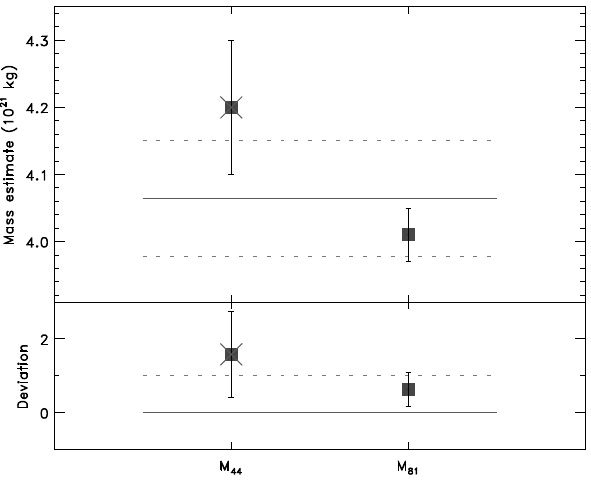}
\caption[Mass estimates for (136108) Haumea]{%
  \label{fap: mass136108}
  Mass estimates for (136108) Haumea.
  The mass estimate from M$_{44}$ was based on few discovery images, and the estimate from M$_{81}$ is preferred.
}
\end{figure}

  \begin{figure}[!ht]
  \centering
  \includegraphics[width=.49\textwidth]{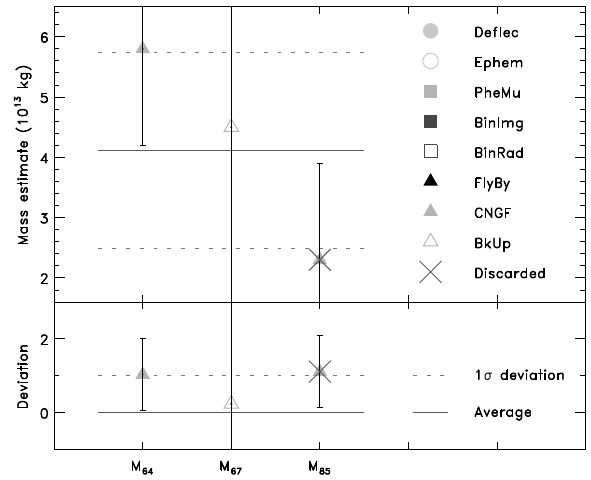}
\caption[Mass estimates for 9P/Tempel1]{%
  \label{fap: mass9P/Tempel1}
  Mass estimates for 9P/Tempel1.
}
\end{figure}


\clearpage
\section{Compilation of volume estimates\label{app: diam}}
  \indent The \numb{1454} volume-equivalent diameter estimates
  gathered in the literature are listed in 
  Table~\ref{tabSM: diam}.
  For objects with more than a single diameter determination, 
  Fig.~\ref{fap: diam000001} to
  Fig.~\ref{fap: diam81P/Wild2}
  presents a comparison of the diameter
  estimates, with additional information on discarded values.
  See~\ref{app: biblio} for the references, and
  Fig.~\ref{fap: diam81P/Wild2} for symbols key.

\setcounter{figure}{0}
\setcounter{table}{0}
\begin{table}[!ht]
\centering
\caption[Volume-equivalent diameter estimates]{%
  Compilation of volume-equivalent diameter estimates ($\phi$, in km) for 232 objects,
  with their associated uncertainty ($\delta \phi$),
  bibliographic references (see~\ref{app: biblio}), 
  and method of analysis: 
  \textsl{H-mag}: crude estimate from absolute magnitude, 
  \textsl{STM}: Standard Thermal Model, 
  \textsl{NEATM}: Near-Earth Asteroid Thermal Model, 
  \textsl{TPM}: Thermophysical Model, 
  \textsl{Occ}: stellar occultation, 
  \textsl{LC+Occ}: lightcurve 3-D model scaled using an occultation, 
  \textsl{PheMu}: mutual eclipsing phenomena in binary systems, 
  \textsl{Img-PSF}: profile deviation from a Point-Spread Function, 
  \textsl{Img}: apprent size in disk-resolved imaging, 
  \textsl{Img-TE}: triaxial ellipsoid model from images, 
  \textsl{KOALA}: combined lightcurves, occultations, and disk-resolved images, 
  \textsl{Radar}: radar imaging, 
  \textsl{FlyBy}: images from spacecraft encounter, 
  \textsl{IAU}: IAU WGCCR consensus value, and 
  \textsl{BkUp}: break-up modeling of comet nucleus. 
  Estimates marked with a dagger ($\dagger$) were rejected from average diameter computation.
\label{tabSM: diam}
}
\begin{tabular}{r@{ }lr@{.}lr@{.}lll}
 \hline
 \hline
 \# & Designation & \multicolumn{2}{c}{$\phi$} & \multicolumn{2}{c}{$\delta \phi$} & Method & Refs.\\
 \hline
     1 & Ceres                & 1014&00 &  101&40 & STM        & $\phi_{96}$           \\ 
     1 & Ceres                &  848&40 &   19&70 & STM        & $\phi_{93}$$^\dagger$ \\ 
     1 & Ceres                &  952&40 &    3&40 & Img-TE     & $\phi_{22}$           \\ 
     1 & Ceres                &  935&20 &    4&40 & Img-TE     & $\phi_{41}$           \\ 
     1 & Ceres                &  950&79 &    7&69 & Img-TE     & $\phi_{48}$           \\ 
     1 & Ceres                &  855&46 &   56&95 & STM        & $\phi_{64}$           \\ 
     1 & Ceres                &  886&47 &   27&29 & NEATM      & $\phi_{64}$           \\
\hline
\end{tabular}
 \end{table}
\setcounter{table}{0}
\begin{table}[!ht]
\caption{Continued}
\begin{tabular}{r@{ }lr@{.}lr@{.}lll}
 \hline
 \hline
 \# & Designation & \multicolumn{2}{c}{$\phi$} & \multicolumn{2}{c}{$\delta \phi$} & Method & Refs.\\
 \hline
     6 & Hebe                 &  204&00 &   20&39 & STM        & $\phi_{96}$           \\ 
     6 & Hebe                 &  185&17 &    2&90 & STM        & $\phi_{93}$           \\ 
     6 & Hebe                 &  180&42 &    8&50 & STM        & $\phi_{64}$           \\ 
     6 & Hebe                 &  214&49 &   10&25 & NEATM      & $\phi_{64}$$^\dagger$ \\ 
     6 & Hebe                 &  180&00 &   40&00 & LC+Occ     & $\phi_{78}$           \\ 
     6 & Hebe                 &  197&14 &    1&83 & STM        & $\phi_{83}$           \\ 
     6 & Hebe                 &  185&00 &   10&68 & NEATM      & $\phi_{72}$           \\
\hline
\end{tabular}
 \end{table}
\setcounter{table}{0}
\begin{table}[!ht]
\caption{Continued}
\begin{tabular}{r@{ }lr@{.}lr@{.}lll}
 \hline
 \hline
 \# & Designation & \multicolumn{2}{c}{$\phi$} & \multicolumn{2}{c}{$\delta \phi$} & Method & Refs.\\
 \hline
    13 & Egeria               &  244&00 &   24&39 & STM        & $\phi_{96}$           \\ 
    13 & Egeria               &  207&63 &    8&30 & STM        & $\phi_{93}$           \\ 
    13 & Egeria               &  223&08 &    3&46 & STM        & $\phi_{64}$           \\ 
    13 & Egeria               &  226&05 &    9&48 & NEATM      & $\phi_{64}$           \\ 
    13 & Egeria               &  203&36 &    2&56 & STM        & $\phi_{83}$           \\ 
    13 & Egeria               &  227&00 &   25&95 & NEATM      & $\phi_{72}$           \\ 
\noalign{\smallskip}
    14 & Irene                &  150&00 &   15&00 & STM        & $\phi_{96}$           \\
\hline
\end{tabular}
 \end{table}
\setcounter{table}{0}
\begin{table}[!ht]
\caption{Continued}
\begin{tabular}{r@{ }lr@{.}lr@{.}lll}
 \hline
 \hline
 \# & Designation & \multicolumn{2}{c}{$\phi$} & \multicolumn{2}{c}{$\delta \phi$} & Method & Refs.\\
 \hline
    21 & Lutetia              &  109&00 &   10&89 & STM        & $\phi_{96}$$^\dagger$ \\ 
    21 & Lutetia              &   95&76 &    4&09 & STM        & $\phi_{93}$$^\dagger$ \\ 
    21 & Lutetia              &  100&00 &   11&00 & Radar      & $\phi_{45}$$^\dagger$ \\ 
    21 & Lutetia              &  110&50 &    3&50 & TPM        & $\phi_{57}$$^\dagger$ \\ 
    21 & Lutetia              &   82&66 &    2&56 & STM        & $\phi_{64}$$^\dagger$ \\ 
    21 & Lutetia              &  104&20 &    7&11 & NEATM      & $\phi_{64}$$^\dagger$ \\ 
    21 & Lutetia              &  105&19 &    7&69 & Img-TE     & $\phi_{62}$$^\dagger$ \\
\hline
\end{tabular}
 \end{table}
\setcounter{table}{0}
\begin{table}[!ht]
\caption{Continued}
\begin{tabular}{r@{ }lr@{.}lr@{.}lll}
 \hline
 \hline
 \# & Designation & \multicolumn{2}{c}{$\phi$} & \multicolumn{2}{c}{$\delta \phi$} & Method & Refs.\\
 \hline
    29 & Amphitrite           &  200&00 &   20&00 & STM        & $\phi_{96}$           \\ 
    29 & Amphitrite           &  212&22 &    6&80 & STM        & $\phi_{93}$           \\ 
    29 & Amphitrite           &  231&21 &    6&55 & STM        & $\phi_{64}$           \\ 
    29 & Amphitrite           &  227&53 &    7&57 & NEATM      & $\phi_{64}$           \\ 
    29 & Amphitrite           &  206&86 &    2&59 & STM        & $\phi_{83}$           \\ 
    29 & Amphitrite           &  227&14 &    3&97 & NEATM      & $\phi_{72}$           \\ 
\noalign{\smallskip}
    30 & Urania               &   94&00 &    9&39 & STM        & $\phi_{96}$           \\
\hline
\end{tabular}
 \end{table}
\setcounter{table}{0}
\begin{table}[!ht]
\caption{Continued}
\begin{tabular}{r@{ }lr@{.}lr@{.}lll}
 \hline
 \hline
 \# & Designation & \multicolumn{2}{c}{$\phi$} & \multicolumn{2}{c}{$\delta \phi$} & Method & Refs.\\
 \hline
    42 & Isis                 &   94&00 &    9&39 & STM        & $\phi_{96}$           \\ 
    42 & Isis                 &  100&19 &    3&40 & STM        & $\phi_{93}$           \\ 
    42 & Isis                 &  102&90 &    2&28 & STM        & $\phi_{64}$           \\ 
    42 & Isis                 &  104&69 &    6&11 & NEATM      & $\phi_{64}$           \\ 
    42 & Isis                 &  104&50 &    1&37 & STM        & $\phi_{83}$           \\ 
\noalign{\smallskip}
    43 & Ariadne              &   85&00 &    8&50 & STM        & $\phi_{96}$$^\dagger$ \\ 
    43 & Ariadne              &   65&87 &    2&50 & STM        & $\phi_{93}$           \\
\hline
\end{tabular}
 \end{table}
\setcounter{table}{0}
\begin{table}[!ht]
\caption{Continued}
\begin{tabular}{r@{ }lr@{.}lr@{.}lll}
 \hline
 \hline
 \# & Designation & \multicolumn{2}{c}{$\phi$} & \multicolumn{2}{c}{$\delta \phi$} & Method & Refs.\\
 \hline
    52 & Europa               &  292&00 &   29&20 & STM        & $\phi_{96}$           \\ 
    52 & Europa               &  302&50 &    5&40 & STM        & $\phi_{93}$           \\ 
    52 & Europa               &  308&00 &    1&50 & Img        & $\phi_{34}$           \\ 
    52 & Europa               &  264&32 &    6&96 & STM        & $\phi_{64}$$^\dagger$ \\ 
    52 & Europa               &  340&75 &   13&36 & NEATM      & $\phi_{64}$           \\ 
    52 & Europa               &  293&00 &   30&00 & LC+Occ     & $\phi_{78}$           \\ 
    52 & Europa               &  350&35 &    5&07 & STM        & $\phi_{83}$$^\dagger$ \\
\hline
\end{tabular}
 \end{table}
\setcounter{table}{0}
\begin{table}[!ht]
\caption{Continued}
\begin{tabular}{r@{ }lr@{.}lr@{.}lll}
 \hline
 \hline
 \# & Designation & \multicolumn{2}{c}{$\phi$} & \multicolumn{2}{c}{$\delta \phi$} & Method & Refs.\\
 \hline
    61 & Danae                &   76&76 &    4&26 & STM        & $\phi_{94}$           \\ 
    61 & Danae                &   82&04 &    4&30 & STM        & $\phi_{93}$           \\ 
    61 & Danae                &   78&12 &    5&11 & STM        & $\phi_{64}$           \\ 
    61 & Danae                &   93&01 &    4&32 & NEATM      & $\phi_{64}$$^\dagger$ \\ 
    61 & Danae                &   83&55 &    1&01 & STM        & $\phi_{83}$           \\ 
    61 & Danae                &   85&12 &    1&96 & NEATM      & $\phi_{72}$           \\ 
\noalign{\smallskip}
    63 & Ausonia              &   93&00 &    9&30 & STM        & $\phi_{96}$           \\
\hline
\end{tabular}
 \end{table}
\setcounter{table}{0}
\begin{table}[!ht]
\caption{Continued}
\begin{tabular}{r@{ }lr@{.}lr@{.}lll}
 \hline
 \hline
 \# & Designation & \multicolumn{2}{c}{$\phi$} & \multicolumn{2}{c}{$\delta \phi$} & Method & Refs.\\
 \hline
    74 & Galatea              &  118&70 &    2&79 & STM        & $\phi_{93}$           \\ 
    74 & Galatea              &   97&83 &    2&94 & Occ        & $\phi_{95}$$^\dagger$ \\ 
    74 & Galatea              &  127&66 &    3&33 & STM        & $\phi_{64}$           \\ 
    74 & Galatea              &  132&02 &    5&32 & NEATM      & $\phi_{64}$           \\ 
    74 & Galatea              &  113&08 &    2&15 & STM        & $\phi_{83}$           \\ 
\noalign{\smallskip}
    76 & Freia                &  117&00 &   11&69 & STM        & $\phi_{96}$$^\dagger$ \\ 
    76 & Freia                &  158&67 &    4&96 & Occ        & $\phi_{95}$           \\
\hline
\end{tabular}
 \end{table}
\setcounter{table}{0}
\begin{table}[!ht]
\caption{Continued}
\begin{tabular}{r@{ }lr@{.}lr@{.}lll}
 \hline
 \hline
 \# & Designation & \multicolumn{2}{c}{$\phi$} & \multicolumn{2}{c}{$\delta \phi$} & Method & Refs.\\
 \hline
    88 & Thisbe               &  210&00 &   21&00 & STM        & $\phi_{96}$           \\ 
    88 & Thisbe               &  200&58 &    5&00 & STM        & $\phi_{93}$           \\ 
    88 & Thisbe               &  207&88 &    5&88 & STM        & $\phi_{64}$           \\ 
    88 & Thisbe               &  223&19 &   10&22 & NEATM      & $\phi_{64}$           \\ 
    88 & Thisbe               &  204&00 &   14&00 & LC+Occ     & $\phi_{78}$           \\ 
    88 & Thisbe               &  220&00 &   16&00 & LC+Occ     & $\phi_{78}$           \\ 
    88 & Thisbe               &  195&58 &    2&72 & STM        & $\phi_{83}$           \\
\hline
\end{tabular}
 \end{table}
\setcounter{table}{0}
\begin{table}[!ht]
\caption{Continued}
\begin{tabular}{r@{ }lr@{.}lr@{.}lll}
 \hline
 \hline
 \# & Designation & \multicolumn{2}{c}{$\phi$} & \multicolumn{2}{c}{$\delta \phi$} & Method & Refs.\\
 \hline
    98 & Ianthe               &  104&44 &    1&79 & STM        & $\phi_{93}$           \\ 
    98 & Ianthe               &  101&97 &    3&67 & STM        & $\phi_{64}$           \\ 
    98 & Ianthe               &  114&30 &    5&23 & NEATM      & $\phi_{64}$           \\ 
    98 & Ianthe               &  104&23 &    1&28 & STM        & $\phi_{83}$           \\ 
    98 & Ianthe               &  110&87 &    2&31 & NEATM      & $\phi_{72}$           \\ 
\noalign{\smallskip}
   105 & Artemis              &  103&65 &    5&26 & STM        & $\phi_{93}$           \\ 
   105 & Artemis              &  119&08 &    2&79 & STM        & $\phi_{64}$           \\
\hline
\end{tabular}
 \end{table}
\setcounter{table}{0}
\begin{table}[!ht]
\caption{Continued}
\begin{tabular}{r@{ }lr@{.}lr@{.}lll}
 \hline
 \hline
 \# & Designation & \multicolumn{2}{c}{$\phi$} & \multicolumn{2}{c}{$\delta \phi$} & Method & Refs.\\
 \hline
   126 & Velleda              &   44&81 &    1&29 & STM        & $\phi_{93}$           \\ 
   126 & Velleda              &   47&47 &    1&53 & STM        & $\phi_{64}$           \\ 
   126 & Velleda              &   49&66 &    2&33 & NEATM      & $\phi_{64}$$^\dagger$ \\ 
   126 & Velleda              &   43&93 &    0&49 & STM        & $\phi_{83}$           \\ 
\noalign{\smallskip}
   127 & Johanna              &  123&33 &    4&36 & STM        & $\phi_{94}$           \\ 
   127 & Johanna              &  112&98 &    8&79 & Occ        & $\phi_{95}$           \\ 
   127 & Johanna              &  109&62 &    0&01 & STM        & $\phi_{64}$$^\dagger$ \\
\hline
\end{tabular}
 \end{table}
\setcounter{table}{0}
\begin{table}[!ht]
\caption{Continued}
\begin{tabular}{r@{ }lr@{.}lr@{.}lll}
 \hline
 \hline
 \# & Designation & \multicolumn{2}{c}{$\phi$} & \multicolumn{2}{c}{$\delta \phi$} & Method & Refs.\\
 \hline
   138 & Tolosa               &   45&50 &    2&09 & STM        & $\phi_{93}$$^\dagger$ \\ 
   138 & Tolosa               &   46&65 &    3&81 & STM        & $\phi_{64}$           \\ 
   138 & Tolosa               &   57&49 &    3&38 & NEATM      & $\phi_{64}$           \\ 
   138 & Tolosa               &   51&61 &    0&84 & STM        & $\phi_{83}$           \\ 
   138 & Tolosa               &   58&93 &    1&64 & NEATM      & $\phi_{72}$$^\dagger$ \\ 
\noalign{\smallskip}
   139 & Juewa                &  172&00 &   17&20 & STM        & $\phi_{96}$           \\ 
   139 & Juewa                &  156&60 &    2&79 & STM        & $\phi_{93}$           \\
\hline
\end{tabular}
 \end{table}
\setcounter{table}{0}
\begin{table}[!ht]
\caption{Continued}
\begin{tabular}{r@{ }lr@{.}lr@{.}lll}
 \hline
 \hline
 \# & Designation & \multicolumn{2}{c}{$\phi$} & \multicolumn{2}{c}{$\delta \phi$} & Method & Refs.\\
 \hline
   154 & Bertha               &  184&92 &    3&59 & STM        & $\phi_{93}$           \\ 
   154 & Bertha               &  189&86 &    5&30 & STM        & $\phi_{64}$           \\ 
   154 & Bertha               &  186&35 &    8&97 & NEATM      & $\phi_{64}$           \\ 
   154 & Bertha               &  185&83 &    2&72 & STM        & $\phi_{83}$           \\ 
   154 & Bertha               &  188&75 &    4&76 & NEATM      & $\phi_{72}$           \\ 
\noalign{\smallskip}
   156 & Xanthippe            &  120&98 &    2&50 & STM        & $\phi_{93}$           \\ 
   156 & Xanthippe            &  122&04 &    3&51 & STM        & $\phi_{64}$           \\
\hline
\end{tabular}
 \end{table}
\setcounter{table}{0}
\begin{table}[!ht]
\caption{Continued}
\begin{tabular}{r@{ }lr@{.}lr@{.}lll}
 \hline
 \hline
 \# & Designation & \multicolumn{2}{c}{$\phi$} & \multicolumn{2}{c}{$\delta \phi$} & Method & Refs.\\
 \hline
   189 & Phthia               &   42&00 &    4&19 & STM        & $\phi_{96}$           \\ 
   189 & Phthia               &   37&65 &    2&00 & STM        & $\phi_{93}$           \\ 
   189 & Phthia               &   32&38 &    1&48 & STM        & $\phi_{64}$$^\dagger$ \\ 
   189 & Phthia               &   40&84 &    3&22 & NEATM      & $\phi_{64}$           \\ 
   189 & Phthia               &   42&47 &    0&67 & STM        & $\phi_{83}$           \\ 
   189 & Phthia               &   40&56 &    0&39 & NEATM      & $\phi_{72}$           \\ 
\noalign{\smallskip}
   192 & Nausikaa             &   99&00 &    9&89 & STM        & $\phi_{96}$           \\
\hline
\end{tabular}
 \end{table}
\setcounter{table}{0}
\begin{table}[!ht]
\caption{Continued}
\begin{tabular}{r@{ }lr@{.}lr@{.}lll}
 \hline
 \hline
 \# & Designation & \multicolumn{2}{c}{$\phi$} & \multicolumn{2}{c}{$\delta \phi$} & Method & Refs.\\
 \hline
   211 & Isolda               &  166&00 &   16&60 & STM        & $\phi_{96}$           \\ 
   211 & Isolda               &  143&19 &    5&09 & STM        & $\phi_{93}$           \\ 
   211 & Isolda               &  142&57 &    4&36 & STM        & $\phi_{64}$           \\ 
   211 & Isolda               &  150&94 &    7&50 & NEATM      & $\phi_{64}$           \\ 
   211 & Isolda               &  153&49 &    1&71 & STM        & $\phi_{83}$           \\ 
   211 & Isolda               &  143&00 &   21&62 & NEATM      & $\phi_{72}$           \\ 
\noalign{\smallskip}
   212 & Medea                &  136&11 &    2&50 & STM        & $\phi_{93}$           \\
\hline
\end{tabular}
 \end{table}
\setcounter{table}{0}
\begin{table}[!ht]
\caption{Continued}
\begin{tabular}{r@{ }lr@{.}lr@{.}lll}
 \hline
 \hline
 \# & Designation & \multicolumn{2}{c}{$\phi$} & \multicolumn{2}{c}{$\delta \phi$} & Method & Refs.\\
 \hline
   240 & Vanadis              &  103&90 &    2&50 & STM        & $\phi_{93}$           \\ 
   240 & Vanadis              &   93&08 &    3&57 & Occ        & $\phi_{95}$           \\ 
   240 & Vanadis              &   99&26 &   10&01 & STM        & $\phi_{64}$           \\ 
   240 & Vanadis              &  112&90 &    4&65 & NEATM      & $\phi_{64}$$^\dagger$ \\ 
   240 & Vanadis              &   90&12 &    1&22 & STM        & $\phi_{83}$           \\ 
   240 & Vanadis              &   91&37 &    2&65 & NEATM      & $\phi_{72}$           \\ 
\noalign{\smallskip}
   241 & Germania             &  200&00 &   20&00 & STM        & $\phi_{96}$           \\
\hline
\end{tabular}
 \end{table}
\setcounter{table}{0}
\begin{table}[!ht]
\caption{Continued}
\begin{tabular}{r@{ }lr@{.}lr@{.}lll}
 \hline
 \hline
 \# & Designation & \multicolumn{2}{c}{$\phi$} & \multicolumn{2}{c}{$\delta \phi$} & Method & Refs.\\
 \hline
   304 & Olga                 &   67&86 &    2&09 & STM        & $\phi_{93}$           \\ 
   304 & Olga                 &   69&19 &    1&79 & Occ        & $\phi_{95}$           \\ 
   304 & Olga                 &   71&13 &    2&54 & STM        & $\phi_{64}$           \\ 
   304 & Olga                 &   77&58 &    5&36 & NEATM      & $\phi_{64}$           \\ 
   304 & Olga                 &   71&26 &    1&29 & STM        & $\phi_{83}$           \\ 
   304 & Olga                 &   68&87 &    2&26 & NEATM      & $\phi_{72}$           \\ 
\noalign{\smallskip}
   306 & Unitas               &   46&70 &    2&29 & STM        & $\phi_{93}$           \\
\hline
\end{tabular}
 \end{table}
 \clearpage
\setcounter{table}{0}
\begin{table}[!ht]
\caption{Continued}
\begin{tabular}{r@{ }lr@{.}lr@{.}lll}
 \hline
 \hline
 \# & Designation & \multicolumn{2}{c}{$\phi$} & \multicolumn{2}{c}{$\delta \phi$} & Method & Refs.\\
 \hline
   345 & Tercidina            &   94&12 &    4&90 & STM        & $\phi_{93}$           \\ 
   345 & Tercidina            &   99&30 &    1&39 & Occ        & $\phi_{95}$           \\ 
   345 & Tercidina            &   93&80 &    5&88 & STM        & $\phi_{64}$           \\ 
   345 & Tercidina            &  106&19 &    7&80 & NEATM      & $\phi_{64}$           \\ 
   345 & Tercidina            &   99&23 &    0&99 & STM        & $\phi_{83}$           \\ 
   345 & Tercidina            &   99&00 &   11&47 & NEATM      & $\phi_{72}$           \\ 
\noalign{\smallskip}
   346 & Hermentaria          &  106&51 &    2&20 & STM        & $\phi_{93}$$^\dagger$ \\
\hline
\end{tabular}
 \end{table}
\setcounter{table}{0}
\begin{table}[!ht]
\caption{Continued}
\begin{tabular}{r@{ }lr@{.}lr@{.}lll}
 \hline
 \hline
 \# & Designation & \multicolumn{2}{c}{$\phi$} & \multicolumn{2}{c}{$\delta \phi$} & Method & Refs.\\
 \hline
   381 & Myrrha               &  120&58 &    2&70 & STM        & $\phi_{93}$           \\ 
   381 & Myrrha               &  130&58 &    2&75 & Occ        & $\phi_{95}$           \\ 
   381 & Myrrha               &  120&30 &    3&45 & STM        & $\phi_{64}$           \\ 
   381 & Myrrha               &  136&57 &    7&13 & NEATM      & $\phi_{64}$           \\ 
   381 & Myrrha               &  117&12 &    1&58 & STM        & $\phi_{83}$           \\ 
   381 & Myrrha               &  129&00 &    9&94 & NEATM      & $\phi_{72}$           \\ 
   381 & Myrrha               &  129&00 &    6&01 & NEATM      & $\phi_{72}$           \\
\hline
\end{tabular}
 \end{table}
\setcounter{table}{0}
\begin{table}[!ht]
\caption{Continued}
\begin{tabular}{r@{ }lr@{.}lr@{.}lll}
 \hline
 \hline
 \# & Designation & \multicolumn{2}{c}{$\phi$} & \multicolumn{2}{c}{$\delta \phi$} & Method & Refs.\\
 \hline
   420 & Bertholda            &  138&85 &    2&41 & STM        & $\phi_{94}$           \\ 
   420 & Bertholda            &  141&25 &    6&90 & STM        & $\phi_{93}$           \\ 
   420 & Bertholda            &  143&97 &    2&51 & Occ        & $\phi_{95}$           \\ 
   420 & Bertholda            &  139&32 &    6&01 & STM        & $\phi_{64}$           \\ 
   420 & Bertholda            &  160&33 &    8&80 & NEATM      & $\phi_{64}$$^\dagger$ \\ 
   420 & Bertholda            &  141&89 &    2&58 & STM        & $\phi_{83}$           \\ 
   420 & Bertholda            &  144&00 &    5&67 & NEATM      & $\phi_{72}$           \\
\hline
\end{tabular}
 \end{table}
\setcounter{table}{0}
\begin{table}[!ht]
\caption{Continued}
\begin{tabular}{r@{ }lr@{.}lr@{.}lll}
 \hline
 \hline
 \# & Designation & \multicolumn{2}{c}{$\phi$} & \multicolumn{2}{c}{$\delta \phi$} & Method & Refs.\\
 \hline
   455 & Bruchsalia           &  105&00 &   10&50 & STM        & $\phi_{96}$           \\ 
   455 & Bruchsalia           &   84&41 &    5&00 & STM        & $\phi_{93}$           \\ 
   455 & Bruchsalia           &   92&08 &    3&36 & STM        & $\phi_{64}$           \\ 
   455 & Bruchsalia           &   99&65 &    4&44 & NEATM      & $\phi_{64}$           \\ 
   455 & Bruchsalia           &   83&45 &    1&00 & STM        & $\phi_{83}$           \\ 
   455 & Bruchsalia           &  112&37 &    4&36 & NEATM      & $\phi_{72}$$^\dagger$ \\ 
\noalign{\smallskip}
   469 & Argentina            &  125&56 &    5&59 & STM        & $\phi_{93}$           \\
\hline
\end{tabular}
 \end{table}
\setcounter{table}{0}
\begin{table}[!ht]
\caption{Continued}
\begin{tabular}{r@{ }lr@{.}lr@{.}lll}
 \hline
 \hline
 \# & Designation & \multicolumn{2}{c}{$\phi$} & \multicolumn{2}{c}{$\delta \phi$} & Method & Refs.\\
 \hline
   508 & Princetonia          &  129&38 &    0&20 & Occ        & $\phi_{95}$$^\dagger$ \\ 
   508 & Princetonia          &  136&14 &    5&44 & STM        & $\phi_{94}$           \\ 
   508 & Princetonia          &  142&35 &    2&59 & STM        & $\phi_{93}$           \\ 
   508 & Princetonia          &  133&75 &    5&71 & STM        & $\phi_{64}$           \\ 
   508 & Princetonia          &  144&99 &    6&77 & NEATM      & $\phi_{64}$           \\ 
   508 & Princetonia          &  139&42 &    2&30 & STM        & $\phi_{83}$           \\ 
   508 & Princetonia          &  120&26 &    2&92 & NEATM      & $\phi_{72}$$^\dagger$ \\
\hline
\end{tabular}%
 \end{table}

\clearpage
    \begin{figure}[!ht]
  \centering
  \includegraphics[width=.49\textwidth]{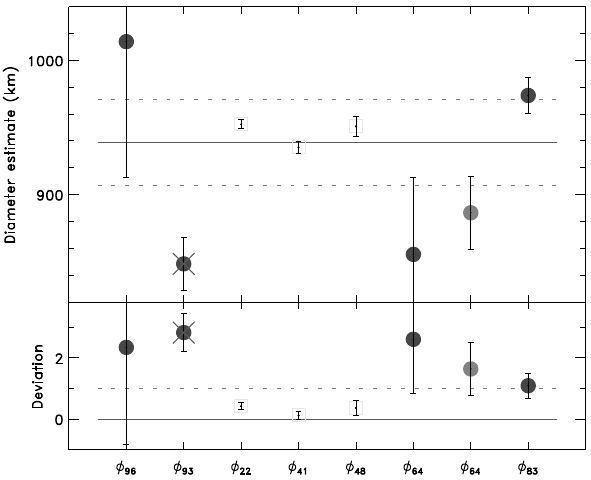}
\caption[Diameter estimates for (1) Ceres]{%
  \label{fap: diam000001}
  Diameter estimates for (1) Ceres.
}
\end{figure}

  \begin{figure}[!ht]
  \centering
  \includegraphics[width=.49\textwidth]{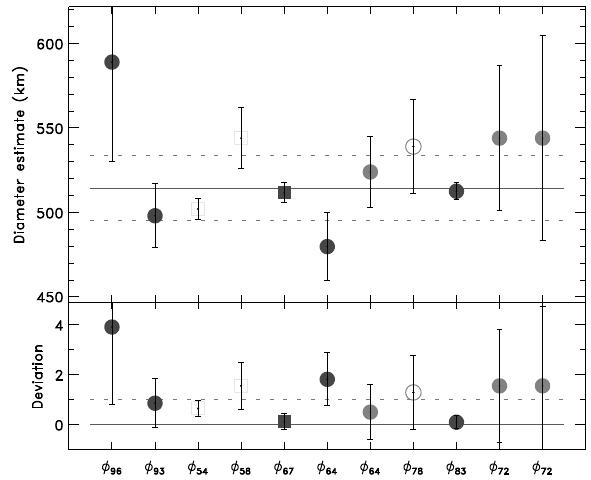}
\caption[Diameter estimates for (2) Pallas]{%
  \label{fap: diam000002}
  Diameter estimates for (2) Pallas.
}
\end{figure}

  \begin{figure}[!ht]
  \centering
  \includegraphics[width=.49\textwidth]{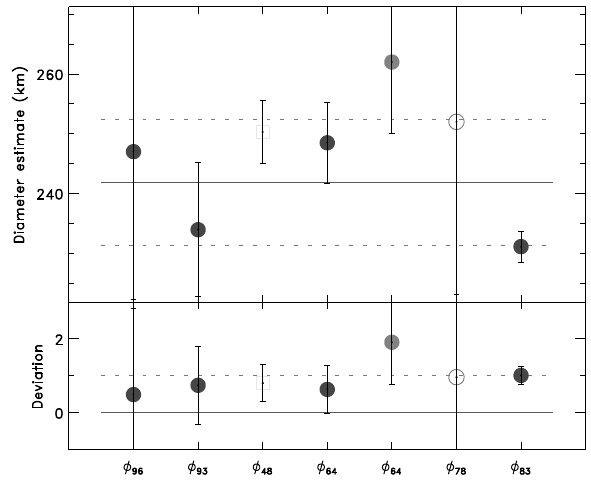}
\caption[Diameter estimates for (3) Juno]{%
  \label{fap: diam000003}
  Diameter estimates for (3) Juno.
}
\end{figure}

  \begin{figure}[!ht]
  \centering
  \includegraphics[width=.49\textwidth]{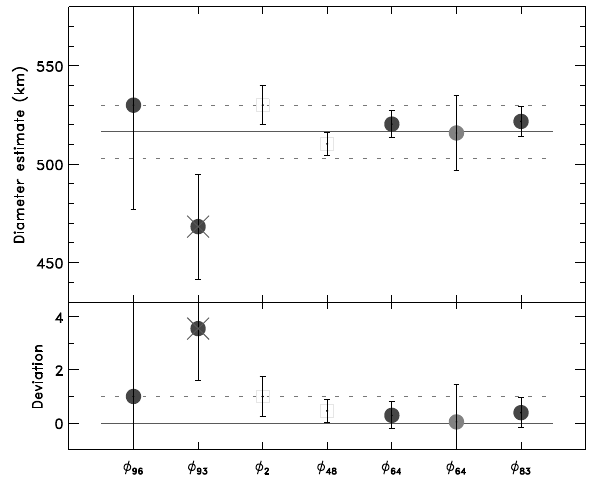}
\caption[Diameter estimates for (4) Vesta]{%
  \label{fap: diam000004}
  Diameter estimates for (4) Vesta.
}
\end{figure}

  \begin{figure}[!ht]
  \centering
  \includegraphics[width=.49\textwidth]{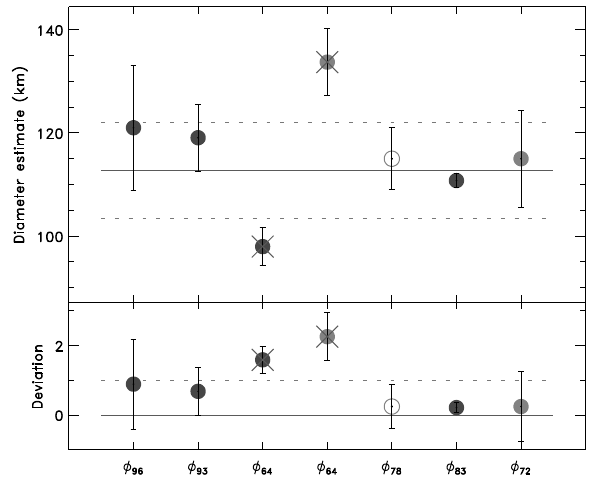}
\caption[Diameter estimates for (5) Astraea]{%
  \label{fap: diam000005}
  Diameter estimates for (5) Astraea.
}
\end{figure}

  \begin{figure}[!ht]
  \centering
  \includegraphics[width=.49\textwidth]{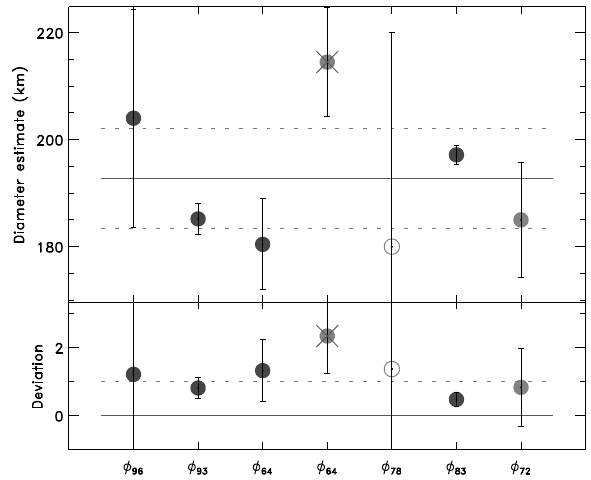}
\caption[Diameter estimates for (6) Hebe]{%
  \label{fap: diam000006}
  Diameter estimates for (6) Hebe.
}
\end{figure}

  \begin{figure}[!ht]
  \centering
  \includegraphics[width=.49\textwidth]{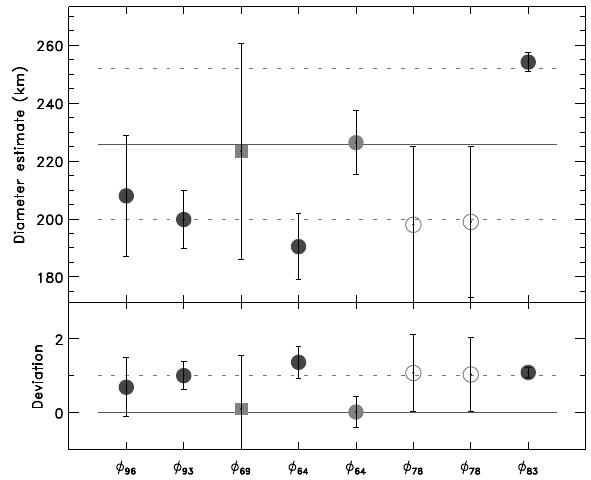}
\caption[Diameter estimates for (7) Iris]{%
  \label{fap: diam000007}
  Diameter estimates for (7) Iris.
}
\end{figure}

  \begin{figure}[!ht]
  \centering
  \includegraphics[width=.49\textwidth]{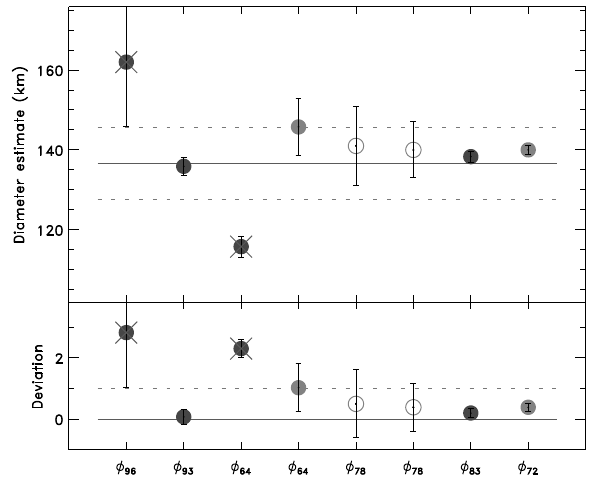}
\caption[Diameter estimates for (8) Flora]{%
  \label{fap: diam000008}
  Diameter estimates for (8) Flora.
}
\end{figure}

  \begin{figure}[!ht]
  \centering
  \includegraphics[width=.49\textwidth]{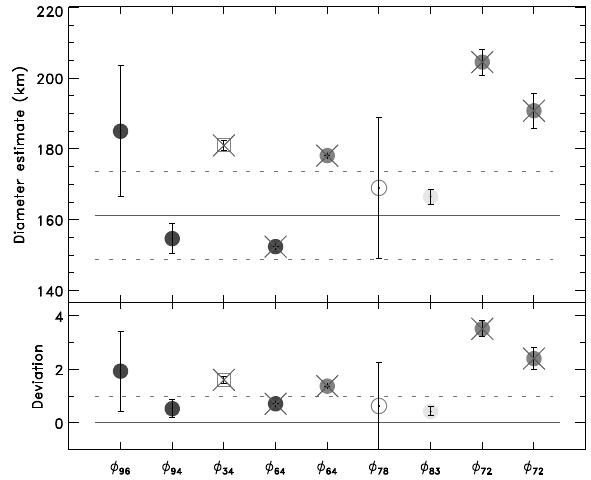}
\caption[Diameter estimates for (9) Metis]{%
  \label{fap: diam000009}
  Diameter estimates for (9) Metis.
  The diameter estimates from $\phi_{64}$ have unrealistic small uncertainties of 0.02\,km.  Using these values strongly biases the average. 
}
\end{figure}

  \begin{figure}[!ht]
  \centering
  \includegraphics[width=.49\textwidth]{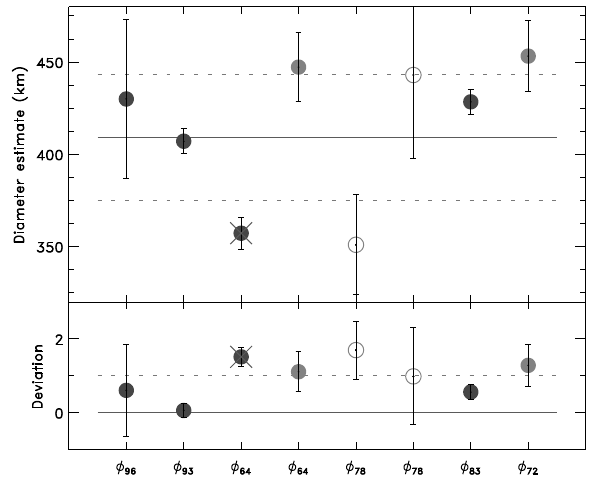}
\caption[Diameter estimates for (10) Hygiea]{%
  \label{fap: diam000010}
  Diameter estimates for (10) Hygiea.
}
\end{figure}

  \begin{figure}[!ht]
  \centering
  \includegraphics[width=.49\textwidth]{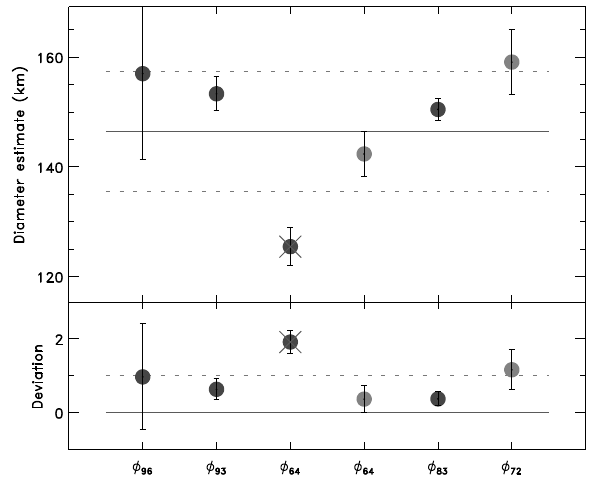}
\caption[Diameter estimates for (11) Parthenope]{%
  \label{fap: diam000011}
  Diameter estimates for (11) Parthenope.
}
\end{figure}

  \begin{figure}[!ht]
  \centering
  \includegraphics[width=.49\textwidth]{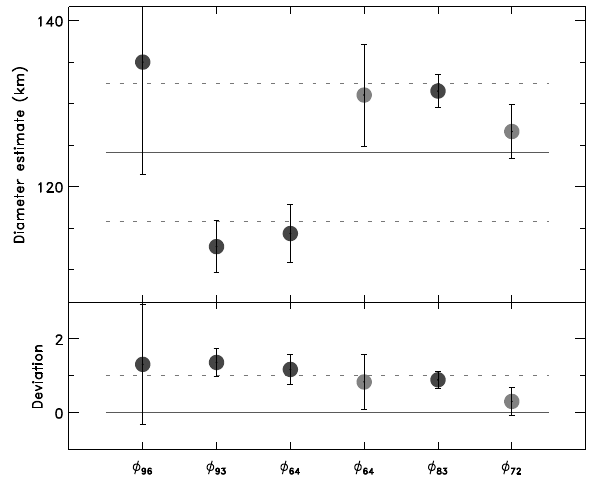}
\caption[Diameter estimates for (12) Victoria]{%
  \label{fap: diam000012}
  Diameter estimates for (12) Victoria.
}
\end{figure}

  \begin{figure}[!ht]
  \centering
  \includegraphics[width=.49\textwidth]{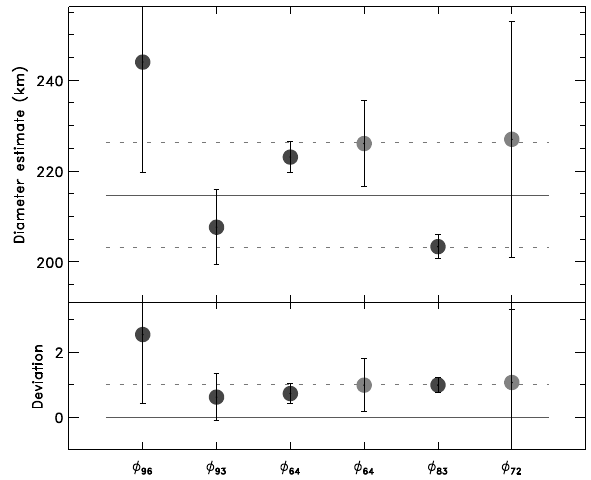}
\caption[Diameter estimates for (13) Egeria]{%
  \label{fap: diam000013}
  Diameter estimates for (13) Egeria.
}
\end{figure}

  \begin{figure}[!ht]
  \centering
  \includegraphics[width=.49\textwidth]{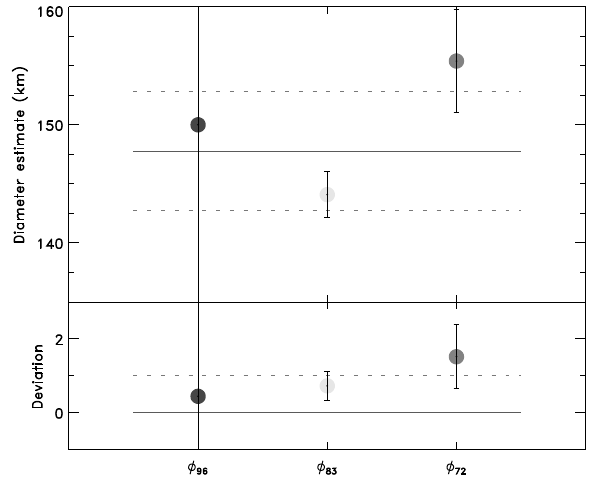}
\caption[Diameter estimates for (14) Irene]{%
  \label{fap: diam000014}
  Diameter estimates for (14) Irene.
}
\end{figure}

  \begin{figure}[!ht]
  \centering
  \includegraphics[width=.49\textwidth]{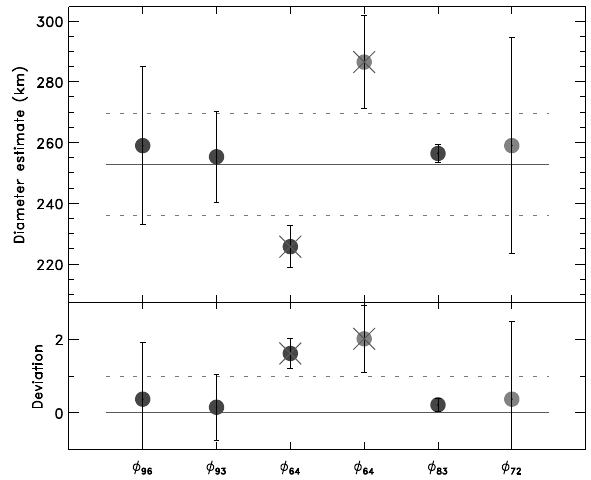}
\caption[Diameter estimates for (15) Eunomia]{%
  \label{fap: diam000015}
  Diameter estimates for (15) Eunomia.
}
\end{figure}

  \begin{figure}[!ht]
  \centering
  \includegraphics[width=.49\textwidth]{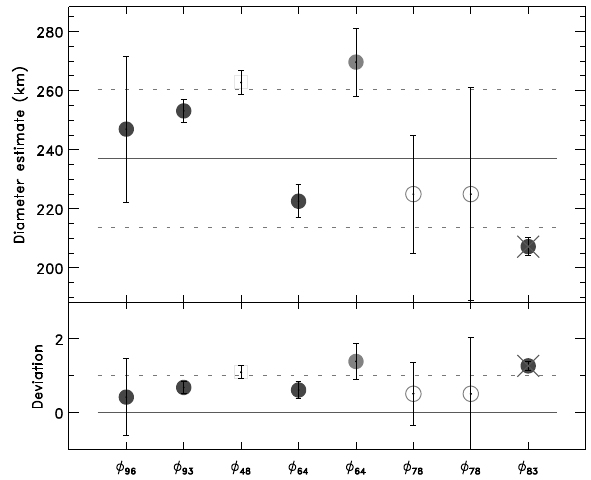}
\caption[Diameter estimates for (16) Psyche]{%
  \label{fap: diam000016}
  Diameter estimates for (16) Psyche.
}
\end{figure}

  \begin{figure}[!ht]
  \centering
  \includegraphics[width=.49\textwidth]{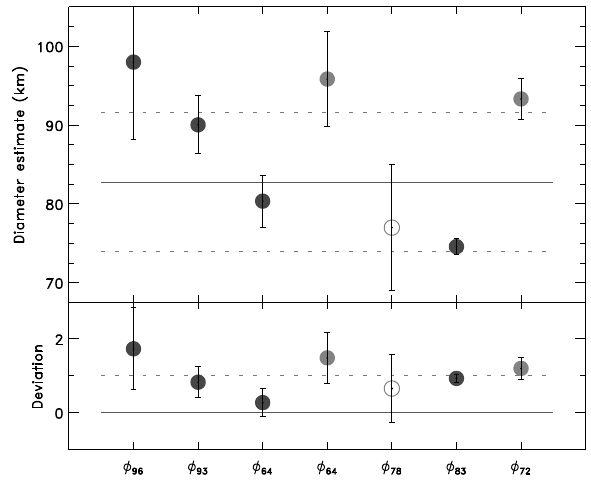}
\caption[Diameter estimates for (17) Thetis]{%
  \label{fap: diam000017}
  Diameter estimates for (17) Thetis.
}
\end{figure}

  \begin{figure}[!ht]
  \centering
  \includegraphics[width=.49\textwidth]{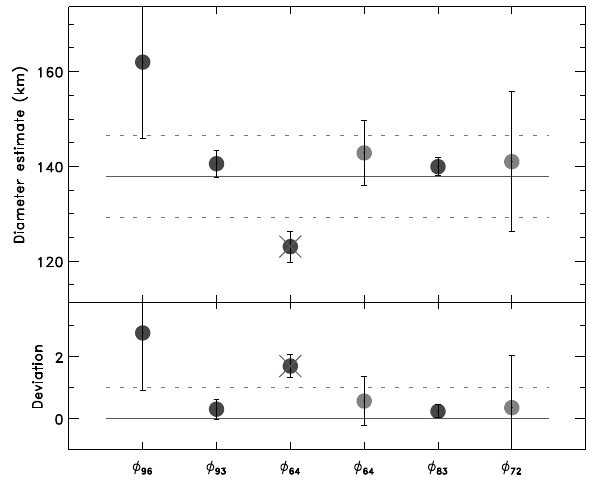}
\caption[Diameter estimates for (18) Melpomene]{%
  \label{fap: diam000018}
  Diameter estimates for (18) Melpomene.
}
\end{figure}

\clearpage
  \begin{figure}[!ht]
  \centering
  \includegraphics[width=.49\textwidth]{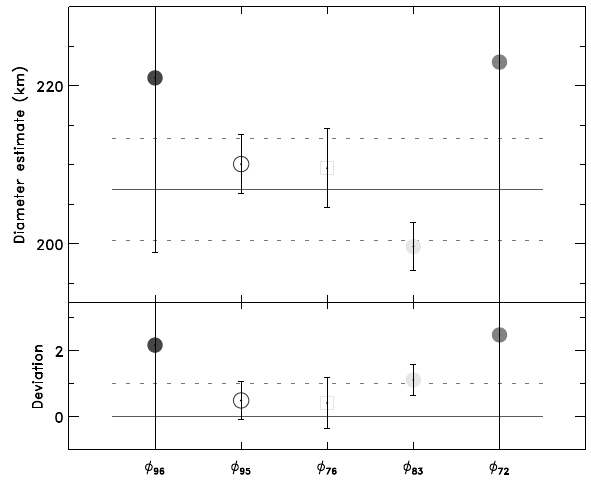}
\caption[Diameter estimates for (19) Fortuna]{%
  \label{fap: diam000019}
  Diameter estimates for (19) Fortuna.
}
\end{figure}

  \begin{figure}[!ht]
  \centering
  \includegraphics[width=.49\textwidth]{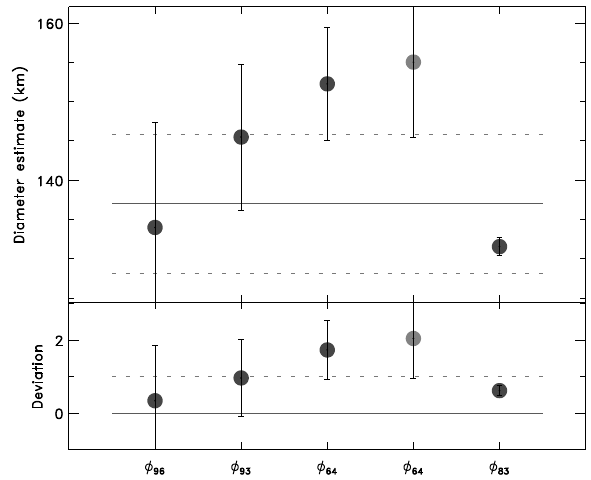}
\caption[Diameter estimates for (20) Massalia]{%
  \label{fap: diam000020}
  Diameter estimates for (20) Massalia.
}
\end{figure}

  \begin{figure}[!ht]
  \centering
  \includegraphics[width=.49\textwidth]{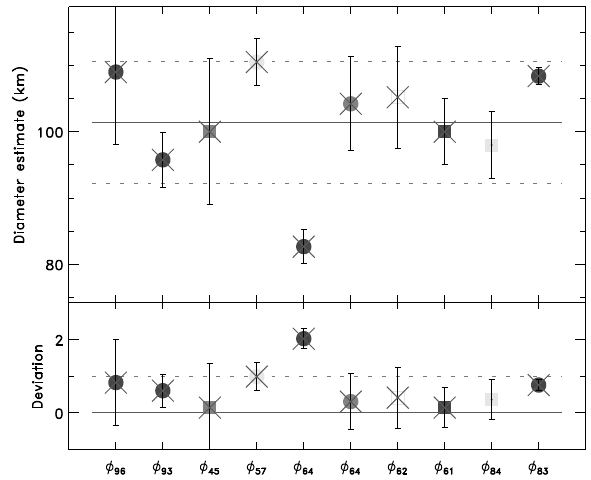}
\caption[Diameter estimates for (21) Lutetia]{%
  \label{fap: diam000021}
  Diameter estimates for (21) Lutetia.
  Only the flyby estimate from $\phi_{61}$$^\dagger$ is used here.
}
\end{figure}

  \begin{figure}[!ht]
  \centering
  \includegraphics[width=.49\textwidth]{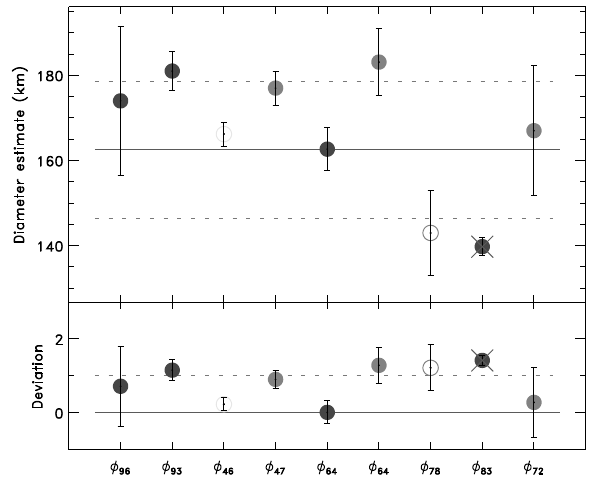}
\caption[Diameter estimates for (22) Kalliope]{%
  \label{fap: diam000022}
  Diameter estimates for (22) Kalliope.
}
\end{figure}

  \begin{figure}[!ht]
  \centering
  \includegraphics[width=.49\textwidth]{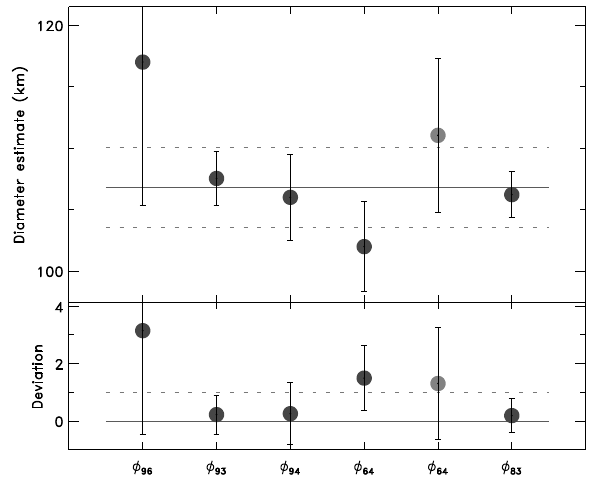}
\caption[Diameter estimates for (23) Thalia]{%
  \label{fap: diam000023}
  Diameter estimates for (23) Thalia.
}
\end{figure}

  \begin{figure}[!ht]
  \centering
  \includegraphics[width=.49\textwidth]{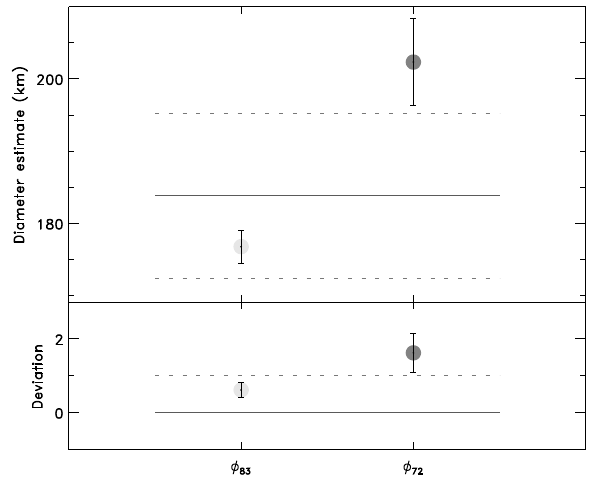}
\caption[Diameter estimates for (24) Themis]{%
  \label{fap: diam000024}
  Diameter estimates for (24) Themis.
}
\end{figure}

  \begin{figure}[!ht]
  \centering
  \includegraphics[width=.49\textwidth]{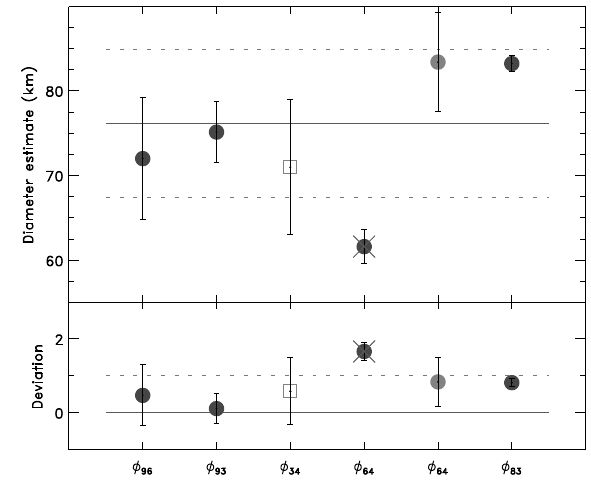}
\caption[Diameter estimates for (25) Phocaea]{%
  \label{fap: diam000025}
  Diameter estimates for (25) Phocaea.
}
\end{figure}

  \begin{figure}[!ht]
  \centering
  \includegraphics[width=.49\textwidth]{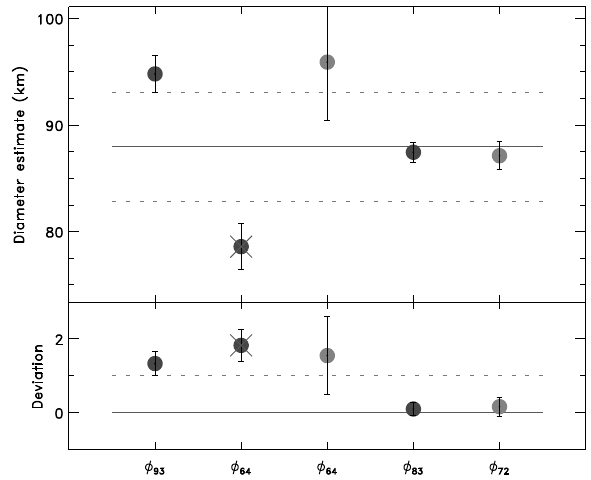}
\caption[Diameter estimates for (26) Proserpina]{%
  \label{fap: diam000026}
  Diameter estimates for (26) Proserpina.
}
\end{figure}

  \begin{figure}[!ht]
  \centering
  \includegraphics[width=.49\textwidth]{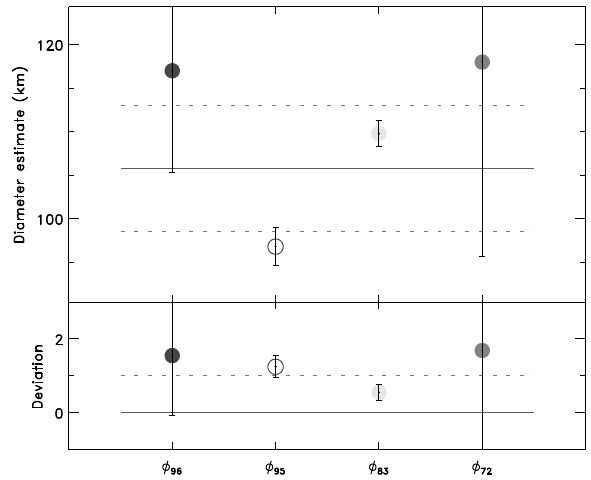}
\caption[Diameter estimates for (27) Euterpe]{%
  \label{fap: diam000027}
  Diameter estimates for (27) Euterpe.
}
\end{figure}

  \begin{figure}[!ht]
  \centering
  \includegraphics[width=.49\textwidth]{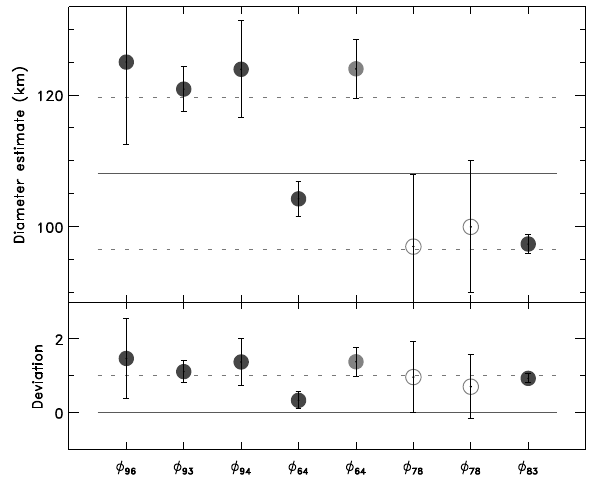}
\caption[Diameter estimates for (28) Bellona]{%
  \label{fap: diam000028}
  Diameter estimates for (28) Bellona.
}
\end{figure}

  \begin{figure}[!ht]
  \centering
  \includegraphics[width=.49\textwidth]{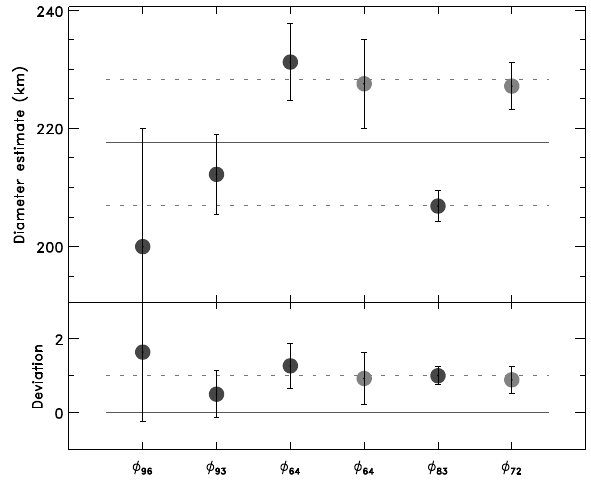}
\caption[Diameter estimates for (29) Amphitrite]{%
  \label{fap: diam000029}
  Diameter estimates for (29) Amphitrite.
}
\end{figure}

  \begin{figure}[!ht]
  \centering
  \includegraphics[width=.49\textwidth]{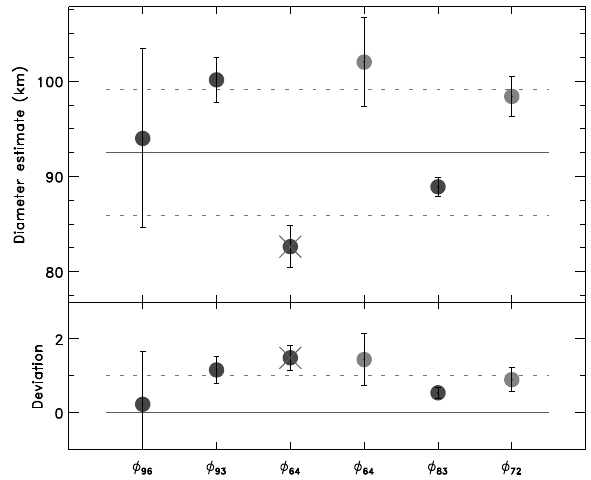}
\caption[Diameter estimates for (30) Urania]{%
  \label{fap: diam000030}
  Diameter estimates for (30) Urania.
}
\end{figure}

  \begin{figure}[!ht]
  \centering
  \includegraphics[width=.49\textwidth]{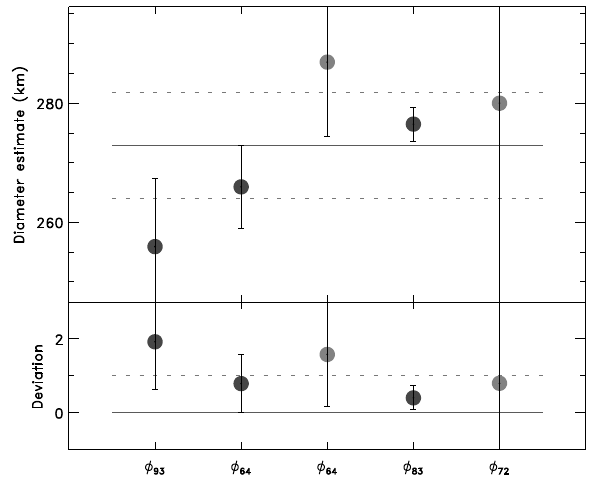}
\caption[Diameter estimates for (31) Euphrosyne]{%
  \label{fap: diam000031}
  Diameter estimates for (31) Euphrosyne.
}
\end{figure}

  \begin{figure}[!ht]
  \centering
  \includegraphics[width=.49\textwidth]{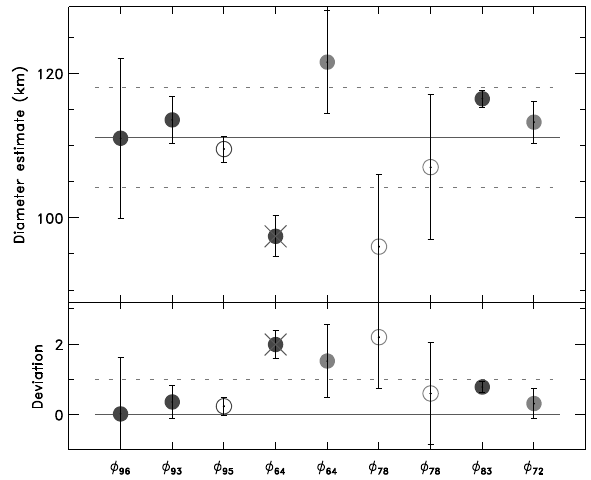}
\caption[Diameter estimates for (34) Circe]{%
  \label{fap: diam000034}
  Diameter estimates for (34) Circe.
}
\end{figure}

  \begin{figure}[!ht]
  \centering
  \includegraphics[width=.49\textwidth]{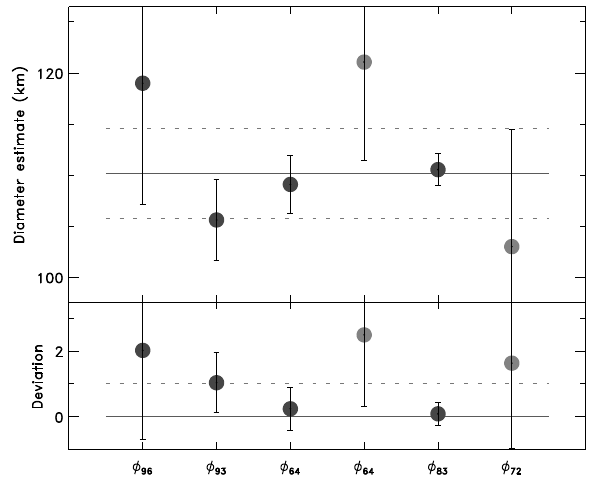}
\caption[Diameter estimates for (36) Atalante]{%
  \label{fap: diam000036}
  Diameter estimates for (36) Atalante.
}
\end{figure}

  \begin{figure}[!ht]
  \centering
  \includegraphics[width=.49\textwidth]{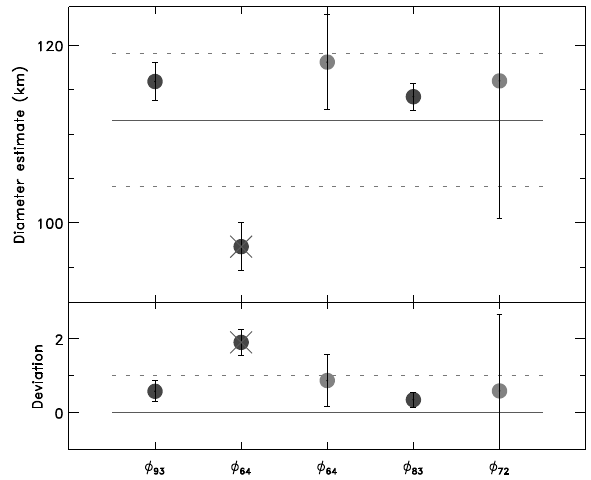}
\caption[Diameter estimates for (38) Leda]{%
  \label{fap: diam000038}
  Diameter estimates for (38) Leda.
}
\end{figure}

  \begin{figure}[!ht]
  \centering
  \includegraphics[width=.49\textwidth]{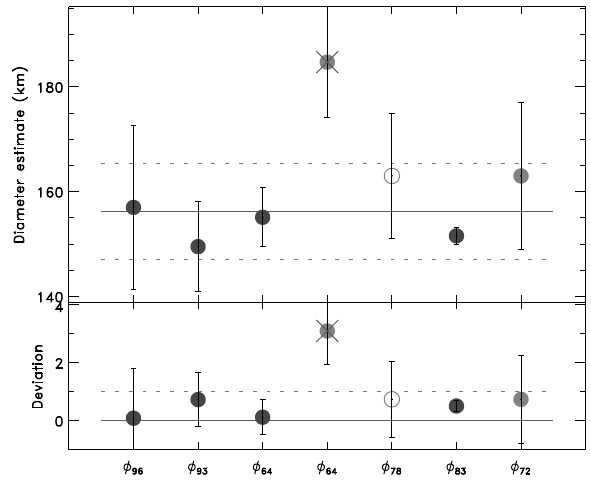}
\caption[Diameter estimates for (39) Laetitia]{%
  \label{fap: diam000039}
  Diameter estimates for (39) Laetitia.
}
\end{figure}

  \begin{figure}[!ht]
  \centering
  \includegraphics[width=.49\textwidth]{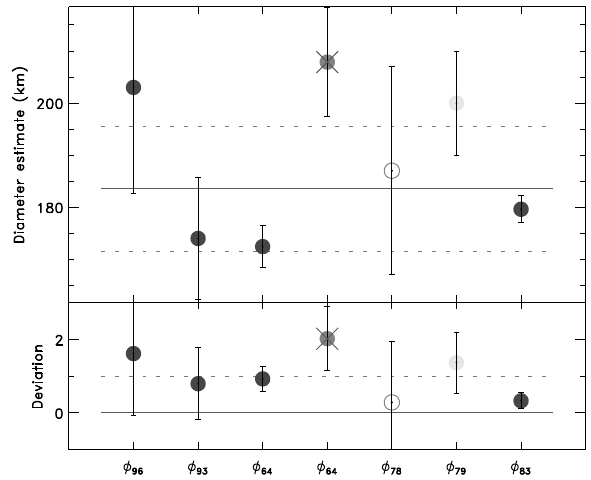}
\caption[Diameter estimates for (41) Daphne]{%
  \label{fap: diam000041}
  Diameter estimates for (41) Daphne.
}
\end{figure}

\clearpage
  \begin{figure}[!ht]
  \centering
  \includegraphics[width=.49\textwidth]{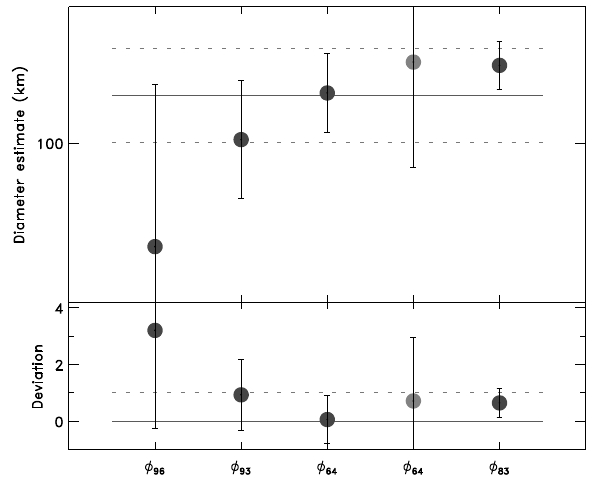}
\caption[Diameter estimates for (42) Isis]{%
  \label{fap: diam000042}
  Diameter estimates for (42) Isis.
}
\end{figure}

  \begin{figure}[!ht]
  \centering
  \includegraphics[width=.49\textwidth]{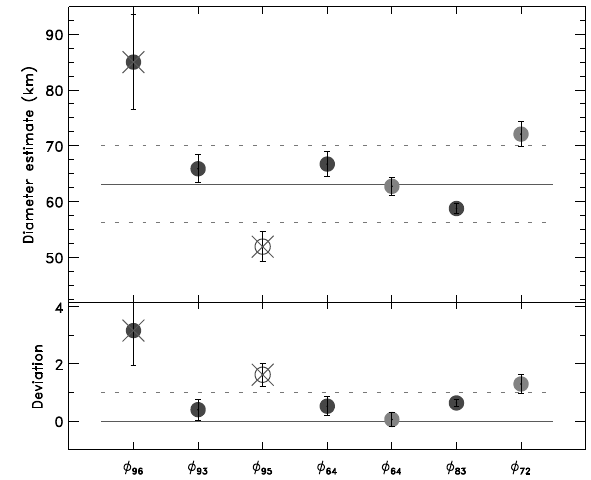}
\caption[Diameter estimates for (43) Ariadne]{%
  \label{fap: diam000043}
  Diameter estimates for (43) Ariadne.
}
\end{figure}

  \begin{figure}[!ht]
  \centering
  \includegraphics[width=.49\textwidth]{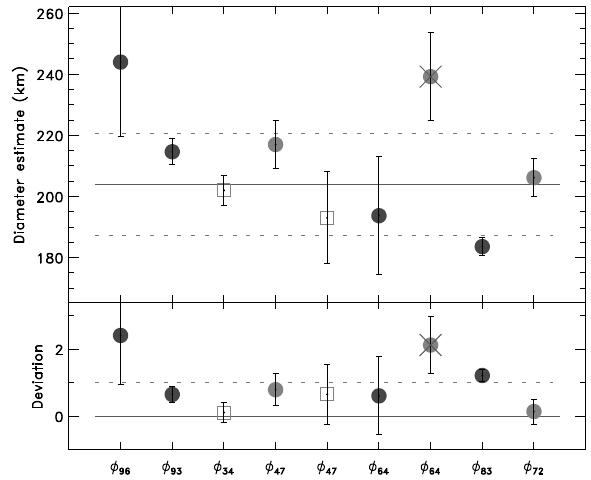}
\caption[Diameter estimates for (45) Eugenia]{%
  \label{fap: diam000045}
  Diameter estimates for (45) Eugenia.
}
\end{figure}

  \begin{figure}[!ht]
  \centering
  \includegraphics[width=.49\textwidth]{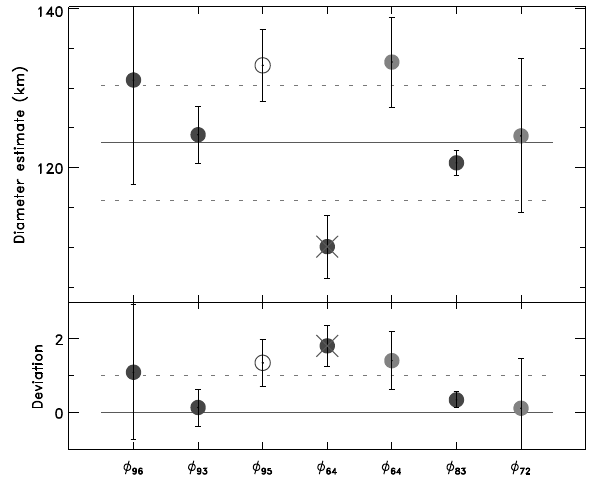}
\caption[Diameter estimates for (46) Hestia]{%
  \label{fap: diam000046}
  Diameter estimates for (46) Hestia.
}
\end{figure}

  \begin{figure}[!ht]
  \centering
  \includegraphics[width=.49\textwidth]{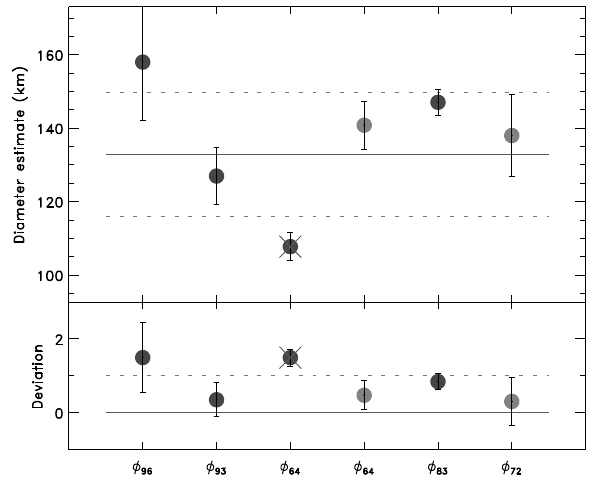}
\caption[Diameter estimates for (47) Aglaja]{%
  \label{fap: diam000047}
  Diameter estimates for (47) Aglaja.
}
\end{figure}

  \begin{figure}[!ht]
  \centering
  \includegraphics[width=.49\textwidth]{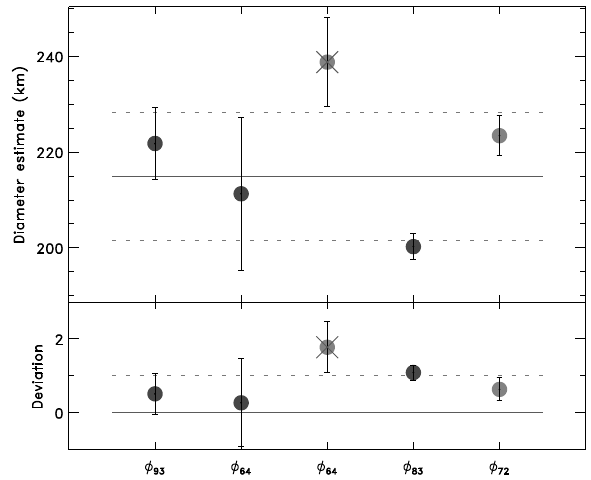}
\caption[Diameter estimates for (48) Doris]{%
  \label{fap: diam000048}
  Diameter estimates for (48) Doris.
}
\end{figure}

  \begin{figure}[!ht]
  \centering
  \includegraphics[width=.49\textwidth]{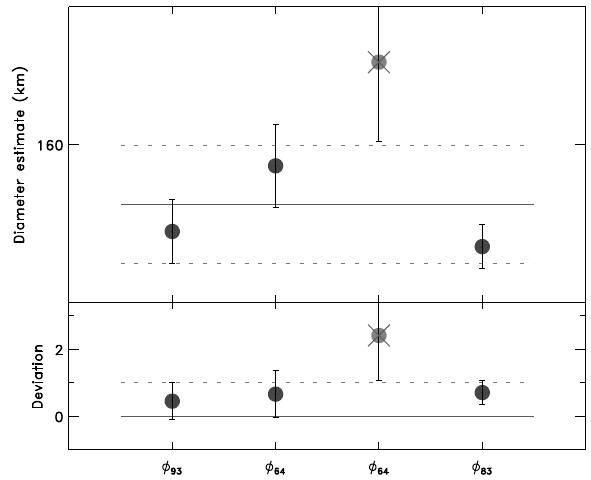}
\caption[Diameter estimates for (49) Pales]{%
  \label{fap: diam000049}
  Diameter estimates for (49) Pales.
}
\end{figure}

  \begin{figure}[!ht]
  \centering
  \includegraphics[width=.49\textwidth]{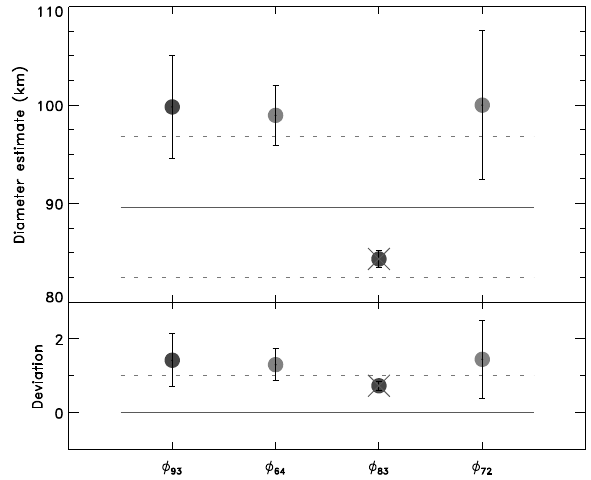}
\caption[Diameter estimates for (50) Virginia]{%
  \label{fap: diam000050}
  Diameter estimates for (50) Virginia.
  The diameter estimate from $\phi_{83}$ raises the density from 7.1\,$\pm$\,5.2 to 9.9\,$\pm$\,7.6 and is therefore discarded.
}
\end{figure}

  \begin{figure}[!ht]
  \centering
  \includegraphics[width=.49\textwidth]{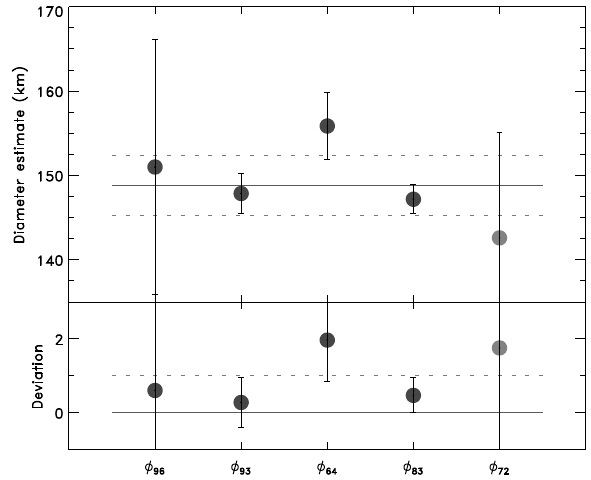}
\caption[Diameter estimates for (51) Nemausa]{%
  \label{fap: diam000051}
  Diameter estimates for (51) Nemausa.
}
\end{figure}

  \begin{figure}[!ht]
  \centering
  \includegraphics[width=.49\textwidth]{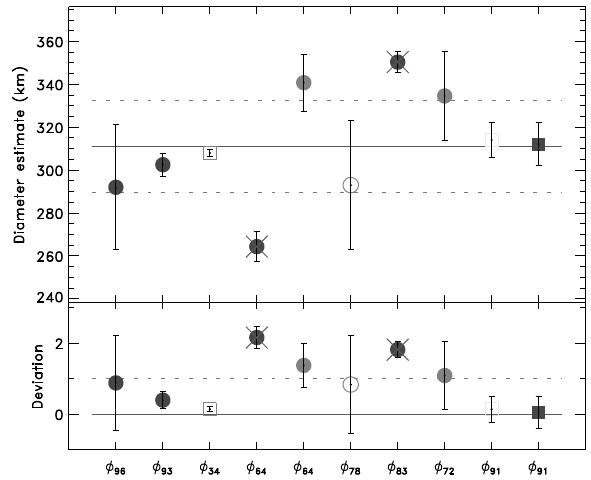}
\caption[Diameter estimates for (52) Europa]{%
  \label{fap: diam000052}
  Diameter estimates for (52) Europa.
}
\end{figure}

  \begin{figure}[!ht]
  \centering
  \includegraphics[width=.49\textwidth]{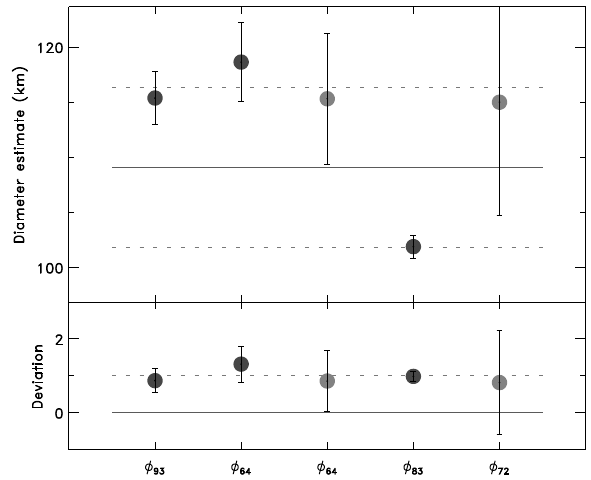}
\caption[Diameter estimates for (53) Kalypso]{%
  \label{fap: diam000053}
  Diameter estimates for (53) Kalypso.
  The large uncertainty on the mass determination forbid to  sort between the diameter estimates. Values from $\phi_{93}$, $\phi_{64}$, and $\phi_{64}$ were obtained using IRAS data \citep{2002-AJ-123-Tedesco-a, 2010-AJ-140-Ryan}, and $\phi_{83}$ with AKARI data \citep{2011-PASJ-63-Usui}. Discrepancy could likely result from different observing geometries (\eg, polar \textsl{vs} equatorial) not taken into account.
}
\end{figure}

  \begin{figure}[!ht]
  \centering
  \includegraphics[width=.49\textwidth]{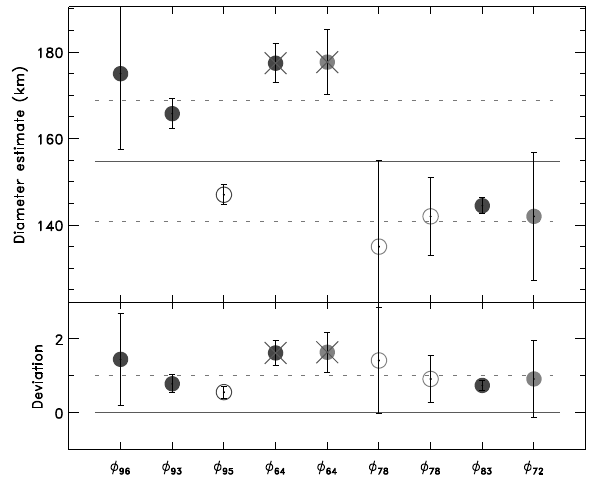}
\caption[Diameter estimates for (54) Alexandra]{%
  \label{fap: diam000054}
  Diameter estimates for (54) Alexandra.
}
\end{figure}

  \begin{figure}[!ht]
  \centering
  \includegraphics[width=.49\textwidth]{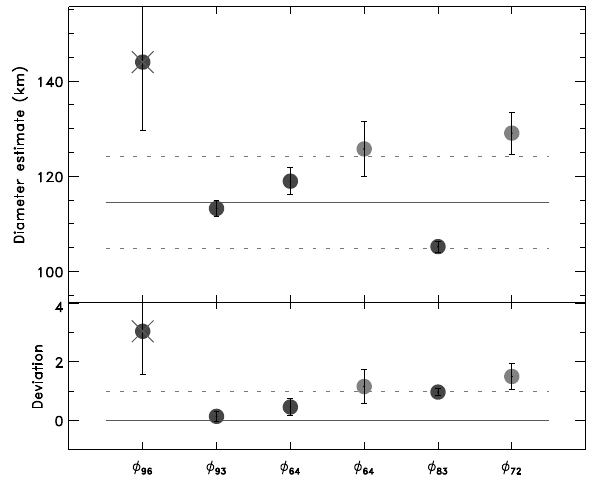}
\caption[Diameter estimates for (56) Melete]{%
  \label{fap: diam000056}
  Diameter estimates for (56) Melete.
}
\end{figure}

  \begin{figure}[!ht]
  \centering
  \includegraphics[width=.49\textwidth]{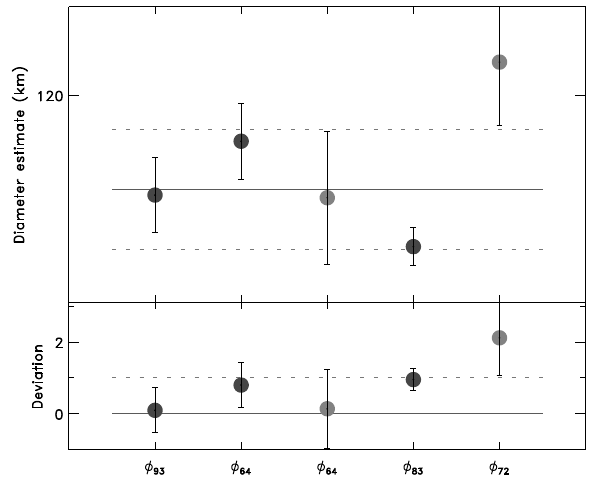}
\caption[Diameter estimates for (57) Mnemosyne]{%
  \label{fap: diam000057}
  Diameter estimates for (57) Mnemosyne.
}
\end{figure}

  \begin{figure}[!ht]
  \centering
  \includegraphics[width=.49\textwidth]{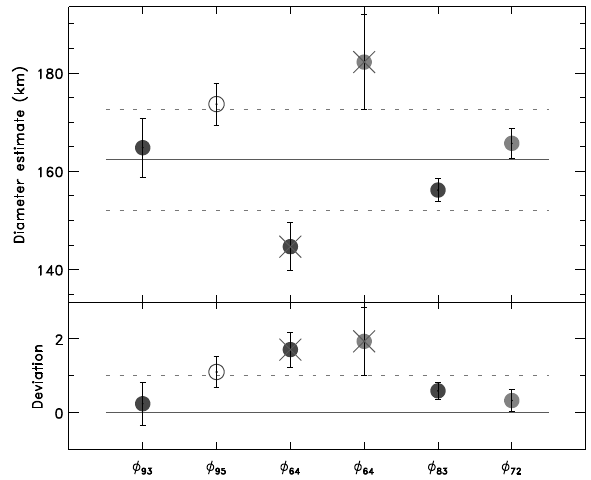}
\caption[Diameter estimates for (59) Elpis]{%
  \label{fap: diam000059}
  Diameter estimates for (59) Elpis.
}
\end{figure}

  \begin{figure}[!ht]
  \centering
  \includegraphics[width=.49\textwidth]{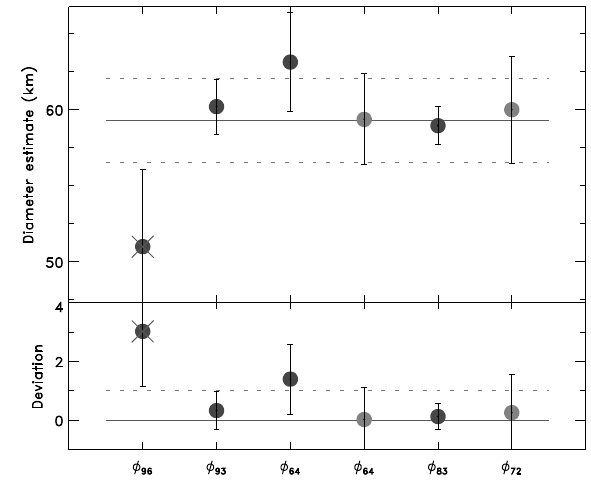}
\caption[Diameter estimates for (60) Echo]{%
  \label{fap: diam000060}
  Diameter estimates for (60) Echo.
}
\end{figure}

  \begin{figure}[!ht]
  \centering
  \includegraphics[width=.49\textwidth]{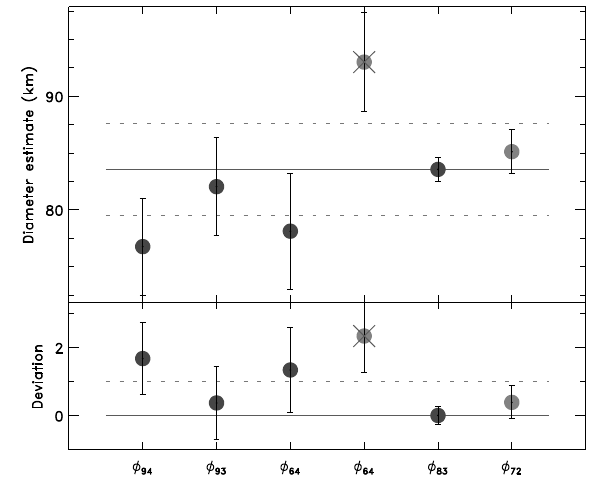}
\caption[Diameter estimates for (61) Danae]{%
  \label{fap: diam000061}
  Diameter estimates for (61) Danae.
}
\end{figure}

  \begin{figure}[!ht]
  \centering
  \includegraphics[width=.49\textwidth]{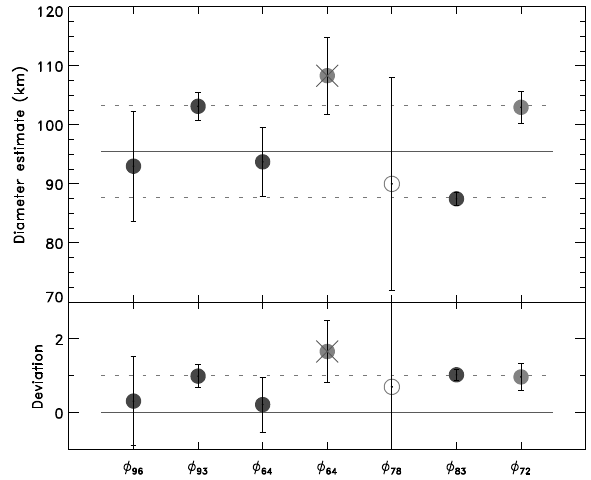}
\caption[Diameter estimates for (63) Ausonia]{%
  \label{fap: diam000063}
  Diameter estimates for (63) Ausonia.
}
\end{figure}

\clearpage
  \begin{figure}[!ht]
  \centering
  \includegraphics[width=.49\textwidth]{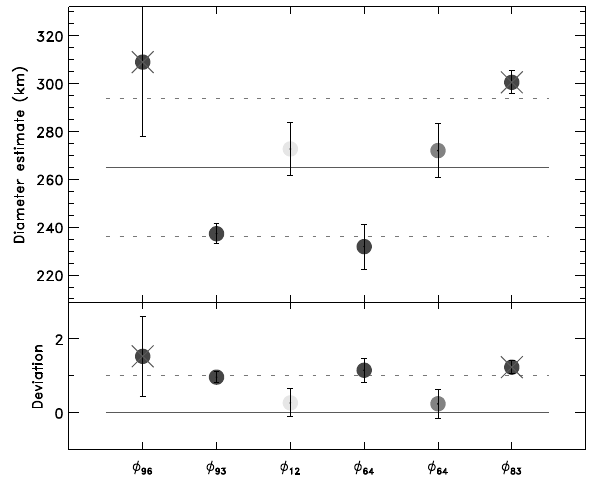}
\caption[Diameter estimates for (65) Cybele]{%
  \label{fap: diam000065}
  Diameter estimates for (65) Cybele.
}
\end{figure}

  \begin{figure}[!ht]
  \centering
  \includegraphics[width=.49\textwidth]{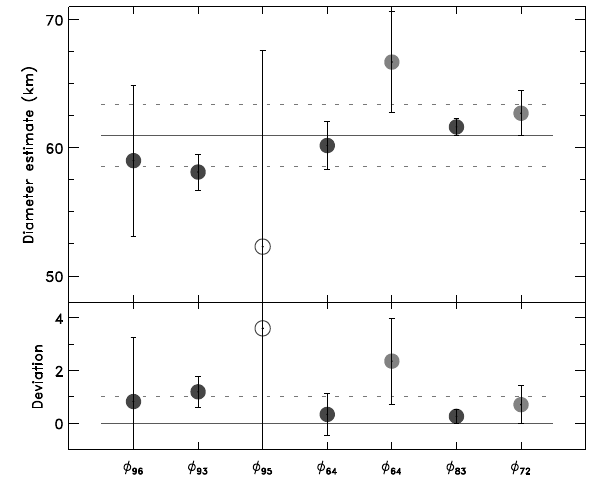}
\caption[Diameter estimates for (67) Asia]{%
  \label{fap: diam000067}
  Diameter estimates for (67) Asia.
}
\end{figure}

  \begin{figure}[!ht]
  \centering
  \includegraphics[width=.49\textwidth]{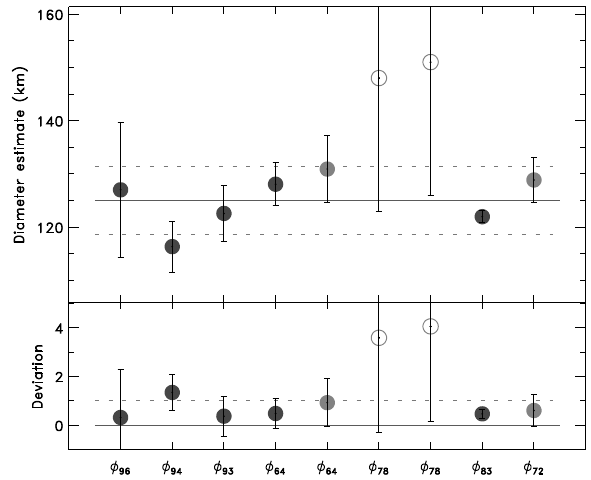}
\caption[Diameter estimates for (68) Leto]{%
  \label{fap: diam000068}
  Diameter estimates for (68) Leto.
}
\end{figure}

  \begin{figure}[!ht]
  \centering
  \includegraphics[width=.49\textwidth]{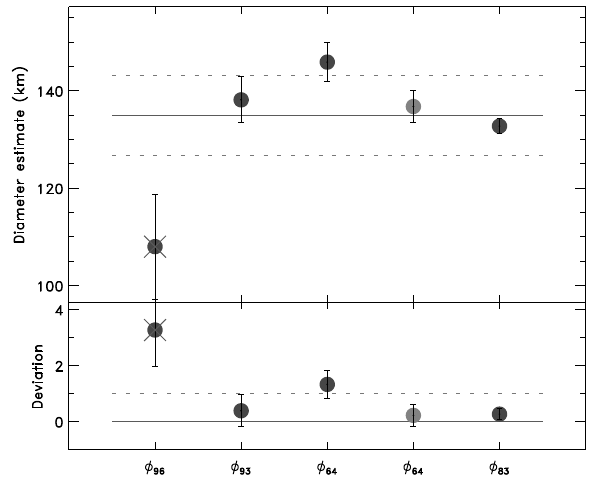}
\caption[Diameter estimates for (69) Hesperia]{%
  \label{fap: diam000069}
  Diameter estimates for (69) Hesperia.
}
\end{figure}

  \begin{figure}[!ht]
  \centering
  \includegraphics[width=.49\textwidth]{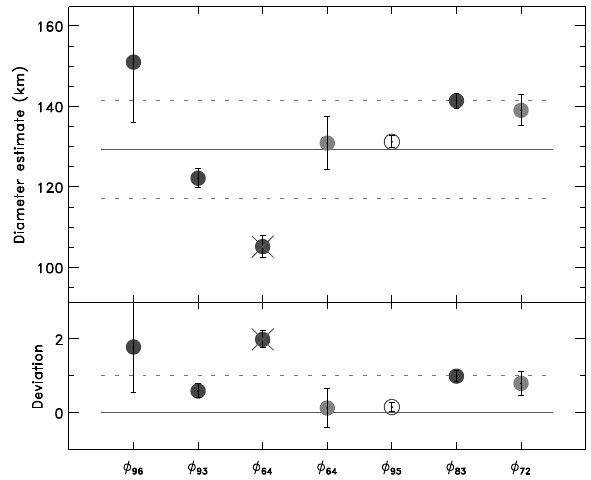}
\caption[Diameter estimates for (70) Panopaea]{%
  \label{fap: diam000070}
  Diameter estimates for (70) Panopaea.
}
\end{figure}

  \begin{figure}[!ht]
  \centering
  \includegraphics[width=.49\textwidth]{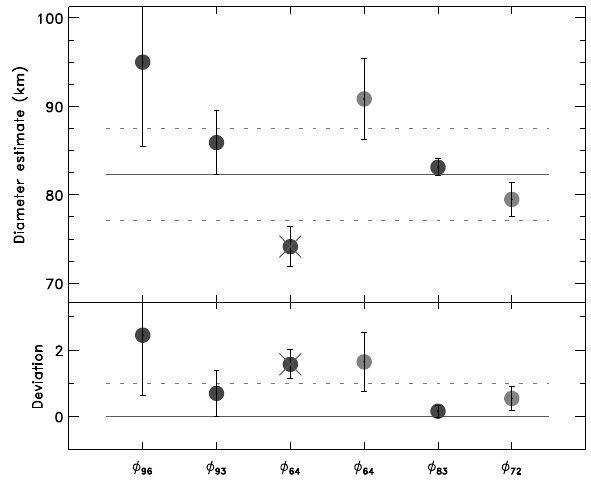}
\caption[Diameter estimates for (72) Feronia]{%
  \label{fap: diam000072}
  Diameter estimates for (72) Feronia.
}
\end{figure}

  \begin{figure}[!ht]
  \centering
  \includegraphics[width=.49\textwidth]{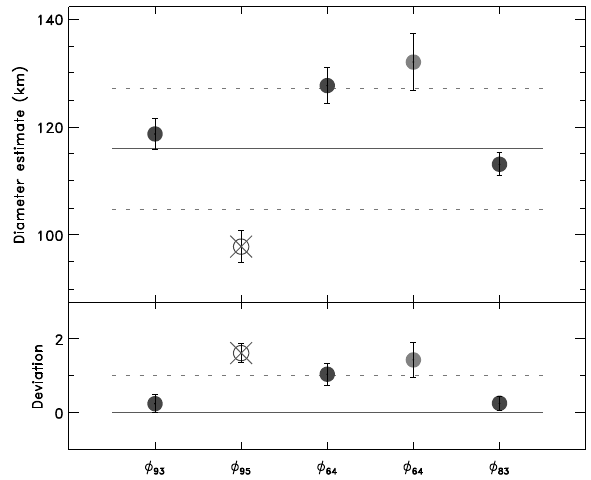}
\caption[Diameter estimates for (74) Galatea]{%
  \label{fap: diam000074}
  Diameter estimates for (74) Galatea.
}
\end{figure}

  \begin{figure}[!ht]
  \centering
  \includegraphics[width=.49\textwidth]{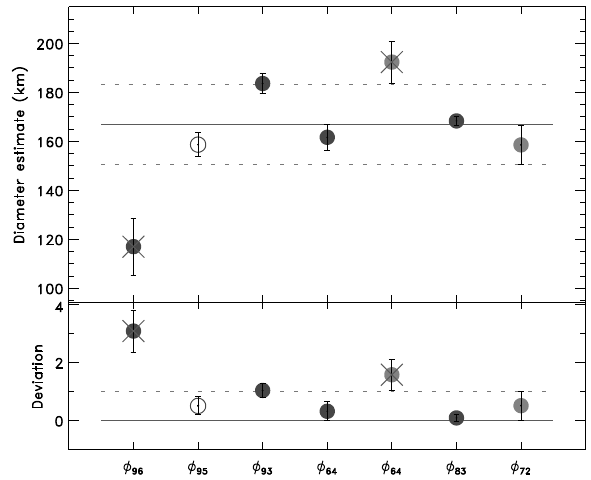}
\caption[Diameter estimates for (76) Freia]{%
  \label{fap: diam000076}
  Diameter estimates for (76) Freia.
}
\end{figure}

  \begin{figure}[!ht]
  \centering
  \includegraphics[width=.49\textwidth]{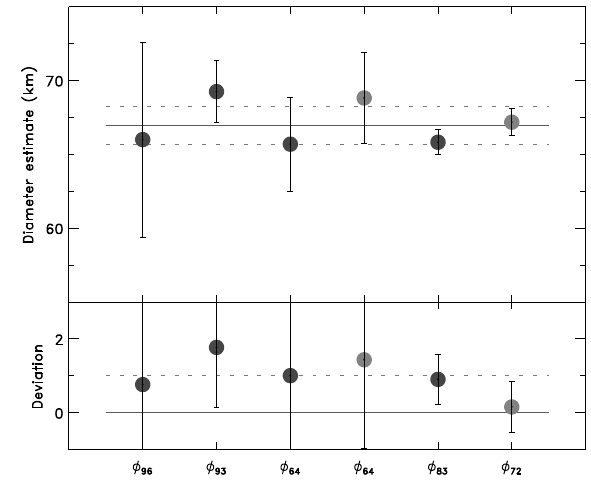}
\caption[Diameter estimates for (77) Frigga]{%
  \label{fap: diam000077}
  Diameter estimates for (77) Frigga.
}
\end{figure}

  \begin{figure}[!ht]
  \centering
  \includegraphics[width=.49\textwidth]{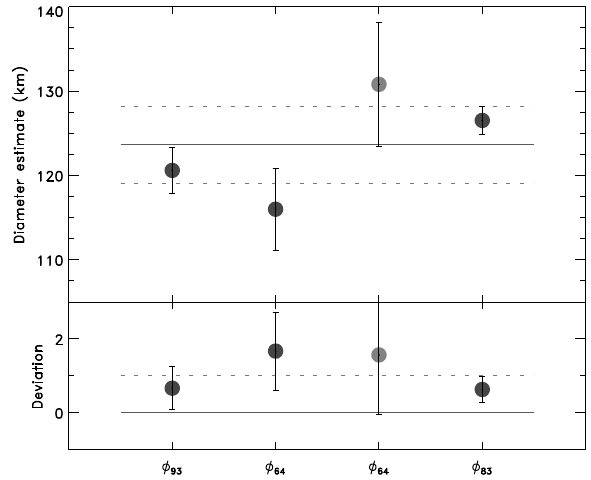}
\caption[Diameter estimates for (78) Diana]{%
  \label{fap: diam000078}
  Diameter estimates for (78) Diana.
}
\end{figure}

  \begin{figure}[!ht]
  \centering
  \includegraphics[width=.49\textwidth]{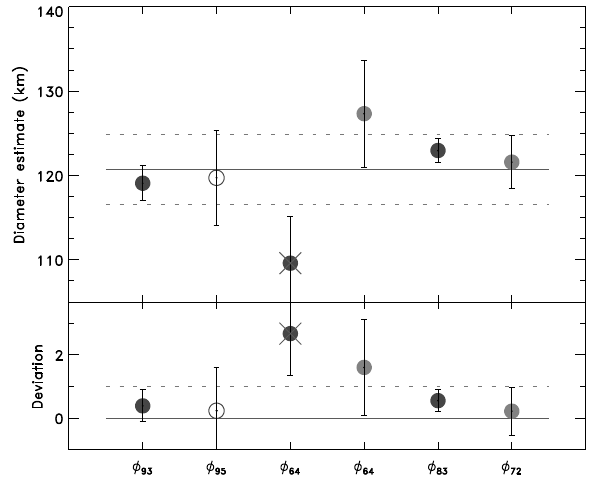}
\caption[Diameter estimates for (81) Terpsichore]{%
  \label{fap: diam000081}
  Diameter estimates for (81) Terpsichore.
}
\end{figure}

  \begin{figure}[!ht]
  \centering
  \includegraphics[width=.49\textwidth]{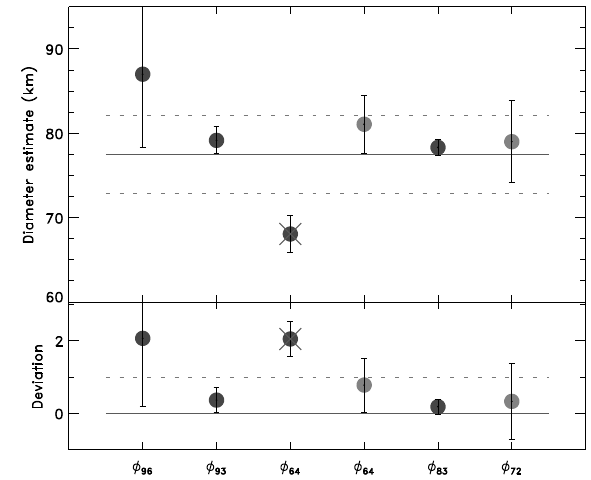}
\caption[Diameter estimates for (84) Klio]{%
  \label{fap: diam000084}
  Diameter estimates for (84) Klio.
}
\end{figure}

  \begin{figure}[!ht]
  \centering
  \includegraphics[width=.49\textwidth]{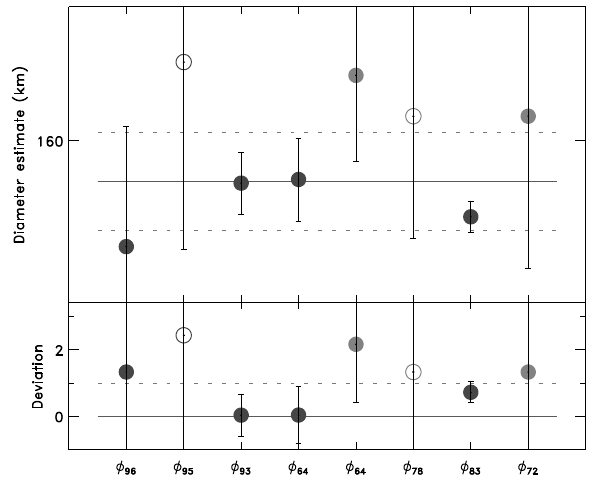}
\caption[Diameter estimates for (85) Io]{%
  \label{fap: diam000085}
  Diameter estimates for (85) Io.
}
\end{figure}

  \begin{figure}[!ht]
  \centering
  \includegraphics[width=.49\textwidth]{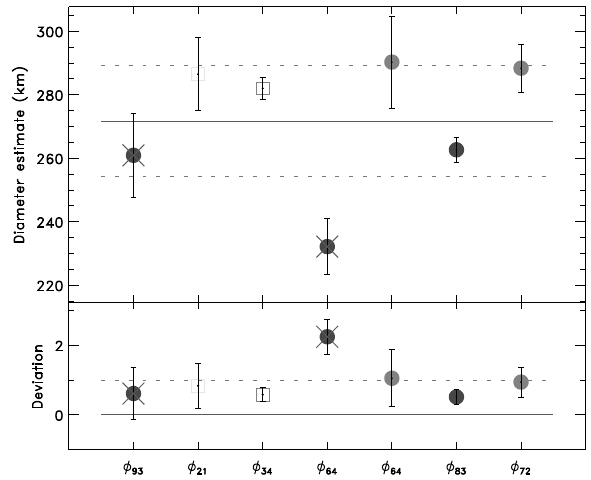}
\caption[Diameter estimates for (87) Sylvia]{%
  \label{fap: diam000087}
  Diameter estimates for (87) Sylvia.
}
\end{figure}

  \begin{figure}[!ht]
  \centering
  \includegraphics[width=.49\textwidth]{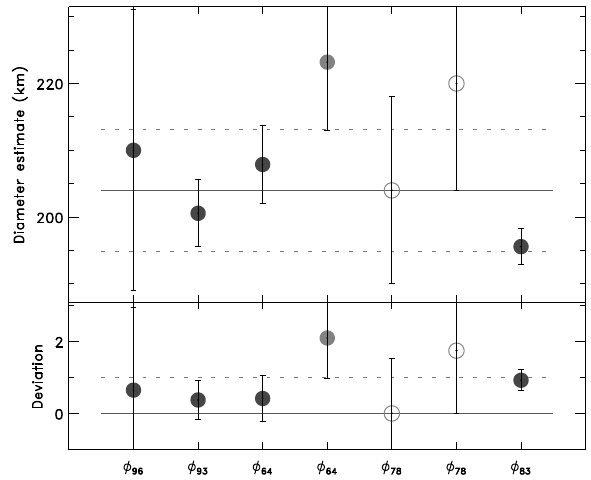}
\caption[Diameter estimates for (88) Thisbe]{%
  \label{fap: diam000088}
  Diameter estimates for (88) Thisbe.
}
\end{figure}

  \begin{figure}[!ht]
  \centering
  \includegraphics[width=.49\textwidth]{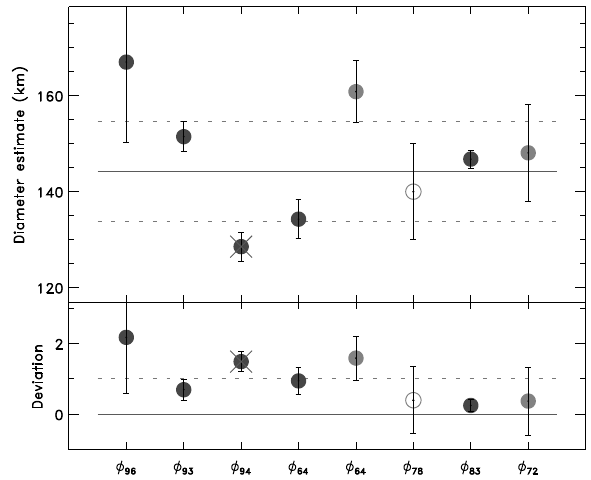}
\caption[Diameter estimates for (89) Julia]{%
  \label{fap: diam000089}
  Diameter estimates for (89) Julia.
}
\end{figure}

  \begin{figure}[!ht]
  \centering
  \includegraphics[width=.49\textwidth]{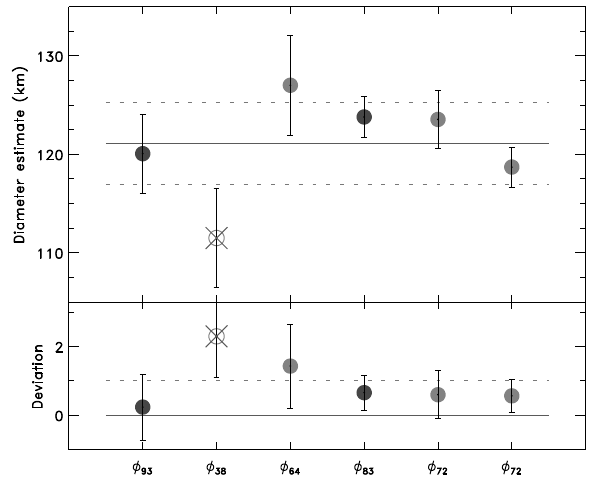}
\caption[Diameter estimates for (90) Antiope]{%
  \label{fap: diam000090}
  Diameter estimates for (90) Antiope.
}
\end{figure}

  \begin{figure}[!ht]
  \centering
  \includegraphics[width=.49\textwidth]{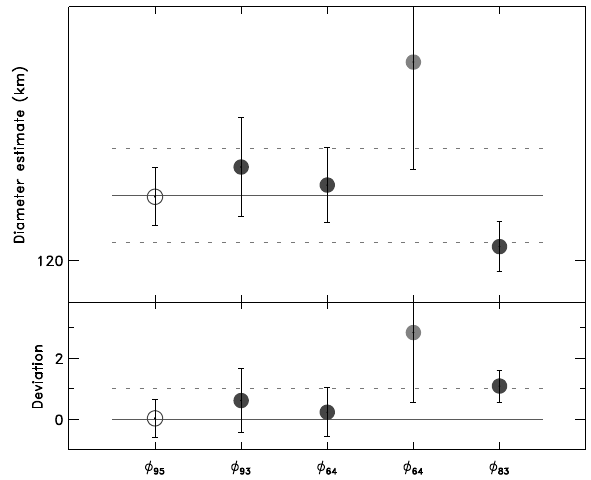}
\caption[Diameter estimates for (92) Undina]{%
  \label{fap: diam000092}
  Diameter estimates for (92) Undina.
}
\end{figure}

\clearpage
  \begin{figure}[!ht]
  \centering
  \includegraphics[width=.49\textwidth]{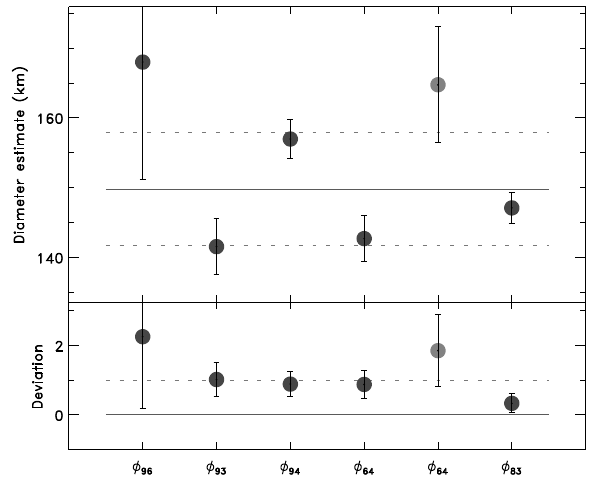}
\caption[Diameter estimates for (93) Minerva]{%
  \label{fap: diam000093}
  Diameter estimates for (93) Minerva.
}
\end{figure}

  \begin{figure}[!ht]
  \centering
  \includegraphics[width=.49\textwidth]{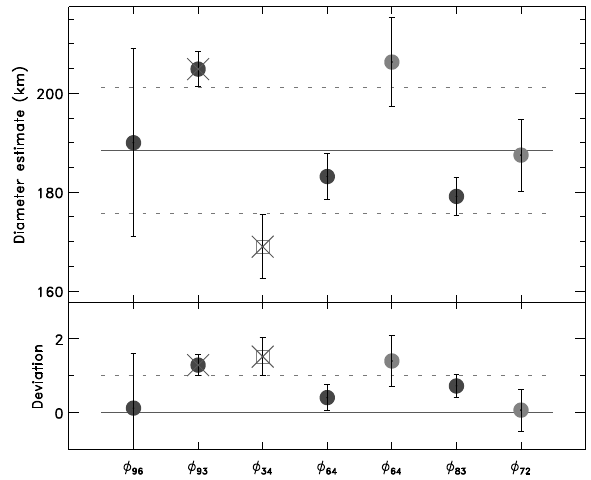}
\caption[Diameter estimates for (94) Aurora]{%
  \label{fap: diam000094}
  Diameter estimates for (94) Aurora.
}
\end{figure}

  \begin{figure}[!ht]
  \centering
  \includegraphics[width=.49\textwidth]{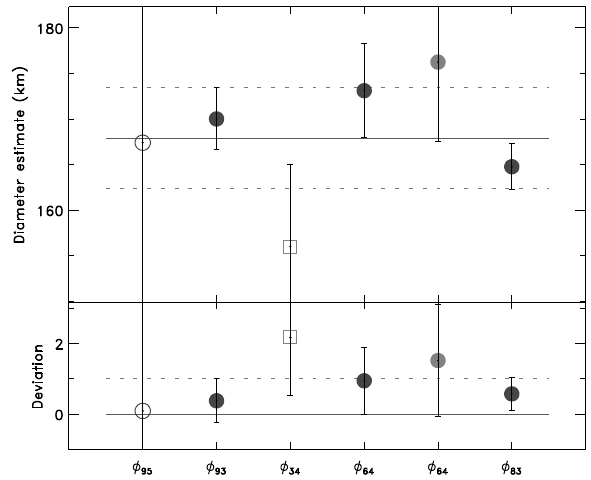}
\caption[Diameter estimates for (96) Aegle]{%
  \label{fap: diam000096}
  Diameter estimates for (96) Aegle.
}
\end{figure}

  \begin{figure}[!ht]
  \centering
  \includegraphics[width=.49\textwidth]{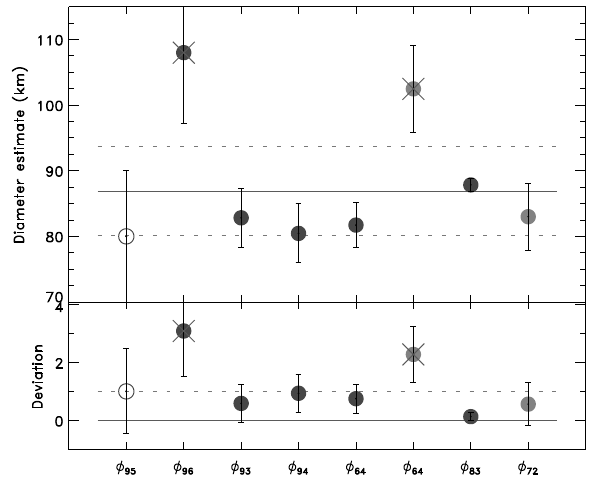}
\caption[Diameter estimates for (97) Klotho]{%
  \label{fap: diam000097}
  Diameter estimates for (97) Klotho.
}
\end{figure}

  \begin{figure}[!ht]
  \centering
  \includegraphics[width=.49\textwidth]{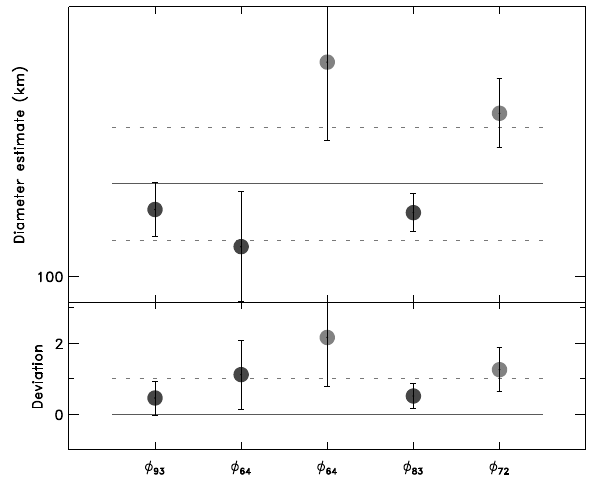}
\caption[Diameter estimates for (98) Ianthe]{%
  \label{fap: diam000098}
  Diameter estimates for (98) Ianthe.
}
\end{figure}

  \begin{figure}[!ht]
  \centering
  \includegraphics[width=.49\textwidth]{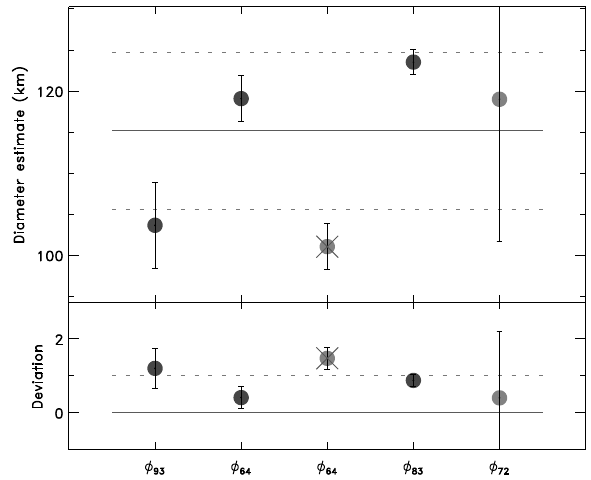}
\caption[Diameter estimates for (105) Artemis]{%
  \label{fap: diam000105}
  Diameter estimates for (105) Artemis.
}
\end{figure}

  \begin{figure}[!ht]
  \centering
  \includegraphics[width=.49\textwidth]{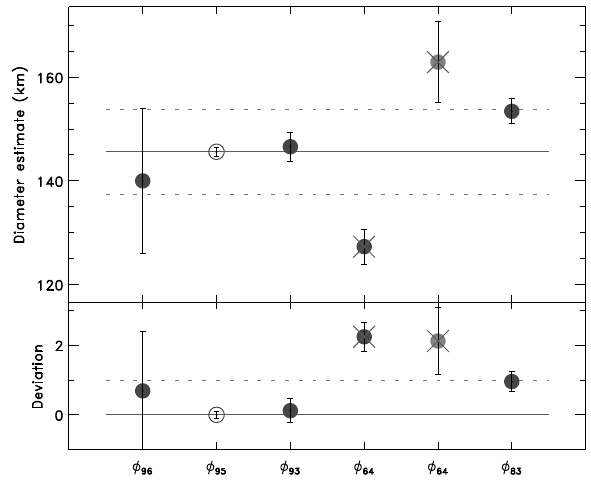}
\caption[Diameter estimates for (106) Dione]{%
  \label{fap: diam000106}
  Diameter estimates for (106) Dione.
}
\end{figure}

  \begin{figure}[!ht]
  \centering
  \includegraphics[width=.49\textwidth]{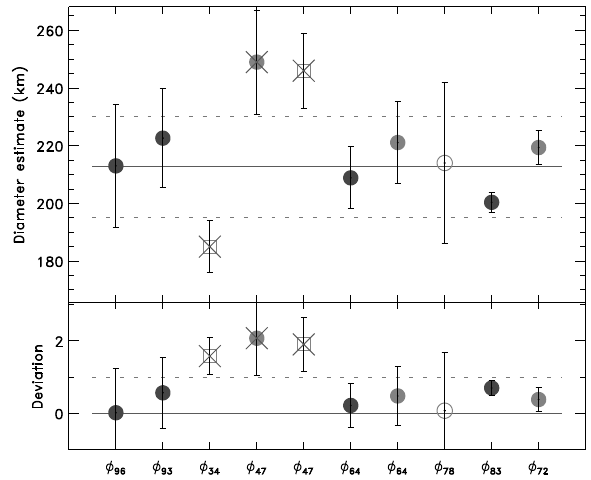}
\caption[Diameter estimates for (107) Camilla]{%
  \label{fap: diam000107}
  Diameter estimates for (107) Camilla.
}
\end{figure}

  \begin{figure}[!ht]
  \centering
  \includegraphics[width=.49\textwidth]{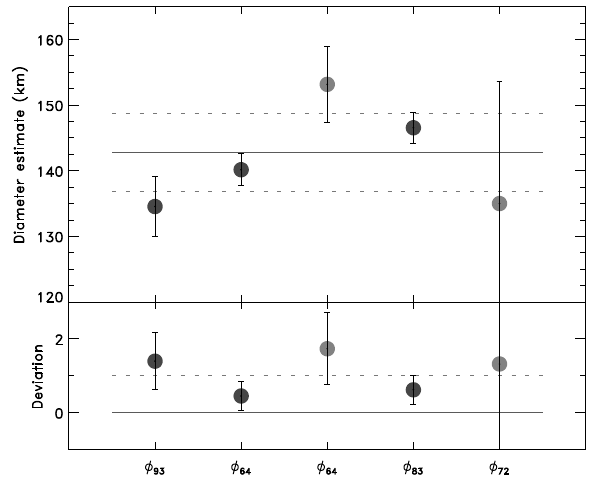}
\caption[Diameter estimates for (111) Ate]{%
  \label{fap: diam000111}
  Diameter estimates for (111) Ate.
}
\end{figure}

  \begin{figure}[!ht]
  \centering
  \includegraphics[width=.49\textwidth]{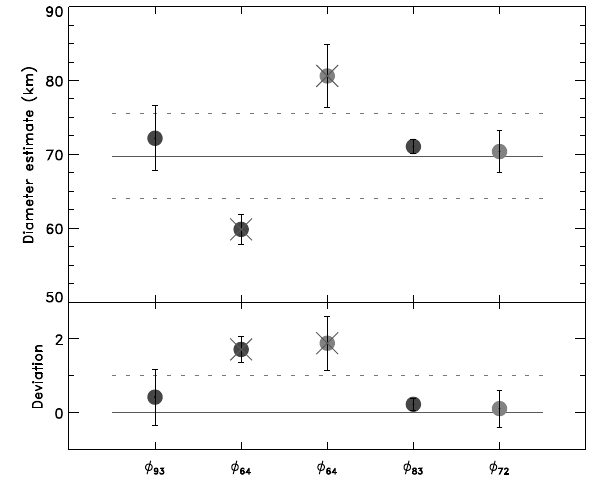}
\caption[Diameter estimates for (112) Iphigenia]{%
  \label{fap: diam000112}
  Diameter estimates for (112) Iphigenia.
}
\end{figure}

  \begin{figure}[!ht]
  \centering
  \includegraphics[width=.49\textwidth]{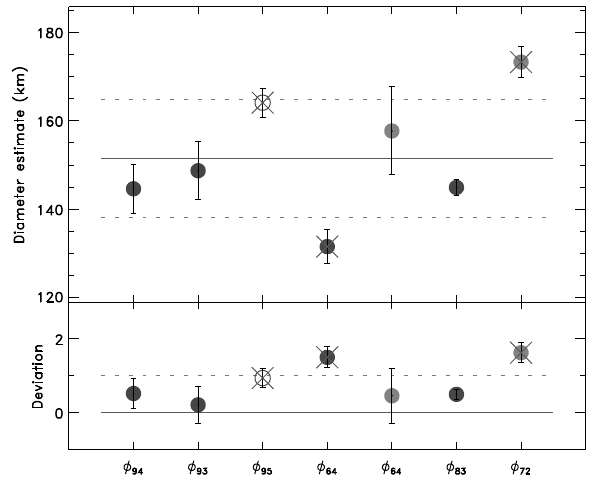}
\caption[Diameter estimates for (117) Lomia]{%
  \label{fap: diam000117}
  Diameter estimates for (117) Lomia.
}
\end{figure}

  \begin{figure}[!ht]
  \centering
  \includegraphics[width=.49\textwidth]{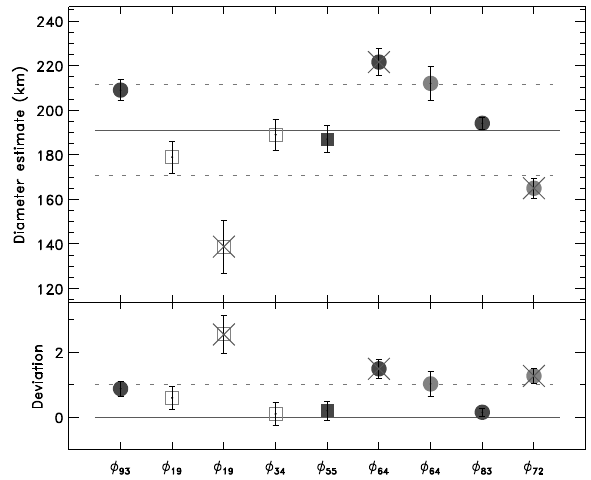}
\caption[Diameter estimates for (121) Hermione]{%
  \label{fap: diam000121}
  Diameter estimates for (121) Hermione.
}
\end{figure}

  \begin{figure}[!ht]
  \centering
  \includegraphics[width=.49\textwidth]{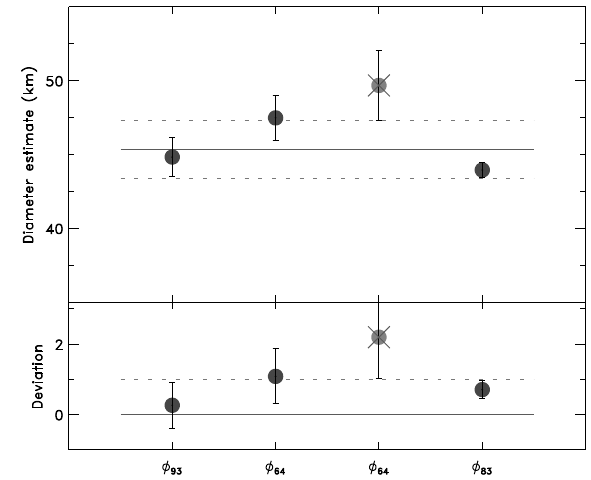}
\caption[Diameter estimates for (126) Velleda]{%
  \label{fap: diam000126}
  Diameter estimates for (126) Velleda.
}
\end{figure}

  \begin{figure}[!ht]
  \centering
  \includegraphics[width=.49\textwidth]{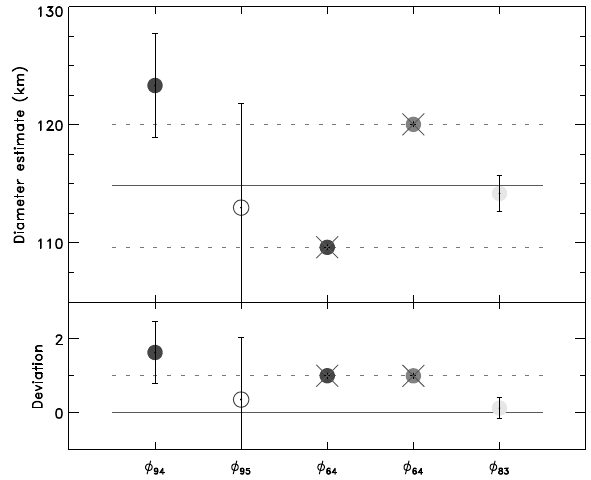}
\caption[Diameter estimates for (127) Johanna]{%
  \label{fap: diam000127}
  Diameter estimates for (127) Johanna.
  The diameter estimates from $\phi_{64}$ have unrealistic small uncertainties of 0.02\,km.  Using these values strongly biases the average. 
}
\end{figure}

  \begin{figure}[!ht]
  \centering
  \includegraphics[width=.49\textwidth]{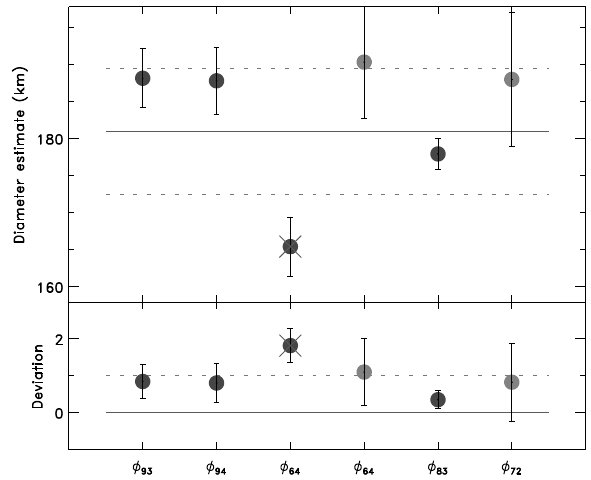}
\caption[Diameter estimates for (128) Nemesis]{%
  \label{fap: diam000128}
  Diameter estimates for (128) Nemesis.
}
\end{figure}

  \begin{figure}[!ht]
  \centering
  \includegraphics[width=.49\textwidth]{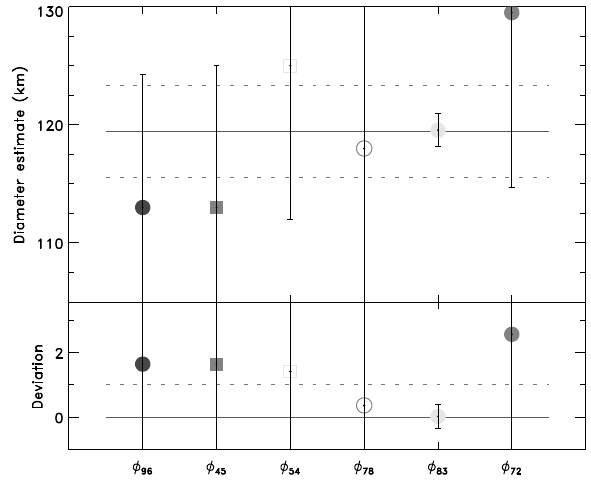}
\caption[Diameter estimates for (129) Antigone]{%
  \label{fap: diam000129}
  Diameter estimates for (129) Antigone.
}
\end{figure}

  \begin{figure}[!ht]
  \centering
  \includegraphics[width=.49\textwidth]{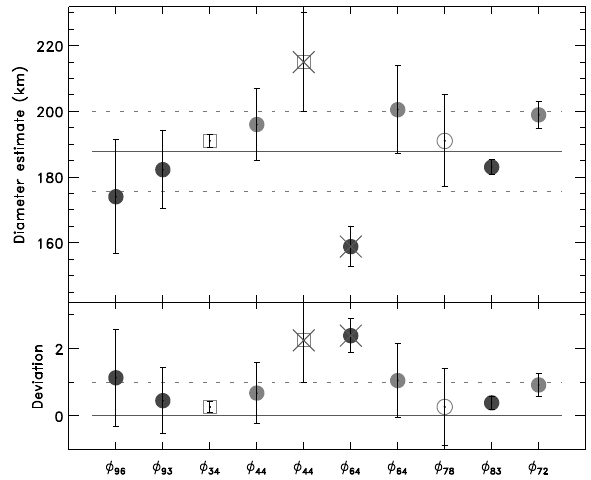}
\caption[Diameter estimates for (130) Elektra]{%
  \label{fap: diam000130}
  Diameter estimates for (130) Elektra.
}
\end{figure}

  \begin{figure}[!ht]
  \centering
  \includegraphics[width=.49\textwidth]{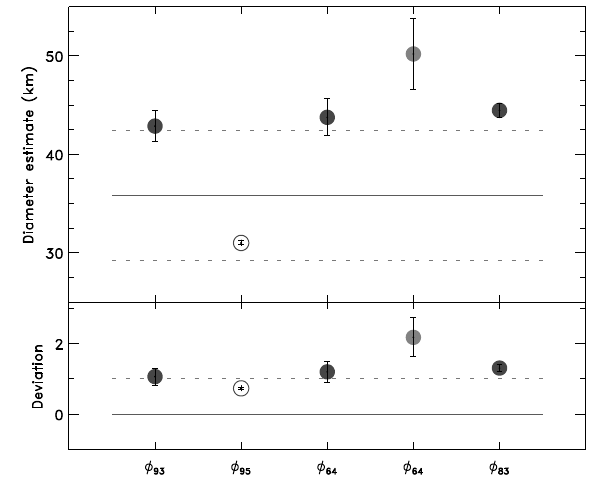}
\caption[Diameter estimates for (132) Aethra]{%
  \label{fap: diam000132}
  Diameter estimates for (132) Aethra.
  The large uncertainty on the mass determination forbid to  sort between the diameter estimates. 
}
\end{figure}

\clearpage
  \begin{figure}[!ht]
  \centering
  \includegraphics[width=.49\textwidth]{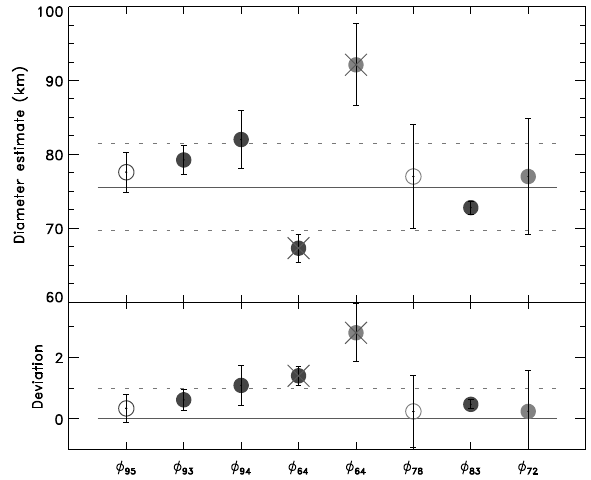}
\caption[Diameter estimates for (135) Hertha]{%
  \label{fap: diam000135}
  Diameter estimates for (135) Hertha.
}
\end{figure}

  \begin{figure}[!ht]
  \centering
  \includegraphics[width=.49\textwidth]{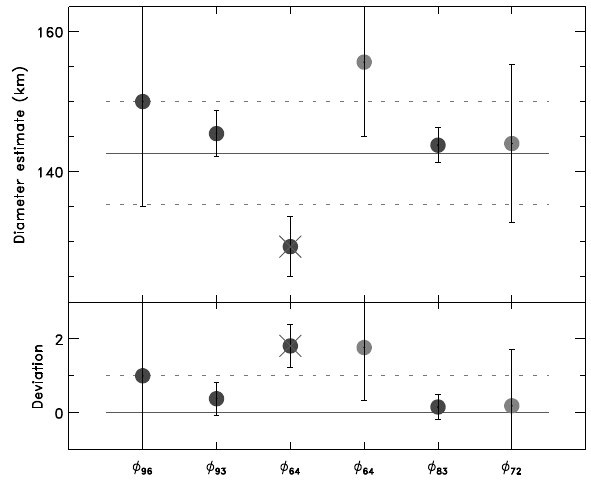}
\caption[Diameter estimates for (137) Meliboea]{%
  \label{fap: diam000137}
  Diameter estimates for (137) Meliboea.
}
\end{figure}

  \begin{figure}[!ht]
  \centering
  \includegraphics[width=.49\textwidth]{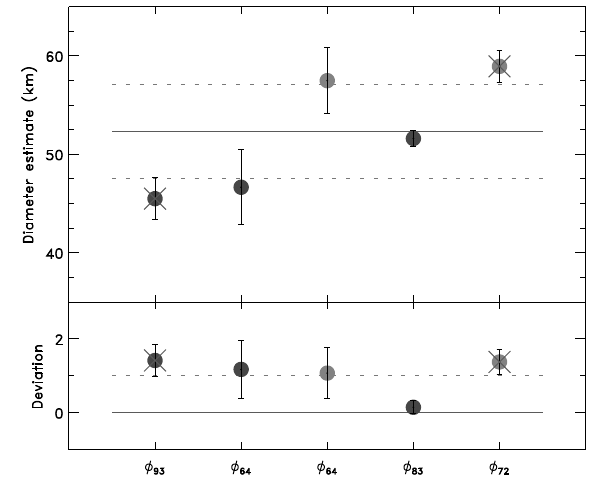}
\caption[Diameter estimates for (138) Tolosa]{%
  \label{fap: diam000138}
  Diameter estimates for (138) Tolosa.
}
\end{figure}

  \begin{figure}[!ht]
  \centering
  \includegraphics[width=.49\textwidth]{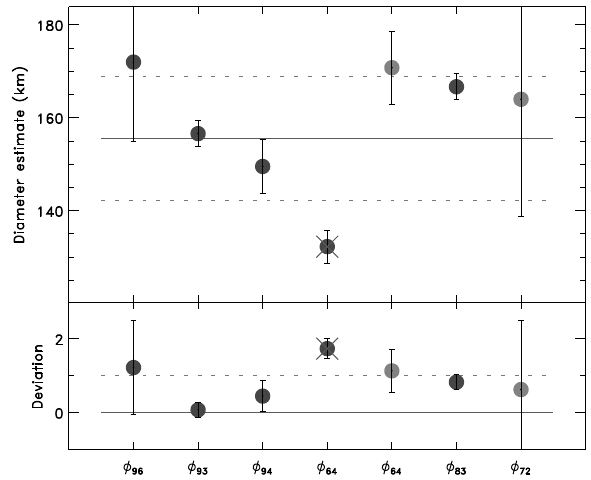}
\caption[Diameter estimates for (139) Juewa]{%
  \label{fap: diam000139}
  Diameter estimates for (139) Juewa.
}
\end{figure}

  \begin{figure}[!ht]
  \centering
  \includegraphics[width=.49\textwidth]{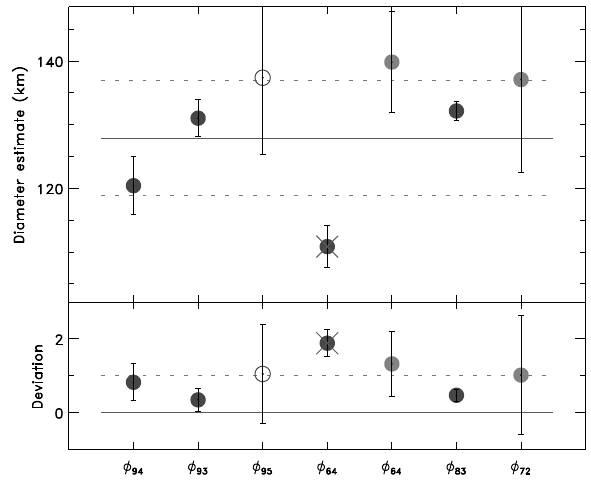}
\caption[Diameter estimates for (141) Lumen]{%
  \label{fap: diam000141}
  Diameter estimates for (141) Lumen.
}
\end{figure}

  \begin{figure}[!ht]
  \centering
  \includegraphics[width=.49\textwidth]{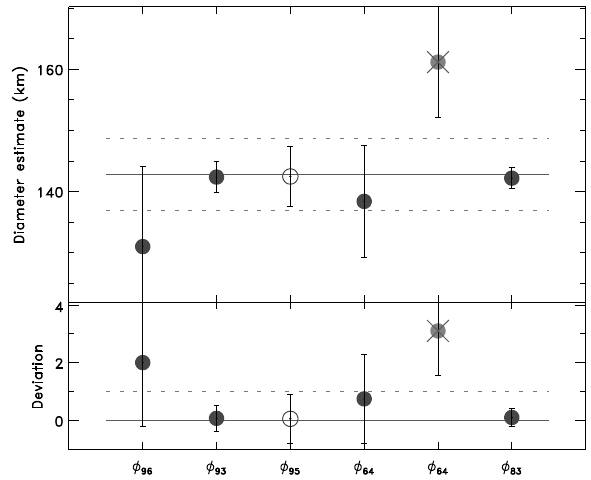}
\caption[Diameter estimates for (144) Vibilia]{%
  \label{fap: diam000144}
  Diameter estimates for (144) Vibilia.
}
\end{figure}

  \begin{figure}[!ht]
  \centering
  \includegraphics[width=.49\textwidth]{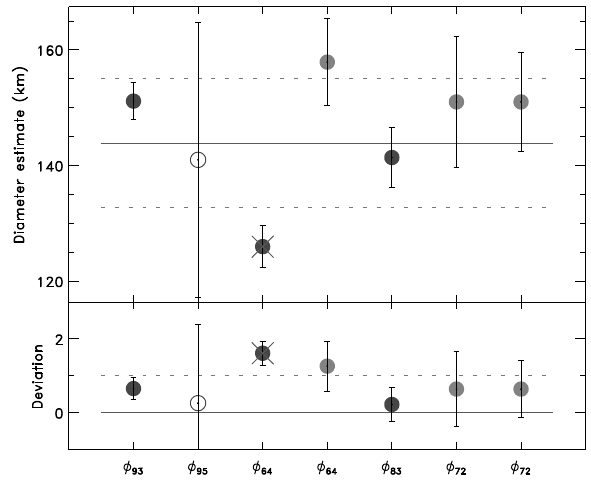}
\caption[Diameter estimates for (145) Adeona]{%
  \label{fap: diam000145}
  Diameter estimates for (145) Adeona.
}
\end{figure}

  \begin{figure}[!ht]
  \centering
  \includegraphics[width=.49\textwidth]{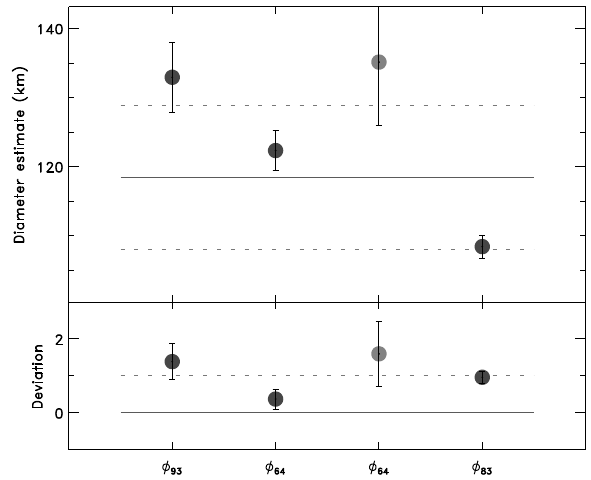}
\caption[Diameter estimates for (147) Protogeneia]{%
  \label{fap: diam000147}
  Diameter estimates for (147) Protogeneia.
}
\end{figure}

  \begin{figure}[!ht]
  \centering
  \includegraphics[width=.49\textwidth]{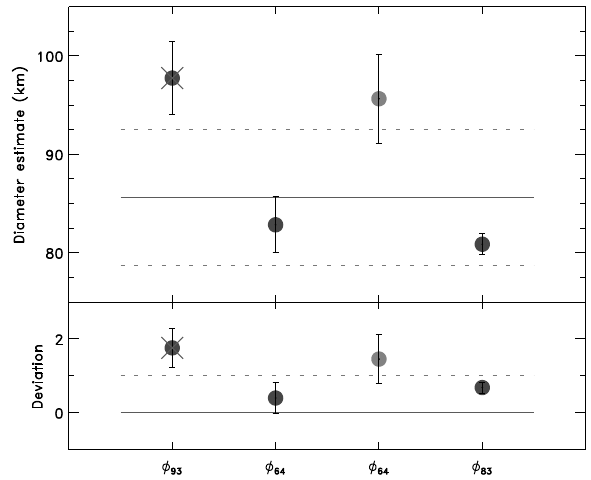}
\caption[Diameter estimates for (148) Gallia]{%
  \label{fap: diam000148}
  Diameter estimates for (148) Gallia.
}
\end{figure}

  \begin{figure}[!ht]
  \centering
  \includegraphics[width=.49\textwidth]{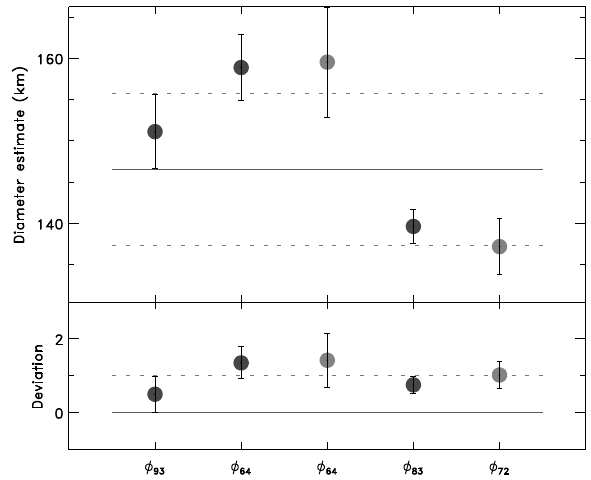}
\caption[Diameter estimates for (150) Nuwa]{%
  \label{fap: diam000150}
  Diameter estimates for (150) Nuwa.
}
\end{figure}

  \begin{figure}[!ht]
  \centering
  \includegraphics[width=.49\textwidth]{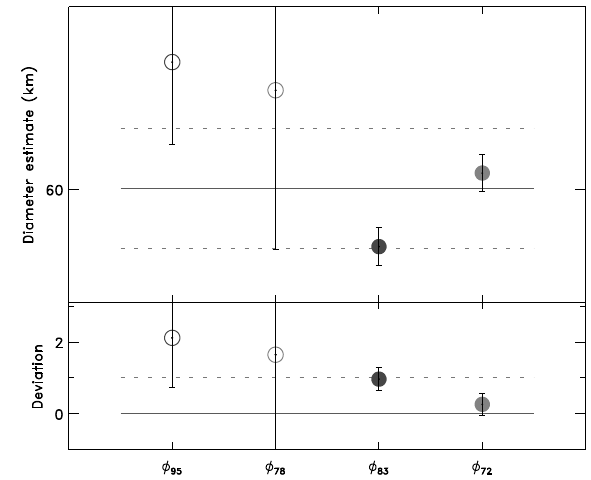}
\caption[Diameter estimates for (152) Atala]{%
  \label{fap: diam000152}
  Diameter estimates for (152) Atala.
}
\end{figure}

  \begin{figure}[!ht]
  \centering
  \includegraphics[width=.49\textwidth]{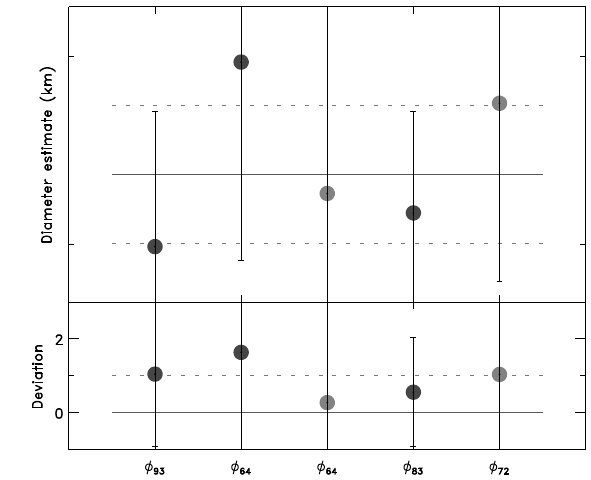}
\caption[Diameter estimates for (154) Bertha]{%
  \label{fap: diam000154}
  Diameter estimates for (154) Bertha.
}
\end{figure}

  \begin{figure}[!ht]
  \centering
  \includegraphics[width=.49\textwidth]{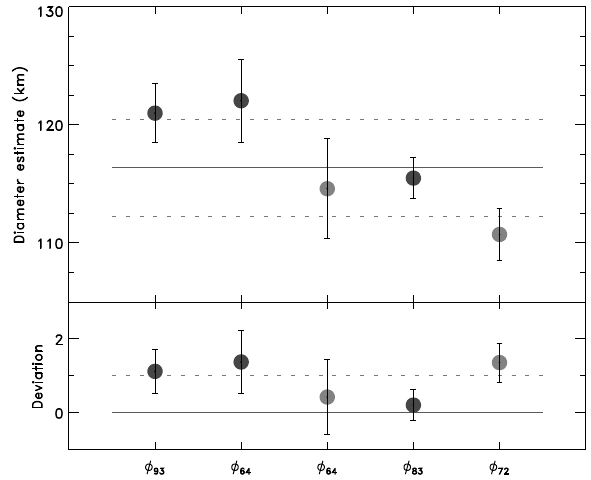}
\caption[Diameter estimates for (156) Xanthippe]{%
  \label{fap: diam000156}
  Diameter estimates for (156) Xanthippe.
}
\end{figure}

  \begin{figure}[!ht]
  \centering
  \includegraphics[width=.49\textwidth]{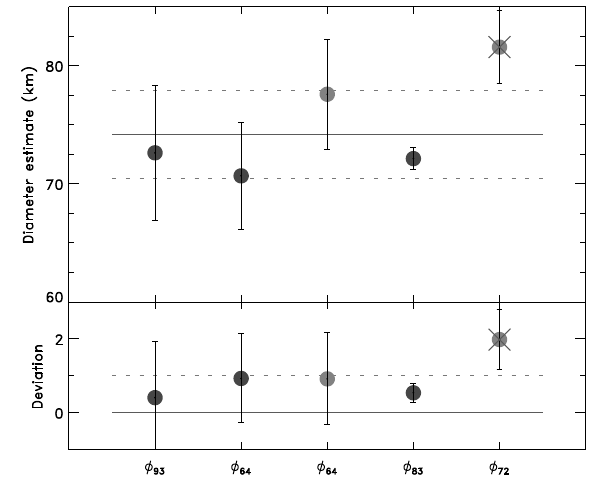}
\caption[Diameter estimates for (163) Erigone]{%
  \label{fap: diam000163}
  Diameter estimates for (163) Erigone.
}
\end{figure}

  \begin{figure}[!ht]
  \centering
  \includegraphics[width=.49\textwidth]{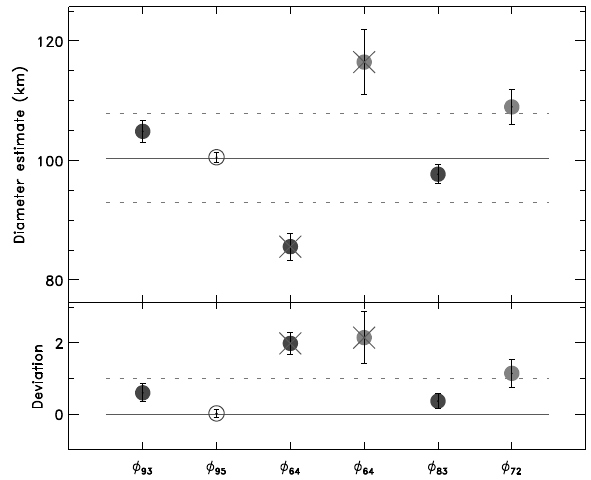}
\caption[Diameter estimates for (164) Eva]{%
  \label{fap: diam000164}
  Diameter estimates for (164) Eva.
}
\end{figure}

  \begin{figure}[!ht]
  \centering
  \includegraphics[width=.49\textwidth]{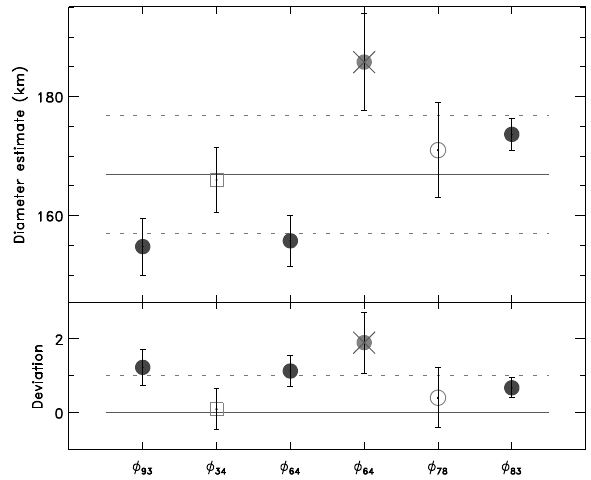}
\caption[Diameter estimates for (165) Loreley]{%
  \label{fap: diam000165}
  Diameter estimates for (165) Loreley.
}
\end{figure}

  \begin{figure}[!ht]
  \centering
  \includegraphics[width=.49\textwidth]{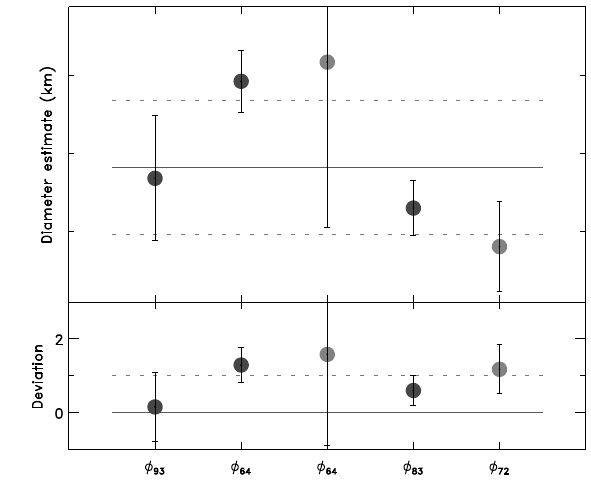}
\caption[Diameter estimates for (168) Sibylla]{%
  \label{fap: diam000168}
  Diameter estimates for (168) Sibylla.
}
\end{figure}

  \begin{figure}[!ht]
  \centering
  \includegraphics[width=.49\textwidth]{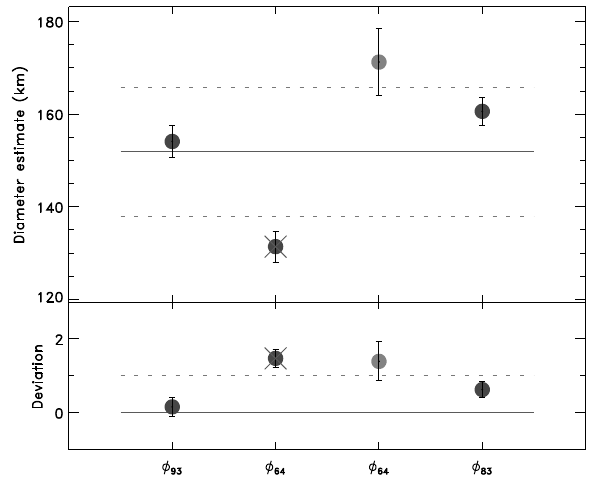}
\caption[Diameter estimates for (173) Ino]{%
  \label{fap: diam000173}
  Diameter estimates for (173) Ino.
}
\end{figure}

\clearpage
  \begin{figure}[!ht]
  \centering
  \includegraphics[width=.49\textwidth]{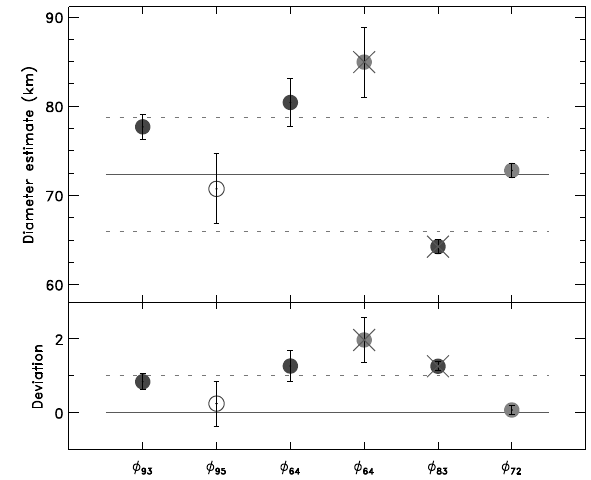}
\caption[Diameter estimates for (179) Klytaemnestra]{%
  \label{fap: diam000179}
  Diameter estimates for (179) Klytaemnestra.
}
\end{figure}

  \begin{figure}[!ht]
  \centering
  \includegraphics[width=.49\textwidth]{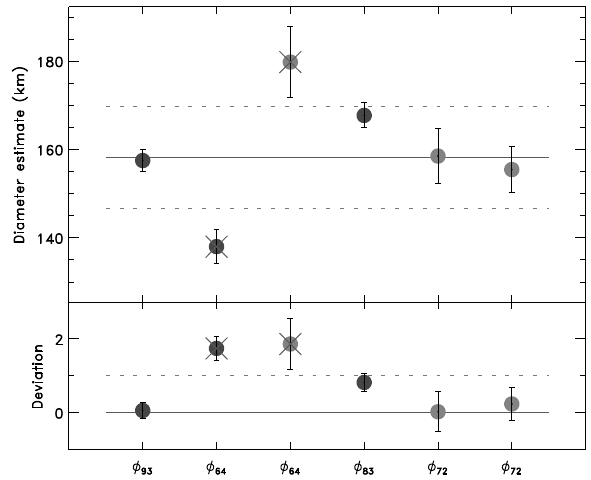}
\caption[Diameter estimates for (185) Eunike]{%
  \label{fap: diam000185}
  Diameter estimates for (185) Eunike.
}
\end{figure}

  \begin{figure}[!ht]
  \centering
  \includegraphics[width=.49\textwidth]{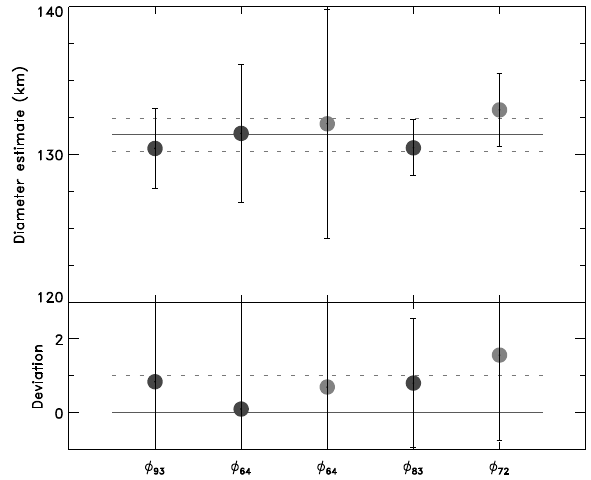}
\caption[Diameter estimates for (187) Lamberta]{%
  \label{fap: diam000187}
  Diameter estimates for (187) Lamberta.
}
\end{figure}

  \begin{figure}[!ht]
  \centering
  \includegraphics[width=.49\textwidth]{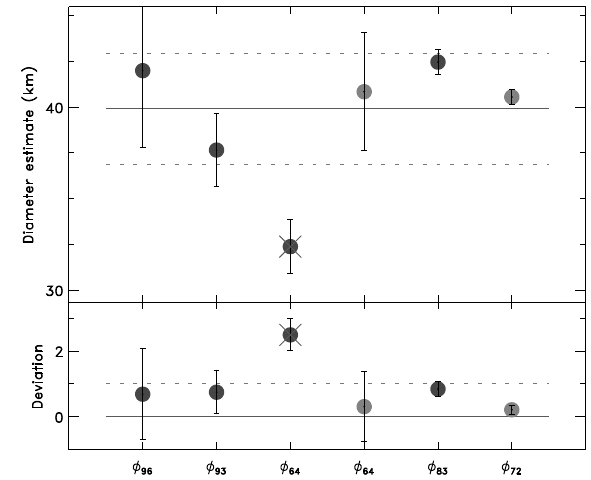}
\caption[Diameter estimates for (189) Phthia]{%
  \label{fap: diam000189}
  Diameter estimates for (189) Phthia.
}
\end{figure}

  \begin{figure}[!ht]
  \centering
  \includegraphics[width=.49\textwidth]{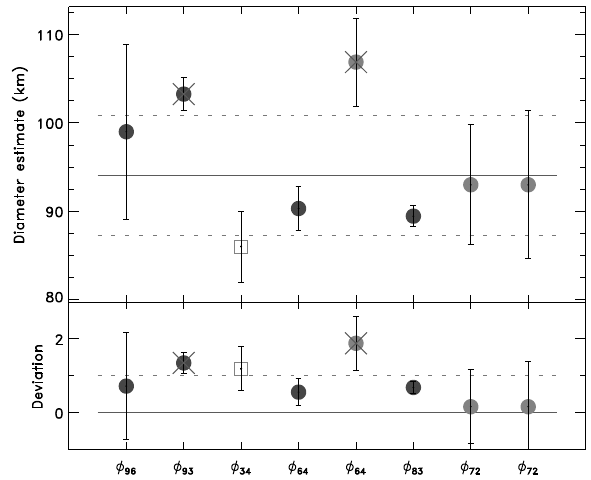}
\caption[Diameter estimates for (192) Nausikaa]{%
  \label{fap: diam000192}
  Diameter estimates for (192) Nausikaa.
}
\end{figure}

  \begin{figure}[!ht]
  \centering
  \includegraphics[width=.49\textwidth]{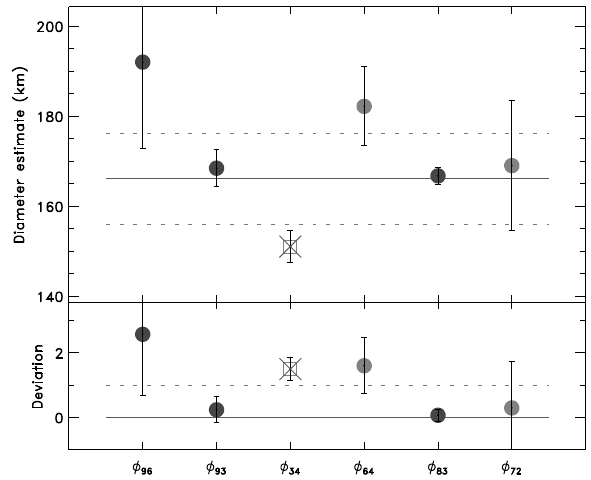}
\caption[Diameter estimates for (194) Prokne]{%
  \label{fap: diam000194}
  Diameter estimates for (194) Prokne.
  The diameter estimated by $\phi_{34}$$^\dagger$ falls outside the  range drawn by the weighted average. It is, however, the unique direct measurement  of Prokne's diameter (although limitated to a single geometry). 
}
\end{figure}

  \begin{figure}[!ht]
  \centering
  \includegraphics[width=.49\textwidth]{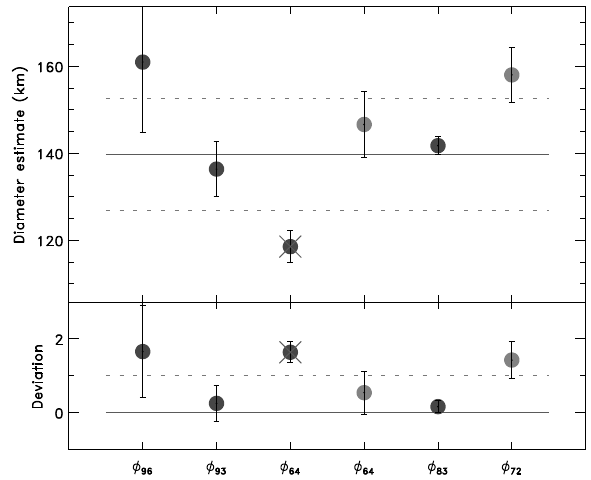}
\caption[Diameter estimates for (196) Philomela]{%
  \label{fap: diam000196}
  Diameter estimates for (196) Philomela.
}
\end{figure}

  \begin{figure}[!ht]
  \centering
  \includegraphics[width=.49\textwidth]{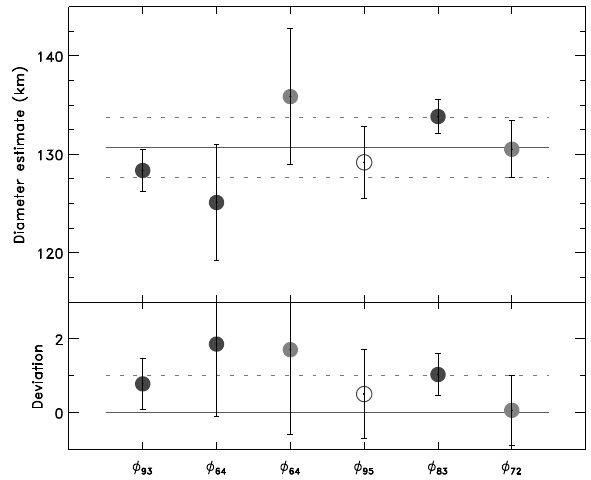}
\caption[Diameter estimates for (200) Dynamene]{%
  \label{fap: diam000200}
  Diameter estimates for (200) Dynamene.
}
\end{figure}

  \begin{figure}[!ht]
  \centering
  \includegraphics[width=.49\textwidth]{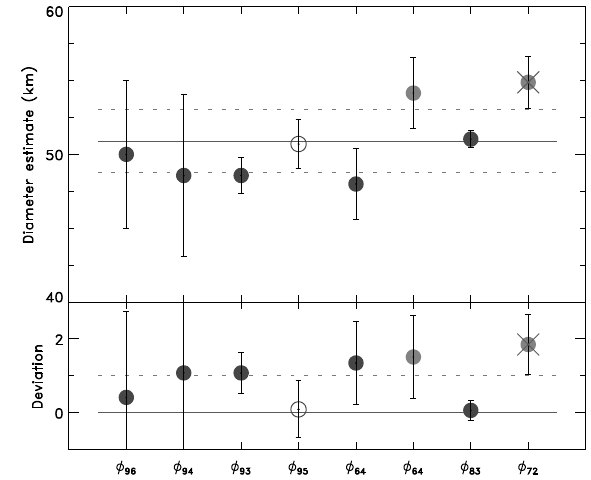}
\caption[Diameter estimates for (204) Kallisto]{%
  \label{fap: diam000204}
  Diameter estimates for (204) Kallisto.
}
\end{figure}

  \begin{figure}[!ht]
  \centering
  \includegraphics[width=.49\textwidth]{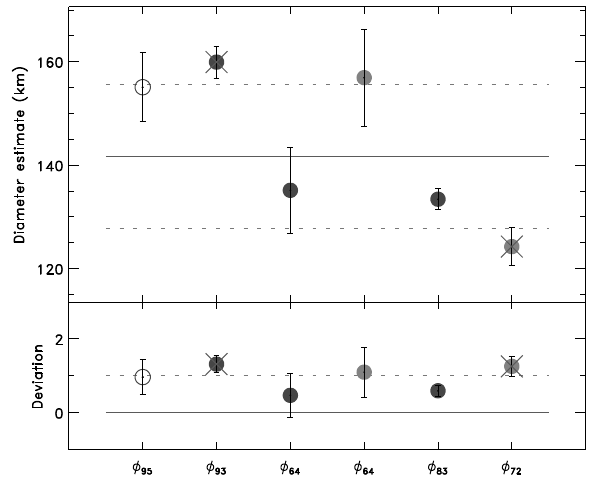}
\caption[Diameter estimates for (209) Dido]{%
  \label{fap: diam000209}
  Diameter estimates for (209) Dido.
}
\end{figure}

  \begin{figure}[!ht]
  \centering
  \includegraphics[width=.49\textwidth]{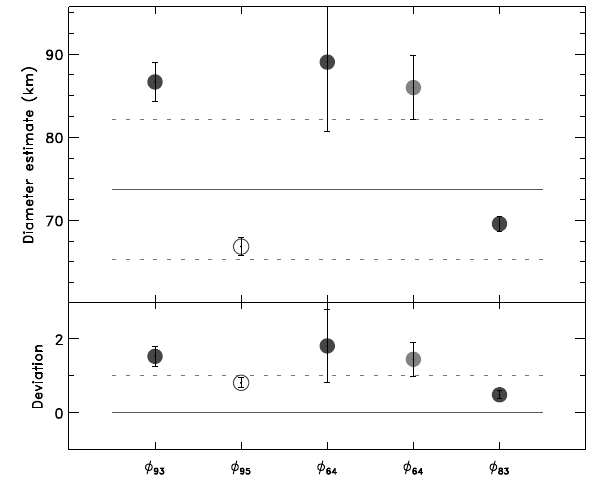}
\caption[Diameter estimates for (210) Isabella]{%
  \label{fap: diam000210}
  Diameter estimates for (210) Isabella.
  The large uncertainty on the mass determination forbid to  sort between the diameter estimates. However,  diameter estimates are all independantly based on single-geometry (\eg, $\phi_{93}$ and $\phi_{64}$ with IRAS, $\phi_{95}$ by occultation)  and their differences most likely result from these different geometries, not properly taken into account.
}
\end{figure}

  \begin{figure}[!ht]
  \centering
  \includegraphics[width=.49\textwidth]{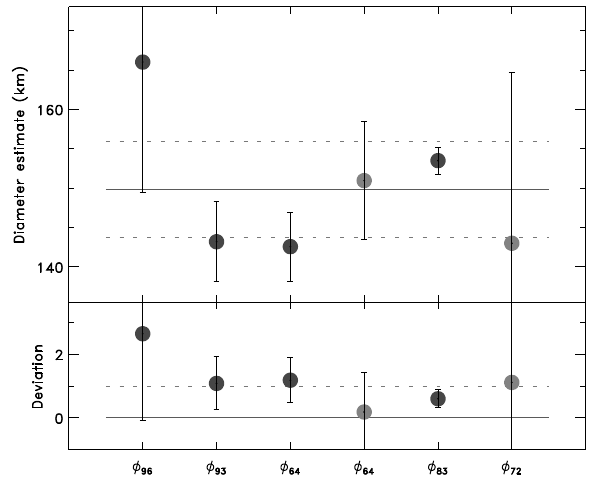}
\caption[Diameter estimates for (211) Isolda]{%
  \label{fap: diam000211}
  Diameter estimates for (211) Isolda.
}
\end{figure}

  \begin{figure}[!ht]
  \centering
  \includegraphics[width=.49\textwidth]{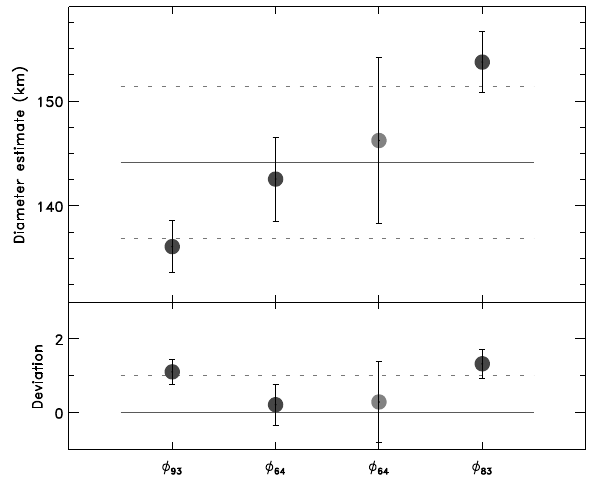}
\caption[Diameter estimates for (212) Medea]{%
  \label{fap: diam000212}
  Diameter estimates for (212) Medea.
}
\end{figure}

  \begin{figure}[!ht]
  \centering
  \includegraphics[width=.49\textwidth]{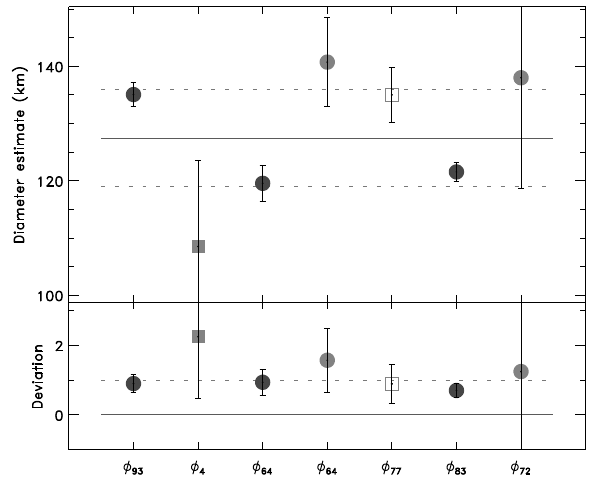}
\caption[Diameter estimates for (216) Kleopatra]{%
  \label{fap: diam000216}
  Diameter estimates for (216) Kleopatra.
}
\end{figure}

  \begin{figure}[!ht]
  \centering
  \includegraphics[width=.49\textwidth]{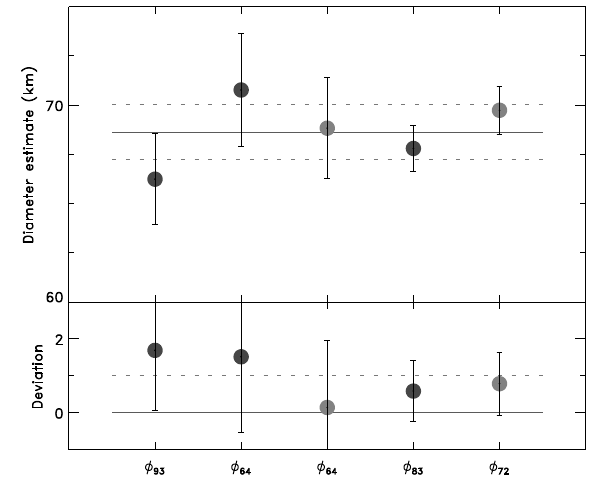}
\caption[Diameter estimates for (217) Eudora]{%
  \label{fap: diam000217}
  Diameter estimates for (217) Eudora.
}
\end{figure}

  \begin{figure}[!ht]
  \centering
  \includegraphics[width=.49\textwidth]{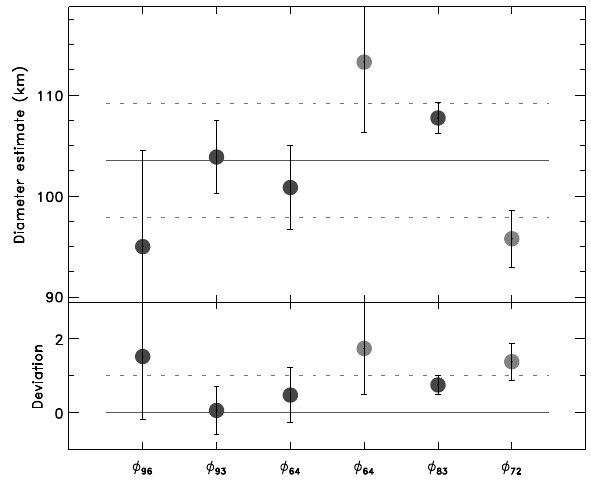}
\caption[Diameter estimates for (221) Eos]{%
  \label{fap: diam000221}
  Diameter estimates for (221) Eos.
}
\end{figure}

  \begin{figure}[!ht]
  \centering
  \includegraphics[width=.49\textwidth]{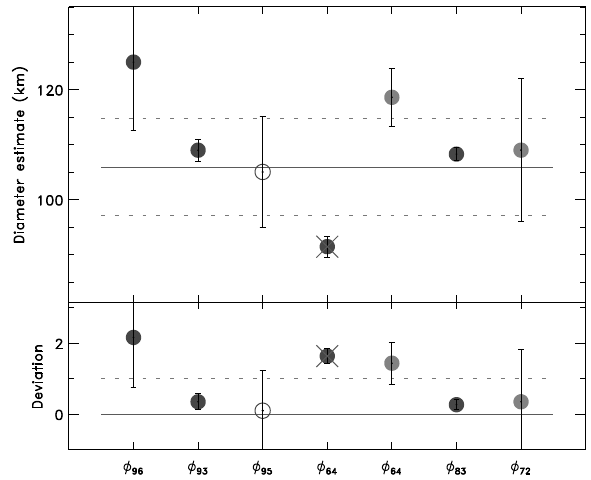}
\caption[Diameter estimates for (230) Athamantis]{%
  \label{fap: diam000230}
  Diameter estimates for (230) Athamantis.
}
\end{figure}

  \begin{figure}[!ht]
  \centering
  \includegraphics[width=.49\textwidth]{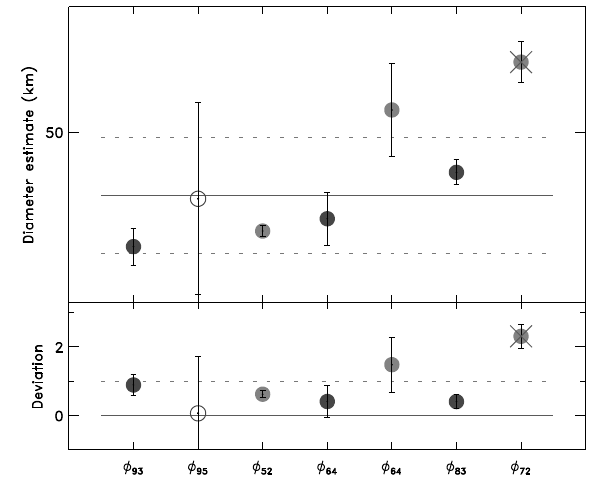}
\caption[Diameter estimates for (234) Barbara]{%
  \label{fap: diam000234}
  Diameter estimates for (234) Barbara.
}
\end{figure}

\clearpage
  \begin{figure}[!ht]
  \centering
  \includegraphics[width=.49\textwidth]{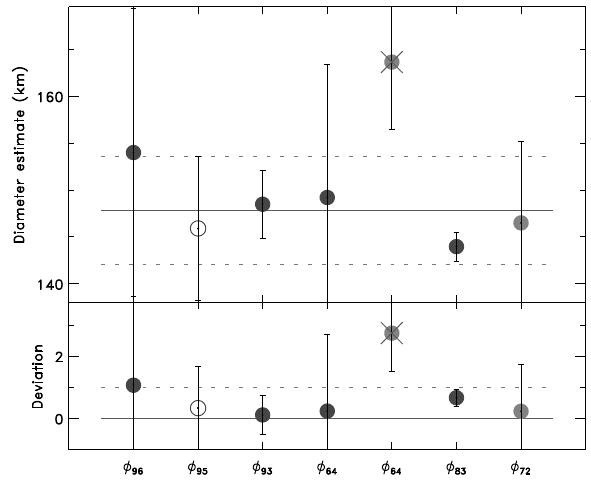}
\caption[Diameter estimates for (238) Hypatia]{%
  \label{fap: diam000238}
  Diameter estimates for (238) Hypatia.
}
\end{figure}

  \begin{figure}[!ht]
  \centering
  \includegraphics[width=.49\textwidth]{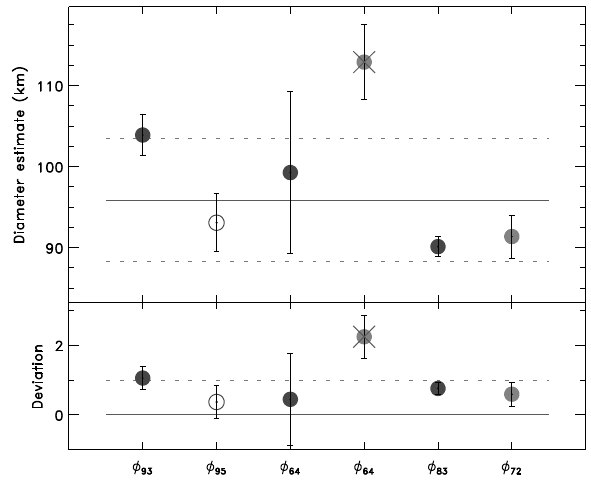}
\caption[Diameter estimates for (240) Vanadis]{%
  \label{fap: diam000240}
  Diameter estimates for (240) Vanadis.
}
\end{figure}

  \begin{figure}[!ht]
  \centering
  \includegraphics[width=.49\textwidth]{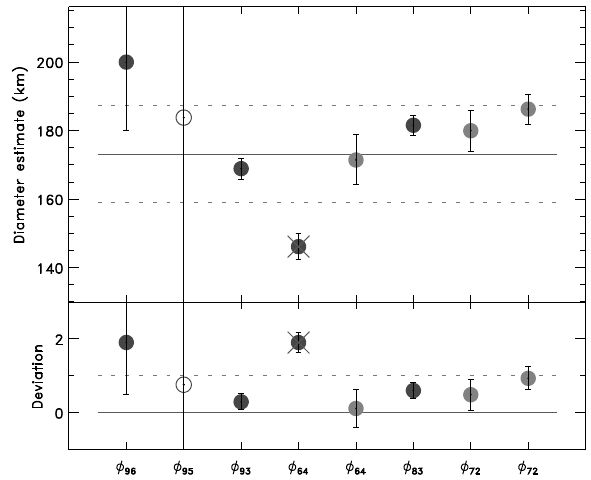}
\caption[Diameter estimates for (241) Germania]{%
  \label{fap: diam000241}
  Diameter estimates for (241) Germania.
}
\end{figure}

  \begin{figure}[!ht]
  \centering
  \includegraphics[width=.49\textwidth]{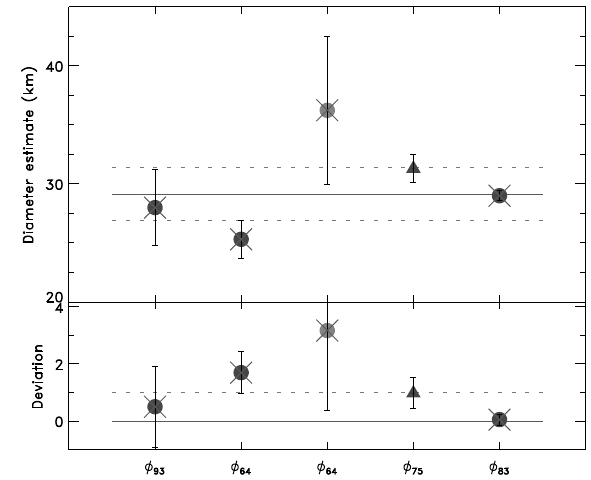}
\caption[Diameter estimates for (243) Ida]{%
  \label{fap: diam000243}
  Diameter estimates for (243) Ida.
  Only the flyby estimate from $\phi_{75}$ is used here. 
}
\end{figure}

  \begin{figure}[!ht]
  \centering
  \includegraphics[width=.49\textwidth]{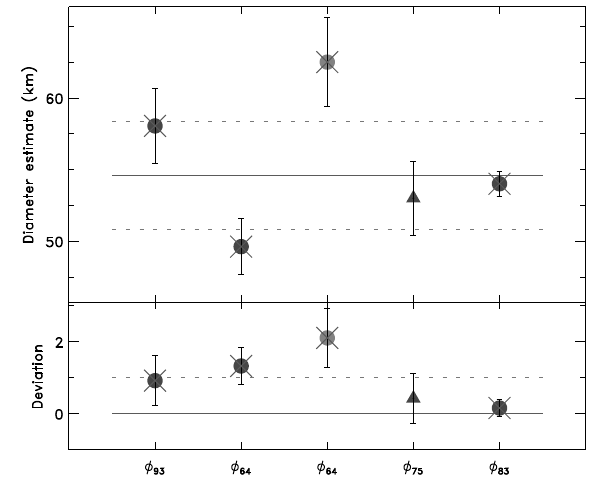}
\caption[Diameter estimates for (253) Mathilde]{%
  \label{fap: diam000253}
  Diameter estimates for (253) Mathilde.
  Only the flyby estimate from $\phi_{75}$ is used here. 
}
\end{figure}

  \begin{figure}[!ht]
  \centering
  \includegraphics[width=.49\textwidth]{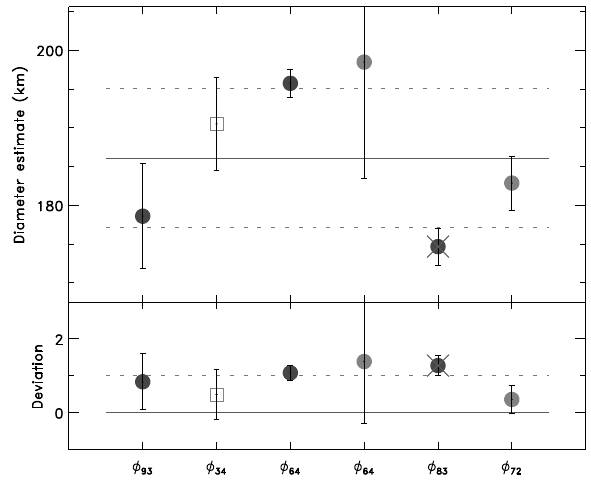}
\caption[Diameter estimates for (259) Aletheia]{%
  \label{fap: diam000259}
  Diameter estimates for (259) Aletheia.
}
\end{figure}

  \begin{figure}[!ht]
  \centering
  \includegraphics[width=.49\textwidth]{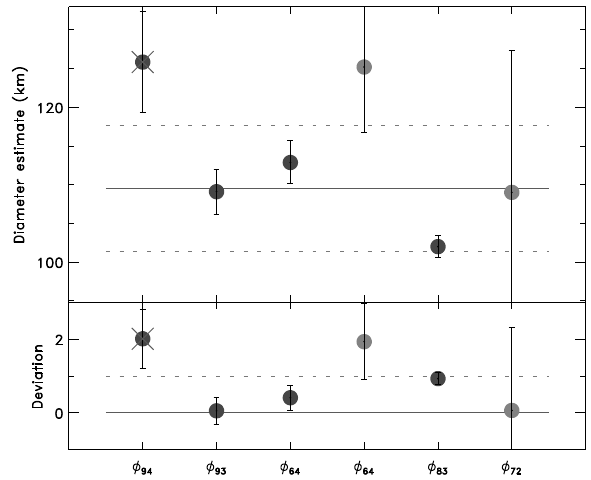}
\caption[Diameter estimates for (266) Aline]{%
  \label{fap: diam000266}
  Diameter estimates for (266) Aline.
}
\end{figure}

  \begin{figure}[!ht]
  \centering
  \includegraphics[width=.49\textwidth]{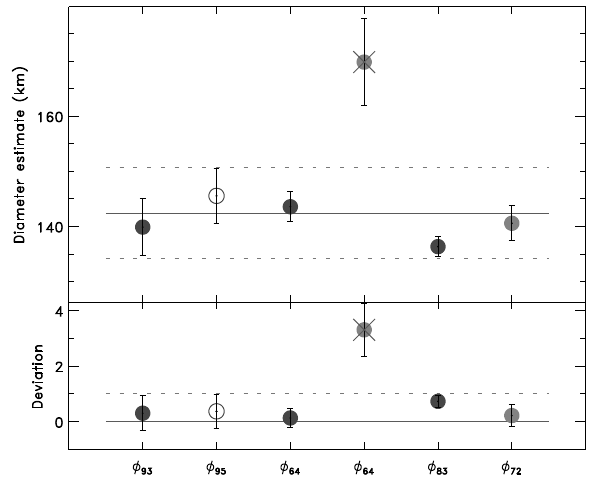}
\caption[Diameter estimates for (268) Adorea]{%
  \label{fap: diam000268}
  Diameter estimates for (268) Adorea.
}
\end{figure}

  \begin{figure}[!ht]
  \centering
  \includegraphics[width=.49\textwidth]{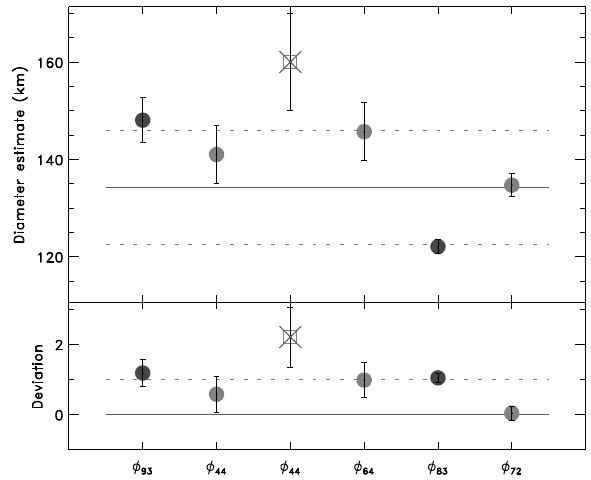}
\caption[Diameter estimates for (283) Emma]{%
  \label{fap: diam000283}
  Diameter estimates for (283) Emma.
}
\end{figure}

  \begin{figure}[!ht]
  \centering
  \includegraphics[width=.49\textwidth]{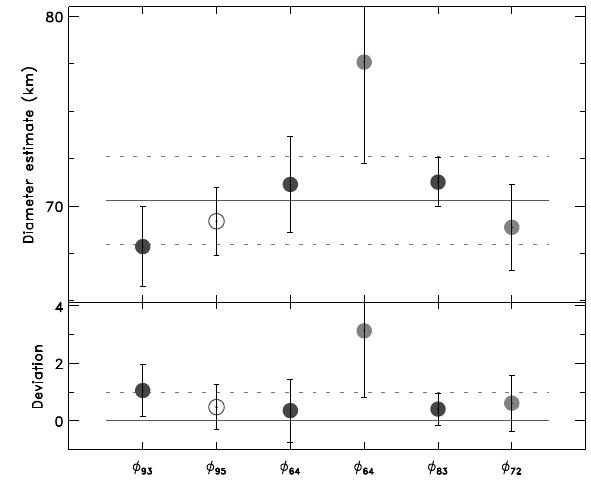}
\caption[Diameter estimates for (304) Olga]{%
  \label{fap: diam000304}
  Diameter estimates for (304) Olga.
}
\end{figure}

  \begin{figure}[!ht]
  \centering
  \includegraphics[width=.49\textwidth]{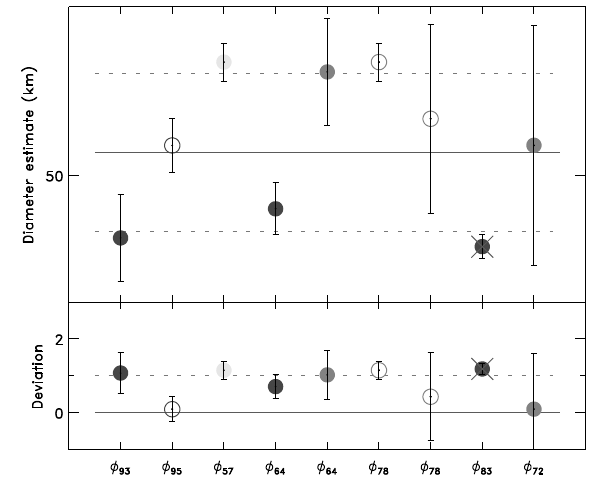}
\caption[Diameter estimates for (306) Unitas]{%
  \label{fap: diam000306}
  Diameter estimates for (306) Unitas.
}
\end{figure}

  \begin{figure}[!ht]
  \centering
  \includegraphics[width=.49\textwidth]{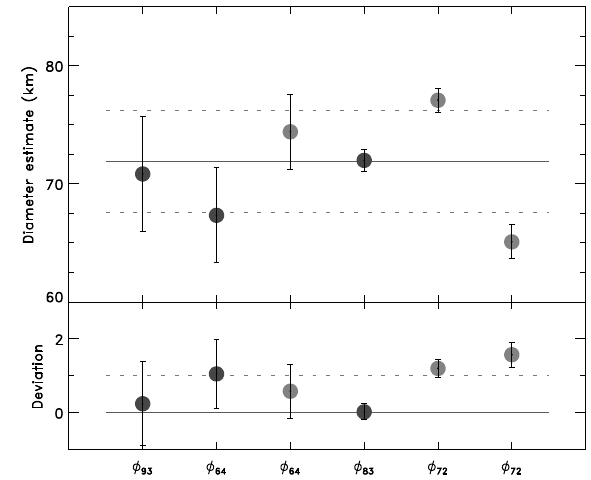}
\caption[Diameter estimates for (322) Phaeo]{%
  \label{fap: diam000322}
  Diameter estimates for (322) Phaeo.
}
\end{figure}

  \begin{figure}[!ht]
  \centering
  \includegraphics[width=.49\textwidth]{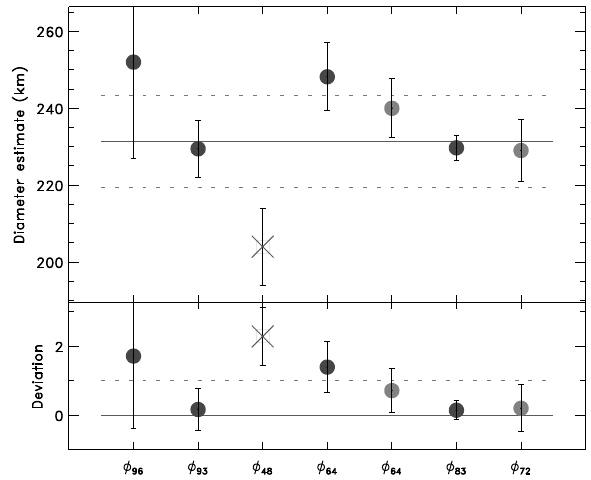}
\caption[Diameter estimates for (324) Bamberga]{%
  \label{fap: diam000324}
  Diameter estimates for (324) Bamberga.
}
\end{figure}

  \begin{figure}[!ht]
  \centering
  \includegraphics[width=.49\textwidth]{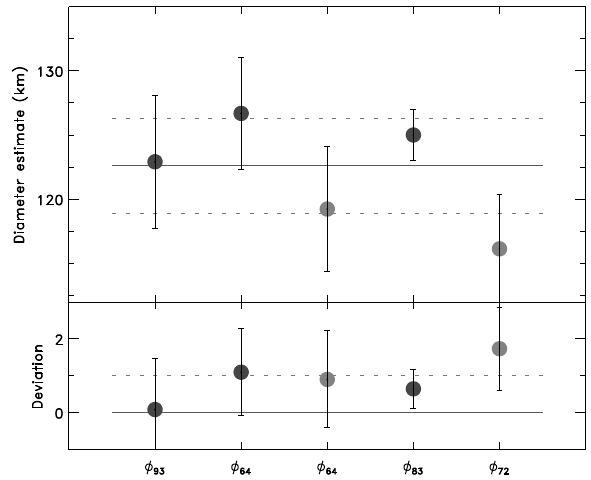}
\caption[Diameter estimates for (328) Gudrun]{%
  \label{fap: diam000328}
  Diameter estimates for (328) Gudrun.
}
\end{figure}

  \begin{figure}[!ht]
  \centering
  \includegraphics[width=.49\textwidth]{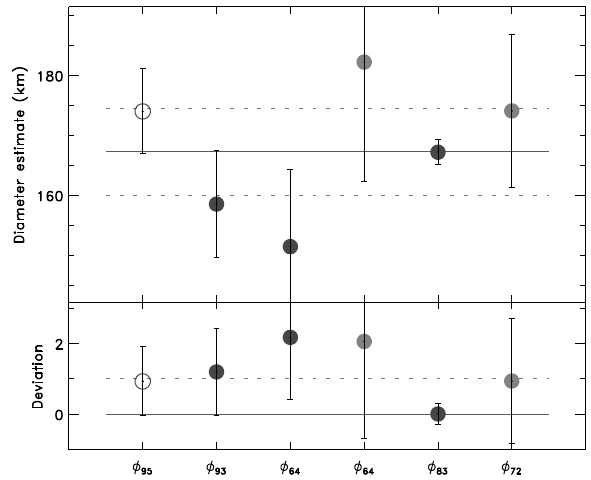}
\caption[Diameter estimates for (334) Chicago]{%
  \label{fap: diam000334}
  Diameter estimates for (334) Chicago.
}
\end{figure}

  \begin{figure}[!ht]
  \centering
  \includegraphics[width=.49\textwidth]{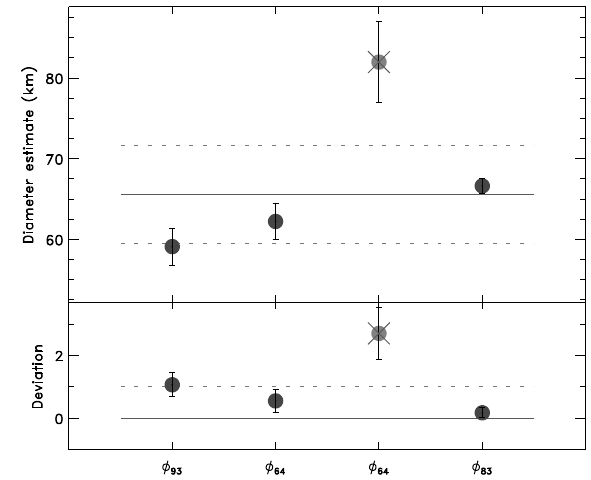}
\caption[Diameter estimates for (337) Devosa]{%
  \label{fap: diam000337}
  Diameter estimates for (337) Devosa.
}
\end{figure}

  \begin{figure}[!ht]
  \centering
  \includegraphics[width=.49\textwidth]{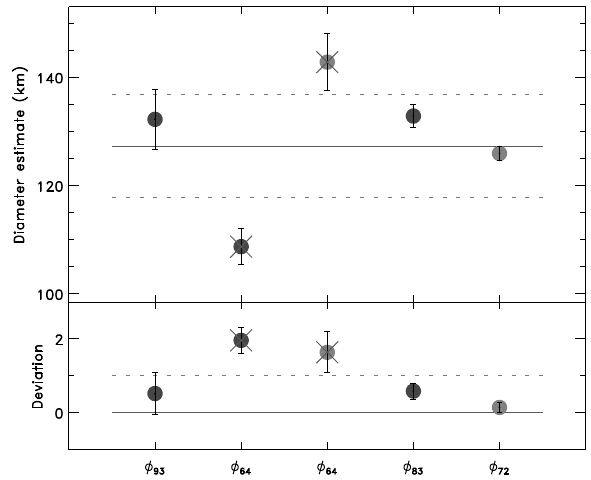}
\caption[Diameter estimates for (344) Desiderata]{%
  \label{fap: diam000344}
  Diameter estimates for (344) Desiderata.
}
\end{figure}

  \begin{figure}[!ht]
  \centering
  \includegraphics[width=.49\textwidth]{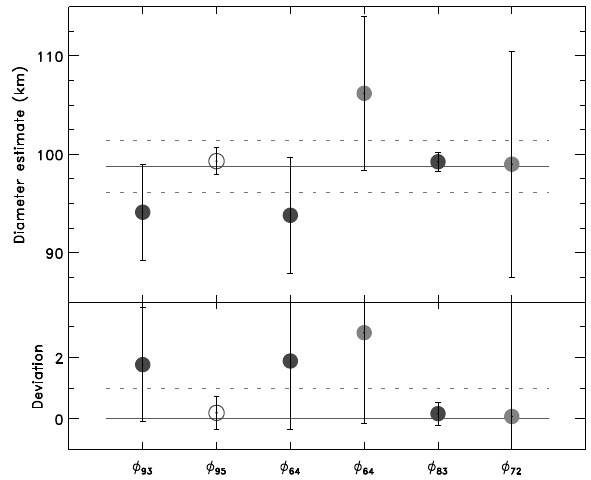}
\caption[Diameter estimates for (345) Tercidina]{%
  \label{fap: diam000345}
  Diameter estimates for (345) Tercidina.
}
\end{figure}

\clearpage
  \begin{figure}[!ht]
  \centering
  \includegraphics[width=.49\textwidth]{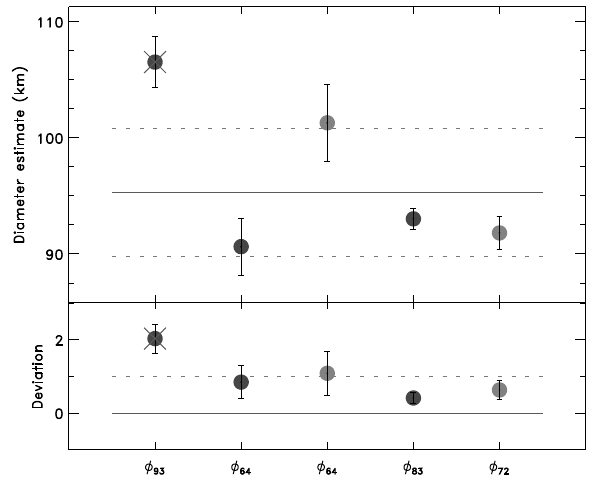}
\caption[Diameter estimates for (346) Hermentaria]{%
  \label{fap: diam000346}
  Diameter estimates for (346) Hermentaria.
}
\end{figure}

  \begin{figure}[!ht]
  \centering
  \includegraphics[width=.49\textwidth]{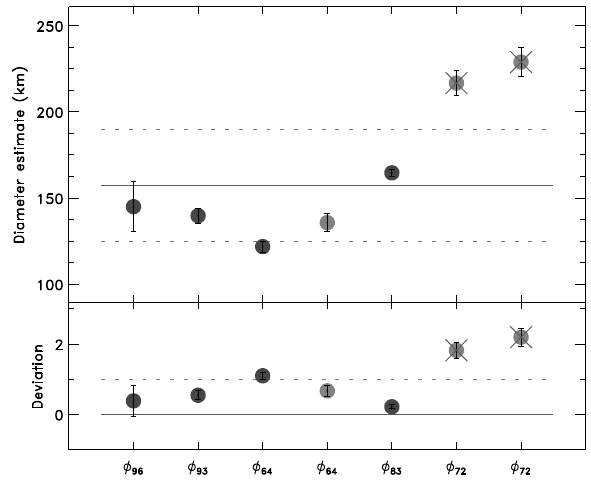}
\caption[Diameter estimates for (349) Dembowska]{%
  \label{fap: diam000349}
  Diameter estimates for (349) Dembowska.
}
\end{figure}

  \begin{figure}[!ht]
  \centering
  \includegraphics[width=.49\textwidth]{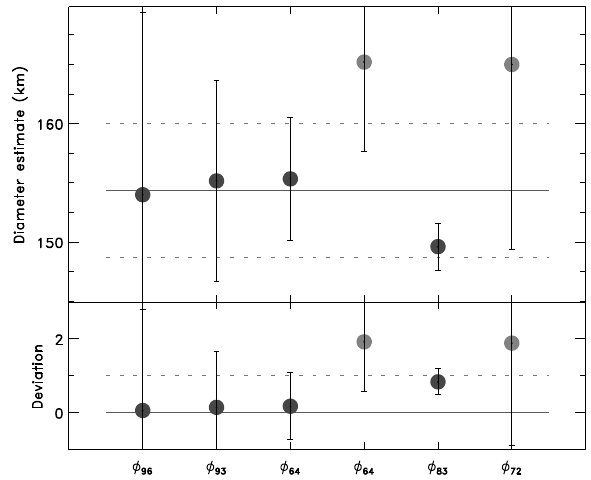}
\caption[Diameter estimates for (354) Eleonora]{%
  \label{fap: diam000354}
  Diameter estimates for (354) Eleonora.
}
\end{figure}

  \begin{figure}[!ht]
  \centering
  \includegraphics[width=.49\textwidth]{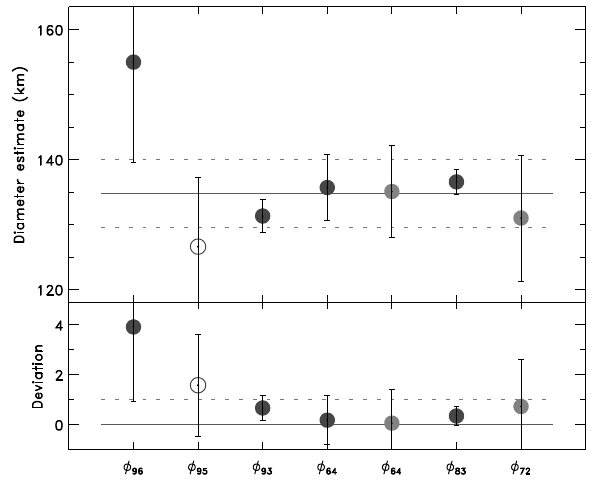}
\caption[Diameter estimates for (356) Liguria]{%
  \label{fap: diam000356}
  Diameter estimates for (356) Liguria.
}
\end{figure}

  \begin{figure}[!ht]
  \centering
  \includegraphics[width=.49\textwidth]{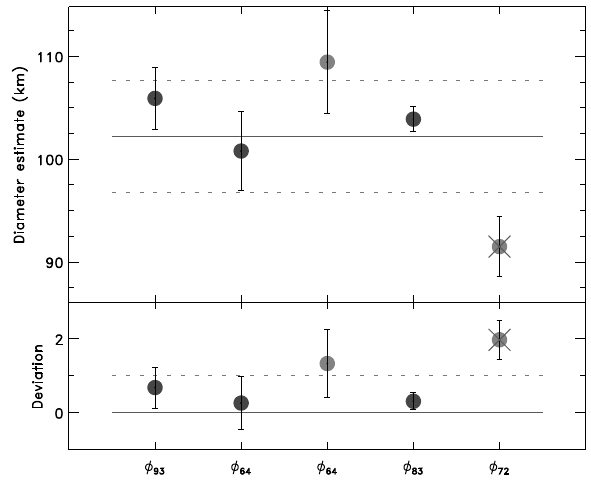}
\caption[Diameter estimates for (365) Corduba]{%
  \label{fap: diam000365}
  Diameter estimates for (365) Corduba.
}
\end{figure}

  \begin{figure}[!ht]
  \centering
  \includegraphics[width=.49\textwidth]{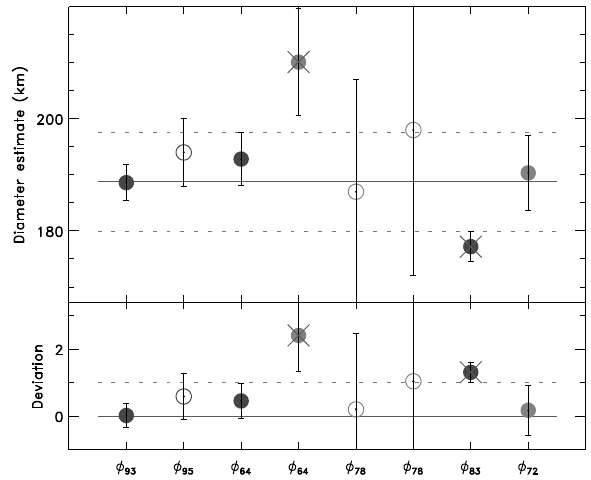}
\caption[Diameter estimates for (372) Palma]{%
  \label{fap: diam000372}
  Diameter estimates for (372) Palma.
}
\end{figure}

  \begin{figure}[!ht]
  \centering
  \includegraphics[width=.49\textwidth]{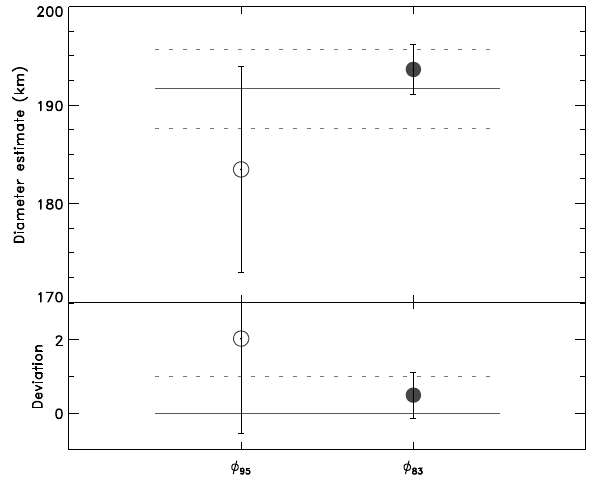}
\caption[Diameter estimates for (375) Ursula]{%
  \label{fap: diam000375}
  Diameter estimates for (375) Ursula.
}
\end{figure}

  \begin{figure}[!ht]
  \centering
  \includegraphics[width=.49\textwidth]{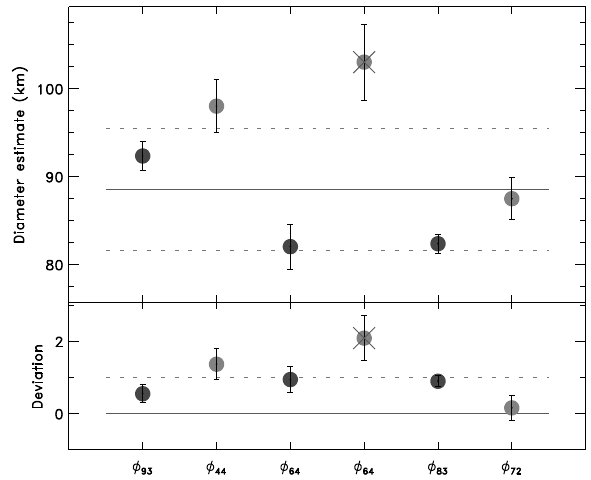}
\caption[Diameter estimates for (379) Huenna]{%
  \label{fap: diam000379}
  Diameter estimates for (379) Huenna.
}
\end{figure}

  \begin{figure}[!ht]
  \centering
  \includegraphics[width=.49\textwidth]{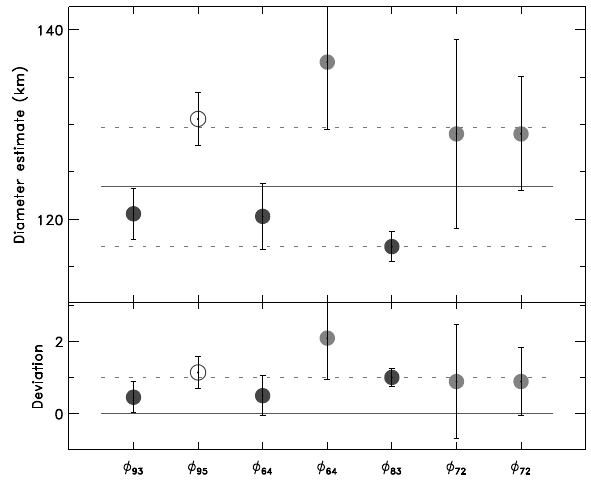}
\caption[Diameter estimates for (381) Myrrha]{%
  \label{fap: diam000381}
  Diameter estimates for (381) Myrrha.
}
\end{figure}

  \begin{figure}[!ht]
  \centering
  \includegraphics[width=.49\textwidth]{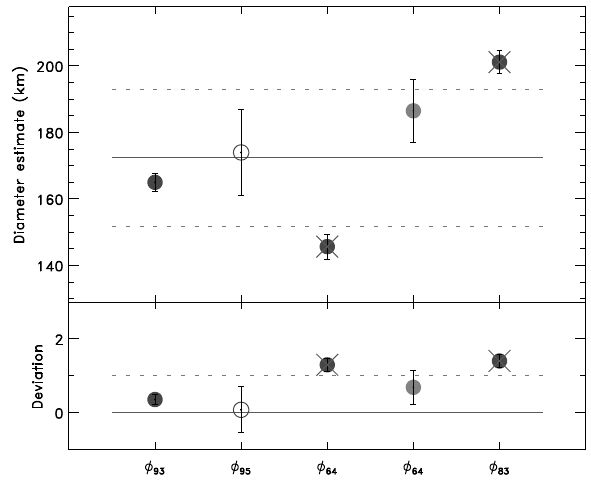}
\caption[Diameter estimates for (386) Siegena]{%
  \label{fap: diam000386}
  Diameter estimates for (386) Siegena.
}
\end{figure}

  \begin{figure}[!ht]
  \centering
  \includegraphics[width=.49\textwidth]{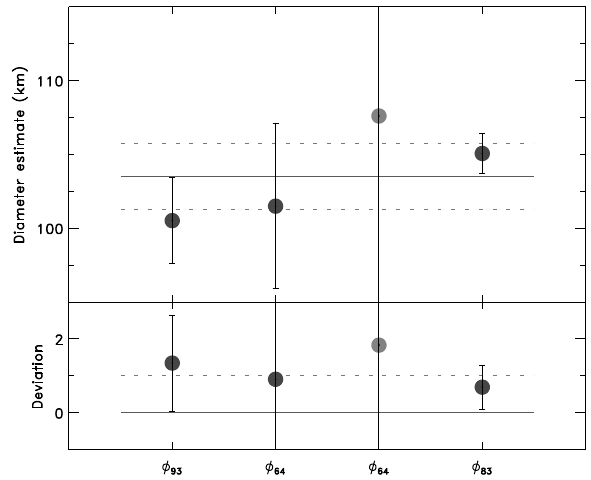}
\caption[Diameter estimates for (387) Aquitania]{%
  \label{fap: diam000387}
  Diameter estimates for (387) Aquitania.
}
\end{figure}

  \begin{figure}[!ht]
  \centering
  \includegraphics[width=.49\textwidth]{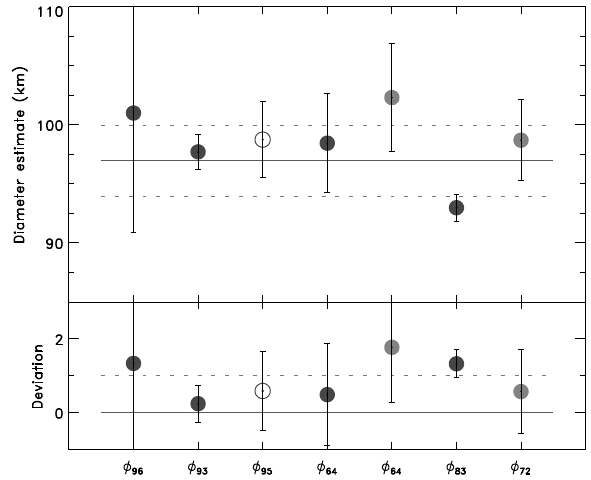}
\caption[Diameter estimates for (404) Arsinoe]{%
  \label{fap: diam000404}
  Diameter estimates for (404) Arsinoe.
}
\end{figure}

  \begin{figure}[!ht]
  \centering
  \includegraphics[width=.49\textwidth]{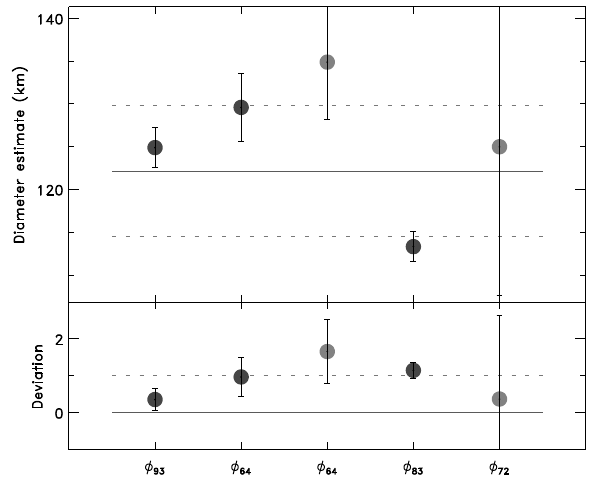}
\caption[Diameter estimates for (405) Thia]{%
  \label{fap: diam000405}
  Diameter estimates for (405) Thia.
}
\end{figure}

  \begin{figure}[!ht]
  \centering
  \includegraphics[width=.49\textwidth]{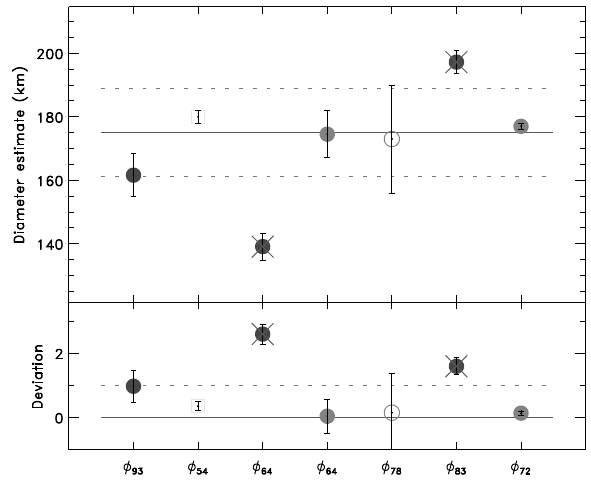}
\caption[Diameter estimates for (409) Aspasia]{%
  \label{fap: diam000409}
  Diameter estimates for (409) Aspasia.
}
\end{figure}

  \begin{figure}[!ht]
  \centering
  \includegraphics[width=.49\textwidth]{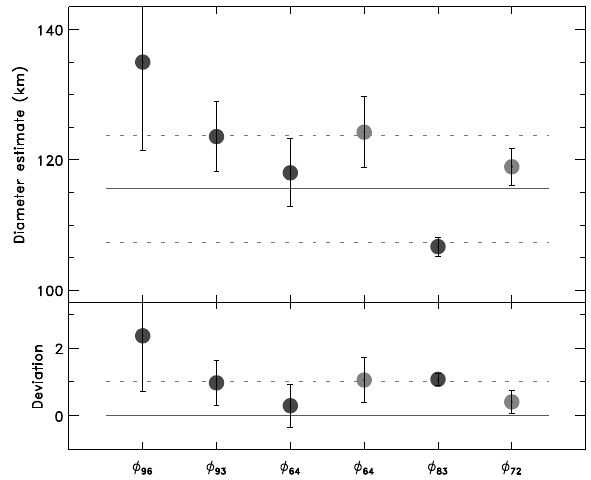}
\caption[Diameter estimates for (410) Chloris]{%
  \label{fap: diam000410}
  Diameter estimates for (410) Chloris.
}
\end{figure}

  \begin{figure}[!ht]
  \centering
  \includegraphics[width=.49\textwidth]{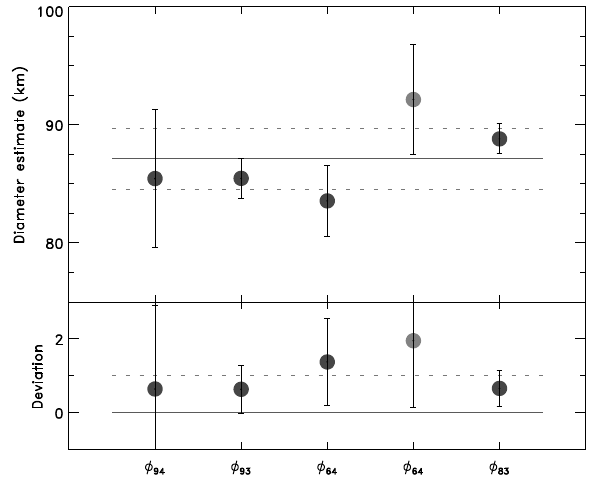}
\caption[Diameter estimates for (416) Vaticana]{%
  \label{fap: diam000416}
  Diameter estimates for (416) Vaticana.
}
\end{figure}

  \begin{figure}[!ht]
  \centering
  \includegraphics[width=.49\textwidth]{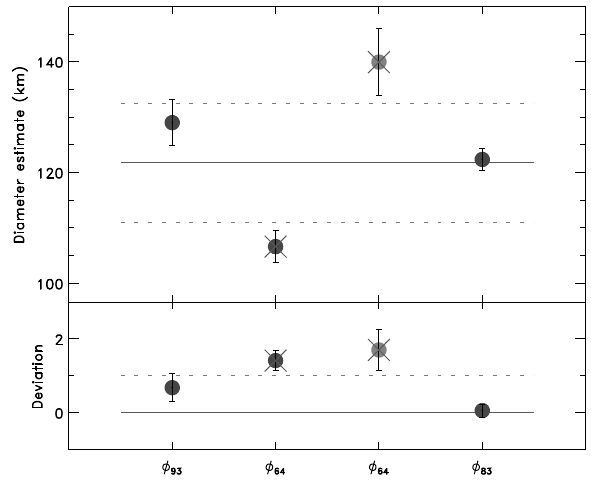}
\caption[Diameter estimates for (419) Aurelia]{%
  \label{fap: diam000419}
  Diameter estimates for (419) Aurelia.
}
\end{figure}

  \begin{figure}[!ht]
  \centering
  \includegraphics[width=.49\textwidth]{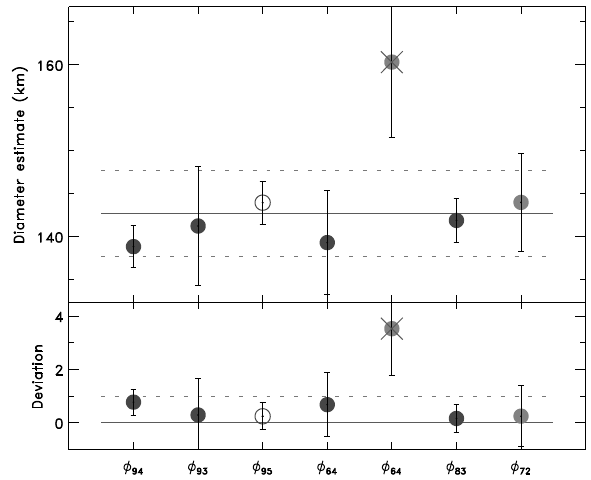}
\caption[Diameter estimates for (420) Bertholda]{%
  \label{fap: diam000420}
  Diameter estimates for (420) Bertholda.
}
\end{figure}

\clearpage
  \begin{figure}[!ht]
  \centering
  \includegraphics[width=.49\textwidth]{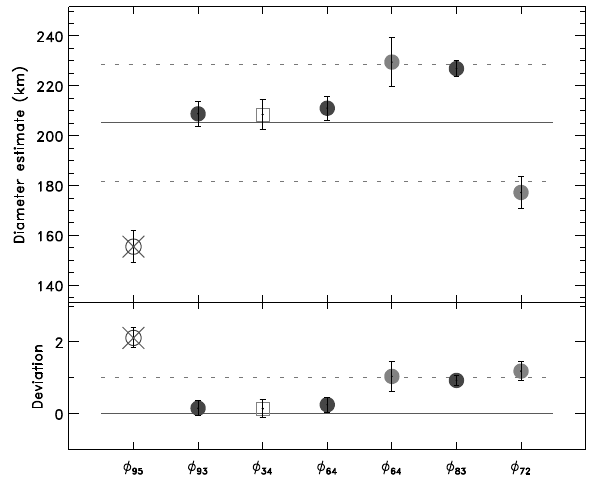}
\caption[Diameter estimates for (423) Diotima]{%
  \label{fap: diam000423}
  Diameter estimates for (423) Diotima.
}
\end{figure}

  \begin{figure}[!ht]
  \centering
  \includegraphics[width=.49\textwidth]{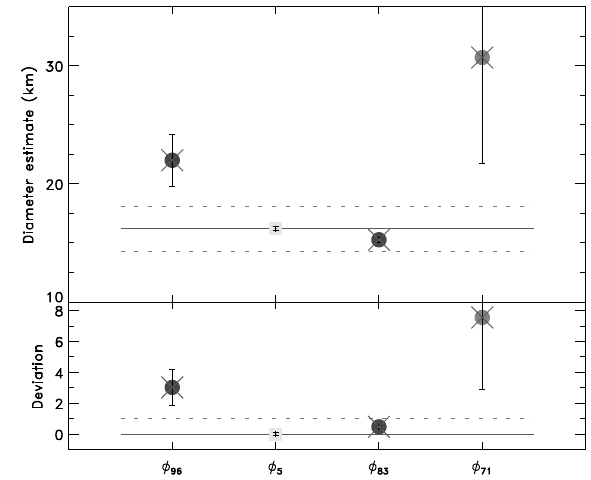}
\caption[Diameter estimates for (433) Eros]{%
  \label{fap: diam000433}
  Diameter estimates for (433) Eros.
  Only the flyby estimate from $\phi_{5}$ is used here. 
}
\end{figure}

  \begin{figure}[!ht]
  \centering
  \includegraphics[width=.49\textwidth]{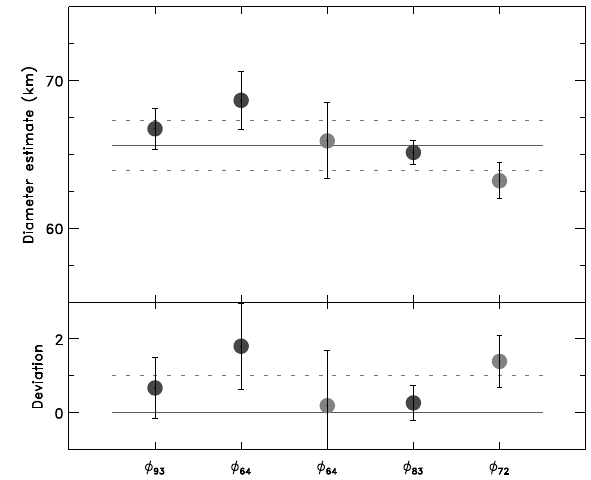}
\caption[Diameter estimates for (442) Eichsfeldia]{%
  \label{fap: diam000442}
  Diameter estimates for (442) Eichsfeldia.
}
\end{figure}

  \begin{figure}[!ht]
  \centering
  \includegraphics[width=.49\textwidth]{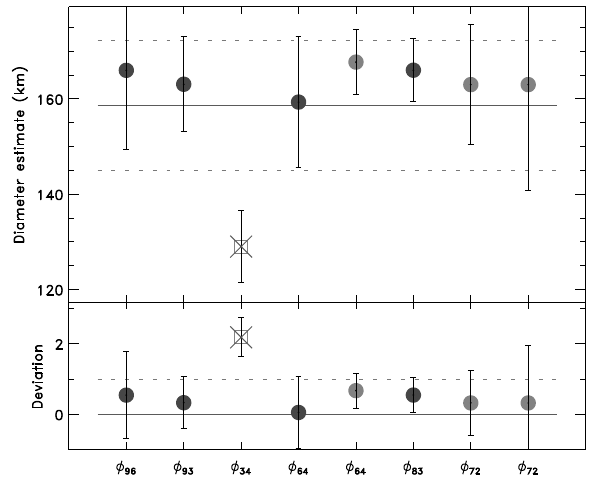}
\caption[Diameter estimates for (444) Gyptis]{%
  \label{fap: diam000444}
  Diameter estimates for (444) Gyptis.
}
\end{figure}

  \begin{figure}[!ht]
  \centering
  \includegraphics[width=.49\textwidth]{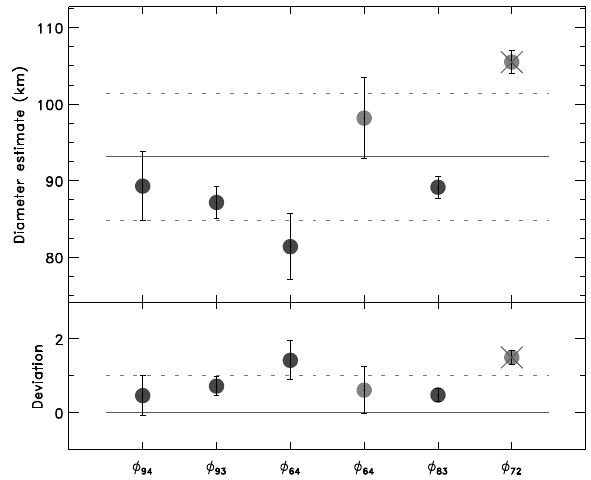}
\caption[Diameter estimates for (445) Edna]{%
  \label{fap: diam000445}
  Diameter estimates for (445) Edna.
}
\end{figure}

  \begin{figure}[!ht]
  \centering
  \includegraphics[width=.49\textwidth]{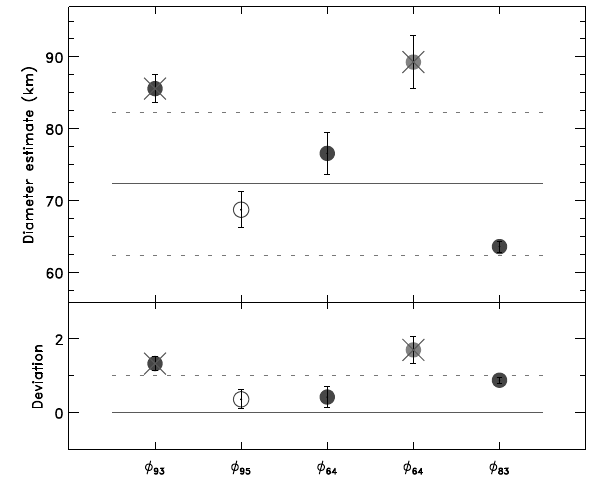}
\caption[Diameter estimates for (449) Hamburga]{%
  \label{fap: diam000449}
  Diameter estimates for (449) Hamburga.
}
\end{figure}

  \begin{figure}[!ht]
  \centering
  \includegraphics[width=.49\textwidth]{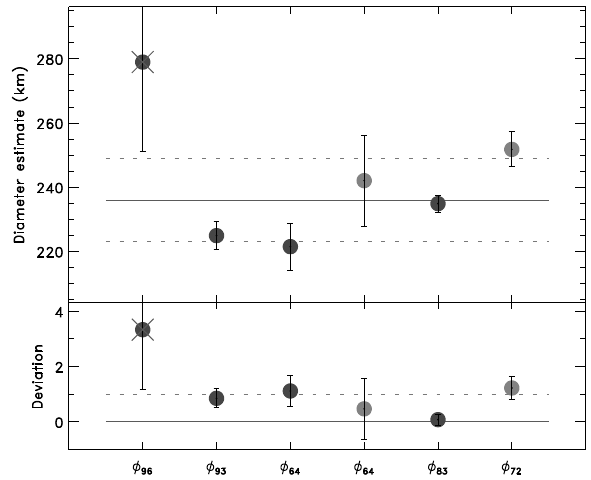}
\caption[Diameter estimates for (451) Patientia]{%
  \label{fap: diam000451}
  Diameter estimates for (451) Patientia.
}
\end{figure}

  \begin{figure}[!ht]
  \centering
  \includegraphics[width=.49\textwidth]{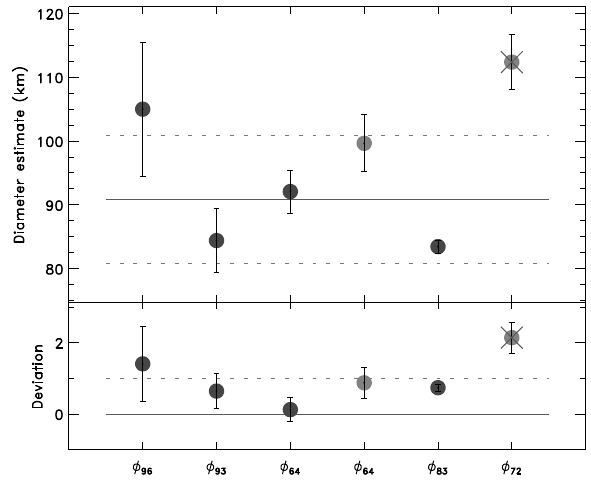}
\caption[Diameter estimates for (455) Bruchsalia]{%
  \label{fap: diam000455}
  Diameter estimates for (455) Bruchsalia.
}
\end{figure}

  \begin{figure}[!ht]
  \centering
  \includegraphics[width=.49\textwidth]{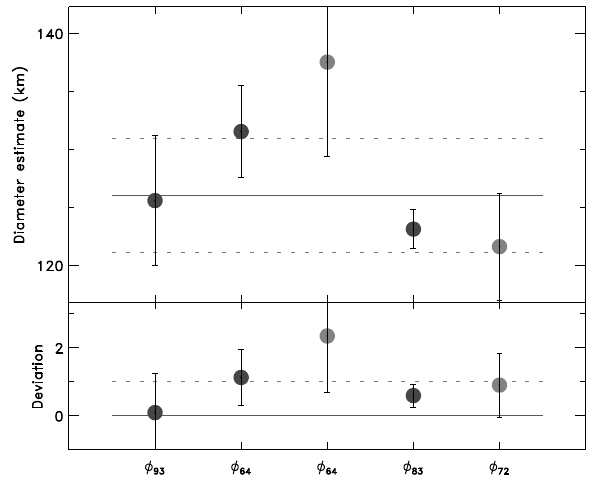}
\caption[Diameter estimates for (469) Argentina]{%
  \label{fap: diam000469}
  Diameter estimates for (469) Argentina.
}
\end{figure}

  \begin{figure}[!ht]
  \centering
  \includegraphics[width=.49\textwidth]{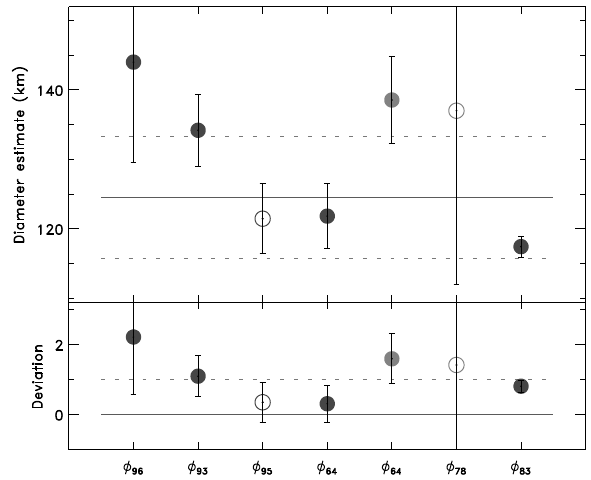}
\caption[Diameter estimates for (471) Papagena]{%
  \label{fap: diam000471}
  Diameter estimates for (471) Papagena.
}
\end{figure}

  \begin{figure}[!ht]
  \centering
  \includegraphics[width=.49\textwidth]{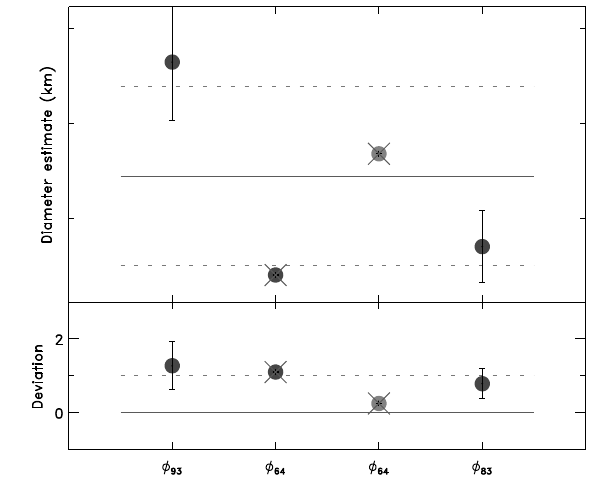}
\caption[Diameter estimates for (481) Emita]{%
  \label{fap: diam000481}
  Diameter estimates for (481) Emita.
  The diameter estimates from $\phi_{64}$ have unrealistic small uncertainties of 0.01\,km.  Using these values strongly biases the average. 
}
\end{figure}

  \begin{figure}[!ht]
  \centering
  \includegraphics[width=.49\textwidth]{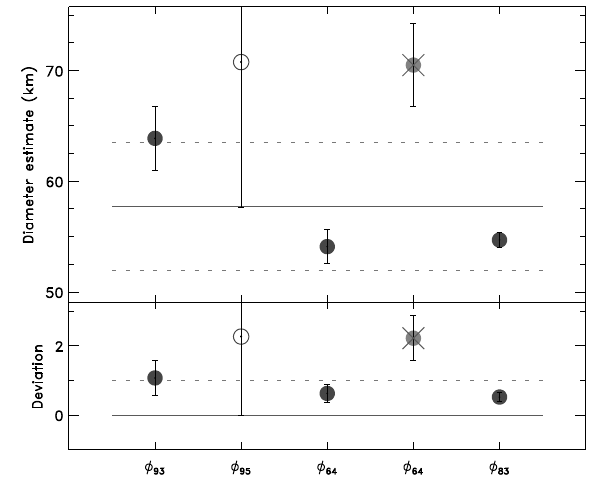}
\caption[Diameter estimates for (485) Genua]{%
  \label{fap: diam000485}
  Diameter estimates for (485) Genua.
}
\end{figure}

  \begin{figure}[!ht]
  \centering
  \includegraphics[width=.49\textwidth]{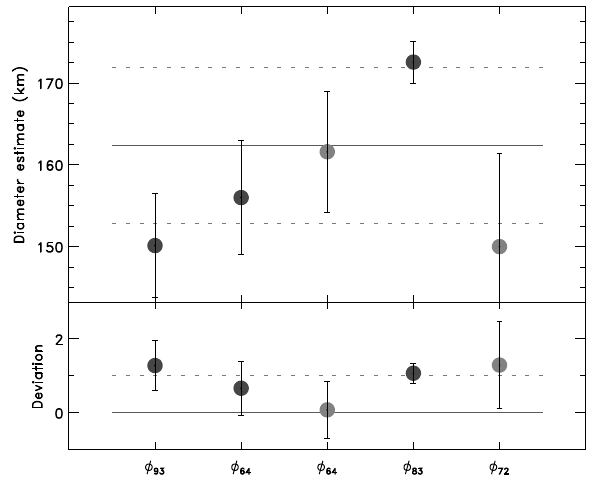}
\caption[Diameter estimates for (488) Kreusa]{%
  \label{fap: diam000488}
  Diameter estimates for (488) Kreusa.
}
\end{figure}

  \begin{figure}[!ht]
  \centering
  \includegraphics[width=.49\textwidth]{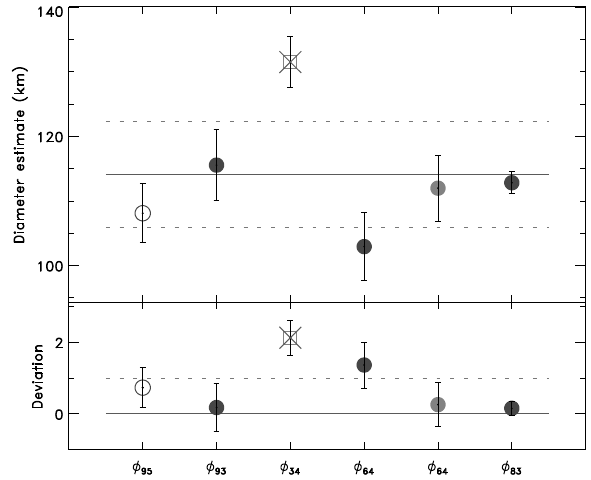}
\caption[Diameter estimates for (490) Veritas]{%
  \label{fap: diam000490}
  Diameter estimates for (490) Veritas.
}
\end{figure}

  \begin{figure}[!ht]
  \centering
  \includegraphics[width=.49\textwidth]{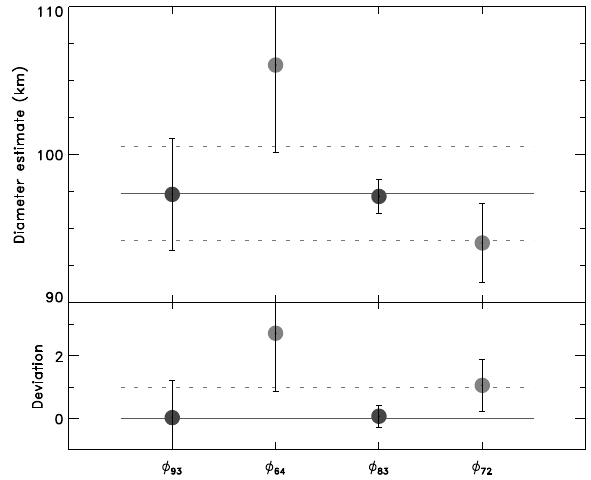}
\caption[Diameter estimates for (491) Carina]{%
  \label{fap: diam000491}
  Diameter estimates for (491) Carina.
}
\end{figure}

  \begin{figure}[!ht]
  \centering
  \includegraphics[width=.49\textwidth]{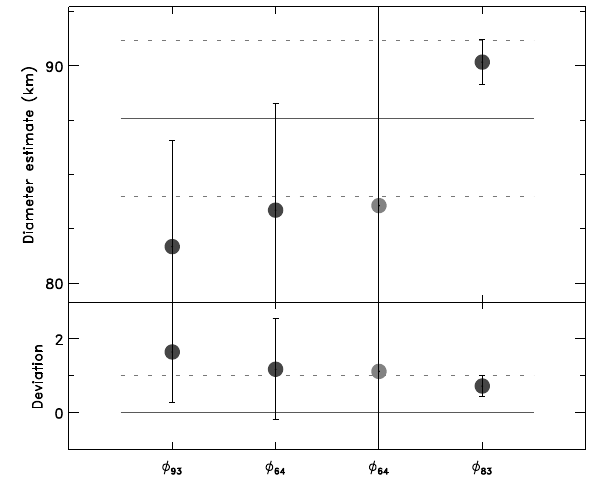}
\caption[Diameter estimates for (503) Evelyn]{%
  \label{fap: diam000503}
  Diameter estimates for (503) Evelyn.
}
\end{figure}

  \begin{figure}[!ht]
  \centering
  \includegraphics[width=.49\textwidth]{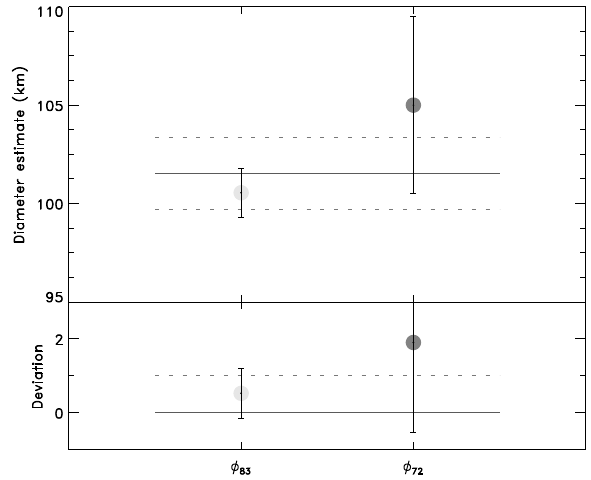}
\caption[Diameter estimates for (505) Cava]{%
  \label{fap: diam000505}
  Diameter estimates for (505) Cava.
}
\end{figure}

  \begin{figure}[!ht]
  \centering
  \includegraphics[width=.49\textwidth]{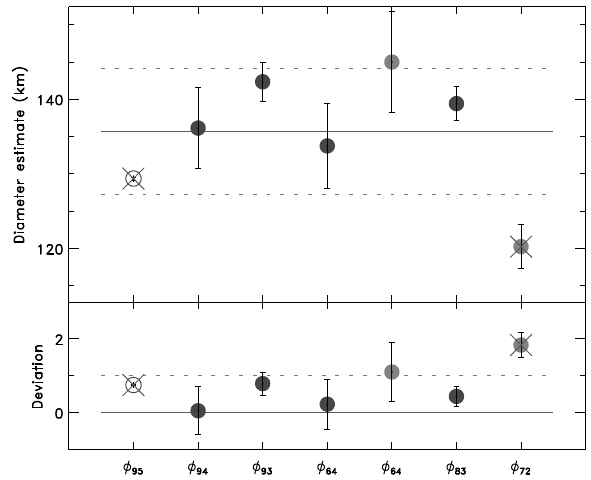}
\caption[Diameter estimates for (508) Princetonia]{%
  \label{fap: diam000508}
  Diameter estimates for (508) Princetonia.
}
\end{figure}

\clearpage
  \begin{figure}[!ht]
  \centering
  \includegraphics[width=.49\textwidth]{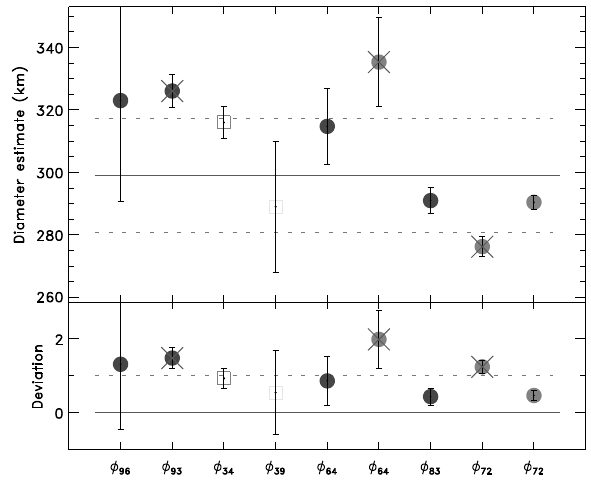}
\caption[Diameter estimates for (511) Davida]{%
  \label{fap: diam000511}
  Diameter estimates for (511) Davida.
}
\end{figure}

  \begin{figure}[!ht]
  \centering
  \includegraphics[width=.49\textwidth]{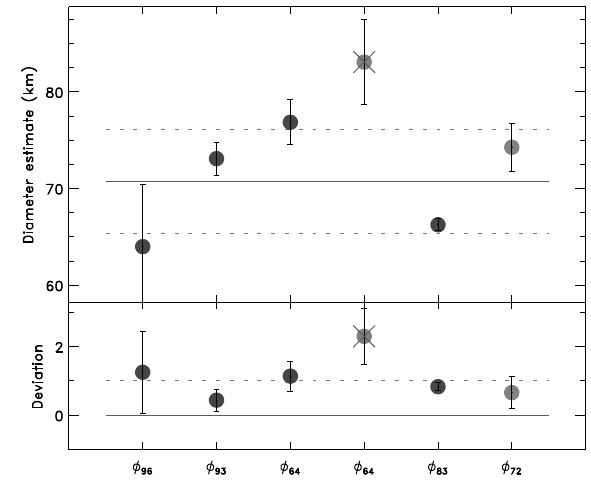}
\caption[Diameter estimates for (516) Amherstia]{%
  \label{fap: diam000516}
  Diameter estimates for (516) Amherstia.
}
\end{figure}

  \begin{figure}[!ht]
  \centering
  \includegraphics[width=.49\textwidth]{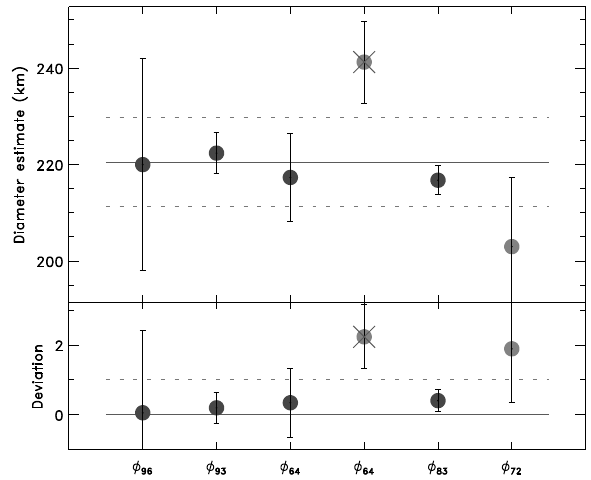}
\caption[Diameter estimates for (532) Herculina]{%
  \label{fap: diam000532}
  Diameter estimates for (532) Herculina.
}
\end{figure}

  \begin{figure}[!ht]
  \centering
  \includegraphics[width=.49\textwidth]{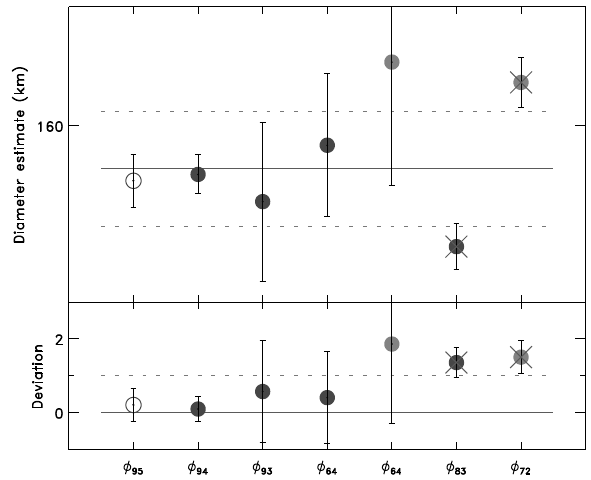}
\caption[Diameter estimates for (536) Merapi]{%
  \label{fap: diam000536}
  Diameter estimates for (536) Merapi.
}
\end{figure}

  \begin{figure}[!ht]
  \centering
  \includegraphics[width=.49\textwidth]{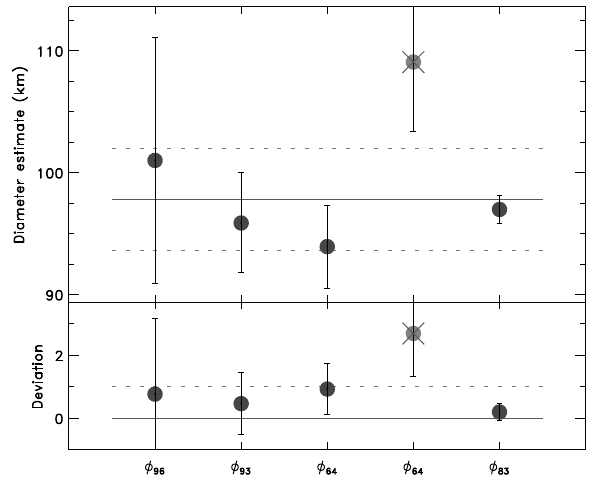}
\caption[Diameter estimates for (554) Peraga]{%
  \label{fap: diam000554}
  Diameter estimates for (554) Peraga.
}
\end{figure}

  \begin{figure}[!ht]
  \centering
  \includegraphics[width=.49\textwidth]{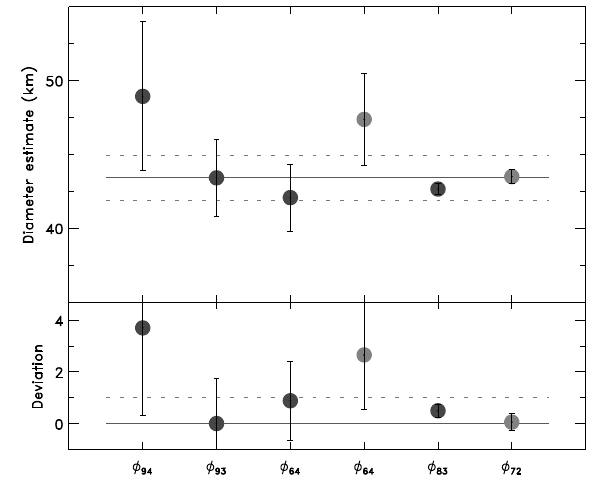}
\caption[Diameter estimates for (582) Olympia]{%
  \label{fap: diam000582}
  Diameter estimates for (582) Olympia.
}
\end{figure}

  \begin{figure}[!ht]
  \centering
  \includegraphics[width=.49\textwidth]{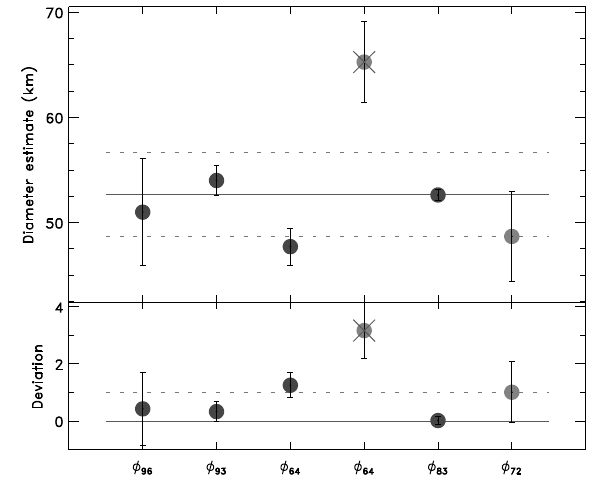}
\caption[Diameter estimates for (584) Semiramis]{%
  \label{fap: diam000584}
  Diameter estimates for (584) Semiramis.
}
\end{figure}

  \begin{figure}[!ht]
  \centering
  \includegraphics[width=.49\textwidth]{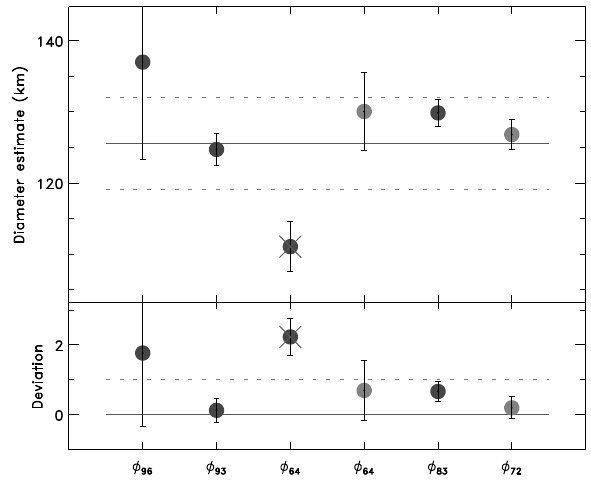}
\caption[Diameter estimates for (602) Marianna]{%
  \label{fap: diam000602}
  Diameter estimates for (602) Marianna.
}
\end{figure}

  \begin{figure}[!ht]
  \centering
  \includegraphics[width=.49\textwidth]{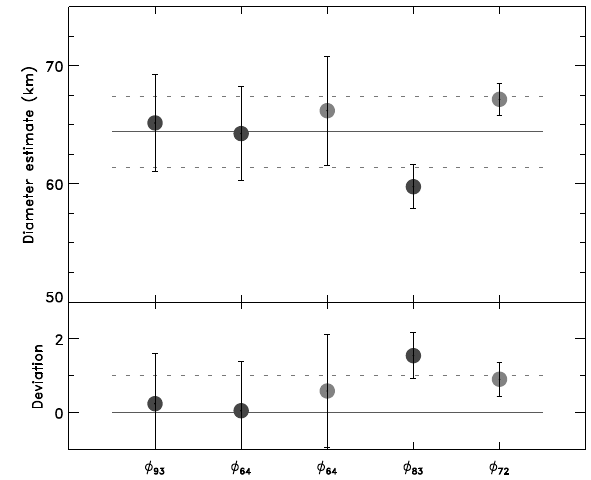}
\caption[Diameter estimates for (604) Tekmessa]{%
  \label{fap: diam000604}
  Diameter estimates for (604) Tekmessa.
}
\end{figure}

  \begin{figure}[!ht]
  \centering
  \includegraphics[width=.49\textwidth]{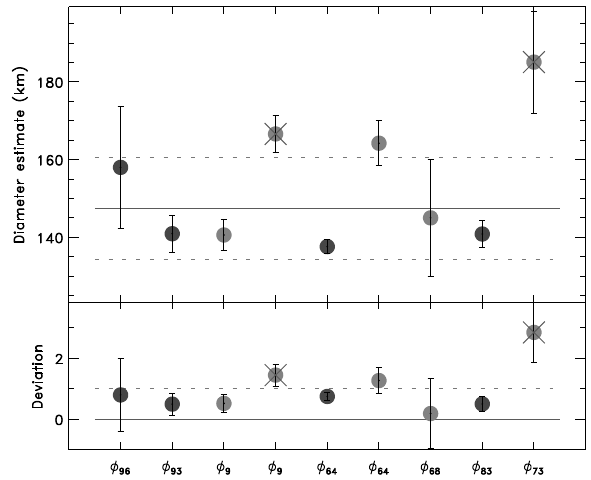}
\caption[Diameter estimates for (617) Patroclus]{%
  \label{fap: diam000617}
  Diameter estimates for (617) Patroclus.
}
\end{figure}

  \begin{figure}[!ht]
  \centering
  \includegraphics[width=.49\textwidth]{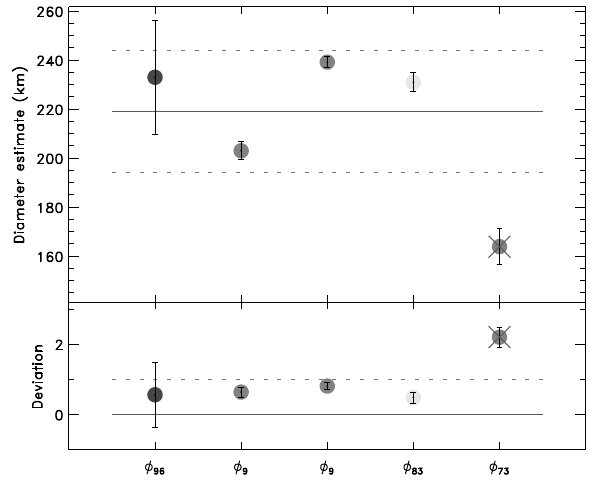}
\caption[Diameter estimates for (624) Hektor]{%
  \label{fap: diam000624}
  Diameter estimates for (624) Hektor.
}
\end{figure}

  \begin{figure}[!ht]
  \centering
  \includegraphics[width=.49\textwidth]{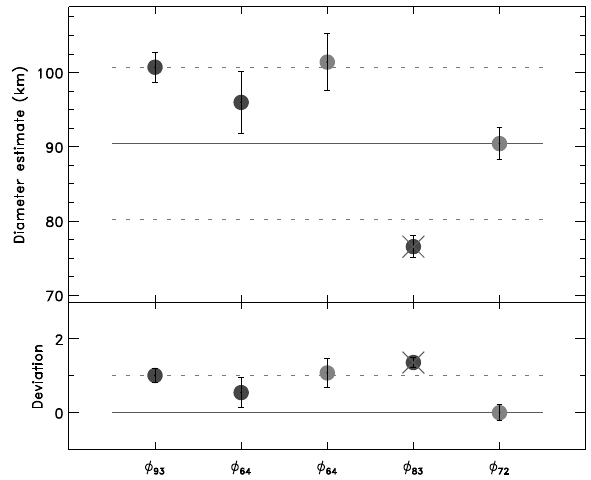}
\caption[Diameter estimates for (626) Notburga]{%
  \label{fap: diam000626}
  Diameter estimates for (626) Notburga.
}
\end{figure}

  \begin{figure}[!ht]
  \centering
  \includegraphics[width=.49\textwidth]{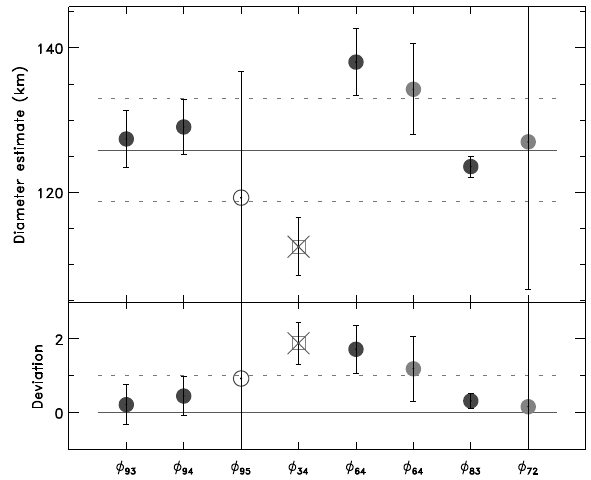}
\caption[Diameter estimates for (654) Zelinda]{%
  \label{fap: diam000654}
  Diameter estimates for (654) Zelinda.
}
\end{figure}

  \begin{figure}[!ht]
  \centering
  \includegraphics[width=.49\textwidth]{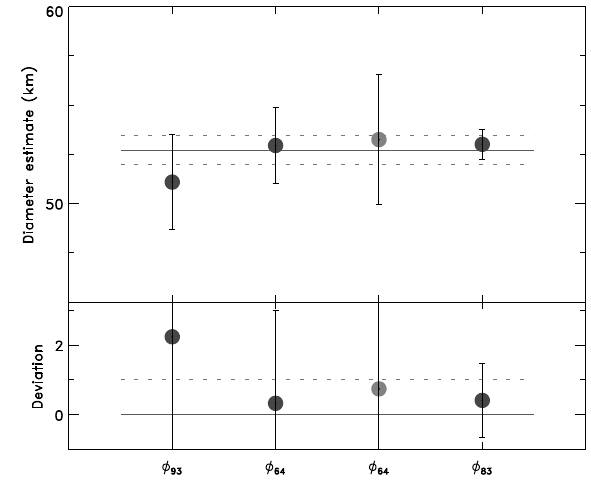}
\caption[Diameter estimates for (665) Sabine]{%
  \label{fap: diam000665}
  Diameter estimates for (665) Sabine.
}
\end{figure}

  \begin{figure}[!ht]
  \centering
  \includegraphics[width=.49\textwidth]{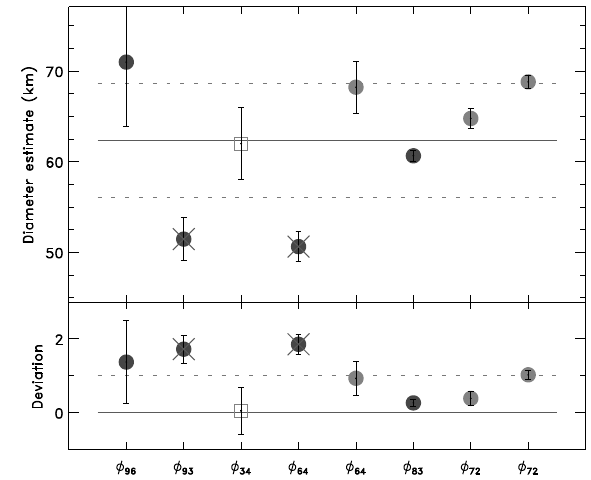}
\caption[Diameter estimates for (679) Pax]{%
  \label{fap: diam000679}
  Diameter estimates for (679) Pax.
}
\end{figure}

  \begin{figure}[!ht]
  \centering
  \includegraphics[width=.49\textwidth]{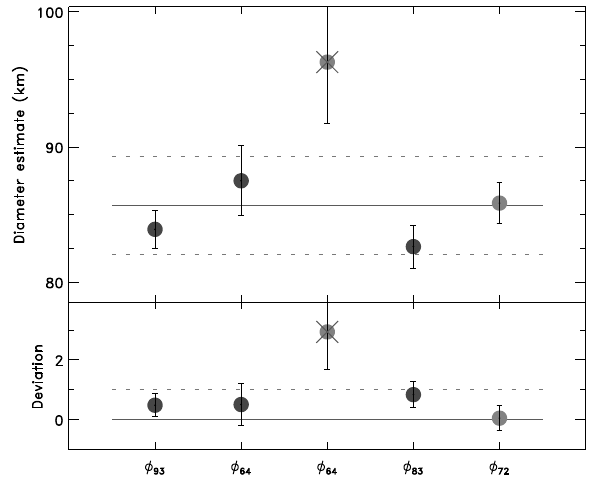}
\caption[Diameter estimates for (680) Genoveva]{%
  \label{fap: diam000680}
  Diameter estimates for (680) Genoveva.
}
\end{figure}

  \begin{figure}[!ht]
  \centering
  \includegraphics[width=.49\textwidth]{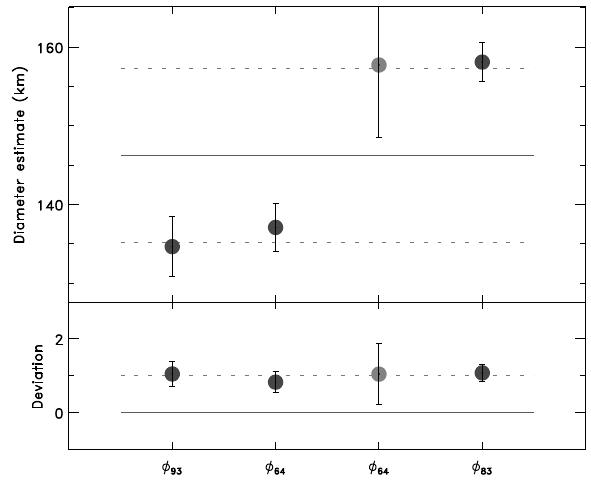}
\caption[Diameter estimates for (690) Wratislavia]{%
  \label{fap: diam000690}
  Diameter estimates for (690) Wratislavia.
}
\end{figure}

  \begin{figure}[!ht]
  \centering
  \includegraphics[width=.49\textwidth]{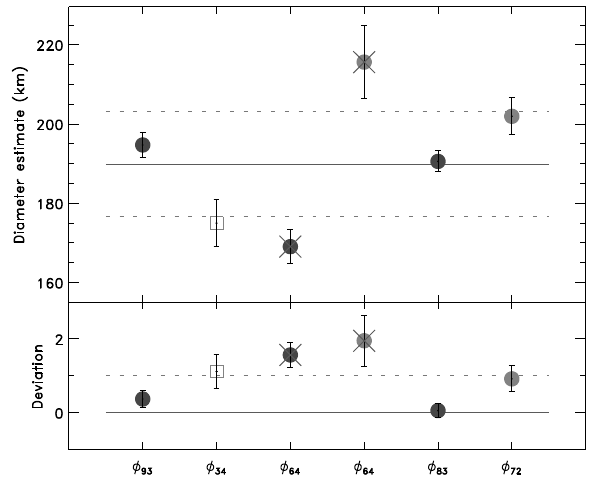}
\caption[Diameter estimates for (702) Alauda]{%
  \label{fap: diam000702}
  Diameter estimates for (702) Alauda.
}
\end{figure}

\clearpage
  \begin{figure}[!ht]
  \centering
  \includegraphics[width=.49\textwidth]{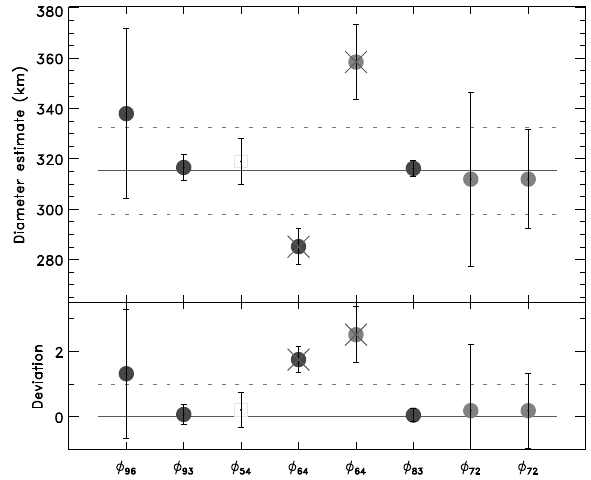}
\caption[Diameter estimates for (704) Interamnia]{%
  \label{fap: diam000704}
  Diameter estimates for (704) Interamnia.
}
\end{figure}

  \begin{figure}[!ht]
  \centering
  \includegraphics[width=.49\textwidth]{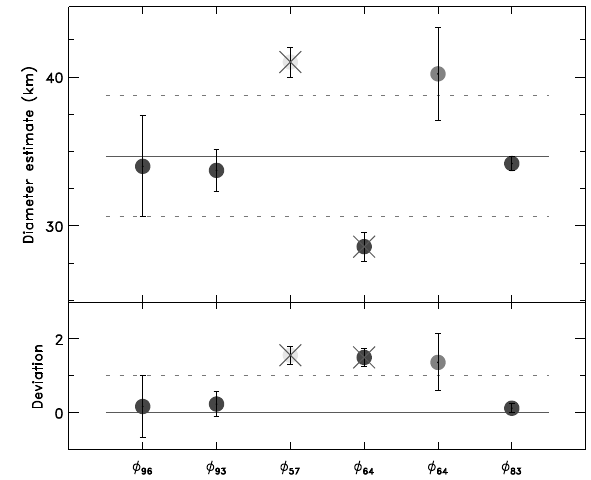}
\caption[Diameter estimates for (720) Bohlinia]{%
  \label{fap: diam000720}
  Diameter estimates for (720) Bohlinia.
}
\end{figure}

  \begin{figure}[!ht]
  \centering
  \includegraphics[width=.49\textwidth]{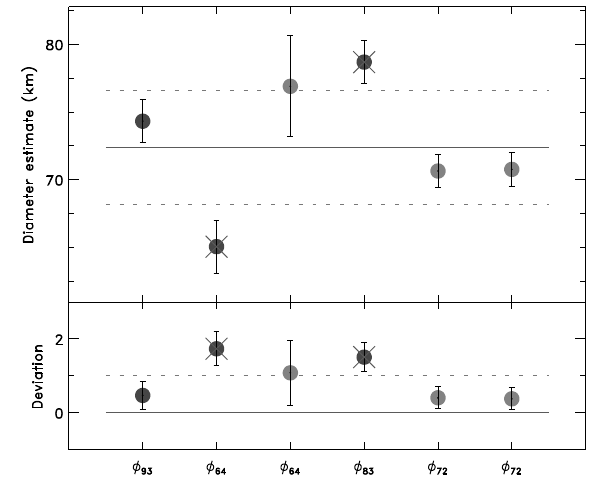}
\caption[Diameter estimates for (735) Marghanna]{%
  \label{fap: diam000735}
  Diameter estimates for (735) Marghanna.
}
\end{figure}

  \begin{figure}[!ht]
  \centering
  \includegraphics[width=.49\textwidth]{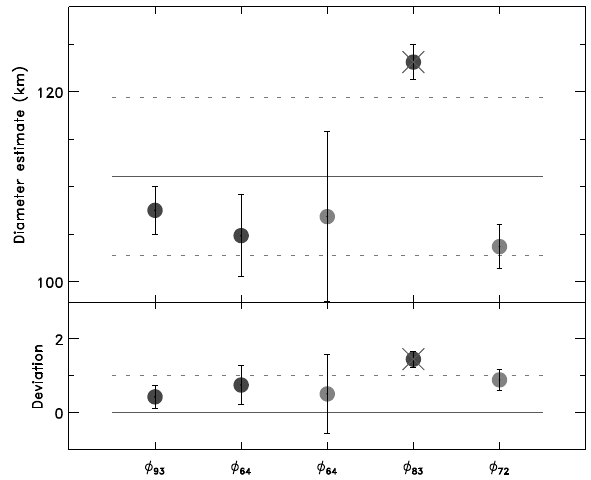}
\caption[Diameter estimates for (739) Mandeville]{%
  \label{fap: diam000739}
  Diameter estimates for (739) Mandeville.
}
\end{figure}

  \begin{figure}[!ht]
  \centering
  \includegraphics[width=.49\textwidth]{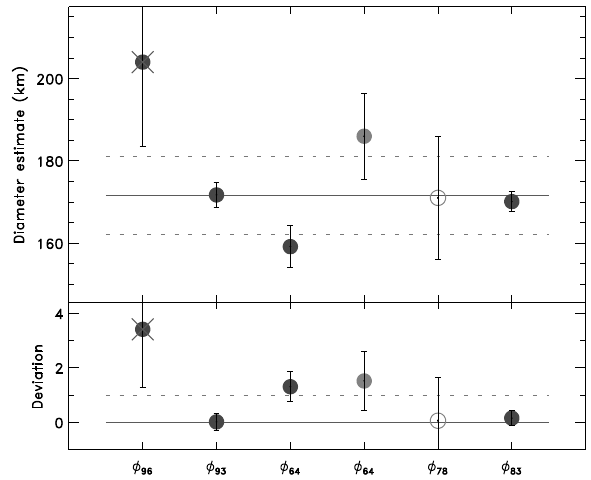}
\caption[Diameter estimates for (747) Winchester]{%
  \label{fap: diam000747}
  Diameter estimates for (747) Winchester.
}
\end{figure}

  \begin{figure}[!ht]
  \centering
  \includegraphics[width=.49\textwidth]{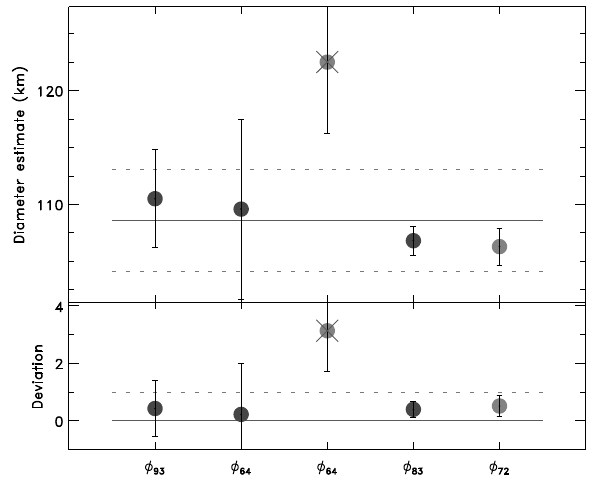}
\caption[Diameter estimates for (751) Faina]{%
  \label{fap: diam000751}
  Diameter estimates for (751) Faina.
}
\end{figure}

  \begin{figure}[!ht]
  \centering
  \includegraphics[width=.49\textwidth]{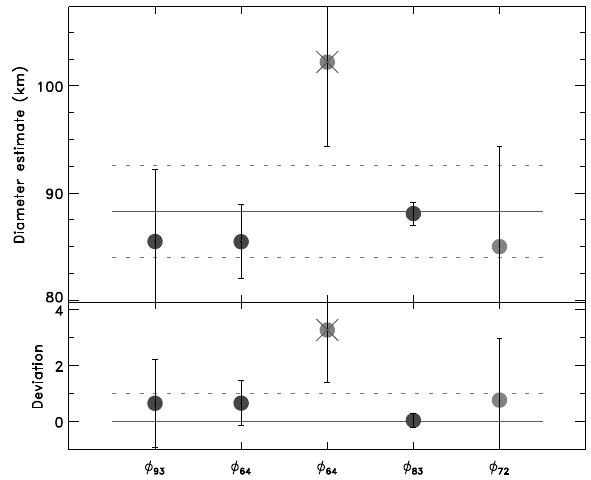}
\caption[Diameter estimates for (758) Mancunia]{%
  \label{fap: diam000758}
  Diameter estimates for (758) Mancunia.
}
\end{figure}

  \begin{figure}[!ht]
  \centering
  \includegraphics[width=.49\textwidth]{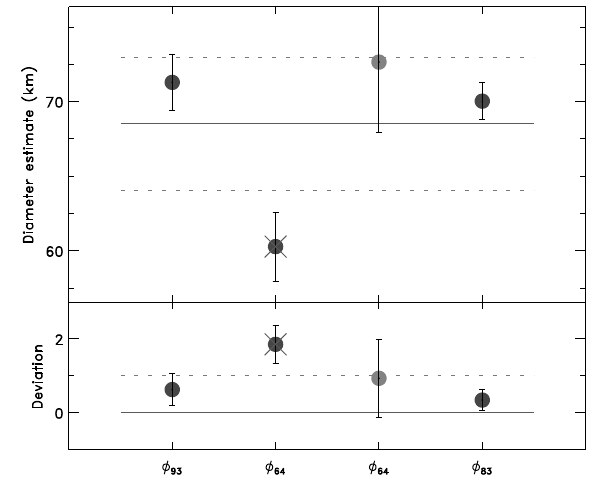}
\caption[Diameter estimates for (760) Massinga]{%
  \label{fap: diam000760}
  Diameter estimates for (760) Massinga.
}
\end{figure}

  \begin{figure}[!ht]
  \centering
  \includegraphics[width=.49\textwidth]{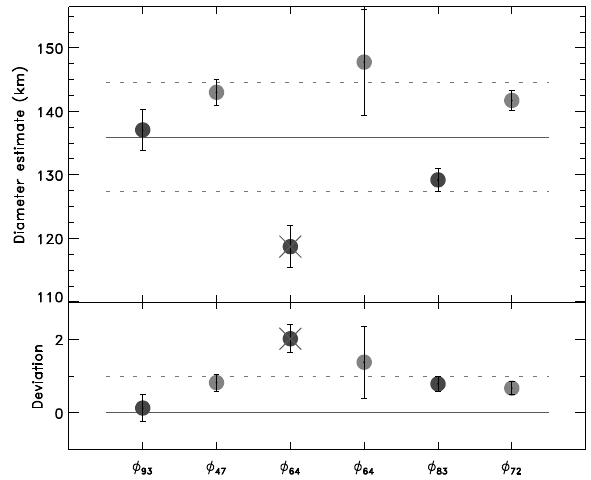}
\caption[Diameter estimates for (762) Pulcova]{%
  \label{fap: diam000762}
  Diameter estimates for (762) Pulcova.
}
\end{figure}

  \begin{figure}[!ht]
  \centering
  \includegraphics[width=.49\textwidth]{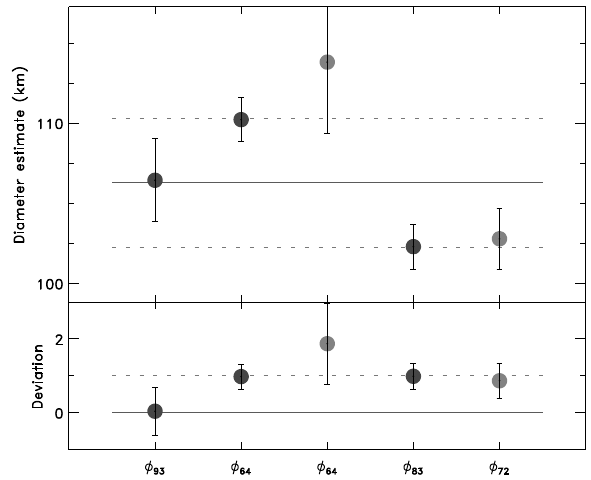}
\caption[Diameter estimates for (769) Tatjana]{%
  \label{fap: diam000769}
  Diameter estimates for (769) Tatjana.
}
\end{figure}

  \begin{figure}[!ht]
  \centering
  \includegraphics[width=.49\textwidth]{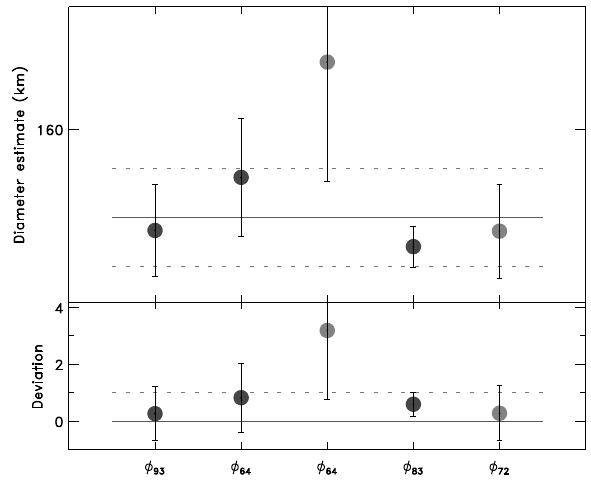}
\caption[Diameter estimates for (776) Berbericia]{%
  \label{fap: diam000776}
  Diameter estimates for (776) Berbericia.
}
\end{figure}

  \begin{figure}[!ht]
  \centering
  \includegraphics[width=.49\textwidth]{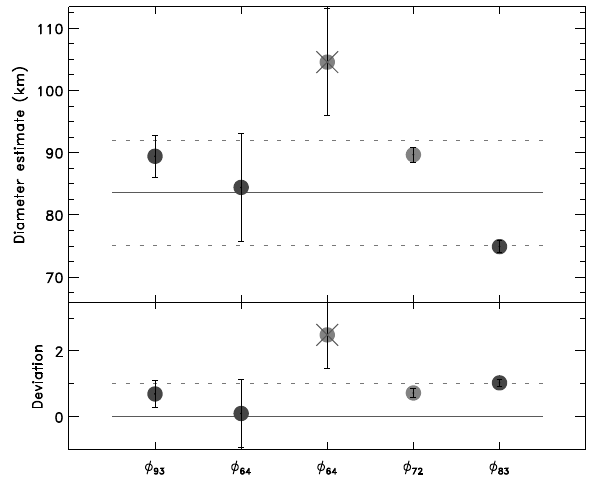}
\caption[Diameter estimates for (784) Pickeringia]{%
  \label{fap: diam000784}
  Diameter estimates for (784) Pickeringia.
}
\end{figure}

  \begin{figure}[!ht]
  \centering
  \includegraphics[width=.49\textwidth]{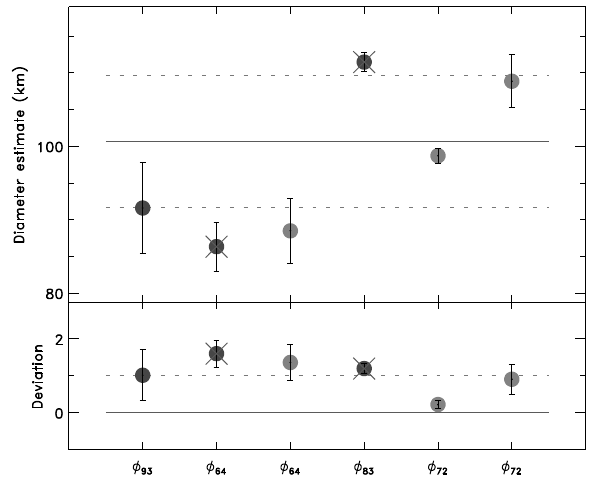}
\caption[Diameter estimates for (786) Bredichina]{%
  \label{fap: diam000786}
  Diameter estimates for (786) Bredichina.
}
\end{figure}

  \begin{figure}[!ht]
  \centering
  \includegraphics[width=.49\textwidth]{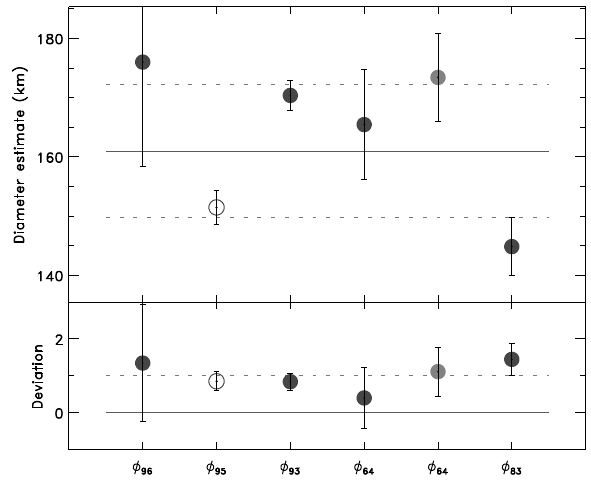}
\caption[Diameter estimates for (790) Pretoria]{%
  \label{fap: diam000790}
  Diameter estimates for (790) Pretoria.
}
\end{figure}

  \begin{figure}[!ht]
  \centering
  \includegraphics[width=.49\textwidth]{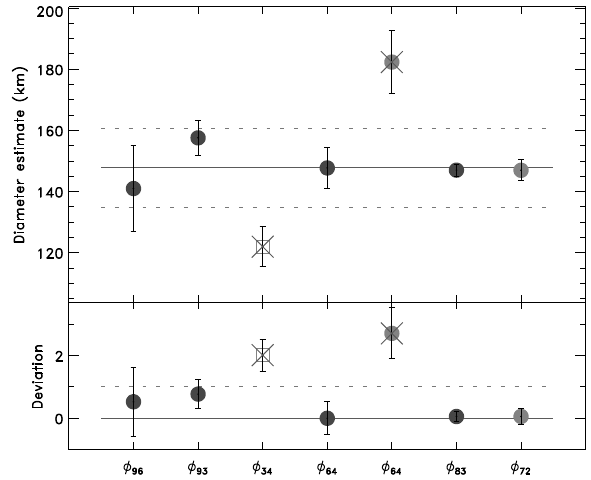}
\caption[Diameter estimates for (804) Hispania]{%
  \label{fap: diam000804}
  Diameter estimates for (804) Hispania.
}
\end{figure}

  \begin{figure}[!ht]
  \centering
  \includegraphics[width=.49\textwidth]{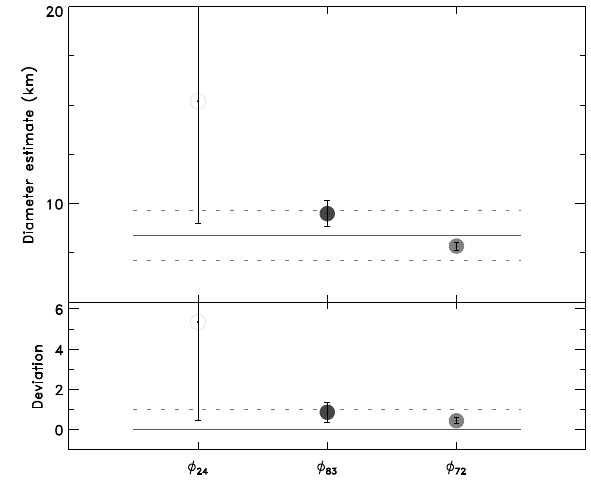}
\caption[Diameter estimates for (854) Frostia]{%
  \label{fap: diam000854}
  Diameter estimates for (854) Frostia.
}
\end{figure}

  \begin{figure}[!ht]
  \centering
  \includegraphics[width=.49\textwidth]{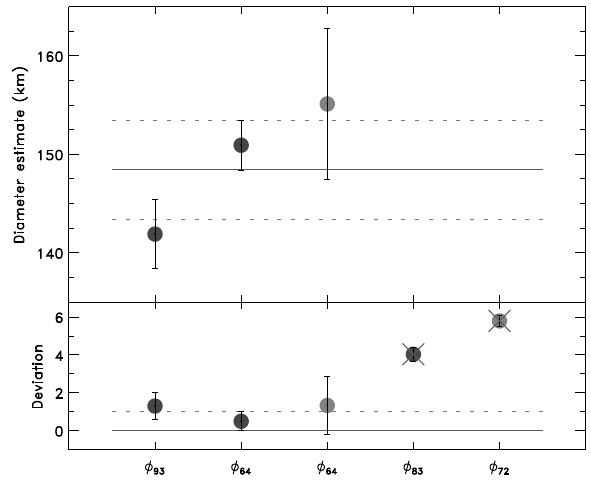}
\caption[Diameter estimates for (895) Helio]{%
  \label{fap: diam000895}
  Diameter estimates for (895) Helio.
}
\end{figure}

  \begin{figure}[!ht]
  \centering
  \includegraphics[width=.49\textwidth]{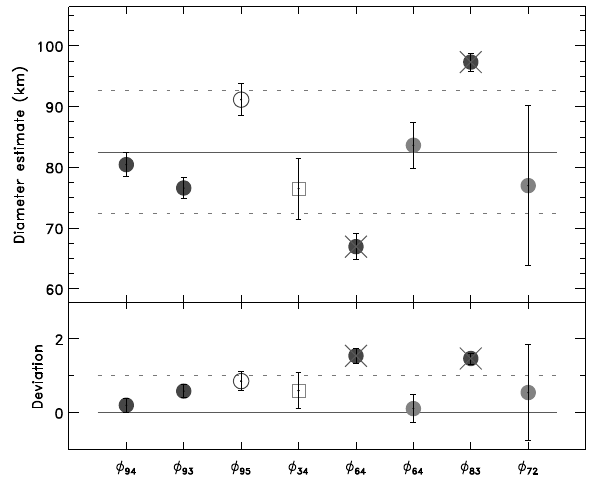}
\caption[Diameter estimates for (914) Palisana]{%
  \label{fap: diam000914}
  Diameter estimates for (914) Palisana.
}
\end{figure}

\clearpage
  \begin{figure}[!ht]
  \centering
  \includegraphics[width=.49\textwidth]{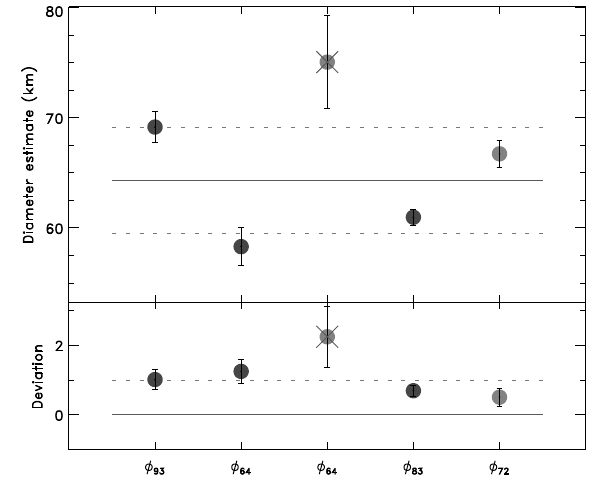}
\caption[Diameter estimates for (949) Hel]{%
  \label{fap: diam000949}
  Diameter estimates for (949) Hel.
}
\end{figure}

  \begin{figure}[!ht]
  \centering
  \includegraphics[width=.49\textwidth]{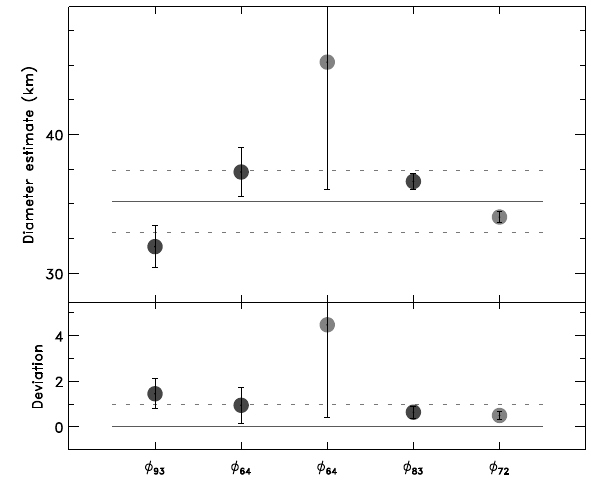}
\caption[Diameter estimates for (1013) Tombecka]{%
  \label{fap: diam001013}
  Diameter estimates for (1013) Tombecka.
}
\end{figure}

  \begin{figure}[!ht]
  \centering
  \includegraphics[width=.49\textwidth]{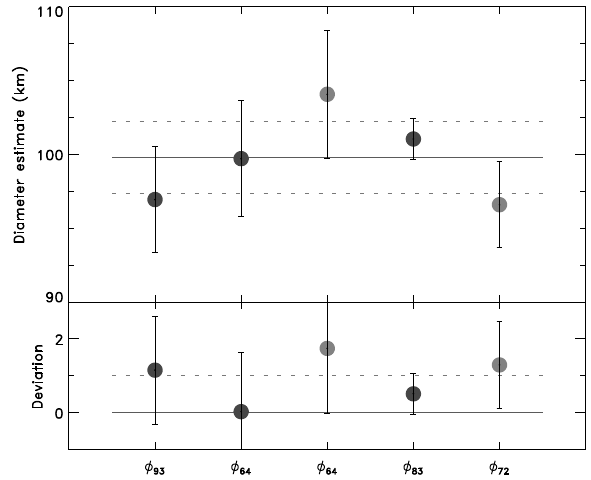}
\caption[Diameter estimates for (1015) Christa]{%
  \label{fap: diam001015}
  Diameter estimates for (1015) Christa.
}
\end{figure}

  \begin{figure}[!ht]
  \centering
  \includegraphics[width=.49\textwidth]{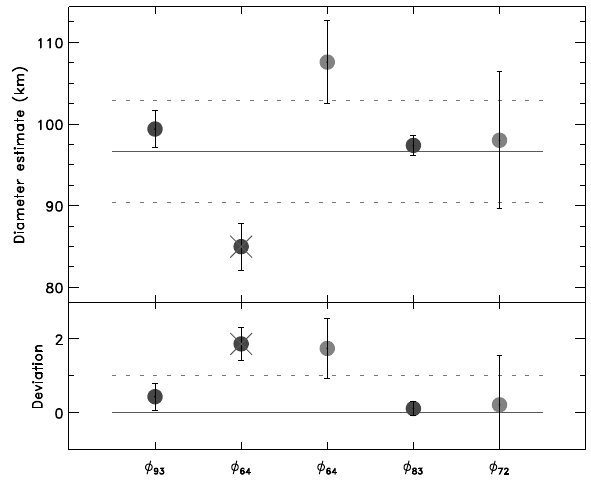}
\caption[Diameter estimates for (1021) Flammario]{%
  \label{fap: diam001021}
  Diameter estimates for (1021) Flammario.
}
\end{figure}

  \begin{figure}[!ht]
  \centering
  \includegraphics[width=.49\textwidth]{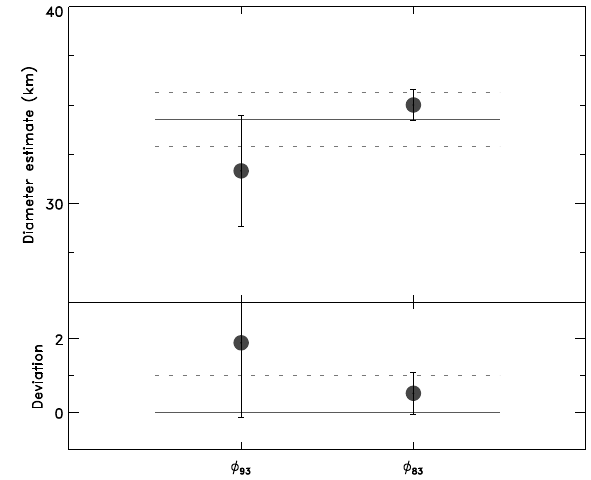}
\caption[Diameter estimates for (1036) Ganymed]{%
  \label{fap: diam001036}
  Diameter estimates for (1036) Ganymed.
}
\end{figure}

  \begin{figure}[!ht]
  \centering
  \includegraphics[width=.49\textwidth]{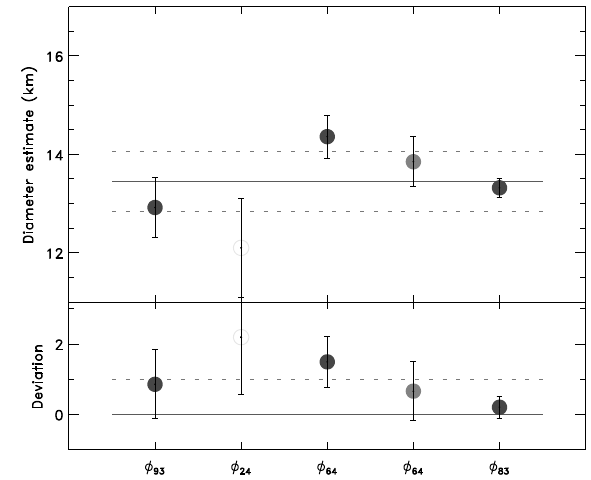}
\caption[Diameter estimates for (1089) Tama]{%
  \label{fap: diam001089}
  Diameter estimates for (1089) Tama.
}
\end{figure}

  \begin{figure}[!ht]
  \centering
  \includegraphics[width=.49\textwidth]{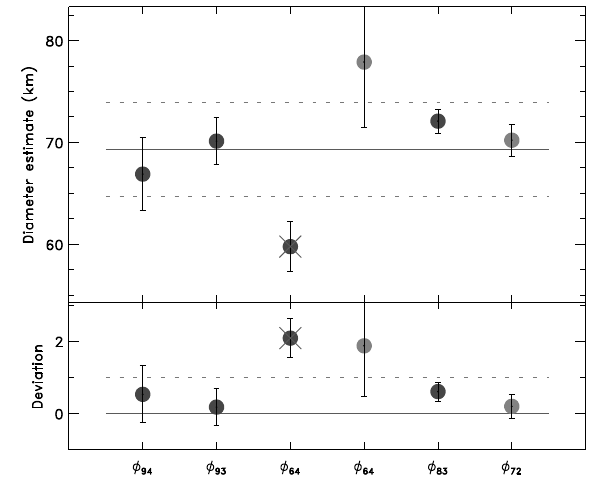}
\caption[Diameter estimates for (1171) Rusthawelia]{%
  \label{fap: diam001171}
  Diameter estimates for (1171) Rusthawelia.
}
\end{figure}

  \begin{figure}[!ht]
  \centering
  \includegraphics[width=.49\textwidth]{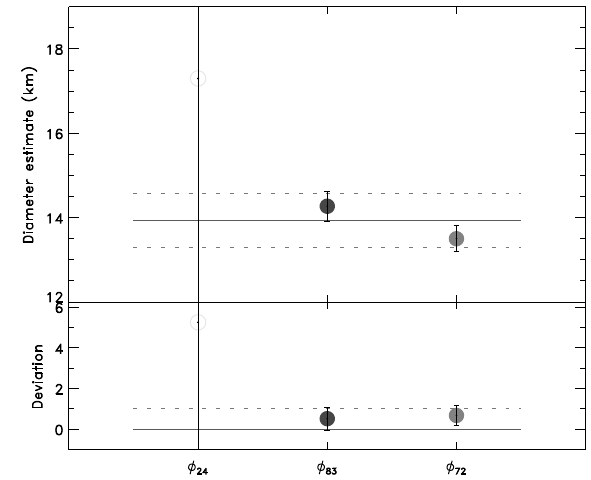}
\caption[Diameter estimates for (1313) Berna]{%
  \label{fap: diam001313}
  Diameter estimates for (1313) Berna.
}
\end{figure}

  \begin{figure}[!ht]
  \centering
  \includegraphics[width=.49\textwidth]{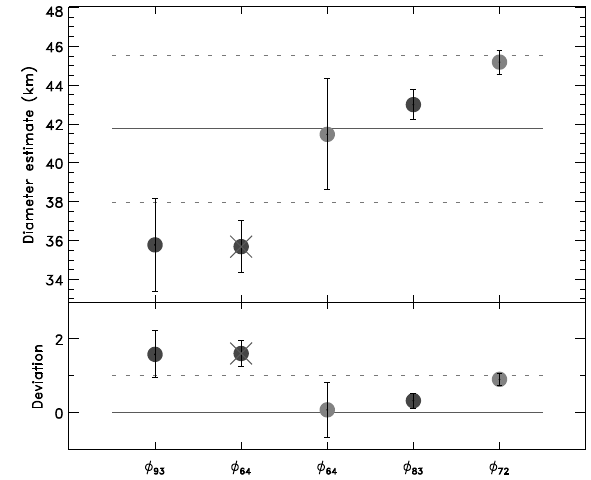}
\caption[Diameter estimates for (1669) Dagmar]{%
  \label{fap: diam001669}
  Diameter estimates for (1669) Dagmar.
}
\end{figure}

  \begin{figure}[!ht]
  \centering
  \includegraphics[width=.49\textwidth]{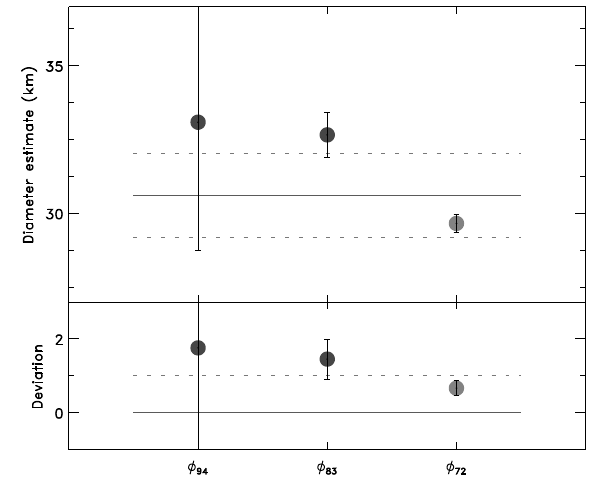}
\caption[Diameter estimates for (1686) De Sitter]{%
  \label{fap: diam001686}
  Diameter estimates for (1686) De Sitter.
}
\end{figure}

  \begin{figure}[!ht]
  \centering
  \includegraphics[width=.49\textwidth]{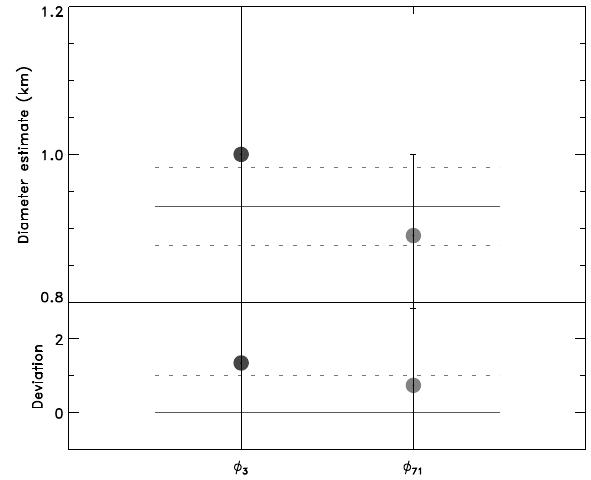}
\caption[Diameter estimates for (3671) Dionysus]{%
  \label{fap: diam003671}
  Diameter estimates for (3671) Dionysus.
}
\end{figure}

  \begin{figure}[!ht]
  \centering
  \includegraphics[width=.49\textwidth]{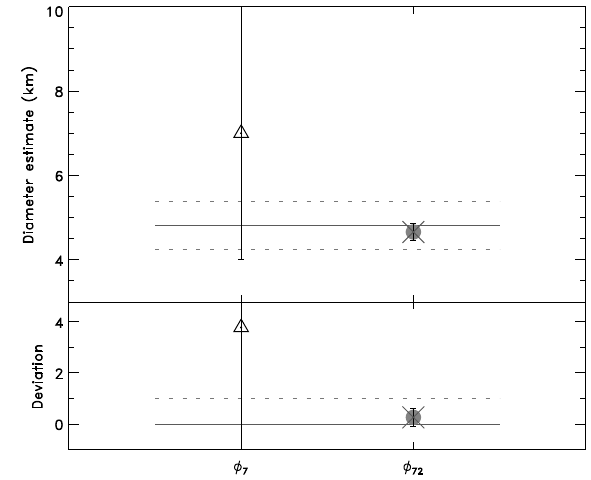}
\caption[Diameter estimates for (3749) Balam]{%
  \label{fap: diam003749}
  Diameter estimates for (3749) Balam.
  The diameter estimate from $\phi_{72}$ give an unrealistic high density of 9.6\,$\pm$\,1.3  if used alone.  Only the estimate from $\phi_{7}$ is used.
}
\end{figure}

  \begin{figure}[!ht]
  \centering
  \includegraphics[width=.49\textwidth]{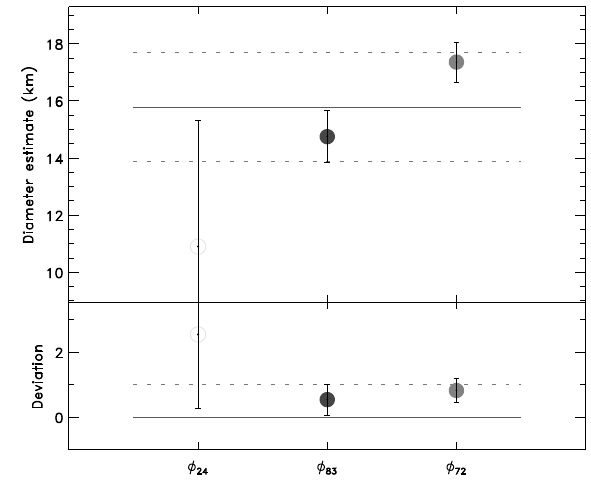}
\caption[Diameter estimates for (4492) Debussy]{%
  \label{fap: diam004492}
  Diameter estimates for (4492) Debussy.
}
\end{figure}

  \begin{figure}[!ht]
  \centering
  \includegraphics[width=.49\textwidth]{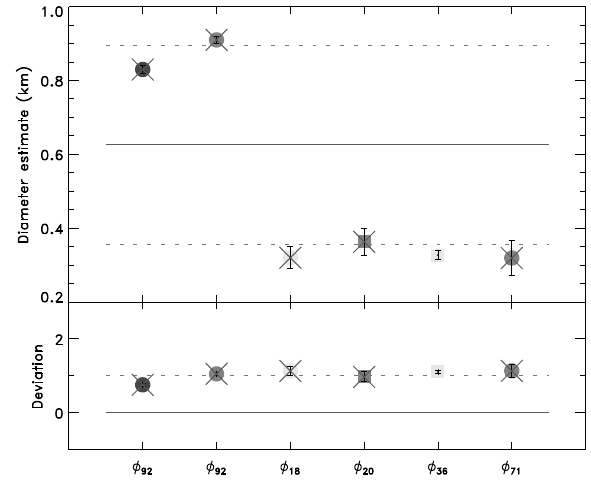}
\caption[Diameter estimates for (25143) Itokawa]{%
  \label{fap: diam025143}
  Diameter estimates for (25143) Itokawa.
  Only the flyby estimate from $\phi_{36}$ is used here. 
}
\end{figure}

  \begin{figure}[!ht]
  \centering
  \includegraphics[width=.49\textwidth]{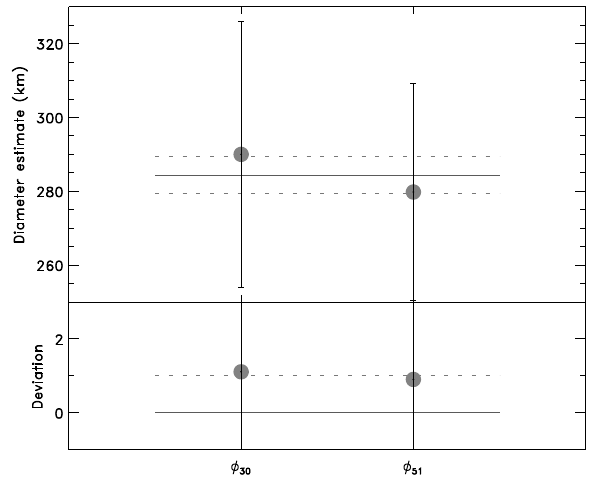}
\caption[Diameter estimates for (26308) 1998 SM165]{%
  \label{fap: diam026308}
  Diameter estimates for (26308) 1998 SM165.
}
\end{figure}

  \begin{figure}[!ht]
  \centering
  \includegraphics[width=.49\textwidth]{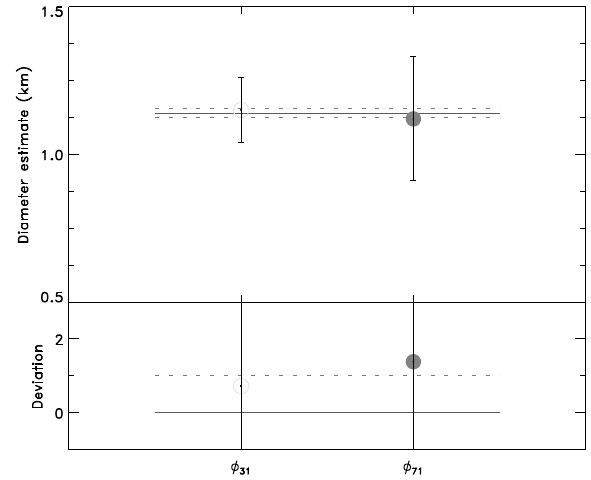}
\caption[Diameter estimates for (35107) 1991 VH]{%
  \label{fap: diam035107}
  Diameter estimates for (35107) 1991 VH.
}
\end{figure}

  \begin{figure}[!ht]
  \centering
  \includegraphics[width=.49\textwidth]{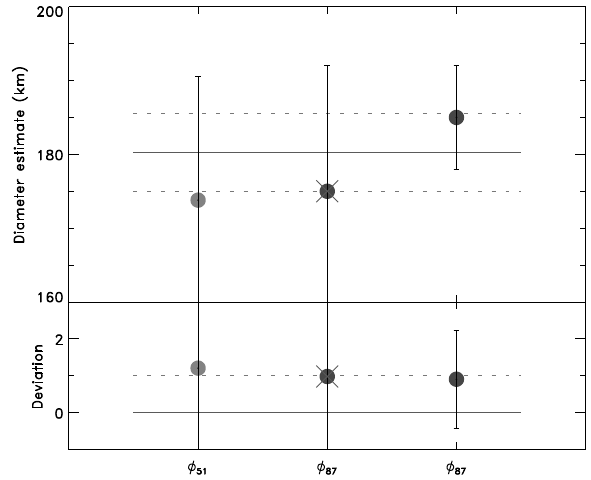}
\caption[Diameter estimates for (42355) Typhon]{%
  \label{fap: diam042355}
  Diameter estimates for (42355) Typhon.
}
\end{figure}

\clearpage
  \begin{figure}[!ht]
  \centering
  \includegraphics[width=.49\textwidth]{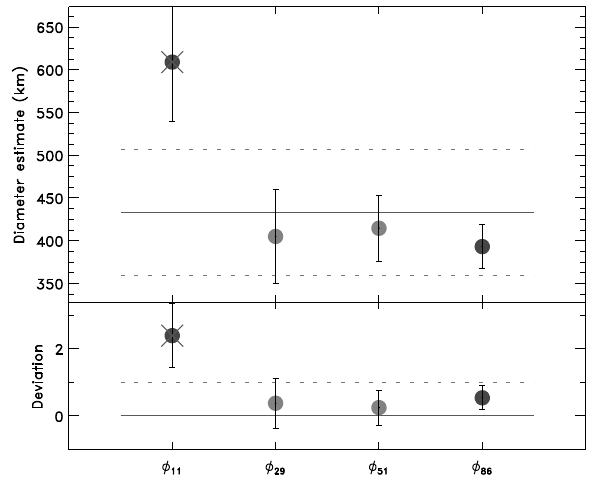}
\caption[Diameter estimates for (47171) 1999 TC36]{%
  \label{fap: diam047171}
  Diameter estimates for (47171) 1999 TC36.
}
\end{figure}

  \begin{figure}[!ht]
  \centering
  \includegraphics[width=.49\textwidth]{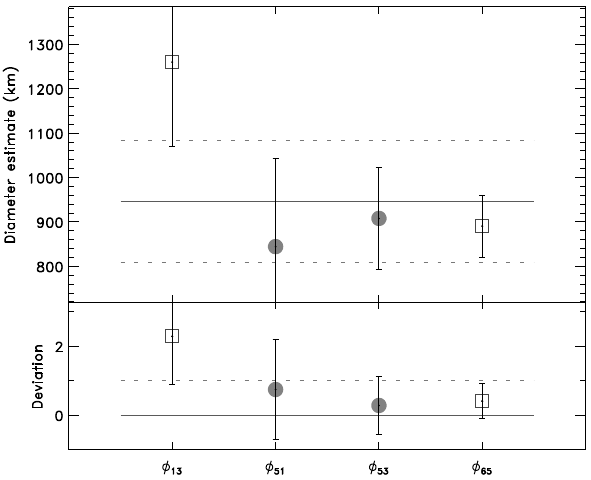}
\caption[Diameter estimates for (50000) Quaoar]{%
  \label{fap: diam050000}
  Diameter estimates for (50000) Quaoar.
}
\end{figure}

  \begin{figure}[!ht]
  \centering
  \includegraphics[width=.49\textwidth]{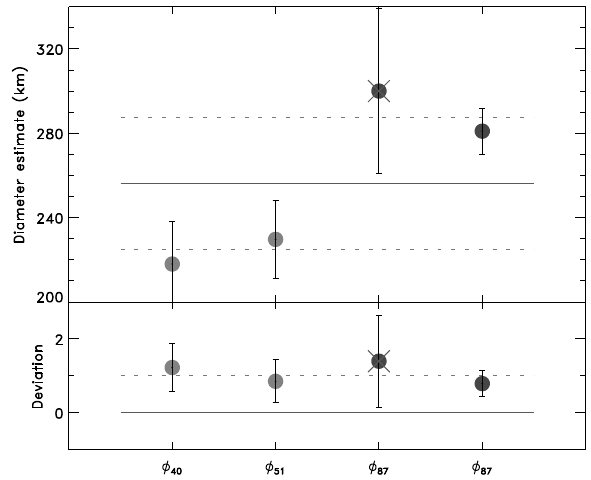}
\caption[Diameter estimates for (65489) Ceto]{%
  \label{fap: diam065489}
  Diameter estimates for (65489) Ceto.
}
\end{figure}

  \begin{figure}[!ht]
  \centering
  \includegraphics[width=.49\textwidth]{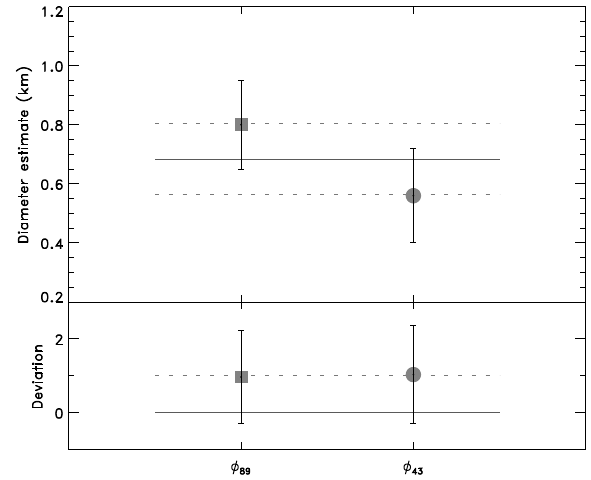}
\caption[Diameter estimates for (66063) 1998 RO1]{%
  \label{fap: diam066063}
  Diameter estimates for (66063) 1998 RO1.
}
\end{figure}

  \begin{figure}[!ht]
  \centering
  \includegraphics[width=.49\textwidth]{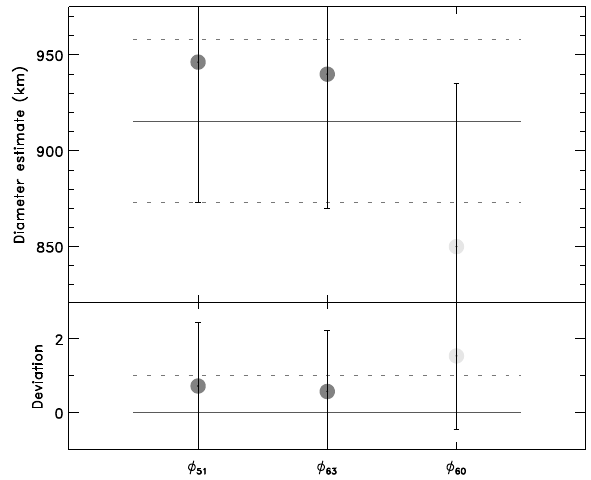}
\caption[Diameter estimates for (90482) Orcus]{%
  \label{fap: diam090482}
  Diameter estimates for (90482) Orcus.
}
\end{figure}

  \begin{figure}[!ht]
  \centering
  \includegraphics[width=.49\textwidth]{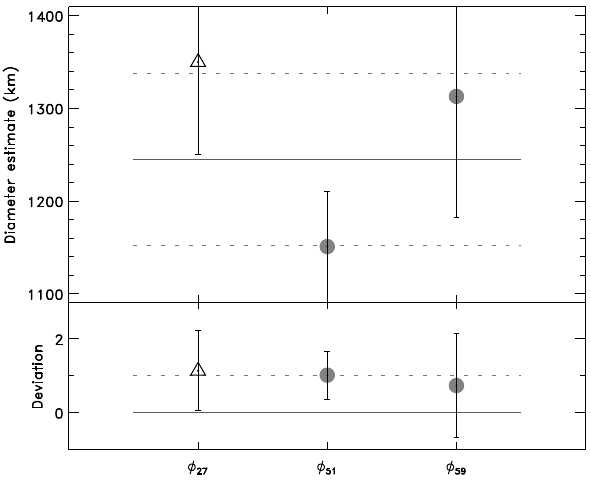}
\caption[Diameter estimates for (136108) Haumea]{%
  \label{fap: diam136108}
  Diameter estimates for (136108) Haumea.
}
\end{figure}

  \begin{figure}[!ht]
  \centering
  \includegraphics[width=.49\textwidth]{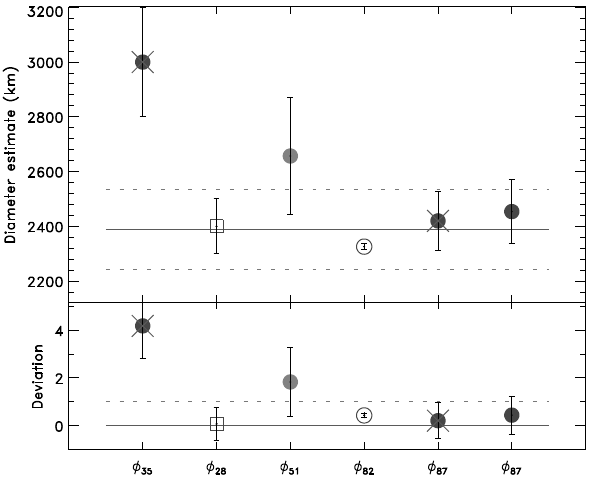}
\caption[Diameter estimates for (136199) Eris]{%
  \label{fap: diam136199}
  Diameter estimates for (136199) Eris.
}
\end{figure}

  \begin{figure}[!ht]
  \centering
  \includegraphics[width=.49\textwidth]{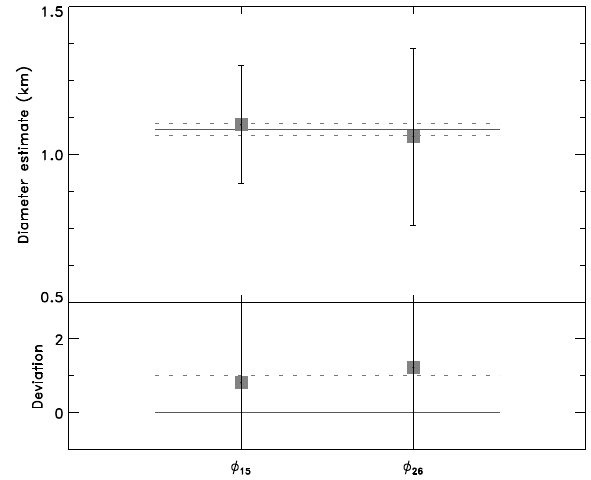}
\caption[Diameter estimates for (164121) 2003 YT1]{%
  \label{fap: diam164121}
  Diameter estimates for (164121) 2003 YT1.
}
\end{figure}

  \begin{figure}[!ht]
  \centering
  \includegraphics[width=.49\textwidth]{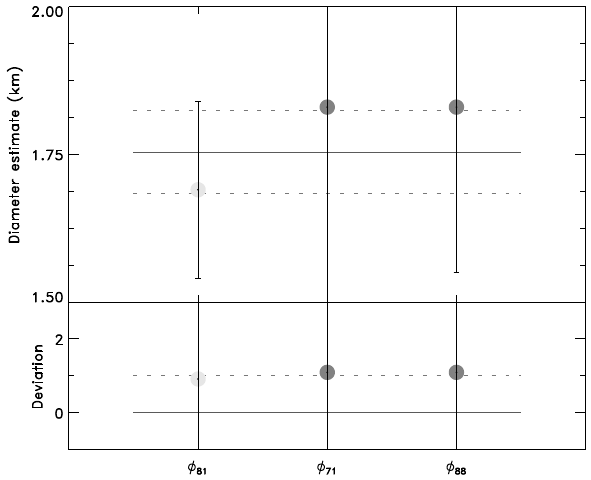}
\caption[Diameter estimates for (175706) 1996 FG3]{%
  \label{fap: diam175706}
  Diameter estimates for (175706) 1996 FG3.
}
\end{figure}

  \begin{figure}[!ht]
  \centering
  \includegraphics[width=.49\textwidth]{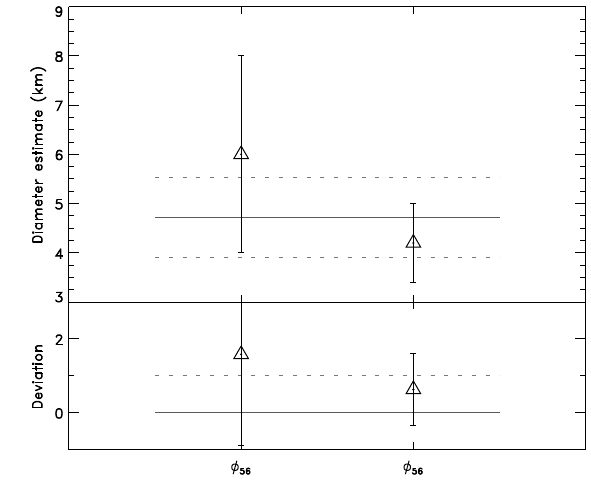}
\caption[Diameter estimates for 2P/Encke]{%
  \label{fap: diam2P/Encke}
  Diameter estimates for 2P/Encke.
}
\end{figure}

  \begin{figure}[!ht]
  \centering
  \includegraphics[width=.49\textwidth]{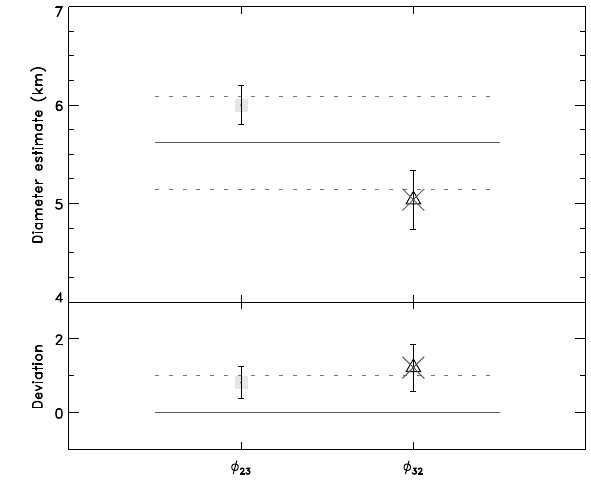}
\caption[Diameter estimates for 9P/Tempel1]{%
  \label{fap: diam9P/Tempel1}
  Diameter estimates for 9P/Tempel1.
}
\end{figure}

  \begin{figure}[!ht]
  \centering
  \includegraphics[width=.49\textwidth]{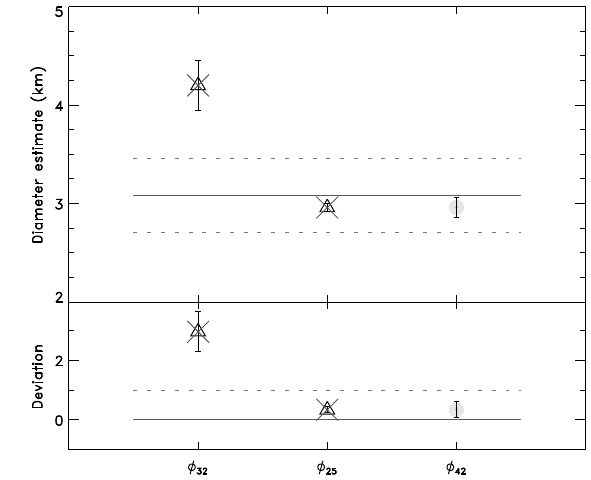}
\caption[Diameter estimates for 67P/C-G]{%
  \label{fap: diam67P/C-G}
  Diameter estimates for 67P/C-G.
}
\end{figure}

  \begin{figure}[!ht]
  \centering
  \includegraphics[width=.49\textwidth]{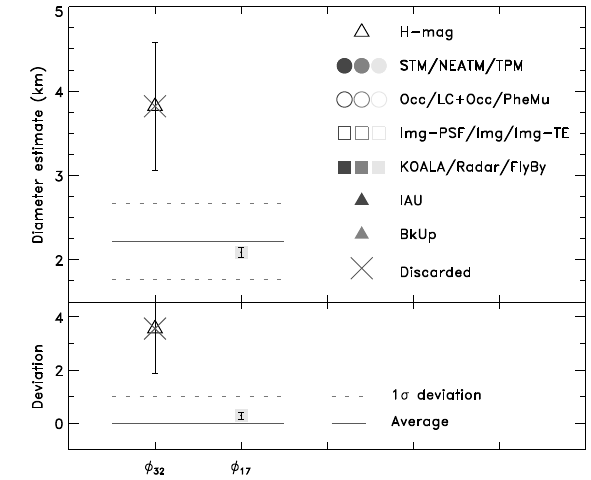}
\caption[Diameter estimates for 81P/Wild2]{%
  \label{fap: diam81P/Wild2}
  Diameter estimates for 81P/Wild2.
}
\end{figure}


\section{Compilation of indirect density estimates\label{app: indens}}
  \indent The \numb{24} indirect density estimates
  gathered in the literature are listed in 
  Table~\ref{tabSM: indirect}.
  See~\ref{app: biblio} for the references.
\setcounter{table}{0}
\begin{table}[!ht]
\centering
\caption[Indirect density estimates]{%
  Compilation of indirect estimates of density ($\rho$) for \numb{24} objects,
  with their associated uncertainty ($\delta \rho$),
  bibliographic references (see~\ref{app: biblio}), 
  and method of analysis: 
  \textsl{PheMu}: mutual eclipsing phenomena in binary systems, 
  \textsl{BinRad}: radar imaging, 
  \textsl{Assum}: assumed, and, 
  for comet nucleii, 
  \textsl{CNGF}: non-gravitational forces, and
  \textsl{BkUp}: break-up modeling. 
  Estimates marked with a dagger ($\dagger$) were used in the
  computation of the average density with other estimates derived from
  the mass and volume measurements. Other estimates are the only
  density estimates available for these targets.
\label{tabSM: indirect}
}
\begin{tabular}{r@{ }lr@{.}lr@{.}lll}
 \hline
 \hline
 \# & Designation & \multicolumn{2}{c}{$\rho$} & \multicolumn{2}{c}{$\delta \rho$} & Method & Refs.\\
 \hline
    809 & Lundia          &  1&64  &  0&1   &  PheMu    & D$_{10}$ \\ 
\noalign{\smallskip}%
    854 & Frostia         &  0&88  &  0&13  &  PheMu    & D$_{8}$ \\ 
\noalign{\smallskip}%
   1089 & Tama            &  2&52  &  0&29  &  PheMu    & D$_{8}$ \\ 
\noalign{\smallskip}%
   1313 & Berna           &  1&21  &  0&14  &  PheMu    & D$_{8}$ \\ 
\noalign{\smallskip}%
   3169 & Ostro           &  2&6   &  0&2   &  PheMu    & D$_{9}$ \\ 
\noalign{\smallskip}%
   3671 & Dionysus        &  1&60  &  0&60  &  PheMu    & D$_{7}$ \\ 
\noalign{\smallskip}%
   4492 & Debussy         &  0&90  &  0&10  &  PheMu    & D$_{8}$ \\ 
\noalign{\smallskip}%
   5381 & Sekhmet         &  1&98  &  0&65  &  BinRad   & D$_{3}$ \\ 
\noalign{\smallskip}%
  35107 & 1991 VH         &  1&40  &  0&50  &  PheMu    & D$_{7}$ \\ 
  35107 & 1991 VH         &  1&60  &  0&50  &  PheMu    & D$_{7}$ \\ 
\noalign{\smallskip}%
  65803 & Didymos         &  1&70  &  0&40  &  PheMu    & D$_{7}$ \\ 
  65803 & Didymos         &  2&10  &  0&60  &  PheMu    & D$_{7}$ \\ 
  65803 & Didymos         &  1&70  &  0&50  &  PheMu    & D$_{11}$ \\ 
  65803 & Didymos         &  2&10  &  0&65  &  PheMu    & D$_{11}$ \\ 
\noalign{\smallskip}%
  66063 & 1998 RO1        &  1&5   &  1&15  &  PheMu    & D$_{11}$ \\ 
  66063 & 1998 RO1        &  4&1   &  1&8   &  PheMu    & D$_{11}$ \\ 
\noalign{\smallskip}%
  66391 & 1999 KW4        &  1&20  &  0&80  &  PheMu    & D$_{11}^\dagger$ \\ 
\noalign{\smallskip}%
 175706 & 1996 FG3        &  1&40  &  0&30  &  PheMu    & D$_{2}$ \\ 
 175706 & 1996 FG3        &  1&30  &  0&60  &  PheMu    & D$_{7}$ \\ 
 175706 & 1996 FG3        &  1&4   &  1&05  &  PheMu    & D$_{11}$ \\ 
\noalign{\smallskip}%
 185851 & 2000 DP107      &  0&8   &  0&6   &  PheMu    & D$_{11}$ \\ 
 185851 & 2000 DP107      &  1&1   &  1&5   &  PheMu    & D$_{11}$ \\ 
\noalign{\smallskip}%
        & 2001 QW322      &  1&0   &  1&0   &  Assum    & D$_{11}$ \\ 
\noalign{\smallskip}%
        & 2003 UN284      &  1&0   &  1&0   &  Assum    & D$_{12}$ \\ 
\noalign{\smallskip}%
        & 2005 EO304      &  1&0   &  1&0   &  Assum    & D$_{12}$ \\ 
\noalign{\smallskip}%
        & 2006 BR284      &  1&0   &  1&0   &  Assum    & D$_{12}$ \\ 
\noalign{\smallskip}%
        & 2006 JZ81       &  1&0   &  1&0   &  Assum    & D$_{12}$ \\ 
\noalign{\smallskip}%
        & 2006 CH69       &  1&0   &  1&0   &  Assum    & D$_{12}$ \\ 
\noalign{\smallskip}%
        & 2007 TY430      &  0&75  &  1&0   &  Assum    & D$_{13}$ \\ 
\noalign{\smallskip}%
        & 19P/Borrely     &  0&24  &  0&06  &  CNGF     & D$_{4}^\dagger$ \\ 
\noalign{\smallskip}%
        & 67P/C-G         &  0&23  &  0&14  &  CNGF     & D$_{5}^\dagger$ \\ 
\noalign{\smallskip}%
        & 81P/Wild2       &  0&70  &  0&10  &  CNGF     & D$_{6}$ \\ 
\noalign{\smallskip}%
        & SL9             &  0&50  &  0&05  &  BkUp     & D$_{1}$ \\ 
\hline
\end{tabular}
\end{table}

\clearpage
\section{Bilbiographic references\label{app: biblio}}
  References for mass estimates:\\
{\scriptsize
  \begin{tabular}{ll}
(M$_{1}$) ~\citet{1992-ACM-Williams} & (M$_{2}$) ~\citet{1992-AcA-42-Sitarski} \\ 
(M$_{3}$) ~\citet{1992-IAUS-152-Landgraf} & (M$_{4}$) ~\citet{1994-Nature-370-Solem} \\ 
(M$_{5}$) ~\citet{1995-AA-111-Viateau} & (M$_{6}$) ~\citet{1995-AcA-45-Sitarski} \\ 
(M$_{7}$) ~\citet{1996-IAUS-172-Carpino} & (M$_{8}$) ~\citet{1996-IAUS-172-Kuzmanoski} \\ 
(M$_{9}$) ~\citet{1997-AA-320-Viateau} & (M$_{10}$) ~\citet{1997-ESAP-402-Bange} \\ 
(M$_{11}$) ~\citet{1997-ESASP-402-Viateau} & (M$_{12}$) ~\citet{1997-IAU-LopezGarcia} \\ 
(M$_{13}$) ~\citet{1997-Icarus-130-Petit} & (M$_{14}$) ~\citet{1997-Science-278-Yeomans} \\ 
(M$_{15}$) ~\citet{1998-AA-334-Viateau} & (M$_{16}$) ~\citet{1998-AA-340-Bange} \\ 
(M$_{17}$) ~\citet{1999-AJ-117-Hilton} & (M$_{18}$) ~\citet{1999-IAA-Vasiliev} \\ 
(M$_{19}$) ~\citet{1999-Nature-401-Merline} & (M$_{20}$) ~\citet{2000-AA-354-Viateau} \\ 
(M$_{21}$) ~\citet{2000-AA-360-Michalak} & (M$_{22}$) ~\citet{2000-Science-289-Yeomans} \\ 
(M$_{23}$) ~\citet{2001-AA-365-Goffin} & (M$_{24}$) ~\citet{2001-AA-370-Viateau} \\ 
(M$_{25}$) ~\citet{2001-AA-371-Pitjeva} & (M$_{26}$) ~\citet{2001-AA-374-Michalak} \\ 
(M$_{27}$) ~\citet{2001-DPS-33-Margot} & (M$_{28}$) ~\citet{2001-IAA-Krasinsky} \\ 
(M$_{29}$) ~\citet{2001-JPL-Standish} & (M$_{30}$) ~\citet{2002-AA-395-Kuzmanoski} \\ 
(M$_{31}$) ~\citet{2002-ACM-Chernetenko} & (M$_{32}$) ~\citet{2002-AsteroidsIII-2.2-Merline} \\ 
(M$_{33}$) ~\citet{2002-Icarus-160-Konopliv} & (M$_{34}$) ~\citet{2002-Science-296-Margot} \\ 
(M$_{35}$) ~\citet{2003-DPS-35-Neish} & (M$_{36}$) ~\citet{2003-EMP-92-Osip} \\ 
(M$_{37}$) ~\citet{2003-Science-300-Margot} & (M$_{38}$) ~\citet{2004-AJ-128-Noll} \\ 
(M$_{39}$) ~\citet{2004-COSPAR-35-Pitjeva} & (M$_{40}$) ~\citet{2004-DPS-36-Margot} \\ 
(M$_{41}$) ~\citet{2004-Icarus-172-Noll} & (M$_{42}$) ~\citet{2004-SoSyR-38-Kochetova} \\ 
(M$_{43}$) ~\citet{2005-AA-430-Kovacevic} & (M$_{44}$) ~\citet{2005-ApJ-632-Brown} \\ 
(M$_{45}$) ~\citet{2005-DDA-36-Chesley} & (M$_{46}$) ~\citet{2005-DPS-37-Kern} \\ 
(M$_{47}$) ~\citet{2005-Icarus-178-Marchis} & (M$_{48}$) ~\citet{2005-Nature-436-Marchis} \\ 
(M$_{49}$) ~\citet{2005-SoSyR-39-Pitjeva} & (M$_{50}$) ~\citet{2006-AA-446-Behrend} \\ 
(M$_{51}$) ~\citet{2006-AA-455-Vitagliano} & (M$_{52}$) ~\citet{2006-AAS-209-Brooks} \\ 
(M$_{53}$) ~\citet{2006-AJ-132-Buie} & (M$_{54}$) ~\citet{2006-ApJ-643-Stansberry} \\ 
(M$_{55}$) ~\citet{2006-Icarus-181-Pravec} & (M$_{56}$) ~\citet{2006-Icarus-182-Konopliv} \\ 
(M$_{57}$) ~\citet{2006-Icarus-183-Kern} & (M$_{58}$) ~\citet{2006-Icarus-184-Shepard} \\ 
(M$_{59}$) ~\citet{2006-Nature-439-Marchis} & (M$_{60}$) ~\citet{2006-Science-312-Fujiwara} \\ 
(M$_{61}$) ~\citet{2006-Science-314-Ostro} & (M$_{62}$) ~\citet{2007-ASPC-370-Aslan} \\ 
(M$_{63}$) ~\citet{2007-EMP-100-Kovacevic} & (M$_{64}$) ~\citet{2007-Icarus-187-Davidsson} \\ 
(M$_{65}$) ~\citet{2007-Icarus-187-Descamps} & (M$_{66}$) ~\citet{2007-Icarus-189-Descamps} \\ 
(M$_{67}$) ~\citet{2007-Icarus-190-Richardson} & (M$_{68}$) ~\citet{2007-Icarus-191-Grundy} \\ 
(M$_{69}$) ~\citet{2007-Science-316-Brown} & (M$_{70}$) ~\citet{2008-AA-477-Fienga} \\ 
(M$_{71}$) ~\citet{2008-CeMDA-100-Baer} & (M$_{72}$) ~\citet{2008-DPS-40-Baer} \\ 
(M$_{73}$) ~\citet{2008-Icarus-195-Marchis} & (M$_{74}$) ~\citet{2008-Icarus-196-Descamps} \\ 
(M$_{75}$) ~\citet{2008-Icarus-196-Marchis} & (M$_{76}$) ~\citet{2008-Icarus-197-Grundy} \\ 
(M$_{77}$) ~\citet{2008-LPI-Taylor} & (M$_{78}$) ~\citet{2008-PSS-56-Ivantsov} \\ 
(M$_{79}$) ~\citet{2009-AA-501-Kryszczynska} & (M$_{80}$) ~\citet{2009-AA-507-Fienga} \\ 
(M$_{81}$) ~\citet{2009-AJ-137-Ragozzine} & (M$_{82}$) ~\citet{2009-Icarus-200-Grundy} \\ 
(M$_{83}$) ~\citet{2009-Icarus-200-Scheirich} & (M$_{84}$) ~\citet{2009-Icarus-203-Descamps} \\ 
(M$_{85}$) ~\citet{2009-MNRAS-393-Sosa} & (M$_{86}$) ~\citet{2009-SciNote-Folkner} \\ 
(M$_{87}$) ~\citet{2010-AJ-139-Brown} & (M$_{88}$) ~\citet{2010-AJ-139-Zielenbach} \\ 
(M$_{89}$) ~\citet{2010-AJ-140-Kuzmanoski} & (M$_{90}$) ~\citet{2010-ApJ-714-Fraser} \\ 
(M$_{91}$) ~\citet{2010-DPS-42-Benner} & (M$_{92}$) ~\citet{2010-PSS-58-Somenzi} \\ 
(M$_{93}$) ~\citet{2010-SciNote-Fienga} & (M$_{94}$) ~\citet{2011-AA-534-Carry} \\ 
(M$_{95}$) ~\citet{2011-AJ-141-Baer} & (M$_{96}$) ~\citet{2011-AJ-141-Fang} \\ 
(M$_{97}$) ~\citet{2011-AJ-142-Zielenbach} & (M$_{98}$) ~\citet{2011-ApJ-727-Rojo} \\ 
(M$_{99}$) ~\citet{2011-ApJ-743-Parker} & (M$_{100}$) ~\citet{2011-DPS-Fienga} \\ 
(M$_{101}$) ~\citet{2011-DPS-Marchis} & (M$_{102}$) ~\citet{2011-Icarus-211-Descamps} \\ 
(M$_{103}$) ~\citet{2011-Icarus-211-Konopliv} & (M$_{104}$) ~\citet{2011-Science-334-Paetzold} \\ 
(M$_{105}$) ~\citet{2011-arXiv-Sheppard} & (M$_{106}$) ~Descamps et al. (pers. com.) \\ 
(M$_{107}$) ~Merline et al. (pers. com.) 
  \end{tabular}
}

References for diameter estimates:\\
{\scriptsize
  \begin{tabular}{ll}
($\phi_{1}$) ~\citet{1994-Nature-370-Solem} & ($\phi_{2}$) ~\citet{1997-Icarus-128-Thomas} \\ 
($\phi_{3}$) ~\citet{1999-Icarus-142-Harris} & ($\phi_{4}$) ~\citet{2000-Science-288-Ostro} \\ 
($\phi_{5}$) ~\citet{2000-Science-289-Veverka} & ($\phi_{6}$) ~\citet{2002-AJ-124-Noll} \\ 
($\phi_{7}$) ~\citet{2002-AsteroidsIII-2.2-Merline} & ($\phi_{8}$) ~\citet{2002-Science-296-Margot} \\ 
($\phi_{9}$) ~\citet{2003-AJ-126-Fernandez} & ($\phi_{10}$) ~\citet{2003-DPS-35-Neish} \\ 
($\phi_{11}$) ~\citet{2004-AA-415-Altenhoff} & ($\phi_{12}$) ~\citet{2004-AA-418-Muller} \\ 
($\phi_{13}$) ~\citet{2004-AJ-127-Brown} & ($\phi_{14}$) ~\citet{2004-AJ-128-Noll} \\ 
($\phi_{15}$) ~\citet{2004-DPS-36-Nolan} & ($\phi_{16}$) ~\citet{2004-Icarus-172-Noll} \\ 
($\phi_{17}$) ~\citet{2004-Science-304-Brownlee} & ($\phi_{18}$) ~\citet{2005-AA-443-Muller} \\ 
($\phi_{19}$) ~\citet{2005-Icarus-178-Marchis} & ($\phi_{20}$) ~\citet{2005-MPS-40-Ostro} \\ 
($\phi_{21}$) ~\citet{2005-Nature-436-Marchis} & ($\phi_{22}$) ~\citet{2005-Nature-437-Thomas} \\ 
($\phi_{23}$) ~\citet{2005-Science-310-AHearn} & ($\phi_{24}$) ~\citet{2006-AA-446-Behrend} \\ 
($\phi_{25}$) ~\citet{2006-AA-458-Lamy} & ($\phi_{26}$) ~\citet{2006-AAS-209-Brooks} \\ 
($\phi_{27}$) ~\citet{2006-ApJ-639-Rabinowitz} & ($\phi_{28}$) ~\citet{2006-ApJ-643-Brown} \\ 
($\phi_{29}$) ~\citet{2006-ApJ-643-Stansberry} & ($\phi_{30}$) ~\citet{2006-DPS-38-Spencer} \\ 
($\phi_{31}$) ~\citet{2006-Icarus-181-Pravec} & ($\phi_{32}$) ~\citet{2006-Icarus-182-Tancredi} \\ 
($\phi_{33}$) ~\citet{2006-Icarus-184-Shepard} & ($\phi_{34}$) ~\citet{2006-Icarus-185-Marchis} \\ 
($\phi_{35}$) ~\citet{2006-Nature-439-Bertoldi} & ($\phi_{36}$) ~\citet{2006-Science-312-Fujiwara} \\ 
($\phi_{37}$) ~\citet{2006-Science-314-Ostro} & ($\phi_{38}$) ~\citet{2007-Icarus-187-Descamps} \\ 
($\phi_{39}$) ~\citet{2007-Icarus-191-Conrad} & ($\phi_{40}$) ~\citet{2007-Icarus-191-Grundy} \\ 
($\phi_{41}$) ~\citet{2008-AA-478-Carry} & ($\phi_{42}$) ~\citet{2008-AA-489-Lamy} \\ 
($\phi_{43}$) ~\citet{2008-Icarus-193-Wolters} & ($\phi_{44}$) ~\citet{2008-Icarus-195-Marchis} \\ 
($\phi_{45}$) ~\citet{2008-Icarus-195-Shepard} & ($\phi_{46}$) ~\citet{2008-Icarus-196-Descamps} \\ 
($\phi_{47}$) ~\citet{2008-Icarus-196-Marchis} & ($\phi_{48}$) ~\citet{2008-Icarus-197-Drummond} \\ 
($\phi_{49}$) ~\citet{2008-Icarus-197-Grundy} & ($\phi_{50}$) ~\citet{2008-LPI-Taylor} \\ 
($\phi_{51}$) ~\citet{2008-SSBN-3-Stansberry} & ($\phi_{52}$) ~\citet{2009-ApJ-694-Delbo} \\ 
($\phi_{53}$) ~\citet{2009-Icarus-201-Brucker} & ($\phi_{54}$) ~\citet{2009-Icarus-202-Drummond} \\ 
($\phi_{55}$) ~\citet{2009-Icarus-203-Descamps} & ($\phi_{56}$) ~\citet{2009-MNRAS-393-Sosa} \\ 
($\phi_{57}$) ~\citet{2009-PSS-57-Delbo} & ($\phi_{58}$) ~\citet{2009-Science-326-Schmidt} \\ 
($\phi_{59}$) ~\citet{2010-AA-518-Lellouch} & ($\phi_{60}$) ~\citet{2010-AA-518-Lim} \\ 
($\phi_{61}$) ~\citet{2010-AA-523-Carry} & ($\phi_{62}$) ~\citet{2010-AA-523-Drummond} \\ 
($\phi_{63}$) ~\citet{2010-AJ-139-Brown} & ($\phi_{64}$) ~\citet{2010-AJ-140-Ryan} \\ 
($\phi_{65}$) ~\citet{2010-ApJ-714-Fraser} & ($\phi_{66}$) ~\citet{2010-DPS-42-Benner} \\ 
($\phi_{67}$) ~\citet{2010-Icarus-205-Carry-a} & ($\phi_{68}$) ~\citet{2010-Icarus-205-Mueller} \\ 
($\phi_{69}$) ~\citet{2010-Icarus-207-Ostro} & ($\phi_{70}$) ~\citet{2011-AJ-141-Fang} \\ 
($\phi_{71}$) ~\citet{2011-AJ-141-Mueller} & ($\phi_{72}$) ~\citet{2011-ApJ-741-Masiero} \\ 
($\phi_{73}$) ~\citet{2011-ApJ-742-Grav} & ($\phi_{74}$) ~\citet{2011-ApJ-743-Parker} \\ 
($\phi_{75}$) ~\citet{2011-CeMDA-109-Archinal} & ($\phi_{76}$) ~\citet{2011-DPS-Drummond} \\ 
($\phi_{77}$) ~\citet{2011-Icarus-211-Descamps} & ($\phi_{78}$) ~\citet{2011-Icarus-214-Durech} \\ 
($\phi_{79}$) ~\citet{2011-Icarus-215-Matter} & ($\phi_{80}$) ~\citet{2011-Icarus-216-Brozovic} \\ 
($\phi_{81}$) ~\citet{2011-MNRAS-418-Wolters} & ($\phi_{82}$) ~\citet{2011-Nature-478-Sicardy} \\ 
($\phi_{83}$) ~\citet{2011-PASJ-63-Usui} & ($\phi_{84}$) ~\citet{2011-Science-334-Sierks} \\ 
($\phi_{85}$) ~\citet{2011-arXiv-Sheppard} & ($\phi_{86}$) ~\citet{2012-AA--Mommert} \\ 
($\phi_{87}$) ~\citet{2012-AA--SantoSanz} & ($\phi_{88}$) ~\citet{2012-ApJ-186-Walsh} \\ 
($\phi_{89}$) ~Benner (pers. com.) & ($\phi_{90}$) ~MPEC \\ 
($\phi_{91}$) ~Merline et al. (pers. com.) & ($\phi_{92}$) ~\citet{PDSSBN-DELBO} \\ 
($\phi_{93}$) ~\citet{PDSSBN-IRAS} & ($\phi_{94}$) ~\citet{PDSSBN-MSX} \\ 
($\phi_{95}$) ~\citet{PDSSBN-OCC} 
  \end{tabular}
}

  References for indirect density estimates:\\
{\scriptsize
  \begin{tabular}{ll}
(D$_{1}$) ~\citet{1994-Nature-370-Solem} & (D$_{2}$) ~\citet{2000-Icarus-146-Mottola} \\ 
(D$_{3}$) ~\citet{2003-DPS-35-Neish} & (D$_{4}$) ~\citet{2004-Icarus-168-Davidsson} \\ 
(D$_{5}$) ~\citet{2005-Icarus-176-Davidsson} & (D$_{6}$) ~\citet{2006-Icarus-180-Davidsson} \\ 
(D$_{7}$) ~\citet{2006-Icarus-181-Pravec} & (D$_{8}$) ~\citet{2006-AA-446-Behrend} \\ 
(D$_{9}$) ~\citet{2007-Icarus-189-Descamps} & (D$_{10}$) ~\citet{2009-AA-501-Kryszczynska} \\ 
(D$_{11}$) ~\citet{2009-Icarus-200-Scheirich} & (D$_{12}$) ~\citet{2011-ApJ-743-Parker} \\
(D$_{13}$)~\citet{2011-arXiv-Sheppard} \\
  \end{tabular}
}


\end{document}